\documentclass[12pt,twoside,a4paper,openany]{book}
\usepackage{graphicx,amsfonts,amsmath,amsthm,appendix,color,amssymb,fancyhdr,lscape,simplewick}

\usepackage[francais]{babel}
\usepackage[latin1]{inputenc}
\usepackage[T1]{fontenc}
\DeclareGraphicsExtensions{.pdf}

\usepackage{caption}
\usepackage{titlesec}
\titleformat{\chapter}[display]{\sffamily\huge}{\hspace{0.62cm}\chaptertitlename\ \thechapter}{20pt}{\sffamily\bfseries\fontsize{32}{38}\selectfont\flushright}
\titleformat{\section}{\sffamily\LARGE\bfseries}{\thesection}{1em}{}
\titleformat{\subsection}{\sffamily\Large\bfseries}{\thesubsection}{1em}{}
\titleformat{\subsubsection}{\sffamily\large\bfseries}{\thesubsubsection}{1em}{}

\renewcommand{\thechapter}{\Roman{chapter}}

\makeatletter
\@addtoreset{equation}{chapter}
\makeatother
\newtheorem{hypo}{Hypoth{\`{e}}se}[chapter]
\newtheorem{theo}{Theor{\`{e}}me}[chapter]
\newtheorem{prop}{Proposition}[chapter]
\newtheorem{lemme}{Lemme}[chapter]
\newtheorem{cor}{Corolaire}[chapter]
\newtheorem{prob}{Probl{\`{e}}me}[chapter]

\newtheorem{conj}{Conjecture}[chapter]

\newcommand{\beq}{\begin{equation}}
\newcommand{\eeq}{\end{equation}}
\newcommand{\bea}{\begin{eqnarray}}
\newcommand{\eea}{\end{eqnarray}}

\newcommand{\z}{{\mathfrak{z}}}
\newcommand{\p}{{\parallel}}
\newcommand{\supp}{{\,\rm supp}}
\newcommand{\Tr}{{\,\rm Tr}\:}
\newcommand{\tr}{{\,\rm tr}\:}
\newcommand{\Res}{\mathop{\,\rm Res\,}}

\newcommand{\ch}{{\mathrm{ch}}}
\newcommand{\sh}{{\mathrm{sh}}}
\newcommand{\On}{{\mathcal{O}(\mathfrak{n})}}
\newcommand{\nn}{\nonumber}
\newcommand{\n}{\mathfrak{n}}
\newcommand{\g}{\mathfrak{g}}
\renewcommand{\Re}{{\mathrm{Re}}\:}
\renewcommand{\Im}{{\mathrm{Im}}\:}

\newcommand{\dd}{\mathrm{d}}
\newcommand{\h}{\mathfrak{h}}

\fancypagestyle{plain}{
\fancyhead{}
\fancyfoot[C]{\thepage}

}

\newcommand{\captionfonts}{\small}
\makeatletter\long\def\@makecaption#1#2{%
  \vskip\abovecaptionskip
  \sbox\@tempboxa{{\captionfonts #1: #2}}%
  \ifdim \wd\@tempboxa >\hsize
    {\captionfonts #1: #2\par}
  \else
    \hbox to\hsize{\hfill\box\@tempboxa\hfill}%
  \fi
  \vskip\belowcaptionskip}
\makeatother

\definecolor{misval}{rgb}{0.851,0.03,0.578}
\definecolor{rouge}{rgb}{0.84,0.18,0.07}
\definecolor{bleu}{rgb}{0.10,0.20,0.35}
\definecolor{vertf}{rgb}{0.08,0.46,0.07}
\usepackage[pdftex]{hyperref}
\hypersetup{colorlinks,urlcolor=vertf,citecolor=black,linkcolor=bleu,filecolor=black}

\begin{document}
\sloppy

\thispagestyle{empty}
\phantom{bubnu}

\newpage

\thispagestyle{empty}

\vspace*{\stretch{1}}
\begin{figure}[h!]
\begin{center}
\includegraphics[width=0.8\textwidth]{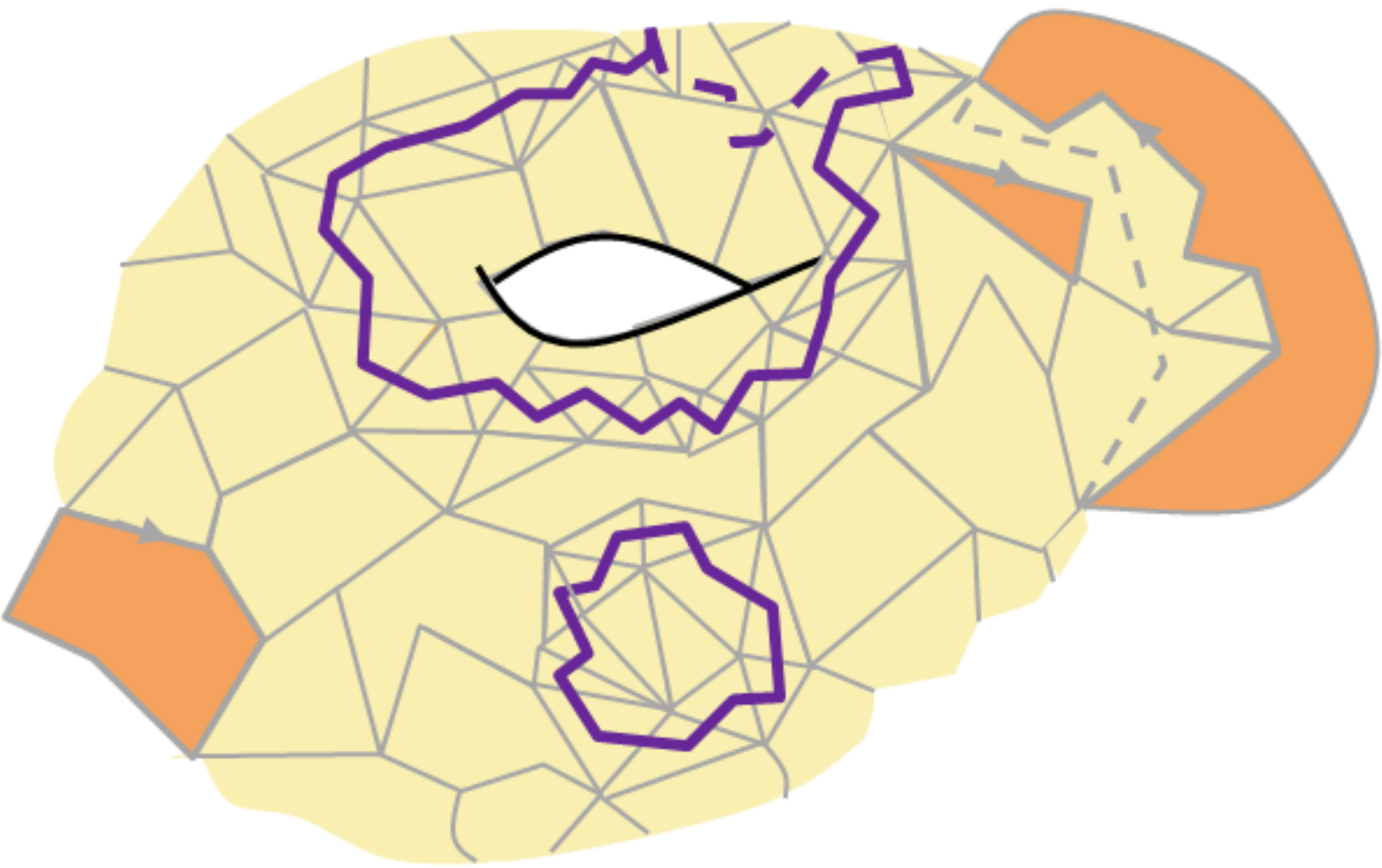}
\end{center}
\end{figure}

\vspace*{\stretch{1}}

{\small \noindent Cette th\`ese a \'et\'e financ\'ee par l'Ecole Normale Sup\'erieure (2008-2009), puis une bourse AMN du Minist\`ere de l'Enseignement Sup\'erieur et de la Recherche (2009-2011). Lors de mes travaux, j'ai b\'en\'efici\'e du soutien financier de l'ANR GranMa (ANR-08-BLAN-0311-01), du CEA Saclay via l'IPhT, et de la fondation CFM pour la recherche.}

\newpage

\thispagestyle{empty}

\subsection*{R\'esum\'e}

{\small La g\'eom\'etrie complexe est un outil puissant pour \'etudier les syst\`emes int\'egrables classiques, la physique statistique sur r\'eseau al\'eatoire, les probl\`emes de matrices al\'eatoires, la th\'eorie topologique des cordes, \ldots{} Tous ces probl\`emes ont en commun la pr\'esence de relations, appel\'ees \'equations de boucle ou contraintes de Virasoro. Dans le cas le plus simple, leur solution compl\`ete a \'et\'e trouv\'ee r\'ecemment \cite{EOFg}, et se formule naturellement en termes de g\'eom\'etrie diff\'erentielle sur une surface de Riemann : la "courbe spectrale", qui d\'epend du probl\`eme. Cette th\`ese est une contribution au d\'eveloppement de ces techniques et de leurs applications.

Pour commencer, nous abordons les questions de d\'eveloppement asymptotique \`a tous les ordres lorsque $N \rightarrow \infty$, des int\'egrales $N$dimensionnelles venant de la th\'eorie des matrices al\'eatoires de taille $N \times N$, ou plus g\'en\'eralement des gaz de Coulomb. Nous expliquons comment \'etablir, dans les mod\`eles de matrice $\beta$ et dans un r\'egime \`a une coupure, le d\'eveloppement asymptotique \`a tous les ordres en puissances de $N$. Nous appliquons ces r\'esultats \`a l'\'etude des grandes d\'eviations du maximum des valeurs propres dans les mod\`eles $\beta$, et en d\'eduisons de fa\c{c}on heuristique des informations sur l'asymptotique \`a tous les ordres de la loi de Tracy-Widom $\beta$, pour tout $\beta > 0$. Ensuite, nous examinons le lien entre int\'egrabilit\'e et \'equations de boucle. En corolaire, nous pouvons d\'emontrer l'heuristique pr\'ec\'edente concernant l'asymptotique de la loi de Tracy-Widom pour les matrices hermitiennes.

Nous terminons avec la r\'esolution de probl\`emes combinatoires en toute topologie. En th\'eorie topologique des cordes, une conjecture de Bouchard, Klemm, Mari\~{n}o et Pasquetti \cite{BKMP} affirme que des s\'eries g\'en\'eratrices bien choisies d'invariants de Gromov-Witten dans les espaces de Calabi-Yau toriques, sont solution d'\'equations de boucle. Nous l'avons d\'emontr\'e dans le cas le plus simple, o\`u ces invariants co\"{i}ncident avec les nombres de Hurwitz simples. Nous expliquons les progr\`es r\'ecents vers la conjecture g\'en\'erale, en relation avec nos travaux. En physique statistique sur r\'eseau al\'eatoire, nous avons r\'esolu le mod\`ele $\On$ trivalent sur r\'eseau al\'eatoire introduit par Kostov, et expliquons la d\'emarche \`a suivre pour r\'esoudre des mod\`eles plus g\'en\'eraux.

Tous ces travaux soulignent l'importance de certaines  "int\'egrales de matrices g\'en\'eralis\'ees" pour les applications futures. Nous indiquons quelques \'el\'ements appelant \`a une th\'eorie g\'en\'erale, encore bas\'ee sur des "\'equations de boucles", pour les calculer.}

\newpage

\thispagestyle{empty}

\subsection*{Abstract}

{\small Complex analysis is a powerful tool to study classical integrable systems, statistical physics on the random lattice, random matrix theory, topological string theory, \ldots{} All these topics share certain relations, called "loop equations" or "Virasoro constraints". In the simplest case, the complete solution of those equations was found recently \cite{EOFg} : it can be expressed in the framework of differential geometry over a certain Riemann surface which depends on the problem : the "spectral curve". This thesis is a contribution to the development of these techniques, and to their applications.

First, we consider all order large $N$ asymptotics in some $N$-dimensional integrals coming from random matrix theory, or more generally from "log gases" problems. We shall explain how to use loop equations to establish those asymptotics in $\beta$ matrix models within a one cut regime. This can be applied in the study of large fluctuations of the maximum eigenvalue in $\beta$ matrix models, and lead us to heuristic predictions about the asymptotics of Tracy-Widom $\beta$ law to all order, and for all $\beta > 0$. Second, we study the interplay between integrability and loop equations. As a corollary, we are able to prove the previous prediction about the asymptotics to all order of Tracy-Widom law for hermitian matrices.

We move on with the solution of some combinatorial problems in all topologies. In topological string theory, a conjecture from Bouchard, Klemm, Mari\~{n}o and Pasquetti \cite{BKMP} states that certain generating series of Gromov-Witten invariants in toric Calabi-Yau threefolds, are solutions of loop equations. We have proved this conjecture in the simplest case, where those invariants coincide with the "simple Hurwitz numbers". We also explain recent progress towards the general conjecture, in relation with our work. In statistical physics on the random lattice, we have solved the trivalent $\On$ model introduced by Kostov, and we explain the method to solve more general statistical models.

Throughout the thesis, the computation of some "generalized matrices integrals" appears to be increasingly important for future applications, and this appeals for a general theory of loop equations.}

\newpage

\thispagestyle{empty}

{\LARGE \textsf{Merci \ldots{}}}

\vspace{0.5cm}

{\small Je voudrais remercier les nombreuses personnes qui m'ont pouss\'e de quelque mani\`ere sur le plan scientifique. Pour commencer, l'ensemble des professeurs qui m'ont form\'e \`a l'\'{E}cole Normale Sup\'erieure. Ensuite, Kay Wiese et Pierre le Doussal m'ont permis, en prolongement d'un stage exp\'erimental de DEA, de participer \`a l'\'ecole d'\'et\'e de Beg Rohu, o\`u j'ai appris gr\^{a}ce aux cours de Patrik Ferrari et Satya Majumdar des rudiments de matrices al\'eatoires. Introduction tr\`es utile lorsque j'ai commenc\'e ma th\`ese, deux mois plus tard.

Bertrand Eynard a \'et\'e un directeur de th\`ese remarquable, par son enthousiasme \`a transmettre ses connaissances, son application \`a pr\'esenter ses id\'ees de mani\`ere accessible, sa patience devant mes tentatives al\'eatoires. Gr\^{a}ce \`a sa culture math\'ematique et ses suggestions, j'ai compris quelques ficelles qui tiennent ensemble des probl\`emes tr\`es diff\'erents. Il a aussi \'et\'e d'un grand soutien pendant la saison moins excitante de la chasse aux post-docs. Enfin, il a r\'euni toutes les conditions pour que j'assiste \`a des conf\'erences, et que je commence \`a discuter avec d'autres scientifiques de nos sujets de recherches.

Je suis tr\`es reconnaissant \`a Boris Dubrovin et Tamara Grava de m'avoir accueilli \`a la SISSA pendant un mois. Ils se sont rendus tr\`es disponibles et j'ai beaucoup appris avec eux sur les syst\`emes int\'egrables. Si ce processus est loin d'\^{e}tre termin\'e, quelques discussions de base ont \'et\'e tr\`es utiles lorsque j'ai commenc\'e \`a travailler sur l'\'equation de Painlev\'e II. J'esp\`ere revenir un jour avec plus d'\'el\'ements pour r\'epondre aux questions que nous avons laiss\'e en suspens.

Je remercie Gabriel Lopes Cardoso \`a Lisbonne, Albrecht Klemm (dont j'ai suivi le cours de th\'eorie topologique des cordes \`a l'IST) et Hartmut Monien \`a Bonn, Jan de Gier \`a Melbourne, pour leurs invitations et les discussions stimulantes qui y sont li\'ees, ainsi que l'ensemble de mes collaborateurs. Enfin, je remercie Philippe Biane et Marcos Mari{\~{n}}o d'avoir accept\'e de rapporter ce manuscrit. Je leur suis redevable, ainsi qu'\`a Pavel Bleher, d'y avoir apport\'e quelques corrections.

\`{A} l'IPhT, j'ai fr\'equemment d\'erang\'e pour de petites questions Michel Bauer, Fran\c{c}ois David, Philippe di Francesco, J\'er\'emie Bouttier, Emmanuel Guitter, Ivan Kostov, St\'ephane Nonnenmacher, que je remercie pour leur accueil et leur r\'eponses. Gr\^{a}ce aux remarques pointilleuses de Michel Berg\`ere, j'ai souvent corrig\'e ma fa\c{c}on de pr\'esenter un probl\`eme pour atteindre plus de rigueur. Les \'echanges avec les autres (post)doctorants, notamment Olivier Marchal, Nicolas Orantin, Jean-\'{E}mile Bourgine, J\'er\^{o}me Dubail, Nicolas Curien, ont \'et\'e enrichissants, ainsi que les discussions avec Gr\'egory Schehr, Satya Majumdar et C\'eline Nadal au LPTMS. Je souhaite aux futurs \'etudiants \`a l'IPhT de b\'en\'eficier encore des excellentes conditions de recherche qui y sont offertes aujourd'hui, et des interactions enrichissantes avec le plus grand nombre de chercheurs.

\vspace{0.2cm}

Les personnages de l'ombre ne sont pas oubli\'e(e)s.}



\subsection*{Avant-propos}
\thispagestyle{empty}
\vspace{0.5cm}

{\small
Je voudrais pour excuser le volume de ce manuscrit, souligner son objectif p\'edagogique (qui sera ou non atteint suivant l'appr\'eciation du lecteur). Plusieurs th\`emes de la physique th\'eorique sont abord\'es : g\'eom\'etrie complexe et alg\'ebrique, th\'eorie des cordes, int\'egrabilit\'e, physique statistique, combinatoire, \ldots{} Chacun requiert un formalisme dont j'ai choisi de pr\'esenter quelques \'el\'ements, pour permettre une lecture sans consultation trop fr\'equente de la bibliographie. Apr\`es une introduction au domaine buissonnant des matrices al\'eatoires (chapitre \textsf{I}), le chapitre \textsf{II} pr\'esente les fils conducteurs de cette th\`ese : le formalisme de la r\'ecurrence topologique et les \'equations de boucles. Les autres chapitres peuvent se lire ind\'ependamment les uns des autres. Les chapitres \textsf{III} et \textsf{IV} sont plut\^{o}t reli\'es \`a des questions d'int\'egrales convergentes de matrices (domaine des probabilit\'es), tandis que les chapitres \textsf{V} et \textsf{VI} utilisent plut\^{o}t des int\'egrales formelles de matrices (domaine de la combinatoire). Un petit formulaire de mod\`eles de matrices (\S~\ref{sec:chainmat} et \ref{sec:correlateurs}) rassemble noms et notations.

Je pr\'esente ceux de mes travaux qui ont aboutis (les r\'esultats sont synth\'etis\'es au \S~\ref{sec:teo}), en essayant de les situer dans un contexte scientifique plus vaste. La plupart ont fait l'objet d'une publication sur l'arXiv. La construction d'un syst\`eme int\'egrable dispersif \`a partir d'une courbe spectrale \'evoqu\'ee au \S~\ref{sec:mono} fait exception : elle est bas\'ee sur un article en pr\'eparation avec Bertrand Eynard, mais nous butons encore sur une preuve technique. J'ai choisi d'en parler pour mieux situer le lien entre nos m\'ethodes et les syst\`emes int\'egrables.

Dans le cours du texte se trouvent quelques r\'esultats secondaires, qui ne sont pas encore apparus \`a ma connaissance dans la litt\'erature :
\begin{itemize}
\item[$\diamond$] La d\'erivation heuristique du d\'eveloppement asymptotique dans un r\'egime \`a plusieurs coupures pour le mod\`ele $\beta$. C'est une g\'en\'eralisation \'evidente du cas hermitien ($\beta = 1$) qui a \'et\'e trait\'e par \cite{Ecv}.
\item[$\diamond$] Dans les mod\`eles de boucles, le calcul des observables avec un nombre fini de changements de conditions de bords (Annexe~\ref{app:bornu}). C'est un r\'esultat relativement simple, qui n'est pas encore assez etoff\'e pour \^{e}tre publi\'e.
\item[$\diamond$] Les \'equations de boucles pour les syst\`emes int\'egrables $d \times d$ et leurs cons\'equences. Seul le cas $2 \times 2$ a \'et\'e trait\'e dans les articles \cite{BEdet} et \cite{BETW}. Cette g\'en\'eralisation et ses applications font l'objet d'un article en pr\'eparation avec Bertrand Eynard.
\end{itemize}
D\`es que possible, j'ai cherch\'e \`a simplifier les raisonnements de base qui se trouvent dans les articles ; j'esp\`ere que le manuscrit leur sera compl\'ementaire. C'est le cas notamment au Chapitre \textsf{V}, o\`u je donne une preuve plus \'el\'egante du lemme "une coupure" dans une large classe de mod\`eles de boucles, inspir\'ee par une discussion avec J\'er\'emie Bouttier et Emmanuel Guitter.

\vspace{0.5cm}

\begin{tabular}{rl}
\textsf{Chapitre I } $-$ & \textsf{\textbf{Introduction}} \\
page$\,\,$ \pageref{sec:debut} $\,$ & \textsf{Les d\'ebuts} \\
\pageref{sec:Mehtapoly} $\,$ & \textsf{Statistique des valeurs propres et polyn\^{o}mes orthogonaux} \\
\pageref{sec:unn} $\,$ & \textsf{Universalit\'e locale} \\
\pageref{sec:ntu} $\,$ & \textsf{Lois de matrices al\'eatoires dans la nature} \\
\pageref{sec:BIPZ} $\,$ & \textsf{Surfaces discr\`etes} \\
\pageref{sec:diss} $\,$ & \textsf{Surfaces discr\`etes \`a la limite continue \ldots} \\
\pageref{sec:intnin} $\,$ & \textsf{\ldots et int\'egrabilit\'e} \\
\pageref{sec:introuniv} $\,$ & \textsf{Physique statistique} \\
\pageref{sec:introgeom} $\,$ & \textsf{G\'eom\'etrie de l'espace des surfaces de Riemann \ldots} \\
\pageref{sec:cordes1} $\,$ & \textsf{\ldots et th\'eorie des cordes} \\
\pageref{sec:boubou} $\,$ & \textsf{\'{E}quations de boucles et g\'eom\'etrie} \\
\pageref{sec:teo} $\,$ & \textsf{R\'esultats obtenus} \\
\pageref{sec:chainmat} $\,$ & \textsf{Description des mod\`eles de matrices \'etudi\'es} \\
\pageref{sec:correlateurs} $\,$ & \textsf{Description des observables} \\
\end{tabular}

\vspace{0.3cm}

\begin{tabular}{rl}
\textsf{Chapitre II} $-$ & \textsf{\textbf{La r\'ecurrence topologique}} \\
\pageref{sec:eqboucl1} $\,$ & \textsf{Les \'equations de Schwinger-Dyson} \\
\pageref{sec:forma} $\,$ & \textsf{Formalisme de la r\'ecurrence topologique} \\
\pageref{sec:ipipi} $\,$ & \textsf{Applications} \\
\pageref{sec:genus} $\,$ & \textsf{G\'en\'eralisations}
\end{tabular}

\vspace{0.3cm}

\begin{tabular}{rl}
\textsf{Chapitre III} $-$ & \textsf{\textbf{Int\'egrales convergentes de matrices}} \\
\pageref{sec:1maj} $\,$ & \textsf{Asymptotique \`a $N \rightarrow \infty$} \\
\pageref{sec:puisn} $\,$ & \textsf{Asymptotique en puissances de $1/N$} \\
\pageref{sec:pluscoup} $\,$ & \textsf{Asymptotique dans un r\'egime \`a plusieurs coupures} \\
\pageref{sec:stata} $\,$ & \textsf{Statistique de $\lambda_{\mathrm{max}}$}
\end{tabular}

\vspace{0.3cm}

\thispagestyle{empty}

\begin{tabular}{rl}
\textsf{Chapitre IV} $-$ & \textsf{\textbf{Syst\`emes int\'egrables et \'equations de boucles}} \\
\pageref{sec:keskao} $\,$ & \textsf{Qu'est-ce qu'un syst\`eme int\'egrable ?} \\
\pageref{sec:isomono} $\,$ & \textsf{D\'eformations isomonodromiques} \\
\pageref{sec:enjo} $\,$ & \textsf{\'{E}quations de boucles et cons\'equences} \\
\pageref{sec:consoK} $\,$ & \textsf{Construction d'un syst\`eme int\'egrable \`a partir d'une courbe spectrale} \\
\pageref{sec:cnuc} $\,$ & \textsf{Conclusion}
\end{tabular}

\vspace{0.3cm}

\begin{tabular}{rl}
\textsf{Chapitre V} $-$ & \textsf{\textbf{Int\'egrales formelles de matrices et combinatoire}} \\
\pageref{sec:foerm} $\,$ & \textsf{Int\'egrales formelles de matrices} \\
\pageref{sec:bouton} $\,$ & \textsf{Les mod\`eles de boucles} \\
\pageref{sec:OnON} $\,$ & \textsf{Le mod\`ele $\On$ trivalent}
\end{tabular}

\vspace{0.3cm}

\begin{tabular}{rl}
\textsf{Chapitre VI} $-$ & \textsf{\textbf{Cordes topologiques et conjecture BKMP}} \\
\pageref{sec:nono} $\,$ & \textsf{Notions de th\'eorie des cordes} \\
\pageref{sec:calcGW} $\,$ & \textsf{R\'ecurrence topologique et th\'eorie topologique des cordes} \\
\pageref{sec:zeroa} $\,$ & \textsf{R\'esultats r\'ecents} \\
\pageref{sec:drop} $\,$ & \textsf{Perspectives}
\end{tabular}

\vspace{0.3cm}
\begin{tabular}{rl}
\textsf{Chapitre VII} $-$ & \textsf{\textbf{Conclusion}} \\
\pageref{eq:conclu} $\,$ &
\end{tabular}

\vspace{0.3cm}

\begin{tabular}{rl}
\textsf{\textbf{Annexes}} & \\
\pageref{app:geomcx} $\,$ & \textsf{G\'eom\'etrie complexe sur les surfaces de Riemann compactes} \\
\pageref{app:bornu} $\,$ & \textsf{Observables \`a bords non uniformes dans les mod\`eles de boucles} \\
\pageref{inde} $\,$ & \textsf{Index} \\
\pageref{app:arti} $\,$ & \textsf{Liste d'articles et bibliographie}
\end{tabular}
}

\thispagestyle{empty}

\newpage

\thispagestyle{empty}
\phantom{bbk}
\newpage

\chapter{Introduction}
\label{chap:inb}
\thispagestyle{plain}
\vspace{-1.5cm}

\noindent\rule{\textwidth}{1.5mm}

\vspace{2.5cm}

\textsf{Au centre de cette th\`ese se trouvent des structures math\'ematiques dont l'\'etude a \'et\'e motiv\'ee, ou encourag\'ee, par la th\'eorie des matrices al\'eatoires. Depuis les ann\'ees 1950, plusieurs probl\`emes ont \'et\'e reli\'es de mani\`ere inattendue, tant en physique qu'en math\'ematiques, \`a des questions de matrices al\'eatoires. Cette introduction r\'esume la progression historique des id\'ees, avec des raccourcis et des omissions qui refl\`etent mon point de vue. En particulier, je vais mettre de c\^{o}t\'e :
\begin{itemize}
\item[$\diamond$] l'analyse statistique (int\'eressant pour certaines applications en biologie, sciences humaines, finance, \ldots). Les premi\`eres matrices al\'eatoires ont \'et\'e introduites \`a cet effet par Wishart \cite{Wishart} en 1928.
\item[$\diamond$] l'emploi de mod\`eles de matrices pour certains r\'egimes de la chromodynamique quantique (cf. la revue d'Akemann \cite{Akerev}).
\end{itemize}
La partie "physique" de cette introduction a \'et\'e inspir\'ee par la lecture de Mehta \cite{Mehtabook} et de la revue de Guhr et al. \cite{Guhrrev}.}

\section{Les d\'ebuts}
\label{sec:debut}
L'histoire commence en physique nucl\'eaire. Pour comprendre le d\'eroulement de certaines r\'eactions nucl\'eaires, il faut pouvoir d\'ecrire comment un neutron projectile interagit avec un atome cible. Lorsque l'\'energie cin\'etique du projectile est assez grande ($E_{\mathrm{neutron}} \succsim 10^{2}-10^{3}$ eV) et le noyau complexe, le neutron va typiquement interagir avec un grand nombre de nucl\'eons \`a l'int\'erieur du noyau. Pour certains $E_{\mathrm{neutron}}$ (les \'energies de r\'esonance), un \'etat quasi li\'e avec le noyau pourra se former avec une grande dur\'ee de vie (par rapport \`a la dur\'ee de collision). Ce compos\'e se d\'esint\'egrera ensuite, avec plusieurs possibilit\'es : fission du noyau, \'ejection d'un nucl\'eon (radioactivit\'e), \ldots{} Les \'energies de r\'esonance $E_n$ et la dur\'ee de vie des \'etats quasi li\'es $\Gamma_n^{-1}$ sont des caract\'eristiques importantes. La complexit\'e des interactions \`a l'int\'erieur du noyau rend la pr\'ediction de ces param\`etres difficile. Il y a de toute fa\c{c}on un grand nombre de r\'esonances (leur espacement moyen est de quelques eV au d\'ebut du spectre, et d\'ecro\^{i}t exponentiellement avec l'\'energie). Mais, dans une r\'eaction nucl\'eaire, la distribution en \'energie des neutrons libres peut \^{e}tre large, donc ce sont les propri\'et\'es statistiques des $E_n$ et $\Gamma_n^{-1}$ qui importent. Cette image \'etait bien \'etablie dans les ann\'ees 1930.

Justement, devant la complexit\'e des interactions, une tentative est de les consid\'erer "typiques" dans un vaste ensemble de mod\`eles d'interactions, donc d'utiliser les outils de la physique statistique.
Wigner franchit ce pas en 1951 \cite{Wign}, et propose comme mod\`ele des propri\'et\'es statistiques des $E_n$, \label{Wignsu}celles des valeurs propres d'une grande matrice al\'eatoire avec entr\'ees ind\'ependantes. Le type de matrice (sym\'etrique, hermitienne, ...) \`a consid\'erer, en l'absence d'autres informations, d\'epend seulement des sym\'etries discr\`etes respect\'ees par les interactions (invariance par renversement du temps, sym\'etrie particule-trou, parit\'e, \ldots). Les matrices $M = (M_{i,j})_{1 \leq i,j \leq N}$, o\`u les entr\'ees $M_{i,j}$ sont des variables al\'eatoires ind\'ependantes, sont appel\'ees \textbf{matrices de Wigner}. Les valeurs propres $\lambda_1 > \ldots > \lambda_N$  d'une matrice de Wigner sont des variables al\'eatoires corr\'el\'ees, qui ont tendance \`a se repousser.  Plus pr\'ecis\'ement, Wigner conjecture que la distribution des espacements $E_{n + 1} - E_{n}$ est bien reproduite par la distribution des espacements $s = \lambda_{i + 1} - \lambda_i$, modulo un changement d'\'echelle tenant compte de la densit\'e locale des $E_n$. Il calcule cette distribution pour une matrice al\'eatoire sym\'etrique de taille $2\times 2$, dont les entr\'ees ont une distribution gaussienne :
\beq
\label{eq:surmise}p_{2}(s)\mathrm{d}s = \frac{\pi s}{2}\,e^{-\frac{\pi s^2}{4}}
\eeq
et cette distribution est proche num\'eriquement de celle obtenue dans une matrice de taille infinie $p_{\infty}(s) = \lim_{N \infty} p_N(s)$.

Vers 1960, cette estim\'ee de Wigner a \'et\'e compar\'ee avec succ\`es aux donn\'ees de la physique nucl\'eaire \cite{RP60} (cf. Fig~\ref{Spectrum}). Naturellement, une matrice $M$ avec des entr\'ees gaussiennes ind\'ependantes ne mod\'elise pas s\'erieusement le hamiltonien d'interaction $H$ d'un syst\`eme quantique \`a $p$ corps ($p \gg 1$) gouvernant la dynamique \`a l'int\'erieur du noyau. De plus, la pr\'esence de r\`egles de s\'election sugg\`ere que la matrice repr\'esentant $H$ a, dans une base adapt\'ee, beaucoup de z\'eros. En ce sens, il n'est pas facile d'envisager un mod\`ele de matrice al\'eatoire r\'ealiste. Cependant, les simulations num\'eriques ont rapidement montr\'e que $p_{\infty}(s)$ d\'ependait peu des d\'etails de la distribution des $M_{i,j}$. Les physiciens esp\'eraient donc que la statistique des valeurs propres de grandes matrices al\'eatoires pr\'esenterait des caract\'eristiques assez universelles, qui se manifesteraient dans les spectres des syst\`emes complexes.

\section{Statistique des valeurs propres et polyn\^{o}mes orthogonaux}
\label{sec:Mehtapoly}
Dans la perspective d'une comparaison plus compl\`ete aux donn\'ees spectroscopiques, un premier enjeu a \'et\'e de calculer analytiquement la statistique des valeurs propres, dans divers mod\`eles de matrices (pas n\'ecessairement de Wigner).
\begin{figure}
\begin{center}
\hspace{-0.5cm}\includegraphics[width = 1.05\textwidth]{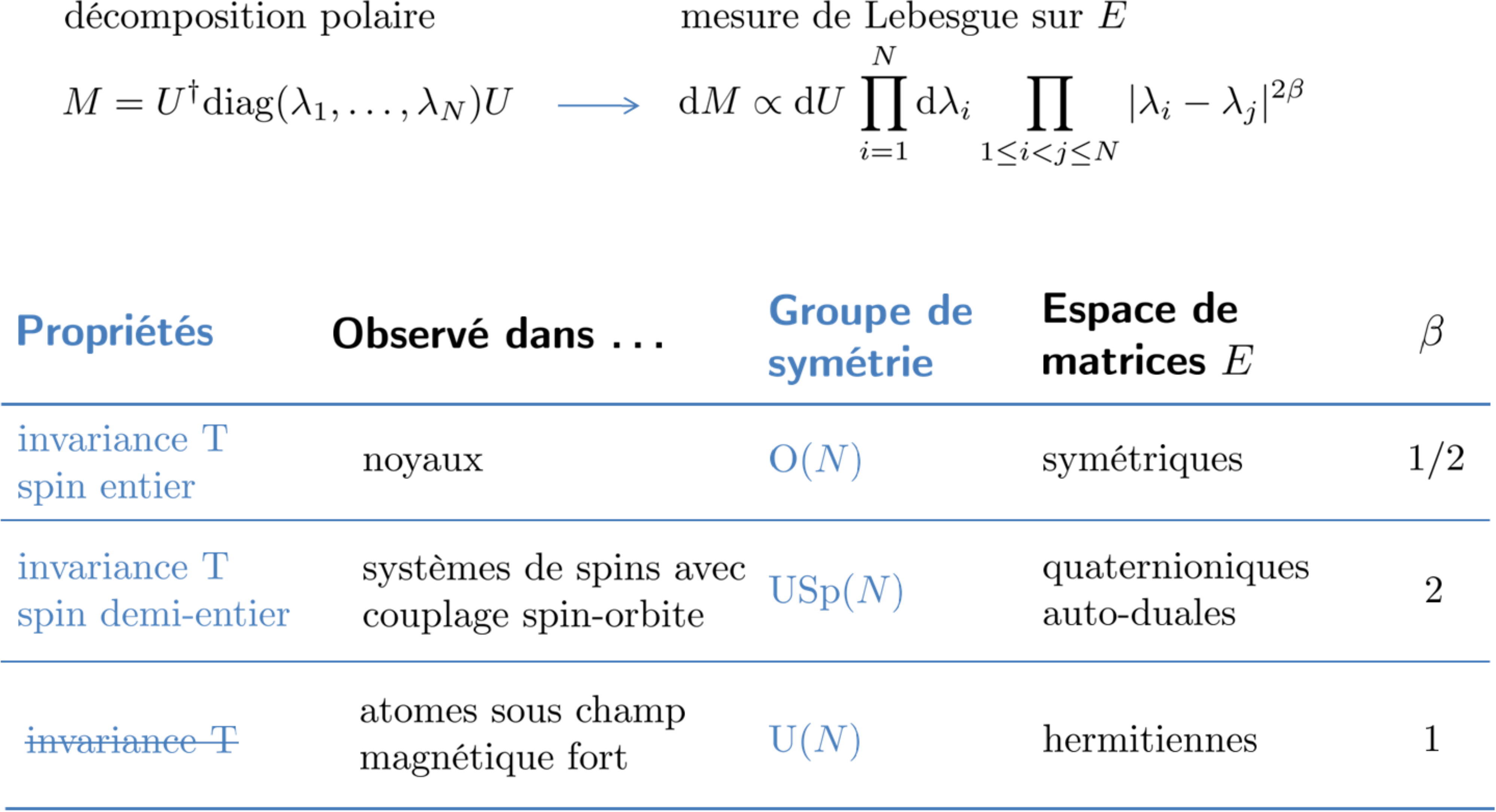}
\caption{\label{Dysonens} {\small Classe de mod\`eles de matrices introduits par Dyson \cite{Dyson}. 'T' repr\'esente l'op\'eration de renversement du temps. $\dd M$ est la mesure canonique sur $E$ consid\'er\'e comme un espace vectoriel, $\dd U$ est la mesure de Haar sur le groupe de sym\'etrie $G$, qui est compact. L'orbite sous l'action de $G$ d'une matrice ayant deux valeurs propres proches a un volume faible : on observe une r\'epulsion effective des valeurs propres, avec un exposant $2\beta$ d\'ependant de $G$ (Dyson utilise par convention l'indice $\overline{\beta} = 2\beta$, de sorte que $\overline{\beta} = 2$ corresponde au cas hermitien). Il y a d'autres combinaisons possibles de sym\'etrie discr\`etes, conduisant \`a d'autres classes de mod\`eles de matrices, plus ou moins pertinents pour la physique. Bernard et LeClair \cite{BLCclass} en ont list\'e $43$ en \'etudiant toutes les repr\'esentations unitaires possibles de la parit\'e et de l'invariance par renversement du temps sur un espace de matrices, \'eventuellement soumises \`a des conditions de r\'ealit\'e. Par exemple, la classe $\mathrm{USp}(N)$ est attendue lorsque les particules ont un spin demi-entier \emph{et} non respect\'e par les interactions (en cas de couplage spin-orbite par exemple).}}
\end{center}
\end{figure}
Les mod\`eles de matrices hermitiennes $M \in \mathcal{H}_N$ sont les plus simples \`a analyser. La mesure de probabilit\'e invariante sous le groupe unitaire $\mathrm{U}(N)$ la plus g\'en\'erale s'\'ecrit :
\beq
\dd\mu_N(M) = \frac{1}{Z_N}\dd M\,F(\Tr f(M)) \nn
\eeq
o\`u $Z_N$ assure la normalisation $\int \dd \mu_N(M) = 1$. Quand on ne s'int\'eresse qu'\`a la statistique des valeurs propres de $M$, on peut utiliser la d\'ecomposition polaire\label{decop} pour faire le changement de variable $M \mapsto (\lambda_1,\ldots,\lambda_N,U)$ o\`u $U \in \mathrm{U}(N)$. Le r\'esultat est donn\'e en Fig.~\ref{Dysonens}, le jacobien du changement de \label{Vande}variable est un d\'eterminant de Vandermonde \`a la puissance $2\beta = 2$. La question est alors de d\'eterminer les corr\'elations de densit\'e entre $n$ valeurs propres :
\bea
\rho_{n|N}(x_1,\ldots,x_n) & = & \frac{\mathrm{Prob}\Big[\bigcap_{j = 1}^n \big\{\exists\:i_j\quad \lambda_{i_j} \in [x_j,x_j + \dd x_j]\big\}\Big]}{\dd x_1\cdots \dd x_n} \nn \\
& = & \frac{N!}{(N - n)!}\,\int_{\mathbb{R}^{N - n}} \prod_{j = n + 1}^N \dd \lambda_i\,|\Delta(\lambda)|^2\,F\Big(\sum_{i = 1}^N f(\lambda_i)\Big)\Big|_{\lambda_1 = x_1,\ldots,\lambda_n = x_n} \nn
\eea
en particulier dans la limite $N \rightarrow \infty$. Une autre quantit\'e int\'eressante est la \label{fonp}fonction de partition :
\beq
Z_N = \int_{\mathcal{H}_N} \dd \mu_N(M)\nn
\eeq

Lorsque la mesure de probabilit\'e est de la forme\footnote{$N$ est la taille des matrices, et sa pr\'esence dans l'exposant est un choix de normalisation.} $\dd\mu_N(M) = \dd M\,e^{-N \Tr V(M)}$, ou apr\`es int\'egration sur la partie angulaire :
\beq
\label{eq:mesl}\dd\mu(\lambda) = \frac{1}{Z_N}\prod_{i = 1}^N \dd\lambda_i\,e^{-N\,V(\lambda_i)}\,|\Delta(\lambda)|^2
\eeq
Mehta et Gaudin \cite{Mehta1,Mehta2} ont d\'ecouvert que les $\rho_n$ pouvaient se calculer en introduisant une famille de polyn\^{o}mes $\big(\pi_n(x)\big)_{n \geq 0}$ orthogonaux :
\beq
\label{eq:ortho}(\pi_n|\pi_m) \equiv \int \mathrm{d}x\,e^{-N\,V(x)}\,\pi_n(x)\,\pi_m(x) = \delta_{n,m}h_n
\eeq
En normalisant $\pi_n(x) = x^n + \cdots$, ils \'ecrivent le d\'eterminant de Vandermonde :
\beq
\Delta(\lambda) = \prod_{1 \leq i < j \leq N} (\lambda_i - \lambda_j) = \det\big(\pi_j(\lambda_i)\big)_{1 \leq i,j \leq N} \nn
\eeq
Apr\`es une alg\`ebre \'el\'ementaire sur les d\'eterminants, et en utilisant l'orthogonalit\'e (\'{E}qn.~\ref{eq:ortho}), ils trouvent :
\beq
\label{eq:det}\rho_{n|N}(x_1,\ldots,x_n) = \det\big(K_N(x_i,x_j)\big)_{1 \leq i < j \leq N}
\eeq
avec le noyau de Christoffel-Darboux :
\beq
K_N(x,y) = \sum_{j = 0}^{N - 1} \frac{\pi_{j}(x)\pi_{j}(y)}{\sqrt{h_j}} = \frac{(\pi_N|\pi_{N - 1})}{\sqrt{h_{N}h_{N - 1}}}\,\frac{\pi_{N}(x)\pi_{N - 1}(y) - \pi_{N - 1}(y)\pi_{N}(x)}{x - y} \nn
\eeq
Pour la fonction de partition, ils retrouvent une formule classique reliant d\'eterminants de Hankel et normes des polyn\^{o}mes orthogonaux :
\bea
Z_N & = & \int_{\mathbb{R}^N} \prod_{i = 1}^N \dd \lambda_i\,e^{-N\,V(\lambda_i)} \nn \\
 \label{eq:Mehta} & = & \mathrm{det}\Big(\int_{\mathbb{R}} \dd x\,x^{i + j - 2}\,e^{-N\,V(x)}\Big)_{1 \leq i,j \leq N} = N!\,\prod_{i = 0}^{N - 1} h_i
\eea
Beaucoup d'autres donn\'ees statistiques, comme la distribution de la plus grande valeur propre (cf.~\ref{sec:largenty}), de la $k^{\textrm{\`{e}me}}$ plus grande valeur propre, d'espacements entre $k$ valeurs propres cons\'ecutives,\ldots{} s'extraient \`a partir du noyau $K_N$. Une strat\'egie analogue peut \^{e}tre appliqu\'ee dans les mod\`eles de matrices invariants sous $\mathrm{O}(N)$ ($\beta = 1/2$) ou $\mathrm{USp}(N)$ ($\beta = 2$), toujours pour une mesure du type $\dd\mu_N(M) = \dd M\,e^{-N\,\Tr V(M)}$. 

Il y a donc un lien profond entre matrices al\'eatoires et th\'eorie des polyn\^{o}mes orthogonaux, \`a double sens. D'une part, \`a la suite de Gaudin et Metha, nombre de r\'esultats exacts ($N$ fini) et asymptotiques ($N \rightarrow \infty$) ont \'et\'e \'etablis sur la statistique des valeurs propres dans les mod\`eles de matrices hermitiennes, notamment ceux associ\'es aux polyn\^{o}mes orthogonaux classiques (Hermite, Laguerre, Jacobi, \ldots). D'autre part, d\'eterminer les propri\'et\'es des polyn\^{o}mes orthogonaux (\'equivalent asymptotique, distribution asymptotique des z\'eros, \ldots) est \'equivalent au calcul d'int\'egrales de matrices dans la limite des grandes tailles. Ce renversement de point de vue a \'et\'e tr\`es fructueux pour les math\'ematiques.

\section{Universalit\'e locale}
\label{sec:unn}
\begin{figure}
\begin{center}
\includegraphics[width = \textwidth]{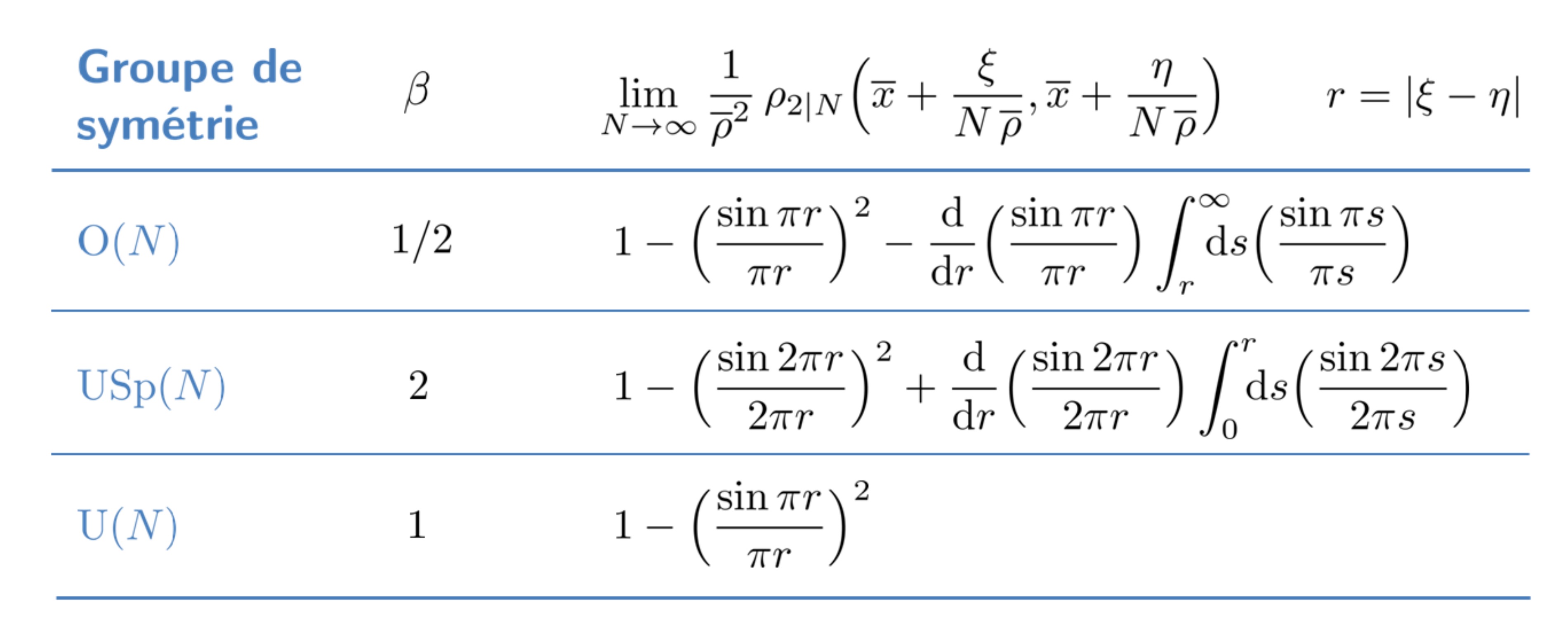}
\caption{\label{2point} {\small $\!\!$ Les corr\'elations locales entre deux valeurs propres sont donn\'ees par des fonctions universelles, qui ne d\'ependent que des sym\'etries du mod\`ele. Les corr\'elations \`a $n$ valeurs propres dans le cas $\beta = 1$ sont donn\'ees par les formules d\'eterminantales (\'{E}qn.~\ref{eq:det}) appliqu\'ees au noyau sinus. Dans les cas $\beta = 1/2$, $\beta = 2$, il existe des formules un peu plus compliqu\'ees, impliquant toujours $K_{\mathrm{sin}}$. En particulier, ces lois sont valides dans les mod\`eles gaussiens $\dd\mu_N(M) = \mathrm{d}M\, e^{-N\Tr M^2/2}$. On fera un abus de langage dans la suite de l'introduction, en appelant statistique EGO ($\beta = 1/2$), EGU ($\beta = 1$) ou EGS ($\beta = 2)$, ces lois locales au milieu du spectre, reli\'ees \`a $K_{\mathrm{sin}}$. EGO pour "Ensemble Gaussien Orthogonal" (GOE en anglais), EGU pour "Ensemble Gaussien Unitaire" (GUE), et EGS pour "Ensemble Gaussien Symplectique" (GSE).}}
\end{center}
\end{figure}

Dans ce paragraphe, la limite $N$ grand sera sous-entendue. Pour des $V$ assez r\'eguliers et croissants \`a l'infini, les $\lambda_i$ suivant \'{E}qn.~\ref{eq:mesl} se situent avec probabilit\'e $\sim 1$ dans une certaine r\'eunion de segments $I = \cup_{i = 1}^r [a_i,b_i]$. La densit\'e de \label{dens}valeurs propres $\rho_1(x) = \lim_{N \infty} N^{-1}\rho_{1|N}(x)$ sur $I$ d\'epend des d\'etails de $V$. Un voisinage born\'e d'un point $\overline{x}$ dans le milieu du spectre, contient une fraction macroscopique des $N$ valeurs propres. Donc, la distance typique entre deux valeurs propres autour de $\overline{x}$ est d'ordre $1/N$. \`{A} cette \'echelle, on peut montrer que les corr\'elations de densit\'e $\rho_n$ ($n \geq 2$) sont universelles\footnote{On sous-entend que la convergence a lieu pour une topologie faible, i.e. au sens des distributions.}. Modulo renormalisation par la densit\'e locale de valeurs propres $\overline{\rho} = \rho_1(\overline{x})$, elles sont donn\'ees par des formules d\'eterminantales\label{dett} (\'{E}qn.~\ref{eq:det}) avec un noyau en sinus :
\beq
K_{\mathrm{sin}}(\xi,\eta) = \overline{\rho}\,\lim_{N \infty}\,K_N\Big(\overline{x} + \frac{\xi}{N\overline{\rho}},\overline{x} + \frac{\eta}{N\overline{\rho}}\Big) \nn
\eeq
o\`u :
\beq
K_{\mathrm{sin}}(\xi,\eta) = \frac{\sin\pi(\xi - \eta)}{\pi(\xi - \eta)} \nn
\eeq
On parle d'\textbf{universalit\'e locale au milieu du spectre} (Fig.~\ref{2point}).

En probabilit\'es, l'\'emergence du th\'eor\`eme central limite pour les sommes de variables al\'eatoires ind\'ependantes est bien comprise, ainsi que ses hypoth\`eses minimales.
En th\'eorie des matrices al\'eatoires, ce programme n'est pas encore compl\'et\'e. Des simulations num\'eriques ont confirm\'e le vaste domaine de validit\'e des propri\'et\'es d'universalit\'e (cf. partie~\ref{sec:introuniv}), bien au-del\`a des mod\`eles solubles comme celui de l'\'{E}qn.~\ref{eq:mesl}. Depuis 10 ans, des progr\`es importants ont \'et\'e faits, auxquels sont associ\'es les noms de K.~Johansson, A.~Soshnikov,\ldots{} puis L.~Erd\"{o}s, J.~Ram\'{i}rez, B.~Schlein, H.-T.~Yau, et T.~Tao et V.~Vu. Tao et Vu ont notamment d\'emontr\'e \cite{TaoVu1} l'universalit\'e locale au milieu du spectre pour les matrices de Wigner hermitiennes, sous une hypoth\`ese sur les trois premiers moments des entr\'ees.

\section{Lois de matrices al\'eatoires dans la nature}
 \label{sec:ntu}
Ces r\'esultats cautionnent la comparaison entre donn\'ees exp\'erimentales et num\'eriques d'une part, et statistique des valeurs propres de matrices al\'eatoires.
Des caract\'eristiques de la statistique EGO ou EGU (suivant les sym\'etries du probl\`eme) ont \'et\'e observ\'ees dans des syst\`emes tr\`es divers. Nous avons \'evoqu\'e les spectres nucl\'eaires des noyaux lourds $Z \lesssim 50-70$, o\`u la statistique EGO a \'et\'e mise en \'evidence. Citons encore le spectre de l'atome hydrog\`ene plac\'e dans un champ magn\'etique fort (o\`u l'on trouve sous certaines conditions une statistique EGO), ou bien les niveaux d'\'energie d'une cavit\'e conductrice, o\`u les \'electrons \'evoluent en r\'egime balistique (o\`u l'on peut trouver une statistique EGO, EGU, ou d'autres classes de sym\'etrie). Dans ces syst\`emes physiques, l'universalit\'e locale des corr\'elations est souvent bien v\'erifi\'ee, jusqu'\`a un certain espacement $r \lesssim r_c$ (notation de la Fig.~\ref{2point}) entre niveaux d'\'energies. Au-del\`a de la ph\'enom\'enologie, on ne comprend pas toujours pourquoi les lois de matrices al\'eatoires apparaissent dans toutes ces situations. Depuis 1960, il y a eu un important travail de documentation de l'apparition de ces lois en physique, qui a inspir\'e des conjectures, \`a leur tour appelant de nouveaux tests et la d\'ecouverte de nouveaux ph\'enom\`enes. Des connections inattendues ont \'et\'e \'etablies avec les syst\`emes dynamiques et la th\'eorie des nombres.

\begin{figure}[h!]
\begin{center}
\includegraphics[width = \textwidth]{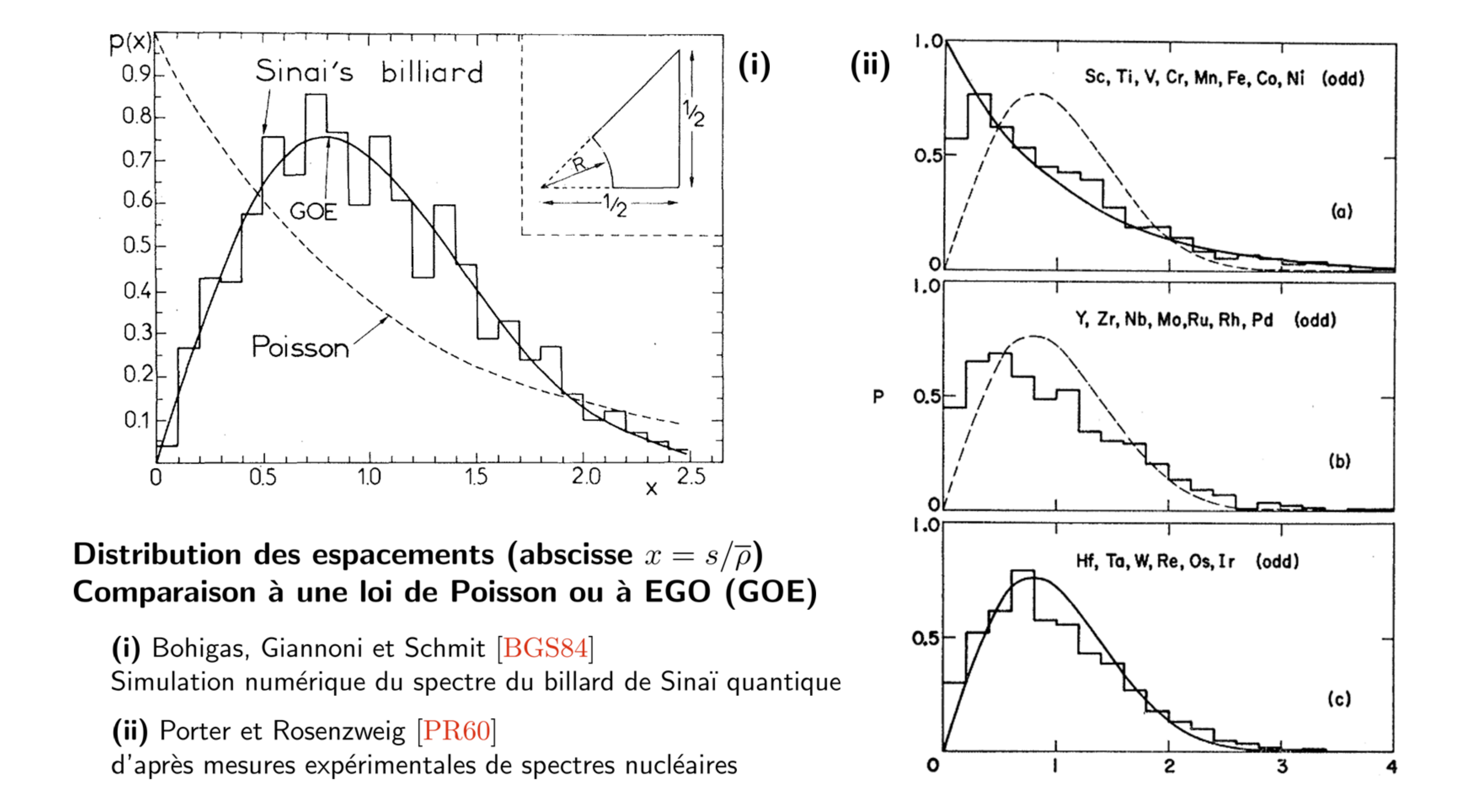}
\caption{\label{Spectrum} {\small Exemple d'application en physique m\'esoscopique $(i)$ et en physique nucl\'eaire $(ii)$ des statistiques de valeurs propres de matrices al\'eatoires. Reproduction avec la permission de l'American Physical Society : $(i)$ \cite{BGS84} \copyright{} APS, 1984 ; $(ii)$ \cite{RP60} \copyright{} APS, 1960.}}
\end{center}
\end{figure}

Le comportement des \'electrons balistiques dans une cavit\'e est la version quantique d'un probl\`eme de trajectoires dans un billard. Il faut trouver les valeurs propres $E_n$ et vecteurs propres $\psi_n$ d'un hamiltonien dans un domaine $\mathcal{D}$ de $\mathbb{R}^d$ (dans les exemples suivants, $d = 2$). Certains syst\`emes sont int\'egrables (par exemple, si $\mathcal{D}$ est un disque ou un rectangle), d'autres sont chaotiques (c'est le cas g\'en\'erique). Dans le cas int\'egrable, la g\'eom\'etrie des $\big(\psi_n(x)\big)_{x \in \mathcal{D}}$ est assez r\'eguli\`ere et la r\'epartition des $E_n$ est assez bien comprise. Dans le cas chaotique, les simulations num\'eriques indiquent que la statistique des $E_n$ est compatible avec EGO et l'estim\'ee de Wigner \'{E}qn.~\ref{eq:surmise}. Citons notamment les simulations de \cite{McDoK79} lorsque $\mathcal{D}$ est un stade, et de \cite{BGS84} lorsque $\mathcal{D}$ est un billard de Sina\"{i}, (Fig.~\ref{Spectrum}). Bohigas\label{boh}, Giannoni et Schmit ont conjectur\'e un ph\'enom\`ene g\'en\'eral :
\begin{conj}
Si le probl\`eme classique est un syst\`eme "assez chaotique\footnote{Bohigas et al. prennent "K-syst\`eme" comme hypoth\`ese.}", la statistique locale des $E_n$ est g\'en\'eriquement\footnote{Merci \`a S.~Nonnenmacher pour cette pr\'ecision. La condition de g\'en\'ericit\'e ne figure pas dans \cite{BGS84}, mais on sait maintenant qu'elle est n\'ecessaire.} celle de EGO ou EGU (suivant le respect ou non de l'invariance par renversement du temps).
\end{conj}
Cette conjecture (ainsi qu'une formulation pr\'ecise des hypoth\`eses n\'ecessaires) reste une question ouverte. Toutefois, il existe aussi une zoologie de syst\`emes dynamiques pr\'esentant un comportement interm\'ediaire entre int\'egrabilit\'e et chaos, o\`u l'on observe (le plus souvent empiriquement) une r\'epulsion des $E_n$ mais pas de loi connue venant des matrices al\'eatoires.

\begin{figure}
\begin{center}
\includegraphics[width = 0.85\textwidth]{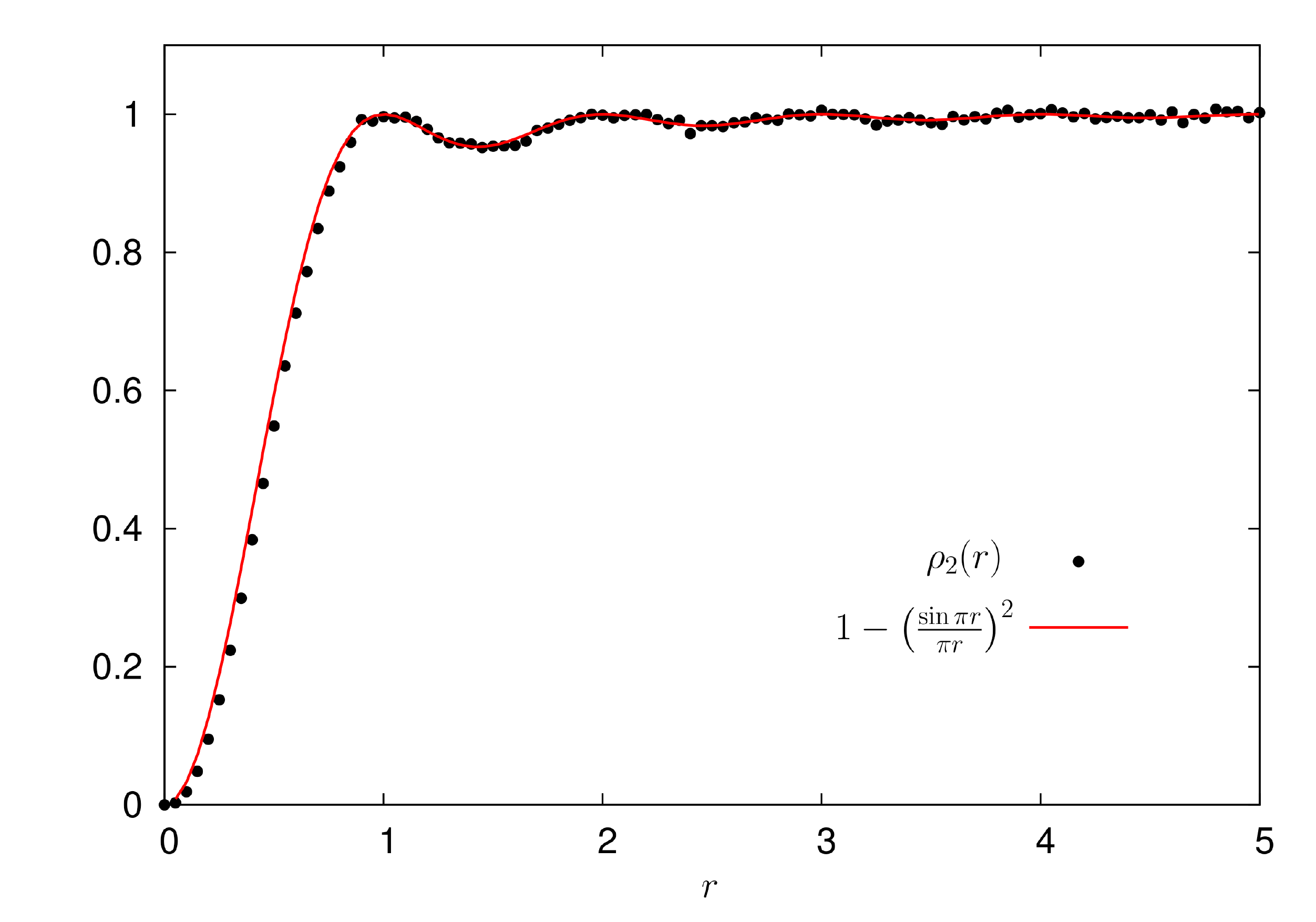}
\caption{\label{fig:Zeta} {\small Corr\'elations de densit\'e entre deux z\'eros de la fonction $\zeta$, pour une s\'equence de $10^{6}$ z\'eros cons\'ecutifs autour du $10^{20}$\`eme. Trac\'e par J.-M.~St\'ephan, d'apr\`es des donn\'ees gracieusement fournies par A.~Odlyzko.}}
\end{center}
\end{figure}

En th\'eorie des nombres, la \label{Zetu2}position des z\'eros de la fonction $\zeta$ de Riemann (le prolongement analytique \`a $\mathbb{C}$ de $\zeta(s) = \sum_{n = 1}^\infty \frac{1}{n^s}$ d\'efini pour $\Re s > 1$) a des cons\'equences importantes sur l'arithm\'etique dans $\mathbb{Z}$, et en particulier la distribution des nombres premiers. L'hypoth\`ese de Riemann (datant de 1859, not\'ee HR) affirme qu'hormis les entiers n\'egatifs pairs, les z\'eros sont de la forme $s_m = \frac{1}{2} + it_m$. On sait d'apr\`es Hardy \cite{Hardy} qu'il existe une infinit\'e de z\'eros de ce type, et il est bien connu que $\zeta(1/2 + it_m) = 0$ implique $\zeta(1/2 - it_m) = 0$, ce qui permet de conserver seulement les $t_m > 0$ et de les lister par ordre croissant. Plusieurs th\'eor\`emes existent sur leur distribution. Montgomery \label{mon}a formul\'e \cite{Montgomery} la conjecture suivante, qu'il a d\'emontr\'e ensuite pour une classe restreinte de fonctions test en supposant HR :
\begin{conj}
On note $w_n = \frac{t_n}{2\pi}\ln t_n$. Pour une classe assez large de fonctions test $f$ :
\beq
\lim_{N \rightarrow \infty} \frac{1}{N} \sum_{1 \leq m' \neq m \leq N} f(w_{m'} - w_m) = \int_{\mathbb{R}} \mathrm{d}w\,f(w)\Big[1 - \Big(\frac{\sin\pi w}{\pi w}\Big)^2\Big] \nn
\eeq
o\`u l'on reconnait la fonction de corr\'elation entre deux valeurs propres de EGU.
\end{conj}
Cette conjecture est support\'ee par les calculs num\'eriques d'Odlyzko \cite{Odly} (cf. Fig.~\ref{fig:Zeta}). Elle en a inspir\'e beaucoup d'autres \cite{Sarn,KeatingS1} concernant $\zeta$ et d'autres fonctions $L$, confirm\'ees par les simulations num\'eriques existantes. Il semble donc, sans compr\'ehension sous-jacente, que l'on puisse importer pour la th\'eorie des nombres certains calculs de la th\'eorie des matrices al\'eatoires.

\section{Surfaces discr\`etes}

\label{sec:BIPZ} Une deuxi\`eme vague d'int\'er\^{e}t pour les matrices al\'eatoires est n\'ee en th\'eorie des champs. Dans ce contexte, on a une grandeur $\phi$ (qui peut ou non avoir un sens physique), \`a laquelle on associe un poids $e^{-S[\phi]}$, et les observables physiques \`a calculer sont les moyennes statistiques $\langle \cdots \rangle$ :
\beq
\label{eq:action} Z = \int_{E} \mathcal{D}\phi\,e^{-S[\phi]},\qquad \langle \mathcal{O}(\phi) \rangle = \frac{1}{Z}\int_{E} \mathcal{D}\phi\,\mathcal{O}(\phi)\,e^{-S[\phi]}
\eeq
L'espace $E$ des configurations possibles est de "tr\`es grande dimension finie" (pour ne pas dire infini en th\'eorie des champs) et on note $\mathcal{D}\phi$ la mesure d'int\'egration. On sait bien \label{Wick2}calculer ces sommes lorsque $S[\phi]$ est quadratique, c'est le th\'eor\`eme de Wick. En g\'en\'eral, on peut essayer de se ramener au cas quadratique en \'etudiant un mod\`ele plus simple que l'\'{E}qn.~\ref{eq:action}. On se place au voisinage d'un minimum de $S$, que l'on d\'ecompose en partie quadratique + le reste : $S[\phi] = S_0[\phi] + \sum_k (-t_k)\mathcal{T}_k[\phi]$. On d\'efinit $\langle \cdots \rangle_0$, la valeur moyenne pour le poids $e^{-S_0[\phi]}$. Puis, on \emph{d\'efinit} le mod\`ele statistique "perturbatif" en d\'eveloppant en s\'erie de Taylor par rapport au 'reste' et en \'echangeant cette s\'erie avec $\int_{E}$.
\bea
\label{eq:pert1}Z_{\mathrm{pert}} & = & \sum_{(v_k)_k} \Big(\prod_{k} \frac{t_k^{v_k}}{v_k!}\Big) \Big\langle  \prod_{k}\mathcal{T}_k^{v_k}[\phi] \Big\rangle_0  \\
\label{eq:pert2}\langle \mathcal{O}[\phi]\rangle_{\mathrm{pert}} & = & \frac{1}{Z_{\mathrm{pert}}}\,\sum_{(v_k)_k} \Big(\prod_{k} \frac{t_k^{v_k}}{v_k!}\Big)\,\Big\langle \mathcal{O}[\phi]\,\prod_{k}\mathcal{T}_k^{v_k}[\phi]\Big\rangle_0
\eea
$Z_{\mathrm{pert}}$ et $\langle \mathcal{O}[\phi]\rangle_{\mathrm{pert}}$ sont d\'efinies comme des s\'eries formelles dans les $t_k$. L'application du th\'eor\`eme de Wick et d'une combinatoire \'el\'ementaire permet d'organiser \'{E}qns.~\ref{eq:pert1}-\ref{eq:pert2} comme s\'eries g\'en\'eratrices de graphes. \`{A} chaque ordre en $\prod_m t_m^{v_m}$, il n'y a qu'un nombre fini de graphes \`a consid\'erer, et ils sont compt\'es avec un certain poids $w(\mathcal{G})$, modulo r\'e-\'etiquetage des sommets et des ar\^{e}tes. Si l'on \'ecrit $Z_{\mathrm{pert}} = \sum_{\mathcal{G}} \frac{w(\mathcal{G})}{|\mathrm{Aut}(\mathcal{G})|}$, alors $\ln Z_{\mathrm{pert}}$ est la m\^{e}me somme restreinte aux graphes connexes, et $\langle \phi_1 \cdots \phi_n \rangle_{\mathrm{pert}}$ est une somme sur des graphes avec $n$ sommets marqu\'es.

En chromodynamique quantique (CQD), $\phi = \big(A(x,t)\big)_{(x,t) \in \mathbb{R}^4}$ est une matrice $N \times N$ de fonctions d\'efinies sur l'espace-temps (quand $A(x,t) \in \mathfrak{su}(3)$, c'est la th\'eorie des interactions fortes), et $S[\phi]$ est une action contenant des termes de degr\'e $2$, $3$ et $4$. t'Hooft observe en 1974 \cite{tHooft} que le poids d'un graphe $\mathcal{G}$ est proportionnel \`a $N^{\chi}$, o\`u $\chi$ est la caract\'eristique d'Euler de $\mathcal{G}$. G\'en\'eralement en th\'eorie des champs et en CQD, faire la somme sur les graphes est difficile, et le nombre de graphes cro\^{i}t exponentiellement avec l'ordre o\`u l'on se place.
\label{caca}
\begin{figure}
\begin{center}
\includegraphics[width=\textwidth]{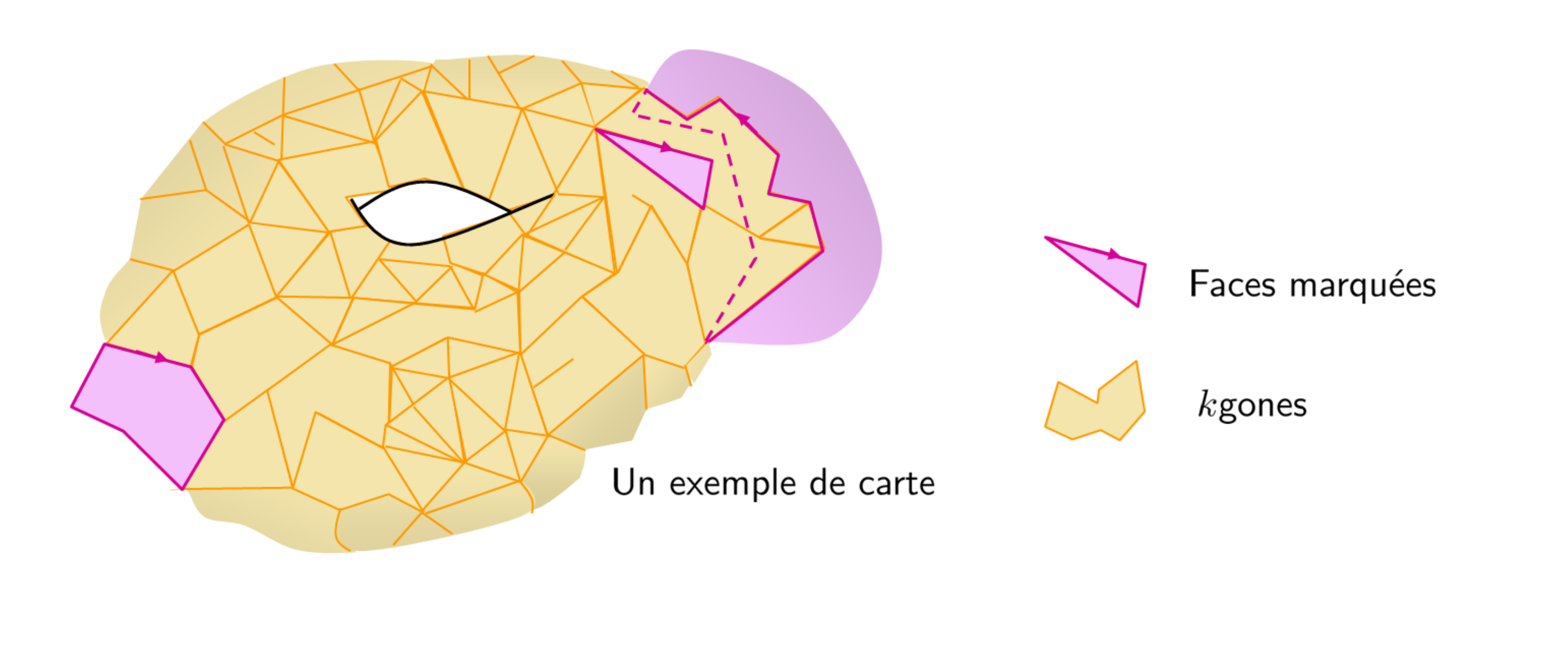}
\caption{\label{fig:carte} Dans les mod\`eles de matrices perturbatifs, les graphes $\mathcal{G}$ sont duaux \`a des surfaces discr\`etes $\Sigma$. Elles sont construites en recollant des $k$gones le long de leurs ar\^{e}tes (les collages de deux ar\^{e}tes appartenant au m\^{e}me polygone sont autoris\'es). Les sommets marqu\'es de $\mathcal{G}$ correspondent \`a des faces marqu\'ees avec une ar\^{e}te marqu\'ee dans $\Sigma$.}
\end{center}
\end{figure}

Mais inversement, on peut obtenir des r\'esultats combinatoires sur les graphes \`a partir d'une th\'eorie perturbative des champs. Cette id\'ee a \'et\'e exploit\'ee astucieusement en 1978 par Br\'ezin, Itzykson, Parisi et Zuber \cite{BIPZ}, avec le mod\`ele tr\`es simple o\`u $\phi$ est une matrice hermitienne de taille $N \times N$, et :
\beq
\label{eq:Scarte} S[\phi] = N\Big(\frac{\Tr \phi^2}{2} - \sum_{k = 3}^d t_k\,\frac{\Tr \phi^k}{k}\Big)\,,
\eeq
 i.e. un mod\`ele \`a une matrice al\'eatoire. Les graphes $\mathcal{G}$ \`a consid\'erer sont alors en bijection avec des surfaces discr\`etes $\Sigma$ (cf. Fig.~\ref{fig:carte}), que l'on appelle \'egalement \textbf{cartes}. Quand $\Sigma$ est constitu\'ee de $v_k$ faces $k$gonales, elle est compt\'ee avec le poids $w(\Sigma) = N^{\chi}\,\prod_k t_k^{v_k}$. Dans la limite $N \rightarrow \infty$ ne survivent que les graphes planaires (les surfaces "sans trous"), et on peut calculer leur nombre, par exemple \`a $v = \sum_{k} v_k$ fix\'e. Le r\'esultat de cette \'enum\'eration avait \'et\'e \'etabli par Tutte dans les ann\'ees 60 \cite{Tutte1,Tutte2} par des m\'ethodes combinatoires, et il est ainsi retrouv\'e par des techniques de matrices al\'eatoires (sans faire de combinatoire). J'exposerai cette m\'ethode au chapitre \ref{chap:formel}, et son application \`a la combinatoire de surfaces discr\`etes d\'ecor\'ees.

De nombreux r\'esultats nouveaux pour la combinatoire ont \'et\'e obtenus \`a la suite de \cite{BIPZ}, en particulier sur les surfaces discr\`etes de genre plus grand ("avec des anses"). Plus r\'ecemment, gr\^{a}ce \`a des bijections invent\'ees par Schaeffer pour les cartes planaires, et g\'en\'eralis\'ees aux genres $g \geq 1$ par Bouttier et Guitter, certains de ces r\'esultats ont \'et\'e retrouv\'es par la combinatoire. Je pr\'ecise toutefois (cf. chapitre~\ref{chap:formel}) que les mod\`eles de matrices ne donnent pas pour les cartes al\'eatoires toutes les informations (notamment, m\'etriques et microscopiques) qui int\'eresseraient les probabilistes et les combinatoristes.

En relation avec la th\'eorie des champs, l'activit\'e de recherche autour des matrices al\'eatoires a \'et\'e renouvel\'ee. Des liens profonds ont \'et\'e \'etablis avec les syst\`emes int\'egrables classiques, domaine florissant depuis les d\'ecouvertes de l'\'ecole russe (Dubrovin, Its, Krichever, Matveev, Novikov, \ldots) au milieu des ann\'ees 1970 \cite{Matveev}. Ces connections ont suscit\'e l'int\'er\^{e}t de nombreux math\'ematiciens pour les matrices al\'eatoires depuis le tournant des ann\'ees 1990, conduisant \`a la preuve (et parfois \`a la correction) de r\'esultats connus des physiciens th\'eoriciens, et \`a de nouvelles math\'ematiques. Les chapitres \ref{chap:conv} et \ref{chap:int} s'inscrivent dans cette direction. Dans les deux paragraphes qui suivent, je retrace le mouvement des ann\'ees 1980, o\`u des probl\`emes de comptage de surfaces, de solutions d'\'equations diff\'erentielles "int\'egrables", de gravit\'e quantique \`a deux dimensions ont \'et\'e simultan\'ement \'etudi\'es.

\section{Surfaces discr\`etes \`a la limite continue \ldots}
\label{sec:diss}
Reprenons le mod\`ele statistique de surfaces discr\`etes g\'en\'er\'e par \'{E}qn.~\ref{eq:Scarte}. Si l'on appelle $F_{g}$, la s\'erie g\'en\'eratrice des cartes connexes de genre $g$, on a une \'egalit\'e entre s\'eries formelles\footnote{\`{A} $v = \sum_k v_k$ fix\'e, la somme dans le membre de droite est finie.} :
\beq
\label{eq:lnZg}\ln Z_{\mathrm{pert}}\big[(t_k)_k\,;\,N\big] = \sum_{g \geq 0} N^{2 - 2g}\,F_g\big[(t_k)_k\big]
\eeq
Le nombre asymptotique de cartes, lorsque $v \rightarrow \infty$, est encod\'e dans les singularit\'es de $F_g\big[(t_k)_k\big]$. En effet\footnote{Il existe un r\'esultat incluant les singularit\'es/asymptotiques logarithmiques. Dans l'\'{E}qn.~\ref{eq:lmp}, $\Leftarrow$ est vrai \`a condition que la s\'erie s'\'etende en une fonction analytique sur un domaine $|t| < t^* + \delta$, priv\'e d'un secteur d'angle $\delta > 0$ autour de $t > t^*$ \cite{Flaj}. Dans la discussion qui suit, je vais mettre de c\^{o}t\'e des subtilit\'es importantes li\'ees \`a la structure des singularit\'es de $Z_{\mathrm{pert}}$.} :
\beq
\label{eq:lmp} \Big(\sum_{v \geq 0} A_v t^v\Big) \mathop{\sim}_{t \rightarrow t^*}  \alpha\,\Big(1 - \frac{t}{t^*}\Big)^{-\lambda} \quad \Longleftrightarrow \quad A_v\,\mathop{\sim}_{v \rightarrow \infty}\, \frac{\alpha\,v^{\lambda - 1}}{\Gamma(\lambda)}
\eeq
On peut \'etudier cette asymptotique dans plusieurs r\'egimes, o\`u l'on privil\'egie certaines esp\`eces de polygones par rapport \`a d'autres : on ajuste simultan\'ement $t_{k_1},\ldots,t_{k_r}$ vers une \label{crti}singularit\'e $(t_{k_1}^*,\ldots,t_{k_r}^*)$. On parle de \textbf{point critique} lorsque $r = 1$, et de \textbf{point multicritique} quand $r \geq 2$. Si $\delta = t_{k_1}^* - t_{k_1}$ mesure une distance au point critique, il est possible de montrer :
\beq
\label{eq:Fgsing}\forall g \geq 0 \qquad (F_g)_{\mathrm{sing}} \sim \delta^{(2 - 2g)(1 - \gamma/2)}\,\widehat{F}_{g}
\eeq
avec un exposant\footnote{Par convention, $\delta^{0} = \ln|\delta|$.} $\gamma$ fix\'e par le r\'egime asymptotique choisi. $\widehat{F}_g$ ne d\'epend plus des param\`etres $t_k$ non ajust\'es. En ce sens, les $\widehat{F}_g$ sont universels.

Si l'on fixe $s = N^{\frac{1}{1 - \gamma/2}}\,\delta$ fini, on peut remarquer que chaque terme dans l'\'{E}qn.~\ref{eq:lnZg} contribue en $O(1)$. Pour approcher un point multicritique, il faut aussi garder certaines variables $s_{k_j} = N^{\sigma_{k_j}}\,(t^*_{k_j} - t_{k_j})$ finies ($j \geq 2$). On devine :
\begin{prop}
\label{heurintro1}
(Heuristique). Il existe une \textbf{double limite d'\'echelle} :
\beq
\label{eq:gravquant}Z^*\big[s,(s_{k_j})_j\big] = \lim_{N \rightarrow \infty} Z_{\mathrm{pert}}\big[t^*_{k_1} + N^{\frac{-1}{1 - \gamma/2}}\,s\,,\,(t_{k_j}^* + N^{-\sigma_{k_j}}\,s_{k_j})_j\,;\,N\big]
\eeq
\end{prop}
On attend aussi que la fonction $Z^*$ se recolle bien avec l'\'{E}qn~\ref{eq:lnZg} et l'\'{E}qn.~\ref{eq:Fgsing} lorsqu'on est assez loin de la singularit\'e :
\begin{prop}
\label{heurintro2}(Heuristique). $\sum_{g \geq 0} s^{(2 - 2g)(1 - \gamma/2)}\,\widehat{F}_g\big[(s_{k_j})_j\big]$ est une s\'erie asymptotique\footnote{Valable seulement dans un certain secteur angulaire autour de $\mathbb{R}_+$.} de $\ln Z^*\big[s,(s_{k_j})_j\big]$ lorsque $s \rightarrow \infty$.
\end{prop}

\label{dens2}Dans le mod\`ele de matrice associ\'e \`a l'\'{E}qn.~\ref{eq:Scarte}, l'approche d'un point critique correspond \`a un changement de comportement de la densit\'e de valeurs propres. G\'en\'eriquement, $\rho_1(x) \propto |x - a|^{1/2}$ pr\`es d'un bord $a$ du spectre. Lorsque $t_{k_1} = t^*_{k_1}$ tandis que les autres $t_{j}$ sont g\'en\'eriques, non critiques, il existe un bord o\`u l'on a plut\^{o}t $\rho_1(x) \propto |x - a|^{3/2}$. En fait, avec des mod\`eles de matrices plus compliqu\'es, comptant des cartes "d\'ecor\'ees" (comme le mod\`ele $\On$ \'etudi\'e au chapitre \ref{chap:formel}), on peut atteindre n'importe quel comportement $\rho_1(x) \propto |x - a|^{\mu}$ ($\mu > -1$). Typiquement, on trouve un nombre fini de valeurs propres autour de $a$ si l'on regarde dans un voisinage de diam\`etre $\sim N^{-\frac{1}{1 + \mu}}$. \`{A} cette \'echelle, les corr\'elations de valeurs propres deviennent universelles (ne d\'ependent pas des param\`etres non ajust\'es). Le cas $\mu = p/q$ a re\c{c}u beaucoup d'attention. Les corr\'elations de \label{noyu}valeurs propres et les $\widehat{F}_g$ dans ce cas peuvent \^{e}tre extraits d'un noyau int\'egrable universel, $K_{\mathrm{(p,q)}}(\xi,\eta)$. Il est justifi\'e de parler de classe d'universalit\'e, que l'on appelle classe $(p,q)$ (cf.~\S~\ref{sec:pq}).

Les cartes avec un grand nombre de polygones, d'un point de vue macroscopique, sont des approximations de surface continues. La th\'eorie des champs o\`u l'on somme sur toutes les surfaces possibles (que l'on interpr\`ete comme des espaces-temps fluctuants) capture la notion de \label{gravq}"gravit\'e quantique \`a deux dimensions" (GQ2d). Les physiciens ont donc propos\'e $Z^*\big[s,(s_{k_j})_j\big]$ (Prop.~\ref{heurintro1}) comme une d\'efinition de la fonction de partition de GQ2d. Cette d\'efinition est non perturbative, car en g\'en\'eral la s\'erie asymptotique dans Prop.~\ref{heurintro2} ne suffit pas pour connaitre $Z^*$. J'insiste sur le fait que $Z^*$ n'est pas une int\'egrale de matrice, mais une limite d'int\'egrale de matrice. Pour des valeurs g\'en\'eriques (assez petites) des $t_k$, lorsque $t_{4} \rightarrow t^*_4$, on tombe dans la classe d'universalit\'e $(4,3)$, et l'on parle de "gravit\'e pure". On peut aussi l'atteindre avec des triangulations : $t_3 \rightarrow t^*_3$ et autres $t_j$ (assez petits) g\'en\'eriques. En cas d'ajustements simultan\'es, plusieurs types de polygones sont en comp\'etition, et l'on peut atteindre les classes d'universalit\'e $(p,q)$.

\section{\ldots et int\'egrabilit\'e}
\label{sec:intnin}
Les polyn\^{o}mes orthogonaux $\pi_n(x)$ du mod\`ele \`a une matrice satisfont une r\'ecurrence (d'ordre $2$) et une \'equation diff\'erentielle (d'ordre $2$) :
\bea
\label{eq:polyrec}&& x\pi_n(x)  =  \pi_{n + 1}(x) + b_n\,\pi_n(x) + \frac{h_n}{h_{n - 1}}\,\pi_{n - 1}(x)  \\
\label{eq:polydiffx}&& \frac{1}{N}\frac{\dd}{\dd x}\left(\begin{array}{c} \pi_{n} \\ \pi_{n - 1} \end{array}\right) = \mathbf{L}\,\left(\begin{array}{c} \pi_n \\ \pi_{n - 1}\end{array}\right)
\eea
o\`u $\mathbf{L}$ est une matrice $2 \times 2$ de polyn\^{o}mes en $x$, de degr\'es $\leq d - 1$ lorsque $t_{j} = 0$ pour tout $j > d$. Si l'on \'etudie la d\'ependance des $\pi_n$ dans les param\`etres $t_k$, on trouve aussi des \'equations diff\'erentielles d'ordre $2$ :
\beq
\label{eq:polydifft}\frac{1}{N}\frac{\partial}{\partial t_k}\left(\begin{array}{c} \pi_{n} \\ \pi_{n - 1}\end{array}\right) = \mathbf{M}_k\,\left(\begin{array}{c} \pi_n \\ \pi_{n - 1}\end{array}\right)
\eeq
o\`u $\mathbf{M}_k$ est une matrice $2 \times 2$ de polyn\^{o}mes en $x$, de degr\'es $\leq k$. Les $\pi_n(x)$ sont donc solutions de syst\`emes compatibles d'\'equations diff\'erentielles (et/ou de r\'ecurrences). Cette propri\'et\'e d\'efinit\footnote{Il y a plusieurs d\'efinitions de l'int\'egrabilit\'e, dont l'\'equivalence n'est pas toujours \'etablie. Nous adopterons celle-ci, attribu\'ee \`a Krichever.} un \textbf{syst\`eme int\'egrable}. En l'occurence, le syst\`eme (S : \'{E}qns.~\ref{eq:polyrec}-\ref{eq:polydiffx}-\ref{eq:polydifft}), est \'equivalent \`a un syst\`eme int\'egrable appel\'e "chaine de Toda". Les \'equations de compatibilit\'e de (S), avec les conditions initiales, d\'eterminent les $b_n$ et $h_n$, donc la fonction de partition $Z\big[(t_k)_k\,;\,N\big]$ (\'{E}qn.~\ref{eq:Mehta}).

Dans la limite continue $n,N \rightarrow \infty$, les \'equations discr\`etes deviennent des \'equations diff\'erentielles dans le param\`etre $t = n/N$. Dans cette limite, on peut chercher des solutions autosimilaires de (S), c'est-\`a-dire des solutions r\'eguli\`eres dans la limite $N \rightarrow \infty$ ne d\'ependant que de variables d'\'echelle comme $s_{k_i}$, et $s$ d\'efini par :
\beq
\frac{n}{N} = t^* + s\,N^{\alpha} \nn
\eeq
On obtient \cite{GrosMig} alors des \'equations diff\'erentielles non lin\'eaires transcendantes remarquables (on les note (P)), qui appartiennent \`a la hi\'erarchie de Painlev\'{e}. Des travaux comme \cite{IFKgravpure} justifient math\'ematiquement que les limites des $b_n$ et $h_n$ tendent vers des fonctions r\'eguli\`eres $b(s)$ et $h(s)$ d\'etermin\'ees par une certaine solution de (P). Ils d\'emontrent ainsi l'existence de la double limite d'\'echelle (Prop.~\ref{heurintro1}), et caract\'erisent $Z^*\big[s_{k_1},(s_{k_j})_j\big]$ comme solution de (P). Par exemple, pour le point critique associ\'e \`a la gravit\'e pure, il existe une variable d'\'echelle, not\'ee $\xi$, de sorte que $u = -\frac{1}{2}\,\frac{\partial^2 \ln Z^*}{\partial^2 \xi}$ soit solution\footnote{Il y a des subtilit\'es importantes sur les "conditions initiales" \`a ajouter pour sp\'ecifier de mani\`ere unique la solution int\'eressante pour la physique.} de l'\'equation de\label{API} Painlev\'e I :
\beq
\frac{\partial^2 u}{\partial \xi^2} = 6 u^2 + \xi \nn
\eeq

Toute une technologie a \'et\'e d\'evelopp\'ee par Bleher, Its, Deift, Venakides, Zhou et leurs successeurs, pour \'etudier rigoureusement la double limite d'\'echelle. Ces m\'ethodes exploitent la structure int\'egrable du mod\`ele de d\'epart. Sans entrer dans les d\'etails, on peut dire que le mod\`ele \`a une matrice hermitienne et les polyn\^{o}mes orthogonaux sont aujourd'hui bien compris \cite{DKMcLVZ1,DKMcLVZ2,DKMcLVZ3,BerMo,BerBou}. En revanche, l'\'etude des autres classes de sym\'etrie ($\beta \neq 1$), ou des chaines de matrices hermitiennes (reli\'ees aux polyn\^{o}mes biorthogonaux), est encore largement ouverte.

\section{Physique statistique}
\label{sec:introuniv}
En physique statistique hors \'equilibre, on rencontre assez fr\'equemment des processus d\'eterminantaux, i.e. dont la loi jointe \`a $n$ variables est du type \'{E}qn.~\ref{eq:det} :
\beq
\rho_n(x_1,\ldots,x_n) = \det\big[K(x_i,x_j)\big]_{1 \leq i,j \leq n} \nn
\eeq
pour un certain noyau $K$. Leur int\'er\^{e}t vient d'un r\'esultat combinatoire de Karlin-McGregor \cite{KarlinMcG} : dans un mod\`ele\label{Karlin} de chemins al\'eatoires avec poids locaux, si l'on note $K(x,y)$ la probabilit\'e d'un chemin de $x$ \`a $y$, $\det\big[K(x_i,y_j)\big]_{1 \leq i,j \leq n}$ est la probabilit\'e d'observer $n$ chemins reliant $\{x_1,\ldots,x_n\}$ \`a $\{y_1,\ldots,y_n\}$ sans s'intersecter. Des processus d\'eterminantaux interviennent donc dans les mod\`eles de marches al\'eatoires sans intersection (MAsI), dans les mod\`eles d'exclusion \`a $1$ dimension, et tous leurs avatars.

\begin{figure}
  \begin{center}
      \hspace{-0.4cm}\includegraphics[width = 1.05\textwidth]{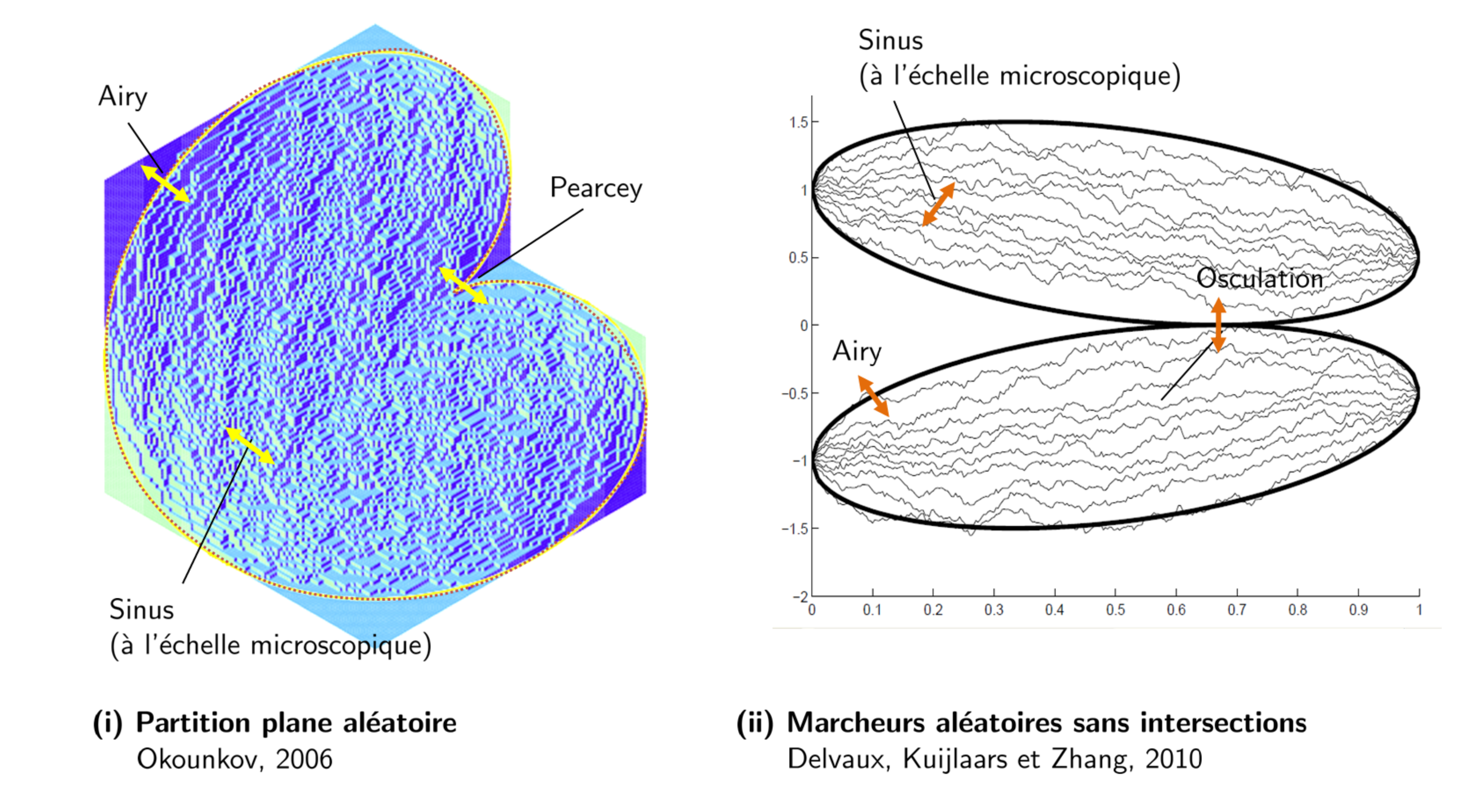}
      \caption{\label{fig:okoun} (Je remercie les auteurs respectifs de ces figures pour leur autorisation de reproduction) $\textbf{\textsf{(i)}}$ Simulation d'une grande partition plane al\'eatoire dans une r\'egion $\mathcal{D}$ de l'espace, par Okounkov \cite{Okounkov06}. Une partition plane d\'etermine une surface $\Sigma$, et apr\`es projection \`a deux dimensions, un pavage de la projection. Une fronti\`ere d\'eterministe apparait entre une r\'egion o\`u $\Sigma$ est rugueuse et fluctue d'un \'echantillon \`a l'autre, et une r\'egion o\`u $\Sigma$ est plate et d\'eterministe. C'est le ph\'enom\`ene du \textbf{cercle arctique}, qui a \'et\'e d\'ecouvert par Logan-Shepp et Kerov-Vershik dans le cas des partitions : les fluctuations sont gel\'ees \`a l'ext\'erieur de la fronti\`ere. Soit $\epsilon \rightarrow 0$, la taille \'el\'ementaire des cubes. Les fluctuations \`a l'\'echelle $\epsilon$ dans la r\'egion rugueuse sont d\'ecrites par le noyau sinus. Les fluctuations autour d'un point r\'egulier sont dans la classe d'universalit\'e $(1,2)$, i.e. d\'ecrites par le noyau d'Airy. En particulier, les fluctuations de la fronti\`ere elle-m\^{e}me suivent la statistique du maximum des valeurs propres al\'eatoires de EGU, que l'on appelle \textbf{distribution de Tracy-Widom} et qui est reli\'ee au noyau d'Airy (cf. chapitre~\ref{chap:conv}). Les fluctuations autour d'un point de rebroussement g\'en\'erique sont d\'ecrites par une autre classe d'universalit\'e, associ\'ee au noyau de Pearcey. $\textbf{\textsf{(ii)}}$ Simulation de marcheurs al\'eatoires sans intersections, par Kuijlaars et al. \cite{Kuijtacnode}. Dans le cas repr\'esent\'e, il y a deux groupes de marcheurs, dont les enveloppes deviennent osculantes en un point. Les auteurs ont \'etudi\'e la classe d'universalit\'e des fluctuations de position autour d'un tel point. Les fluctuations locales \`a l'int\'erieur des enveloppes sont bien s\^{u}r d\'ecrites par le noyau sinus, et les fluctuations autour d'un bord r\'egulier par le noyau d'Airy.}
\end{center}
\end{figure}

Dans les MAsI (cf. Fig.~\ref{fig:okoun}), on peut faire correspondre la position $x_i(t)$ \`a la $i$\`eme valeur propre d'une matrice $M(t)$, et \'ecrire un mod\`ele de matrices hermitiennes encodant la dynamique des $x_i(t)$. Lorsque $n \rightarrow \infty$, en invoquant l'universalit\'e locale, les corr\'elations locales entre les marcheurs "du milieu" sont d\'ecrites par le noyau sinus. Autour de marcheurs extr\^{e}mes, on doit trouver la classe d'universalit\'e $(1,2)$, valable g\'en\'eriquement au bord du spectre d'une matrice al\'eatoire, et d\'ecrite par le noyau d'Airy\label{aaa}. On peut inventer des mod\`eles o\`u l'enveloppe $\mathcal{E}$ des positions des marcheurs a une forme plus compliqu\'ee. Chaque type de g\'eom\'etrie locale pour $\mathcal{E}$ correspond \`a une classe d'universalit\'e, qui d\'ecrit au voisinage de ce point les corr\'elations des valeurs propres/des positions des marcheurs.

\begin{figure}[h!]
  \begin{center}
      \includegraphics[width = \textwidth]{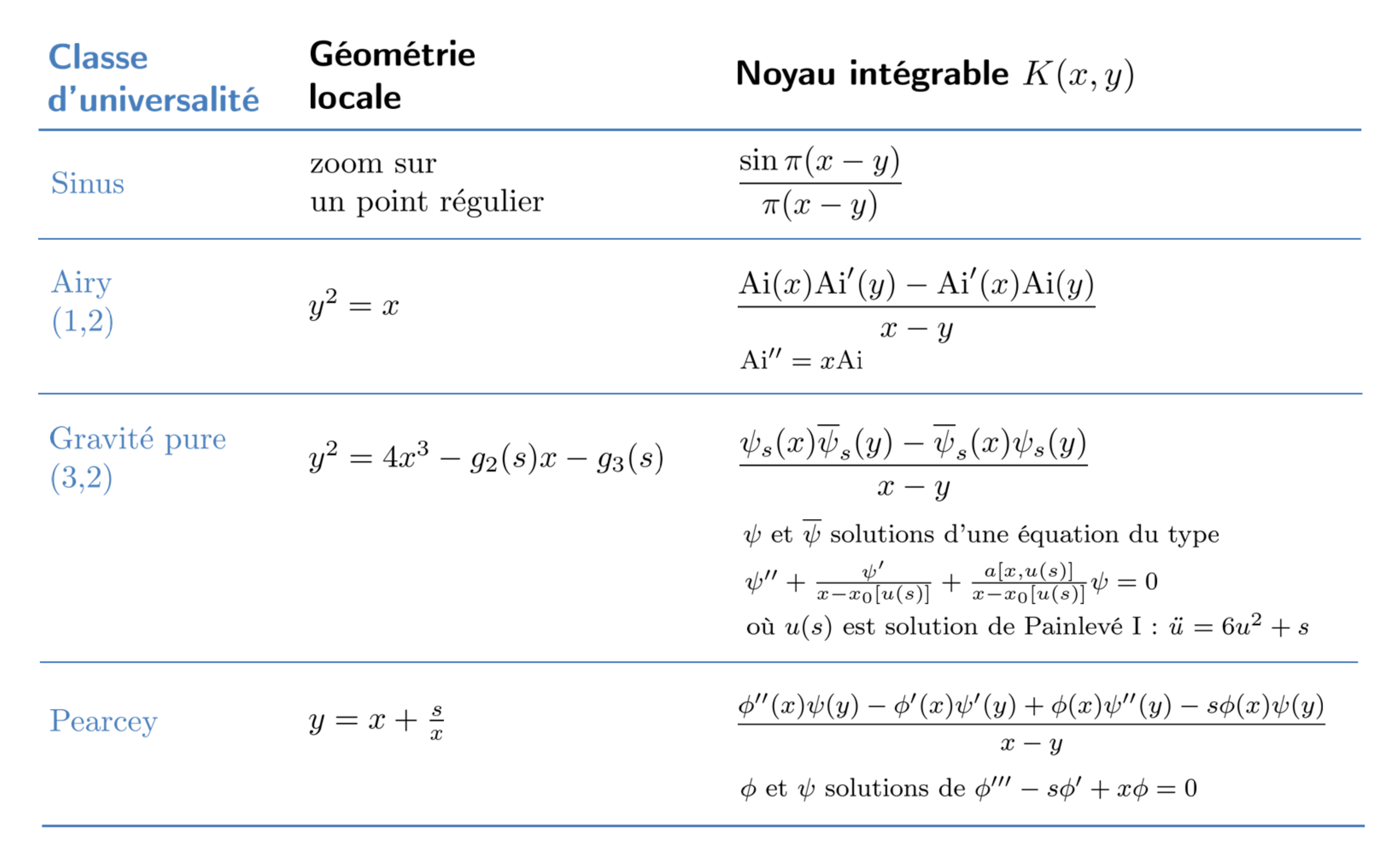}
      \caption{\label{fig:univ} Exemples de classes d'universalit\'e de matrices al\'eatoires hermitiennes.}
      \end{center}
\end{figure}

D'un point de vue fondamental, de nombreux\label{bbb} mod\`eles de partition al\'eatoire (empilement de carr\'es) ou de partition plane al\'eatoire (empilement de cubes, cf. Fig.~\ref{fig:okoun}) peuvent se repr\'esenter comme des int\'egrales multidimensionnelles qui sont des mod\`eles de matrices hermitiennes. On verra une application de cette technique avec le calcul des nombres de Hurwitz au chapitre~\ref{chap:cordes}. Ces probl\`emes de partitions sont duaux \`a des mod\`eles de permutations al\'eatoires, de pavages al\'eatoires, de dim\`eres, de particules en interactions, d'o\`u l'ubiquit\'e des lois de matrices al\'eatoires hermitiennes en physique statistique. Quelques exemples de classes d'universalit\'e sont donn\'es en Fig.~\ref{fig:univ}.

L'\'etude rigoureuse et la classification de ces classes d'universalit\'e (souvent anticip\'ee par les physiciens) a beaucoup progress\'e ces vingt derni\`eres ann\'ees, et a servi la physique statistique. Dans un mod\`ele donn\'e, cela revient \`a prouver l'existence d'une double limite d'\'echelle. Lorsque $K$ a une repr\'esentation assez explicite, des m\'ethodes \'el\'ementaires suffisent. En g\'en\'eral, la m\'ethode developp\'ee par Bleher-Its et Deift-Zhou a longtemps \'et\'e le seul angle d'attaque. Parmi les derniers r\'esultats obtenus par cette m\'ethode, mentionnons la d\'etermination d'une classe d'universalit\'e associ\'ee \`a une singularit\'e de type "osculation" \cite{Ferrtacnode,Kuijtacnode}. Le chapitre~\ref{chap:int} peut \^{e}tre interpr\'et\'e comme une autre approche rigoureuse de l'id\'ee (g\'eom\'etrie locale) $\leftrightarrow$ (classe d'universalit\'e associ\'ee \`a un syst\`eme int\'egrable), bas\'ee sur la "r\'ecurrence topologique" introduite au chapitre~\ref{chap:toporec}. Cela est illustr\'e en particulier \`a la page \pageref{sec:TWTW}.

Enfin, il y a d'autres mod\`eles comme les processus d'exclusion (TASEP, ASEP), des mod\`eles de croissance d'interfaces \cite{SpohnKPZ} ou de cristaux, \ldots{} o\`u l'on observe des lois de matrices al\'eatoires (diverses, pas n\'ecessairement hermitiennes) mais le lien est plus fragile. On y trouve aussi des d\'eformations de processus d\'eterminantaux, des transitions entre lois de matrices al\'eatoires et d'autres classes d'universalit\'e, \ldots{} dont les structures math\'ematiques restent \`a \'elucider.

\section{G\'eom\'etrie de l'espace des surfaces de Riemann \ldots}
\label{sec:introgeom}
Enfin, je vais parler des d\'eveloppements des techniques \'evoqu\'ees en partie~\ref{sec:BIPZ}, pour la combinatoire des surfaces de Riemann.

\begin{figure}[h!]
\begin{center}
\includegraphics[width = \textwidth]{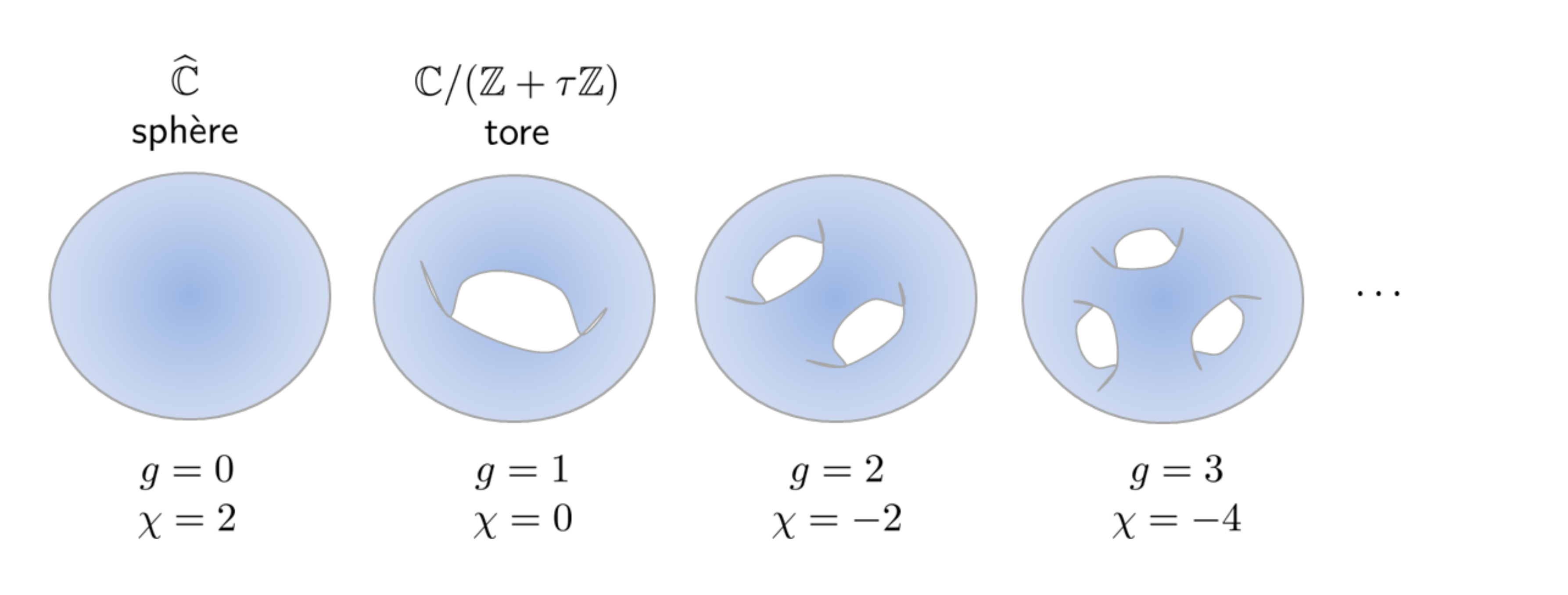}
\caption{\label{fig:genus} Classification topologique des surfaces compactes connexes orientables. $g$ est le genre, $\chi = 2 - 2g$ la caract\'eristique d'Euler.}
\end{center}
\end{figure}

Une \textbf{surface de Riemann} $\Sigma$ est une surface (un objet de dimension $2$) lisse, orientable et connexe. \`{A} d\'eformation continue pr\`es, les surfaces de Riemann compactes sont classifi\'ees par leur genre $g$, i.e. leur nombre d'anses (Fig.~\ref{fig:genus}).

\begin{figure}
  \begin{center}
      \includegraphics[width = 1.1\textwidth]{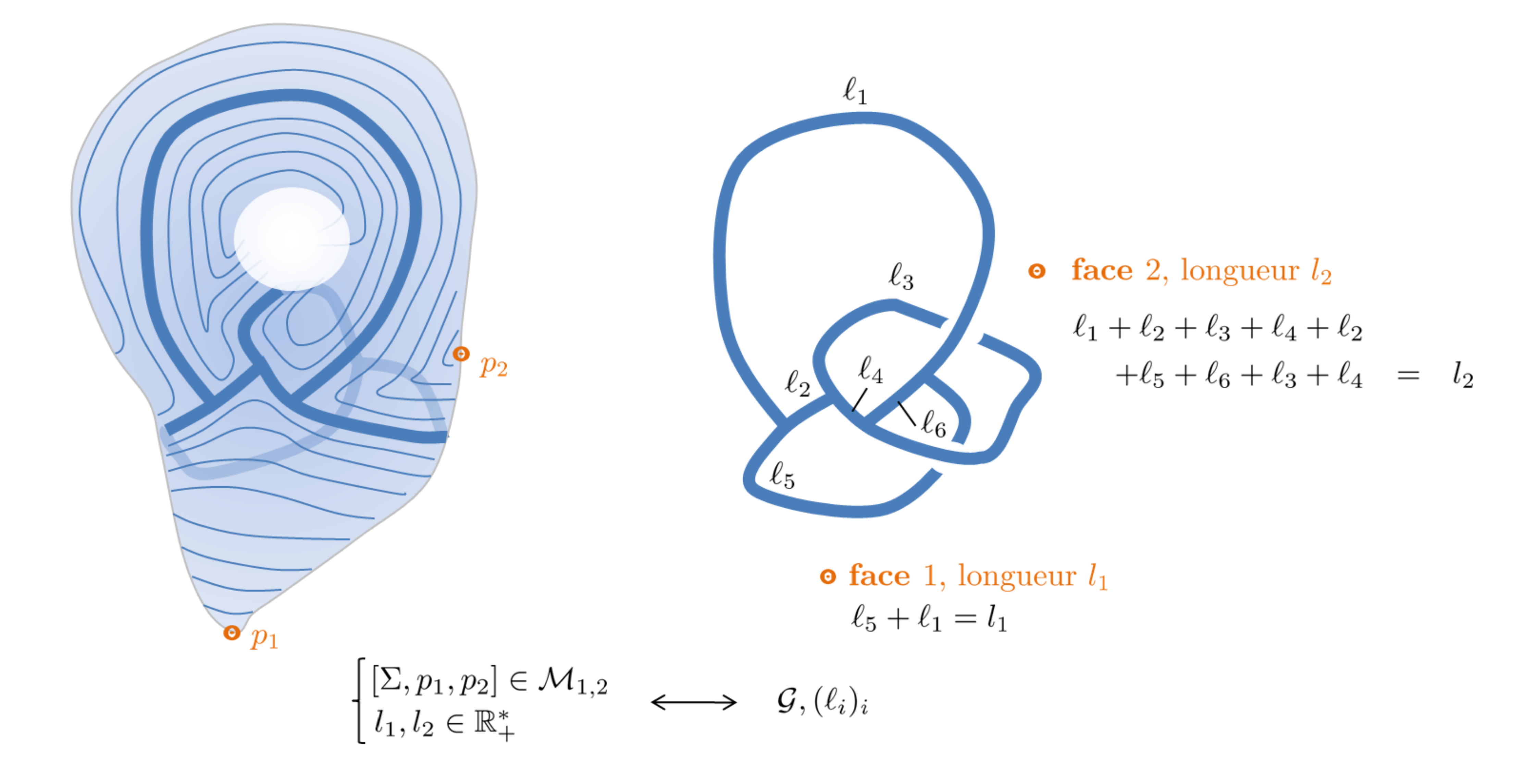}
      \caption{\label{fig:Streb} On suppose $2 - 2g - n < 0$. Pour toute surface de Riemann $\Sigma$ avec $n$ points marqu\'es $p_1,\ldots,p_n$, si l'on se donne des longueurs $l_1,\ldots,l_n$, d'apr\`es un th\'eor\`eme de Strebel \cite{Strebel} : il existe une unique fonction hauteur $h$, de sorte que $(i)$ $p_1,\ldots,p_n$ ait une hauteur extr\'emale ; $(ii)$ les trajectoires horizontales critiques forment un graphe \'epais $\mathcal{G} \subseteq \Sigma$, qui a $n$ faces centr\'ees sur les $p_i$. $(iii)$ les faces (composantes connexes de $\Sigma\setminus\mathcal{G}$) aient la topologie d'un disque. $\mathcal{G}$ est un graphe dont les sommets ont une valence $\geq 3$, et les longueurs des ar\^{e}tes ne d\'ependent pas que de la classe conforme de $\Sigma,p_1,\ldots,p_n$. R\'eciproquement, on peut reconstruire une surface de Riemann $\Sigma,p_1,\ldots,p_n$ en recollant des disques de mani\`ere conforme le long d'un graphe \'epais $\mathcal{G}$ \`a $n$ faces. Notant $s$ le nombre de sommets et $a$ le nombre d'ar\^{e}tes de $\mathcal{G}$, la dimension de la famille de surfaces reconstruite \`a partir de $\mathcal{G}$ est $d_{\mathcal{G}} = a$, et l'on sait $s - a + n = 2 - 2g$. En d\'eformant un peu les graphes \'epais, on voit que $\mathcal{G}$ est g\'en\'eriquement trivalent. Dans ce cas, $3s = 2a$, et on obtient $d_{\mathcal{G}} = 3(2g - 2 + n)$. Dans l'exemple ci-dessus, $g = 1$, $s = 4$, $a = 6$ et $n = 2$, et on a repr\'esent\'e les lignes de niveaux de $h$ sur $\Sigma$.}
\end{center}
\end{figure}

Mais on peut aussi faire de la g\'eom\'etrie complexe sur $\Sigma$, auquel cas on s'int\'eresse \`a $\Sigma$ \`a bijection holomorphe pr\`es. On utilise parfois le mot \textbf{conforme} comme synonyme d'holomorphe. L'ensemble des classes d'\'equivalence conforme de surfaces de Riemann de genre $g$, avec $n$ points marqu\'es, est not\'e $\mathcal{M}_{g,n}$. Plusieurs bijections ont \'et\'e invent\'ees pour d\'ecrire l'espace $\mathcal{M}_{g,n}$, et je pr\'esente en Fig.~\ref{fig:Streb} la d\'ecomposition\label{fig:Streb22} utilisant la th\'eorie de Strebel pour en donner une image intuitive :
\beq
\mathcal{M}_{g,n} \times (\mathbb{R}_+^*)^n \simeq \coprod_{\substack{\mathcal{G}\,\mathrm{graphe}\,\textrm{\'{e}pais} \\ \mathrm{\`a}\,n\,\mathrm{faces}\,\mathrm{et}\,s\,\mathrm{sommets}}} \big(\mathbb{R}_+^*\big)^{2g - 2 + n + s} \nn
\eeq
$\mathcal{M}_{g,n}$ est donc un espace \textbf{stratifi\'e}. Les strates sont localement model\'ees sur un espace $\mathbb{R}^{m}$ quotient\'e par un groupe discret (on parle d'\textbf{orbivari\'et\'e}), et\label{orbi2} ont pour dimension maximale :
\beq
d_{g,n} = 3(2g - 2 + n) - n = 6g - 6 + 2n \nn
\eeq
La limite d'une famille de surfaces de Riemann lisses n'est pas forc\'ement un objet lisse. Par exemple, on peut pincer une surface en plusieurs points. Donc $\mathcal{M}_{g,n}$ n'est pas \label{em}compact, mais on peut lui ajouter des surfaces pinc\'ees pour obtenir un espace compact $\overline{\mathcal{M}}_{g,n}$ (c'est la compactification de Deligne-Mumford).
On peut alors donner un sens\footnote{Les difficult\'es techniques sont cach\'ees dans la construction des $\nu$ contre lesquels on veut int\'egrer.} \`a $\int_{\overline{\mathcal{M}}_{g,n}} \nu$ et ainsi "compter des surfaces continues". Des facteurs de sym\'etrie sont naturellement inclus car $\overline{\mathcal{M}}_{g,n}$ est une orbivari\'et\'e.

Ces d\'ecompositions peuvent \^{e}tre utilis\'ees en principe pour ramener un probl\`eme de "comptage de surfaces continues" \`a un probl\`eme de combinatoire de graphes. Associ\'ee \`a des calculs de matrices al\'eatoires, cette strat\'egie a produit des th\'eor\`emes nouveaux en g\'eom\'etrie alg\'ebrique. En 1986, dans un article fondateur, Harer et Zagier \cite{HarerZagier} ont calcul\'e entre autres la caract\'eristique d'Euler de $\mathcal{M}_{g,0}$ en \'etudiant les moments $\langle\mathrm{Tr} M^{2n}\rangle_0$ dans l'int\'egrale \`a une matrice hermitienne avec un poids gaussien (cf. partie~\ref{sec:BIPZ}) :
\beq
\chi(\mathcal{M}_{g,0}) = \frac{B_{2g}}{2g(2g - 2)},\qquad B_{m} = m^{\textrm{\`{e}me}}\,\textrm{nombre}\,\mathrm{de}\,\mathrm{Bernoulli} \nn
\eeq
Quelques ann\'ees plus tard, partant des m\^{e}mes id\'ees, Kontsevich a prouv\'e un r\'esultat remarquable, conjectur\'e par Witten, conduisant \`a un calcul effectif de la th\'eorie de l'intersection \label{WW} dans $\mathcal{M}_{g,n}$. Je dois expliquer quelques notions avant de l'\'enoncer.

Un enjeu en g\'eom\'etrie alg\'ebrique est de comprendre la g\'eom\'etrie d'un espace comme $Y = \overline{\mathcal{M}}_{g,n}$, i.e. comment s'intersectent des sous-espaces de $Y$. Pour $Y = \widehat{\mathbb{C}}$, c'est tr\`es simple : on a des courbes $\mathcal{C}$ et des points $p$, avec les r\`egles $\mathcal{C}_1\cap\mathcal{C}_2 = \bigcup_j\{p_j\}$,  et g\'en\'eriquement $\mathcal{C}\cap \{p\} = \emptyset$, $\{p_1\} \cap \{p_2\} = \emptyset$. Il y a aussi des formules int\'egrales pour le nombre d'intersections de deux courbes $\mathcal{C}_1$ et $\mathcal{C}_2$. Pour $Y = \overline{\mathcal{M}}_{g,n}$ qui est une orbivari\'et\'e stratifi\'ee, la th\'eorie de l'intersection est plus compliqu\'ee. Il faut d'abord d\'efinir les objets naturels (les "classes") \`a intersecter. \label{cls}Dans cette th\`ese (chapitre~\ref{chap:cordes}), nous rencontrerons les classes $\psi_i$ et les classes $\kappa_{r}$ construites par Mumford \cite{Mum}. Informellement, une classe $\psi_i$ se comporte comme l'\'el\'ement d'aire de la surface de $\overline{\mathcal{M}}_{g,n}$ engendr\'ee en d\'epla\c{c}ant le point $p_i$ sur $\Sigma$. En particulier, les \textbf{nombres d'intersections} :
\beq
\label{eq:psiclass} \langle \tau_{d_1}\cdots \tau_{d_n}\rangle \equiv \int_{\overline{\mathcal{M}}_{g,n}} \prod_{i = 1}^{n} \psi_i^{d_i} \quad \in \mathbb{Q_+}
\eeq
peuvent \^{e}tre d\'efinis math\'ematiquement et sont non nuls lorsque $\sum_{i} 2d_i = d_{g,n}$. Quant aux classes $\kappa_r$, elles se comportent plut\^{o}t comme des \'el\'ements de volume $2r$-dimensionnels. Par construction, $\kappa_r$ intervient lorsque l'on veut d\'ej\`a int\'egrer sur un point $p_1$ dans l'\'{E}qn.~\ref{eq:psiclass} :
\beq
\forall d_1 \in \mathbb{N}^*,\qquad \int_{\overline{\mathcal{M}}_{g,n}}\!\!\!\!\!\! \psi_1^{d_1}\,\prod_{i = 2}^{n} \psi_i^{d_i} = \int_{\overline{\mathcal{M}}_{g,n - 1}} \!\!\!\!\!\!\!\!\!\kappa_{d_1 - 1}\,\prod_{i = 2}^n \psi_i^{d_i} \nn
\eeq

Kontsevich \cite{Kontsevich} a d\'emontr\'e en construisant un mod\`ele de matrice ad\'equat :
\begin{theo}
\label{thKont}La s\'erie formelle :
\beq
F_{\mathrm{K}}(t_0,t_1,\ldots) = \sum_{d_1,\ldots,d_k,\ldots, \geq 0} \Big[\prod_{k} \frac{\big((2k + 1)!!\,t_{2k + 1}\big)^{d_k}}{d_k !}\Big]\,\Big\langle \prod_{k}\tau_k^{d_k}\Big\rangle \nn
\eeq
est le logarithme d'une fonction tau\footnote{La notion de fonction tau pour un syst\`eme int\'egrable sera introduite au chapitre~\ref{chap:int}. Disons seulement qu'une fonction tau satisfait certaines \'equations diff\'erentielles non lin\'eaires.} de la hi\'erarchie int\'egrable KdV, pour les temps $t_i$. En particulier, $u = \frac{\partial^2 F}{\partial t_1^2}$ satisfait l'\'equation de \label{KDV}Korteweg-de Vries :
\beq
\frac{\partial u}{\partial t_3} = u\,\frac{\partial u}{\partial t_1} + \frac{1}{12}\frac{\partial^3 u}{\partial t_1^3} \nn
\eeq
On utilise la notation $(2k + 1)!! = 1\cdot 3\cdot \cdots \cdot (2k + 1)$.
\end{theo}
Comme les classes $\psi$ encodent des familles de surfaces avec des points marqu\'es fluctuants, on a appel\'e $Z_{\mathrm{K}}(t_0,t_1,\ldots) = e^{F_{\mathrm{K}}(t_0,t_1,\ldots)}$ la fonction de partition de la gravit\'e topologique \`a deux dimensions. En fait, les $Z^*_{\mathrm{(2m + 1,2)}}$ de la gravit\'e quantique (\'{E}qn.~\ref{eq:gravquant}) sont des sp\'ecialisations de $Z_{\mathrm{K}}$ \`a $t_j = 0$ pour $j \geq (2m + 1)$.

\section{\ldots et th\'eorie des cordes}
\label{sec:cordes1}
Plus g\'en\'eralement, l'\'etude des familles de surfaces (modulo reparam\'etrage conforme) plong\'ees dans un espace cible $X$, s'est consid\'erablement d\'evelopp\'ee depuis les ann\'ees 90, avec des travaux de Kontsevich et bien d'autres. Il s'agit l\`a encore de comprendre la th\'eorie de l'intersection, de r\'esoudre des probl\`emes d'\'enum\'eration de configurations g\'eom\'etriques dans un "espace $\mathcal{M}_{g,n}(X)$". On souhaite aussi d\'efinir des invariants, calculables de fa\c{c}on algorithmique \`a partir de la d\'efinition de $X$, peut-\^{e}tre applicables pour mieux distinguer (voire classifier) des objets $X$. Comprendre $\mathcal{M}_{g,n}(X)$ permet de mieux comprendre la g\'eom\'etrie de l'espace $X$.

En fait, explorer l'ensemble des surfaces $[\Sigma]$ plong\'ees dans $X$, les compter avec certains poids (au sens de \'{E}qn.~\ref{eq:action}), est exactement ce que l'on entend par une th\'eorie des cordes dans $X$. On parle de \textbf{th\'eorie des cordes topologiques} lorsque les observables $\langle\mathcal{O}\rangle$ ne d\'ependent que de donn\'ees topologiques de $\Sigma \hookrightarrow X$ : le genre $g$, le nombre de bords/points marqu\'es $n$, leurs nombres d'enroulements $\vec{\beta}$ sur certains sous-espaces de $X$, \ldots{} Ces observables, lorsqu'elles sont proprement d\'efinies, sont des s\'eries g\'en\'eratrices d'invariants g\'eom\'etriques de $X$. On a vu un exemple au Th\'eor\`eme~\ref{thKont} avec $Z_{\mathrm{K}}$, la s\'erie g\'en\'eratrice des nombres d'intersections dans $\mathcal{M}_{g,n} = \mathcal{M}_{g,n}(\{\mathrm{pt}\})$.

\begin{figure}
\begin{center}
\includegraphics[width = 0.88\textwidth]{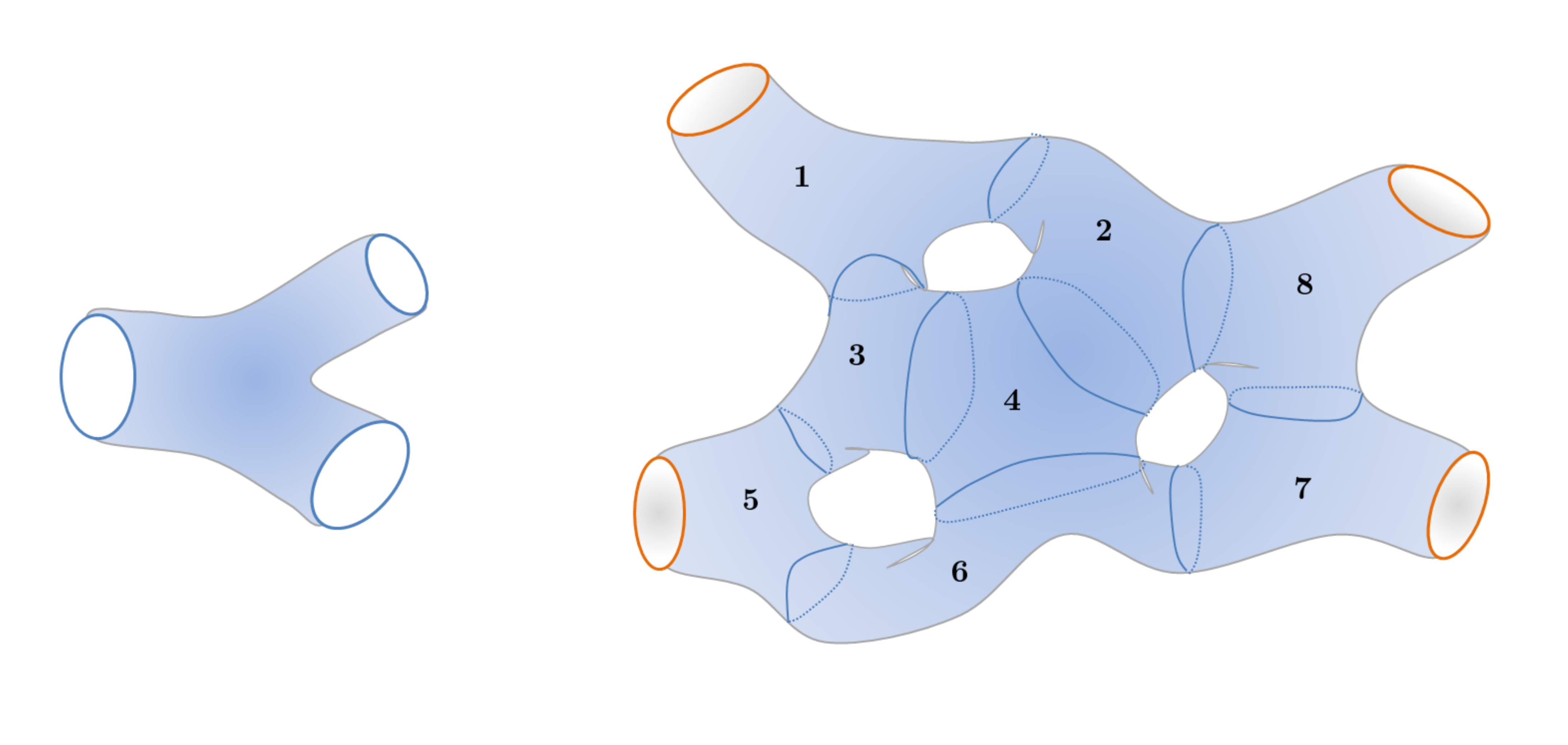}
\caption{\label{fig:cordes} "Th\'eorie des cordes = g\'eom\'etrie alg\'ebrique". Avec un vocabulaire de physique, $\Sigma \in \mathcal{M}_{g,n}$ s'interpr\`ete comme la surface engendr\'ee dans $X$ par la propagation de cordes. Un point marqu\'e $p_i$ (que l'on peut \'elargir et repr\'esenter comme un bord de $\Sigma$) correspond \`a une corde entrante ou sortante. Comme en physique des particules, la dynamique permet des interactions \'el\'ementaires (fusions ou scissions) entre cordes, repr\'esent\'ee par un "pantalon". $\Sigma \in \mathcal{M}_{g,n}$ d\'ecrit alors un processus impliquant $n$ cordes, avec $2g - 2 + n = -\chi$ interactions \'el\'ementaires (ce nombre est ind\'ependant de la d\'ecomposition en pantalons choisie). Dans l'exemple ci-dessus, $n = 4$, $g = 3$, d'o\`u $-\chi = 8$.}
\end{center}
\end{figure}

Cette pr\'esentation informelle cache des aspects math\'ematiques non triviaux pour d\'efinir proprement $\overline{\mathcal{M}}_{g,n}(X,\vec{\beta})$ et des classes sur lesquelles on peut int\'egrer. Lorsque $X$ est une vari\'et\'e complexe lisse et compacte, Li et Tian \cite{LiTian} ont ainsi construit\footnote{$\overline{\mathcal{M}}_{g,n}(X,\vec{\beta})$ n'est pas un ensemble, mais plut\^{o}t une pile, en utilisant le langage des cat\'egories.} $\overline{\mathcal{M}}_{g,n}(X,\vec{\beta})$ et une bonne classe de comptage $[\mathbf{1}]$ : les nombres $\int_{\overline{\mathcal{M}}_{g,n}(X,\vec{\beta})} [\mathbf{1}]$ ont un sens, et sont des invariants g\'eom\'etriques de $X$. Ce sont les\label{TGWH} \textbf{invariants de Gromov-Witten}, dont nous reparlerons au chapitre~\ref{chap:cordes}. Ils comptent le nombre de fa\c{c}ons in\'equivalentes de plonger $[\Sigma] \in \mathcal{M}_{g,n}$ dans $X$. Ces travaux ont \'et\'e progressivement g\'en\'eralis\'es \`a des $X$ moins sympathiques.

Cette construction ne donne pas de r\`egle pour obtenir ces invariants de Gromov-Witten. Depuis leur invention, des progr\`es importants ont \'et\'e r\'ealis\'es pour les calculer effectivement dans des espaces $X$ particuliers, gr\^{a}ce aux \'echanges f\'econds entre th\'eoriciens des cordes et math\'ematiciens.  Notamment, lorsque l'espace $X$ poss\`ede assez de sym\'etries, l'int\'egrale sur $\mathcal{M}_{g,n}(X,\vec{\beta})$ peut \^{e}tre exactement localis\'ee, et devient une somme discr\`ete sur certaines configurations g\'eom\'etriques\label{instan} (appel\'ees instantons). Derri\`ere ces probl\`emes de "g\'eom\'etrie \'enum\'erative", on retrouve des syst\`emes int\'egrables, des calculs d'int\'egrales sur $\mathcal{M}_{g,n}$, des sommes sur des partitions ou des partitions planes \ldots{} et \textit{in fine}, les structures math\'ematiques qui apparaissent dans les mod\`eles de matrices, si ce n'est directement les mod\`eles de matrices.

\section{\'{E}quations de boucles et g\'eom\'etrie}
\label{sec:boubou}
Au cours de cette th\`ese, nous allons voir que ces structures math\'ematiques gravitent toutes autour des \textbf{\'equations de boucles}, ou de la \textbf{hi\'erarchie des \'equations de boucles}, introduites au \S~\ref{sec:heqb}. Les observables des mod\`eles de matrices, les s\'eries g\'en\'eratrices de cartes, les s\'eries g\'en\'eratrices de nombres d'intersections sont (ou peuvent \^{e}tre extraites) des solutions de ces \'equations. Mon directeur de th\`ese Bertrand Eynard a d\'etermin\'e en 2004 \cite{E1MM}  la solution g\'en\'erale de la hi\'erarchie d'\'equations de boucles, en \'etendant et reformulant la \textbf{m\'ethode des moments} initi\'ee par Ambj{\o}rn et Makeenko en 1990 \cite{AM90}. Ces solutions d\'erivent d'un formalisme uniquement bas\'e sur de la g\'eom\'etrie complexe, que l'on appelle \textbf{r\'ecurrence topologique} et qu'il a d\'evelopp\'e avec son \'etudiant \`a l'\'epoque, Nicolas Orantin. Dans sa version la plus simple, la r\'ecurrence topologique prend une courbe plane $\mathcal{E} : E(x,y) = 0$ comme donn\'ee, et d\'efinit des nombres $F_g \in \mathbb{C}$, et des formes diff\'erentielles $\omega_n^{(g)}(z_1,\ldots,z_n)$ ($z_i \in E$). Vu les pr\'ec\'edents math\'ematiques des matrices al\'eatoires, il \'etait naturel d'esp\'erer que ces objets aient de nombreuses propri\'et\'es, et c'est le cas. En particulier, Eynard et Orantin ont d\'emontr\'e que les $F_g$ sont des \textbf{invariants symplectiques} de $\mathcal{E}$. En suivant une pr\'esentation axiomatique, je r\'eexpliquerai la r\'ecurrence topologique au chapitre \ref{chap:toporec}, en signalant les propri\'et\'es connues utiles \`a cette th\`ese, et les propri\'et\'es conjectur\'ees. Pendant ma th\`ese, j'ai r\'epondu \`a (ou progress\'e vers) certaines de ces conjectures, notamment en relation avec l'int\'egrabilit\'e et la g\'eom\'etrie \'enum\'erative.

Les travaux sur la r\'ecurrence topologique (2004-2011), et notamment les th\`eses de Nicolas Orantin \cite{TheseNico}, d'Olivier Marchal \cite{TheseOliv}, et la pr\'esente contribution,  laissent penser que les \'equations de boucles forment la structure math\'ematique universelle qui sous-tend toutes les questions \'evoqu\'ees dans cette introduction. Cette structure est essentiellement g\'eom\'etrique, tant par sa formulation (g\'eom\'etrie complexe sur les surfaces de Riemann) que ses propri\'et\'es (invariance symplectique) et applications (en g\'eom\'etrie alg\'ebrique, en int\'egrabilit\'e). Quant \`a son interpr\'etation : s'il est clair que $F_g[\mathcal{E}]$ et $\omega_n^{(g)}[\mathcal{E}]$ pour n'importe quel $\mathcal{E}$ s'expriment toujours \`a partir des nombres d'intersections dans $\mathcal{M}_{g,n}$, la relation pr\'ecise n'est bien connue que dans certains cas (\S~\ref{sec:interFg}).

\section{R\'esultats obtenus}
\label{sec:teo}
Mes travaux ont \'et\'e r\'ealis\'es en collaboration \'etroite avec mon directeur de th\`ese, et s'orientent suivant deux axes. En premier lieu, le d\'eveloppement de la th\'eorie associ\'ee \`a la r\'ecurrence topologique (en abr\'eg\'e, RT) et aux \'equations de boucles, et ses g\'en\'eralisations.  En second lieu, le d\'eveloppement de nouvelles applications de cette th\'eorie. Voici les principaux r\'esultats que j'ai obtenus pendant la th\`ese. Je mentionnerai leur contexte le moment venu.

\vspace{0.2cm}

\noindent $\diamond\,$ L'\'enum\'eration des cartes d\'ecor\'ees par des boucles auto\'evitantes (le mod\`ele $\mathcal{O}(\mathfrak{n}))$ dans toutes les topologies (chapitre~\ref{chap:formel}, article \cite{BEMS}). En corolaire, la preuve des exposants KPZ pour les grandes cartes dans ce mod\`ele. La r\'eponse est donn\'ee par la r\'ecurrence topologique appliqu\'ee \`a une courbe $\mathcal{E}$ non alg\'ebrique.

\vspace{0.2cm}

\noindent $\diamond\,$ La preuve de la conjecture de Bouchard et Mari\~{n}o : les nombres de Hurwitz sont donn\'es par la r\'ecurrence topologique appliqu\'es \`a la courbe de Lambert $\mathcal{E}\,:\, e^{x} = ye^{-y}$. Ce travail a \'et\'e r\'ealis\'e en collaboration avec Motohico Mulase (UC Davis) et Brad Safnuk (Central Michigan University). C'est le premier cas d\'emontr\'e de la conjecture BKMP, qui affirme que les invariants de Gromov-Witten d'une vari\'et\'e torique de Calabi-Yau $X$, sont donn\'es par la r\'ecurrence topologique appliqu\'ee \`a une courbe $\mathcal{E}_X$, appel\'ee courbe miroir et facilement calculable \`a partir de $X$ (chapitre~\ref{chap:cordes}).

\vspace{0.2cm}

\noindent $\diamond\,$ Pour la loi de Tracy-Widom $\mathsf{TW}_{\beta}(s)$ avec $\beta > 0$ quelconque, la d\'etermination heuristique de l'asymptotique $s \rightarrow -\infty$ \`a tous les ordres : elle est donn\'ee par la r\'ecurrence topologique version $\beta$ appliqu\'ee \`a $\mathcal{E}_{\mathrm{TW}}\,:\, y^2 = x + \frac{1}{x} - 2$ (chapitre~\ref{chap:conv}). Ce travail s'inscrit dans une collaboration (article~\cite{BEMN}) avec C\'eline Nadal et son directeur de th\`ese Satya Majumdar (LPTMS Orsay).

\vspace{0.2cm}

\noindent $\diamond\,$ La d\'emonstration du d\'eveloppement en $1/N$ dans les mod\`eles $\beta$ sur l'axe r\'eel, dans un r\'egime \`a une coupure hors d'un point critique (chapitre~\ref{chap:conv}). C'est un travail en collaboration avec Alice Guionnet (ENS Lyon), qui sera bient\^{o}t publi\'e (article~\cite{BG11})

\vspace{0.2cm}

\noindent $\diamond\,$ Pour $\beta = 1$, la d\'emonstration de l'heuristique ci-dessus, \`a savoir que la resommation des $F_g(\mathcal{E}_{\mathsf{TW}})$ est asymptotique \`a la solution de Painlev\'e II qui permet d'exprimer $\mathsf{TW}_1$ d'apr\`es les travaux originaux de Tracy et Widom \cite{TW92} (chapitre~\ref{chap:conv}, article \cite{BETWc}).

\vspace{0.2cm}

\noindent $\diamond\,$ L'association d'\'equations de boucles \`a tout syst\`eme int\'egrable classique. L'article \cite{BETW} et sa version courte \cite{BETWc} sont en fait une application de ces travaux aux syst\`emes $2 \times 2$ associ\'es \`a l'\'equation de Painlev\'e II. Le cas des syst\`emes $2 \times 2$ a \'et\'e fait pour l'essentiel par Berg\`ere et Eynard \cite{BEdet}. J'ai trouv\'e une solution technique qui permet d'affaiblir leurs hypoth\`eses, et de g\'en\'eraliser aux syst\`emes $d \times d$. Elle est expliqu\'ee au chapitre~\ref{chap:int}, et devrait figurer dans un prochain article, avec une application \`a la double limite d'\'echelle dans les mod\`eles minimaux $(p,q)$.

\vspace{0.2cm}

\noindent $\diamond\,$ Des r\'esultats partiels sur la r\'eciproque, i.e. la construction d'un syst\`eme int\'egrable \`a partir d'\'equations de boucles. Ils seront expliqu\'es au chapitre~\ref{chap:int} mais ne sont pas encore publi\'es.

\section{Description des mod\`eles de matrices \'etudi\'es}
\label{sec:chainmat}
\label{Vande2} On utilise la notation $\Delta(\lambda) = \mathrm{det}\big(\lambda_i^{j - 1})_{1 \leq i,j \leq N} = \prod_{1 \leq i <  j \leq} (\lambda_i - \lambda_j)$ pour le d\'eterminant de Vandermonde.

\subsubsection*{Le mod\`ele \`a une matrice hermitienne}

L'ensemble $\mathcal{H}_N$ des matrices hermitiennes est un espace vectoriel de dimension$_{\mathbb{R}}$ $N^2$, que l'on munit de la mesure canonique $
\dd M = \prod_{i = 1}^N\dd M_{ii}\prod_{1 \leq i < j \leq N} \dd(\Re M_{ij})\,\dd(\Im M_{ij})$.
\beq
\label{eq:1mm} Z = \frac{1}{C}\int_{\mathcal{H}_N} \!\!\!\!\!\dd M\,e^{-\frac{N}{t}\,\Tr V(M)}
\eeq
$C$ est une constante de normalisation. Gr\^{a}ce \`a la d\'ecomposition polaire :
\bea
\mathcal{H}_N & \simeq &\frac{\mathbb{R}^N \times \mathrm{U}(N)}{\mathfrak{S}_N \times \big(\mathrm{U}(1)\big)^N} \nn \\
M & \mapsto & \big[(\lambda_1,\ldots,\lambda_N,U)\quad \mathrm{tel}\,\mathrm{que}\quad M = U\cdot\mathrm{diag}(\lambda_1,\ldots,\lambda_N)\cdot U^{\dagger}\big] \nn
\eea
on peut changer de variable et int\'egrer sur $U$. Le jacobien de ce changement de variable est $|\Delta(\lambda)|^2$ :
\beq
\label{eq:1mml}Z = \frac{1}{C}\,\frac{\mathrm{vol}\big(\mathrm{U}(N)\big)}{N!\,(2\pi)^N}\int_{\mathbb{R}^N} \dd\lambda_1\cdots\dd\lambda_N\,|\Delta(\lambda)|^2\,e^{-N\sum_{i = 1}^N V(\lambda_i)}
\eeq
Remarquons que les int\'egrales sur les \'el\'ements de matrice $M_{i,j}$ ne se factorisent dans \'{E}qn~\ref{eq:1mm} que si $V(x)$ est un polyn\^{o}me de degr\'e $2$. Autrement dit, les matrices $M$ distribu\'ee avec le poids de l'\'{E}qn.~\ref{eq:1mm} sont des matrices de Wigner ssi $V(x)$ est un polyn\^{o}me de degr\'e $2$.

\subsubsection*{Remarque : volume de $\mathrm{U}(N)$}

Une m\'ethode classique pour calculer le volume de $\mathrm{U}(N)$ (pour la mesure induite par la mesure canonique sur $\mathrm{GL}_N(\mathbb{C})$) est justement de sp\'ecialiser avec le potentiel quadratique $\frac{N}{t}V(x) = x^2$. Dans ce cas, \'{E}qn~\ref{eq:1mm} est un produit d'int\'egrales gaussiennes :
\beq
Z = \pi^{N^2/2} \nn
\eeq
tandis que l'int\'egrale sur les $\lambda_i$ peut \^{e}tre calcul\'ee avec la m\'ethode de Gaudin et Mehta \'evoqu\'ee \`a la partie~\ref{sec:Mehtapoly} (\'{E}qn.~\ref{eq:Mehta}). Les polyn\^{o}mes orthogonaux en jeu sont les polyn\^{o}mes de Hermite \mbox{$\pi_n(x) = e^{x^2}\Big(-\frac{\dd}{\dd x}\Big)^{n}e^{-x^2}$}, de norme $h_n = n!2^n$. D'o\`u :
\beq
\mathrm{vol}\big(\mathrm{U}(N)\big) = \frac{2^{-N(N - 3)/2}\,\pi^{N(N + 1)/2}}{\prod_{n = 0}^{N - 1} n!} \nn
\eeq

\subsubsection*{Le mod\`ele \`a une matrice normale}

 Une matrice carr\'ee $(M_{i,j})_{1 \leq i,j \leq N}$ de taille $N \times N$ est normale ssi elle est diagonalisable dans une base orthonorm\'ee. On veut \'etendre la d\'efinition \'{E}qn.~\ref{eq:1mml} en pla\c{c}ant les valeurs propres sur des chemins (ferm\'es ou ouverts) $\gamma_1,\ldots,\gamma_r$ dans $\widehat{\mathbb{C}}$. On d\'efinit donc des fractions de remplissage $\epsilon_i = n_i/N$ ($n_i \in \mathbb{N}$)  telles que $\sum_{i = 1}^r \epsilon_i = 1$, et on appelle $\mathcal{H}(\gamma_1^{n_1}\times\cdots\times\gamma_r^{n_r})$ l'ensemble des matrices normales ayant $n_i$ valeurs propres sur $\gamma_i$. On se donne aussi un \textbf{potentiel} $V\,:\,\cup_{i} \gamma_i \rightarrow \widehat{\mathbb{C}}$.
    \beq
   \label{eq:1mmn} Z(\epsilon_1,\ldots,\epsilon_r) = \frac{N!}{\prod_{i = 1}^r (N\epsilon_i)!} \int_{\gamma_1^{N\epsilon_1}\times \cdots\times\gamma_r^{N\epsilon_r}}\!\!\!\!\!\!\!\!\!\!\!\!\!\!\!\!\!\!\!\!\! \mathrm{d}\lambda_1\cdots\mathrm{d}\lambda_N\,|\Delta(\lambda)|^2\, e^{-\frac{N}{t}\Big(\sum_{i = 1}^N V(\lambda_i)\Big)}
    \eeq
     Par rapport \`a l'\'{E}qn.~\ref{eq:1mml}, on a fix\'e le pr\'efacteur. Avec cette notation, $\mathcal{H}(\mathbb{R}^N)$ est l'ensemble $\mathcal{H}_N$ des matrices hermitiennes. Par la suite, on note g\'en\'eriquement $\Gamma$, un produit cart\'esien de contours du plan complexe, et on appelle $\mathrm{Ombre}(\Gamma)$ la r\'eunion de ces contours. Cette g\'en\'eralisation du mod\`ele hermitien intervient naturellement quand on doit consid\'erer l'asymptotique des int\'egrales de matrices dans le r\'egime \`a plusieurs coupures.

\subsubsection*{Les mod\`eles $\beta$ \`a une matrice}

C'est la g\'en\'eralisation de l'int\'egrale \'{E}qn.~\ref{eq:1mmn} sur les valeurs propres, en prenant un d\'eterminant de Vandermonde \`a une puissance $2\beta$ :
\beq
\label{eq:1mmbeta} Z = \int_{\Gamma} \dd\lambda_1\cdots\dd\lambda_N\,|\Delta(\lambda)|^{2\beta}\,e^{-\frac{N\beta}{t}\Big(\sum_{i = 1}^N V(\lambda_i)\Big)}
\eeq
Quand $\beta = 1$, c'est l'int\'egrale \'{E}qn.~\ref{eq:1mmn} sur un espace de matrices normales. Les cas $\beta = 1/2$ et $2$ sont atteints par des int\'egrales du type \'{E}qn.~\ref{eq:1mm}, mais plut\^{o}t sur un espace de matrices diagonalis\'ees par des transformations orthogonales ($\beta = 1/2$), ou symplectiques unitaires ($\beta = 2)$ (cf. aussi Fig.~\ref{Dysonens}). Il n'y a pas d'int\'egrales sur un ensemble de matrices connu qui reproduisent, apr\`es passage aux valeurs propres, l'int\'egrale \'{E}qn.~\ref{eq:1mmbeta} en toute g\'en\'eralit\'e. Le seul r\'esultat en ce sens, d\^{u} \`a Dumitriu et Edelman \cite{DE02} et g\'en\'eralis\'e r\'ecemment, est un mod\`ele de matrices tridiagonales qui reproduit \'{E}qn.~\ref{eq:1mmbeta} pour les potentiels polynomiaux pairs. Il y a bien des mani\`eres de d\'eformer le mod\`ele \`a une matrice, dont les exemples donn\'es au chapitre \ref{chap:formel}. Historiquement, les mod\`eles $\beta$ ont retenu l'attention \`a cause de leur lien avec le hamiltonien de Calogero-Sutherland et les polyn\^{o}mes de Jack, et plus g\'en\'eralement leur relation suppos\'ee aux th\'eories conformes d\'ecrivant des effets Hall quantiques fractionnaires : la mesure d'int\'egration dans l'\'{E}qn.~\ref{eq:1mmbeta} est la fonction d'onde \`a $N$ corps propos\'ee par R. Laughlin \cite{BobLaugh} pour rendre compte de particules \`a statistique fractionnaire (\`a un op\'erateur de vertex $e^{-\frac{N\beta}{t} V(\lambda_i)}$ pr\`es).

\subsubsection*{Remarque : convergence des int\'egrales}
\label{concons}
Dans les chapitres \ref{chap:int} et \ref{chap:conv}, on s'int\'eressera \`a des \textbf{int\'egrales de matrices convergentes}, en particulier dans la limite o\`u $N$ est grand. $\Re \beta \geq 0$ est une premi\`ere condition, et d'autres conditions portent sur $V$ et $\Gamma$. Par exemple, si un des $\gamma_i$ va \`a l'infini, $\Re\:V$ doit cro\^{i}tre assez rapidement \`a l'infini le long de $\gamma_i$. Une condition suffisante est l'existence de $\alpha > 1$, tel que $|x|^{2\beta(N - 1)}e^{-\frac{N\beta}{t}\Re V(x)} \in O(|x|^{-\alpha})$. Comme on s'int\'eressera au comportement de $N$ grand de l'\'{E}qn.~\ref{eq:1mmn}, il est suffisant de supposer :
    \beq
    \liminf_{x \mathop{\rightarrow}_{\gamma_i} \infty} \frac{\Re V(x)}{2t\ln|x|} \geq 1 \nn
    \eeq
On rencontrera dans les chapitres~\ref{chap:formel} et \ref{chap:cordes} des \textbf{int\'egrales de matrices formelles}, dont on donnera une d\'efinition pr\'ecise. Elle est analogue \`a la d\'efinition des th\'eories des champs "perturbatives" donn\'ee dans l'introduction (partie~\ref{sec:BIPZ}), et les conditions sur $V$ sont de nature diff\'erentes.

\subsubsection*{Remarque : les int\'egrales de Selberg}
\label{Seljk}
Pour des potentiels tr\`es particuliers, il existe des formules closes dues \`a Selberg\footnote{Elles sont rassembl\'ees par exemple au Chapitre 17 de \cite{Mehtabook}, et leurs implications ont fait l'objet d'une revue \cite{FoSel}.} \cite{Sel} pour la fonction de partition des mod\`eles $\Re\,\beta > 0$. Notamment :

\vspace{0.2cm}

\noindent $\diamond\,$ Pour l'ensemble de Laguerre version $\beta$, i.e. $V_{\mathrm{L}}(x) = x$ :
{\small \bea
\int_{\mathbb{R}_+^N} \prod_{i = 1}^N \dd\lambda_i\,e^{-\frac{N\beta}{t}\,\lambda_i}\,\prod_{1 \leq i < j \leq N} |\lambda_i - \lambda_j|^{2\beta} & = & \Big(\frac{t}{\beta N}\Big)^{\beta N^2 - (\beta - 1)N}\,\frac{\prod_{n = 1}^N \big(\Gamma(1 + n\beta)\big)^2}{\Gamma(1 + \beta N)\big(\Gamma(1 + \beta)\big)^N} \nn \\
\label{eq:SelL} &&
\eea}

\vspace{-0.2cm}

\noindent $\diamond\,$ Pour l'ensemble gaussien version $\beta$, i.e. $V_{\mathrm{G}}(x) = x^2/2$ :
{\small \bea
\int_{\mathbb{R}^N} \prod_{i = 1}^N \dd\lambda_i\,e^{-\frac{N\beta}{t}\,\frac{\lambda_i^2}{2}}\,\prod_{1 \leq i < j \leq N} |\lambda_i - \lambda_j|^{2\beta} = (2\pi)^{N/2}\Big(\frac{t}{N\beta}\Big)^{\frac{\beta N^2}{2} - \frac{(1 - \beta)N}{2}}\,\frac{\prod_{n = 1}^N \Gamma(1 + n\beta)}{\big(\Gamma(1 + \beta)\big)^N} \nn \\
\label{eq:SelG} &&
\eea}

\subsubsection*{Le mod\`ele \`a une matrice avec champ ext\'erieur}

Dans ce mod\`ele, on souhaite appliquer un potentiel diff\'erent \`a chaque $\lambda_i$. Ici, on ne s'int\'eressera qu'\`a une mani\`ere minimale de le faire : on se donne une matrice diagonale $\mathbf{R}$ (le \textbf{champ ext\'erieur}), et on d\'efinit
\beq
\label{eq:1mmchpext}    Z = \frac{1}{C}\int_{\mathcal{H}(\Gamma)} \!\!\!\!\!\!\dd M\,e^{-\frac{N}{t}\Tr\big(V(M) - M\,\mathbf{R}\big)} \nn
\eeq
C'est une version "formelle" de ce mod\`ele que l'on utilisera pour calculer les nombres de Hurwitz au chapitre \ref{chap:cordes}.

\subsubsection*{L'int\'egrale d'Harish-Chandra}

Soit $\mathbf{X} = \mathrm{diag}(x_1,\ldots,x_N)$ et $\mathbf{Y} = \mathrm{diag}(y_1,\ldots,y_N)$. On munit $\mathrm{U}(N)$ de la mesure de Haar $\mathrm{d}U/\mathrm{vol}\big(\mathrm{U}(N)\big)$. Harish-Chandra \cite{HarishChandra}, puis Itzykson et Zuber \cite{IZint} ont calcul\'e l'int\'egrale :
\beq
\label{eq:HarishChandra} I(\mathbf{X},\mathbf{Y}) \equiv \frac{1}{\mathrm{vol}\big(\mathrm{U}(N)\big)}\int_{\mathrm{U}(N)}\!\!\!\!\!\!\! \dd U\,e^{\Tr \mathbf{X}U\mathbf{Y}U^{\dagger}} = \frac{\det\big(e^{x_iy_j}\big)}{\Delta(\mathbf{X})\,\Delta(\mathbf{Y})}
\eeq

\subsubsection*{D\'eterminant de Cauchy}
\label{detCa}
Le d\'eterminant de Cauchy de $\mathbf{X}$ et $\mathbf{Y}$ est d\'efini par :
\beq
\Delta(\mathbf{X},\mathbf{Y}) = \mathrm{det}_{1 \leq j,k \leq N} \Big(\frac{1}{x_j + y_k}\Big) = \frac{\Delta(\mathbf{X})\Delta(\mathbf{Y})}{\prod_{j,k = 1}^N (x_j + y_k)} \nn
\eeq
Un calcul \'el\'ementaire \`a partir de l'int\'egrale de Harish-Chandra permet de le repr\'esenter par une int\'egrale de matrice :
\bea
\Delta(\mathbf{X},\mathbf{Y}) & = & \frac{(2\pi)^N}{\big(\mathrm{vol}\,\mathrm{U}(N)\big)^3} \int_{\mathrm{U}(N)}\!\!\!\!\!\!\!\dd U\int_{\mathcal{H}(\mathbb{R}_+^N)}\!\!\!\!\!\!\dd M \int_{\mathrm{U}(N)} \!\!\!\!\!\!\!\dd V\,e^{-\Tr(M U\mathbf{X}U^{\dagger} + MV\mathbf{Y}V^{\dagger})} \nn \\
\label{eq:Cauint} &&
\eea

\subsubsection*{La chaine de matrices}

C'est un mod\`ele \`a plusieurs matrices de m\^{e}me taille $N \times N$, coupl\'ees en chaine avec une interaction lin\'eaire.
    \bea
    Z_{\textrm{chaine}} & = & \frac{1}{C}\int_{\mathcal{H}(\Gamma_1)}\!\!\!\!\!\!\!\!\!\dd M_1 \cdots \int_{\mathcal{H}(\Gamma_P)}\!\!\!\!\!\!\!\!\! \dd M_P\,e^{-\frac{N}{t}\Tr\Big(\sum_{p = 1}^P V_p(M_p) + \sum_{p = 1}^{P - 1} c_{p,p + 1}\,M_p\,M_{p + 1}\Big)} \nn \\
    \label{sec:chain}&&
    \eea
Gr\^{a}ce \`a la formule d'Harish-Chandra, on peut diagonaliser les matrices $M_p = U_p\cdot\mathrm{diag}(\lambda_{1}^{(p)},\ldots,\lambda_{N}^{(p)})\cdot U_p^{\dagger}$, et int\'egrer sur les variables angulaires $U_p \in \mathrm{U}(N)$.
\bea
Z_{\mathrm{chaine}} & = & \int_{\Gamma_1\times\cdots\times\Gamma_P} \Big[\prod_{p = 1}^P \prod_{i = 1}^N \dd \lambda_i^{(p)}\Big]\,|\Delta(\lambda^{(1)})|\,|\Delta(\lambda^{(P)})| \nn \\
&& \phantom{\int_{\Gamma_1}}\,\prod_{p = 1}^{P - 1} \mathop{\det}_{1\leq k,l \leq N}\big(e^{-\frac{N}{t}\,c_{p,p + 1} \lambda_k^{(p)}\,\lambda_{l}^{(p + 1)}}\big)\,\prod_{p = 1}^P\prod_{i = 1}^N e^{-\frac{N}{t}\,V_p(\lambda_i^{(p)})} \nn
\eea
\begin{figure}[h!]
\begin{center}
\includegraphics[width = 0.6\textwidth]{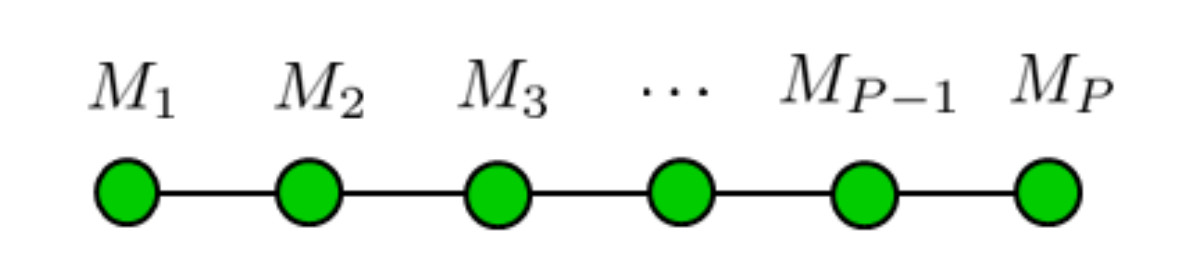}
\end{center}
\end{figure}
Lorsque $\Gamma_p = \gamma_p^N$ (les valeurs propres ne vivent que sur un contour), et lorsque les $V_p$ sont des polyn\^{o}mes, $Z_{\mathrm{chaine}}$ et les fonctions de corr\'elations de ce mod\`ele peuvent \^{e}tre \'etudi\'es en introduisant deux bases de polyn\^{o}mes duales pour le produit scalaire :
\bea
&& \big(f|g\big) = \int_{\gamma_1 \times \cdots \times \gamma^P}\!\!\!\!\!\!\!\!\!\!\!\!\!\!\!\!\dd x^{(1)}\cdots \dd x^{(P)}\,f(x^{(1)})\,g(x^{(P)})\Big[\prod_{p = 1}^P e^{-\frac{N}{t}V_p(x^{(p)})}\Big]\Big[\prod_{p = 1}^{P - 1} e^{-\frac{N}{t}\,c_{p,p + 1}\,x^{(p)}\,x^{(p + 1)}}\Big] \nn \\
\label{eq:poyh} &&
\eea
La limite des grandes chaines ($P \rightarrow \infty$) peut \^{e}tre vue, en normalisant correctement les couplages, comme un mod\`ele \`a une matrice $M(p)$ d\'ependant d'un temps $p$ continu, avec une dynamique dict\'ee par un terme cin\'etique quadratique $\Big(\frac{\dd M}{\dd p}\Big)^2$.

 Les chaines de matrices n'interviendront pas directement dans cette th\`ese. Je les d\'efinis car ce sont actuellement les mod\`eles les plus g\'en\'eraux\footnote{On peut aussi ajouter comme dans \'{E}qn.~\ref{eq:1mmchpext} un champ ext\'erieur \`a une des extr\'emit\'es.} \`a plusieurs matrices o\`u l'on sait obtenir des \'equations de boucles ferm\'ees, et les r\'esoudre. Par ailleurs, on voit que pour des mod\`eles avec couplages entre matrices plus compliqu\'es, l'astuce d'Harish-Chandra ne permet pas d'int\'egrer sur les variables angulaires pour ne garder que les valeurs propres. Du point de vue probabiliste, on peut penser que la chaine de matrices est un mod\`ele particulier. Cependant, ils sont singularis\'es du point de vue de l'int\'egrabilit\'e, et interviennent souvent en physique statistique ou en th\'eorie des cordes (probl\`emes de partitions planes al\'eatoires). On peut aussi garder \`a l'esprit, pour des travaux futurs, qu'\'etudier l'asymptotique des chaines de matrices, revient \`a \'etudier l'asymptotique des polyn\^{o}mes biorthogonaux.

\subsubsection{Les mod\`eles de matrices g\'en\'eralis\'es}

Nous appellerons \textbf{int\'egrales de matrices g\'en\'eralis\'ees}, les mod\`eles du type :
\beq
\label{eq:intqh} Z = \int_{\Gamma} \Big[\prod_{i = 1}^N \dd \lambda_i\,e^{-\frac{N}{t}V(\lambda_i)}\Big]\,\prod_{1 \leq i < j \leq N} K_{\mathrm{sing}}(\lambda_i,\lambda_j)\,\prod_{i,j = 1}^N K_{\mathrm{reg}}(\lambda_i,\lambda_j)
\eeq
et leurs extensions \`a plusieurs matrices. L'interaction entre deux "valeurs propres" $\lambda_i$ et $\lambda_j$ n'est plus d\'ecrite par un d\'eterminant de Vandermonde, mais une fonction \`a deux points arbitraire. Ici, on a s\'epar\'e artificiellement un facteur $K_{\mathrm{sing}}(\lambda_i,\lambda_j)$ qui s'annule \`a points co\"{i}ncidants, d'un facteur r\'egulier $K_{\mathrm{reg}}$. Le mod\`ele $\beta$ \`a une matrice $K_{\mathrm{sing}}(\lambda,\mu) = |\lambda - \mu|^{2\beta}$, et \label{Ontri}le mod\`ele $\On$ trivalent $K_{\mathrm{reg}}(\lambda,\mu) = (\lambda + \mu)^{-\n/2}$ en sont deux exemples. Comme les mod\`eles de matrices, cette g\'en\'eralisation a de nombreuses applications en physique th\'eorique, dont quelques-unes sont \'evoqu\'ees dans cette th\`ese.

\section{Description des observables}
\label{sec:correlateurs}
\subsubsection*{R\'esolvantes et corr\'elateurs}

Dans cette th\`ese, nous nous int\'eresserons le plus souvent aux valeurs moyennes d'observables d'un mod\`ele \`a une matrice, qui sont sym\'etriques par changement de base (i.e. par permutation des valeurs propres) : elles s'\'ecrivent $\big\langle\Tr \mathcal{O}_1(M)\cdots \Tr \mathcal{O}_n(M)\big\rangle$. Pour cette raison, on les appelle parfois \textbf{observables bosoniques}. Les \textbf{r\'esolvantes} $\mathcal{O}_x(M) = \frac{1}{x - M}$ engendrent une base de ces observables bosoniques ($x \in \mathbb{C}\setminus\mathrm{Ombre}(\Gamma)$). L'exp\'erience montre que c'est la plus commode du point de vue de la g\'eom\'etrie complexe. On d\'efinit alors les \textbf{corr\'elateurs non connexes} :
\beq
\label{eq:nonon1}\overline{W}_n(x_1,\ldots,x_n) \equiv \Big\langle \prod_{j = 1}^n \Tr \frac{1}{x_j - M} \Big\rangle = \Big\langle \prod_{j = 1}^{n} \Big(\sum_{i_j = 1}^N \frac{1}{x_j - \lambda_{i_j}}\Big) \Big\rangle
\eeq
et les \textbf{corr\'elateurs connexes} :
\bea
W_n(x_1,\ldots,x_n) & \equiv & \partial_{t_1}\cdots\partial_{t_n}\ln\Big\langle e^{\sum_{j = 1}^n t_j\,\mathcal{O}_{x_j}(M)}\Big\rangle\Big|_{t_1 = \ldots = t_n = 0} \nn \\
& \equiv & \Big\langle \prod_{j = 1}^n \Tr \frac{1}{x_j - M} \Big\rangle_c \nn
\eea
La notation $\langle\cdots\rangle_c$ rappelle que $W_n$ est d\'efini \`a partir des cumulants\label{cum}. Les corr\'elateurs connexes ou pas \'etant sym\'etriques dans les $x_i$, on utilisera la notation $W_n(x_I)$ pour d\'esigner $W_n(\{x_i,\,\,i\in I\})$. La relation entre $\overline{W}_n$ et $W_n$ est \`a l'origine du vocabulaire non connexe/connexe :
\bea
\overline{W}_n(x_I) & = & \sum_{m = 1}^{|I|}\,\, \sum_{\substack{\{J_1,\ldots,J_m\},\,J_l \neq \emptyset \\ \mathrm{partition}\,\mathrm{de}\,I}}\, \,\,\prod_{l = 1}^m W_{|J_l|}(x_{J_l}) \nn \\
W_n(x_I) & = &   \sum_{m = 1}^{|I|} \frac{(-1)^{m + 1}}{m}\sum_{\substack{\{J_1,\ldots,J_m\},\,J_l \neq \emptyset \\ \mathrm{partition}\,\mathrm{de}\,I}} \,\,\,\prod_{l = 1}^m W_{|J_l|}(x_{J_l}) \nn
\eea
Par exemple, pour $n = 2$ et $n = 3$ :
\bea
\overline{W}_2(x_1,x_2) & = & W_2(x_1,x_2) + W_1(x_1)W_1(x_2) \nn \\
\overline{W}_2(x_1,x_2,x_3) & = & W_3(x_1,x_2,x_3) + W_1(x_1)W_1(x_2)W_1(x_3)  \nn \\
& & + W_2(x_1,x_2)W_1(x_3)  + W_2(x_1,x_3)W_1(x_2) + W_2(x_2,x_3)W_1(x_3) \nn
\eea
Par la suite, corr\'elateur signifiera en l'absence de pr\'ecision "corr\'elateur connexe". Les observables bosoniques pour une matrice $M$ peuvent \^{e}tre engendr\'ees en perturbant le potentiel $V$ dans les int\'egrales de matrices. Si l'on explicite cette d\'ependance dans le potentiel :
\bea
\overline{W}_n[V](x_1,\ldots,x_n) & = & \frac{1}{Z[V]}\,\partial_{\varepsilon_1}\cdots\partial_{\varepsilon_n}\,Z\Big[V - \frac{t}{\beta N}\sum_{j = 1}^n \frac{\varepsilon_j}{x_j - \bullet}\Big]\Big|_{\varepsilon_1,\ldots,\varepsilon_n = 0} \nn \\
\label{eq:Wnnconnexe} & & \\
W_n[V](x_1,\ldots,x_n) & = & \partial_{\varepsilon_1}\cdots\partial_{\varepsilon_n}\ln Z\Big[V - \frac{t}{\beta N} \sum_{j = 1}^n \frac{\varepsilon_j}{x_j - \bullet}\Big]\Big|_{\varepsilon_1 = \cdots = \varepsilon_n = 0} \nn \\
\label{eq:Wnconnexe} & &
\eea
La d\'eriv\'ee $\partial_{\varepsilon = 0}$ le long de la d\'eformation $V_{\varepsilon}(\xi) = V(\xi) - \frac{t}{N\beta}\frac{\varepsilon}{x - \xi}$ est appel\'ee \textbf{op\'erateur d'insertion}, ou parfois op\'erateur de vertex infinit\'esimal.

\subsubsection*{Ratios de polyn\^{o}mes caract\'eristiques}

Du point de vue des syst\`emes int\'egrables, ce sont les \textbf{observables fermioniques} qui ont \'et\'e introduites \`a l'origine, notamment les valeurs moyennes de ratios des polyn\^{o}mes caract\'eristiques :
\beq
K_{n;m}\big(y_1,\ldots,y_n;x_1,\ldots,x_m\big) = \Big\langle \frac{\prod_{l = 1}^n \det(y_l - M)}{\prod_{j = 1}^m \det(x_j - M)}\Big\rangle \nn
\eeq
Les $K_{n;m}$ sont engendr\'es par l'insertion d'op\'erateurs de vertex finis. Par cons\'equent, on peut ais\'ement \'ecrire une formule exponentielle exprimant les $K_{n;m}$ en fonction de $W_n$ ou $\overline{W}_n$, et un op\'erateur de d\'erivation \`a points co\"{i}ncidant pour faire l'op\'eration inverse. Par exemple :
\bea
K_{1;1}(x,y) & = & \Big\langle \exp\Big(\sum_{k = 1}^{\infty} \int_{y}^x \!\!\!\!\dd\xi_1\cdots\int_y^x \dd\xi_l \prod_{k' = 1}^k \Tr \frac{1}{\xi_{k'} - M}\Big)\Big\rangle \nn \\
\overline{W}_n(x_1,\ldots,x_n) & = & \partial_{y_1}\cdots\partial_{y_n} K_{n;n}(y_1,\ldots,y_n;x_1,\ldots,x_n) \Big|_{y_i = x_i} \nn
\eea

\newpage
\thispagestyle{empty}
\phantom{bbk}

\newpage
\thispagestyle{empty}
\phantom{bbk}

\newpage

\chapter{La r\'ecurrence topologique}
\label{chap:toporec}
\thispagestyle{plain}
\vspace{-1.5cm}

\noindent \rule{\textwidth}{1.5mm}
\addtolength{\baselineskip}{0.20\baselineskip}

\vspace{2.5cm}
{\textsf{ Je commence par une discussion \'el\'ementaire sur les \'equations de Schwinger-Dyson, afin de montrer les \'equations de boucles satisfaites par les mod\`eles de matrices, et de justifier leur \'etude dans un cadre plus g\'en\'eral. Puis, je pr\'esente l'axiomatique de la r\'ecurrence topologique (RT), d\'evelopp\'ee par Eynard et Orantin : \`a partir d'une courbe spectrale (notion qui sera d\'efinie), RT exhibe la solution g\'en\'erale d'une hi\'erarchie d'\'equations de boucles. Je rappelle ses propri\'et\'es utiles pour la suite de la th\`ese, en particulier son application aux chaines de matrices hermitiennes. Je discute aussi des extensions (connues ou encore \`a explorer) de RT.
Le contenu de ce chapitre n'\'etant pas nouveau, je donnerai peu de d\'emonstrations d\'etaill\'ees, et renvoie \`a \cite{EORev} pour une revue plus compl\`ete datant de 2008 et des r\'ef\'erences. Le lecteur qui n'est pas familier avec la g\'eom\'etrie complexe sur les surfaces de Riemann pourra consulter l'Annexe~\ref{app:geomcx} en pr\'eliminaire.}}

\newpage

\addtolength{\baselineskip}{-0.20\baselineskip}

\section{Les \'equations de Schwinger-Dyson}
\label{sec:eqboucl1}
\subsection{Principes}

Reprenons les notations de la partie~\ref{sec:BIPZ}. Imaginons une th\'eorie des champs (ou une loi de probabilit\'e), avec une fonction de partition $Z = \int_E \mathcal{D}\phi\,e^{-S[\phi]}$. Imaginons aussi $E$ et $\mathcal{D}\phi$ construits de sorte que les th\'eor\`emes habituels sur l'invariance de l'int\'egration par changement de variable s'appliquent, et que l'on dispose d'une famille diff\'erentiable de changements de variable $\phi \mapsto \Upsilon_{\varepsilon}(\phi)$, qui envoie $E$ sur $E_{\varepsilon}$ :
\beq
\int_{E_\varepsilon} \mathcal{D}\Upsilon_{\varepsilon}(\phi)\,e^{-S[\Upsilon_{\varepsilon}(\phi)]} = \int_E \mathcal{D}\phi\,e^{-S[\phi]} \nn
\eeq
Ce qui donne, pour des changements de variables infinit\'esimaux, une \'equation \`a trois termes :
\beq
\label{eq:exSD}\big(\partial_{\varepsilon} \ln Z_{\varepsilon}\big)\big|_{\varepsilon = 0} + \big\langle \mathcal{J}[\phi] \big\rangle - \big\langle (\delta S)[\phi]\big\rangle = 0
\eeq
Le premier vient de la variation du domaine d'int\'egration, on a not\'e $Z_{\varepsilon} = \int_{\mathcal{E}_{\varepsilon}} \mathcal{D}\phi\,e^{-S[\phi]}$. Le second est le jacobien infinit\'esimal du changement de variable :
\beq
\mathcal{D}\Upsilon_{\varepsilon}(\phi) = \mathcal{D}\phi\cdot\big(1 + \varepsilon\,\mathcal{J}[\phi] + o(\varepsilon)\big) \nn
\eeq
Le troisi\`eme est la variation du poids d'int\'egration :
\beq
S[\Upsilon_\varepsilon(\phi)] = S[\phi] + \varepsilon(\delta S)[\phi] + o(\varepsilon) \nn
\eeq
L'\'{E}qn.~\ref{eq:exSD} donne une relation entre valeurs moyennes de certaines observables. On appelle \textbf{\'equation de Schwinger-Dyson}, toute relation d\'ecoulant du principe d'invariance par changement de variable (ou d'une int\'egration par partie, ce qui est \'equivalent). On peut \'ecrire une \'equation de Schwinger-Dyson pour toutes les familles de changements de variable possibles dans $E$ (elles sont param\'etr\'ees par l'espace tangent de $E$, si tant est que cela ait un sens). On peut appliquer le m\^{e}me principe pour les int\'egrales donnant les valeurs moyennes $\int_{E}\mathcal{D}\phi\,\mathcal{O}[\phi]\,e^{-S[\phi]}$. Apr\`es diverses manipulations, on peut esp\'erer obtenir des \'equations ferm\'ees pour certaines observables $\mathcal{O}_k$, et peut-\^{e}tre ainsi calculer $\big\langle\mathcal{O}_k\big\rangle$. De plus, si $S = \sum_{k} t_k\,\mathcal{O}_k$, on peut remonter \`a $Z$ modulo un facteur constant en int\'egrant $\partial_{t_k}\ln Z = -\big\langle \mathcal{O}_k \big\rangle$ (comme en probabilit\'es, lorsque l'on connait bien les moments d'une loi, on peut remonter \`a la loi).

\subsection{Exemples}
\label{sec:beta1}
Nous allons illustrer la d\'erivation des \'equations de Schwinger-Dyson. Pour bien montrer que la m\'ethode est \'el\'ementaire, nous allons d'abord le faire sur une int\'egrale unidimensionnelle. Nous passerons ensuite aux \'equations de Schwinger-Dyson du mod\`ele $\beta$ \`a une matrice pour la fonction de partition $Z$ et toutes les observables $W_n$. Enfin, je pr\'esenterai l'\'equation de Schwinger-Dyson pour le mod\`ele \`a une matrice hermitienne en champ externe. Ces deux cas nous seront utiles par la suite.

\subsubsection{Exemple trivial}

On se propose de calculer les moments pour la mesure de probabilit\'e $\dd \nu = \frac{1}{Z}\,\dd x\,e^{-\alpha x^2/2}$ sur $\mathbb{R}$ ($\Re\,\alpha > 0$), en suivant la strat\'egie que l'on vient d'\'evoquer. Pour des questions de convergence, on restreint d'abord $\nu$ \`a $[-m,m]$ : $\dd\nu_m = \frac{1}{Z_m}\mathbf{1}_{[-m,m]} \dd \nu$. Toute fonction $h\,:\,\mathbb{R} \rightarrow \mathbb{R}$ qui est $\mathcal{C}^1$ d\'efinit une famille diff\'erentiable de changements de variable $\Upsilon_{\varepsilon}(x) = x + \varepsilon\,h(x)$, pourvu que $\varepsilon$ soit assez petit. L'\'equation de Schwinger-Dyson s'\'ecrit :
\beq
\big(h(m) - h(-m)\big)\frac{e^{-\alpha m^2/2}}{Z_m} + \big\langle \big(h'(x) - \alpha h(x)\big)\rangle_{m} = 0 \nn
\eeq
Pour $h$ \`a croissance mod\'er\'ee \`a l'infini, le terme de bord disparait quand $m \rightarrow \infty$, et l'on trouve :
\beq
\big\langle h'(x) - \alpha x h(x)\big\rangle = 0 \nn
\eeq
En prenant $h(x) = x^p$ pour $p \in \mathbb{N}$, on trouve par r\'ecurrence :
\beq
\forall p \in \mathbb{N},\qquad \langle x^{2p}\rangle = \alpha^{-p}\prod_{k = 1}^{p} (2k - 1)\,,\qquad \langle x^{2p + 1} \rangle = 0 \nn
\eeq
Enfin, $\partial_{\alpha} \ln Z = -\frac{\langle x^2 \rangle}{2} = - \frac{1}{2\alpha}$. Il existe donc une constante $C$ telle que :
\beq
Z = \frac{C}{\sqrt{\alpha}} \nn
\eeq
$C = \sqrt{\pi}$ doit \^{e}tre calcul\'e par d'autres moyens.

\subsubsection{Mod\`ele $\beta$ \`a une matrice}

Le mod\`ele $\beta$ \`a une matrice est d\'efini en \'{E}qn.~\ref{eq:1mmbeta} :
\beq
\label{eq:Zbeta2} Z_N = \int_{\Gamma}\prod_{i = 1}^N \dd \lambda_i\,|\Delta(\lambda)|^{2\beta}\,e^{-\frac{N\beta}{t}\sum_{i = 1}^N V(\lambda_i)}
\eeq
Etudions les \'equations de Schwinger-Dyson g\'en\'er\'ees par les changements de variable de la forme :
\beq
\Upsilon_{\varepsilon}(\lambda_i) = \lambda_i + \varepsilon\,h(\lambda_i) \nn
\eeq
Si l'on choisit $h$ de sorte qu'il n'y ait pas de termes de bord, on obtient :
{\small \beq\label{eq:laier}
 \Big\langle\sum_{i = 1}^N h'(\lambda_i) + \beta \sum_{1 \leq i \neq j \leq N} \frac{h(\lambda_i) - h(\lambda_j)}{\lambda_i - \lambda_j} - \frac{N}{t}\sum_{i = 1}^N V'(\lambda_i)\,h(\lambda_i)\Big\rangle = 0
 \eeq}
$\!\!\!$Cette \'equation est lin\'eaire en $h$, donc on peut en principe choisir une base quelconque de $h$, comme la base des polyn\^{o}mes $\big(h(\lambda) = \lambda^p\big)_{p \in \mathbb{N}}$, puis en faire des combinaisons lin\'eaires pour tenter de fermer les \'equations. On ne reproduira pas ces t\^{a}tonnements, et on choisit $h_i(\lambda) = \frac{1}{x - \lambda}$ qui donne efficacement la r\'eponse. L'\'{E}qn.~\ref{eq:laier} s'\'ecrit :
\beq
\Big\langle (1 - \beta)\sum_{i = 1}^N \frac{1}{(x - \lambda_i)^2} + \beta \Big(\sum_{i = 1}^N \frac{1}{x - \lambda_i}\Big)^2 + \frac{N\beta}{t}\sum_{i = 1}^N \Big(-\frac{V'(x)}{x - \lambda_i} + \frac{V'(x) - V'(\lambda_i)}{x - \lambda_i}\Big)\Big\rangle = 0 \nn
\eeq
ou en termes de corr\'elateurs (cf. partie~\ref{sec:correlateurs}) :
\bea
-(1 - \beta)W_1'(x) + \beta\,W_2(x,x) + \beta\big(W_1(x)\big)^2  && \nn \\
\label{eq:masterloop} + \frac{N\beta}{t}\big(-V'(x)W_1(x) + P_1(x)\big) & = & 0
\eea
avec :
\beq
P_1(x) = \Big\langle \sum_{i = 1}^N \frac{V'(x) - V'(\lambda_i)}{x - \lambda_i}\Big\rangle \nn
\eeq
On peut aussi \'ecrire des \'equations de Schwinger-Dyson \`a partir de l'int\'egrale qui repr\'esente le corr\'elateur \`a $n - 1$ points. En fait, elles se d\'eduisent de l'\'{E}qn.~\ref{eq:masterloop} par des perturbations infinit\'esimales du potentiel (cf. \'{E}qn.~\ref{eq:Wnconnexe}) $V(\lambda) \rightarrow V(\lambda) - \frac{t}{\beta N} \sum_{j = 2}^{n} \frac{\epsilon_j}{x_j -  \lambda}$. Il en r\'esulte :
\bea
\label{eq:masterloopn} -(1 - \beta)\frac{\dd}{\dd x}\big(W_{n}(x,x_I)\big) + \beta\,W_{n + 1}(x,x,x_I) && \\
\beta \sum_{J \subseteq I} W_{|J| + 1}(x,x_J)\,W_{n - |J|}(x,x_{I\setminus J}) - \frac{N\beta}{t}\,V'(x)\,W_n(x,x_I)  && \nn \\
+ \sum_{i = 2}^{n} \frac{\dd}{\dd x_i}\Big(\frac{W_{n - 1}(x,x_{I\setminus\{i\}}) - W_{n - 1}(x_I)}{x - x_i}\Big) + \frac{N\beta}{t}\,P_n(x;x_I) & = & 0 \nn
\eea
o\`u l'on a introduit :
\beq
P_n(x;x_2,\ldots,x_n) = \Big\langle \Big(\sum_{i = 1}^N\frac{V'(x) - V'(\lambda_i)}{x - \lambda_i}\Big) \prod_{j = 2}^n\Big( \sum_{i_j = 1}^{N} \frac{1}{x_j - \lambda_{i_j}}\Big)\Big\rangle_c \nn
\eeq

Ces \'equations sont quadratiques, \`a cause du terme d'interaction entre deux valeurs propres (le d\'eterminant de Vandermonde) dans l'\'{E}qn.~\ref{eq:Zbeta2}.

\subsubsection{Mod\`ele \`a une matrice hermitienne en champ ext\'erieur}

Le\label{chuk} mod\`ele est d\'efini en \'{E}qn.~\ref{eq:1mmchpext} :
\beq
Z = \int_{\mathcal{H}_N(\Gamma)} \!\!\!\!\!\!\!\!\dd M\,e^{-\frac{N}{t}\Tr\big(V(M) - M \mathbf{R}\big)} \nn
\eeq
\'{E}crivons l'\'equation de Schwinger-Dyson pour $Z$, en utilisant le changement de variable $\Upsilon_{\varepsilon}(M) = M + \varepsilon\,\frac{1}{x - M}\frac{1}{\zeta - \mathbf{R}}$ :
{\small \bea
\Big\langle\Tr\frac{1}{x - M}\Tr\Big(\frac{1}{x - M}\frac{1}{\zeta - \mathbf{R}}\Big) - \frac{N}{t}\,\Tr\Big(\frac{V'(M)}{x - M}\frac{1}{\zeta - \mathbf{R}}\Big) + \frac{N}{t}\Tr\Big(\frac{1}{x - M}\frac{\mathbf{R}}{\zeta - \mathbf{R}}\Big)\Big\rangle = 0 && \nn \\
\label{eq:fdsf}& &
\eea}
$\!\!\!$Il est utile d'introduire :
\bea
U_1(x,\zeta) & = & \Big\langle\Tr\frac{1}{x - M}\frac{1}{\zeta - \mathbf{R}}\Big\rangle \nn \\
U_2(x_1;x,\zeta) & = & \Big\langle\Tr\frac{1}{x_1 - M}\Tr\Big(\frac{1}{x - M}\frac{1}{\zeta - \mathbf{R}}\Big)\Big\rangle_c \nn \\
P_1(x;\zeta) & = & \Big\langle\Tr\Big(\frac{V'(x) - V'(M)}{x - M}\frac{1}{\zeta - \mathbf{R}}\Big)\Big\rangle \nn
\eea
En particulier, notons que $P_1(x;\zeta)$ est r\'egulier lorsque $x \in \mathrm{Ombre}(\Gamma)$, et a des p\^{o}les simples lorsque $\zeta \rightarrow R_i$. Nous pouvons mettre en forme l'\'{E}qn.~\ref{eq:fdsf} :
\bea
U_2(x;x,\zeta) + \Big[W_1(x) + \frac{N}{t}\big(\zeta - V'(x)\big)\Big]U_1(x,\zeta) & & \nn \\
-\frac{N}{t}\,W_1(x) + \frac{N}{t}\,P_1(x;\zeta) & = & 0 \nn
\eea
Cette \'equation est valable pour tout $\zeta$, en particulier, on peut faire disparaitre $U_1(x,\zeta)$ en sp\'ecialisant $\zeta = V'(x) - \frac{t}{N}W_1(x)$ :
\beq
\label{eq:fdogih}
\left\{\begin{array}{l} U_2\big(x;x,Y(x)\big) - \frac{N}{t}W_1(x) + P_1\big(x;Y(x)\big)  = 0 \\ Y(x) = V'(x) - \frac{t}{N}W_1(x) \end{array}\right.
\eeq

\section{Formalisme de la r\'ecurrence topologique}
\label{sec:forma}
Je vais maintenant introduire la r\'ecurrence topologique, qui exhibe l'unique solution des \'equations de Schwinger-Dyson des mod\`eles de matrices hermitiennes (\'{E}qns.~\ref{eq:masterloop} et \ref{eq:masterloopn} avec $\beta = 1$)  prescrite par certaines conditions analytiques. L'extension au cas $\beta \neq 1$ sera mentionn\'ee au \S~\ref{sec:ha}.

\subsection{Notion de courbe spectrale}
\label{sec:defc}
\noindent Pour nous, une \textbf{courbe spectrale} sera la donn\'ee :
\begin{itemize}
\item[$\diamond$] d'une surface de Riemann $\Sigma$.
\item[$\diamond$] d'un couple $(x,y)$ de fonctions m\'eromorphes sur $\Sigma$. On suppose que $\qquad x(\{X\})$ est un ensemble fini pour tout $X \in \Sigma$.
\item[$\diamond$] On note $a_i \in \Sigma$, les z\'eros de $\dd x$, on suppose qu'ils sont en nombre fini, et on demande que $y\dd x$ soit r\'egulier aux $a_i$.
\item[$\diamond$] d'un\label{Berg} \textbf{noyau de Bergman} $B \in \mathcal{T}^*(\Sigma)\otimes \mathcal{T}^*(\Sigma)$, i.e. d'un objet $B(z_1,z_2)$ qui est une forme diff\'erentielle en $z_1$ et en $z_2$, sym\'etrique en $(z_1,z_2)$, et qui est m\'eromorphe avec pour seule singularit\'e un p\^{o}le double avec coefficient $1$ en $z_1 = z_2$, sans r\'esidu, i.e. dans n'importe quelle coordonn\'ee locale $\xi$ :
    \beq
    B(z_1,z_2) \mathop{=}_{z_1 \rightarrow z_2} \frac{\dd\xi(z_1)\dd\xi(z_2)}{\big(\xi(z_1) - \xi(z_2)\big)^2} + O(1) \nn
    \eeq
\end{itemize}
Les deux premiers points sont \'equivalents \`a la donn\'ee d'un atlas $\{(x(z),y(z))\,\,z \in U_i\}_{i}$, ou encore d'une courbe plane $\Sigma_E \subseteq \mathbb{C}^2$ d'\'equation $E(x,y) = 0$. On peut \'eventuellement se restreindre \`a une courbe plane $\Sigma \subseteq \Sigma_E$ (pour enlever un ensemble de singularit\'es par exemple), o\`u rajouter des points \`a $\Sigma_E$ (pour la rendre compacte par exemple). Il n'est pas n\'ecessaire que $E$ soit une fonction alg\'ebrique.

Les $a_i \in U$ sont les points de ramification de $x$, et les valeurs $X_i = x(a_i) \in \widehat{\mathbb{C}}$ sont les points de branchement\label{brancha}. On dit qu'une courbe spectrale est \textbf{simple} si les $a_i$ sont des z\'eros simples de $\dd x$ et les $X_i$ sont deux \`a deux distincts. Dans ce cas, il existe une \textbf{involution locale} $z \mapsto \overline{z}$, d\'efinie sans ambig\"{u}it\'e au voisinage de chaque $a_i$ par $x(z) = x(\overline{z})$ et $z \neq \overline{z}$.

Outre l'\'equivalence g\'eom\'etrique (libert\'e de param\'etrage, de choix de coordonn\'ees locales,\ldots), nous utiliserons deux notions d'\'equivalence de courbes planes, qui se transmettent aux courbes spectrales.

\vspace{0.2cm}

$\diamond$ $\mathcal{S}_1$ est \textbf{faiblement \'equivalente} \`a $\mathcal{S}_2$ si elles peuvent \^{e}tre\label{faible} repr\'esent\'ees avec la m\^{e}me surface de Riemann $\Sigma$, et qu'il existe une constante $C$, une fonction $f$ m\'eromorphe au voisinage des points de branchement de $\mathcal{S}_1$, telles que :
\beq
x_1 = x_2 + C,\qquad y_1 = y_2 + f(x_1) \nn
\eeq

\vspace{0.2cm}

$\diamond$ $\mathcal{S}_1$ est \textbf{symplectiquement \'equivalente} \`a $\mathcal{S}_2$ si elles se d\'eduisent l'une de l'autre par composition de transformations d'\'equivalence faible, d'\'echanges $x \leftrightarrow y$, et de $x \rightarrow -x$. Informellement, ce sont les transformations qui pr\'eservent $|\dd x\wedge\dd y|$, d'o\`u le nom "symplectique".

\subsection{Exemples}

\noindent $\diamond\,$ Une courbe spectrale alg\'ebrique de genre $0$, avec deux points de ramification $z = \pm 1$ :
\beq
\Sigma = \mathbb{C},\qquad\left\{\begin{array}{l} x(z) = \alpha + \gamma(z + 1/z) \\ y(z) = \sum_{k = 1}^{d} u_k(z^k - z^{-k}) \end{array}\right.,\qquad  B(z_1,z_2) = \frac{\dd z_1\dd z_2}{(z_1 - z_2)^2} \nn
\eeq
L'involution locale est en fait globale : $z \mapsto 1/z$. Cette courbe spectrale intervient dans le d\'enombrement des cartes construites avec des $j$gones ($3 \leq j \leq d$).

\vspace{0.2cm}

\noindent $\diamond\,$ La \textbf{courbe spectrale d'Airy}, avec un seul point de ramification $y = 0$ :
\beq
\label{eq:Airy}\Sigma = \widehat{\mathbb{C}},\qquad y^2 = x,\qquad B(y_1,y_2) = \frac{\dd y_1\dd y_2}{(y_1 - y_2)^2}
\eeq
L'involution locale est en fait globale : $y \mapsto -y$.

\vspace{0.2cm}

\noindent $\diamond\,$ La courbe plane la plus g\'en\'erale qui est de genre $0$ et poss\`ede un seul point de ramification $z = 0$ s'\'ecrit :
\beq
\label{eq:courbeK}\Sigma \subseteq \mathbb{C},\qquad \left\{\begin{array}{l} x(z) = z^2 \\ y(z) = \sum_{k \geq 0} t_{k + 1}\,z^{k - 1} \end{array}\right.
\eeq
L'involution locale est en fait globale : $z \mapsto -z$. Par des transformations d'\'equivalence faible, on peut toujours se ramener \`a $\forall k \geq 0,\,t_{2k} \equiv 0$. Si $t_0 = 0$, et si l'on adjoint \`a $[\Sigma,x,y]$ le choix du noyau de Bergman :
 \beq
\label{eq:BK} B_{\mathrm{K}}(x_1,x_2) = \frac{\dd\sqrt{x_1}\,\dd\sqrt{x_2}}{(\sqrt{x_1} - \sqrt{x_2})^2}
\eeq
on obtient une courbe spectrale symplectiquement \'equivalente \`a celle du mod\`ele de Kontsevich (cf. \S~\ref{sec:interFg} ou le Th\'eor\`eme~\ref{thKont} de l'introduction)

\vspace{0.2cm}

\noindent $\diamond\,$ Une courbe spectrale non alg\'ebrique, de genre $0$, avec un point de ramification en $(x,y) = (1,1)$ :
\beq
\Sigma = \mathbb{C}^*,\qquad  e^{x} = ye^{-y},\qquad B(y_1,z_2) = \frac{\dd y_1\dd y_2}{(y_1 - y_2)^2} \nn
\eeq
Ici, $y$ fournit\footnote{Attention, les transformations d'\'equivalence faible et symplectiques ne touchent pas au noyau de Bergman. Ainsi, cette courbe spectrale est symplectiquement \'equivalente \`a la courbe  $e^{y} = xe^{-x}$ munie de $B(x_1,x_2) = \frac{\dd x_1\dd x_2}{(x_1 - x_2)^2}$.} une coordonn\'ee globale, et $y(x) = - \mathcal{W}(-x)$ o\`u $\mathcal{W}$ est la fonction de Lambert\label{Lambert}. Cette courbe spectrale intervient dans le calcul des nombres de Hurwitz. Ce n'est pas un cas particulier de la courbe spectrale de Kontsevich. Certes, on peut mettre la courbe plane $[\Sigma,x,y]$ sous la forme \'{E}qn.~\ref{eq:courbeK}, avec des coefficients $t_k$ explicitement calculables, mais le noyau de Bergman $B = \frac{\dd y_1\dd y_2}{(y_1 - y_2)^2}$ est diff\'erent de $B_{\mathrm{K}}$ donn\'e \`a l'\'{E}qn.~\ref{eq:BK}.

\vspace{0.2cm}

\noindent $\diamond\,$ Une courbe spectrale alg\'ebrique elliptique, avec un point de ramification $z = (1 + \tau)/2$ :
\beq
\Sigma = \mathbb{C}/(\mathbb{Z} \oplus \tau\mathbb{Z}),\qquad \left\{\begin{array}{c} x(z) = \wp(z|\tau) \\ y(z) = \wp'(z|\tau) \end{array}\right.,\qquad B(z_1,z_2) = \dd z_1\dd z_2\,\wp(z_1 - z_2) \nn
\eeq
o\`u $\Im \tau > 0$ et $\wp(z|\tau)$ est la fonction\label{Wei} de Weierstra\ss. Cette courbe spectrale est bas\'ee sur la courbe plane d'\'equation $y^2 = 4x^3 - g_2(\tau)x - g_3(\tau)$. Depuis les travaux de Seiberg et Witten \cite{SW94}, elle intervient pour les calculs semiclassiques dans les th\'eories de Yang-Mills supersym\'etriques $\mathcal{N} = 2$.

\vspace{0.2cm}

\noindent $\diamond\,$ Les courbes spectrales hyperelliptiques\label{hyper}. L'\'equation :
\beq
y^2 = \frac{\big(\mathrm{Pol}(x)\big)^2}{\prod_{i = 1}^{d} (x - X_i)} \nn
\eeq
d\'efinit une surface de Riemann compacte $\Sigma$, qui est g\'en\'eriquement de genre $g = \lfloor d/2 \rfloor$. Les involutions locales se recollent en fait en une involution globale : $y \mapsto -y$. Cette propri\'et\'e caract\'erise les courbes \textbf{hyperelliptiques}. Avec un noyau de Bergman appropri\'e, c'est la courbe spectrale la plus g\'en\'erale que l'on rencontre dans le mod\`ele \`a une matrice hermitienne avec potentiel polynomial.

\subsection{Axiomatique}
\label{sec:axiom}
\vspace{0.2cm}
\`{A} partir d'une courbe spectrale simple, on d\'efinit des nombres $\mathcal{F}^{(g)} = \omega_0^{(g)} \in \widehat{\mathbb{C}}$ pour $g \geq 2$, et des objets $\omega_n^{(g)}(z_1,\ldots,z_n) \in \bigotimes_{j = 1}^n \mathcal{T}^*(\Sigma)$. Pour initialisation :
\beq
\omega_1^{(0)}(z) = -y(z)\dd x(z),\qquad \omega_2^{(0)}(z_1,z_2) = B(z_1,z_2) \nn
\eeq
Ensuite, on introduit un \textbf{noyau de r\'ecurrence} $\mathcal{K}(z_0,z)$, qui est une forme diff\'erentielle en $z_0 \in \Sigma$, et l'inverse d'une forme diff\'erentielle en $z$ d\'efinie au voisinage de chaque $a_i$.
\beq
\label{eq:noyrecu} \mathcal{K}(z_0,z) = -\frac{1}{2}\frac{\int_{\overline{z}}^z \omega_2^{(0)}(z_0,\cdot)}{\big(y(z) - y(\overline{z})\big)\dd x(z)}
\eeq
La \textbf{formule des r\'esidus} d\'efinit $\omega_n^{(g)}$ par r\'ecurrence sur $|\chi| = 2g - 2 + n > 0$ :
{\small \bea
\omega_n^{(g)}(z,z_I) = \sum_{i} \Res_{z \rightarrow a_i} \mathcal{K}(z_0,z)\Big[\omega_{n + 1}^{(g - 1)}(z,\overline{z},z_I) + \sum_{\substack{J \subseteq I \\ \,0 \leq h \leq g}}^{'} \omega_{|J| + 1}^{(h)}(z,z_J)\,\omega_{n - |J|}^{(g - h)}(\overline{z},z_{I\setminus J}) \Big] && \nn\\
\label{eq:res} &&
\eea \small}
$\sum^{'}$ signifie que l'on exclut de la somme les termes o\`u $\omega_1^{(0)}$ apparait. Enfin, on d\'efinit pour $g \geq 2$ des nombres $\mathcal{F}_g$, par l'\textbf{\'equation de dilatation} :
\beq
\label{eq:dilat}\mathcal{F}^{(g)} = \frac{1}{2 - 2g}\sum_{i} \Res_{z \rightarrow a_i} \Big(\int^{z}\!\! y\dd x\Big)\,\omega_1^{(g)}(z)
\eeq
Comme un noyau de Bergman n'a pas de r\'esidus, on peut montrer \`a partir de la formule des r\'esidus que $\Res_{z \rightarrow a_i} \omega_n^{(g)}(z,z_I) = 0$, sauf pour $(n,g) = (1,0)$. Par cons\'equent, la d\'efinition de $\mathcal{F}^{(g)}$ ne d\'epend pas de la primitive de $y\dd x$ choisie.

\subsection{Repr\'esentation diagrammatique}

En d\'ebobinant la r\'ecurrence, il est possible d'\'ecrire $\omega_n^{(g)}(z_1,\ldots,z_n)$ comme une somme sur une certaine classe de graphes fl\'ech\'es trivalents $\mathcal{G}$, de genre $g$, et avec une ar\^{e}te externe pour chaque variable $z_i$. $\omega_n^{(g)}$ est calcul\'e par un emboitement de $2g - 2 + n$ r\'esidus. Le poids de $\mathcal{G}$ est obtenu en multipliant les poids des ar\^{e}tes internes $B(z_i,z_j)$, les poids des sommets trivalents $\mathcal{K}(z_0,z)$ (o\`u $z_0,z,\overline{z}$ sont les variables associ\'ees aux ar\^{e}tes incidentes), et en prenant les r\'esidus dans l'ordre des fl\`eches. \`{A} cause du fl\'echage, ces poids sont non locaux. Cette diagrammatique est d\'evelopp\'ee dans \cite{EORev}.

Il est aussi possible de repr\'esenter $\omega_n^{(g)}$ comme une fonction g\'en\'eratrice de surfaces orientables de genre $g$, \`a $n$ bords, avec une variable $z_i$ vivant sur chaque bord (Fig.~\ref{fig:toporec}). Ici, j'indique seulement une fa\c{c}on graphique de d'\'ecrire/lire la formule des r\'esidus, sans pr\'etention bijective. C'est plut\^{o}t un choix de vocabulaire, qui a une contrepartie combinatoire pr\'ecise d\'evelopp\'ee dans \cite{EOGeo} mais que je n'aborderai pas.

\begin{landscape}
\addtolength{\footskip}{30pt}
\addtolength{\linewidth}{50pt}
\begin{figure}
\begin{center}
\hspace{-0.9cm}\begin{minipage}[c]{0.70\linewidth}
\raisebox{-12cm}{\includegraphics[width = \textwidth]{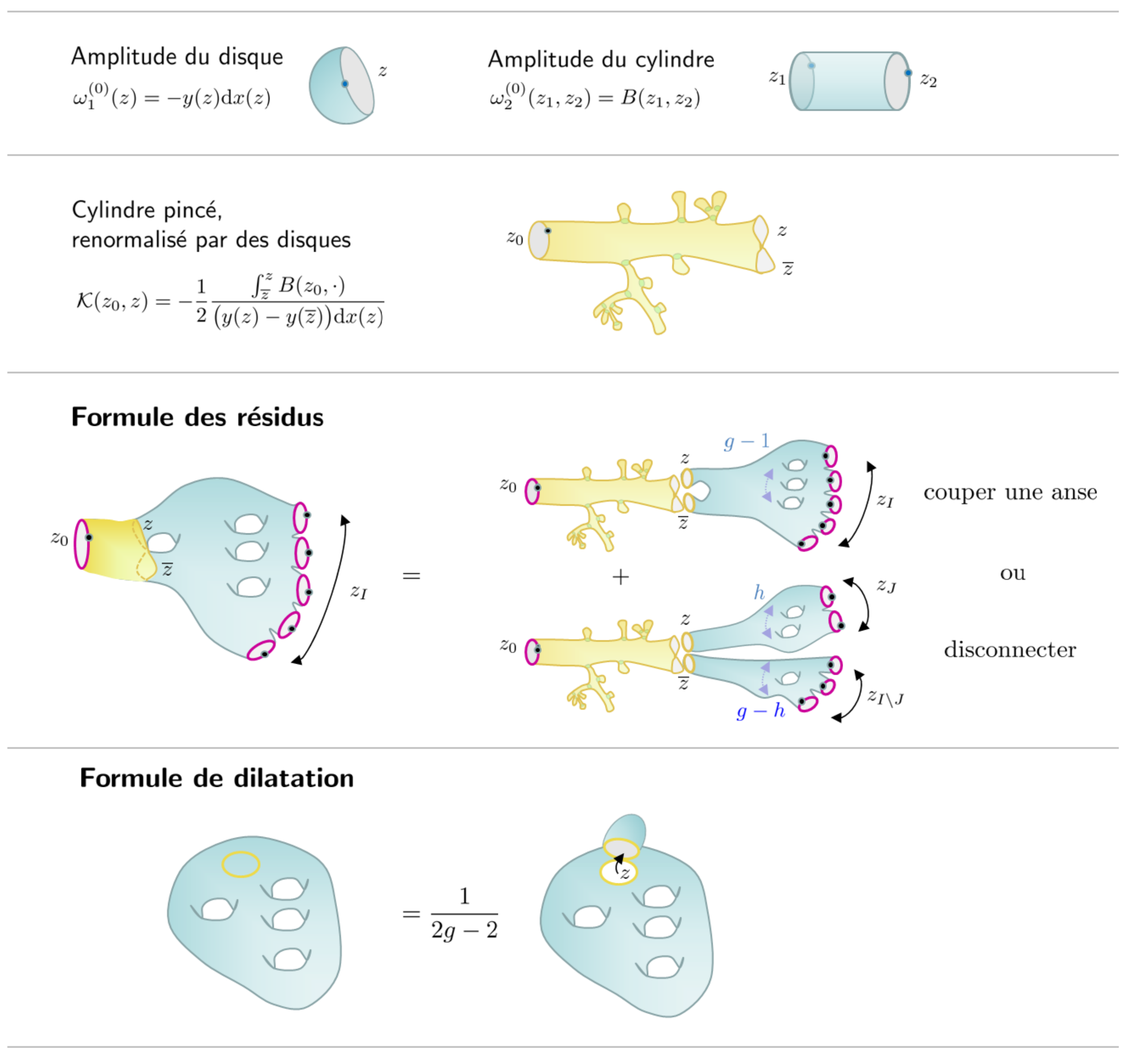}}
\end{minipage} \hfill \begin{minipage}[c]{0.295\linewidth}
\caption{\label{fig:toporec} {\small $\omega_n^{(g)}$ est repr\'esent\'e par une surface orientable $\Sigma$ de genre $g$ \`a $n$ bords. Pour calculer $\omega_n^{(g)}$, il faut choisir une variable $z_0$/un bord $C_0$. On parcourt $\Sigma$ jusqu'\`a rencontrer un branchement si bien que la surface restant \`a parcourir $\Sigma'$ change de topologie. Il y a deux possibilit\'es pour passer de $\Sigma$ \`a $\Sigma'$ : soit on a coup\'e une anse, soit on a disconnect\'e $\Sigma$ (et les bords restants sont r\'epartis entre les deux composantes connexes de $\Sigma'$). La surface parcourue avant ce d\'ecoupage \'etait un cylindre pinc\'e, sur lequel pouvaient aussi se brancher un nombre arbitraire de disques. On en tient compte en "renormalisant les cylindres pinc\'es".}}
\end{minipage}
\end{center}
\end{figure}
\end{landscape}

\subsection{D\'efinition des termes instables}
\label{stbsu}
En g\'eom\'etrie alg\'ebrique, on montre que le groupe d'automorphismes holomorphes d'une surface de Riemann $\Sigma$ de genre $g$ avec $n$ points marqu\'es est discret ssi $\chi = 2 - 2g - n < 0$. Dans ce cas, on dit que $\Sigma$ est \textbf{stable}, sinon elle est \textbf{instable}. Les objets stables sont aussi ceux qui peuvent \^{e}tre munis d'une m\'etrique hyperbolique \`a courbure constante strictement n\'egative et rendant les $n$ bords g\'eod\'esiques. Ces deux faits sont li\'es et rendent les objets stables plus faciles \`a \'etudier que les objets instables. Par exemple, l'espace des modules $\overline{\mathcal{M}}_{g,n}$ n'est bien d\'efini que pour $2 - 2g - n < 0$, et la d\'ecomposition de Strebel (partie~\ref{sec:introgeom}) existe aussi \`a cette condition. Toutefois, les topologies instables sont peu nombreuses : les sph\`eres $(n,g) = (0,0)$, les disques $(n,g) = (1,0)$, les cylindres $(n,g) = (2,0)$, et le tore $(n,g) = (0,1)$. On peut donc les traiter au cas par cas. C'est un ph\'enom\`ene analogue qui rend la d\'efinition d'un $\mathcal{F}^{(0)}$ et $\mathcal{F}^{(1)}$, ayant de bonnes propri\'et\'es, plus compliqu\'ee que celle des $\mathcal{F}^{(g)}$ ou $\omega_n^{(g)}$ stables. "Bonnes propri\'et\'es" signifie que les d\'eriv\'ees de $\mathcal{F}^{(0)}$ et $\mathcal{F}^{(1)}$ par rapport \`a un maximum de param\`etres doivent \^{e}tre d\'ecrites par la g\'eom\'etrie sp\'eciale (\S~\ref{sec:geomspec}). Par cons\'equent, $\mathcal{F}^{(0)}$ et $\mathcal{F}^{(1)}$ ne sont jamais d\'efinis qu'\`a une constante pr\`es. Dans une famille donn\'ee de courbes spectrales, il peut exister des choix naturels de cette constante.

Ici, on va se contenter de donner des d\'efinitions pour une famille de courbes spectrales simples alg\'ebriques. Elles ont l'avantage d'\^{e}tre bas\'ees sur une surface de Riemann compacte, sur laquelle la g\'eom\'etrie complexe est bien connue (cf. Annexe~\ref{app:geomcx}).

\subsubsection{D\'efinition de $\mathcal{F}^{(1)}$}

Lorsque $\Sigma$ est une surface de Riemann compacte, de genre $\g$, l'ensemble des noyaux de Bergman sur $\Sigma$ est un espace affine de dimension$_{\mathbb{C}}$ $\g(\g + 1)$ (cf. Annexe~\ref{app:geomcx}) :
\beq
\label{eq:Bkappa}B(z_1,z_2) = B_0(z_1,z_2) + 2i\pi\sum_{\h,\h' = 1}^{\g} \kappa_{\h\h'}\,\dd u_{\h}(z_1)\dd u_{\h'}(z_2),\qquad \kappa_{\h\h'} = \kappa_{\h'\h}
\eeq
$B_0$ est enti\`erement d\'etermin\'e par le choix d'une base symplectique $(\mathcal{A}_{\h},\mathcal{B}_{\h})$ de $H_1(\Sigma,\mathbb{Z})$. Si $\mathcal{S}$ est une courbe spectrale simple, on peut d\'efinir sans ambig\"{u}it\'e (je ne le justifie pas ici) une famille\footnote{Cette famille est parfois appel\'ee "espace de Hurwitz" dans la litt\'erature touchant aux syst\`emes int\'egrables et aux vari\'et\'es de Frobenius.} $\big(\mathcal{S}_{\kappa,X}\big)$ de courbes spectrales simples param\'etr\'ee par $\kappa_{\h,\h'}$ et les positions des points de \label{branch2}branchement $X_i = x(a_i)$. La structure conforme de $\Sigma$ change avec $\kappa$ et $\textbf{X} = \{X_i\}$. Cependant, on peut s'int\'eresser aux variations infinit\'esimales d'une fonction $f_{\kappa,\mathbf{X}}(z)$ ou d'une forme diff\'erentielle $\omega_{\kappa,\mathbf{X}}(z)$, lorsqu'une coordonn\'ee locale $\xi(z)$ est fix\'ee et $X_i$ varie. Le r\'esultat est not\'e $(\partial_{X_i}\:f_{\kappa,X})(z)$ ou $(\partial_{X_i} \omega)(z)$, et vit encore\footnote{Une petite illustration : $\partial_{X_i}\Big(\frac{\dd x}{\sqrt{\prod_{j = 1}^{d} (x - X_j)}}\Big) = \frac{1}{2(x - X_i)}\,\frac{\dd x}{\sqrt{\prod_{j = 1}^d (x- X_j)}}$.} sur la surface de Riemann $\Sigma_{\kappa,\mathbf{X}}$.

Les noyaux de Bergman $B$ dans cette famille satisfont la \textbf{formule variationnelle de Rauch} \cite{Rauchvar} :
\beq
\label{eq:Rauch}\frac{\partial}{\partial X_i} B(z_1,z_2) = \Res_{z \rightarrow a_i} \frac{B(z_1,z)\,B(z,z_2)}{\dd x(z)}
\eeq
Elle assure l'existence d'une fonction $\mathcal{T}_{B,x}(X_i)$ telle que \cite{EKK2} :
\beq
\label{eq:TauBerg}\partial_{X_i} \ln \mathcal{T}_{B,x} = \Res_{z \rightarrow a_i} \frac{B(z,\overline{z})}{\dd x(z)}
\eeq
On utilise la notation :
\beq
y(z) - y(\overline{z}) \mathop{\sim}_{z \rightarrow a_i} y_i\,\xi_i^{2m_i - 1}(z) \nn
\eeq
Comme $y$ multipli\'e par n'importe quelle fonction rationnelle de $x$, vit toujours sur la m\^{e}me surface de Riemann, $\tau_{B,x}$ ne d\'epend pas des $m_i$. Par exemple, pour la famille des \label{hyper2} courbes hyperelliptiques $y^2 = \prod_{i} (x - X_i)^{2m_i - 1}$, apr\`es le choix d'une base symplectique $(\mathcal{A}_{\h},\mathcal{B}_{\h})$ sur une courbe origine dans cette famille :
\beq
\mathcal{T}_{B,x}(\mathbf{X}) = \Delta(\mathbf{X})^{1/4}\cdot\det \mathbf{P},\qquad P_{\h,\h'} = \oint_{\mathcal{A}_{\h}} \frac{x^{\h' - 1}\dd x}{y} \nn
\eeq
$\Delta(\mathbf{X})$ est le d\'eterminant de Vandermonde des $X_i$, et $\mathbf{P}$ la matrice des p\'eriodes canonique pour les courbes hyperelliptiques. Le choix d'un pr\'efacteur $1$ s'impose naturellement.

On d\'efinit \`a une constante pr\`es :
\beq
\label{eq:F1def} \mathcal{F}^{(1)} =  - \frac{1}{2}\ln\mathcal{T}_{B,x}(\mathbf{X}) - \frac{1}{24}\ln\Big|\!\!\prod_{i\,/\,m_i = 0} \!\!y_i^3\cdot\!\!\prod_{i\,/\,m_i = 1} \!\!y_i\Big| - \ln\det\kappa
\eeq
Les points de branchement avec $m_i \geq 2$ ne contribuent pas au second terme.

\subsubsection{D\'efinition de $\mathcal{F}^{(0)}$}
\label{pre}
Notons $p$, les p\^{o}les de $y$, et $\xi_p$ la coordonn\'ee locale au voisinage de $p$ fix\'ee par la fonction $x$ (Annexe~\ref{app:geomcx}). D\'ecomposons :
\beq
y(z)\dd x(z) \mathop{=}_{z \rightarrow p} \dd V_p(z) + t_{p;0}\frac{\dd\xi_p(z)}{\xi_p(z)} + O(1),\qquad V_p = \sum_{j = 1}^{d_p} \frac{t_{p;j}}{j}\,\frac{1}{\xi_{p}^{j}} \nn
\eeq
Avec un choix de point base $o \in \Sigma$, on introduit le \textbf{potentiel chimique} :
\beq
\mu_{p} = \lim_{z \rightarrow p} \Big(\int_{o}^{z} y\dd x - \dd V_p(z) - t_{p;0}\dd\ln\xi_p\Big) \nn
\eeq
\label{cycsym} On choisit \'egalement une base symplectique de cycles $(\mathcal{A}_{\h},\mathcal{B}_{\h})$ sur $\Sigma$. Alors, on d\'efinit $\mathcal{F}^{(0)}$, qui est aussi appel\'e \textbf{le pr\'epotentiel} :
\bea
 && \mathcal{F}^{(0)} = - \frac{1}{2}\Big(\sum_{p} t_{0,p}\,\mu_p + \Res_{z \rightarrow p} y(z)\dd x(z)\,V_p(z)\Big) + \frac{1}{4i\pi}\sum_{\h = 1}^{\g} \oint_{\mathcal{A}_\h}\!\!y\dd x\oint_{\mathcal{B}_{\h}}\!\!y\dd x \nn \\
 \label{eq:F0def} &&
\eea
Cette expression ne d\'epend pas du choix de point base $o$ car $\sum_{p} \Res_p y\dd x = 0$.

\subsubsection{Remarque}

Il n'est pas facile d'\'etendre cette d\'efinition de $\mathcal{F}^{(1)}$ a des courbes spectrales $\mathcal{S}$ non alg\'ebriques. Il faut s'assurer que l'on peut d\'eformer $\mathcal{S}$ en param\'etrant par la position des points de branchement, tout en pr\'eservant la formule variationnelle de Rauch (\'{E}qn.~\ref{eq:Rauch}). Dans le cas alg\'ebrique, $\mathcal{F}^{(1)}$ est reli\'e au d\'eterminant r\'egularis\'e du laplacien sur $\Sigma$ muni de la m\'etrique $|y\dd x|^2$, et on pourrait aussi essayer de g\'en\'eraliser cette relation \`a des cas non alg\'ebriques.

Le pr\'epotentiel a une interpr\'etation g\'eom\'etrique simple, li\'ee \`a l'aire r\'egularis\'ee de $\Sigma$ munie de la m\'etrique $|y\dd x|^2$. 
Cette r\'egularisation est transparente : il est moins difficile de g\'en\'eraliser la d\'efinition de $\mathcal{F}^{(0)}$ \`a des courbes spectrales non alg\'ebriques, au cas par cas. On peut consid\'erer que $\mathcal{F}^{(0)}$ est d\'efini pour toute courbe spectrale (pas n\'ecessairement simple). En fait, l'\'{E}qn.~\ref{eq:F0def} ne fait pas intervenir le noyau de Bergman $B$, autrement dit, on peut parler du pr\'epotentiel d'une courbe plane $[\Sigma,x,y]$ quelconque.

\subsection{Propri\'et\'es \'el\'ementaires}
\label{sec:propele}

On pose $\xi_i = \sqrt{x - X_i}$, la coordonn\'ee locale\footnote{Si $X_i = \infty$, il faut plut\^{o}t utiliser $\xi_i = \frac{1}{\sqrt{x}}$.} au voisinage de $a_i$ telle que $\xi_i(z) = -\xi_i(\overline{z})$.
On introduit $m_i$, l'entier positif tel que $y(z) - y(\overline{z}) \propto \xi_i^{2m_i - 1}$, et on distingue les points de ramification \textbf{durs} (quand $m_i = 0$) et \textbf{mous} (quand $m_i > 0$).

\vspace{0.2cm}

\noindent \emph{Analyticit\'e} $\diamond\,$ Les $\omega_n^{(g)}(z_1,\ldots,z_n)$ stables sont des formes diff\'erentielles m\'eromorphes dans chacun des $z_j$, dont les p\^{o}les sont n\'ecessairement situ\'es aux $a_i$, et n'ont pas de r\'esidus.

\vspace{0.2cm}

\noindent \emph{Sym\'etrie} $\diamond\,$ Bien que la formule des r\'esidus privil\'egie une variable, tous les $\omega_n^{(g)}(z_1,\ldots,z_n)$ sont invariants par permutation des $z_j$. Cela peut \^{e}tre explicitement prouv\'e en utilisant la repr\'esentation diagrammatique.

\vspace{0.2cm}

\noindent  \emph{Degr\'e des p\^{o}les} $\diamond\,$ Les $\omega_n^{(g)}(z,z_I)$ stables ont un p\^{o}le de degr\'e maximal $\mathrm{max}\big(0 ; 2m_i(2g - 2 + n) + 2g\big)$ lorsque $z \rightarrow a_i$. En particulier, si $a_i$ est un point de ramification dur, tous les $\omega_n^{(0)}(z,z_I)$ sont r\'eguliers en $z = a_i$. Ces propri\'et\'es se prouvent par r\'ecurrence.

\vspace{0.2cm}

\noindent \emph{Localit\'e} $\diamond\,$ On consid\`ere $\Sigma$ et $B$ donn\'es. Les $\omega_n^{(g)}$ stables ne d\'ependent que du d\'eveloppement de Taylor \`a l'ordre $2m_i(2g + n) + 2g - 2$ de $\big(y(z) - y(\overline{z})\big)\dd x(z)$ lorsque $z \rightarrow a_i$. En revanche, les objets instables $\omega_1^{(0)}$, $\mathcal{F}^{(1)}$ et $\mathcal{F}^{(0)}$ d\'ependent des propri\'et\'es globales de la courbe spectrale.

\vspace{0.2cm}

\noindent \emph{Normalisation} $\diamond\,$ Pour la courbe spectrale d'Airy \label{Airy}$\mathcal{S}_{\mathrm{Airy}}$ (\'{E}qn.~\ref{eq:Airy}) :
\beq
\forall g,\qquad \mathcal{F}^{(g)}[\mathcal{S}_{\mathrm{Airy}}] = 0 \nn
\eeq

\vspace{0.2cm}

\noindent \emph{Invariance faible} $\diamond\,$ Tous les $\omega_n^{(g)}$ (sauf $\omega_1^{(0)}$) sont invariants lors de transformations d'\'equivalence faible de la courbe spectrale. La preuve est tr\`es facile.

\vspace{0.2cm}

\noindent \emph{Invariance symplectique} $\diamond\,$ Tous les $\mathcal{F}^{(g)}$ sont invariants lors de transformations symplectiques \label{inini}de la courbe spectrale. En revanche, les $\omega_n^{(g)}$ ne sont pas invariants, ils changent par un signe $(-1)^n$ ou/et par l'ajout d'une forme exacte, i.e. d'une quantit\'e $\dd_{z_1}\cdots\dd_{z_n}f(z_1,\ldots,z_n)$. La preuve \cite{EO2MM} demande des outils conceptuels que nous n'aborderons pas ici\footnote{Le point difficile est l'invariance par \'echange $x \leftrightarrow y$. On peut le v\'erifier \`a la main pour $\mathcal{F}^{(0)}$ et $\mathcal{F}^{(1)}$. La preuve g\'en\'erale consiste \`a d\'efinir des quantit\'es mixtes $\omega_{n;m}^{(g)}$ o\`u $x$ et $y$ jouent des r\^{o}les sym\'etriques, qui satisfont des \'equations de boucles. Il faut enfin montrer que l'on trouve le m\^{e}me r\'esultat en calculant $\mathcal{F}^{(g)} = \omega_{0;0}^{(g)}$, \`a partir de $\omega_{1;0}^{(g)}$ ou de $\omega_{0;1}^{(g)}$ par une \'equation de dilatation.}.

\vspace{0.2cm}

\noindent\emph{Homog\'en\'eit\'e} $\diamond\,$ Tous les $\omega_n^{(g)}$ et les $\mathcal{F}^{(g)}$ sont homog\`enes en $y\dd x$ de degr\'e $2 - 2g - n$. I.e., lorsque $y\dd x \rightarrow \lambda\,y\dd x$ pour $\lambda \in \mathbb{C}^*$ :
\beq
\label{eq:homoh}\omega_n^{(g)} \rightarrow \lambda^{2 - 2g - n}\,\omega_n^{(g)}\,,\qquad \mathcal{F}^{(g)} \rightarrow \lambda^{2 - 2g}\,\mathcal{F}^{(g)}
\eeq

\vspace{0.2cm}

\noindent \emph{\'{E}quation de dilatation} $\diamond\,$ $\omega_n^{(g)}$ se d\'eduit de $\omega_{n + 1}^{(g)}$ par :
\beq
(2g - 2 + n)\omega_n^{(g)}(z_I) = \sum_{i} \Res_{z \rightarrow a_i}\,\Big(\int^{z} \!\!y\dd x\Big)\,\omega_{n + 1}^{(g)}(z,z_I) - \delta_{n,1}\delta_{g,0}\,y\dd x(z_1) \nn
\eeq
Cette \'equation pour $n = 0$ d\'efinit les $\mathcal{F}^{(g)} = \omega_0^{(g)}$ pour $g \geq 2$.

\subsection{Hi\'erarchie d'\'equations de boucles}
\label{sec:heqb}

On indexe par $\alpha$, les diff\'erents feuillets de $x\,:\,\Sigma \rightarrow \widehat{\mathbb{C}}$, i.e. pour tout $z \in \Sigma$, $\{{}^{\alpha}z\} = x^{-1}(\{x(z)\})$. La r\'ecurrence topologique est \'equivalente au syst\`eme suivant d'\'equations de boucles :

\vspace{0.2cm}

\noindent $\diamond$ \emph{Hi\'erarchie d'\'equations de boucles lin\'eaires}.
\bea
\sum_{\alpha} \omega_{n}^{(g)}({}^{\alpha}z,z_2,\ldots,z_n) & = & \delta_{n,1}\delta_{g,0}V'(x(z)) + \delta_{n,2}\delta_{g,0} \frac{\dd x(z)\dd x(z_2)}{\big(x(z) - x(z_2)\big)^2} \nn \\
\label{eq:VVVV1} && \eea
o\`u $V'$ est une fonction de $x(z)$ est holomorphe au voisinage de $x(z) = X_i$.

\vspace{0.2cm}

\noindent $\diamond$ \emph{Hi\'erarchie d'\'equations de boucles quadratiques}.
{\small \bea
\sum_{\alpha < \beta} \omega_{n + 1}^{(g - 1)}({}^{\alpha}z,{}^{\beta}z,z_I) + \sum_{J \subseteq I} \sum_{h = 0}^{g} \omega_{|J| + 1}^{(h)}({}^{\alpha}z,z_J)\,\omega_{n - |J|}^{(g - h)}({}^{\beta}z,z_{I\setminus J}) = B_k^{(g)}(x(z);z_I)\big(\dd x(z)\big)^2 && \nn \\
\label{eq:VVVV2}  &&
\eea}
$\!\!\!$o\`u $B_k^{(g)}$ est une fonction de $x(z)$ qui est holomorphe au voisinage $x(z) = X_i$.

\vspace{0.2cm}

En pratique, lorsque $z \rightarrow a_i$, seuls les feuillets qui se ramifient en $a_i$ donnent lieu \`a des termes singuliers dans ces \'equations. Ainsi, si la courbe spectrale est simple, seuls les termes impliquant $z$ et $\overline{z}$ sont singuliers quand $z \rightarrow a_i$. On appelle parfois le cas $(n,g) = (1,0)$ de l'\'{E}qn.~\ref{eq:VVVV1} l'\textbf{\'equation de boucle maitresse}, car si l'on part d'un feuillet donn\'e, elle prescrit la structure de $\omega_1^{(0)}$ dans les autres feuillets, et donc la structure de la courbe plane $[\Sigma,x,y]$. On peut \'egalement remarquer que la composante $(n,g) = (2,0)$ de l'\'{E}qn.~\ref{eq:VVVV1} :
\beq
\label{eq:BBBB}\sum_{\alpha} \omega_2^{(0)}(z_1^{\alpha},z_2) = \frac{\dd x_1\dd x_2}{(x_1 - x_2)^2}
\eeq
est \'equivalente \`a dire que $\omega_2^{(0)}(z_1,z_2)$ est un noyau de Bergman.

\subsection{Limite singuli\`ere de courbes spectrales}
\label{sec:limitos}
Soit une famille lisse de courbes spectrales simples $(S_{\epsilon})_{0 < \epsilon \leq \epsilon_0}$. Il y a trois fa\c{c}ons pour $\mathcal{S}_{\epsilon}$ de devenir singuli\`ere lorsque $\epsilon \rightarrow 0$, qui peuvent \^{e}tre combin\'ees simultan\'ement.

\vspace{0.2cm}

\noindent \emph{Pincement} $\diamond\,$ Deux points de branchement $x(a_i)$ et $x(a_{i'})$ se rapprochent jusqu'\`a fusionner \`a $\epsilon = 0$.

\vspace{0.2cm}

\noindent \emph{Saut de $m_i$} $\diamond\,$ $y(z) - y(\overline{z}) \sim y_i\,\xi_i^{2m_i - 1}(z)$ avec $y_i \rightarrow 0$ lorsque $\epsilon \rightarrow 0$, donc $m_i(\mathcal{S}_0) \leq m_i(\mathcal{S}) + 1$.

\vspace{0.2cm}

\noindent \emph{Collision avec une singularit\'e essentielle} $\diamond\,$ $y$ a une singularit\'e essentielle qui s'approche d'un $a_i$.

\vspace{0.2cm}

Si $(\mathcal{S}_{\epsilon})_{\epsilon}$ est singuli\`ere \`a la limite $\epsilon \rightarrow 0^+$, m\^{e}me dans l'hypoth\`ese o\`u on peut appliquer la r\'ecurrence topologique \`a $\mathcal{S}_{\epsilon = 0}$, les propri\'et\'es analytiques des $\omega_n^{(g)}[\mathcal{S}_{\epsilon = 0}]$ stables sont diff\'erentes de celles des $\omega_n^{(g)}[\mathcal{S}_{\epsilon}]$ stables pour $\epsilon > 0$. Par exemple, si $m_i$ augmente brusquement, le degr\'e des p\^{o}les \`a $z = a_i$ dans $\omega_n^{(g)}(z,z_I)$ augmente aussi brusquement. C'est un signe de divergence des $\omega_n^{(g)}[\mathcal{S}_{\epsilon}]$ stables lorsque $\epsilon \rightarrow 0^+$.

Imaginons que, par un zoom \`a une certaine \'echelle autour des $a_i$ incrimin\'es, l'on voit une courbe spectrale simple $\mathcal{S}^* = [\Sigma^*,x^*,y^*,B^*]$. Plus pr\'ecis\'ement, on suppose qu'il existe une bijection locale $\xi_{\epsilon}$ telle que :
\beq
\left\{\begin{array}{l}  x_{\epsilon}\big(\xi_{\epsilon}(z)\big) =  \alpha_{x}\,x^*(z) + o\big(\alpha_{x}(\epsilon)\big) \\ y_{\epsilon}\big(\xi_{\epsilon}(z)\big) = R\big(x_{\epsilon}) + \alpha_{y}\,y^*(z) + o\big(\alpha_{y}(\epsilon)\big) \end{array}\right. \nn
\eeq
avec un facteur d'\'echelle $\alpha = \alpha_{x}\alpha_{y} \rightarrow 0$ quand $\epsilon \rightarrow 0$. Par r\'ecurrence, on peut alors prouver pour les objets stables :
\bea
\omega_n^{(g)}[\mathcal{S}_{\epsilon}]\big(\xi_{\epsilon}(z_1),\cdots,\xi_{\epsilon}(z_n)\big) & \sim & \alpha^{2 - 2g - n}\,\omega_n^{(g)}[\mathcal{S}^*](z_1,\ldots,z_n) \nn \\
\mathcal{F}^{(g)}[\mathcal{S}_{\epsilon}] & \sim & \alpha^{2 - 2g}\,\mathcal{F}^{(g)}[\mathcal{S}^*] \nn
\eea
Ces formules sont aussi valables avec quelques pr\'ecautions pour les objets instables :
\bea
\mathcal{F}^{(1)}[\mathcal{S}_{\epsilon}] & \sim& \ln|\alpha|\,\,\mathcal{F}^{(1)}[\mathcal{S}^*] \nn \\
\mathcal{F}^{(0)}[\mathcal{S}_{\epsilon}] & = & \mathcal{F}^{(0)}[\mathcal{S}_{\epsilon = 0}] + \alpha^{2}\,\mathcal{F}^{(0)}[\mathcal{S}^*] + o(\alpha^2) \nn
\eea
Chaque formule est valide tant que l'\'equivalent n'est pas nul, ce qui pourrait ponctuellement arriver avec certains $\mathcal{S}^*$.

Dans ce r\'esultat, il est remarquable que la contribution dominante \`a $\omega_n^{(g)}[\mathcal{S}_{\epsilon}]$ \`a l'approche d'une singularit\'e, ne d\'epende plus des d\'etails de la famille $\mathcal{S}_{\epsilon}$, mais seulement du type de singularit\'e, encod\'e par $\mathcal{S}^*$ \`a \'equivalence(s) pr\`es. C'est une propri\'et\'e d'universalit\'e, dont on rencontrera un exemple au \S~\ref{sec:conn}.

\subsection{G\'eom\'etrie sp\'eciale}
\label{sec:geomspec}

\subsubsection{D\'eformations m\'eromorphes}

Soit $\mathcal{S} = [\Sigma,x,y,B]$ une courbe spectrale simple. On peut chercher \`a changer le couple $(x,y)$ (ce qui en g\'en\'eral change la position des points de branchement donc la classe conforme de $\Sigma$). On veut donc calculer la variation de $\omega_n^{(g)}$ le long de transformations infinit\'esimales \mbox{$(x,y) \rightarrow (x + s\,\delta x,y + s\,\delta y)$} avec $s \rightarrow 0$, o\`u $\delta x$ et $\delta y$ sont des fonctions analytiques sur $\Sigma$.  Comme les $\omega_n^{(g)}$ stables ne d\'ependent que de la courbe spectrale locale autour des $a_i$, il est suffisant de supposer $\delta x$ et $\delta y$ m\'eromorphes au voisinage des $a_i$. Si l'on veut calculer les variations d'objets instables, on supposera que $\delta x$ et $\delta y$ sont m\'eromorphes sur tout $\Sigma$.

\`{A} cause de la libert\'e de reparam\'etrage, la variation des $\omega_n^{(g)}$ ne d\'epend que de la forme diff\'erentielle $\Omega = (\delta x)\dd y - (\delta y)\dd x$. Par dualit\'e forme-cycle, on peut la repr\'esenter :
\beq
\Omega(z) = \int_{z' \in \Omega^*}\!\!\!\!\!\!\!\! (\delta \Lambda_{\Omega})(z')\,B(z,z') \nn
\eeq
Alors, on peut montrer par r\'ecurrence pour tout $g,n$ :
\beq
\label{eq:geomspec0}\delta_{\Omega} \omega_{n}^{(g)}(z_I) = \int_{z' \in \Omega^*} \!\!\!\!\!\!\!\!(\delta\Lambda_{\Omega})(z')\,\omega_{n + 1}^{(g)}(z',z_I)
\eeq
Cette famille de propri\'et\'es est appel\'ee \textbf{g\'eom\'etrie sp\'eciale}. Pour les objets instables, il faut parfois r\'egulariser le membre de droite pour que cette formule ait un sens et soit valide, mais il existe une mani\`ere canonique de le faire, prescrite par l'atlas de coordonn\'ees locales fix\'ees en tout point de $\Sigma$ par la fonction $x$.

\subsubsection{Exemples}

\noindent \emph{Ajouter un p\^{o}le simple} $\diamond\,$ Comme $\sum_p \Res_p \Omega = 0$ (au moins pour une forme m\'eromorphe sur une surface compacte), une mani\`ere minimale est d'ajouter deux p\^{o}les simples $p$ et $q$ avec r\'esidus $1$ et $-1$. Cela revient \`a faire varier $t_{q;0} - t_{p;0} \equiv \Res_{p} y\dd x - \Res_{q}y\dd x$.
\beq
\Omega_{p,q;0}(z) = \mathop{\dd S}_{p,q}(z) \longleftrightarrow \Omega^*_{p,q;0} = \int_{p}^q\cdot \nn
\eeq

\vspace{0.2cm}

\noindent \emph{Ajouter un p\^{o}le multiple sans r\'esidu} $\diamond\,$ Si l'on choisit un point $z_0\in \Sigma$ et la coordonn\'ee locale $\xi$ dict\'ee par $x$ au voisinage de $z_0$, on peut ajouter une forme m\'eromorphe $\Omega(z)$ qui a un unique p\^{o}le, de degr\'e $k \geq 2$ en $z = z_0$, tel que $\Omega(z) = \frac{\dd\xi(z)}{\xi(z)^{k + 1}} + O(1)$ lorsque $z \rightarrow z_0$. Cela revient \`a changer $t_{z_0;k} \equiv \Res_{p} \xi_{z_0}^{k}\,y\dd x$.
\beq
\Omega_{z_0;k}(z) = \Res_{z' \rightarrow z_0} B(z,z')\big(\xi_{z_0}(z')\big)^{-k} \longleftrightarrow \Omega^*_{z_0;k} = \Res_{p} \xi_{z_0}^{-k}\cdot \nn
\eeq
En particulier, on note $\frac{\delta_{z_0}}{\dd x(z_0)}$ la d\'eformation par ajout d'un p\^{o}le double $(k = 2)$ en $z_0$ : $\Omega_{z_0;2}(z) = B(z_0,z)/\dd x(z_0)$. La relation de g\'eom\'etrie sp\'eciale s'\'ecrit :
\beq
\delta_{z_0}\omega_n^{(g)}(z_I) = \omega_{n + 1}^{(g)}(z_0,z_I) \nn
\eeq
\label{eq:opin}Pour cette raison, on appelle $\delta_{z_0}$ l'\textbf{op\'erateur d'insertion}. Il permet d'exprimer n'importe quelle d\'eformation $\Omega$. En effet, par construction :
\beq
\delta_{\Omega} = \int_{z \in \Omega^*}\!\!\!\!\! \delta_{z} \nn
\eeq

\vspace{0.2cm}

\noindent \emph{Ajouter une forme holomorphe} $\diamond\,$ Lorsque $\Sigma$ est une surface de Riemann compacte de genre $\g$, le choix d'une base symplectique \label{cycsym2} de cycles $(\mathcal{A}_{\h},\mathcal{B}_{\h})$ d\'efinit une base de $\g$ formes holomorphes $\dd u_{\h}$. D\'eformer $y\dd x$ en ajoutant $\dd u_{\h}$ revient \`a faire varier la \textbf{fraction de remplissage} $\epsilon_{\h} \equiv -\frac{1}{2i\pi}\oint_{\mathcal{A}_h} y\dd x$.
\beq
\Omega_{\h}(z) = 2i\pi\,\dd u_{\h}(z) \longleftrightarrow \Omega^*_{\h} = \oint_{\mathcal{B}_{\h}}\cdot \nn
\eeq
Notamment, \label{tauto}la matrice $\tau$ des p\'eriodes de $\Sigma$ v\'erifie :
\beq
\tau_{\h,\h'} \equiv \frac{1}{2i\pi}\oint_{z\in\mathcal{B}_{h}}\oint_{z'\in\mathcal{B}_{\h'}}\!\!\!\!\!\! B(z,z') = \frac{1}{2i\pi}\frac{\partial^2 \mathcal{F}^0}{\partial \epsilon_{\h}\partial\epsilon_{\h'}} \nn
\eeq
En fait, le pr\'epotentiel $\mathcal{F}^{(0)}$ a \'et\'e d\'efini (\'{E}qn.~\ref{eq:F0def}) afin que ses d\'eriv\'ees secondes soient donn\'ees par la relation de g\'eom\'etrie sp\'eciale, i.e. par des p\'eriodes de $B(z,z')$.

\subsubsection{D\'eformations du noyau de Bergman}

\label{bjs}On peut aussi s'int\'eresser aux variations des $\omega_n^{(g)}$ le long d'une famille de courbes spectrales o\`u le noyau de Bergman change \'egalement, via une d\'eformation $\frac{\partial}{\partial \kappa}$ dans l'\'{E}qn.~\ref{eq:Bkappa}. C'est le cas par exemple lorsque l'on change de base symplectique de cycles. Sans donner plus de d\'etails, disons lorsque $\Sigma$ est compacte, que les formules de g\'eom\'etrie sp\'eciale (\'{E}qn.~\ref{eq:geomspec0}) sont encore valables si l'on remplace $\delta_{\Omega}$ par :
\beq
D_{\Omega} = \delta_{\Omega} + \tr\Big(\kappa\:\delta_{\Omega}\tau\:\kappa\:\frac{\partial}{\partial \kappa}\Big) \nn
\eeq
o\`u la variation de la matrice des p\'eriodes $\tau_{\h,\h'}$ se calcule par :
\beq
\delta_{\Omega}\tau_{\h,\h'} = \int_{z \in\Omega^*}\!\!\!\!\!\!\!\!(\delta\Lambda_{\Omega})(z)\oint_{z_1 \in \mathcal{B}_{\h}}\oint_{z_1' \in \mathcal{B}_{\h'}}\!\!\!\!\!\! \omega_3^{(0)}(z,z_1,z_1') \nn
\eeq

\subsubsection{$\omega_3^{(0)}$ et la formule de Rauch}
\label{Rauchh}
En supposant $m_i \equiv 1$, on peut calculer $\omega_3^{(0)}$ et l'\'ecrire sous la forme :
\beq
\label{eq:W30v}\omega_3^{(0)}(z_0,z_1,z_2) = \sum_{i} \frac{B(z_0,\cdot)}{\dd y}(a_i)\, \Res_{z \rightarrow a_i} \frac{B(z_1,a_i)B(z_2,a_i)}{\dd x(z)}
\eeq
Par ailleurs, lors d'une d\'eformation $\Omega$ de sorte que $\delta y$ soit r\'egulier aux $a_i$, la position des points de branchement $X_i$ varie de $\delta X_i = \frac{\dd \Omega}{\dd y}(a_i)$.  D'apr\`es la formule variationnelle de Rauch (\'{E}qn.~\ref{eq:Rauch}), on peut \'ecrire :
\beq
\omega_3^{(0)}(z_0,z_1,z_2) = \sum_{i} \delta_{z_0} X_i\,\frac{\partial}{\partial X_i} B(z_1,z_2) \nn
\eeq
On voit donc que la formule de Rauch est \'equivalente \`a la relation de g\'eom\'etrie sp\'eciale $\delta_{z_0}\omega_2^{(0)}(z_1,z_2) = \omega_3^{(0)}(z_0,z_1,z_2)$.

\section{Applications}
\label{sec:ipipi}
Dans cette partie, je rappelle le lien entre r\'ecurrence topologique et mod\`eles de matrices, et avec les nombres d'intersections dans $\overline{\mathcal{M}}_{g,n}$. Plusieurs applications seront discut\'ees plus loin dans la th\`ese : \`a l'\'etude des classes d'universalit\'e (chapitre~\ref{chap:conv}), aux syst\`emes int\'egrables classiques  (chapitre~\ref{chap:int}), \`a la combinatoire des cartes (chapitre~\ref{chap:formel}), \`a la th\'eorie topologique des cordes (chapitre~\ref{chap:cordes}). J'ai choisi de les reporter, afin de les pr\'esenter en relation avec mes travaux.

\subsection{Mod\`eles de matrices}
\label{sec:bout}
\subsubsection{D\'eveloppement topologique}
\label{sec:devM}
Dans la chaine de matrices hermitiennes d\'efinie en \'{E}qn.~\ref{sec:chain}, supposons qu'il \label{eneli}existe un param\`etre $g_s$ de sorte que l'\'energie libre $\ln Z$ et les corr\'elateurs\footnote{Lorsqu'il y a plusieurs matrices, il faut aussi supposer que tous les corr\'elateurs mixtes, i.e. impliquant plusieurs $M_i$, ont un d\'eveloppement en puissances de $g_s$.} $W_ {n;p}$ (le $p$ indique qu'ils sont d\'efinis \`a partir de la matrice $M_p$ de la chaine) aient un \label{devo}"d\'eveloppement topologique" :
\bea
\ln Z = \sum_{g \geq 0} g_s^{2g - 2}\,F^{g},\qquad W_{n;p}(x_1,\ldots,x_n) = \sum_{g \geq 0} g_s^{2g - 2 + n}\,W_{n;p}^{(g)}(x_1,\ldots,x_n) && \nn \\
\label{eq:dev} &&
\eea
Peu importe sa nature, asymptotique (lorsque $Z$ est une int\'egrale de matrices convergente) ou formel (lorsque $Z$ est une int\'egrale de matrices d\'efinie perturbativement). On demande surtout que $W_n$ commence en $g_s^{n - 2}$, ce qui contraint dans un exemple concret le choix d'un bon param\`etre $g_s$. Souvent $g_s = 1/N$ o\`u $N$ est la taille des matrices, est un bon param\`etre, car les grandes matrices al\'eatoires ont une propri\'et\'e de factorisation :
\beq
\label{eq:factor}\langle\Tr f_1(M)\cdot \Tr f_2(M)\rangle \sim \langle\Tr f_1(M)\rangle\langle\Tr f_2(M)\rangle
\eeq
qui est le signe d'une forte corr\'elation entre valeurs propres de $M$.

Ce d\'eveloppement peut \^{e}tre ins\'er\'e dans les \'equations de Schwinger-Dyson, qui sont satisfaites ordre par ordre en $g_s$. On trouve alors que les objets :
\beq
\mathcal{W}_n^{(g)}(x_1,\ldots,x_n)  = W_{n;p}^{(g)}(x_1,\ldots,x_n)\dd x_1\cdots\dd x_n + \delta_{n,2}\delta_{g,0}\,\frac{\dd x_1\dd x_2}{(x_1 - x_2)^2} \nn
\eeq
satisfont les hi\'erarchies d'\'equations de boucles lin\'eaires et quadratiques (\S~\ref{sec:heqb}). En gardant les notations de l'\'{E}qn.~\ref{eq:VVVV1}, $V = V_p$ est le potentiel pour la matrice $M_p$ figurant dans l'\'{E}qn.~\ref{sec:chain}. La preuve est facile pour le mod\`ele \`a une matrice, \`a partir des \'equations de Schwinger-Dyson \'{E}qn.~\ref{eq:masterloopn}. En revanche, mettre les \'equations de Schwinger-Dyson de la chaine de matrice sous la forme \'{E}qn.~\ref{eq:VVVV1}-\ref{eq:VVVV2} demande un tour de force, d\'ej\`a n\'ecessaire dans la chaine \`a deux matrices. Cela a \'et\'e fait dans \cite{CEO06}. Par cons\'equent, $\mathcal{W}_n^{(g)} = \omega_n^{(g)}[\mathcal{S}]$ et $F^{g} = \mathcal{F}^{(g)}[\mathcal{S}]$ si l'on peut trouver une courbe spectrale $\mathcal{S} = [\Sigma,x,y,B]$ telle que :
\beq
\mathcal{W}_1^{(0)} = -y\dd x,\qquad \mathcal{W}_2^{(0)} = B \nn
\eeq

\subsubsection{Calcul de la courbe spectrale}
\label{jjjj}
Par construction, le corr\'elateur $W_{1;p}^{(0)}(x)$ est une fonction holomorphe dans $\mathbb{C}\setminus\gamma^{(0)}_p$, discontinue sur $\gamma^{(0)}_p \subseteq \mathrm{Ombre}(\Gamma)$. Ce $\gamma^{(0)}_p$ est le support des valeurs propres de $M_p$ dans la limite $g_s \rightarrow 0$, et est a priori inconnu. En premier lieu, il faut d\'eterminer de mani\`ere consistante :

\vspace{0.2cm}

\begin{itemize}
\item[$\diamond$] Le support $\gamma^{(0)}_p$, qui est en g\'en\'eral une r\'eunion d'arcs.
\item[$\diamond$] Une surface de Riemann $\Sigma$ sur laquelle $W_{1;p}^{(0)}(x)$ admet un prolongement analytique. Le plongement de $\widehat{\mathbb{C}}\setminus\gamma^{(0)}_p$ dans $\Sigma$ est appel\'e \textbf{feuillet physique} \label{fph}pour la matrice $M_p$.
\item[$\diamond$] \label{ffra}Pour chaque fraction de remplissage $\epsilon_{\h;p}$ fix\'ee pour la matrice $M_p$, des cycles $\mathcal{A}_{\h} \subseteq \Sigma$ sur lesquels :
\beq
\frac{1}{2i\pi}\oint_{\mathcal{A}_{\h}} W_{1;p}^{(0)}(x)\,\dd x = \frac{N\epsilon_{\h;p}}{g_s} \nn
\eeq
\item[$\diamond$]  $\mathcal{W}_1^{(0)} = W_{1;p}^{(0)}\dd x$ d\'efini sur $\Sigma$ et qui est la solution de l'\'equation de boucle maitresse :
\beq
\label{eq:probRHW10}\sum_{\alpha} \mathcal{W}_{1;p}^{(0)}({}^{\alpha}z) = \dd V_p(x(z))\,\qquad z \in \Sigma
\eeq
La notion de feuillet et la d\'efinition pr\'ecise de $\sum_{\alpha}$ est prescrite par le mod\`ele de matrice consid\'er\'e. Cependant, l'\'equation \`a r\'esoudre pour $\mathcal{W}_1^{(0)}$ prend toujours la forme \'{E}qn.~\ref{eq:probRHW10}.
\end{itemize}

\vspace{0.2cm}

En d'autres termes, $[\Sigma,x,y]$ est la solution d'un probl\`eme de type Riemann-Hilbert scalaire, qu'il est parfois difficile de r\'esoudre. Si certaines fractions de remplissage $\epsilon'_k$ sont libres dans le mod\`ele de matrice, il faut les d\'eterminer de sorte que $[\Sigma,x,y]$ soit un extremum pour le pr\'epotentiel $\mathcal{F}^{(0)}$ (par stricte convexit\'e de $\mathcal{F}^0$, c'est n\'ecessairement un minimum). Les conditions de minimisation sont :
\beq
\forall \mathcal{C} \subseteq \Sigma,\quad \Re\Big(\oint_{\mathcal{C}} y\dd x\Big) = 0 \nn
\eeq
\label{boubou}Cette propri\'et\'e est la d\'efinition des \textbf{courbes de Boutroux}. Gr\^{a}ce \`a une bonne compr\'ehension des courbes de Boutroux hyperelliptiques, il est prouv\'e que l'ensemble de ce probl\`eme pour le mod\`ele \`a une matrice admet une solution, qui est unique \cite{BerBou}. L'existence de courbes de Boutroux dans le cas g\'en\'eral (ou d\'ej\`a dans le cas alg\'ebrique non hyperelliptique) est une question difficile. Empiriquement l'ensemble de ce probl\`eme a une solution unique, mais on n'en connait pas de preuve.

\subsubsection{Noyau de Bergman}

Une fois la g\'eom\'etrie complexe trouv\'ee (surface $\Sigma$, feuillets, cycles  $\mathcal{A}_{\h}$,\ldots), il faut d\'eterminer $W_2^{(0)}$. On se contente de dire que le noyau de Bergman appropri\'e dans les mod\`eles de matrices satisfait :
\beq
\label{eq:sjdq}\oint_{z \in \mathcal{A}_{\h}} \!\!\!\!\!\!\!\! B(z,z') = 0
\eeq
Les $\mathcal{A}$-cycles sont les cycles non contractibles tels que $x(\mathcal{A})$ entourent les arcs du support $\gamma^{(0)}$ dans le feuillet physique. Lorsque $\Sigma$ est compact, l'\'{E}qn.~\ref{eq:sjdq} prescrit $B$ de mani\`ere unique.

\subsubsection{Exemples}

\noindent \emph{Le mod\`ele \`a une matrice hermitienne} $\diamond\,$ \'{E}crivons l'\'equation de Schwinger-Dyson au rang $1$ (\'{E}qn.~\ref{eq:masterloop} avec $\beta = 1$) \`a l'ordre dominant en $g_s$ :
 \beq
 \label{eq:dsgts}\big(W_1^{(0)}(x)\big)^2 + \frac{Ng_s}{t}\big(-V'(x)W_1^{(0)}(x) + P_1^{(0)}(x)\big) = 0
 \eeq
 C'est une \'equation du second degr\'e, dont on peut \'ecrire la solution. Comme $P_1^{(0)}(x)$ est r\'eguli\`ere sur $\gamma^{(0)} \subseteq \mathrm{Ombre}(\Gamma)$, on apprend que $W_1^{(0)}(x)$ a une discontinuit\'e sur $\gamma^{(0)}$ de type 'racine carr\'ee', et reste finie sur $\gamma^{(0)}$. On peut donc la prolonger analytiquement sur une surface de Riemann \label{hyper3} hyperelliptique $\Sigma$. $\Sigma$ a deux feuillets, un pour chaque d\'etermination de la racine carr\'ee. On a \'egalement a priori, gr\^{a}ce aux fractions de remplissage, des informations sur les int\'egrales de cycles (dans le plan $x\in \widehat{\mathbb{C}}\setminus\gamma^{(0)}$) de $W_1^{(0)}(x)\dd x$. Calculons maintenant la discontinuit\'e de \'{E}qn.~\ref{eq:dsgts} de part et d'autre de $\gamma^{(0)}$ :
\beq
\label{eq:discW10}\forall x \in \gamma^{(0)},\qquad  W_1^{(0)}(x + i0) + W_1^{(0)}(x - i0) = \frac{Ng_s}{t}\,V'(x)
\eeq
L'\'equation est maintenant lin\'eaire. Comme $W_1^{(0)}(x)\dd x$ peut \^{e}tre \'etendue en une forme diff\'erentielle analytique sur $\Sigma$, l'\'{E}qn.~\ref{eq:discW10} est une relation entre fonctions analytiques valable sur un continuum de points $\gamma^{(0)}\hookrightarrow\Sigma$, donc sur $\Sigma$ tout entier. On retrouve l'\'equation de boucle maitresse. . Par ailleurs, en perturbant le potentiel \`a partir de l'\'{E}qn.~\ref{eq:discW10}, on trouve :
\bea
&& \forall (x_1,x_2)\in\gamma^{(0)}\times\widehat{\mathbb{C}}\setminus\gamma^{(0)},\qquad W_2^{(0)}(x_1 + i0,x_2) + W_2^{(0)}(x_1 - i0,x_2) = -\frac{1}{(x_1 - x_2)^2} \nn \\
\label{eq:Bergame} &&
\eea
En traduisant \`a nouveau cette relation en \'egalit\'e de formes diff\'erentielles analytiques sur $\Sigma$, on d\'eduit que la combinaison :
\beq
\label{eq:W2H} \mathcal{W}_2^{(0)}(x_1,x_2) = \Big(W_2^{(0)}(x_1,x_2) + \frac{1}{(x_1 - x_2)^2}\Big)\dd x_1\dd x_2
\eeq
est bien un noyau de Bergman sur $\Sigma$  (cf. \'{E}qn.~\ref{eq:BBBB}).

\vspace{0.2cm}

\noindent \emph{Le mod\`ele \`a une matrice avec champ ext\'erieur} $\diamond\,$ Dans l'\'equation de Schwinger-Dyson~\ref{eq:fdogih}, $U_2(x;x,y)$ est une corr\'elation connexe \`a deux points et donc n\'egligeable \`a l'ordre dominant. On obtient alors une \'equation sur $W_1^{(0)}$ :
\beq
\label{eqoug}\frac{t}{Ng_s}\,W_1^{(0)}(x) = P_1^{(0)}\big(x,Y^{(0)}(x)\big),\qquad Y^{(0)}(x) = V'(x) - \frac{t}{Ng_s}W_1^{(0)}(x)
\eeq
En pratique, les propri\'et\'es analytiques de $P_1^{(0)}$ permettent de deviner $\Sigma$ et d'exhiber un param\'etrage $z \mapsto (x(z),Y^{(0)}(z))$ de la courbe d'\'equation~\ref{eqoug}. Lorsque ce probl\`eme peut avoir plusieurs solutions, il faut s\'electionner la bonne en minimisant le pr\'epotentiel.

\vspace{0.2cm}

\subsection{Nombres d'intersection dans $\overline{\mathcal{M}}_{g,n}$}
\label{sec:interFg}
Nous pr\'esentons dans ce paragraphe les r\'esultats de \cite{Ekappa}. Il ne faut pas faire confiance aux facteurs $2$ et aux signes dans cet article, mais les calculs ont \'et\'e d\'etaill\'es et corrig\'es dans \cite{Ebook}.
Je vais utiliser les notions de th\'eorie de l'intersection dans $\overline{\mathcal{M}}_{g,n}$ introduites \`a la partie~\ref{sec:introgeom}. Kontsevich \cite{Kontsevich} a montr\'e pour la s\'erie g\'en\'eratrice des nombres d'intersections :
\bea
\label{eq:modsa} Z_{\mathrm{K}} & = & \frac{1}{Z_0}\,\int_{\mathcal{H}_N} \dd X\,e^{\Tr\big(\frac{iX^3}{6} - \frac{\mathbf{T}X^2}{2}\big)},\qquad  \\
& = & \exp\Big\{\sum_{n \geq 1} \sum_{d_1,\ldots,d_n \geq 0} \Big[\prod_{j = 1}^n (2d_j + 1)!!\,t_{2d_j + 1}\Big]\,\int_{\overline{\mathcal{M}}_{g(\mathbf{d}),n}} \prod_{j = 1}^n \psi_j^{d_j}\Big\} \nn
\eea
avec les temps $(t_k)_{k \geq 1}$ qui sont en correspondance avec une matrice diagonale de taille assez grande par $t_{k} = \frac{(-1)^k}{k}\,\Tr \mathbf{T}^{-k}$, et $Z_0$ la constante de normalisation :
\beq
Z_0 = \int_{\mathcal{H}_N} \dd X\,e^{-\Tr \frac{\mathbf{T}\,X^2}{2}} \nn
\eeq
$g(\mathbf{d})$ est le genre fix\'e par la dimension $d_{g,n} = 3g - 3 + n = \sum_{j = 1}^n d_j$. Nous allons \'ecrire $\mathbf{S} = N^{-1/3}\,\mathbf{T}$ et $s_k = N^{-1}\,\Tr\mathbf{S}^{-k}$ afin de le mettre en valeur :
\bea
 Z_{\mathrm{K}} & = & \exp\Big\{\sum_{n \geq 1} \sum_{g \geq 0} N^{2 - 2g}\sum_{d_1 + \cdots + d_n  = d_{g,n}} \Big[\prod_{j = 1}^n (2d_j + 1)!!\,s_{2d_j + 1}\Big]\,\int_{\overline{\mathcal{M}}_{g(\mathbf{d}),n}} \prod_{j = 1}^n \psi_j^{d_j}\Big\} \nn \\
\label{eq:konkon} &&
\eea
Ici, il est commode de fixer $s_1 = 0$, ce qui permet de supprimer les termes redondants contenant $\psi^{0} \equiv 1$, donc de restreindre la somme \`a des entiers $d_i$ strictement positifs. Cette convention revient \`a choisir une normalisation pour la matrice des temps : $\Tr \mathbf{S}^{-1} = 0$.

\subsubsection{$\mathcal{F}^{(g)}$ et classes $\psi_i$}
\label{psipsi}
L'int\'egrale de matrice de Kontsevich peut se r\'e\'ecrire comme une int\'egrale formelle \`a une matrice $M = (2N)^{-1/3}X + 2^{-1/3}\mathbf{S}$ dans un champ externe $\mathbf{R} = 2^{-2/3}\mathbf{S}^2$ :
\beq
Z_{\mathrm{K}} = \frac{\big(i(2N)^{1/3}\big)^{N^2}}{Z_0}\,e^{N\Tr\mathbf{S}^3/6}\,\int_{\mathcal{H}(\Gamma)} \dd M\,e^{N\Tr\big(M^3/3 - \mathbf{R} M\big)} \nn
\eeq
Nous sommes dans le champ d'application du paragraphe pr\'ec\'edent, avec $1/N$ qui joue le r\^{o}le\footnote{Ce passage peut \^{e}tre rendu rigoureux (les manipulations d'int\'egrales formelles seront justifi\'ees au chapitre~\ref{chap:formel}). Nous invoquons ici le cas du mod\`ele \`a une matrice en champ externe pour justifier l'application de la r\'ecurrence topologique, car nous l'avons d\'ej\`a rencontr\'e. On pourrait tout aussi bien r\'esoudre les \'equations de Schwinger-Dyson du mod\`ele de l'\'{E}qn.~\ref{eq:modsa} et arriver aux m\^{e}mes conclusions.} de $g_s$.
\beq
Z_{\mathrm{K}} = \exp\Big\{\sum_{g \geq 0} N^{2 - 2g}\,\mathcal{F}^{(g)}[\mathcal{S}_{\mathrm{K}}]\Big\} \nn
\eeq
La courbe spectrale satisfait l'\'equation maitresse :
{\small \bea
W_1^{(0)}(x) & = & P_1^{(0)}(x,Y(x)), \nn \\
\label{eq:dsda} Y(x) & = & - x^2 - W_1^{(0)}(x),\qquad P_1^{(0)}(x;\zeta) = \Big\langle \Tr \frac{x + M}{\zeta - \mathbf{R}}\Big\rangle^{(0)}
\eea}
$\!\!\!$\`{A} la limite $t_j \rightarrow 0$, l'int\'egrale de Kontsevich est un mod\`ele de matrice gaussien, dont la courbe spectrale est de genre $0$. Comme les $t_j$ sont des variables formelles, la courbe spectrale $\mathcal{S}_{\mathrm{K}}$ en est une perturbation, en particulier doit aussi \^{e}tre de genre $0$. Autrement dit, $\mathcal{S}_{\mathrm{K}}$ admet un param\'etrage rationnel. Ceci d\'etermine une unique solution de l'\'{E}qn.~\ref{eq:dsda}. Il suffit d'\'etudier le comportement aux p\^{o}les pour v\'erifier que :\label{courbeK2}
\beq
\Sigma_{\mathrm{K}}\: : \: \left\{\begin{array}{l} x(z) = z - \frac{1}{N}\Tr \frac{1}{2 (2^{-1/3}\mathbf{S})(2^{-1/3}\mathbf{S} - z)} = z(1 - s_3) - \sum_{k \geq 2} 2^{(k - 1)/3}s_{k + 2}\,z^{k} \\ y(z) = z^2 \end{array}\right. \nn
\eeq
est la solution d\'esir\'ee. Nous donnons aussi le r\'esultat pour le noyau de Bergman :
\beq
B(z_1,z_2) = \frac{\dd z_1\,\dd z_2}{(z_1 - z_2)^2} \nn
\eeq
Par comparaison avec l'\'{E}qn.~\ref{eq:konkon}, pour $g \geq 2$ :
\beq
\label{eq:jonkon}\mathcal{F}^{(g)}[\mathcal{S}_{\mathrm{K}}] = \sum_{n \geq 1} \sum_{d_1 + \cdots + d_n = d_{g,n}} \prod_{j = 1}^{n} (2d_j + 1)!!\,s_{2d_j + 1}\,\int_{\overline{\mathcal{M}}_{g,n}} \prod_{j = 1}^n \psi_j^{d_j}
\eeq
Par invariance symplectique, ce sont aussi les $\mathcal{F}^{(g)}$ de la courbe spectrale :
{\small \beq
\widetilde{\mathcal{S}}_{\mathrm{K}}\: : \: x(z) = z^2,\qquad y(z) = z(1 - s_3) - \sum_{k \geq 2} 2^{(k - 1)/3}s_{k + 2}\,z^{k},\qquad B(z_1,z_2) = \frac{\dd z_1\dd z_2}{(z_1 - z_2)^2} \nn
\eeq}
$\!\!\!$Par une nouvelle transformation symplectique, on peut se d\'ebarasser des $s_{2k}$ pour $k \geq 1$. Cela est \'evident car le membre de droite de \'{E}qn~\ref{eq:jonkon} ne d\'epend pas de $s_{2k}$, et on le retrouve a posteriori gr\^{a}ce \`a la propri\'et\'e d'invariance symplectique des $\mathcal{F}^{(g)}$. On pourrait absorber le facteur $(1 - s_3)$ en faisant une dilatation de $y$. D'apr\`es l'\'{E}qn.~\ref{eq:homoh}, cela signifie que l'\'{E}qn.~\ref{eq:jonkon} d\'epend de $s_3$ uniquement via un pr\'efacteur $(1 - s_3)^{2g - 2}$ et les variables renormalis\'ees $s_k/(1 - s_3)$.

\subsubsection{R\'e\'ecriture avec les classes $\kappa_r$}
\label{kaka}
Les int\'egrales de classes $\psi$ sur $\overline{\mathcal{M}}_{g,n}$ peuvent \^{e}tre r\'eduites \`a des int\'egrales sur $\overline{\mathcal{M}}_{g,0}$ en introduisant les classes $\kappa_r$. Cela permet de r\'e\'ecrire l'\'{E}qn.~\ref{eq:jonkon} de fa\c{c}on compacte. Le r\'esultat est :
\beq
\label{eq:hujs} \mathcal{F}^{(g)}[\mathcal{S}_{\mathrm{K}}] = \int_{\overline{\mathcal{M}}{g,0}} K^{\vee}[\mathcal{S}],\qquad K^{\vee}[\mathcal{S}] = e^{\sum_{k \geq 0} \widehat{s}_k\,\kappa_k}
\eeq
Il faut comprendre $K^{\vee}$ comme une s\'erie g\'en\'eratrice de classes $\kappa_r$ (qui sont des formes $2r$-dimensionnelles), et $\int_{\overline{\mathcal{M}}_{g,0}}$ est nulle par convention lorsque l'on int\`egre une forme de dimension totale diff\'erente de $d_{g,0} = 3g - 3$. Les $\widehat{s}_k$ sont les param\`etres modifi\'es :
\beq
\widehat{s}_k = 2^{2k/3} \sum_{n \geq 1} \frac{1}{n} \sum_{\widehat{d}_1 + \cdots + \widehat{d}_n = k} \prod_{i = 1}^n (2\widehat{d}_i + 1)!!\,s_{2\widehat{d}_i + 3} \nn
\eeq
Ils sont reli\'es \`a la transform\'ee de Laplace de $y \dd x$ sur la courbe spectrale :
\bea
 && \exp\Big(-\sum_{k \geq 0} \widehat{s}_k\,u^k\Big) = \frac{1}{4\sqrt{\pi} u^{3/2}}\int_{\Re x(z) < 0} \dd x(z)\,e^{-\frac{x(z)}{2u}}\,\big(y(z) - y(\overline{z})\big) \nn\\
\label{eq:Kvee} &&
\eea

\subsubsection{$\omega_n^{(g)}$, classes $\psi$ et $\kappa_r$}

Les $\omega_n^{(g)}$ sont engendr\'es par les variations des $\mathcal{F}^{(g)}$ lorsque l'on fait des variations infinit\'esimales de la courbe spectrale. La variation $\mathbf{S} \rightarrow \mathbf{S} + \epsilon 2^{1/3}N\,\mathbf{V}$ r\'ealise l'op\'erateur d'insertion $\sum_{i = 1}^N v_i \delta_{z_i}$. Ici, $\mathbf{V}$ est une matrice diagonale\footnote{Pour pr\'eserver la condition $\Tr \mathbf{S}^{-1} = 0$, il faut supposer $\Tr \mathbf{V}\mathbf{S}^{-2} = 0$.}, on note $\sigma_i$ (resp. $v_i$) les valeurs propres de $\mathbf{S}$ (de $\mathbf{V})$, et les $z_i$ sont les p\^{o}les de $y(z)$, i.e. $z_i = 2^{-1/3}\sigma_i$. En effet, un calcul simple montre :
\bea
\delta_{\epsilon}\Omega & = & (\delta_{\epsilon}) x \dd y - (\delta_{\epsilon} y)\dd x =  \Tr \frac{\dd z}{(z - 2^{-1/3}\mathbf{S})^2} \nn \\
\int_{(\delta_{\epsilon} \Omega)^*} \bullet & = & \sum_{i = 1}^N \Res_{z \rightarrow z_i} \frac{\bullet}{z - z_i} \nn
\eea
Les $\omega_n^{(g)}$ doivent donc s'exprimer \'egalement en termes d'int\'egrales de classes $\psi_i$. Nous allons seulement donner le r\'esultat :
\beq
\label{eq:omomom}\omega_n^{(g)}[\widetilde{\mathcal{S}}](z_1,\ldots,z_n)  =  \sum_{d_1 + \cdots d_n = d_{g,n}} \Big[\prod_{j = 1}^n \dd\varsigma_d(z_j)\Big] \int_{\overline{\mathcal{M}}_{g,n}} K^{\vee}[\mathcal{S}]\,\prod_{j = 1}^n \psi_i^{d_i}
\eeq
o\`u l'on a introduit les formes diff\'erentielles :
\beq
\dd\varsigma_d(z) = (2d + 1)!!\,\frac{\dd z}{z^{2d + 2}} \nn
\eeq
Cette formule est valable pour les topologies stables, i.e. $2 - 2g  - n < 0$. Elle est aussi valable pour $n = 0$, o\`u il n'y a plus d'insertions de classes $\psi_i$ et l'on retrouve \'{E}qn.~\ref{eq:hujs}.
Elle donne une interpr\'etation des $\mathcal{F}^{(g)}$ et $\omega_n^{(g)}$ en termes de nombres d'intersections\footnote{Cela permet aussi de calculer de fa\c{c}on r\'ecursive les nombres d'intersections en utilisant la formule des r\'esidus. Historiquement, c'est Mirzakhani \cite{Mirza1} qui a trouv\'e la premi\`ere formule r\'ecursive pour calculer des intersections de classes $\psi$ et de $\kappa_1$ que l'on appelle \textbf{forme volume de Weil-Petersson}. Une autre preuve a \'et\'e propos\'ee par \cite{Mul1,LiuXu1}, et l'\'equivalence avec la r\'ecurrence topologique a \'et\'e d\'emontr\'ee dans \cite{EOwp}. Ces r\'esultats ont \'et\'e \'etendus pour inclure les autres classes $\kappa_r$ dans \cite{LiuXu2}, et du point de vue de la r\'ecurrence topologique dans \cite{Ekappa}.} dans $\overline{\mathcal{M}}_{g,n}$ pour toutes les courbes spectrales qui n'ont qu'un seul point de branchement mou $x = a$, et dont le noyau de Bergman est :
\beq
B_{\mathrm{K}}(x_1,x_2) = \frac{\dd\sqrt{x_1 - a}\,\dd\sqrt{x_2 - a}}{\big(\sqrt{x_1 - a} - \sqrt{x_2 - a}\big)^2} \nn
\eeq

R\'esumons de mani\`ere informelle. $K^{\vee}[\mathcal{S}]$ est un habillage gravitationnel, qui d\'epend des temps, et est d\'efini via une transform\'ee de Laplace dans la courbe plane $[\Sigma,x,y]$. $\psi_i$ est l'\'el\'ement de surface associ\'ee \`a la famille de surfaces de Riemann de $\mathcal{M}_{g,n}$ o\`u le point marqu\'e $p_i$ fluctue. $\omega_n^{(g)}$ calcule une fonction de corr\'elation \`a $n$ points parmi les surfaces de Riemann de genre $g$ pond\'er\'ees par $K^{\vee}[\mathcal{S}]$. En utilisant un vocabulaire de physique, ce sont les fonctions de \label{gravq2}corr\'elation de la gravit\'e quantique (\S~\ref{sec:introgeom}).

\section{G\'en\'eralisations}
\label{sec:genus}
\subsection{D\'eformations $\beta$ de la r\'ecurrence topologique}
\label{sec:ha}
\subsubsection{Le param\`etre $\hbar$}

Nous avons \'ecrit les \'equations de Schwinger-Dyson du mod\`ele $\beta$ \`a une matrice au \S~\ref{sec:beta1}. Lorsque $\beta = 1$, elles sont de nature alg\'ebrique, et sont \'equivalentes apr\`es d\'eveloppement topologique \`a la hi\'erarchie d'\'equations de boucles que r\'esoud la r\'ecurrence topologique (\S~\ref{sec:heqb}). Lorsque $\beta \neq 1$, elles font apparaitre une d\'eriv\'ee, par exemple :
\bea
\label{eq:eqdg}-(1 - \beta)W_1'(x) + \beta\,W_2(x,x) + \beta\big(W_1(x)\big)^2  && \\
 + \frac{N\beta}{t}\big(-V'(x)W_1(x) + P_1(x)\big) & = & 0 \nn
\eea
On veut \'etudier ces \'equations en pr\'esence d'un petit param\`etre tel que $W_n \sim g_s^{n - 2}\,W_{n}^{(0)}$. \`{A} l'ordre dominant :
\beq
\label{eq:xgf} \hbar\,\frac{\dd}{\dd x}\big(W_1^{(0)}(x)\big)  + \big(W_1^{(0)}(x)\big)^2  + \frac{Ng_s}{t}\big(- V'(x)W_1^{(0)}(x) + P_1^{(0)}(x)\big) = 0
\eeq
avec $\hbar = g_s\big(1 - \frac{1}{\beta}\Big)$. Je vais \'ecarter le cas singulier o\`u le terme quadratique est n\'egligeable devant les deux autres, et distingue deux r\'egimes int\'eressants :

\vspace{0.2cm}

\emph{$\hbar \ll 1$} $\diamond\,$ Le terme d\'eriv\'ee est n\'egligeable \`a l'ordre dominant, $W_1^{(0)}$ v\'erifie la m\^{e}me \'equation que pour $\beta = 1$ (\'{E}qn.~\ref{eq:dsgts}). Les \'equations de Schwinger-Dyson, ordre par ordre en puissances de $g_s$, forment encore un syst\`eme r\'ecursif d'\'equations alg\'ebriques. Elles peuvent \^{e}tre r\'esolues en termes de g\'eom\'etrie complexe sur la courbe spectrale $[\Sigma,x,-W_1^{(0)},W_2^{(0)}]$ du mod\`ele hermitien ($\beta = 1$). Par exemple, ce r\'egime s'applique d\`es que $g_s = \beta^{\kappa}/N$ et $\beta$ reste fini et loin de $0$, ce qui recouvre beaucoup de cas int\'eressants pour la th\'eorie des matrices al\'eatoires.

\vspace{0.2cm}

\emph{$\hbar$ fini} $\diamond\,$ Tous les termes de l'\'{E}qn.~\ref{eq:xgf} sont du m\^{e}me ordre, c'est une \'equation diff\'erentielle de type Ricatti. En posant $W_1^{(0)} = \frac{Ng_s}{t}\,\frac{V'(x)}{2} + \hbar\frac{\psi'}{\psi}$, elle est \'equivalente \`a une \'equation de Schr\"{o}dinger :
\beq
\label{eq:Scvhro}\hbar^2\frac{\dd^2\psi}{\dd x^2} + \frac{Ng_s}{t}\Big(\hbar\frac{V''(x)}{2} - \frac{Ng_s}{t}\,\frac{\big(V'(x)\big)^2}{4} + P_1^{(0)}(x)\Big)\psi(x) = 0
\eeq
avec une fonction inconnue $P_1^{(0)}(x)$ \`a d\'eterminer par consistance avec les propri\'et\'es analytiques que l'on requiert pour $W_1^{(0)}$. Les autres \'equations de Schwinger-Dyson, ordre par ordre en puissances de $g_s$, sont des \'equations diff\'erentielles lin\'eaires du premier ordre pour $W_n^{(g)}(x,x_I)$, avec des inconnues $P_n^{(g)}(x,x_I)$ \`a d\'eterminer encore par consistance. Dans ce r\'egime, il est n\'ecessaire (mais pas suffisant) que $\beta \rightarrow 0$ ou $\infty$. Par exemple, il serait pertinent du point de vue probabiliste, pour \'etudier la transition entre un r\'egime de fortes corr\'elations ($\beta > 0$) et de d\'ecorr\'elation ($\beta = 0$) pour $N$ variables al\'eatoires $\lambda_1,\ldots,\lambda_N$.

\vspace{0.2cm}

 R\'esoudre les \'equations de Schwinger-Dyson \`a $\hbar$ fini requiert une nouvelle technologie, principalement d\'evelopp\'ee par Bertrand Eynard et Olivier Marchal \cite{EMq0,TheseOliv}, rejoints par Leonid Chekhov \cite{CEMq1,CEMq2}. Ils sont parvenus \`a mettre la solution sous la forme d'une r\'ecurrence topologique, o\`u toutes les notions de g\'eom\'etrie complexe (noyau de Bergman, r\'esidus, formule de Cauchy, cycles non contractibles, identit\'e bilin\'eaire de Riemann, \ldots) sont remplac\'ees par des notions de la th\'eorie des \'equations diff\'erentielles. En quelque sorte, ils ont \'etabli l'analogue de la g\'eom\'etrie complexe classique d'une courbe plane $E(x,y) = 0$, pour un $\mathcal{D}$-module $E(x,y)\cdot\psi = 0$, o\`u $x$ et $y$ sont des objets non commutatifs : $[x,y] = \hbar$. Disons que ce programme est compl\'et\'e au moins dans le "cas hyperelliptique" $E(x,y) = y^2 - \mathrm{Pol}(x)$. Il est conjectur\'e que les $\mathcal{F}^{(g)}$ sont des invariants symplectiques, i.e. sont pr\'eserv\'es par toute transformation $(x,y) \mapsto (x_o,y_o)$ telle que $[x,y] = [x_o,y_o] = \hbar$. L'extension \`a des $E(x,y)$ plus g\'en\'eraux (atteints par exemple dans la d\'eformation $\beta$ de la chaine \`a deux matrices) permettrait d'\'etudier cette conjecture. Cette th\'eorie est prometteuse, je l'ai peu abord\'ee pendant ma th\`ese, mais j'ai l'intention d'y participer dans le futur.

Dans cette th\`ese, nous rencontrerons seulement $\hbar \ll 1$. L'extension de la r\'ecursion topologique est assez facile car on reste dans le domaine de la g\'eom\'etrie complexe. Elle a \'et\'e d\'evelopp\'ee par Chekhov et Eynard depuis 2006 \cite{CE06,C10}. Les nombres $\mathcal{F}^{(g;l)}[\mathcal{S}]$ et des formes diff\'erentielles $\omega_n^{(g;l)}[\mathcal{S}]$ sur $\Sigma$ peuvent \^{e}tre d\'efinis axiomatiquement en termes de g\'eom\'etrie complexe sur $\mathcal{S}$. Ils satisfont outre la hi\'erarchie d'\'equations de boucles ($l = 0$), toutes leurs descendantes ($l = 1,2,\ldots$). Au niveau $l$ apparait la d\'eriv\'ee de l'objet de niveau $l - 1$ (cf. \'{E}qn.~\ref{eq:M2} un peu plus loin). En particulier, $\omega_n^{(g;0)}[\mathcal{S}] = \omega_n^{(g)}[\mathcal{S}]$ pour tout $n,g$. Les objets $\omega_n^{(g;l)}[\mathcal{S}]$ v\'erifient la g\'eom\'etrie sp\'eciale, sont homog\`enes de degr\'e $2 - 2g - l - n$, et se comportent bien par rapport aux limites singuli\`eres de courbes spectrales\footnote{L'indice 'sing' rappelle que les objets instables $2g - 2 + n + l > 0$ ont une partie non divergente. Les nouveaux objets instables sont $\mathcal{F}^{(0;1)}$, $\mathcal{F}^{(0;2)}$, $\omega_1^{(0;1)}$.}
\beq
\big(\omega_n^{(g;l)}[\mathcal{S}_{\epsilon}]\big)_{\mathrm{sing}} \sim \alpha^{2 - 2g - n - l}\omega_n^{(g;l)}[\mathcal{S}^*] \nn
\eeq
L'invariance symplectique des $\mathcal{F}^{(g;l)}$ pour $l > 0$ est conjectur\'ee mais n'a pas encore \'et\'e prouv\'ee.

\subsubsection{Application au mod\`ele $\beta$ \`a une matrice ($\hbar \propto 1/N$)}

Revenons au mod\`ele $\beta$ (fini loin de $0$) \`a une matrice, avec un potentiel $V$ ind\'ependant de $\beta$ et $N$. Ici, je voudrais justifier la d\'ependance en $\beta$ et en $N$ du d\'eveloppement \`a $N \rightarrow \infty$, moyennant quelques hypoth\`eses qu'il faut juger sur des exemples pratiques (cf. chapitre ~\ref{chap:conv}). Supposons qu'il existe un d\'eveloppement :
\beq
\label{eq:ansat}\ln Z = \sum_{k \geq 0} N^{2 - k}\,F^{[k]},\qquad W_n = \sum_{k \geq 0} N^{2 - n - k}\,W_n^{[k]}
\eeq
En particulier, on demande que l'ordre dominant de $W_n$ soit $N^{2 - n}$, et par convention $W_n^{[k]} \equiv 0$ si $k < 0$. L'\'equation de Schwinger-Dyson \ref{eq:eqdg} s'\'ecrit ordre par ordre en $1/N$ :
\bea
\Big(\frac{1}{\beta} - 1\Big)(W_1^{[k - 1]})'(x)  + W_2^{[k - 2]}(x,x) + \sum_{m = 0}^{k} W_1^{[k]}(x)\,W_1^{[k - m]}(x) && \nn \\
+ \frac{1}{t}\big(-V'(x)W_1^{[k]}(x) + P_1^{[k]}(x)\big) & = & 0 \nn \\
\label{eq:M1}  &&
 \eea
et plus g\'en\'eralement l'\'{E}qn.~\ref{eq:masterloopn} :
\bea
\Big(\frac{1}{\beta} - 1\Big)\frac{\dd}{\dd x}\big(W_{n}^{[k - 1]}(x,x_I)\big) + W_{n + 1}^{[k - 2]}(x,x,x_I) && \nn \\
\sum_{J \subseteq I} \sum_{m = 0}^{k} W_{|J| + 1}^{[m]}(x,x_J)\,W_{n - |J|}^{[k - m]}(x,x_{I\setminus J}) - \frac{1}{t}\,V'(x)\,W_n^{[k]}(x,x_I)  && \nn \\
 + \frac{1}{\beta}\sum_{i = 2}^{n} \frac{\dd}{\dd x_i}\Big(\frac{W_{n - 1}^{[k]}(x,x_{I\setminus\{i\}}) - W_{n - 1}^{[k]}(x_I)}{x - x_i}\Big) + \frac{1}{t}\,P_n^{[k]}(x;x_I) & = & 0 \nn \\
\label{eq:M2} &&
\eea
Ces \'equations sont r\'ecursives lin\'eaires : on peut d\'eterminer $W_n^{[k]}$ (niveau $p = n + k$) en r\'esolvant une \'equation lin\'eaire, o\`u interviennent seulement des termes de niveau inf\'erieur $p' \leq p - 1$. Lorsque $\beta = 1$, on remarque que la premi\`ere correction g\'en\'er\'ee \`a partir des termes $k = 0$, n'est pas $k = 1$, mais $k = 2$. Autrement dit, la substitution du d\'eveloppement \'{E}qn.~\ref{eq:ansat} dans les \'equations de Schwinger-Dyson produit naturellement un d\'eveloppement en $1/N^2$ lorsque $\beta = 1$. En revanche, pour $\beta = 1$, tous les ordres $N^{-k}$ apparaissent. $W_n^{[k]}$. L'ordre dominant de $W_1^{[0]}$ v\'erifie l'\'equation (\'{E}qn.~\ref{eq:xgf}) :
\beq
\big(W_1^{[0]}(x)\big)^2 - \frac{1}{t}\big(V'(x)W_1^{[0]}(x) - P_1^{[0]}(x)\big) = 0 \nn
\eeq
donc ne d\'epend pas de $\beta$. Avec cette initialisation, une analyse rapide montre que la d\'ependance en $W_n^{[k]}$ est polynomiale en $\frac{1}{\beta}$ et $\Big(1 - \frac{1}{\beta}\Big)$, et plus pr\'ecis\'ement :
\beq
W_n^{[k]} = \beta^{1 - n}\sum_{g = 0}^{\lfloor k/2 \rfloor} \frac{1}{\beta^m}\Big(1 - \frac{1}{\beta}\Big)^{k - n - 2m}\,W_n^{[k;m]} \nn
\eeq
avec $W_n^{[k;m]}$ ind\'ependant de $\beta$. Par int\'egration de $W_1^{[k]}$, on trouve aussi :
\beq
F^{[k]} = \beta \sum_{m = 0}^{\lfloor k/2 \rfloor} \frac{1}{\beta^m}\Big(1 - \frac{1}{\beta}\Big)^{k - 2m}\,F^{[k;m]} \nn
\eeq
avec $F^{[k;m]}$ ind\'ependant de $\beta$.

En particulier, on peut d\'efinir une courbe spectrale $\mathcal{S}$ qui est essentiellement celle du mod\`ele hermitien :
\beq
y(x) = -W_1^{[0]}(x),\qquad B(x_1,x_2) = \beta\:W_2^{[0]}(x_1,x_2)\dd x_1\dd x_2 \nn
\eeq
$\mathcal{S}$ est une courbe spectrale hyperelliptique, ind\'ependante de $\beta$. La version $\beta$ de la r\'ecurrence topologique appliqu\'ee \`a $\mathcal{S}$ est pr\'ecis\'ement d\'efinie afin de donner la solution de la hi\'erarchie d'\'equations de Schwinger-Dyson (\'{E}qn.~\ref{eq:M1} et \ref{eq:M2}), avec la\label{devo6} correspondance :
\beq
W_n^{[k;m]}(x_1,\ldots,x_n) = \frac{\omega_n^{(m;k - 2m)}[\mathcal{S}](x_1,\ldots,x_n)}{\dd x_1\cdots\dd x_n} - \frac{\delta_{n,2}\delta_{k,0}\delta_{m,0}}{(x_1 - x_2)^2} \nn
\eeq
et :
\beq
F^{[k;m]} = \mathcal{F}^{(m;k - 2m)}[\mathcal{S}] \nn
\eeq

\subsection{Qu'est-ce qu'une courbe spectrale ?}
\label{sec:defsq}
J'ai fix\'e une notion de courbe spectrale au \S~\ref{sec:defc}, afin de pr\'esenter une th\'eorie g\'en\'erale. Mais, d'un point de vue empirique, on peut d\'efinir des objets $\mathcal{F}^{(g)}[\mathcal{S}]$ et $\omega_n^{(g)}[\mathcal{S}]$ d\`es que la formule des r\'esidus (\'{E}qn.~\ref{eq:res}) et la formule de dilatation (\'{E}qn.~\ref{eq:dilat}) font sens. Notamment, on a vu que les objets stables ($2g - 2 + n > 0$) ne font intervenir que les propri\'et\'es locales de $\mathcal{S}$ autour des points de ramification. Certaines propri\'et\'es de la r\'ecurrence topologique (comme une hi\'erarchie d'\'equations de boucles, le comportement \`a l'approche d'une limite singuli\`ere) seront automatiquement satisfaites, d'autres (comme la g\'eom\'etrie sp\'eciale) demanderont quelques v\'erifications.

Si $x\,:\,\Sigma \rightarrow \widehat{\mathbb{C}}$ est donn\'ee, on peut chercher \`a appliquer la r\'ecurrence topologique avec une fonction $y$ poss\'edant des monodromies autour des $a_i$. Il suffit de passer \`a un recouvrement universel  $\widetilde{\Sigma}$ de $\Sigma$, trouver un noyau de Bergman $\widetilde{B}$ dans $\widetilde{\Sigma}$ et \'ecrire une formule des r\'esidus aux $a_i \hookrightarrow\widetilde{\Sigma}$. Notamment, le cas o\`u $\Sigma$ est compact alors que $\widetilde{\Sigma}$ est non compact est int\'eressant.

\label{Ontri2}Un exemple est donn\'e par le mod\`ele de matrice $\On$, o\`u les corr\'elateurs \`a l'ordre dominant $W_1^{(0)}$ et $W_2^{(0)}$ sont connus depuis longtemps, et sont des fonctions analytiques avec monodromies, d\'efinies sur une surface $\Sigma$ elliptique. Pour des valeurs $\n \neq 2\cos(\pi\mathfrak{p}/\mathfrak{q})$, $\widetilde{\Sigma} \simeq \mathbb{C}$ est non compacte ($W_1^{(0)}$ et $W_2^{(0)}$ ont une singularit\'e essentielle au point $\infty$). On peut chercher \`a appliquer la r\'ecurrence topologique \`a ces donn\'ees, mais \`a ce stade, ce n'est qu'une d\'efinition axiomatique. J'ai montr\'e \cite{BEOn} que les $\omega_n^{(g)}$ construits de cette mani\`ere donnent bien le d\'eveloppement topologique des corr\'elateurs du mod\`ele de matrice $\On$, si l'on d\'efinit correctement l'involution locale sur $\widetilde{\Sigma}$ et si l'on choisit le bon noyau de Bergman $\widetilde{B}$. La situation est similaire dans plusieurs autres mod\`eles de matrices intervenant en combinatoire/physique statistique sur r\'eseau al\'eatoire : dans le mod\`ele de Potts \cite{BE99}, le mod\`ele \`a six vertex \cite{Kos6}, certains mod\`eles ABAB \cite{PZinnAB}, \ldots{} on connait l'ordre dominant, qui d\'efinit une courbe spectrale poss\'edant des monodromies. On conjecture que le d\'eveloppement topologique de tous ces mod\`eles est donn\'e par une r\'ecurrence topologique appropri\'ee. Cela signifierait que les \'equations de Schwinger-Dyson dans ces mod\`eles de matrices peuvent se mettre sous la forme d'\'equations de boucles (\'{E}qn.~\ref{eq:VVVV1} et \ref{eq:VVVV2}) avec une notion de feuillet appropri\'ee. Je reviendrai sur ce point au chapitre~\ref{chap:formel}.

Enfin, j'ai mentionn\'e au \S~\ref{sec:ha} l'extension de la r\'ecurrence topologique \`a une g\'eom\'etrie alg\'ebrique non commutative. Plus g\'en\'eralement, on voit que la surface de Riemann $\Sigma$ munie de la fonction $x$ sert uniquement \`a donner une r\'ealisation g\'eom\'etrique \`a l'\'equation de boucle lin\'eaire (\'{E}qn.~\ref{eq:VVVV1}), o\`u la notion de feuillet correspond aux diff\'erents rel\`evements de $x\,:\,\Sigma \rightarrow \widehat{\mathbb{C}}$. On peut chercher \`a \'etudier l'\'{E}qn.~\ref{eq:VVVV1} :

\vspace{0.2cm}

\begin{itemize}
\item[$\diamond$] dans un cadre g\'eom\'etrique plus g\'en\'eral. Par exemple, on peut dire que dans la d\'eformation $\beta$ des \'equations de boucles, on a $f(z^{a}) = f'(z)$ pour un certain feuillet $a$. Cela conduit \`a \'etudier les courbes spectrales non pas comme une surface de Riemann avec structure additionnelle, mais plut\^{o}t comme un module $\mathrm{Ker}\,\mathcal{L}$, o\`u $\mathcal{L}$ est un certain op\'erateur lin\'eaire. Par exemple $\mathcal{L} = \hbar^2\frac{\dd^2}{\dd x^2} - \mathrm{Pol}(x)$ (cf. \'{E}qn.~\ref{eq:Scvhro}). Pour le mod\`ele $\On$, on rencontrera \`a la partie~\ref{sec:OnON} :
    \beq
    \mathcal{L} = e^{2\tau\,\partial_u} - \n\,e^{\tau\,\partial u} + \mathrm{Id} \nn
    \eeq
    o\`u $u(x)$ est une fonction elliptique. Plus g\'en\'eralement, on pourrait s'int\'eresser \`a :
    \beq
    \mathcal{L} = \mathrm{Pol}\big(x,e^{x},\frac{\dd}{\dd f_{\alpha}(x)},e^{\tau_{\alpha}\frac{\dd}{\dd f_{\alpha}(x)}}\big)  \nn
    \eeq
C'est ce que les auteurs de la d\'eformation $\beta$ de la r\'ecurrence topologique ont initi\'e.
\item[$\diamond$] dans un cadre alg\'ebrique. On appelle $\mathbf{T}^{a,b}$ l'op\'erateur de passage de passage du feuillet $a$ au feuillet $b$, et on se donne une structure abstraite de groupe pour les $\mathbf{T}^{a,b}$. On cherche alors \`a trouver les solutions de la hi\'erarchie d'\'equations de boucles. \label{Galois}Cela nous am\`enerait vers les th\'eories de Galois classiques et diff\'erentielles.
\end{itemize}

\vspace{0.2cm}

Les deux aspects sont tr\`es li\'es. Ce programme est largement ouvert. Il est motiv\'e par les progr\`es r\'ecents fait sur la d\'eformation $\beta$ de la r\'ecurrence topologique et le mod\`ele $\On$, et les nombreuses similarit\'es entre ph\'enom\`enes math\'ematiques observ\'es dans les mod\`eles de matrices, et les th\'eories conformes, les th\'eories de jauge, les probl\`emes \`a $N$ corps, \ldots{} En fait, le lien entre tous ces probl\`emes se trouve peut-\^{e}tre dans les \'equations de boucles (dont les solutions repr\'esentables par des int\'egrales de matrices ne sont qu'un cas particulier). Leur formulation \'{E}qn.~\ref{eq:VVVV1} et l'\'{E}qn.~\ref{eq:VVVV2} est assez universelle.
Il est bien connu (e.g. \cite{deBoer}) que ces \'equations encodent une partie de l'alg\`ebre de \label{confor}Virasoro\footnote{Dans les travaux d'Ivan Kostov, les mod\`eles de matrices apparaissent comme fonction de partition de la th\'eorie conforme d'un boson chiral vivant sur la courbe spectrale \cite{Kostov1}. Les termes du d\'eveloppement topologique apparaissent comme des corrections quantiques, et peuvent \^{e}tre calcul\'es par des techniques diagrammatiques. Kostov et Orantin ont montr\'e r\'ecemment \cite{KostovNico} que ces diagrammes sont une resommation de ceux de la r\'ecurrence topologique. Ces aspects ne seront pas abord\'es dans cette th\`ese.} ou plus g\'en\'eralement des contraintes issues d'une alg\`ebre $\mathcal{W}$ \cite{RevSchou}. Par cons\'equent, de nouveaux r\'esultats concernant la r\'esolution des \'equations de boucles auraient des cons\'equences en th\'eorie conforme ou pour les syst\`emes int\'egrables.

\newpage
\thispagestyle{empty}
\phantom{bbk}

\newpage

\chapter{Int\'egrales convergentes \mbox{de matrices}}
\label{chap:conv}
\thispagestyle{plain}
\vspace{-1.5cm}

\rule{\textwidth}{1.5mm}

\vspace{2.5cm}
\addtolength{\baselineskip}{0.20\baselineskip}

{\textsf{Cette partie traite de l'asymptotique \`a $N \rightarrow \infty$ dans les mod\`eles $\beta$ \`a une matrice, voire les mod\`eles de matrices g\'en\'eralis\'es. Tout d'abord dans le r\'egime "une coupure", o\`u le d\'eveloppement asymptotique ne contient que des puissances de $1/N$. Apr\`es un rappel des th\'eor\`emes sur l'ordre dominant, j'explique comment les \'equations de boucles permettent de d\'emontrer l'existence de l'asymptotique \`a tous les ordres (article \cite{BG11}). Leur expression est alors donn\'ee par la r\'ecurrence topologique. Dans le r\'egime "plusieurs coupures", je pr\'esente une d\'erivation heuristique de l'asymptotique, due \`a Eynard, qui justifie l'apparition de termes $\frac{\textrm{oscillant}(N)}{N^k}$ \`a tous les ordres $k$. La derni\`ere partie est d\'evolue \`a l'\'etude de la statistique de la valeur propre maximale dans les mod\`eles $\beta$, comme application des techniques de r\'ecurrence topologique. Dans une double limite d'\'echelle, cela permet d'obtenir des informations pr\'ecises sur la queue gauche des lois de Tracy-Widom $\beta$, et de comprendre leur universalit\'e (article \cite{BEMN}).}}

\addtolength{\baselineskip}{-0.20\baselineskip}

\section{Asymptotique \`a $N \rightarrow \infty$}
\label{sec:1maj}
Une int\'egrale convergente de matrice est simplement une loi de probabilit\'e $Z^{-1}\dd\nu$ sur un ensemble de matrices, ou bien la donn\'ee de la loi jointe sur leurs valeurs propres $\lambda_i$. Nous serons aussi int\'eress\'es par les mesures de probabilit\'e provenant de mod\`eles de matrices g\'en\'eralis\'es (\'{E}qn.~\ref{eq:intqh}). Pour nous, la question principale sera de d\'eterminer la statistique des $\lambda_i$, dans la limite $N \rightarrow \infty$, et de calculer l'\'energie libre $F = \ln Z = \ln\big(\int \dd \nu\big)$.

Pour le mod\`ele \`a une matrice hermitienne, pour la chaine de matrices, pour les mod\`eles $\beta = 1/2,1,2$, il existe des m\'ethodes de polyn\^{o}mes orthogonaux, qui reviennent \`a r\'ealiser ces mod\`eles comme des syst\`emes int\'egrables classiques. Une technologie puissante, appel\'ee \textbf{approche de Riemann-Hilbert}, a \'et\'e d\'evelopp\'ee pour \'etudier leur asymptotique lorsque $N \rightarrow \infty$, et l'on pourra consulter comme bonne introduction la s\'erie de cours de Deift \cite{Defcours}. Dans cette th\`ese, nous nous int\'eressons au cas o\`u cette m\'ethode ne peut pas \^{e}tre appliqu\'ee. Cela concerne la plupart des mod\`eles de matrices g\'en\'eralis\'es (\'{E}qn.~\ref{eq:intqh}), et notamment le mod\`ele $\beta \notin 1/2,1,2$ (\'{E}qn.~\ref{eq:1mmbeta}).

Dans ce chapitre, nous pr\'esentons une approche bas\'ee sur les \'equations de Schwinger-Dyson (cf. partie~\ref{sec:eqboucl1}). Ces \'equations sont exactes pour $N$ fini, et contraignent la structure du d\'eveloppement asymptotique lorsque $N \rightarrow \infty$. Elles permettent aussi de contr\^{o}ler les erreurs successives lorsque l'on tronque le d\'eveloppement. Pastur et ses collaborateurs sont les pr\'ecurseurs de cette m\'ethode.

\subsection{Cadre}
\label{bebeb}
Notre propos sera illustr\'e en d\'etail pour le mod\`ele $\beta$ sur l'axe r\'eel, avec bords $a,b$. Il est d\'efini par la mesure de probabilit\'e sur $\mathbb{R}^N$ :
\beq
\label{eq:lames} \dd\nu^{[b,a];V}_{N} = \frac{1}{Z^{[b,a];V}_{N}}\,\prod_{i = 1}^N \dd \lambda_i\,\mathbf{1}_{[b,a]}(\lambda_i)\,e^{-\frac{N\beta}{t}\,V(\lambda_i)}\prod_{1 \leq i < j \leq N} |\lambda_i - \lambda_j|^{2\beta}
\eeq
Le d\'eveloppement asymptotique lorsque $N \rightarrow \infty$ de ce mod\`ele est l'objet de l'article \cite{BG11} \'ecrit avec Alice Guionnet. Le but n'est pas ici d'\'enoncer des hypoth\`eses minimales ou de pr\'esenter les preuves compl\`etes qui apparaissent dans \cite{BG11} et ses r\'ef\'erences. Nous voulons plut\^{o}t d\'elimiter la d\'emarche g\'en\'erale. Nous indiquerons au \S~\ref{sec:paraa} et \ref{sec:gene} ce qui se g\'en\'eralise sans difficult\'e, et ce qui exigerait des travaux futurs. Pour fixer les id\'ees, nous supposerons que :

\vspace{0.2cm}

\noindent $\diamond\,$ $\beta \in ]0,\infty[$ est ind\'ependant de $N$ (r\'egime $\hbar \ll 1$ du \S~\ref{sec:ha}).

\vspace{0.2cm}

\noindent $\diamond\,$ $V$ est une fonction \`a valeurs r\'eelles, qui admet un d\'eveloppement asymptotique
\beq
\label{eq:Vhyp} V(x) = \sum_{k \geq 0} N^{-k}\,V^{[[k]]}(x)
\eeq
o\`u chaque $V^{[[k]]}$ se prolonge en une fonction holomorphe (\'eventuellement nulle) sur un voisinage de $[b,a]$.

\vspace{0.2cm}

\noindent $\diamond\,$ Si $a_i \in \{a,b\}$ est infini, il faut supposer pour assurer la convergence :
\beq
\liminf_{x \rightarrow a_i} \frac{V(x)}{2 \ln |x|}  > 1 \nn
\eeq

\vspace{0.2cm}

Cette derni\`ere hypoth\`ese dit aussi que $V$ cro\^{i}t assez vite \`a l'infini, pour que les valeurs propres vivent dans une r\'egion finie, \`a des configurations de poids exponentiellement petit pr\`es quand $N$ est grand :
\begin{lemme}
\label{lfini} Il existe un segment fini $[b',a']$, inclus dans $[b,a]$, et une constante $C > 0$ tels que :
{\small \beq
Z^{[b,a];V}_{N} = Z^{[b',a'];V}_{N}\big(1 + O(e^{-NC})\big)\,\quad\,F^{[b,a];V}_{N} = F^{[b',a'];V}_{N} + O(e^{-NC})\nn
\eeq}
\end{lemme}
Si l'on ne s'int\'eresse pas \`a ces termes exponentiellement petits, on peut donc supposer d\`es le d\'epart $a$ et $b$ finis. L'avantage est que les corr\'elateurs $W_n(x_1,\ldots,x_n)$ d\'efinis en \'{E}qn.~\ref{eq:nonon1} :
\beq
W_n(x_1,\ldots,x_n) = \Big\langle \prod_{j = 1}^n \sum_{i_j = 1}^N \frac{1}{x_j - \lambda_{i_j}}\Big\rangle_c \nn
 \eeq
sont automatiquement des fonctions holomorphes pour $x_i \in \mathbb{C}\setminus[b,a]$. Cela va nous permettre d'utiliser les outils de l'analyse complexe, et de d\'eplacer des contours d'int\'egration dans $\mathbb{C}\setminus[b,a]$.

\subsubsection{Quelle notion de convergence ?}

\begin{figure}[h!]
\begin{center}
\includegraphics[width=0.9\textwidth]{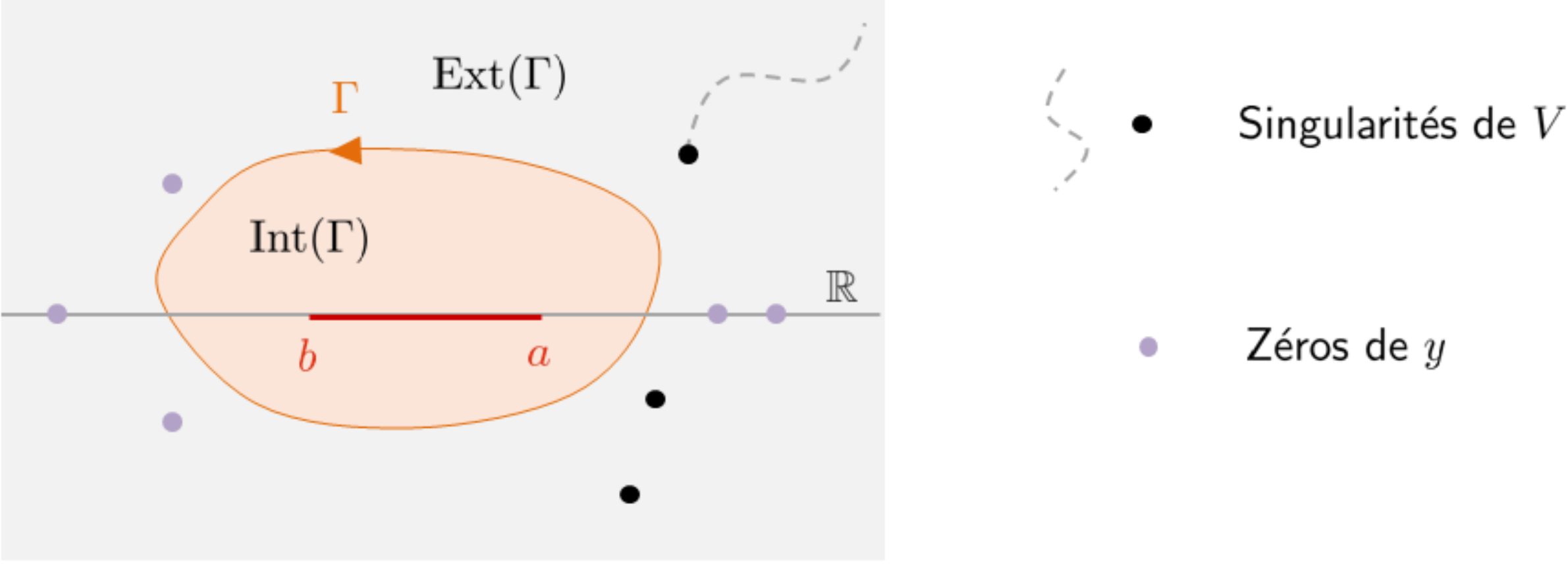}
\caption{\label{fig:contours} $\Gamma$ est choisi assez proche de $[a,b]$ de sorte que les singularit\'es de $V$, et les z\'eros de $y$ (cf. \S~\ref{sec:paraa}), se situent \`a l'ext\'erieur de $\Gamma$.}
\end{center}
\end{figure}

Plus pr\'ecis\'ement, $W_n$ appartient \`a l'espace $\mathcal{H}_{n;[b,a]}$ des fonctions \`a $n$ variables, holomorphes sur $\mathbb{C}\setminus[b,a]$ au moins, et qui ont une limite \`a l'infini (ici, $W_n(x,x_I) \rightarrow 0$ lorsque $x \rightarrow \infty$). Si l'on veut parler de d\'eveloppement asymptotique, il faut pr\'eciser pour quelle notion de convergence. Choisissons une courbe de Jordan $\Gamma$ qui entoure $[b,a]$, et notons $\mathrm{Ext}(\Gamma)$ la composante connexe de $\widehat{\mathbb{C}}\setminus\Gamma$ contenant $\infty$, et $\mathrm{Int}(\Gamma)$ celle qui contient $[b,a]$ (Fig.~\ref{fig:contours}). D\'efinissons pour $f \in \mathcal{H}_{n;[b,a]}$ :
\beq
\label{eq:norm}\p f \p_{\Gamma} = \sup_{(\xi_1,\ldots,\xi_n) \in \mathrm{Ext}(\Gamma)^n} |f(\xi_1,\ldots,\xi_n)| = \sup_{(\xi_1,\ldots,\xi_n) \in \Gamma^{n}} |f(\xi_1,\ldots,\xi_n)|
\eeq
Gr\^{a}ce au principe des z\'eros isol\'es, $\p \cdot \p_{\Gamma}$ est une norme. Puisque $f(\xi,\ldots)$ n'a pas de singularit\'e lorsque $\xi$ vit dans $\widehat{\mathbb{C}}\setminus\mathrm{Int}(\Gamma)$, le sup est atteint pour $(\xi_1,\ldots,\xi_n) \in \Gamma$ en appliquant $n$ fois le principe du maximum. L'espace $(\mathcal{H}_{n;[b,a]},\p\cdot\p_{\Gamma})$ est complet. La notion de convergence dans cet espace revient \`a contr\^{o}ler uniform\'ement la convergence de tous les coefficients de la s\'erie de Laurent de $f$ au voisinage de $\infty$ ; autrement dit, en termes de $W_n$, \`a contr\^{o}ler la \label{momen}convergence de tous les moments impliquant $n$ valeurs propres distinctes.

Remarquons que, pour tout $f \in \mathcal{H}_{1;[b,a]}$, il existe une constante $A_{\Gamma,\Gamma'}$ telle que :
\beq
\label{eq:rq} \p \frac{f(\bullet) - f(x_0)}{\bullet - x_0}\p_{\Gamma} \leq \p f' \p_{\Gamma} \leq A_{\Gamma,\Gamma'}\,\p f \p_{\Gamma'}
\eeq
o\`u $\Gamma' \subset \mathrm{Int}(\Gamma)$. Comme d'habitude avec les fonctions analytiques, contr\^{o}ler $f$ uniform\'ement suffit \`a contr\^{o}ler sa d\'eriv\'ee uniform\'ement.

\subsubsection{\'{E}quations de Schwinger-Dyson}
\label{SDD}
Les \'equations de Schwinger-Dyson pour $\dd\nu^{[b,a];V}_N$ ont \'et\'e essentiellement \'ecrites \`a l'\'{E}qn.~\ref{eq:laier}, o\`u il faut choisir des changements de variable $\lambda_i \rightarrow \lambda_i + \varepsilon\,h(\lambda_i)$ qui pr\'eservent le domaine d'int\'egration $[b,a]$. Ils sont tous g\'en\'er\'es par $h(\lambda) = \frac{(\lambda - a)(\lambda - b)}{x - \lambda}$. Apr\`es quelques calculs, on trouve l'\'equation au rang $1$ :
{\small \bea
W_2(x,x) + \big(W_1(x)\big)^2  + \Big(1 - \frac{1}{\beta}\Big)\frac{\dd}{\dd x}\big(W_1(x)\big) && \nn \\
+ \frac{(1 - 1/\beta)N - N^2}{(x - a)(x - b)} - \frac{N}{t}\,\frac{1}{2i\pi}\oint_{\mathcal{C}([b,a])} \frac{\dd \xi}{x - \xi}\,\frac{(\xi - a)(\xi - b)}{(x - a)(x - b)}\,V'(\xi)\,W_1(\xi) && = 0 \nn \\
\label{eq:cvf} &&
\eea}
$\!\!\!$Et en perturbant le potentiel, les \'equations au rang $n$ :
\begin{footnotesize}
\bea
W_{n + 1}(x,x,x_I) + \sum_{J \subseteq I} W_{|J| + 1}(x,x_J)\,W_{n - |J|}(x,x_{I\setminus J}) + \Big(1 - \frac{1}{\beta}\Big)\frac{\dd}{\dd x}\big(W_n(x,x_I)\big) && \nn \\
- \frac{N}{t}\,\frac{1}{2i\pi}\oint_{\mathcal{C}([b,a])} \frac{\dd \xi}{x - \xi}\,\frac{(\xi - a)(\xi - b)}{(x - a)(x - b)}\,V'(\xi)\,W_{n}(\xi,x_I) && \nn \\
 + \frac{1}{\beta} \sum_{i \in I} \frac{\dd}{\dd x_i}\Big(\frac{W_{n - 1}(x,x_{I\setminus\{i\}}) - \frac{(x_i - a)(x_i - b)}{(x - a)(x - b)}\,W_{n - 1}(x_I)}{x - x_i}\Big) & = & 0 \nn \\
\label{eq:ooo} &&
\eea
\end{footnotesize}
$\!\!\!\mathcal{C}([b,a])$ est un contour entourant $[b,a]$ mais laissant $x$ \`a l'ext\'erieur. En le d\'epla\c{c}ant pour inclure le p\^{o}le \`a $\xi = x$, la deuxi\`eme ligne s'\'ecrit :
\beq
\frac{N}{t}\Big(-V'(x)W_n(x,x_I) + \frac{\widetilde{P}_n(x,x_I)}{(x - a)(x - b)}\Big) \nn
\eeq
o\`u $\widetilde{P}_n(x,x_I)$ est holomorphe au voisinage de $[b,a]$. Les \'equations \'ecrites sous cette forme se rapprochent des \'{E}qns.~\ref{eq:masterloop} et \ref{eq:masterloopn} qui sont valables en l'absence de bords $[b,a]$ : le seul effet des bords $a,b$ est d'autoriser a priori un p\^{o}le simple en $x \rightarrow a,b$ dans la quantit\'e $P_n(x,x_I)$.

\subsection{Mesure empirique des valeurs propres}
\label{dens3}
La \textbf{mesure empirique} est d\'efinie comme la mesure de comptage des valeurs propres :
\beq
\dd L_N = \sum_{i = 1}^N \delta_{\lambda_i} \nn
\eeq
Il existe depuis longtemps des th\'eor\`emes assez g\'en\'eraux qui garantissent que $\frac{1}{N}\,L_N$ converge vers une mesure de probabilit\'e $L^*$ pour la topologie faible, i.e. pour toute fonction test $f$ assez r\'eguli\`ere :
\beq
\frac{1}{N}\sum_{i = 1}^N f(\lambda_i) \rightarrow \int \dd L^*(\xi)\,f(\xi) \nn
\eeq
Par exemple, pour $\dd\nu^{[b,a];V}_{N}$ \cite{Johan}
\begin{theo}
$L^*$ existe, et est l'unique mesure de probabilit\'e sur $[b,a]$ qui minimise :
\beq
\label{eq:45}\mathcal{E}[\mu] = -\beta\iint_{[b,a]^2}\!\!\!\!\! \dd\mu(\xi)\dd\mu(\eta)\,\ln|\xi - \eta| + \frac{\beta}{t} \int_{b}^a \dd\mu(\xi)\,V(\xi)
\eeq
On note $\mathrm{supp}\,L^* \subseteq [b,a]$ le support de $L^*$. Si $a = +\infty$ ou $b = -\infty$, pr\'ecisons que $\mathrm{supp}\,L^*$ est compact.
\end{theo}
Avec nos hypoth\`eses de r\'egularit\'e sur $V$, $\dd L^*$ est absolument continue par rapport \`a la mesure de Lebesgue, i.e. c'est une densit\'e $\rho(\xi)\dd\xi$.

L'existence d'un $\mu = \mu^*$ qui minimise $\mathcal{E}$ vient de consid\'erations g\'en\'erales de th\'eorie de la mesure. Le point essentiel est l'unicit\'e. La fonctionnelle \'{E}qn.~\ref{eq:45} est \label{convex}en fait strictement convexe, donc ne peut avoir qu'un seul minimum. La propri\'et\'e de convexit\'e :
\beq
\forall s \in [0,1] \qquad \mathcal{E}[(1 - s)\mu + s\mu'] \leq (1 - s)\mathcal{E}[\mu] + s\mathcal{E}[\mu'] \nn
\eeq
ne voit pas le terme lin\'eaire et se r\'esume \`a :
\beq
\label{eq:insee}\Delta[\mu_0] = -2\beta\iint_{[b,a]^2}\!\!\!\!\! \dd\mu_0(\xi)\dd\mu_0(\eta)\,\ln|\xi - \eta| \leq 0
\eeq
o\`u $\mu_0 = \mu - \mu'$ est la diff\'erence de deux mesures de probabilit\'e $\mu,\mu'$ quelconques sur $[b,a]$. L'indice $0$ est l\`a pour rappeler que $\mu_0$ est une mesure sign\'ee telle que $\int \dd\mu_0 = 0$. Avec cette notation, et la repr\'esentation :
\beq
\label{eq:astc}\frac{1}{2}\ln(s^2 + \epsilon^2) = \ln \epsilon - \mathrm{Re}\,\int_{0}^\infty\!\!\! \dd u\,e^{-\epsilon u}\,\frac{e^{isu} - 1}{u}
\eeq
un petit calcul suivant \cite[p142-143]{Defcours} montre que :
\beq
\Delta[\mu_0] = -2\beta\int_{0}^{\infty} \frac{|\widehat{\mu_0}(u)|^2}{u} \nn
\eeq
$\widehat{\mu}_0$ est la transform\'ee de Fourier de la mesure $\mu_0$. L'int\'egrand est r\'egulier en $u = 0$, car $\int \dd\mu_0 = 0$ implique $\widehat{\mu}(u) \propto u$ lorsque $u \rightarrow 0$. Par cons\'equent, $\Delta[\mu_0] \leq 0$, avec \'egalit\'e ssi $\widehat{\mu_0} \equiv 0$, ssi $\mu_0 = 0$. Ceci \'etablit la stricte convexit\'e de $\mathcal{E}$.

\subsubsection{Caract\'erisation de $L^*$}

Les \'equations de minimisation sont :
\bea
\label{eq:vk}\forall x\in\mathrm{supp}\,L^*\,&& 2\int_{b}^a \dd L^*(\xi)\ln |x - \xi| = \frac{1}{t}\,V^{[[0]]}(x) + C \\
\forall x \notin \mathrm{supp}\,L^* && 2 \int_{b}^a \dd L^*(\xi)\ln |x - \xi| \geq \frac{1}{t}\,V^{[[0]]}(x) + C \nn
\eea
o\`u $C$ est une constante. En d\'erivant l'\'egalit\'e par rapport \`a $x$ :
\beq
\label{eq:inqin}\forall x\in \mathrm{supp}\,L^*,\quad 2\,\mathrm{v.p} \int_{b}^a \frac{\dd L^*(\xi)}{x - \xi} = (V^{[[0]]})'(x)
\eeq
v.p. d\'esigne la partie principale de l'int\'egrale, i.e. l'expression $\frac{1}{2}\big(I(x + i0) + I(x - i0)\big)$. L'utilisation de la transform\'ee de Stieltjes :
\beq
W_1^{(0)}(x) = \int_{b}^a \frac{\dd L^*(\xi)}{x - \xi} \nn
\eeq
clarifie la nature g\'eom\'etrique du probl\`eme :
\beq
\label{eq:hhsd}\forall x \in \mathrm{supp}\,L^*,\quad W_1^{(0)}(x + i0) + W_1^{(0)}(x - i0) = \frac{1}{t}\,(V^{[[0]]})'(x)
\eeq

Cette \'equation est bien s\^{u}r compatible avec l'\'equation de Schwinger-Dyson au rang $1$ (\'{E}qn.~\ref{eq:cvf}), \`a l'ordre dominant lorsque $N \rightarrow \infty$. En effet, l'\'{E}qn.~\ref{eq:hhsd} assure que $\big(W_1^{(0)}(x)\big)^2 - (V^{[[0]]})'(x)W_1^{(0)}(x)$ n'a pas de discontinuit\'e sur $\supp\,L^*$. Cette quantit\'e a donc des singularit\'es uniquement \`a celles de $V^{[[0]]}(x)$, \'eventuellement en $x = a_i \in \{a,b\}$ o\`u $W_1^{(0)}$ peut se comporter comme $O\big((x - a_i)^{-1/2})$, ce qui la d\'etermine de mani\`ere unique. La solution v\'erifie :
\begin{small}\bea
\label{eq:joop} && \big(W_1^{(0)}(x)\big)^2 - (V^{[[0]]})'(x)W_1^{(0)}(x) \\
& = & \frac{1}{(x - a)(x - b)} + \frac{1}{t}\,\frac{1}{2i\pi}\oint_{\mathcal{C}([b,a],x)} \frac{\dd\xi}{x - \xi}\,\frac{(\xi - a)(\xi -b)}{(x - a)(x - b)}\,(V^{[[0]]})'(\xi)\,W_1^{(0)}(\xi) = 0 \nn
\eea
\end{small}
En comparant avec l'\'{E}qn.~\ref{eq:cvf}, on apprend au passage que $W_2(x,x) \in o(N^2)$.

L'\'{E}qn.~\ref{eq:hhsd} est ind\'ependante de $\beta$ (c'est la m\^{e}me \'equation que pour le mod\`ele hermitien) et de la pr\'esence des bords. En fait, les bords influent sur les propri\'et\'es analytiques a priori de $W_1^{(0)}(x)$. Si l'on note $a_i \in \{a,b\}$ un des bords, on sait d'apr\`es l'\'{E}qn.~\ref{eq:cvf} \`a l'ordre dominant, que $W_1^{(0)}(x) \in O\big((x - a_i)^{-1/2}\big)$. En revanche, en l'absence de bord, $W_1^{(0)}(x)$ reste fini partout sur l'axe r\'eel. Trouver $L^*$ revient \`a d\'eterminer la solution $W_1^{(0)}$ de l'\'{E}qn.~\ref{eq:hhsd} qui satisfait ces propri\'et\'es analytiques, et (si l'on a le choix) qui minimise $\mathcal{E}$.

Les techniques d'analyse complexe sont efficaces pour r\'esoudre l'\'{E}qn.~\ref{eq:hhsd}, elles sont r\'eexpliqu\'ees en d\'etail dans \cite{BEMN}. Elles ont aussi l'avantage de se g\'en\'eraliser \`a des \'equations lin\'eaires plus compliqu\'ees, comme celle du mod\`ele $\On$ (\S~\ref{sec:OnON}). Dans des cas plus restreints (dont l'\'{E}qn.~\ref{eq:inqin} fait partie), on peut trouver ind\'ependamment dans la litt\'erature la solution de l'\'equation int\'egrale : c'est une formule int\'egrale due \`a Plemelj et red\'eriv\'ee par Tricomi \cite{Tricomi} pour \ref{eq:inqin}, et une formule int\'egrale due \`a B¸ckner \cite{Buckner} pour l'\'equation du mod\`ele $\On$ dans le cas particulier $\mathrm{supp}\,L^* = [0,b]$.

Rappelons que la courbe plane $[\Sigma,x,y]$ associ\'ee au mod\`ele de matrice est reli\'e \`a $W_1^{(0)}$ par :
\beq
W_1^{(0)}(x) = \frac{(V^{[[0]]})'(x)}{2t} - y(x) \nn
\eeq
et $\Sigma$ est une surface de Riemann sur laquelle $y(x)$ se prolonge analytiquement. En particulier, $y(x)$ est aussi le prolongement analytique de la quantit\'e $\frac{1}{2}\big(W_1^{(0)}(x - i0) - W_1^{(0)}(x + i0)\big)$, initialement d\'efinie sur $\mathrm{supp}\,L^*$.

\subsection{L'op\'erateur lin\'eaire $\mathcal{K}$}

L'existence de $L^*$ indique que $W_1 = NW_1^{(0)} + \delta W_1(x)$ avec $\delta W_1 \in o(N)$, sous-entendu pour la norme $\p \cdot \p_{\Gamma}$. On peut aussi en d\'eduire que $\ln Z = -N^2 \mathcal{E}[L^*] + o(N^2)$. Que se passe-t-il aux ordres suivants ?

\vspace{0.2cm}

\'{E}crivons \'egalement $V = V^{[[0]]} + \frac{1}{N}\,\delta V$. $W_1^{(0)}$ satisfait l'\'{E}qn.~\ref{eq:cvf} \`a l'ordre dominant (\'{E}qn.~\ref{eq:joop}), et le reste de l'\'equation s'\'ecrit :
\begin{footnotesize}
\bea
&& \Big(\mathcal{K} + \frac{1}{N}\,\delta W_1(x) + \frac{1}{N}\,\mathcal{N}_{\delta V'} + \frac{1}{N}(1 - 1/\beta)\frac{\dd}{\dd x}\Big)[\delta W_1](x) \nn \\
&& = \mathcal{N}_{\delta V'}[W_1^{(0)}](x) - (1 - 1/\beta)\frac{\dd}{\dd x}\big(W_1^{(0)}(x)\big) - \frac{(1 - 1/\beta)}{(x - a)(x - b)} - \frac{1}{N}\,W_2(x,x) \nn \\
\label{eq:bjud}  & &
\eea
\end{footnotesize}
$\!\!\!\mathcal{K}$ et $\mathcal{N}_{g}[f]$ sont des endomorphismes de $\mathcal{H}_{1;[b,a]}$.
\bea
\mathcal{K}[f](x) & = & 2W_1^{(0)}(x)f(x) - \frac{1}{2i\pi t}\oint_{\mathcal{C}([b,a])}\frac{\dd \xi}{x - \xi}\,\frac{(\xi - a)(\xi - b)}{(x - a)(x - b)}\,(V^{[[0]]})'(\xi)\,f(\xi) \nn \\
\label{eq:NG} \mathcal{N}_{g}[f](x) & = & \frac{1}{2i\pi{}t}\oint_{\mathcal{C}([b,a])} \frac{\dd \xi}{x - \xi}\,\frac{(\xi - a)(\xi - b)}{(x - a)(x - b)}\,g(\xi)\,f(\xi)
\eea
L'op\'erateur $\mathcal{K}$ surtout joue un r\^{o}le important, car il faudrait l'inverser pour d\'eterminer $W_1$ \`a l'ordre sous-dominant. Il intervient de m\^{e}me dans toutes les \'equations de Schwinger-Dyson de rang $n \geq 2$ (\'{E}qn.~\ref{eq:ooo}), que l'on peut \'ecrire :
\begin{footnotesize}\bea
&& \Big(\mathcal{K} + \frac{2}{N}\delta W_1(x) - \frac{1}{N}\,\mathcal{N}_{\delta V'} + \frac{1}{N}(1 - 1/\beta)\frac{\dd}{\dd x}\Big)[W_n](x,x_I) = A_{n + 1} + B_n + C_{n - 1}  \nn \\
&& A_{n + 1} = -\frac{1}{N}\,W_{n + 1}(x,x,x_I) \nn \\
&& B_n = -\frac{1}{N}\sum_{\substack{n_1,n_2 \geq 2 \\ n_1 + n_2 = n + 1}} \sum_{J \subseteq I,\,\,|J| = n_1} W_{n_1}(x,x_J)\,W_{n_2}(x,x_{I\setminus J}) \nn \\
\label{eq:bjud1}  && C_{n - 1} = - \frac{1}{N}\,\frac{1}{\beta}\sum_{i \in I} \frac{\dd}{\dd x_i}\Big(\frac{W_{n - 1}(x,x_{I\setminus\{i\}}) - \frac{(x_i - a)(x_i - b)}{(x - a)(x - b)}\,W_{n - 1}(x_I)}{x - x_i}\Big)
\eea\end{footnotesize}
$\!\!\!$o\`u les $x_I$ sont spectateurs et les op\'erateurs agissent sur la variable $x$. On aurait besoin \`a nouveau de l'inverse de $\mathcal{K}$, pour d\'eterminer $W_n$ \`a l'ordre dominant. Deux cas se pr\'esentent.

\vspace{0.2cm}
\label{coucou}\label{bobo}\label{bobo3}
\noindent \emph{R\'egime \`a une coupure} $\diamond\,$ Si $\mathrm{supp}\,L^*$ est un segment $[b^*,a^*]$, $\mathcal{K}$ est bien inversible. Ici, nous allons simplifier notre discussion en supposant que les bords $a$ et $b$ sont \textbf{durs}, i.e. $a^* = a$ et $b^* = b$. En pr\'esence de bords mous, la m\'ethode que nous allons d\'ecrire demande d'\'etablir a priori un "principe de d\'ependance faible dans les bords mous", i.e. que les observables du mod\`ele d\'efini sur $[a,b]$ co\"{i}ncident \`a un $O(e^{-N\eta})$, avec celles du mod\`ele d\'efini sur tout intervalle $[a',b']$ qui est un petit \'elargissement de $[a^*,b^*]$. Ceci sera fait dans l'article en cours d'\'ecriture avec Alice Guionnet, o\`u nous d\'emontrons le r\'esultat final (Propositions~\ref{props} et \ref{orps}) avec des hypoth\`eses l\'eg\`erement plus faibles, et quelque soit la nature des bords. Ainsi, en supposant que les bords $a$ et $b$ sont durs, un argument classique conduit \`a la formule de Tricomi \cite{Tricomi} pour l'inverse de $\mathcal{K}$:
\beq
\label{eq:TriKA} \mathcal{K}^{-1}[g](x) = \frac{1}{2i\pi t}\oint_{\mathcal{C}([b,a])} \frac{\dd \xi}{x - \xi}\,\sqrt{\frac{(\xi - a^*)(\xi - b^*)}{(x - a^*)(x - b^*)}}\,\frac{g(\xi)}{2y(\xi)}
\eeq
Dans ce cas, $[\Sigma,x,y]$ est une courbe de genre $0$.

\vspace{0.2cm}

\noindent \emph{R\'egime \`a plusieurs coupures} $\diamond\,$ Si $\mathrm{supp}\,L^*$ est une r\'eunion de $\mathfrak{g} + 1$ segments  $[b^*_j,a^*_j]$, $\mathcal{K}$ n'est pas inversible, son noyau est de dimension$_{\mathbb{C}}$ $\mathfrak{g}$. En effet, les \'el\'ements du noyau sont les $\dd u/\dd x$, o\`u $\dd u$ est une forme holomorphe sur la surface de Riemann $\overline{\Sigma}$ d'\'equation $\sigma^2 = \prod_{j = 1}^{\mathfrak{g} + 1}(x - a^*_j)(x - b^*_j)$ :
\beq
\frac{\dd u}{\dd x} = \frac{\mathrm{Pol}(x)}{\sqrt{\prod_{j = 1}^{\mathfrak{g} + 1} (x - a^*_j)(x - b^*_j)}},\quad  \mathrm{deg}\,\mathrm{Pol} \leq \mathfrak{g} - 1 \nn
\eeq
On peut en choisir une base $\dd u_{\mathfrak{h}}$ duale aux contours\footnote{Le contour $\mathcal{A}_{\mathfrak{g} + 1}$ n'est pas ind\'ependant des autres, puisque $\sum_{\mathfrak{h} = 1}^{\mathfrak{g} + 1} \mathcal{A}_{\mathfrak{h}}$ est contractible sur $\overline{\Sigma}$.} $\mathcal{A}_{\mathfrak{h}}$ qui entourent $[b^*_{\mathfrak{h}},a^*_{\mathfrak{h}}]$ :
\beq
\forall \mathfrak{h} \in \{1,\ldots,\mathfrak{g}\},\qquad  \oint_{\mathcal{A}_{\mathfrak{h}}} \dd u_{\mathfrak{h}'} = \delta_{\mathfrak{h}\mathfrak{h'}} \nn
\eeq
Le mieux que l'on puisse dire est :
{\small \bea
\mathcal{K}[f] = g  &  \Leftrightarrow &\exists c_1,\ldots,c_{\mathfrak{g}} \in \mathbb{C} \quad f(x) = \sum_{\mathfrak{h} = 1}^{\mathfrak{g}} c_{\mathfrak{h}}\frac{\dd u_{\mathfrak{h}}}{\dd x} + f_0(x) \nn \\
 &&  f_0(x) = \frac{1}{2i\pi t}\oint_{\mathcal{C}([b,a])} \frac{\dd\xi}{x - \xi}\,\sqrt{\prod_{j = 1}^{\mathfrak{g} + 1} \frac{(\xi - a^*_j)(\xi - b^*_j)}{(x - a^*_j)(x - b^*_j)}}\,\frac{g(\xi)}{2y(\xi)} \nn
\eea}
$\!\!\!$N\'eanmoins, $\mathcal{K}$ peut \^{e}tre invers\'e si l'on se fixe a priori les int\'egrales de cycles $\oint_{\mathcal{A}_{\mathfrak{h}}} f(x)\dd x$.

\vspace{0.2cm}

Dans un r\'egime \`a une coupure, on peut esp\'erer que $\ln Z$ et $W_n$ aient un d\'eveloppement en puissances de $1/N$, en r\'esolvant r\'ecursivement les \'equations de Schwinger-Dyson. C'est ce que nous allons faire. Ce n'est jamais vrai dans un r\'egime \`a plusieurs coupures, car des oscillations rapides avec $N$ apparaissent dans la partie de $W_n$ qui est annihil\'ee par $\mathcal{K}$. Leur origine sera expliqu\'ee en partie~\ref{sec:pluscoup}.

\begin{figure}[h!]
\begin{center}
\includegraphics[width=1.05\textwidth]{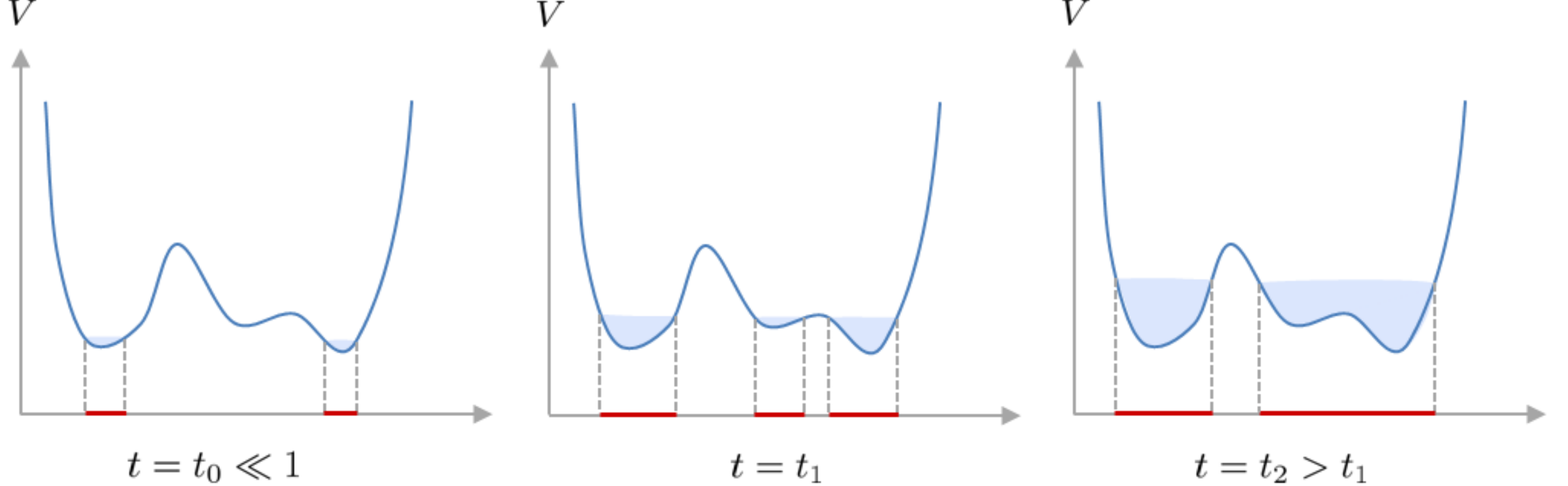}
\caption{\label{fig:vete} Croissance de $\supp\,L^*$ \`a partir des minima de $V$, lorsque $t$ augmente en partant de $t \ll 1$.}
\end{center}
\end{figure}

Pour $t$ assez petit, il est facile de d\'eterminer qualitativement le r\'egime dans lequel on se trouve (Fig.~\ref{fig:vete}). \`{A} $t \rightarrow 0$, les valeurs propres se concentrent au voisinage des minima absolus de $V^{[[0]]}$. Lorsque $t$ augmente, $\supp\,L^*$ est constitu\'e de segments croissants autour des minima globaux, puis d'autres segments au voisinage de minima locaux peuvent apparaitre. Lorsque $t$ augmente encore, les valeurs propres s'\'etalent de plus en plus et certains segments peuvent fusionner. Si $V^{[[0]]}$ est strictement convexe sur $[b,a]$ (donc admet un unique minimum), nous serons toujours dans un r\'egime \`a une coupure. En effet, la fonction
 \beq
x \mapsto \int_b^a \dd L^*(\xi)\ln|x - \xi| - \frac{1}{t}V^{[[0]]}(x) - C \nn
\eeq
 est maintenant strictement concave sur $\mathrm{supp}\,L^*$. Comme elle est nulle sur $\mathrm{supp}\,L^*$ et continue strictement n\'egative sur $\mathbb{R}\setminus\mathrm{supp}\,L^*$ (\'{E}qn.~\ref{eq:vk}), $\mathrm{\supp}\,L^*$ doit \^{e}tre connexe.

\section{Asymptotique en puissances de $1/N$}
\label{sec:puisn}
\subsection{Ce qui est important}
\label{sec:paraa}
\subsubsection{Continuit\'e de $\mathcal{K}^{-1}$}

Notre approche est bas\'ee sur l'hypoth\`ese fondamentale :
\begin{hypo}
\label{hypo1}$\mathcal{K}$ a un inverse, qui est un op\'erateur lin\'eaire continu.
\end{hypo}

\begin{figure}[h!]
\begin{center}
\includegraphics[width=0.9\textwidth]{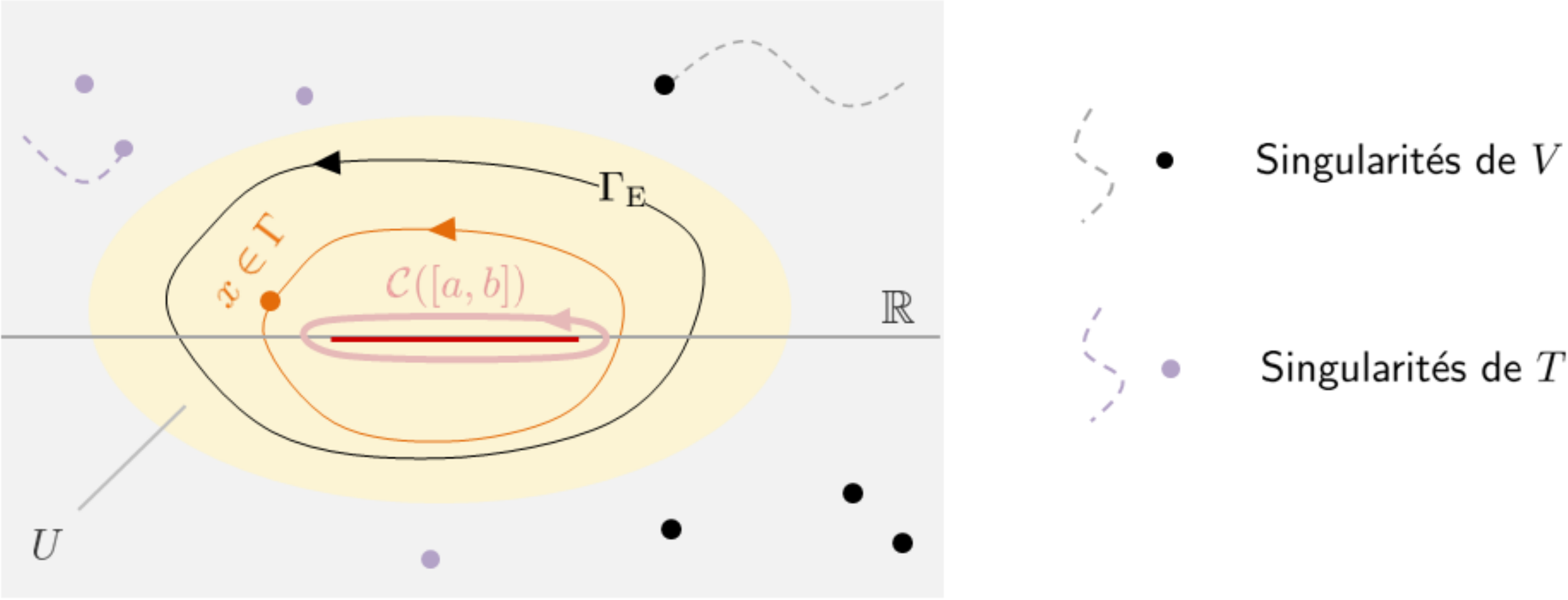}
\caption{\label{fig:c2} Situation des contours.}
\end{center}
\end{figure}

Cette hypoth\`ese est v\'erifi\'ee d\`es que $\mathcal{K}^{-1}$ est donn\'e par une formule de type Tricomi :
\beq
\label{eq:invinv} \forall g \in \mathcal{H}_{1;[b,a]},\quad \mathcal{K}^{-1}[g](x) = \frac{1}{2i\pi}\oint_{\mathcal{C}([b,a])} \frac{\dd\xi}{\xi - x}\,T(x,\xi)\,g(\xi)
\eeq
o\`u $T(x,\xi)$ est une fonction holomorphe dans $U \times U$, o\`u $U$ est un voisinage ouvert de $[b,a]$ priv\'e du segment $[b,a]$ lui-m\^{e}me, et $\mathcal{C}([b,a]) \subseteq U$ entoure $[b,a]$ mais pas $x$. En effet, pla\c{c}ons $x$ sur un contour $\Gamma \subseteq U$ qui d\'efinit la norme $\p \cdot \p_{\Gamma}$ (\'{E}qn.~\ref{eq:norm}), et d\'epla\c{c}ons $\mathcal{C}([b,a])$ vers un contour $\Gamma_{\mathrm{E}} \subseteq U \cap \mathrm{Ext}(\Gamma)$ (Fig.~\ref{fig:c2}).
\beq
\forall x \in \Gamma,\qquad \mathcal{K}^{-1}[g](x) = -T(x,x)g(x) + \frac{1}{2i\pi}\oint_{\Gamma_{\mathrm{E}}} \frac{\dd \xi}{x - \xi}\,T(x,\xi)\,g(\xi) \nn
\eeq
On peut maintenant borner le membre de droite :
\bea
\p \mathcal{K}^{-1}[g] \p & \leq & \p T \p_{\Gamma^2}\p g \p_{\Gamma} + \p T \p_{\Gamma \times\Gamma_{\mathrm{E}}}\,\frac{\ell(\Gamma_{\mathrm{E}})}{2\pi\,d(\Gamma,\Gamma_{\mathrm{E}})}\,\p g \p_{\Gamma_{\mathrm{E}}} \nn \\
& \leq & \Big(1 + \frac{\ell(\Gamma_{\mathrm{E}})}{2\pi\,d(\Gamma,\Gamma_{\mathrm{E}})}\Big)\,\p T \p_{\Gamma_{\mathrm{E}}}\,\p g \p_{\Gamma} \nn
\eea
donc $\mathcal{K}^{-1}$ est un op\'erateur lin\'eaire continu dans l'espace $(\mathcal{H}_{1;[b,a]},\p \cdot \p_{\Gamma})$. Sa norme $\p \mathcal{K}^{-1} \p_{\Gamma}$ est contr\^{o}l\'ee par la distance entre les singularit\'es de $T$ qui sont \`a l'ext\'erieur de $[b,a]$, et le segment $[b,a]$ lui-m\^{e}me. Par exemple, pour le mod\`ele $\beta$ dans le r\'egime \`a une coupure :
\beq
T(\xi,x) = \frac{1}{2y(\xi)}\,\sqrt{\frac{(\xi - a^*)(\xi - b^*)}{(x - a^*)(x - b^*)}} \nn
\eeq
et la norme de $\mathcal{K}^{-1}$ est control\'ee par la distance entre $[b,a]$ et les z\'eros de $y$.

\subsubsection{Ordre de grandeur de $W_n$}

En second lieu, nous allons nous baser sur un contr\^{o}le a priori des $W_n$, que nous \'enon\c{c}ons d'abord sous une forme contraignante :
\begin{hypo}
\label{hypo2} $\forall n \geq 1,\qquad W_n \in O(N^{2 - n})$.
\end{hypo}

Hyp.~\ref{hypo1} garantit que l'on peut appliquer $\mathcal{K}^{-1}$ \`a l'\'{E}qn.~\ref{eq:bjud} ou l'\'{E}qn.~\ref{eq:bjud1}, en contr\^{o}lant la norme \`a une constante multiplicative pr\`es.
Si l'on sait que $W_n \in O(c_n(N))$ pour certaines constantes $c_n(N)$  ($c_1(N) = N$), les \'equations de Schwinger-Dyson (\'{E}qn.~\ref{eq:bjud1}) impliquent :
\beq
c_{n} \leq \frac{1}{N}\,\mathrm{max}(c_{n + 1},c_{n - 1},\{c_{j}c_{n + 1 - j}\quad 2 \leq j \leq n - 1 \}) \nn
\eeq
L'analyse de cette relation montre que des bornes a priori grossi\`eres sur $W_n$, sont suffisantes pour conclure que $W_n$ est un $O(N^{2 - n})$. Par exemple :
\begin{lemme}
\label{laime} Si il existe $\gamma \in [0,1[$ et $\delta \in [0,\infty[$ tel que pour tout $n \geq 2$, $W_n \in O(N^{\gamma n - \delta})$, alors l'hypoth\`ese~\ref{hypo2} est v\'erifi\'ee.
 \end{lemme}

Pour le mod\`ele $\dd\nu_{N}^{[b,a];V}$, en supposant que $V$ est une fonction \`a valeurs r\'eelles, lipschitzienne sur $[b,a]$, mais sans hypoth\`eses sur $\mathrm{supp}\,L^*$, Boutet de Monvel, Pastur et Shcherbina \cite{BPSmat} ont \'etabli des bornes dont on peut d\'eduire $W_n \in O\big((\ln N)^n)$ \cite[Lemme 4.6]{BG11}. Si l'on dispose de l'hypoth\`ese~\ref{hypo1}, le Lemme~\ref{laime} s'applique et justifie dans ce cas l'hypoth\`ese~\ref{hypo2}.

Voyons les premi\`eres cons\'equences de l'hypoth\`ese \ref{hypo2}, en commen\c{c}ant avec l'\'{E}qn.~\ref{eq:bjud}. Dans le membre de gauche, $\mathcal{K}[\delta W_1]$ est le terme dominant quoiqu'il arrive ; dans le membre de droite, il y a des termes d'ordre $1$ donc $-\frac{1}{N}\,W_2(x,x) \in O(1/N)$ est n\'egligeable devant eux\footnote{On a seulement besoin de $W_2 \in o(N)$ \`a ce stade.}. Par cons\'equent\footnote{On utilise la continuit\'e de $\mathcal{K}^{-1}$, ainsi que la continuit\'e des op\'erateurs $\frac{\dd}{\dd x}$ et de $\mathcal{N}_g$ qui est \'evidente d'apr\`es la remarque faite \`a l'\'{E}qn.~\ref{eq:rq}.}, $\delta W_1$ a une limite $\delta^{[[0]]}W_1$ quand $N \rightarrow \infty$ :
\begin{footnotesize}
\bea
\delta^{[[0]]} W_1(x) & = & \mathcal{K}^{-1}\Big[\mathcal{N}_{(V^{[[1]]})'}[W_1^{(0)}](x) - (1 - 1/\beta)\frac{\dd}{\dd x}\big(W_1^{(0)}(x)\big) - \frac{(1 - 1/\beta)}{(x - a)(x - b)}\Big] \nn \\
\label{eq:Wbh} &&
\eea
\end{footnotesize}
$\!\!\!$Les m\^{e}mes arguments s'appliquent \`a l'\'{E}qn.~\ref{eq:bjud1}, qui contient a priori des termes qui sont $O(N^{2 - n})$, tandis que $A_{n + 1} = - \frac{1}{N}W_{n + 1}(x,x,x_I) \in O(N^{-n})$ est d'un ordre de grandeur inf\'erieur. L'\'equation qui d\'etermine l'ordre dominant de $W_n$ ne fait plus intervenir que l'ordre dominant des $W_j$ ($1 \leq j \leq n - 1$). Par cons\'equent, une r\'ecurrence montre que $W_n/N^{2 - n}$ admet une limite $W_n^{(0)}$ pour tout $n \geq 2$, et :
\begin{footnotesize} \bea
&& W_n^{(0)}(x,x_I) = \mathcal{K}^{-1}\Big[-\sum_{\substack{n_1, n_2 \geq 2 \\ n_1 + n_2 = n + 1}} \sum_{J \subseteq I,\,\,|J| = n_1} W_{n_1}^{(0)}(x,x_J)\,W_{n_2}^{(0)}(x,x_{I\setminus J}) + \nn \\
\label{eq:Whb} && \phantom{W_n^{(0)}} - \frac{1}{\beta}\sum_{i \in I} \frac{\dd}{\dd x_i}\Big(\frac{W_{n - 1}^{(0)}(x,x_{I\setminus\{i\}}) - \frac{(x_i - a)(x_i - b)}{(x - a)(x - b)}W_{n - 1}^{(0)}(x_I)}{x - x_i}\Big)\Big]
\eea
\end{footnotesize}

Remarquons que les \'equations de Schwinger-Dyson, m\^{e}me pour $\beta = 1$, contiennent des d\'eriv\'ees de $W_n$. \`{A} une \'etape $l$ de la r\'ecurrence, si les corr\'elateurs sont control\'es pour une norme $\p \cdot \p_{\Gamma}$, le contr\^{o}le \`a l'\'etape $l + 1$ sera valable pour une norme $\p \cdot \p_{\Gamma'}$, o\`u $\Gamma'$ est l\'eg\`erement plus grand\footnote{Je remercie Pavel Bleher pour une correction sur ce point.}, i.e. $\Gamma \subset \mathrm{Int}(\Gamma)$. il faut \'elargir l\'eg\`erement le contour. Ceci peut \^{e}tre pris en compte en introduisant une famille de contours emboit\'es, ce qui est expliqu\'e dans \cite{BG11}. Le prix \`a payer pour cela est que les $O(\cdots)$ concernant $W_n$ ne peuvent \^{e}tre uniformes par rapport \`a $n$.


\subsection{Th\'eor\`eme central limite et cons\'equences}
\label{thcl}
L'existence de $\delta^{[[0]]} W_1 = \lim_{N \rightarrow \infty} W_1 - NW_1^{(0)}$ et la propri\'et\'e $W_2 \in o(N)$ est suffisante pour \'etablir un \textbf{th\'eor\`eme central limite} pour les valeurs propres $\lambda_i$. Pour le mod\`ele $\beta$, c'est un r\'esultat d\'ej\`a connu de Johansson \cite{Johan}, essentiellement par la m\^{e}me m\'ethode.
\begin{prop}
\label{thoqh} Soit $V$ de sorte que l'hypoth\`ese ~\ref{hypo1} soit v\'erifi\'ee ainsi que $W_2 \in o(N)$. Soit une fonction $h\,:\,[b,a] \rightarrow \mathbb{R}$ qui se prolonge analytiquement sur un voisinage de $[b,a]$. Posons :
\bea
X[h] & = & \frac{\sum_{i = 1}^N h(\lambda_i) - N\int_b^a \dd L^*(\xi)\,h(\xi) - M[h]}{\sigma[h]\,\sqrt{N}} \nn \\
M[h] & = & \frac{1}{2i\pi}\oint_{\mathcal{C}([b,a])} \!\!\!\!\!\!\!\!\dd\xi\,(\delta^{[[0]]} W_1)(\xi)\,h(\xi)\nn \\
\sigma[h] & = & -\frac{1}{2i\pi\beta}\,\oint_{\mathcal{C}([b,a])} \!\!\!\!\!\!\!\!\dd \xi\,(\mathcal{K}^{-1}\circ\mathcal{N}_{th'})[W_1^{(0)}](\xi) \nn
\eea
La loi de la variable al\'eatoire $X[h]$, induite par $\dd\nu_{N}^{[b,a];V}$, converge vers la loi normale lorsque $N \rightarrow \infty$.
\end{prop}

Cela vient d'un petit calcul avec le potentiel modifi\'e $V_s(x) = V - \frac{ts}{N\beta}\,h(x)$. Nous allons noter $W_{1;s}$ le corr\'elateur \`a un point pour le potentiel $V_s$ sur $[b,a]$. Les hypoth\`eses assurent la validit\'e de l'\'{E}qn.~\ref{eq:Wbh}, et l'on a :
\bea
\ln\nu_{N}^{[b,a];V}[e^{\sum_{i = 1}^N h(\lambda_i)}] & = & \int_0^1 \dd s \,\partial_s \ln Z_{N}^{[b,a];V_s} \nn \\
& = &  \frac{1}{2i\pi}\oint_{\mathcal{C}([b,a])} \!\!\!\!\!\!\!\!\dd\xi\,W_{1;s}(\xi)\,h(\xi) \nn \\
& = & \frac{1}{2i\pi}\oint_{\mathcal{C}([b,a])} \!\!\!\!\!\!\!\!\dd \xi\,\big(N\,W_{1}^{(0)}(\xi) + \delta^{[[0]]} W_{1;s}(\xi)\big)h(\xi) + o(1) \nn
\eea
Le premier terme est $N\int_b^a \dd L^*(\xi)\,h(\xi)$ d'apr\`es la d\'efinition de la transform\'ee de Stieltjes. En utilisant explicitement l'\'{E}qn.~\ref{eq:Wbh}, $\delta^{[[0]]} W_{1;s} = \delta^{[[0]]} W_1$ + un terme lin\'eaire en $h$, et l'on trouve :
\beq
\label{eq:com}\ln \nu_{N}^{[b,a];V}[e^{\sum_{i = 1}^N h(\lambda_i)}] = N\,\int_{b}^a \dd L^*(\xi)\,h(\xi) + M[h] + \frac{\sigma[h]}{2} + o(1)
\eeq
Notons que $M[h]$ est lin\'eaire et $\sigma[h]$ est quadratique en $h$. L'\'{E}qn.~\ref{eq:com} indique que la transform\'ee de Fourier de la loi de $X[h]$ co\"{i}ncide converge vers celle de la loi normale, d'o\`u la convergence en loi \'enonc\'ee \`a la Prop.~\ref{thoqh}.

\subsection{D\'eveloppement asymptotique complet}
\label{sec:Amj}
Il est temps de syst\'ematiser la strat\'egie initi\'ee \`a la fin du \S~\ref{sec:paraa}. Prenons comme hypoth\`ese de r\'ecurrence au niveau $k \geq -1$ l'assertion \textsf{HR}$(k)$ :
\beq
\forall n \geq 1\,\qquad W_n = \sum_{m = n - 2}^{k} N^{-m} W_{n}^{[[m]]} + N^{-(k + 1)}\delta^{[[k + 1]]}W_{n} \nn
\eeq
o\`u $W_n^{[[m]]} \in \mathcal{H}_{n;\mathrm{supp}\,L^*}$ est connu pour $m \leq k$, et le reste $\delta^{[[k + 1]]}W_n \in \mathcal{H}_{n;[b,a]}$ est un $O(1)$ a priori inconnu. Nous utiliserons \'egalement la notation $\delta^{[[l]]} V$ pour :
\beq
V = \sum_{m = 0}^{l - 1} N^{-m}\,V^{[[m]]} + N^{-l}\delta^{[[l]]} V \nn
\eeq
Pour relier ces notations aux pr\'ec\'edentes, $W_1^{[[-1]]} = W_1^{(0)}$.

\textsf{HR}$(-1)$ est v\'erifi\'ee d'apr\`es \S~\ref{sec:paraa}. Soit $k \geq -1$ et supposons \textsf{HR}$(k)$. Les \'equations de Schwinger-Dyson ~\ref{eq:bjud} et \ref{eq:bjud1} tronqu\'ees \`a un $o(N^{-k})$ pr\`es donnent des relations entre $W_{n}^{[[m]]}$ d\'ej\`a connues. De nouvelles informations se trouvent dans le $o(N^{-k})$.

\vspace{0.2cm}

\noindent $\diamond\,$ Pour $n = 1$ :
{\small \bea
&& N^{-(k + 1)}\Big(\mathcal{K} + \sum_{m = 0}^{k} N^{-(m + 1)}\,W_1^{[[m]]}(x) + N^{-(k + 2)}\,\delta^{[[k + 1]]}W_1(x) \nn \\
&& + \frac{1}{N}(1 - 1/\beta)\frac{\dd}{\dd x} - \frac{1}{N}\,\mathcal{N}_{\delta V'}\Big)[\delta^{[[k + 1]]}W_1](x) \nn \\
&& =  N^{-(k + 1)}\big(E_{1}^{[[k]]}(x) + \Delta_1^{[[k]]}(x)\big) \nn
\eea}
$\!\!\!\diamond\,$ Pour $n \geq 2$ :
{\small \bea
&& N^{-(k + 1)}\Big(\mathcal{K} + \frac{2}{N}\,\delta W_1(x) + \frac{1}{N}(1 - 1/\beta)\frac{\dd}{\dd x} - \frac{1}{N}\,\mathcal{N}_{\delta V'}\Big)[\delta^{[[k + 1]]}W_n](x,x_I)  \nn \\
& = & N^{-(k + 1)}\big(E_{n}^{[[k]]}(x,x_I) + \Delta_n^{[[k]]}(x,x_I)\big) \nn
\eea}

Dans ces expressions, nous avons isol\'e $E_{n}^{[[k]]}$, qui est une expression ind\'ependante de $N$ o\`u interviennent uniquement les $W_{n}^{[[m]]}$ pour $m \in \{0,\ldots,k\}$, donc qui n'a de discontinuit\'e que sur $\mathrm{supp}\,L^*$ :
\begin{footnotesize}
\bea
E_n^{[[k]]}(x,x_I) & = & - W_{n + 1}^{[[k]]}(x,x,x_I) - \sum_{J \subseteq I,\,0 \leq m \leq k} W_{|J| + 1}^{[[m]]}(x,x_J)\,W_{n - |J|}^{[[k - m]]}(x,x_{I\setminus J}) \nn \\
& & - (1 - 1/\beta)\frac{\dd}{\dd x}\big(W_n^{[[k]]}(x)\big) + \sum_{m = 0}^{k} \mathcal{N}_{V^{[[k + 1 - m]]})'}[W_n^{[[m]]}](x,x_I) \nn \\
&& - \frac{1}{\beta}\sum_{i \in I}\frac{\dd}{\dd x_i}\Big(\frac{W_{n - 1}^{[[k]]}(x,x_{I\setminus\{i\}}) - \frac{(x_i - a)(x_i - b)}{(x - a)(x - b)}W_{n -1}^{[[k]]}(x_I)}{x - x_i}\Big)\nn
\eea
\end{footnotesize}
$\!\!\!$Le reste $\Delta_{n}^{[[k]]}$ est un $O(1/N)$ (d'apr\`es \textsf{HR}$(k)$) que nous n'allons pas expliciter, mais qui est dans $\mathcal{H}_{n;[b,a]}$. En cons\'equence, en utilisant la continuit\'e de $\mathcal{K}^{-1}$, $\delta^{[[k + 1]]}W_n$ a une limite $W_n^{[[k + 1]]}$ :
\beq
\label{eq:invj} W_n^{[[k + 1]]}(x,x_I) = \mathcal{K}^{-1}\big[E_{n}^{[[k]]}\big](x,x_I)
\eeq
 qui appartient \`a $\mathcal{H}_{n;\mathrm{supp}\,L^*}$, et le reste $\delta^{[[k + 1]]}W_n - W_{n}^{[[k + 1]]}$ est un $O(1/N)$ dans $\mathcal{H}_{1;[b,a]}$. Ce qui prouve \textsf{HR}$(k + 1)$. L\`a encore, il faut \'elargir le contour $\Gamma$ d\'efinissant la norme \`a chaque \'etape de la r\'ecurrence afin de bien contr\^{o}ler les d\'eriv\'ees intervenant dans $\Delta_n^{[[k]]}$. Finalement, $\Delta_n^{[[k]]}$ est un bien un $O(1/N)$, mais qui n'est pas uniforme par rapport \`a $n$ et \`a $k$

 Ainsi, par r\'ecurrence :
\begin{prop}
\label{props}
Sous les hypoth\`eses~\ref{hypo1} et \ref{hypo2}, $W_n$ admet un d\'eveloppement asymptotique en puissances de $1/N$ :
\beq
W_n = \sum_{k \geq n - 2} N^{-k}\,W_n^{[[k]]} \nn
\eeq
avec $W_n^{[[k]]}$ holomorphe sur $\big(\mathbb{C}\setminus\mathrm{supp}\,L^*\big)^n$.
\end{prop}

En principe, l'\'energie libre $\ln Z$ se d\'eduit de $W_1$ \`a une constante pr\`es, en int\'egrant sur les param\`etres du mod\`ele.
\begin{prop}
\label{orps} Soit $(V_s)_{s \in [0,1]}$ une famille diff\'erentiable de potentiels qui satisfont les hypoth\`eses~\ref{hypo1} et \ref{hypo2}. Alors :
\beq
\ln Z_{N}^{[b,a];V_1} - \ln Z_{N}^{[b,a];V_0} = -\frac{N\beta}{t} \frac{1}{2i\pi}\oint_{\mathcal{C}([b,a])} (\partial_s V_s)(\xi)\,W_{1}^{V_s}(\xi) \nn
\eeq
Par cons\'equent, il existe un d\'eveloppement asymptotique :
\beq
\ln Z_{N}^{[b,a];V_1} -\ln Z_{N}^{[b,a];V_0} = \sum_{k \geq -2} N^{-k}\,F^{[[k]]} \nn
\eeq
\end{prop}

Ces d\'eveloppements asymptotiques sont uniformes tant que les param\`etres du mod\`ele prennent des valeurs \`a distance born\'ee loin de $0$ de celles o\`u $\p \mathcal{K}^{-1} \p_{\Gamma}$ explose ou l'hypoth\`ese \ref{hypo2} est prise en d\'efaut.

\subsection{Lien avec la version $\beta$ de la r\'ecurrence topologique}
\label{sec:lor}
Supposons d'abord que $V$ est ind\'ependant de $N$ et $\beta$. Nous savons maintenant que $\ln Z$ et $W_n$ admettent un d\'eveloppement asymptotique en $1/N$ lorsque $N \rightarrow \infty$ :
\beq
\label{eq:asyn}\ln Z = C_{N,\beta} + \sum_{k \geq -2} N^{-k}\,F^{[[k]]},\qquad W_n = \sum_{k \geq n - 2} N^{-k}\,W_n^{[[k]]}
\eeq
pour une constante $C_{N,\beta}$ ind\'ependante de $V$. Nous pouvons donc appliquer le formalisme de la r\'ecurrence topologique\footnote{C'est simplement une mise en forme des coefficients $F^{[[k]]}$, et $W_n^{[[k]]}$ calcul\'es p\'edestrement avec l'op\'erateur $\mathcal{K}^{-1}$ \`a l'\'{E}qn.~\ref{eq:invj}.} pr\'esent\'e au \S~\ref{sec:ha}. La courbe spectrale $\mathcal{S} = [\Sigma,x,y,B]$ s'\'ecrit\footnote{$W_2^{[[0]]} \propto 1/\beta$ donc $\mathcal{S}$ est ind\'ependante de $\beta$ contre les apparences.} :
\beq
y(x) = W_1^{[[-1]]}(x) - \frac{V'(x)}{2t},\qquad B(x_1,x_2) = \Big(\beta W_2^{[[0]]}(x_1,x_2) + \frac{1}{(x_1 - x_2)^2}\Big)\dd x_1\dd x_2 \nn
\eeq
\label{devo3}et les autres coefficients du d\'eveloppement valent :
\bea
W_n^{[[k]]} & = & \sum_{g = 0}^{\lfloor \frac{k - n}{2}\rfloor + 1} \!\!\beta^{1 - g - n}\Big(1 - \frac{1}{\beta}\Big)^{k - n + 2 - 2g}\,\frac{\omega_n^{(g;k - n + 2 - 2g)}[\mathcal{S}]}{\dd x_1\cdots\dd x_n} \nn \\
 F^{[[k]]} & = & \sum_{g = 0}^{\lfloor \frac{k}{2} \rfloor + 1} \beta^{1 - g}\Big(1 - \frac{1}{\beta}\Big)^{k + 2 - 2g}\,\mathcal{F}^{(g;k + 2 - 2g)}[\mathcal{S}] \nn \\
 \label{eq:asynfu} &&
\eea

Imaginons \`a pr\'esent que l'on s'int\'eresse \`a un potentiel $V(x) =v_{N,\beta}(x)$, qui admet un d\'eveloppement asymptotique quelconque. Une astuce est d'introduire des nouveaux param\`etres $N'$ et $\beta'$, qui n'ont rien \`a avoir avec $N$, le nombre de valeurs propres, et $\beta$, l'exposant de r\'epulsion entre valeurs propres. Si l'on suppose que le mod\`ele avec potentiel $v_{N',\beta'}(x)$ satisfait les hypoth\`eses~\ref{hypo1} et \ref{hypo2} pour toutes valeurs de $N'$ (finie ou infinie) et $\beta'$, il admet pour une courbe spectrale $\mathcal{S} = \mathcal{S}_{N',\beta'}$ le d\'eveloppement ~\ref{eq:asynfu}, qui est uniforme par rapport \`a $N'$. On peut \'egalement calculer le d\'eveloppement asymptotique lorsque $N' \rightarrow \infty$ de  $\omega_n^{(g;l)}[\mathcal{S}_{N',\beta'}]$ et $\mathcal{F}^{(g;l)}[\mathcal{S}_{N',\beta'}]$. Gr\^{a}ce \`a l'uniformit\'e, il suffit de substituer $N' = N$, et de regrouper les termes de m\^{e}me ordre en $N$ dans $\ln Z$ et $W_n$ : le mod\`ele avec potentiel $v_{N,\beta}(x)$ admet lui aussi un d\'eveloppement asymptotique quand $N \rightarrow \infty$. Naturellement, si le d\'eveloppement de $v_{N,\beta}(x)$ contient des termes qui ne sont pas des puissances de $N$, il en sera de m\^{e}me du d\'eveloppement de $\ln Z$ et $W_n$. Il se calcule nonobstant en regroupant les termes donn\'es par la version $\beta$ de la r\'ecurrence topologique.

\subsection{Mod\`eles de matrices g\'en\'eralis\'es}
\label{sec:gene}
Ces r\'esultats peuvent \^{e}tre \'etendus \`a des mod\`eles de matrices g\'en\'eralis\'es :
\beq
\dd\nu(\lambda) = \prod_{i = 1}^N \dd \lambda_i\,\mathbf{1}_{[b,a]}(\lambda_i)\,e^{-\frac{N}{t}V(\lambda_i)}\,\prod_{1 \leq i < j \leq N} |K(\lambda_i,\lambda_j)| \nn
\eeq
Le th\'eor\`eme d'unicit\'e de $L^*$ et les bornes a priori de \cite{BPSmat} sont toujours valides pourvu que la fonctionnelle $\mathcal{E}_{\mathrm{K}}$ soit strictement convexe :
\beq
\label{eq:fonce}\mathcal{E}_{\mathrm{K}}[\mu] = - \iint_{[b,a]^2} \!\!\!\!\!\dd\mu(\xi)\dd\mu(\eta) \ln |K(\xi,\eta)| + \frac{1}{t}\int_{b}^a \dd\mu(\xi)\,V(\xi)
\eeq
Les hypoth\`eses \ref{hypo1} et \ref{hypo2} seront alors valides dans un r\'egime \`a une coupure. La condition de convexit\'e s'\'ecrit toujours :
\beq
\Delta_{\mathrm{K}}[\mu_0] = - \iint_{[b,a]^2} \!\!\!\!\!\dd\mu_0(\xi)\dd\mu_0(\eta)\,\ln |K(\xi,\eta)| \leq 0 \nn
\eeq
 pour toute mesure sign\'ee $\mu_0$ telle que $\int\dd\mu_0 = 0$. Si l'on d\'ecompose $K(\xi,\eta) = \prod_{d} K_d(\xi,\eta)^{\alpha_d}$, on peut utiliser \`a nouveau l'astuce \'{E}qn.~\ref{eq:astc} et calculer :
\beq
\Delta_{\mathrm{K}}[\mu_0] = -\sum_{d} \int_{0}^{\infty} \frac{\dd u}{u}\,\mathrm{Re}\Big(\alpha_d\,\int_{[b,a]^2}\!\!\!\!\! \dd\mu_0(\xi)\dd\mu_0(\eta)\,e^{iK_d(\xi,\eta)u}\Big) \nn
\eeq

Le mod\`ele $\On$ trivalent (et sa d\'eformation $\beta$) est un bon exemple pour voir cette condition \`a l'{\oe}uvre :
\beq
\label{eq:Krk1} K(\xi,\eta) = (\xi - \eta)^{2\beta}\,(\xi + \eta)^{-\mathfrak{n}}
\eeq
Pour $\beta = 1$, ce mod\`ele est \'etudi\'e du point de vue des int\'egrales formelles de matrices en partie~\ref{sec:OnON}, en particulier dans le r\'egime $|\mathfrak{n}| < 2$. Ici, consid\'erons l'int\'egrale convergente associ\'ee \`a \'{E}qn.~\ref{eq:Krk1}. On trouve :
\beq
\Delta_{\mathrm{K}}[\mu_0] = \int_0^{\infty} \frac{\dd u}{u}\Big[(\mathfrak{n} - 2\beta)\big(\widehat{\mu_0}_\mathrm{R}(u)\big)^2 + (-\mathfrak{n} - 2\beta)\big(\widehat{\mu_0}_{\mathrm{I}}(u)\big)^2\Big] \nn
\eeq
o\`u $\widehat{\mu_0}_{\mathrm{R,I}}$ sont les parties r\'eelles et imaginaires de $\widehat{\mu_0}$. Si $\mathfrak{n} < 2\beta$, on peut conclure que $\mathcal{E}_{\mathrm{K}}$ est strictement convexe. De plus,  lorsque le potentiel $V$ conduit \`a un r\'egime \`a une coupure, l'op\'erateur $\mathcal{K}^{-1}$ a \'et\'e construit dans \cite{TheseBE,EKOn} et prend la forme \'{E}qn.~\ref{eq:invinv}.

\begin{prop}
Les Propositions~\ref{props} et \ref{orps} sont aussi valides pour l'int\'egrale de matrice convergence du mod\`ele $\On$ trivalent (et de sa d\'eformation $\beta)$, lorsque $\mathfrak{n} < 2\beta$ et dans un r\'egime \`a une coupure.
\end{prop}

En cons\'equence des r\'esultats de la partie~\ref{sec:OnON}, les coefficients du d\'eveloppement seront donn\'es par la version $\beta$ de la r\'ecurrence topologique, appliqu\'ee \`a la courbe spectrale du mod\`ele\footnote{Comme d'habitude, la courbe spectrale d'un mod\`ele \`a $\beta$ fini ne d\'epend pas de $\beta$.} $\On$ (\'{E}qn.~\ref{eq:Wms}).

\section{Asymptotique dans un r\'egime \`a plusieurs coupures}
\label{sec:pluscoup}
Dans cette partie, nous allons suivre la m\'ethode propos\'ee par \cite{Ecv} pour obtenir de fa\c{c}on heuristique l'asymptotique des int\'egrales de matrices convergentes, dans un r\'egime \`a plusieurs coupures.
La formule finale (\'{E}qn.~\ref{eq:resul}) sera utile au Chapitre~\ref{chap:int}. Au passage, nous verrons quels probl\`emes importants il faudrait r\'esoudre pour d\'emontrer cette asymptotique.

\subsection{Une bonne base de contours}

\begin{figure}[h!]
\begin{center}
\includegraphics[width=0.98\textwidth]{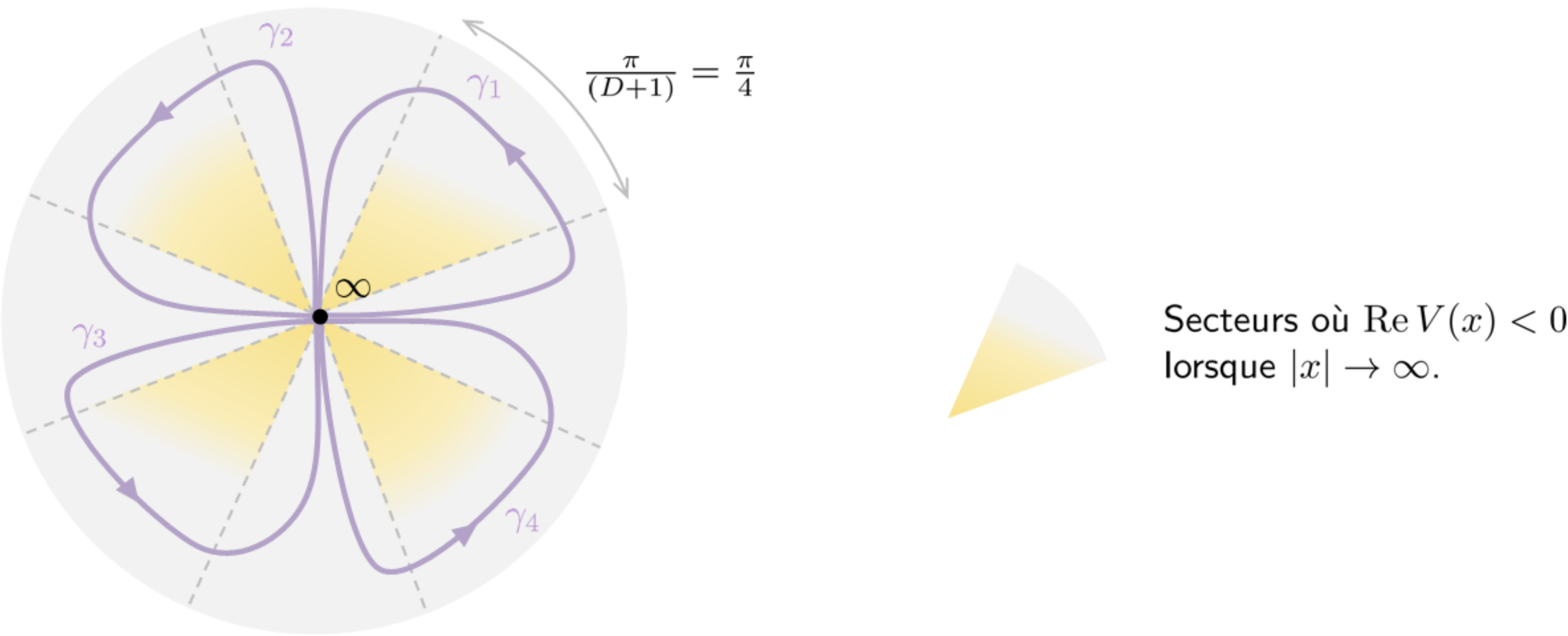}
\caption{\label{fig:homolo} Pour un potentiel quartique ($D = 3$), les valeurs propres peuvent \^{e}tre plac\'ees sur trois contours ind\'ependants parmi $\gamma_1,\ldots,\gamma_4$. Comme $\sum_{j} \gamma_i$ est contractible dans $\mathbb{C}$ et $\dd\nu(\lambda)$ enti\`ere, $\int_{(\sum_j \gamma_j)^N} \dd\nu(\lambda) = 0$.}
\end{center}
\end{figure}

Nous nous int\'eresserons ici \`a une mesure \`a valeurs complexes :
\beq
\dd\nu_N(\lambda) = \prod_{i = 1}^N \dd \lambda_i\,e^{-\frac{N}{t}V(\lambda_i)}\,\prod_{1 \leq i < j \leq N} K(\lambda_i,\lambda_j) \nn
\eeq
o\`u $K(\xi,\eta)$ est une fonction enti\`ere de $\xi$ et $\eta$. Par exemple, $K(\xi,\eta) = (\xi - \eta)^{2\beta}$ est admissible ssi $\beta$ est un entier. Nous supposerons \'egalement, par souci de simplicit\'e, que $V$ est un polyn\^{o}me de degr\'e $D + 1$ \`a coefficients complexes, ind\'ependants de $N$. Au lieu de placer les $\lambda_i$ par d\'efaut sur l'axe r\'eel, nous allons les placer sur des cycles $\gamma_{j_i}$ du plan complexe. $\dd\lambda_i$ d\'esigne alors une mesure curviligne quelconque sur $\gamma_{j_i}$. Puisque $\frac{\dd\nu_N(\lambda)}{\dd\lambda_1\cdots\dd\lambda_N}$ est une fonction enti\`ere des $\lambda_i$, $\int_{\prod_i \gamma_{j_i}}\!\! \dd\nu(\lambda)$ est inchang\'e si l'on d\'eforme continument les $\gamma_{j_i}$. Les seuls contours qui donnent une contribution non nulle sont ceux qui passent par l'infini. Et, afin que l'int\'egrale soit convergente, ils doivent le faire dans la r\'egion o\`u $\Re V(x) > 0$ lorsque $|x| \rightarrow \infty$. Par cons\'equent, il existe $D$ contours ind\'ependants $\gamma_{1},\ldots,\gamma_{D}$ sur lesquels on peut placer les valeurs propres (Fig.~\ref{fig:homolo}). \label{fonp2}D\'efinissons une fonction de partition en pla\c{c}ant $p_j = N\epsilon_j$ valeurs propres sur $\gamma_{j}$ :
\beq
Z_{N}(\epsilon_1,\ldots,\epsilon_D) = \frac{1}{\prod_{j = 1}^D (N\epsilon_j)!}\,\int_{\Gamma} \dd\nu_N(\lambda),\quad \Gamma = \prod_{j = 1}^{D} \gamma_{j}^{N\epsilon_j} \nn
\eeq
Les $\epsilon_j$ sont les \textbf{fractions de remplissage}. \label{ffra2}Comme $\sum_{j = 1}^{D} \epsilon_j = 1$, il n'y a que $D - 1$ fractions de remplissage ind\'ependantes.

Puisque $V$ cro\^{i}t assez vite \`a l'infini le long des $\gamma_{j}$, les \'equations de Schwinger-Dyson associ\'ees \`a $Z_N(\epsilon_1,\ldots,\epsilon_D)$ ne contiennent pas de termes de bord. De plus, elles sont ind\'ependantes des $\epsilon_j$. L'exemple du mod\`ele \`a une matrice hermitienne $M = \mathrm{diag}(\lambda_i)$ permet de bien comprendre cette libert\'e. L'\'equation de Schwinger-Dyson au rang $1$ fait intervenir la quantit\'e a priori inconnue :
\beq
P_1(x) = \Big\langle \Tr \frac{V'(x) - V'(M)}{x - M}\Big\rangle \nn
\eeq
Il est clair que $P_1(x)$ est un polyn\^{o}me de degr\'e $D - 1$, dont le coefficient dominant vaut $Nt_{D}$ si l'on a not\'e $\frac{t_{D}}{D + 1}$ celui de $V(x)$. $P_1$ contient donc $D - 1$ coefficients inconnus, i.e. autant que de fractions de remplissage ind\'ependantes. Ce d\'ecompte est rassurant, mais ne prouve pas que les coefficients de $P_1$ sont en bijection avec les fractions de remplissage (la situation est en g\'en\'eral plus compliqu\'ee).

Nous allons supposer :
\begin{hypo}
\label{hypo3} Il existe une base d'homologie de contours $\gamma_{1},\ldots,\gamma_{D}$, de sorte que, \`a fractions de remplissage fix\'ees $\epsilon = (\epsilon_1,\ldots,\epsilon_{D'})$, $Z_N(\epsilon)$ admette un d\'eveloppement en puissances de $1/N$ :
\beq
\ln Z_{N}(\epsilon) = \sum_{k \geq -2} N^{-k}\,F^{[[k]]}(\epsilon) \nn
\eeq
\end{hypo}

\`{A} l'heure actuelle, cette hypoth\`ese n'a re\c{c}u de preuve compl\`ete que dans le mod\`ele \`a une matrice hermitienne avec potentiel polyn\^{o}mial. M\^{e}me dans ce mod\`ele, la preuve est indirecte. En effet, les techniques de minoration/majoration utilis\'ees pour la partie~\ref{sec:puisn} sont valables pour des mesures de probabilit\'e, et il n'est pas facile de les \'etendre au cas o\`u $\dd\nu$ est une mesure complexe. D\'emontrer le Lemme~\ref{lfini}, qui assure que les valeurs propres vivent dans une r\'egion born\'ee de $\mathbb{C}$ \`a des configurations de poids exponentiellement petit pr\`es, demanderait d\'ej\`a de nouvelles astuces. Qualitativement, on s'attend \`a ce que l'hypoth\`ese \ref{hypo3} soit vraie, avec pour $\gamma_{j}$ une base de chemins de plus grande pente pour $V$. Dans ce cas, une valeur propre sur $\gamma_j$ ne voit alors qu'un potentiel avec un seul minimum global $x_{0;j}$, donc les $N\epsilon_j$ valeurs propres vivant sur $\gamma_j$ devraient se condenser sur une partie connexe $\gamma^{(0)}_{j}$ de $\gamma_{(j)}$ autour de $x_{0;j}$. Si c'est le cas, on a alors une \'egalit\'e, valable ordre par ordre en $1/N$ :
\beq
\frac{1}{2i\pi}\oint_{\mathcal{C}(\gamma^{(0)}_{j})} \dd x\,W_n(x,x_I) = \delta_{n,1}\,N\epsilon_{j} \nn
\eeq
Connaissant ces int\'egrales de cycles, on pourrait inverser l'op\'erateur $\mathcal{K}$ qui intervient dans les \'equations de Schwinger-Dyson, et d\'emontrer l'existence d'un d\'eveloppement en $1/N$.
Les coefficients $F^{[[k]]}(\epsilon)$ seront automatiquement calcul\'es par la r\'ecurrence topologique appliqu\'ee \`a la courbe spectrale $\mathcal{S}_{\epsilon}$ du mod\`ele. C'est une courbe spectrale de genre maximal $D - 1$\footnote{On rappelle que les fractions de remplissage sont en correspondance avec les formes holomorphes sur la courbe spectrale : son genre est le nombre de fractions de remplissage ind\'ependantes et non nulles.}. Elle est caract\'eris\'ee, en sus des \'equations de boucles lin\'eaires, par les int\'egrales de cycles (cf. \S~\ref{sec:bout}) :
\beq
\label{eq:bui}\frac{1}{2i\pi}\oint_{\mathcal{A}_j} y\dd x = -\epsilon_j,\qquad \oint_{\mathcal{A}_j} B = 0,\qquad \mathcal{A}_j = \mathcal{C}(\gamma^{(0)}_j)
\eeq

\subsection{Somme sur les fractions de remplissage}
\label{sec:fraco}
Cherchons \`a calculer l'asymptotique de :
\beq
Z_N^{\gamma} = \frac{1}{N!}\int_{\gamma^n} \dd\nu(\lambda)\nn
\eeq
o\`u $\gamma$ est un contour sans bords du plan complexe, en acceptant l'hypoth\`ese~\ref{hypo3}. $\gamma$ se d\'ecompose sur un sous-ensemble des chemins de la base $\gamma = \sum_{j = 1}^{D'} c_j\,\gamma_j$ o\`u $c_j \in \mathbb{Z}^*$, dans le sens o\`u :
\beq
\label{somu} Z_{N}^{\gamma} = \sum_{p_1 + \cdots + p_{D'} = N} c_1^{p_1}\cdots c_{D'}^{p_{D'}}\,Z_{N}(p_1/N,\ldots,p_{D'}/N,0,\ldots,0)
\eeq
On peut d'ailleurs donner un sens au membre de gauche pour $c_j \in \mathbb{C}^*$, que l'on va repr\'esenter $c_j = e^{2i\pi\nu_j}$. \'{E}qn.~\ref{somu} est une somme toutes les configurations possibles de fractions de remplissage. Chaque terme admet un d\'eveloppement asymptotique $\exp\big(\sum_{k \geq -2} N^{-k}\,F^{[[k]]}(\epsilon)\big)$, et poss\`ede une phase, car il a fallu introduire des chemins $\gamma_1,\ldots,\gamma_{D'}$ passant par le plan complexe, m\^{e}me si $\gamma$ est au d\'epart sur l'axe r\'eel. L'interf\'erence de tous ces termes donne naissance \`a des oscillations rapides avec $N$, et l'asymptotique de la somme ne peut \^{e}tre en g\'en\'eral de la forme $\exp\big(\sum N^{-k}\,f^{[[k]]}\big)$.

On s'attend \`a ce que les configurations $\mathbf{p}$ voisines d'un maximum global de $\Re\,F^{[[-2]]}(\epsilon)$ dominent la somme. Ceci sugg\`ere, heuristiquement, que relever la contrainte $p_1 + \cdots p_{D'} = N$ dans la somme n'a qu'un effet exponentiellement petit. On peut m\^{e}me \`a ce prix \'etendre la somme \`a $p_j \in \mathbb{Z}$, et l'on conjecture $Z_N^{\gamma} = \widetilde{Z}_N^{\gamma} + O(e^{-NC})$, avec :
\beq
\label{eq:jko}\widetilde{Z}_{N}^{\gamma} = \sum_{p_1,\ldots,p_{D'} \in \mathbb{Z}} c_1^{p_1}\cdots c_{D'}^{p_{D'}}\,\exp\Big(\sum_{k \geq -2} N^{-k}\,F^{[[k]]}(\mathbf{p}/N)\Big)
\eeq
La fonction de partition modifi\'ee $\widetilde{Z}_{N}^{\gamma}$ est plus facile \`a manier. Nous allons la mettre en forme pour arriver au r\'esultat final \'{E}qn.~\ref{eq:resul}.

\subsubsection{Maxima et condition de Boutroux}

Comme nous avons utilis\'e plusieurs notations suivant le contexte, rappelons que l'ordre dominant de $\ln Z(\epsilon)$, le pr\'epotentiel de $\mathcal{S}_{\epsilon}$, et la valeur $\mathcal{E}(L^*_{\epsilon})$ de la fonctionnelle \'{E}qn.~\ref{eq:45} pour la mesure d'\'equilibre \`a fractions de remplissage fix\'ee, d\'esignent le m\^{e}me objet.
\beq
F^{[[-2]]}(\epsilon) = \beta\mathcal{F}^{(0)}[\mathcal{S}_{\epsilon}] = -\mathcal{E}(L^*_{\epsilon}) \nn
\eeq
\label{cycsym3}Les cycles $(\mathcal{B}_j)_j$ qui forment avec les cycles $(\mathcal{A}_j)_j$ une base symplectique de $\mathcal{S}_{\epsilon}$ sont duaux aux variations des fractions de remplissage (\S~\ref{sec:geomspec}) :
\beq
\frac{\partial F^{[[-2]]}}{\partial \epsilon_j} = -\oint_{\mathcal{B}_j} y\dd x,\quad \frac{\partial^2 F^{[[-2]]}}{\partial \epsilon_j\partial\epsilon_{j'}} = \oint_{\mathcal{B}_j}\oint_{\mathcal{B}_{j'}} B = 2i\pi\,\tau_{jj'}\nn
\eeq
o\`u $\tau_{jj'}$ est la matrice des p\'eriodes de $\mathcal{S}_{\epsilon}$.

Si $\epsilon_*$ est r\'eel et est un minimum local de $\Re\,F^{[[-2]]}(\epsilon)$, $\mathcal{S}_{\epsilon^*}$ v\'erifie\footnote{L'existence d'un minimum local dans le domaine $\sum \epsilon_j = 1$, $\epsilon_j \geq 0$ est assur\'ee par $\Im \tau > 0$. Il y a des subtilit\'es car certains $\epsilon^*_j$ pourraient \^{e}tre nuls, et nous admettrons que la restriction initiale \`a un $D'$uple de fractions de remplissages non nulles, avec $D' \leq D$, permet d'en tenir compte. Il y a aussi des difficult\'es importantes pour identifier les cycles $\mathcal{A}_j$ lorsque $\supp\,L^*_{\epsilon^*}$ n'est pas une r\'eunion d'arcs (mais plut\^{o}t une r\'eunion d'arbres). Pour cette raison, le probl\`eme de Boutroux n'est pas r\'esolu en g\'en\'eral. Seul le cas des courbes spectrales hyperelliptiques est bien compris \cite{BerBou}.} :
\beq
\Re\Big(\oint_{\mathcal{A}_j} y \dd x\Big) = 0,\quad \Re\Big(\oint_{\mathcal{B}_j} y\dd x\Big) = 0 \nn
\eeq
\label{boubou2} Cela signifie que $\Re \int_{\mathcal{C}} y\dd x$ est nul pour tout cycle $\mathcal{C}$. Le \textbf{probl\`eme de Boutroux} consiste \`a trouver de tels $\mathcal{S}_{\epsilon^*}$ sous la contrainte \'{E}qn.~\ref{eq:bui}.

\subsubsection{Mise en forme}

On choisit un $\epsilon^*$ quelconque, pas n\'ecessairement un minimum pour $\Re\,F^{[[-2]]}(\epsilon)$, ce qui sera comment\'e plus tard. Nous allons d\'evelopper $F^{[[k]]}(\mathbf{p}/N)$ en s\'erie de Taylor au voisinage de $\mathbf{p}/N = \epsilon_*$, en mettant de c\^{o}t\'e les contributions divergentes quand $N \rightarrow \infty$ (\`a $\epsilon^*$ fix\'e) dans chaque terme exponentiel. Quelques notations pr\'eliminaires :

\vspace{0.2cm}

\noindent $\diamond\,$ On note $Z^* = \exp\Big(\sum_{k \geq 2} N^{-k}F^{[[k]]}(\epsilon^*)\Big)$, et $\mathbf{p}^* = N\,\epsilon^*$.

\vspace{0.2cm}

\noindent $\diamond\,$ Si $\mathbf{q}$ et $\mathbf{r}$ sont deux $D'$uples, $\mathbf{q}\cdot\mathbf{r} = \sum_{j = 1}^{D'} q_jr_j$ d\'esigne leur produit scalaire.

\vspace{0.2cm}

\noindent $\diamond\,$ On note $F^{[[k]],(\ell)}_*$ la diff\'erentielle d'ordre $\ell$ de $F^{[[k]]}$ par rapport au $D'$uple $\epsilon$, \'evalu\'ee en $\epsilon = \epsilon^*$. Autrement dit, pour tout $D'$uple $\mathbf{q}_1,\ldots,\mathbf{q}_{\ell}$ :
\beq
 F^{[[k]],(\ell)}_*\cdot(\mathbf{q}_1,\ldots,\mathbf{q}_{\ell})= \sum_{j_1 = 1}^{D'}\cdots\sum_{j_\ell = 1}^{D'} \frac{\partial^{\ell} F^{[[k]]}}{\partial \epsilon_{j_1}\cdots\partial \epsilon_{j_{\ell}}}(\epsilon^*)\,q_{j_1}\cdots q_{j_{\ell}} \nn
 \eeq

\noindent $\diamond\,$ $\mathbf{q}^{\otimes \ell}$ d\'esigne la suite $(\mathbf{q},\ldots,\mathbf{q})$ de longueur $\ell$.

\vspace{0.2cm}

\begin{footnotesize} \bea
\widetilde{Z}_N^{\gamma} & = & \sum_{p_1,\ldots,p_{D'} \in \mathbb{Z}} Z^*\,\exp\Big\{2i\pi\nu\cdot\mathbf{p}^* + \big(NF_*^{[[-2]];(1)} + F_*^{[[-1]];(1)} + 2i\pi\nu\big)\cdot(\mathbf{p} - \mathbf{p}^*) \nn \\
&& \phantom{\sum_{p_1,\ldots,p_{D'} \in \mathbb{Z}} Z^*\,\exp} + \frac{F^{[[-2]];(2)}_*}{2}\cdot(\mathbf{p} - \mathbf{p}^*)^{\otimes 2}\Big\} \nn \\
&& \times \Big\{\sum_{r = 0}^{\infty} \frac{1}{r!}\sum_{\substack{k_j \geq -2,\,\,\ell_j \geq 1 \\ k_j + \ell_j \geq 1}} N^{-\sum_{j} k_j + \ell_j} \Big[\prod_{j = 1}^r \frac{F^{[[k_j]],(\ell_j)}_*}{\ell_j !} \cdot(\mathbf{p} - \mathbf{p}^*)^{\otimes \ell_j}\Big]\Big\} \nn
\eea
\end{footnotesize}

Si la seconde ligne n'existait pas, nous aurions une somme qui ressemble \`a une fonction th\^{e}ta de genre $D'$ :
\beq
\label{theq}\Theta(\mathbf{w}|\tau) = \sum_{p_1,\ldots,p_{D'}\in \mathbb{Z}} e^{2i\pi \nu \cdot \mathbf{p}^* + 2i\pi \sum_{j} w_j (p_j - p_{j'}^*) + i\pi\sum_{j,j'}\tau_{jj'} (p_j - p_j^*)(p_{j'} - p_{j'}^*)}
\eeq
avec $\mathbf{w}$ et $\tau$ prenant les valeurs :
\beq
\mathbf{w}^* = \frac{1}{2i\pi}\big(NF_*^{[[-2]];(1)} + F^{[[-1]];(1)}_*\big) + \nu,\qquad \tau^* = \frac{F^{[[-2]];(2)}_*}{2i\pi}  \nn
\eeq
Cette s\'erie converge car $\mathrm{Im}\,\tau^*$ est d\'efinie positive. Remarquons qu'il suffit de d\'eriver $\ell$ fois \'{E}qn.~\ref{theq} par rapport \`a $\mathbf{w}$ afin de calculer la m\^{e}me somme avec une insertion de $(\mathbf{p} - \mathbf{p}^*)^{\otimes \ell}$. Finalement :
{\small
\bea
\label{eq:resul} \widetilde{Z}_N^{\gamma} & = & \exp\Big(\sum_{k \geq -2} N^{-k}\,F^{[[k]]}_*\Big) \\
& & \times\,\Big\{\sum_{r = 0}^{\infty} \frac{1}{r!} \sum_{\substack{k_j \geq -2,\,\,\ell_j \geq 1 \\ k_j + \ell_j \geq 1}}\,N^{-\sum_{j} (k_j + \ell_j)} \Big[\prod_{j = 1}^{r} \frac{F^{[[k_j]];(\ell_j)}_*}{\ell_j !}\cdot\frac{\nabla_{\mathbf{w}}^{\otimes \ell_j}}{(2i\pi)^{\ell_j}}\Big]\Big\}\Theta(\mathbf{w} = \mathbf{w}^*|\tau = \tau^*) \nn
\eea}
$\!\!\!$Le premier facteur est simplement le d\'eveloppement asymptotique lorsque $N \rightarrow \infty$ de $Z_N(\epsilon^*)$. Le second produit toute une s\'erie de d\'eriv\'ees de fonctions th\^{e}ta, avec des pr\'efacteurs $N^{-K}$ ($K \geq 0$), et dont la variable d\'epend lin\'eairement de $N$. \'{E}crivons explicitement le d\'ebut de cette s\'erie, en notant $\Theta^{(\ell)}_* = (2i\pi)^{-\ell_j}\,\nabla_{\mathbf{w}}^{\otimes \ell}\,\Theta(\mathbf{w} = \mathbf{w}^*,\tau = \tau^*)$ :
{\small \bea
\widetilde{Z}_N^{\gamma} & = & \exp\Big(N^{2}F^{[[-2]]}_* + NF^{[[-1]]}_* + F^{[[0]]}_*\Big) \nn \\
&& \times \left\{\Theta_* + \frac{1}{N}\Big(F^{[[1]]}_*\,\Theta_* + F^{[[0]];(1)}_*\,\Theta'_* + \frac{F^{[[-1]];(2)}_*}{2}\,\Theta''_{*} + \frac{F^{[[-2]];(3)}_*}{6}\,\Theta'''_*\Big) + O(1/N^2)\right\} \nn
\eea}

\subsubsection{Commentaire}

La d\'efinition standard de la fonction th\^{e}ta de Riemann est :
\beq
\theta(\mathbf{w}|\tau) = \sum_{\mathbf{p} \in \mathbb{Z}^{D'}} e^{i\pi\mathbf{p}\cdot \tau\mathbf{p} + 2i\pi \mathbf{w}\cdot\mathbf{p}} \nn
\eeq
C'est une fonction p\'eriodique suivant $\mathbb{Z}^{D'}$ et quasi \label{qausi}p\'eriodique suivant $\tau(\mathbb{Z}^{D'})$ :
\beq
\forall \mathbf{m},\mathbf{n} \in \mathbb{Z}^{D'},\quad \theta\big(\mathbf{w} + \mathbf{m} + \tau\mathbf{n}|\tau\big) = e^{-i\pi\big(2\mathbf{w} + \tau\mathbf{n}\big)\cdot\mathbf{n}}\,\theta(\mathbf{w}|\tau) \nn
\eeq
La fonction $\Theta$ lui est reli\'ee par :
\beq
\Theta(\mathbf{w}|\tau) = e^{i\pi \mathbf{p}^*\cdot(\tau\mathbf{p}^* + 2\nu - 2\mathbf{w})}\,\theta(\mathbf{w} - \tau\mathbf{p}^*|\tau) \nn
\eeq
et en cons\'equence est quasi p\'eriodique dans les directions $\mathbb{Z}^{D'}$ et $\tau(\mathbb{Z}^{D'})$.

L'\'{E}qn.~\ref{eq:resul} est valable pour n'importe quel $D'$uple $\epsilon^*$ choisi, et $\widetilde{Z}_N^{\gamma}$ est ind\'ependant de ce choix. Si l'on choisit $\epsilon^*$ qui maximise $\Re\,F^{[[-2]]}$ :
\beq
\mathbf{w}^* = \frac{N\,\Im F^{[[-2]];(1)}_*}{2\pi} + \Big(\frac{F^{[[-1]];(1)}_*}{2i\pi} + \nu\Big) \nn
\eeq
Lorsque $N$ d\'efile, le pr\'efacteur de $N^{-K}$ dans l'\'{E}qn.~\ref{eq:resul} a donc un comportement quasi p\'eriodique avec $N$.

\subsection{Conclusion}

Les m\'ethodes bas\'ees sur les \'equations de Schwinger-Dyson permettent d'\'etablir le d\'eveloppement asymptotique d'une grande classe de mod\`eles \`a une matrice dans un r\'egime \`a une coupure.
Des ajouts seraient n\'ecessaires pour d\'emontrer avec ces outils l'asymptotique dans un r\'egime \`a plusieurs coupures. Cela passe par une g\'en\'eralisation de la partie~\ref{sec:puisn} au cas o\`u la mesure $\dd\nu(\lambda)$ est complexe. Une extension aux mod\`eles \`a plusieurs matrices est aussi souhaitable. N\'eanmoins, le r\'esultat attendu (\'{E}qn.~\ref{eq:resul}) est bien compris, il s'exprime uniquement en termes d'objets d\'efinis par la r\'ecurrence topologique du Chapitre~\ref{chap:toporec} appliqu\'ee \`a la courbe spectrale du mod\`ele de matrice.

Ces m\'ethodes sont bien s\^{u}r applicables aux mod\`eles de matrices int\'egrables, comme les mod\`eles $\beta \in \{1/2,1,2\}$. On peut esp\'erer que ces outils pourront suppl\'eer l'approche de Riemann-Hilbert, l\`a o\`u elle n'a pas encore \'et\'e appliqu\'ee en raison de difficult\'es techniques. Un enjeu important serait :
\begin{prob}
\label{rpob}\'{E}tablir le d\'eveloppement asymptotique \'{E}qn.~\ref{eq:resul} pour la chaine de matrice \'{E}qn.~\ref{sec:chain} avec potentiels polyn\^{o}miaux (voire plus g\'en\'eraux).
\end{prob}
Cela reviendrait \`a \'etablir le d\'eveloppement asymptotique des polyn\^{o}mes biorthogonaux. Les courbes de Boutroux jouent un r\^{o}le important pour cet asymptotique. Une autre interpr\'etation de $L^*$ assure que $\mathrm{supp}\,L^*$ est le lieu d'accumulation des z\'eros des polyn\^{o}mes (bi)orthogonaux lorsque $N \rightarrow \infty$, et que $L^*$ est leur densit\'e \cite{Defcours}. Le probl\`eme inverse est tout aussi int\'eressant, et intimement reli\'e au probl\`eme de Boutroux :
\begin{prob}
Construire des familles de polyn\^{o}mes (bi)orthogonaux \`a coefficients complexes, dont les z\'eros s'accumulent d'une fa\c{c}on prescrite sur une r\'eunion d'arcs ou d'arbres dans le plan complexe.
\end{prob}

\section{Statistique de $\lambda_{\mathrm{max}}$}
\label{sec:stata}
La th\'eorie des matrices al\'eatoires s'int\'eresse \`a toutes les questions imaginables sur la statistique des $\lambda_i$. Apr\`es le choix d'une bonne \'echelle, quelle est la loi de la $i$-\`eme valeur propre (par ordre d\'ecroissant), de l'espacement entre deux valeurs propres, \ldots{} dans la limite $N \rightarrow \infty$ ? Et, est-ce que ces lois sont robustes lorsque les param\`etres du mod\`ele changent, se rangent-elles dans des classes d'universalit\'e ? Dans l'article \cite{BEMN} \'ecrit avec Bertrand Eynard, Satya Majumdar et C\'eline Nadal, nous avons illustr\'e l'int\'er\^{e}t des m\'ethodes bas\'ees sur la r\'ecurrence topologique, en \'etudiant la distribution du maximum des valeurs propres $\lambda_{\mathrm{max}} = \mathrm{max}_{1 \leq i \leq N} \lambda_i$ dans un mod\`ele $\beta$ \`a une matrice (\'{E}qn.~\ref{eq:lames}) :
\beq
P_{N,\beta}(a) = \nu_{N,\beta}^{\mathbb{R};V}[\lambda_{\mathrm{max}} < a] = \frac{Z_{N,\beta}^{[-\infty,a];V}}{Z_{N,\beta}^{\mathbb{R};V}} \nn
\eeq
Nous supposerons que $V$ est un potentiel polyn\^{o}mial ind\'ependant de $N$ et $\beta$, et qui conduit pour la mesure $\dd\nu^{\mathbb{R};V}_{N,\beta}$ \`a un r\'egime \`a une coupure. On note $D + 1$ son degr\'e. Par convention $t \equiv 1$.

Les r\'esultats essentiels de \cite{BEMN} seront expos\'es aux \S~\ref{sec:ben} et \ref{sec:conn}. Cet article prend l'existence de l'asymptotique en puissances de $1/N$ dans les mod\`eles $\beta \neq 1$ comme une hypoth\`ese, car il est ant\'erieur \`a \cite{BG11} o\`u cette asymptotique est \'etablie.

\subsection{Description qualitative}

Dans la situation \`a une coupure que nous \'etudions, les valeurs propres se condensent dans la limite $N$ grand, sur un segment $[b^*,a^*]$ pour la mesure $\dd\nu_{N,\beta}^{\mathbb{R};V}$. En particulier, $\lambda_{\mathrm{max}}$ vaut $a^*$ en moyenne. Un argument classique, valable pour tout $\beta > 0$, permet d'\'evaluer l'\'echelle $\delta\lambda_{\mathrm{max}}$ de \label{dens4}ses fluctuations autour de $a^*$. La densit\'e limite de valeurs propres sur $[a^*,b^*]$ ne d\'epend pas de $\beta$, elle est de la forme :
\beq
\forall x \in [a^*,b^*]\quad \rho(x)\dd x = \frac{M(x)}{2\pi}\sqrt{(a^* - x)(x - b^*)} \nn
\eeq
o\`u $M$ est un polyn\^{o}me de degr\'e $\mathrm{deg}\,V - 2$. Si $V$ est g\'en\'erique, $\rho(x)$ se comporte en $(a^* - x)^{1/2}$ pr\`es de $a^*$. Le nombre de valeurs propres sup\'erieures \`a $a$ est en moyenne $\overline{n}(a) = N\int_{a}^{a^*} \rho(x)\dd x$. On s'attend \`a $\overline{n}(a^* - \delta\lambda_{\mathrm{max}}) \sim 1$, d'o\`u l'estimation $\delta\lambda_{\mathrm{max}} \sim N^{-2/3}$.

\begin{figure}[h!]
\begin{center}
\begin{minipage}[c]{0.60\linewidth}
\raisebox{-4cm}{\includegraphics[width=\textwidth]{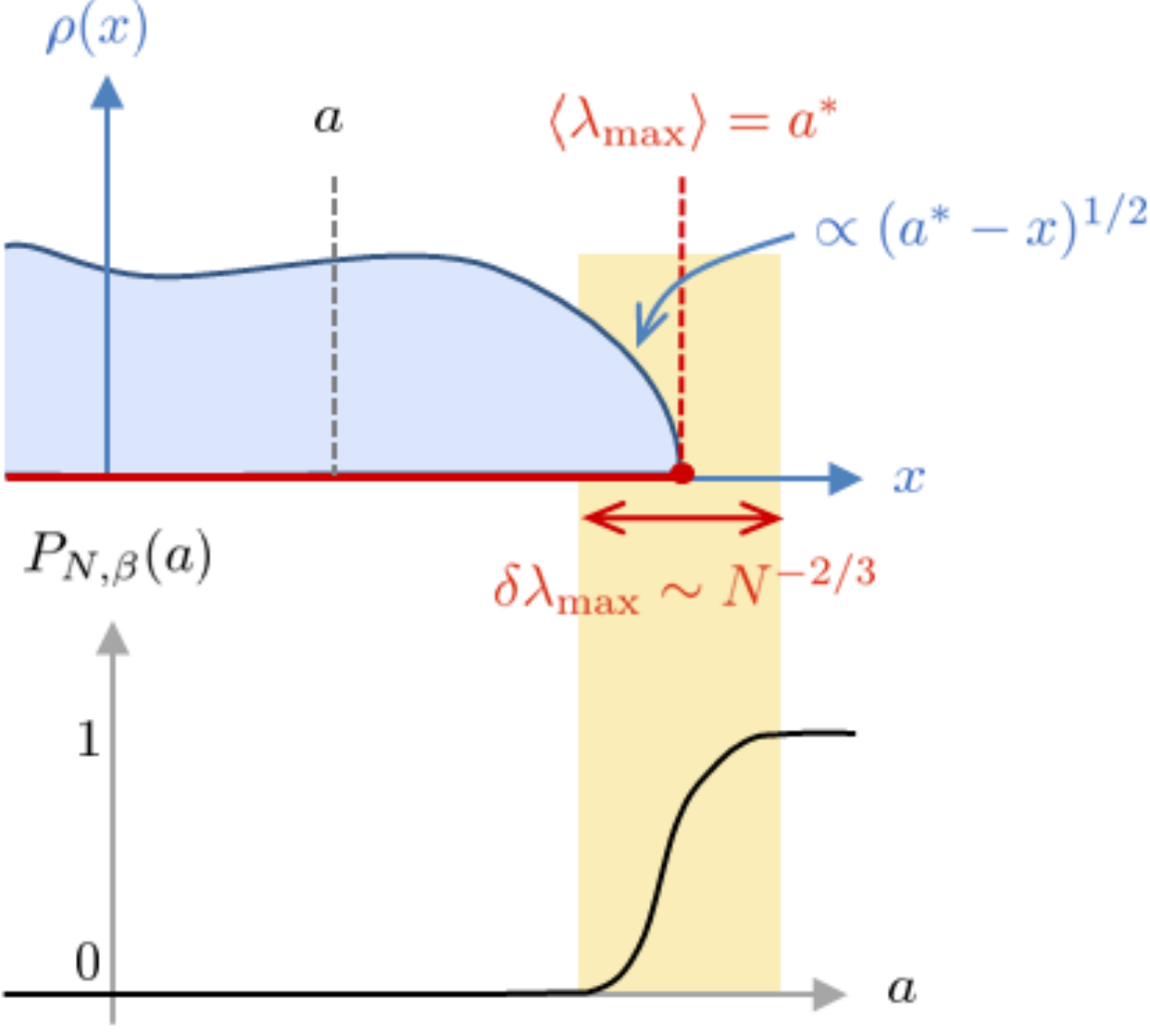}}
\end{minipage} \hfill \begin{minipage}{0.39\linewidth}
\caption{\label{fig:tra} Fonction de r\'epartition de $\lambda_{\mathrm{max}}$ : transition entre grandes d\'eviations et \'ev\`enements typiques.}
\end{minipage}
\end{center}
\end{figure}

Si l'on choisit $a < a^*$, les \'ev\`enements "$\lambda_{\mathrm{max}} < a$" sont rares. En fait, la \textbf{mesure conditionn\'ee} $\dd\nu^{[-\infty,a];V}_{N,\beta}\!\!$ reste dans un r\'egime \`a une coupure tant que $a < a^*$, et les r\'esultats de la partie~\ref{sec:puisn} sont applicables\footnote{Ici, le bord $a$ est dur, tandis que le bord $b$ est mou. La justification des r\'esultats de la partie~\ref{sec:puisn} dans ce cas se trouve dans \cite{BG11}.} Par cons\'equent, la probabilit\'e de "$\lambda_{\mathrm{max}} < a$" est de l'ordre de $e^{-N^2\Phi(a)}$ lorsque $N$ est grand.
Plus pr\'ecis\'ement, en notant $\mathcal{S}_a$ la courbe spectrale du mod\`ele conditionn\'e :
{\small \bea
P_{N,\beta}(a) & = & \frac{C_{N,\beta}}{Z_{N,\beta}^{\mathbb{R};V}}\,\exp\Big\{\sum_{k \geq -2} N^{-k}\,\sum_{g = 0}^{\lfloor \frac{k}{2}\rfloor + 1} \beta^{1 - g}(1 - 1/\beta\big)^{k + 2 - 2g}\,\mathcal{F}^{(g;k + 2 - 2g)}[\mathcal{S}_{a}]\Big\} \nn \\
\label{eq:devv}&&
\eea}
$\!\!\!$Les premiers termes de cette s\'erie sont :
\bea
P_{N,\beta}(a) & = & \frac{C_{N,\beta}}{Z_{N,\beta}^{\mathbb{R};V}}\exp\Big\{N^2\beta\,\mathcal{F}^{(0;0)}[\mathcal{S}_{a}] + N(\beta - 1)\mathcal{F}^{(0;1)}[\mathcal{S}_a] \nn \\
\label{eq:Instable} && + (\beta + 1/\beta - 2)\mathcal{F}^{(0;2)}[\mathcal{S}_a] + \mathcal{F}^{(1;0)}[\mathcal{S}_a] + O(1/N)\Big\}
\eea
On identifie la \textbf{fonction de grande d\'eviation} $\Phi(a) = -\mathcal{F}^{(0;0)}[\mathcal{S}_a] + \mathrm{cte}$, et les autres $\mathcal{F}^{(g;l)}[\mathcal{S}_a]$ en calculent les corrections de taille finie.

Si l'on zoome autour de $a^*$ \`a l'\'echelle $N^{-2/3}$, on s'attend \`a voir la transition entre $P_{N,\beta}(a) \ll 1$ lorsque $a < a^*$, et $P_{N,\beta}(a) \sim 1$ lorsque $a > a^*$ (Fig.~\ref{fig:tra}). Il devrait exister une limite :
\beq
\label{eq:Pbeta} \widehat{P}_{\beta}(s) = \lim_{N \rightarrow \infty} P_{N,\beta}(a^* + sN^{-2/3})
\eeq
On s'attend \'egalement \`a ce que cette limite soit universelle pour la classe des potentiels $V$ (ne conduisant pas n\'ecessairement \`a un r\'egime une coupure) tel que $\rho(x) \propto (a^* - x)^{1/2}$ au voisinage du maximum $a^*$ du spectre. Autrement dit, si $V_0$ et $V_1$ sont de tels potentiels :
\beq
\label{eq:uninv}\widehat{P}_{\beta}^{V_0}(s) = \widehat{P}_{\beta}^{V_1}(s/A_{\beta}^{V_0,V_1})
\eeq
pour une constante non universelle $A_{\beta}^{V_0,V_1}$.

\subsection{Lois de Tracy-Widom}
\label{sec:largenty}
Ce paragraphe r\'esume ce qui est connu sur les distributions $\widehat{P}_{\beta}$, et qui nous int\'eresse au premier chef pour situer les r\'esultats de \cite{BEMN}.

\subsubsection{$\beta \in \{1/2,1,2\}$}

Le mod\`ele $\beta$ pour ces valeurs est un mod\`ele de matrice associ\'e \`a un des groupes de sym\'etrie classique (Fig.~\ref{Dysonens}), soluble par les techniques de polyn\^{o}mes orthogonaux (partie~\ref{sec:Mehtapoly}). Les fonctions de corr\'elation de densit\'e entre valeurs propres \label{dens5}s'\'ecrivent  :
\beq
\rho_{n|N}(x_1,\ldots,x_n) = \mathrm{det}\big(K_N(x_i,x_j)\big) \nn
\eeq
Un calcul utilisant cette repr\'esentation d\'eterminantale montre alors que la probabilit\'e de $\lambda_{\mathrm{max}}$ soit inf\'erieure \`a $a$ est un d\'eterminant de Fredholm\label{Fredo} :
\beq
P_{N,\beta}(a) = \mathrm{Det}(1 - \mathbf{K}_N)_{\mathrm{L}^{2}[a,+\infty[} \nn
\eeq

Pour le potentiel $V_{\mathrm{G}}(x) = x^2/2$ (et $t \equiv 1$ par convention), les noyaux $K_N$ pertinents sont reli\'es au noyau d'Airy (Fig.~\ref{fig:univ}). En \'etudiant leur limite, Tracy et Widom \cite{TW92,TW95} ont \'etabli l'existence de la limite \'{E}qn.~\ref{eq:Pbeta} pour tout $s \in \mathbb{R}$. Nous noterons $\mathsf{TW}_{\beta}$ les fonctions de Tracy-Widom :
\beq
\label{eq:normee}\mathsf{TW}_{1/2}(s) = \widehat{P}_{1/2}^{V_{\mathrm{G}}}(s),\quad \mathsf{TW}_{1}(s) = \widehat{P}_{1}^{V_{\mathrm{G}}}(s),\quad \mathsf{TW}_{2}(s) = \widehat{P}_{2}^{V_{\mathrm{G}}}(2^{-2/3}s)
\eeq
Ces auteurs ont surtout \'etabli des \'equations diff\'erentielles pour les d\'eterminants de Fredholm sur $\mathrm{L}^{2}\big(\bigcup_{j} [b_j,a_j]\big)$, qui leur ont permis notamment de repr\'esenter $\textsf{TW}_{\beta}(s)$ en fonction de la solution d'une \'equation de Painlev\'e II.
\bea
&& \mathsf{TW}_{1/2}(s) = \mathsf{E}(s)\mathsf{F}(s),\quad \mathsf{TW}_{1}(s) = \big(\mathsf{F}(s)\big)^2,\quad \mathsf{TW}_{2}(s) = \big(\mathsf{E}(s) + \mathsf{E}^{-1}(s)\big)\mathsf{F}(s) \nn  \\ && \mathsf{E}(s) = \exp\Big(-\frac{1}{2}\int_s^{\infty} q(\sigma)\dd\sigma\Big),\qquad \mathsf{F}(s) = \exp\Big(-\frac{1}{2}\int_{s}^{\infty}(s - \sigma)q^2(\sigma)\Big) \nn \\
\label{eq:jbhh} &&
\eea
et $q$ \label{APII}est l'unique solution de $q'' = 2q^3 + sq$ telle que $q(s) \sim -s/2$ lorsque $s \rightarrow -\infty$. Son existence et unicit\'e ont \'et\'e prouv\'ee par Hastings et McLeod \cite{HMcL}, et elle se comporte aussi comme
\beq
\label{eq:jud} q(s) \sim \mathrm{Ai}(s) \sim \frac{\exp(-\frac{2s^{3/2}}{3})}{2\sqrt{\pi}s^{1/4}}\qquad s \rightarrow +\infty
\eeq
Cette \'equation de Painlev\'e II apparait comme la condition de compatibilit\'e dans un syst\`eme int\'egrable $2 \times 2$, nous y reviendrons au \S~\ref{sec:PII}.

\label{orhu}L'universalit\'e des lois de Tracy-Widom a \'et\'e d\'emontr\'ee par Deift et Gioev \cite{DG07}. Gr\^{a}ce aux asymptotiques des polyn\^{o}mes orthogonaux \'etablis par les m\'ethodes de Riemann-Hilbert \cite{DKMcLVZ3}, ces auteurs contr\^{o}lent la convergence du noyau $K_N$ vers le noyau d'Airy, de mani\`ere suffisamment pr\'ecise pour \'etablir l'universalit\'e au bord du spectre.

\begin{figure}[h!]
\begin{center}
\includegraphics[width=0.9\textwidth]{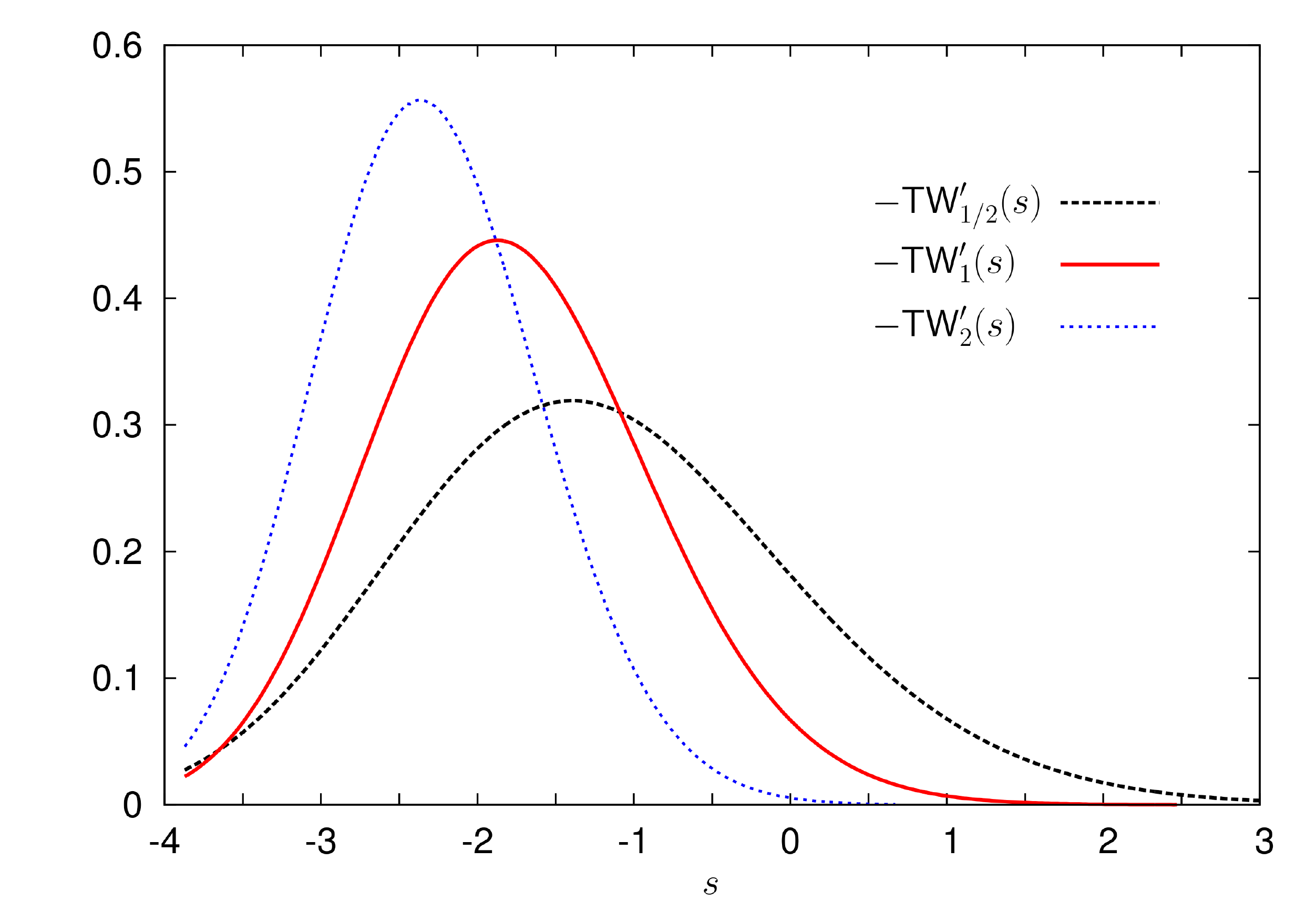}
\caption{Graphe des densit\'es de probabilit\'e de Tracy-Widom pour $\beta \in \{1/2,1,2\}$. La moyenne est toujours n\'egative et se d\'ecale vers la gauche lorsque $\beta$ augmente. L'\'ecart-type augmente avec $\beta$. Trac\'e par J.-M.~St\'ephan, d'apr\`es les donn\'ees num\'eriques de A.Iu.~Bejan \cite{Bejan}.}
\end{center}
\end{figure}

Le d\'eveloppement de la \textbf{queue} gauche \textbf{des lois de Tracy-Widom}, i.e. lorsque $s \rightarrow -\infty$, se d\'eduit r\'ecursivement de l'\'{E}qn.~\ref{eq:jbhh} et de l'\'equation de \label{cteTW}Painlev\'e II, sauf le pr\'efacteur constant $\tau_{\beta}$ qui est difficile \`a obtenir.
\bea
\textsf{TW}_{1/2}(s) & = & \tau_{1/2}\,\exp\Big\{-\frac{|s|^3}{24} - \frac{|s|^{3/2}}{3\sqrt{2}} - \frac{\ln|s|}{16} + O(|s|^{-3/2})\Big\} \nn \\
\textsf{TW}_{1}(s) & = & \tau_1\,\exp\Big\{-\frac{|s|^{3}}{12} - \frac{\ln|s|}{8} + O(s^{-3})\Big\} \nn \\
\textsf{TW}_{2}(s) & = & \tau_2\,\exp\Big\{-\frac{|s|^{3}}{24} + \frac{|s|^{3/2}}{3\sqrt{2}} - \frac{\ln|s|}{16} + O(|s|^{-3/2})\Big\} \nn
\eea
La valeur de la constante pour $\beta = 1$ a \'et\'e conjectur\'ee par Tracy et Widom \cite{TW92} d'apr\`es une co\"{i}ncidence num\'erique, et \'etablie par Deift, Its et Krasovsky \cite{DIK}. En \'etudiant plus g\'en\'eralement les int\'egrales de $q$, Baik, Buckingham et DiFranco \cite{BBdiF} ont retrouv\'e $\tau_1$ et d\'etermin\'e la valeur de $\tau_{1/2}$ et $\tau_{2}$ :
\beq
\tau_{1/2} = 2^{-11/48}\,e^{\zeta'(-1)/2},\quad \tau_1 = 2^{1/24}e^{\zeta'(-1)},\quad \tau_2 = 2^{-35/48}\,e^{\zeta'(-1)/2} \nn
\eeq
$\zeta$ est le prolongement analytique de la\label{zetu} fonction z\^{e}ta de Riemann $\zeta(s) = \sum_{n = 1}^{\infty} n^{-s}$. Toutes ces preuves utilisent les techniques de Riemann-Hilbert.

Ce genre de constante a pour origine l'asymptotique de la fonction de Barnes \cite{Voros} :
{\small \beq
G(N + 1) \equiv \prod_{n = 1}^{N - 1} n! \sim \exp\Big(\frac{N^2\ln N}{2} - \frac{3N^2}{4} + \frac{N\ln(2\pi)}{2} - \frac{\ln N}{12} + \zeta'(-1)\Big) \nn
\eeq}
\label{Seljk2}$\!\!\!$qui intervient dans les int\'egrales de Selberg (\'{E}qn.~\ref{eq:SelL} et \ref{eq:SelG}). En fait, c'est la version $\beta$ de la fonction de Barnes qui apparaissait :
\beq
\label{eq:Geta}G_{\beta}(N + 1) \equiv \prod_{n = 1}^{N - 1} \Gamma(1 + n\beta)
\eeq
Mais, pour $\beta = 2$, ou $\beta = 1/2$ gr\^{a}ce \`a la formule de duplication pour $\Gamma$, $G_{\beta}(N + 1)$ s'exprime en termes de $G(N + 1)$.

\subsubsection{$\beta > 0$ quelconque}
\label{bebeb2}
Les informations pour $\beta$ quelconque sont plus rares, et datent principalement des cinq derni\`eres ann\'ees. \'{E}qn.~\ref{eq:Pbeta} a \'et\'e d\'emontr\'ee pour un potentiel $V_{\mathrm{G}}(x) = x^2/2$ par Rider, Ram\'{i}rez et Vir\'{a}g \cite{RRV}. Leur point de d\'epart est la repr\'esentation du mod\`ele $\beta$ comme un mod\`ele \`a une matrice tridiagonale al\'eatoire de taille $N \times N$ : ils d\'emontrent que, dans la limite $N \rightarrow \infty$, les premi\`eres plus grandes valeurs propres $\lambda_{1} > \lambda_{2} > \cdots > \lambda_k > \cdots$ convergent en loi, apr\`es une mise \`a l'\'echelle, vers celles d'un op\'erateur de Schr\"{o}dinger perturb\'e par le processus brownien $B$ et agissant sur $\mathrm{L}^2(\mathbb{R}_+)$ :
\beq
\mathcal{H}_{\beta} = -\frac{\dd^2}{\dd \tau^2} + \tau + \frac{2}{\sqrt{\beta}}\frac{\dd B}{\dd \tau} \nn
\eeq
L'\'equation de Schr\"{o}dinger stochastique $\mathcal{H}_{\beta} \psi = \lambda \psi$ est \'equivalente \`a une \'equation de Ricatti stochastique v\'erifi\'ee par $w = \psi'/\psi$ :
\beq
\dd w_{\tau} = (\tau - w^2_{\tau})\dd\tau + \frac{2}{\sqrt{\beta}}\dd B_{\tau} \nn
\eeq
La probabilit\'e que $\lambda_1 = \lambda_{\mathrm{max}}$ soit plus petite que $a = a^* + sN^{-2/3}$ s'interpr\`ete alors, dans la limite $N$ grand, comme la probabilit\'e que le processus $(w_{\tau})_{\tau \geq s}$, avec la condition initiale $w_{s} = + \infty$, ne rejoigne pas $-\infty$ en temps fini. Cela d\'efinit une "loi de Tracy-Widom $\beta$", not\'ee\footnote{Par la suite, nous noterons g\'en\'eriquement $\textsf{TW}_{\beta}(s) = \widehat{P}^{V_{\mathrm{G}}}_{\beta}(s)$, et cela sous-entendra qu'il faut remplacer $s \rightarrow 2^{-2/3}s$ lorsque $\beta = 2$ pour coller \`a la d\'efinition standard (\'{E}qn.~\ref{eq:normee}).} $\textsf{TW}_{\beta}(s)$. Par ailleurs, Rider \cite{RMSRI} a annonc\'e une preuve de l'\'{E}qn.~\ref{eq:Pbeta} pour tous les potentiels polyn\^{o}miaux pairs, ainsi qu'un r\'esultat d'universalit\'e.

Bloemendal et Vir\'{a}g \cite{VirBloe} ont d\'eriv\'e de cette interpr\'etation stochastique une \'equation diff\'erentielle maitresse pour calculer $\textsf{TW}_{\beta}$. Leur r\'esultat dit que $\mathsf{TW}_{\beta}(s) = \lim_{w \rightarrow + \infty} f(w,s)$, o\`u $f$ est l'unique solution born\'ee de
\bea
&& \frac{\partial f}{\partial s} + \frac{2}{\beta}\frac{\partial^2 f}{\partial w^2} + (s - w^2)\frac{\partial f}{\partial w} = 0 \nn \\
\label{eq:carara} && \lim_{\substack{w \rightarrow +\infty \\ s \rightarrow +\infty}} f(w,s) = 1,\quad \forall s_0 \in \mathbb{R},\,\,\lim_{\substack{w \rightarrow -\infty \\ s \rightarrow \tau_0}} f(w,s) = 0
\eea
Pour $\beta = 1,2$, ces auteurs ont v\'erifi\'e que les repr\'esentations \'{E}qn.~\ref{eq:jbhh} sont bien solutions de ce probl\`eme, mais le cas $\beta = 1/2$ leur a r\'esist\'e. Cette caract\'erisation donne un point de vue nouveau sur les lois de Tracy-Widom, dont il faudrait explorer les liens (s'il y en a) avec les syst\`emes int\'egrables.

En travaillant sur l'interpr\'etation stochastique, \cite{RRV} ont pu \'etablir le premier terme de la queue gauche :
\beq
\textsf{TW}_{\beta}(s) = \exp\Big(-\frac{\beta|s|^3}{12} + o(|s|^3)\Big)\qquad s \rightarrow -\infty \nn
\eeq
Sans connaitre ces travaux, \`a la m\^{e}me \'epoque, Dean et Majumdar \cite{DeanMaj} ont obtenu ce r\'esultat heuristiquement, en calculant la fonction de grande d\'eviation $\Phi(a) = - \mathcal{E}[L^*_a]$ pour $a < a^*$, et en ins\'erant $a = a^* + sN^{-2/3}$ dans cette expression. $L^*_a$ d\'esigne ici la densit\'e limite des valeurs propres dans le mod\`ele conditionn\'e $\dd\nu^{[-\infty,a];V_{\mathrm{G}}}_{N,\beta}$. Notre article \cite{BEMN} est en fait une extension de cette d\'emarche \`a tous les ordres en $N$, puisque la r\'ecurrence topologique donne tous les coefficients et leurs propri\'et\'es.

Toujours avec l'interpr\'etation stochastique, signalons enfin que \cite{VirDum} ont \'etabli les deux premiers termes de la queue droite :
\beq
1 - \textsf{TW}_{\beta}(s) = \exp\Big(-\frac{4\beta s^{3/2}}{3} - \frac{3\beta}{2}\ln s + O(\sqrt{\ln s})\Big)\qquad s \rightarrow +\infty \nn
\eeq

\subsection{Grandes d\'eviations}
\label{sec:ben}

\subsubsection{Courbe spectrale $\mathcal{S}_a$}

\begin{figure}[h!]
\begin{center}
\includegraphics[width=1.05\textwidth]{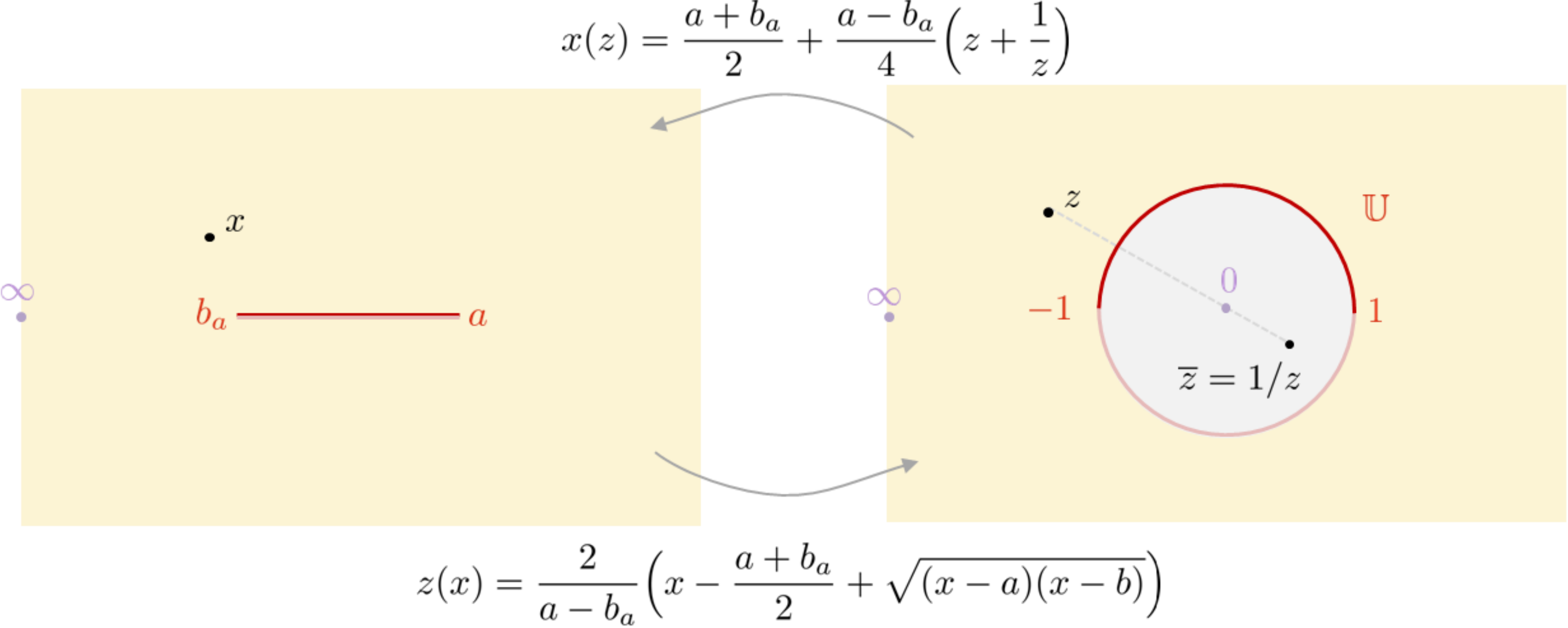}
\caption{\label{fig:Joukov} L'application de Joukovski met $\Sigma$ en bijection conforme avec $\widehat{\mathbb{C}}$. La coupure est envoy\'ee sur le cercle unit\'e $\mathbb{U}$, le domaine $\widehat{\mathbb{C}}\setminus[b_a,a]$ sur $\mathrm{Ext}(\mathbb{U})$ (le feuillet physique). L'involution $z \mapsto \overline{z} = 1/z$ est d\'efinie globalement sur $\Sigma$. $z = 1$ (image de $a$) est un point de branchement dur, $z = -1$ (image de $b_a$) est un point de branchement mou.}
\end{center}
\end{figure}

Si $a < a^*$, $a$ est un \label{bobo2} bord dur pour le mod\`ele conditionn\'e $\dd\nu_{N,\beta}^{]-\infty,a]}$. La densit\'e limite de valeurs propres $\rho_a(x)$ dans ce mod\`ele est support\'ee par un segment $[b_a,a]$. D'apr\`es les \'equations de Schwinger-Dyson, $\rho_a(x)$ dans ce mod\`ele diverge en $(a - x)^{-1/2}$ lorsque $x \rightarrow a$, et reste finie \label{dens6}partout ailleurs sur $[b_a,a[$ :
 \beq
\label{eq:Msji} \rho_a(x) = -\frac{M(x)}{2\pi}\sqrt{\frac{x - b_a}{a - x}},\qquad \mathrm{deg}\,M = D
 \eeq
 Qualitativement, une fraction macroscopique de valeurs propres voudraient se condenser aussi dans la r\'egion $[a,a^*]$ qui leur est interdite, donc marquent leur pr\'ef\'erence en s'accumulant \`a gauche de $a$. La mani\`ere la plus efficace de r\'esoudre l'\'{E}qn.~\ref{eq:hhsd} passe par le param\'etrage de \label{Joukov} Joukovski (Fig.~\ref{fig:Joukov}), qui ouvre la coupure $[b_a,a]$. Gr\^{a}ce \`a lui, on peut calculer la courbe spectrale $\mathcal{S}_a = [\Sigma,x,y,B]$, et r\'esoudre r\'ecursivement et "\`a la main" les premi\`eres \'equations de Schwinger-Dyson : les $\omega_n^{(g;l)}[\mathcal{S}_a](z_1,\ldots,z_n)$ sont des fractions rationnelles des $z_i$, et cette r\'esolution requiert seulement des calculs de d\'ecompositions en \'el\'ements simples. Cela est expliqu\'e en d\'etail dans \cite{BEMN}, mais ce n'est qu'une application de la version $\beta$ de la r\'ecurrence topologique d\'evelopp\'ee dans \cite{CE06}, \`a une courbe spectrale de genre $0$.

Si l'on d\'ecompose $V$ sur les polyn\^{o}mes de Tchebychev $T_k(x(z)) = \frac{1}{2}(z^k + z^{-k})$ :
\beq
V(x(z)) = v_0 + \sum_{k = 1}^{D + 1} v_k(z^k + z^{-k}) \nn\qquad  \rho_a(x)\dd x = \frac{\dd z}{i\pi z}\Big(-1 + \sum_{k = 1}^{D + 1} kv_k\,T_k(x)\Big)\nn
\eeq
Il est commode d'\'ecrire la fonction $y = \frac{V'}{2t} - W_1^{(0)} = i\pi\rho_a$ d\'efinissant la courbe spectrale sous une forme \'equivalente :
\beq
\label{eq:dhu} y(z) = \frac{1}{2}\frac{z + 1}{z - 1}\,M(x(z)),\quad M(x(z)) = t_{D + 1}\,\Big(\frac{a - b_a}{4z}\Big)^{D} \prod_{j = 1}^{D} (z - s_j)(z - 1/s_j)
\eeq
Cela met en \'evidence les z\'eros $s_j$ de $y$ dans le feuillet physique. C'est leur distance \`a $[b_a,a]$ qui contr\^{o}le la norme de l'op\'erateur $\mathcal{K}^{-1}$ (\S~\ref{sec:paraa}). L'extr\'emit\'e $b_a$ est d\'etermin\'ee par la condition $W_1^{(0)}(x(z)) \in O(1/x(z))$ lorsque $z \rightarrow \infty$, qui s'\'ecrit :
\beq
\sum_{k = 1}^{D + 1} (-1)^k\,kv_k = 1 \nn
\eeq
Le noyau de Bergman est $B(z_1,z_2) = \frac{\dd z_1\dd z_2}{(z_1 - z_2)^2}$ (cf.~\ref{sec:bout}).

\subsubsection{Quelques r\'esultats explicites}

Comme $\mathcal{S}_a$ est de genre $0$, les calculs peuvent \^{e}tre men\'es \`a terme avec des fonctions \'el\'ementaires. Voici les expressions des termes de $\ln P_{N,\beta}(a)$ qui subsistent dans la limite $N \rightarrow \infty$ (\'{E}qn.~\ref{eq:Instable}).

\vspace{0.2cm}

\noindent $\diamond\,$ La fonction de grande d\'eviation $\Phi(a) = -\beta \mathcal{F}^{(0;0)}[\mathcal{S}_a] = -\beta\mathcal{F}^{(0)}[\mathcal{S}_a]$, avec le pr\'epotentiel d\'efini \`a l'\'{E}qn.~\ref{eq:F0def}. Plus explicitement :
\beq
\label{eq:hy1}\mathcal{F}^{(0;0)}[\mathcal{S}_a] = \sum_{k = 1}^{D + 1} \frac{kv_k^2}{2} - v_0 + \ln\Big(\frac{a - b_{a}}{4}\Big)
\eeq
Cette expression est un avatar du th\'eor\`eme de la limite forte de Szeg{\"{o}} \cite{Szegofort}.

\vspace{0.2cm}

\noindent $\diamond\,$ La premi\`ere correction, $\mathcal{F}^{(0;1)}[\mathcal{S}_a]$, est au signe pr\`es l'entropie de la distribution $\rho_a$.
{\small \bea
\label{eq:hy2} \mathcal{F}^{(0;1)}[\mathcal{S}_a] & = & \int_{b_a}^a \rho_a(x)\ln\rho_a(x)\dd x  \\
& = & \ln\Big[\frac{t_{D + 1}}{2\pi}\Big(\frac{a - b_a}{4}\Big)^{D}\prod_{j = 1}^{D} |s_j|\Big] - \sum_{k\,\,\mathrm{impair}} v_k + \sum_{k = 1}^{D + 1} v_k\Big(\sum_{j = 1}^{D} \frac{1}{s_j^k}\Big) \nn \eea}

\noindent $\diamond\,$ $\mathcal{F}^{(1;0)} = \mathcal{F}^{(1)}$ a \'et\'e d\'efini \`a l'\'{E}qn.~\ref{eq:F1def}. Puisque $\mathcal{S}_a$ est de genre $0$, il n'y a pas d'int\'egrales de cycles dans la fonction tau de Bergman :
\beq
\label{eq:hy3}\mathcal{F}^{(1;0)}[\mathcal{S}_a] = -\frac{1}{24}\ln\big[M^3(a)M(b_a)(a - b_a)^{4}\big]
\eeq

\noindent $\diamond\,$ $\mathcal{F}^{(0;2)}$ est reli\'ee \`a certains d\'eterminants d'op\'erateurs sur la courbe spectrale \cite{WZ06}, comme $\mathcal{F}^{(1;0)}$.
{\small \bea
\label{eq:hy4} \mathcal{F}^{(0;2)}[\mathcal{S}_a] & = & \frac{1}{12}\ln\big[M^{3}(a)M^{-1}(b_a)(a - b_a)^2\big] + \mathcal{F}_{\mathrm{I}}^{(0;2)}[\mathcal{S}_a] \\
\mathcal{F}^{(0;2)}_{\mathrm{I}}[\mathcal{S}_a] & = & \frac{1}{16\pi^2} \int_{z_1 \in \mathcal{C}_{\mathrm{ext}}}\!\!\frac{\dd\rho_a(z_1)}{\rho_a(z_1)}\int_{z_2 \in \mathcal{C}_+} \!\!\!\!\!\ln\big(\rho_a(z_2)\big)\Big(\int_{z_1}^{\overline{z_1}}\!\!\! B(z_2,\cdot)\Big) \nn \\
& = & \frac{1}{4}\ln\Big[\prod_{j = 1}^{D} \frac{s_j^4}{(s_j - 1)(s_j + 1)^3} \prod_{1 \leq j < l \leq D} \frac{1}{1 - 1/s_js_l}\Big]  \nn
\eea}

Les expressions ci-dessus sont des formules closes pour les termes instables de la r\'ecurrence topologique (les plus difficiles \`a calculer). Pour tous les autres $\mathcal{F}^{(g;l)}$, il y a un algorithme uniforme via la formule d'int\'egration :
\beq
\label{eq:Fgl} \mathcal{F}^{(g;l)}[\mathcal{S}_a] = \frac{1}{2 - 2g - l}\,\Res_{z \rightarrow \pm 1} \Big(\int_{1}^{z} \!\!y\dd x\Big)\omega_1^{(g;l)}[\mathcal{S}_a](z)
\eeq
Pour les obtenir de fa\c{c}on explicite, il suffit de r\'esoudre m\'ecaniquement les \'equations de Schwinger-Dyson pour $\omega_n^{(g;l)}$, qui forment une r\'ecurrence sur $2g - 2 + n + l$ \cite{CE06}. Le r\'esultat de l'\'{E}qn.~\ref{eq:Fgl} sera une fonction rationnelle sym\'etrique des $s_j$.

\subsection{Connection avec les lois de Tracy-Widom}
\label{sec:conn}
\subsubsection{Universalit\'e}

La famille de courbes $(\mathcal{S}_a)_{a < a^*}$ est singuli\`ere dans la limite $a \rightarrow a^*$, car $a$ devient brusquement un point de branchement mou lorsqu'il atteint $a^*$. Cette situation advient lorsque l'un des z\'eros de $y$ dans le feuillet physique tend vers $1 = z(a)$ (appelons-le $s_1$). G\'en\'eriquement, on trouve $(s_1 - 1) \sim \sigma\,(a^* - a)^{1/2}$ pour une constante positive $\sigma$, tandis que les autres $s_j$ tendent vers $s_j^* \neq \pm 1$. D'apr\`es \S~\ref{sec:limitos}, il faut calculer la r\'esolution de la famille de courbes spectrales au voisinage de $a$. L'effet de $s_1$ se fait sentir \`a l'\'echelle o\`u $\zeta$ est d'ordre $1$, si l'on pose $z = 1 + \sigma(a^* - a)^{1/2} \zeta$. L'\'{E}qn.~\ref{eq:dhu} devient :
\beq
\label{eq:AVSS}\left\{\begin{array}{l} x(z) = a + \frac{a^* - b^*}{4}\,\sigma^2 (a^* - a)\,\widehat{x}(\zeta) + \cdots \\ y(z) \sim M_0\,\sigma\,(a^* - a)^{1/2}\,\widehat{y}(\zeta) \end{array}\right.\qquad B(z_1,z_2) \sim \widehat{B}(\zeta_1,\zeta_2)
\eeq
\label{TWaa2}La courbe spectrale $\widehat{\mathcal{S}} = [\mathbb{C},\widehat{x},\widehat{y},\widehat{B}]$ est universelle :
\beq
\label{eq:courbun}\widehat{x}(\zeta) = \zeta^2,\qquad \widehat{y}(\zeta) = \zeta - \frac{1}{\zeta},\qquad \widehat{B}(\zeta_1,\zeta_2) = \frac{\dd\zeta_1\dd\zeta_2}{(\zeta_1 - \zeta_2)^2}
\eeq
et d\'ecrit la r\'esolution de $(\mathcal{S}_a)_{a < a^*}$ \`a une constante de proportionnalit\'e (ind\'ependante de $\beta$) non universelle pr\`es :
\beq
\label{eq:AVF} A_{V} = \frac{a^* - b^*}{4}\,M_0\,\sigma^3 =  t_{D + 1}\,\sigma^3\Big(\frac{a^* - b^*}{4}\Big)^{D + 1}\prod_{j = 2}^{D} \big(2 - s_j^* + \frac{1}{s_j^*}\big)
\eeq

Par cons\'equent, si l'on note $\widehat{\mathcal{F}}^{(g;l)}$ les invariants symplectiques de $\widehat{\mathcal{S}}$ :
\bea
\mathcal{F}^{(0;0)}[\mathcal{S}_a] & = & \mathcal{F}^{(0;0)}[\mathcal{S}_{a^*}] + \big[A_{V}\,(a^* - a)^{3/2}\big]^2\,\widehat{\mathcal{F}}^{(0;0)} + o\big((a^* - a)^{3}\big) \nn \\
\mathcal{F}^{(0;1)}[\mathcal{S}_a] & = & \mathcal{F}^{(0;1)}[\mathcal{S}_{a^*}] + A_{V}\,(a^* - a)^{3/2}\,\widehat{\mathcal{F}}^{(0;1)} + o\big((a^* - a)^{3/2}\big) \nn \\
\mathcal{F}^{(1;0)}[\mathcal{S}_a] & = & \ln\big[A_{V}(a^* - a)^{3/2}\big]\,\widehat{\mathcal{F}}^{(1;0)} + f_{V}^{(1;0)} + o(1)  \nn \\
\mathcal{F}^{(0;2)}[\mathcal{S}_a] & = &  \ln\big[A_{V}(a^* - a)^{3/2}\big]\,\widehat{\mathcal{F}}^{(0;2)} + f_{V}^{(0;2)} + o(1) \nn \\
\mathcal{F}^{(g;l)}[\mathcal{S}_a] & \sim & \big[A_V\,(a^* - a)^{3/2}\big]^{2 - 2g - l}\,\widehat{\mathcal{F}}^{(g;l)} \quad\quad  \mathrm{si}\,\,2 - 2g - l < 0 \nn
\eea
Puisque les $\widehat{\mathcal{F}}^{(g;l)}$ sont universels, leur valeur pour $2 - 2g - l \geq 0$ se d\'eduit ais\'ement des formules closes \ref{eq:hy1}-\ref{eq:hy4} en choisissant le \label{UPO}cas simple d'un potentiel gaussien $V_{\mathrm{G}}(x) = x^2/2$ :
\beq
\widehat{\mathcal{F}}^{(0;0)} = -\frac{2}{3},\quad \widehat{\mathcal{F}}^{(0;1)} = \frac{4}{3},\quad \widehat{\mathcal{F}}^{(1;0)} = -\frac{1}{12},\quad \widehat{\mathcal{F}}^{(0;2)} = \frac{1}{12} \nn
\eeq
Et tous les autres $\widehat{\mathcal{F}}^{(g;l)}$ se calculent avec la (version $\beta$ de la) r\'ecurrence topologique appliqu\'ee \`a $\widehat{\Sigma}$ \cite{CE06}. Voici les premiers:
\beq
\begin{array}{llll}
\widehat{\mathcal{F}}^{(1;1)} = \frac{31}{2^{7}\cdot 3} &\: \widehat{\mathcal{F}}^{(0;3)} = -\frac{9}{2^{6}} & & \\
\widehat{\mathcal{F}}^{(2;0)} = \frac{3}{2^9} &\: \widehat{\mathcal{F}}^{(1;2)} = \frac{487}{2^{12}\cdot 3} &\: \widehat{\mathcal{F}}^{(0;4)} = -\frac{595}{2^{11}\cdot 3} & \\
\widehat{\mathcal{F}}^{(2;1)} = \frac{8831}{2^{17}\cdot 3} & \: \widehat{\mathcal{F}}^{(1;3)} = \frac{9281}{2^{17}\cdot 3^2}&\: \widehat{\mathcal{F}}^{(0;5)} = -\frac{19977}{2^{16}\cdot 3}& \\
\widehat{\mathcal{F}}^{(3;0)} = \frac{63}{2^{14}} &\: \widehat{\mathcal{F}}^{(2;2)} = \frac{62761}{2^{20}}&\: \widehat{\mathcal{F}}^{(1;4)} = -\frac{539193}{2^{23}} &\: \widehat{\mathcal{F}}^{(0;6)} = -\frac{577879}{2^{22}}
\end{array} \nn
\eeq
Les constantes $f_{V}^{(1;0)}$ et $f_{V}^{(0;2)}$ sont a priori non universelles\footnote{A posteriori, on v\'erifie que $f_{V}^{(1;0)} = (\ln 2)\widehat{\mathcal{F}}^{(1;0)} + \mathcal{F}^{(1;0)}[\mathcal{S}_{a^*}]$. Il est probable que $f_{V}^{(2;0)} = c\,\widehat{\mathcal{F}}^{(0;2)} + c' + \mathcal{F}^{(0;2)}[\mathcal{S}_{a^*}]$ pour des constantes universelles $c$ et $c'$, mais je n'ai pas fait le calcul.}, il faut les calculer \`a partir de \ref{eq:hy3}-\ref{eq:hy4}.

\subsubsection{Double limite d'\'echelle}
\label{dls}
Comme attendu, chaque terme du d\'eveloppement~\ref{eq:devv} devient d'ordre $1$ lorsque la variable d'\'echelle $s = N^{2/3}(a - a^*) < 0$ est d'ordre $1$, et les $o(\cdots)$ pr\'ec\'edents contribuent pour $o(1)$ lorsque $N \rightarrow \infty$. En g\'en\'eral, la somme de tous ces termes d'ordre $1$ n'est pas convergente. Il faut plut\^{o}t lui donner, heuristiquement, la signification de d\'eveloppement asymptotique lorsque $s \rightarrow -\infty$. Ainsi, nous avons une pr\'ediction pour la queue gauche de $\textsf{TW}_{\beta}(s)$ \`a tous les ordres (formules~\ref{eq:TWde} et \ref{eq:taube} ci-dessous). De plus, cette pr\'ediction v\'erifie la propri\'et\'e d'universalit\'e (\'{E}qn.~\ref{eq:uninv}) : les coefficients du d\'eveloppement sont les invariants symplectiques de la courbe spectrale universelle $\widehat{\Sigma}$, qui d\'ecrit $\rho_a(x)$ lorsque $a$ approche le bord du spectre (\'{E}qn.~\ref{eq:AVSS}). La constante de normalisation $A_V$ est explicitement calculable (\'{E}qn.~\ref{eq:AVF}).

Pour avoir les normalisations correctes, il faut calculer les constantes qui apparaissent avec le potentiel gaussien $V_{\mathrm{G}}(x) = x^2/2$ qui d\'efinit $\textsf{TW}_{\beta}$.
\bea
&& a^* = -b^* = 2,\quad \sigma = 2^{-1/2},\quad A_{V_{\mathrm{G}}} = 2^{-3/2} \nn \\
\label{eq:hyu} && \mathcal{F}^{(0;0)}[\mathcal{S}_{a^*}] = -3/4,\quad \mathcal{F}^{(0;1)}[\mathcal{S}_{a^*}] = 1/2 - \ln(2\pi) \\
&& f_{V_{\mathrm{G}}}^{(1;0)} = -(5\ln 2)/12,\quad f_{V_{\mathrm{G}}}^{(2;0)} = -(7\ln 2)/12 \nn
\eea
Par cons\'equent :
\bea
\textsf{TW}_{\beta}(s) & = & \tau_{\beta}\exp\Big\{-\frac{\beta}{12}\,|s|^3 + \frac{\sqrt{2}(\beta - 1)}{3}\,|s|^{3/2} + \frac{\beta + 1/\beta - 3}{8}\ln|s| \nn \\
& & + \sum_{k \geq 1} (|s|/2)^{-3k/2} \sum_{g = 0}^{\lfloor \frac{k}{2} + 1 \rfloor}  \beta^{1 - g}(1 - 1/\beta)^{k + 2 - 2g}\,\widehat{\mathcal{F}}^{(g;k + 2 - 2g)}\Big\} \nn \\
\label{eq:TWde} && \eea
avec la constante :
{\small \bea
\tau_{\beta} & = & \lim_{N \rightarrow \infty} C_{N,\beta} \nn \\
& & \times \big(Z_{N,\beta}^{V_{\mathrm{G}};\mathbb{R}}\big)^{-1}\,\exp\Big\{N^2 \mathcal{F}^{(0;0)}[\mathcal{S}_{a^*}] + N(\beta - 1)\mathcal{F}^{(0;1)}[\mathcal{S}_{a^*}]\Big\} \nn \\
&& \times \exp\Big\{\big[f_{V_{\mathrm{G}}}^{(1;0)} -\widehat{\mathcal{F}}^{(1;0)}\ln N\big] + (\beta + 1/\beta - 2)\big[f_{V_{\mathrm{G}}}^{(0;2)} - \widehat{\mathcal{F}}^{(0;2)}\ln N\big]\Big\} \nn
\eea}
$\!\!\!$En analogie avec \cite{BBdiF}, cette constante se compose de trois morceaux :

\vspace{0.2cm}

\noindent \emph{Morceau exact} $\diamond\,$ $C_{N,\beta}$ est le facteur de normalisation entre l'int\'egrale sur les valeurs propres avec bords durs $\int\dd\nu_{\beta,N}^{[a,b];V}$, et la d\'efinition des invariants symplectiques $\mathcal{F}^{(g;l)}[\mathcal{S}]$ de la courbe spectrale (\'{E}qn.~\ref{eq:asyn}). Il ne d\'epend pas de $V$, $a$ et $b$ : on peut le d\'eterminer lorsque l'on a des formules closes, d'une part pour l'int\'egrale sur les valeurs propres, d'autre part pour tous les $\mathcal{F}^{(g;l)}$. Le choix $a \rightarrow -\infty$ et $V(x) = V_{\mathrm{G}}(x)$ fait l'affaire. D'une part, $\int \dd \nu$ se ram\`ene lorsque $a \rightarrow -\infty$ \`a une int\'egrale de Selberg \label{Seljk3}pour l'ensemble $\beta$ de Laguerre (\'{E}qn.~\ref{eq:SelL}). D'autre part tous les $\mathcal{F}^{(g;l)}$ pour $2 - 2g - l < 0$ sont des $o(1)$ dans cette limite, et se d\'eduisent des formules~\ref{eq:hy1}-\ref{eq:hy4} pour $2 - 2g - l \geq 0$.

\noindent \emph{Morceau d'Airy} $\diamond\,$ La deuxi\`eme ligne est essentiellement le facteur de normalisation entre l'int\'egrale sur les valeurs propres sans bords durs $\int \dd\nu_{\beta,N}^{\mathbb{R};V}$, et la d\'efinition des invariants symplectiques. Il se calcule \`a partir de l'\'{E}qn.~\ref{eq:hyu} et de $Z_{N,\beta}^{V_{\mathrm{G}}}$, qui est une int\'egrale de Selberg avec poids gaussien (\'{E}qn.~\ref{eq:SelG}).

\vspace{0.2cm}

\noindent \emph{Choix de la variable d'\'echelle} $\diamond\,$ Le d\'eveloppement asymptotique de $\ln \textsf{TW}_{\beta}$ contient un logarithme $\ln |s|$, donc son terme constant est sensible \`a la normalisation choisie pour la variable d'\'echelle $s$. La troisi\`eme ligne est li\'ee \`a ce ph\'enom\`ene.

\vspace{0.2cm}

Le calcul montre que la limite $N \rightarrow \infty$  existe bien, ce qui est rassurant pour la consistance de ce paragraphe.
\beq
\label{eq:taube}\tau_{\beta} = \frac{\kappa_{\beta}}{\sqrt{2\pi\beta}}\,2^{\frac{5 - 2(\beta + 1/\beta)}{6}}
\eeq
Le morceau d'Airy et le morceau exact font intervenir (via les int\'egrales de Selberg) la d\'eformation $\beta$ de la fonction de Barnes (\'{E}qn.~\ref{eq:Geta}), et plus pr\'ecis\'ement sa partie finie $\kappa_{\beta}$ :
\begin{footnotesize}
\bea
G_{\beta}(N + 2) & \sim  &\exp\Big\{\frac{\beta N^2\ln N}{2} + \beta\Big(-\frac{3}{4} + \frac{\ln\beta}{2}\Big)N^2 + \frac{\beta + 1}{2}\ln N  \nn \\
&& + \Big(\frac{\beta + 1}{2}(\ln\beta - 1) + \frac{\ln(2\pi)}{2}\Big)N + \frac{\beta + 1/\beta - 3}{12}\ln N + \ln\kappa_{\beta} + o(1)\Big\} \nn
\eea
\end{footnotesize}

Les formules~\ref{eq:TWde} et \ref{eq:taube} sont en accord avec tous les r\'esultats connus \'evoqu\'es au \S~\ref{sec:largenty}. Il faut souligner que la d\'erivation de ce paragraphe est heuristique, elle suppose que l'on peut prolonger les r\'esultats obtenus rigoureusement \`a $a < a^*$ fix\'e, jusqu'\`a $a_* - a \in O(N^{-2/3})$. En fait, il suffirait pour le justifier de d\'emontrer l'existence d'un d\'eveloppement en $|s|^{-3k/2}$ pour $\ln \textsf{TW}_{\beta}(s)$ lorsque $s \rightarrow -\infty$. C'est un probl\`eme \`a explorer \`a la lumi\`ere du d\'ebut de ce chapitre.

\subsubsection{Commentaires}

Cette m\'ethode ne donne malheureusement pas de caract\'erisation de la loi de Tracy-Widom $\beta$, ni ne permet de v\'erifier \`a la main si $\mathsf{TW}_{\beta}(s)$ satisfait une \'equation diff\'erentielle comme \'{E}qn.~\ref{eq:carara}. Pour le cas int\'egrable $\beta = 1$ n\'eanmoins, il existe une relation pr\'ecise entre r\'ecurrence topologique et syst\`emes int\'egrables\footnote{On ne dispose pas actuellement d'une th\'eorie analogue pour les cas $\beta = 1/2$ ou $2$, dont on sait pourtant qu'ils sont int\'egrables.}, qui sera pr\'esent\'ee au Chapitre~\ref{chap:int}. Cela a \'et\'e exploit\'e dans l'article \cite{BETW} pour d\'emontrer que :
\beq
\tau_{1}\,\exp\Big\{-\frac{|s|^{3}}{12} + \frac{\ln|s|}{8} + \sum_{g \geq 2} (|s|/2)^{3 - 3g}\,\widehat{\mathcal{F}}^{(g)}\Big\} \nn
\eeq
est bien le d\'eveloppement asymptotique de $\mathsf{TW}_{1}(s)$ repr\'esent\'e en termes de la \label{fhua}solution d'Hastings-McLeod de Painlev\'e II (\'{E}qn.~\ref{eq:jbhh}).

\`{A} cause de la structure particuli\`ere en $\beta$ du d\'eveloppement~\ref{eq:TWde}, on peut remarquer une dualit\'e $\beta \leftrightarrow 1/\beta$ au niveau des d\'eveloppements asymptotiques :
\beq
\mathsf{TW}_{1/\beta} = \frac{\beta\kappa_{1/\beta}}{\kappa_{\beta}}\,\widetilde{\mathsf{TW}}_{\beta}(\beta^{-2/3}s) \nn
\eeq
o\`u $\widetilde{\cdots}$ d\'esigne le choix du signe oppos\'e pour la racine carr\'ee. Une relation de ce type existe par exemple entre deux solutions ind\'ependantes d'une \'equation de Schr\"{o}dinger.
Cela sugg\`ere que $\mathsf{TW}_{\beta}(s)$ et $\mathsf{TW}_{1/\beta}(s)$ correctement normalis\'es sont solutions d'une m\^{e}me \'equation diff\'erentielle d'ordre deux. \`{A} l'heure de l'\'ecriture, cette dualit\'e n'est pas \'evidente \`a partir de la caract\'erisation~\ref{eq:carara}.

La queue droite de $\textsf{TW}_{\beta}(s)$ \`a tous les ordres pourrait \^{e}tre \'etudi\'ee par les m\^{e}mes m\'ethodes. Il faudrait plut\^{o}t employer le mod\`ele de matrice pour la densit\'e de probabilit\'e de $\lambda_{\mathrm{max}}$ :
\beq
\nu_{N,\beta}[a < \lambda_{\mathrm{max}} < a + \dd a] = \frac{NZ_{N - 1,\beta}^{[-\infty,a];U_{N,a}}\,\dd a}{Z_{N,\beta}^{V;\mathbb{R}}} \nn
\eeq
avec le potentiel d\'ependant de $N$ et de $a$ :
\beq
U_{N,a}(x) = \frac{N}{N - 1} V(x) - \frac{t}{2(N - 1)}\ln |x - a| \nn
\eeq
Qualitativement, lorsque $a > a^*$, ce mod\`ele sera dans un r\'egime \`a une coupure, avec un support $[b',a']$ tel que $a^* < a' < a$. Les r\'esultats de la partie \ref{sec:puisn} s'appliqueront, avec la proc\'edure \'evoqu\'ee au \S~\ref{sec:lor} pour traiter la d\'ependance en $N$ du potentiel.

D'autres statistiques de valeurs propres (et leur r\'egime d'\'echelle pour $N$ grand) seraient calculables dans les mod\`eles $\beta$ quelconque par les m\^{e}mes m\'ethodes, parall\`element \`a \cite[Chapitre 9]{Loggas} qui expose les m\'ethodes bas\'ees sur les d\'eterminants de Fredholm pour $\beta \in \{1/2,1,2\}$. De notre point de vue, il suffirait d'\'etudier un mod\`ele de matrice appropri\'e avec bords (et sa double limite d'\'echelle), de pr\'ef\'erence dans un r\'egime une coupure car les calculs sont plus tractables. Deux exemples parmi d'autres :

\vspace{0.2cm}

\noindent $\diamond\,$ La loi jointe des $r$ plus grandes valeurs propres $\lambda_1 > \cdots > \lambda_r$.
{\small\bea
\frac{\nu_{N,\beta}\big[\lambda_1 \in [a_1,a_1 + \dd a_1],\ldots,\lambda_r \in [a_r,a_r + \dd a_r]\big]}{\dd a_1\cdots\dd a_r} & = & \frac{N!}{(N - r)!}\,|\Delta(\mathbf{a})|^{2\beta}\,\frac{Z_{N - r,\beta}^{[-\infty,a_r];U_{N,\mathbf{a}}}}{Z_{N,\beta}^{\mathbb{R};V}} \nn \\
U_{N,\mathbf{a}}(x) = \frac{N}{N - r}V(x) - \frac{t}{2(N - r)}\sum_{j = 1}\ln|x - a_j| \nn
\eea}
$\!\!\!$Qualitativement, ce mod\`ele devrait avoir une coupure dans le r\'egime o\`u $a_1 > \ldots > a_r > a^*$.

\vspace{0.2cm}

\noindent $\diamond\,$ La probabilit\'e qu'un segment $[b,a]$ fix\'e ne contienne aucune valeur propre :
\beq
\nu_{N,\beta}\Big[\bigcap_i \lambda_i \notin [b,a]\Big] = \frac{Z_{N,\beta}^{]-\infty,b]\cup[a,\infty[;V}}{Z_{N,\beta}^{\mathbb{R};V}} \nn
\eeq
Lorsque $[a,b]\subset [a^*,b^*]$, i.e. au milieu du spectre, c'est un mod\`ele de matrice \`a deux coupures.

\vspace{0.2cm}

Cela conduirait de fa\c{c}on syst\'ematique \`a des r\'esultats nouveaux pour $\beta \notin \{1/2,1,2\}$, mais un peu frustrants \`a l'heure actuelle en l'absence de m\'ethode pour caract\'eriser ces lois. Inversement, pour $\beta = 1$, les lois sont connues et l'on \'etablirait leurs relations avec la r\'ecurrence topologique appliqu\'ee \`a certaines courbes spectrales.

\newpage
\thispagestyle{empty}
\phantom{bbk}

\newpage
\thispagestyle{empty}
\phantom{bbk}

\newpage

\chapter{Syst\`emes int\'egrables et \'equations de boucles}
\label{chap:int}
\thispagestyle{plain}
\vspace{-1.5cm}

\rule{\textwidth}{1.5mm}

\vspace{2.5cm}

\addtolength{\baselineskip}{0.20\baselineskip}

\textsf{Nous commen\c{c}ons par fixer les d\'efinitions et les objets associ\'es \`a un syst\`eme int\'egrable classique : probl\`eme lin\'eaire, noyau int\'egrable, fonction tau, \'equations de Hirota, courbe spectrale, \ldots{}  Nous d\'efinissons aussi des "corr\'elateurs" \`a la suite de Berg\`ere et Eynard, et proposons une m\'ethode plus efficace pour montrer qu'ils satisfont des \'equations de boucles. Lorsque la courbe spectrale est de genre $0$, le Th\'eor\`eme~\ref{thsu} \'etablit le lien entre asymptotique BKW (\`a tous les ordres) d'un syst\`eme int\'egrable, et r\'ecurrence topologique. R\'eciproquement, nous rappelons et reformulons la construction alg\'ebro-g\'eom\'etrique de Krichever, qui associe un syst\`eme int\'egrable \`a une courbe spectrale fix\'ee. On peut chercher \`a construire un syst\`eme int\'egrable dispersif, qui est une d\'eformation du pr\'ec\'edent o\`u la courbe spectrale \'evolue lentement au cours du temps. Guid\'e par le r\'esultat heuristique pour l'asymptotique des mod\`eles de matrices, nous exhibons des formules pour la fonction d'onde, la fonction tau, \ldots{} \`a tous les ordres dans le param\`etre lent. Leur validit\'e repose sur une conjecture \'equivalente \`a l'\'equation de Hirota, qui reste en suspens \`a l'heure de l'\'ecriture.}

\addtolength{\baselineskip}{-0.20\baselineskip}

\section{Qu'est-ce-qu'un syst\`eme int\'egrable ?}
\label{sec:keskao}
\subsection{Introduction informelle}
\label{sec:avant}
Il existe plusieurs d\'efinitions, pas toujours \'equivalentes, des syst\`emes int\'egrables. Dans cette th\`ese, il sera question de \textbf{syst\`emes int\'egrables classiques} d\'efinis par la donn\'ee d'une famille compatible de syst\`emes diff\'erentiels lin\'eaires. Cela doit \^{e}tre compris au sens large. Voici une revue des notions associ\'ees \`a un syst\`eme int\'egrable, pour situer le vocabulaire que nous allons employer. Nous utiliserons souvent le livre \cite{BBT} comme r\'ef\'erence.

Chaque syst\`eme diff\'erentiel d\'efinit un flot, et l'on appelle \textbf{temps} le param\`etre d'\'evolution $t_{j_0}$ correspondant. En g\'en\'eral, une des variables d'\'evolution joue un r\^{o}le particulier, on la note $x$, et la famille des autres temps est not\'ee $\mathbf{t}$. L'ensemble de ces syst\`emes diff\'erentiels est appel\'e \textbf{probl\`eme lin\'eaire} : il se traduit typiquement en \'equations diff\'erentielles lin\'eaires de type Schr\"{o}dinger (et leur g\'en\'eralisation d'ordre $> 2$) \`a une dimension, dont les coefficients $a(x,\mathbf{t})$ d\'ependent de tous les param\`etres d'\'evolution. Une solution de ce probl\`eme est g\'en\'eriquement not\'ee $\Psi(x,\mathbf{t})$, et appel\'ee \textbf{fonction de Baker-Akhiezer}. Les conditions de compatibilit\'e du probl\`eme lin\'eaire donnent par \'elimination des \'equations aux d\'eriv\'ees partielles, non lin\'eaires, v\'erifi\'ees par les coefficients $a(x,\mathbf{t})$. Elles forment une \textbf{hi\'erarchie int\'egrable}, caract\'eris\'ee par l'existence d'une \textbf{quantit\'e conserv\'ee} $Q_j$ par flot $\partial_{t_j}$.

La compatibilit\'e des syst\`emes diff\'erentiels implique l'existence d'une \textbf{fonction tau} $\tau(\mathbf{t})$, telle que $\Psi(x,\mathbf{t})$ soit le ratio de deux fonctions tau. C'est la \textbf{formule de Sato} :
\beq
\Psi(x,\mathbf{t}) = \frac{\tau(\mathbf{t} - [x])}{\tau(\mathbf{t})}\,e^{V(x,\mathbf{t})} \quad \begin{array}{l} V(x,\mathbf{t})\,\,\textrm{lin\'{e}aire}\,\,\textrm{en}\,\,\mathbf{t} \\ \big[x\big]\,\,\textrm{un}\,\,\textrm{certain}\,\,\textrm{vecteur}\,\,\textrm{de}\,\,\textrm{temps} \end{array} \nn
\eeq
Une caract\'eristique importante des syst\`emes int\'egrables est que, une fois la fonction tau construite, l'\'evolution de $\Psi(x,\mathbf{t})$ est lin\'eaire dans tous les temps : \label{flot}c'est la manifestation de la commutativit\'e des flots $\partial_{t_j}$. L'\'etude des propri\'et\'es analytiques dans la variable $x$ de $\Psi(x,\mathbf{t})$ permettent d'\'etablir une \textbf{identit\'e bilin\'eaire} pour les solutions du probl\`eme lin\'eaire. Cela implique pour la fonction $\tau$ une \'equation fonctionnelle bilin\'eaire, appel\'ee \textbf{\'equation de Hirota}. Celle-ci est \'equivalente \`a la hi\'erarchie int\'egrable. L'\'equation de Hirota peut \^{e}tre \'ecrite de plusieurs fa\c{c}ons, mais c'est essentiellement la m\^{e}me \'equation dans tous les syst\`emes int\'egrables. Sa solution g\'en\'erale peut toujours s'\'ecrire comme un d\'eterminant infini (qui se r\'eduit parfois \`a un d\'eterminant fini).

\`{A} partir des fonctions de Baker-Akhiezer, on peut construire un \textbf{noyau int\'egrable} $K(x_i,x_j)$ qui a des propri\'et\'es remarquables. L'\'equation de Hirota \'equivaut notamment \`a une propri\'et\'e d'\textbf{autor\'eplication} pour le noyau int\'egrable.

\`{A} partir de la fonction tau, on peut d\'efinir\footnote{Cette d\'efinition a \'et\'e propos\'ee par Berg\`ere et Eynard \cite{BEdet}. Toute mention d'une r\'ef\'erence ant\'erieure serait bienvenue.} des \textbf{corr\'elateurs non connexes \`a $n$ points} $\overline{\mathcal{W}}_n(z_1,\ldots,z_n)$ qui encodent les d\'eriv\'ees $n^{\textrm{\`{e}me}}$ de la fonction $\tau$ par rapport aux temps. La formule de Sato se traduit par une \textbf{formule exponentielle}, exprimant le noyau int\'egrable en fonction de tous les corr\'elateurs. R\'eciproquement, l'autor\'eplication de $K$ entraine des \textbf{formules d\'eterminantales}, exprimant le corr\'elateur \`a $n$ points comme des d\'eterminants $n \times n$ du noyau int\'egrable :
\beq
\overline{\mathcal{W}}_n(z_1,\ldots,z_n) = "\mathrm{det}"\big(K(z_i,z_j)\big)_{1 \leq i,j \leq n} \nn
\eeq

\subsection{Exemples brefs}
\label{sec:pq}
\subsubsection{La hi\'erarchie KP}
\label{KPKP}
La hi\'erarchie KP est un exemple fondamental de syst\`eme int\'egrable. Les hi\'erarchies de Painlev\'e, les mod\`eles minimaux que nous rencontrerons plus tard, sont en fait des r\'eductions ou des troncations de KP (voir la d\'efinition un peu plus loin). Il n'est pas question de la pr\'esenter ici (cf. \cite{BBT}), mais plut\^{o}t d'illustrer \`a partir de l'\'equation de Hirota pour KP :
\beq
\label{eq:Hirhi}\forall \mathbf{s},\quad  \Res_{\xi \rightarrow \infty} \dd\xi\,e^{\sum_{k \geq 1} 2\,s_k\,\xi^k}\,e^{-\sum_{k \geq 1} \frac{\xi^{-k}}{k} \frac{\partial}{\partial s_k}}\, \tau(\mathbf{t} + \mathbf{s})\tau(\mathbf{t} - \mathbf{s}) = 0
\eeq
comment apparait la hi\'erarchie d'\'equations diff\'erentielles non lin\'eaires. Ici, les temps $t_j$ sont index\'es par $j \in \mathbb{N}^*$. La premi\`ere \'etape est de d\'evelopper en s\'erie de Taylor :
\beq
\frac{\tau(\mathbf{t} + \mathbf{s})\tau(\mathbf{t} - \mathbf{s})}{\big(\tau(\mathbf{t})\big)^2}  = 1 + \sum_{n \geq 1} \sum_{j_1\,\cdots\,j_{2n}} \frac{2}{(2n!)}\,\Big[\prod_{i} s_{j_i}\Big]\,T_{j_1\,\ldots\, j_{2n}}(\mathbf{t}) \nn
\eeq
Le premier ordre de l'\'{E}qn.~\ref{eq:Hirhi} qui contient une information est $s_1^3$ :
\beq
T_{1111} + 3T_{22} - 4T_{13} = 0 \nn
\eeq
o\`u :
\beq
T_{1111} = \partial_{t_1}^4 \ln \tau + 6\big(\partial_{t_1}^2\ln \tau)^2,\qquad T_{jk} = \partial_{t_j}\partial_{t_k} \ln \tau \nn
\eeq
En d\'erivant deux fois\label{KPeq} par rapport \`a $t_1$, on trouve que $u = \partial_{t_1}^2 \ln \tau$ satisfait :
\beq
3\frac{\partial^2 u}{\partial t_2^2} + \frac{\partial}{\partial t_1}\Big(-4 \frac{\partial u}{\partial t_3} + \frac{\partial^3 u}{\partial t_1^3} + 12 u\,\frac{\partial u}{\partial t_1}\Big) = 0 \nn
\eeq
\`{A} des coefficients num\'eriques pr\`es que l'on peut absorber dans la d\'efinition des $t_j$, c'est l'\'equation originale de Kadomtsev et Petviashvili \cite{KPorig}. Elle d\'ecrit la propagation selon $x = t_1$ d'une perturbation $u$ dans un milieu faiblement dispersif \`a deux dimensions $(x,y) = (t_1,t_2)$. Elle peut \^{e}tre d\'eriv\'ee par des arguments physiques en supposant l'\'echelle de variation de $u$ dans la direction $y$ assez grande pour que le caract\`ere dispersif ne s'y manifeste pas. Dans la limite d'un probl\`eme unidimensionnel ($\partial_{t_2} \equiv 0$), cet argument conduit \`a l'\'equation de Korteweg-de Vries\label{KDV2} \cite{KdV} :
\beq
-4\frac{\partial u}{\partial t_3} + \frac{\partial^3 u}{\partial t_1^3} + 12 u \,\frac{\partial u}{\partial t_1} = 0 \nn
\eeq
qui a \'et\'e utilis\'ee historiquement pour mod\'eliser les mascarets.

\textbf{Tronquer} la hi\'erarchie revient \`a imposer $t_{j} \equiv 0$ lorsque $j > p$ pour un certain $p$. \textbf{R\'eduire} la hi\'erarchie signifie rechercher des solutions qui ne d\'ependent que de certaines variables d'\'echelle construites \`a partir des temps.

\subsubsection{Les mod\`eles minimaux}
\label{minimo}
Il existe des troncations, r\'eductions de la hi\'erarchie KP, qui se ram\`enent au probl\`eme suivant : trouver deux op\'erateurs diff\'erentiels $\mathrm{P}$ et $\mathrm{Q}$, de degr\'es $p$ et $q$ premiers entre eux, tels que
\beq
\label{eq:PPO} \mathrm{P} = \partial^p + \sum_{j = 0}^{p - 2} u_{p - 2 - j}\partial^{j},\quad \mathrm{Q} = \partial^q + \sum_{j = 0}^{q - 2} v_{q - 2 - j} \partial^j,\quad [\mathrm{P},\mathrm{Q}] = 1
\eeq
Cela d\'efinit les \textbf{mod\`eles minimaux $(p,q)$}, et $[\mathrm{P},\mathrm{Q}] = 1$ est appel\'ee \textbf{\'equation de corde}. La solution g\'en\'erale pour $\mathrm{P}$ s'\'ecrit \cite{dFZJ} :
\beq
\label{eq:Phu}\mathrm{P} = -\frac{1}{q}\sum_{k = 1}^{p} (k + q)t_{k + q}\,\big(\mathrm{Q}^{k/q}\big)_+,\qquad t_{p + q} = -\frac{q}{p + q}
\eeq
L'\'equation de corde d\'etermine alors les coefficients $u_{p - 2 - j}$ et $v_{q - 2 - j}$ \`a des constantes d'int\'egration pr\`es. On d\'efinit le temps $t_1$ tel que $\partial = \partial_{t_1}$. La relation pr\'ecise avec la fonction $\tau$ de la hi\'erarchie KP est alors $\partial_{t_1}^2 \ln \tau(\mathbf{t}) = -u_0(\mathbf{t})/q$.

Ces mod\`eles minimaux interviennent dans la chaine \`a deux matrices hermitiennes (\S~\ref{sec:chainmat}). $\mathrm{P}$ et $\mathrm{Q}$ sont les op\'erateurs repr\'esent\'es par les matrices infinies :
\bea
\mathrm{P}_{nm} & = & \iint \dd x_1\dd x_2\, x_1\,\pi_n(x_1)\,\widetilde{\pi}_m(x_2)\,e^{-\frac{N}{t}\big(V_1(x_1) + V_2(x_2) + c_{1,2} x_1x_2\big)} \nn \\
\mathrm{Q}_{nm} & = & \iint \dd x_1\dd x_2\,\frac{\partial \pi_n}{\partial x_1}(x_1)\,\widetilde{\pi}_m(x_2)\,e^{-\frac{N}{t}\big(V_1(x_1) + V_2(x_2) + c_{1,2}\,x_1x_2\big)} \nn
\eea
dans les bases $(\pi_n)_{n \in \mathbb{N}}$ et $(\widetilde{\pi}_m)_{m \in \mathbb{N}}$ des polyn\^{o}mes \label{orhu2}biorthogonaux pour le produit scalaire d\'efini \`a l'\'{E}qn.~\ref{eq:poyh}. Lorsque $V_1$ et $V_2$ sont de degr\'es respectifs $p$ et $q$, $\mathrm{P}$ et $\mathrm{Q}$ sont des matrices \`a bande finie de largeur $p + 1$ et $q + 1$. Dans une double limite d'\'echelle, $\mathrm{P}$ et $\mathrm{Q}$ deviennent au moins heuristiquement des op\'erateurs diff\'erentiels de degr\'e $p$ et $q$, et l'on est amen\'e \`a \'etudier l'\'{E}qn.~\ref{eq:PPO}.

En particulier, les mod\`eles $(2m + 1,2)$ d\'ecrivent le $m^{\textrm{\`{e}me}}$ point multicritique pour le mod\`ele \`a une matrice hermitienne (polyn\^{o}mes orthogonaux cette fois). $m = 1$, i.e. le mod\`ele $(3,2)$, d\'ecrit le point critique correspondant \`a la gravit\'e pure (\S~\ref{sec:diss}). En \'ecrivant l'\'equation de corde pour $\mathrm{P} = \partial^3 - 3u\partial + w$ et $\mathrm{Q} = \partial^2 - 2u$, on trouve en effet l'\'equation de Painlev\'e I :
\beq
\label{eq:PIA} \frac{\partial w}{\partial t_1} = -\frac{3}{2}\,\frac{\partial^2 u}{\partial t_1^2},\quad \frac{\partial}{\partial t_1}\Big(\frac{\partial^2 u}{\partial t_1^2} + 6u^2  - t_1\Big) = 0
\eeq
Naturellement, le mod\`ele $(2m + 1,2)$ est inclus dans $(2m' + 1,2)$ pour $m$ croissant, il suffit de geler certains des temps de l'\'{E}qn.~\ref{eq:Phu}. L'\'equation de Painlev\'e I est donc contenue dans la hi\'erarchie d'\'equations du mod\`ele minimal $(2m + 1,2)$, que l'on appelle \textbf{hi\'erarchie de Painlev\'e I}.

\section{D\'eformations \mbox{isomonodromiques}}
\label{sec:isomono}
\subsection{Syst\`emes de Lax et probl\`emes de Riemann-Hilbert}
\label{sec:su}
Dans la suite de ce chapitre, notre point de d\'epart sera un autre type de syst\`emes int\'egrables, que l'on appellera ici \textbf{syst\`emes de Lax}. Il est temps de passer de la pr\'esentation du \S~\ref{sec:avant} \`a des formules concr\`etes. Initialement, on se pose le probl\`eme lin\'eaire :
\beq
\partial_x \Psi(x) = \mathbf{L}(x)\,\Psi(x) \nn
\eeq
o\`u $\mathbf{L}(x)$ est maintenant une matrice donn\'ee de taille $d \times d$, et $\Psi(x)$ une matrice $d \times d$ inversible. La litt\'erature se place souvent dans le cas $\Sigma = \widehat{\mathbb{C}}$, o\`u les coefficients de $L$ sont des fonctions rationnelles de $x$, c'est aussi l'hypoth\`ese que nous allons prendre. Il me semble que la plupart des \'enonc\'es de ce paragraphe sont g\'en\'eralisables (avec quelques subtilit\'es) lorsque $x$ vit sur une surface de Riemann $\Sigma$, et les coefficients de $\mathbf{L}(x)$ sont des formes m\'eromorphes : cela pourrait faire l'objet d'un article dans le futur.

\subsubsection{Flots et probl\`eme lin\'eaire}

\begin{figure}[h!]
\begin{center}
\begin{minipage}[c]{0.3\linewidth}
\raisebox{-2.5cm}{\includegraphics[width=\textwidth]{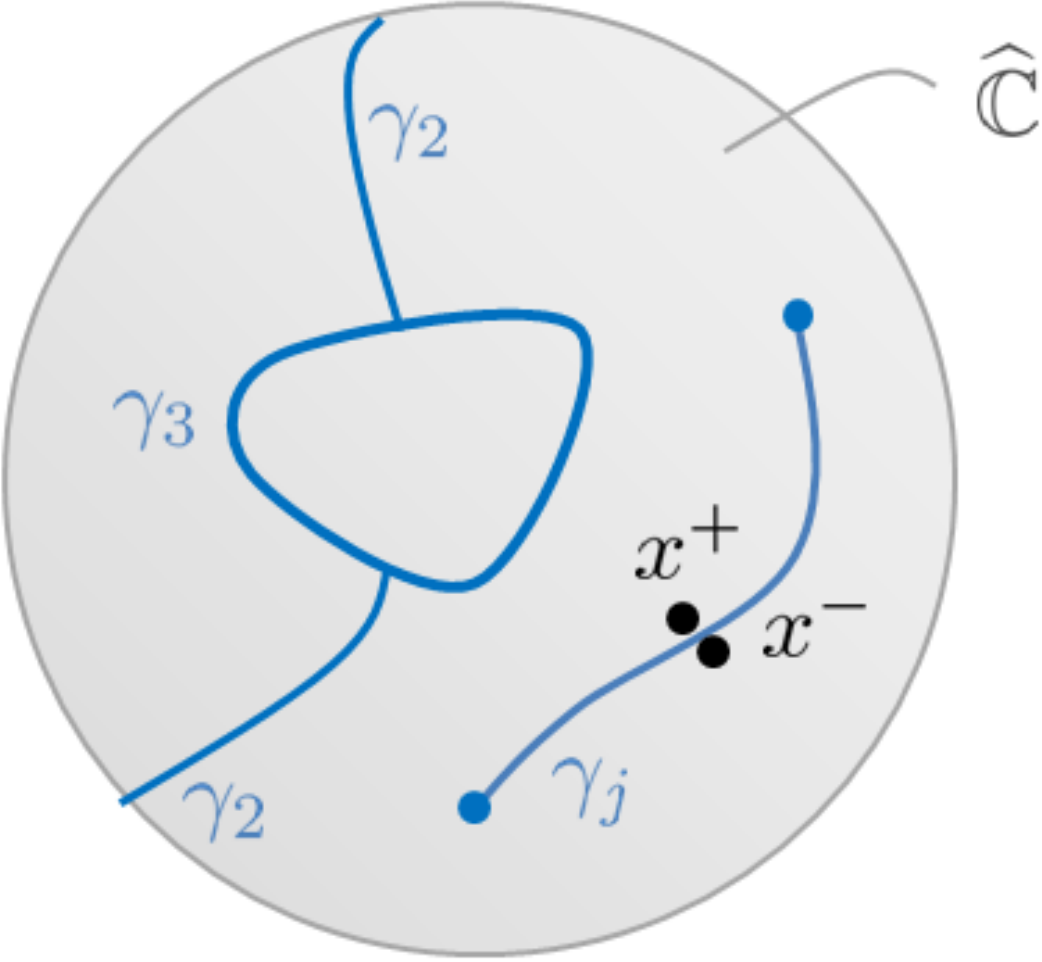}}
\end{minipage} \hfill
\begin{minipage}[l]{0.67\linewidth}
\caption{\label{fig:RH} Dans un probl\`eme de Riemann-Hilbert, on recherche une matrice $\Psi$ de taille $d \times d$, m\'eromorphe sur $\Sigma \setminus \cup_{j} \gamma_j$, dont les sauts sur les contours $\gamma_j$ sont fix\'es : $\Psi(x^+) = \mathcal{M}_j\cdot\Psi(x^-)$ ; et qui v\'erifie \'eventuellement des propri\'et\'es analytiques suppl\'ementaires (p\^{o}les ou z\'eros en certains points) et des conditions de croissance aux extr\'emit\'es des $\gamma_j$. Ce type de probl\`eme peut-\^{e}tre d\'efini sur une surface de Riemann $\Sigma$ quelconque.}
\end{minipage}
\end{center}
\end{figure}

Pr\`es des p\^{o}les $p$ de $\mathbf{L}(x)$, toute solution $\Psi(x)$ a une singularit\'e essentielle. Localement en $p$, on peut la d\'ecomposer en une partie r\'eguli\`ere $\widetilde{\Psi}_p(x)$, et une partie singuli\`ere $\Xi_{p}(x)$ :
\beq
\Psi(x) = \widetilde{\Psi}_p(x)\,\Xi_p(x) \nn
\eeq
En\label{RHH} g\'en\'eral, une solution $\Psi(x)$ poss\`ede des \textbf{monodromies} : si l'on prolonge analytiquement $\Psi(x)$ le long d'un cycle qui fait le tour de $p$, $\Psi(x)$ ne revient pas \`a sa valeur initiale apr\`es un tour complet. De mani\`ere \'equivalente, $\Psi(x)$ est la solution d'un \textbf{probl\`eme de Riemann-Hilbert} : cela signifie qu'elle admet des discontinuit\'es prescrites $\Psi_{-}(x) = \mathcal{M}_{j}(x)\Psi_+(x)$ de part et d'autre de certains contours $\gamma^j \subseteq \widehat{\mathbb{C}}$, o\`u $\mathcal{M}_j$ sont des matrices $d \times d$ (Fig.~\ref{fig:RH}). Ici, $\Psi^{-1}(x)\partial_x \Psi(x)$ est une forme rationnelle en $x$, donc sans monodromie, par cons\'equent les $\mathcal{M}_j$ ne d\'ependent pas de $x$.

D'apr\`es une construction classique \cite[Chapitre 8]{BBT}, il existe une infinit\'e de flots qui commutent $\partial_{t_{p;a;j}}$ ($a \in \{1,\ldots,d\}, j \in \mathbb{N}^*$), avec lesquels $\Psi(x)$ \'evolue en $\Psi(x,\mathbf{t})$ tout en gardant les m\^{e}mes monodromies $\mathcal{M}_j$. D'o\`u leur nom : \textbf{d\'eformations isomonodromiques}. Ces flots perturbent le comportement de $\Psi$ au voisinage de la singularit\'e essentielle :
\bea
\Xi_p(x,\mathbf{t}) = \Xi_p(x)\,e^{\mathbf{T}(x,\mathbf{t})} \qquad \mathbf{T}(x,\mathbf{t}) = \sum_{a = 1}^{d} 2i\pi\,t_{p;a;0} \ln \xi_p(x) + \sum_{j \geq 1} \frac{t_{p;a;j}}{\xi_p(x)^j}\,\mathbf{E}_{a,a}&& \nn \\
\label{eq:Tap}&&
\eea
$\mathbf{E}_{a,a}$ est la matrice $d \times d$ qui ne contient que des z\'eros, \`a l'exception d'un $1$ en position $(a,a)$. $\xi_p(x)$ d\'esigne une coordonn\'ee locale sur $\Sigma$ centr\'ee en $p$. Puisque l'on travaille sur $\Sigma =\widehat{\mathbb{C}}$, on peut prendre $\xi_p(x) = x - p$ si $p \neq \infty$, et $\xi_{\infty}(x) = 1/x$.

Les temps $t_{p;a;0}$ sont sp\'eciaux, car seules les transformations discr\`etes $t_{p;a;0} \rightarrow t_{p;a;0} + \mathbb{Z}$ pr\'eservent les monodromies : \label{Schl}elles sont appel\'ees \textbf{transformations de Schlesinger}. Les flots $\partial_{t_{p;a;j}}$ pour $j \geq 1$ sont eux d\'efinis de mani\`ere consistante par :
\bea
\partial_{x} \Psi(x,\mathbf{t}) & = & \mathbf{L}(x,\mathbf{t})\Psi(x,\mathbf{t}),\quad \partial_{t_I} \Psi(x,\mathbf{t}) = \mathbf{M}_I(x,\mathbf{t}) \Psi(x,\mathbf{t}) \nn \\
\mathbf{M}_I(x,\mathbf{t}) & = & \big[\widetilde{\Psi}_p(x,\mathbf{t})\Xi_p(x)\,\partial_{t_I} \mathbf{T}(x,\mathbf{t}) \widetilde{\Psi}_p^{-1}(x,\mathbf{t})\Xi_p^{-1}(x)\big]_{-,p} \nn
\eea
car ils v\'erifient les conditions de compatibilit\'e :
\bea
\partial_x \mathbf{M}_I(x,\mathbf{t}) - \partial_{t_I} \mathbf{L}(x,\mathbf{t}) + [\mathbf{M}_I(x,\mathbf{t}),\mathbf{L}(x,\mathbf{t})] = 0 && \nn \\
\partial_{t_J} \mathbf{M}_{I}(x,\mathbf{t}) - \partial_{t_I} \mathbf{M}_J(x,\mathbf{t}) + [\mathbf{M}_{I}(x,\mathbf{t}),\mathbf{M}_J(x,\mathbf{t})] = 0 && \nn
\eea
$[\cdots]_{-,p}$ d\'esigne la partie divergente en $p$ de la d\'ecomposition en \'el\'ements simples de $\cdots$. Ainsi, $\mathbf{M}_I(x,\mathbf{t})$ et $\mathbf{L}(x,\mathbf{t})$ restent rationnelles. Par un argument d\'ej\`a employ\'e, cela assure que les monodromies $\mathcal{M}_j$ ne d\'ependent ni de $x$, ni de $\mathbf{t}$. Les \textbf{temps diagonaux} $t_{p;a;j}$ engendrent donc des transformations isomonodromiques continues pour $j \geq 1$, qui forment un sous-groupe commutatif maximal de d\'eformations isomonodromiques. En fait, toutes les matrices \'el\'ementaires $\mathbf{E}_{a,b}$ g\'en\`erent des d\'eformations isomonodromiques, et l'on pourrait tr\`es bien choisir un autre sous-groupe commutatif de flots pour travailler.

Pour collecter tous les flots continus associ\'es \`a un p\^{o}le $p$ et une direction $a$, on d\'efinit l'\textbf{op\'erateur d'insertion} :
\beq
\label{eq:inser} D_{p;a}^{(x)} = \sum_{j \geq 1} \xi_p^{j - 1}(x)\,\partial_{t_{p;a;j}}
\eeq
Le probl\`eme lin\'eaire se r\'e\'ecrit alors :
\beq
\label{eq:problin} D_{p;a}^{(x)} \Psi(x_0,\mathbf{t}) = \Psi(x_0)\,\mathbf{E}_{a,a}\,\frac{\Psi^{-1}(x_0,\mathbf{t})\Psi(x,\mathbf{t})}{\xi_p(x) -\xi_p(x_0)}
\eeq

\subsubsection{Fonction tau et noyau}

\noindent $\diamond\,$ \`{A} toute famille de solutions $\Psi(x,\mathbf{t})$, Jimbo, Miwa et Ueno ont montr\'e\footnote{Le cas o\`u $\mathbf{T}(x,\mathbf{t})$ n'est pas une matrice diagonale est un ajout de Bertola et Marchal \cite{BMtau}.} dans leur article fondateur \cite{JM81} que l'on peut associer une fonction $\tau(\mathbf{t})$ v\'erifiant :
\beq
\label{eq:Jibu} \partial_{t_I} \ln\tau(\mathbf{t}) = - \sum_{p} \Res_{x \rightarrow p}  \mathrm{Tr}\,\big(\widetilde{\Psi}_p^{-1}(x,\mathbf{t})\,\dd_x \widetilde{\Psi}_p(x,\mathbf{t})\,\partial_{t_I} \Xi_{p}(x,\mathbf{t})\,\Xi_p^{-1}(x,\mathbf{t})\big)
\eeq
R\'eciproquement, la solution du probl\`eme lin\'eaire se d\'eduit de $\tau(\mathbf{t})$ par une formule de Sato (\'{E}qn.~\ref{eq:Sato}, \cite{Sato2}). Cette fonction tau est un objet important pour les syst\`emes int\'egrables. Par exemple, $\tau$ s'annule en $\mathbf{t} = \mathbf{t}_0$ ssi le probl\`eme de Riemann-Hilbert que $\Psi(x,\mathbf{t}_0)$ devrait v\'erifier avec les conditions \'{E}qn.~\ref{eq:Tap} n'a pas de solution. C'est un th\'eor\`eme d\^{u} \`a Malgrange \cite{Malgrange} lorsque $p$ est un p\^{o}le simple de $\mathbf{L}$, et \`a Palmer \cite{Palmer} en toute\label{RHH2} g\'en\'eralit\'e.

\vspace{0.2cm}

\noindent $\diamond\,$ Le \textbf{noyau int\'egrable}, parfois\label{noyu2} appel\'e noyau de Christoffel-Darboux, est la matrice $d \times d$ d\'efinie par :
\beq
\mathbf{K}(x_1,x_2) = \frac{\Psi^{-1}(x_1)\Psi(x_2)}{x_1 - x_2} \nn
\eeq
Le fait que $\Psi$ soit solution du probl\`eme lin\'eaire \'{E}qn.~\ref{eq:problin} se traduit par une propri\'et\'e\footnote{Elle est \'equivalente \`a l'\'equation de Ricatti de \cite[Section 8.7]{BBT}.} d'\textbf{autor\'eplication} \label{autorep} :
\beq
\label{eq:amto} D_{p;a_2}^{(x_2)} \mathbf{K}_{a_1a_3}(x_1,x_3) = -\mathbf{K}_{a_1a_2}(x_1,x_2)\mathbf{K}_{a_2a_3}(x_2,x_3)
\eeq
La formule de Sato exprime $\mathbf{K}(x_1,x_2)$ en fonction de $\tau$ \cite{BBT} :
\beq
\label{eq:Sato} \mathbf{K}_{a_1a_2}(x_1,x_2) = \sum_{1 \leq b_1,b_2 \leq d} \big(\Xi_{p_1}^{-1}(x_1,\mathbf{t})\big)_{a_1b_1}\,\frac{\tau(\mathbf{t} + [x_1]_{p_1;b_1} - [x_2]_{p_2;b_2})}{\tau(\mathbf{t})}\,\big(\Xi_{p_2}(x_2,\mathbf{t})\big)_{b_2a_2}
\eeq
$[x]_{p;b}$ d\'esigne la famille de temps :
\beq
t_{q;c;0} = \delta_{q,p}\,\delta_{c,b},\qquad t_{q;c;j} = \delta_{q,p}\,\delta_{c,b}\,\frac{1}{j\,\xi_p^j(x)} \nn
\eeq
Le d\'ecalage dans l'argument de la fonction $\tau$ dans Eqn~\ref{eq:Sato} comporte \`a la fois une \'evolution des temps diagonaux, et des transformations de Schlesinger. L'autor\'eplication pour $\mathbf{K}(x_1,x_2)$ (\'{E}qn.~\ref{eq:amto}) est alors \'equivalente \`a l'\'equation de Hirota pour $\tau(\mathbf{t})$.

\subsubsection{Remarque}

Dans la suite, nous commencerons souvent avec deux syst\`emes diff\'erentiels compatibles :
\beq
\label{eq:2dis}\partial_x \Psi(x,s) = \mathbf{L}(x,s)\Psi(x,s),\qquad \partial_s \Psi(x,s) = \mathbf{M}(x,s)\Psi(x,s)
\eeq
dont l'\'equation de compatibilit\'e est une \'equation diff\'erentielle $(\star)$ par rapport \`a $s$, non lin\'eaire. Avec la construction \'evoqu\'ee ci-dessus, on peut toujours d\'efinir une infinit\'e de flots $\partial_{t_j}$ qui sont compatibles avec l'\'{E}qn.~\ref{eq:2dis} et perturbent le comportement de $\Psi$ au voisinage d'une singularit\'e essentielle ($\partial_s$ lui m\^{e}me s'identifiera peut-\^{e}tre avec une combinaison de ces flots). Cela garantit l'existence d'un op\'erateur d'insertion (\'{E}qn.~\ref{eq:inser}), qui sera cruciale pour la suite. On peut ensuite oublier l'existence des $\partial_{t_j}$ si on le souhaite et se concentrer sur le syst\`eme initial \'{E}qn.~\ref{eq:2dis}.


\subsection{Exemples}
\label{sec:PIIex}
\subsubsection{Hi\'erarchie de Painlev\'e I}
\label{hiePI}
Il existe un syst\`eme de Lax de taille $2 \times 2$ dont les \'equations de compatibilit\'e sont \'equivalentes \`a celles du mod\`ele minimal $(2m + 1,2)$, i.e. \`a la hi\'erarchie de Painlev\'e I. Il est construit comme suit, avec un temps $s$ pour commencer :
{\small \beq
 \label{eq:sysy}\mathbf{L}(x,s) = \sum_{k = 0}^m \widetilde{t}_k \left(\begin{array}{cc} A_k & B_k \\ C_k & -A_k \end{array}\right),\qquad \mathbf{M}(x,s) = \left(\begin{array}{cc} 0 & 1 \\ x + 2u(s) & 0 \end{array}\right)\qquad
\eeq}
$\!\!\!$avec :
\beq
B_k = -\sum_{j = 0}^{k} x^{k - j} R_j[u](s),\quad A_k = -\frac{\partial_s B_k}{2},\quad C_k = (x + 2u)B_k + \partial_s A_k \nn
\eeq
$R_j$ est le $j^{\textrm{\`{e}me}}$ \textbf{polyn\^{o}me diff\'erentiel de Gelfand-Dikii}, \label{GD}d\'efini r\'ecursivement \`a des constantes d'int\'egration pr\`es par :
\beq
R_0 = 1,\quad \frac{\partial R_{j + 1}}{\partial s} = -2u\,\frac{\partial R_j}{\partial s} - \frac{\partial u}{\partial s}\,R_j + \frac{1}{4}\frac{\partial^3 R_j}{\partial s^3} \nn
\eeq
Les premiers termes sont :
\beq
R_1 = -u\,,\qquad R_2 = \frac{3u^2}{2} - \frac{\partial_s^2 u}{4}\,,\qquad \ldots \nn
\eeq
L'\'equation de compatibilit\'e $[\partial_x - \mathbf{L},\partial_s - \mathbf{M}] = 0$ s'\'ecrit alors :
\beq
\label{eq:sysy2}\frac{\partial}{\partial s}\Big(\sum_{k = 0}^{m} 2\,\widetilde{t}_k R_{k + 1}[u](s)\Big) = 1
\eeq
Pour $m = 1$, $(\widetilde{t}_0,\widetilde{t}_1) = (0,1)$, on retrouve bien l'\'equation de Painlev\'e I (\'{E}qn.~\ref{eq:PIA}), et le probl\`eme lin\'eaire associ\'e se sp\'ecialise \`a :
\beq
\mathbf{L}_{(3,2)}(x,s) = \left(\begin{array}{cc} \frac{x + \partial_s u}{2} & -x + \partial_s u \\ (x + 2u)(-x + \partial_s u) - \frac{\partial^2_s u}{2} & -\frac{x + \partial_s u}{2} \end{array}\right) \nn
\eeq
$\partial_s$ est en quelque sorte le flot $\partial_{t_1}$. On peut bien s\^{u}r construire tous les autres flots $\partial_{t_j}$ ($j \geq 2$) en suivant le \S~\ref{sec:su}.

\subsection{\'{E}quation de Painlev\'e II}
\label{sec:PII}
Un exemple important en relation avec les lois de Tracy-Widom est \textbf{la hi\'erarchie de Painlev\'e II}. L'\'equation de Painlev\'e II :
\beq
\label{eq:PII}\frac{\partial^2 u}{\partial s^2} = 2u^3 + s\,u
\eeq
en est le premier repr\'esentant. C'est l'\'equation de compatibilit\'e des deux syst\`emes d\'efinis par les matrices de Lax \cite{AKNS} :
{\small \beq
\label{eq:poq}\mathbf{L}(x,s) = \left(\begin{array}{cc} -4ix^2 - i(s + 2u^2) & 4xu + 2i\partial_s u \\ 4xu - 2i\partial_s u & 4ix^2 + i(s + 2u^2)\end{array}\right),\quad \mathbf{M}(x,s) = \left(\begin{array}{cc} -ix & u \\ u & ix \end{array}\right)
\eeq}
$\!\!\!$Le probl\`eme de Riemann-Hilbert associ\'e est pr\'esent\'e en d\'etail dans l'article original \cite{FN80}, et dans le livre \cite{FIKN} qui est une revue de la m\'ethode g\'en\'erale. Il est important de garder \`a l'esprit l'\'equivalence entre syst\`emes \label{sejio}isomonodromiques et probl\`emes de Riemann-Hilbert pour les applications potentielles de la partie~\ref{sec:enjo}, mais nous n'en aurons pas besoin ici. Signalons surtout :
\begin{theo}
\label{theu} Lorsque $u(s)$ est la solution de Hastings-McLeod de Painlev\'e II, la fonction $\tau(s)$ de ce syst\`eme co\"{i}ncide avec le d\'eterminant du noyau d'Airy, qui est aussi la loi de Tracy-Widom pour $\beta = 1$ (\S~\ref{sec:largenty}) :
\beq
\tau(s) = \mathsf{TW}_{1}(s) \nn
\eeq
\end{theo}
Ceci est la manifestation d'un ph\'enom\`ene g\'en\'eral, \'etudi\'e dans \cite{BD02} : grosso modo, le \label{Fredo2}d\'eterminant de Fredholm d'un noyau int\'egrable est une fonction tau.

La hi\'erarchie compl\`ete de Painlev\'e II et le syst\`eme de Lax correspondant sont d\'ecrits par exemple\footnote{Afin que la comparaison avec le syst\`eme~\ref{eq:poq} soit plus facile, les matrices donn\'ees par \cite{BlE} ont \'et\'e conjugu\'ees par $\left(\begin{array}{cc} 1 & i \\ i & 1 \end{array}\right)$.} par Bleher et Eynard \cite{BlE} :
{\small \bea
\mathbf{L}(x,s) & = & \left(\begin{array}{cc} -is & 0 \\ 0 & is \end{array}\right) + \sum_{k = 1}^{m} \widetilde{t}_k \left(\begin{array}{cc} -iC_k & \frac{i}{2}\,\partial_s B_k + xB_k \\ -\frac{i}{2}\,\partial_s B_k + xB_k & iC_k \end{array}\right) \nn \\
\label{eq:h1} \quad \mathbf{M}(x,s) & = & \left(\begin{array}{cc} -ix & u \\ u & ix \end{array}\right)
\eea}
$\!\!\!$avec :
\beq
B_k = \sum_{j = 0}^{k - 1} x^{2(k - 1) - 2j}\,\hat{R}_j,\qquad C_{k} = x^{2k} + \sum_{j = 0}^{k - 1} x^{2(k - 1) - 2j}\,\check{R_j} \nn
\eeq
$\hat{R}$ et $\check{R}$ sont \label{GD2}d\'efinis par une relation de r\'ecurrence \`a la Gelfand-Dikii :
\bea
\hat{R}_0 = \frac{u^2}{2} \quad && \hat{R}_{k + 1} = u\check{R}_k - \frac{1}{4}\,\frac{\partial^2 \hat{R}_k}{\partial s^2} \nn \\
\check{R}_0 = \frac{u}{2}\quad && \check{R}_{k + 1} = \int_0^{s} u\,\frac{\partial \hat{R}_k}{\partial s} \nn
\eea
L'\'equation de compatibilit\'e $[\partial_x - \mathbf{L},\partial_s - \mathbf{M}] = 0$ s'\'ecrit alors :
\beq
\label{eq:h2}  s\,u(s) + \sum_{k = 1}^{m} \widetilde{t}_k \hat{R}_k = 0
\eeq
On retrouve le syst\`eme~\ref{eq:poq} lorsque $m = 1$ et $\widetilde{t}_1 = 4$.

\section{\'{E}quations de boucles et cons\'equences}
\label{sec:enjo}

\subsection{Notion de courbe spectrale semiclassique}
\label{semi}
Lorsque que l'on veut \'etudier l'asymptotique d'une solution $\Psi$ d'un probl\`eme lin\'eaire, il est commode de red\'efinir les diff\'erents param\`etres de $\Psi$ ($x$, les temps $t_j$, d'autres param\`etres libres \ldots), en introduisant un param\`etre redondant $N$, pour avoir un probl\`eme lin\'eaire \'equivalent :
\beq
\label{eq:probN} \frac{1}{N}\,\partial_x\Psi = \mathbf{L} \Psi,\qquad \frac{1}{N}\,\partial_{t_j}\Psi = \mathbf{M} \Psi
\eeq
Maintenant, chaque d\'eriv\'ee vient avec un pr\'efacteur $1/N$, et $\Psi$, $\mathbf{L}$ et $\mathbf{M}$ d\'ependent de $N$. On garde cependant les d\'efinitions :
\bea
D_{(p;a)}^{(x)} & = & \sum_{j \geq 1} \xi_p(x)^{j - 1}\,\partial_{t_{p;a;j}} \nn \\
\mathbf{T}(x,\mathbf{t}) & = & \sum_{a = 1}^{d} \Big(2i\pi t_{p,0}\,\ln\xi_p(x) + \sum_{j \geq 1} \frac{t_{p;a;j}}{\xi_p^{j}(x)}\Big)\mathbf{E}_{a,a} \nn
\eea
Cette mise \`a l'\'echelle est pertinente lorsque le r\'egime asymptotique que l'on veut \'etudier correspond \`a $N \rightarrow +\infty$, et lorsque $\mathbf{L}$ et $\mathbf{M}$ ont une limite $\mathbf{L}^{[[0]]}$ et $\mathbf{M}^{[[0]]}$ lorsque $N \rightarrow +\infty$.

La \textbf{courbe spectrale semiclassique} associ\'ee\footnote{Cette d\'efinition sera compl\'et\'ee plus tard par un noyau de Bergman.} \`a \ref{eq:probN} est par d\'efinition la courbe alg\'ebrique :
\beq
\mathcal{E}(\mathbf{t})\: : \: \mathrm{det}\big(y - \mathbf{L}^{[[0]]}(x,\mathbf{t})\big) = 0  \nn
\eeq
Elle est toujours d\'efinie \`a partir de la limite $N \rightarrow \infty$ du syst\`eme diff\'erentiel par rapport \`a $x$. C'est une courbe de degr\'e $d$ en $y$, il y a donc $d$ valeurs ${}^{1}y(x),\ldots,{}^{d}y(x)$ sur $\mathcal{E}$ correspondant \`a une valeur de $x$ donn\'ee. La raison de cette d\'efinition est que l'asymptotique de $\Psi$ lorsque $N \rightarrow +\infty$ est de la forme :
\beq
\label{eq:nh}\Psi = \widetilde{\Psi}\,\exp\Big\{N\sum_{a = 1}^{d}\Big(\int^{x}\!\!{}^{a}y(x')\dd x'\Big)\mathbf{E}_{a,a}\Big\}
\eeq
Cette \'equation est simplement l'ordre dominant d'une approximation BKW, $\mathcal{E}$ donnant la relation de dispersion semiclassique o\`u $y \propto \textrm{impulsion}$. Elle joue un r\^{o}le important lorsque l'on souhaite d\'eterminer les coefficients d'un d\'eveloppement asymptotique complet de $\Psi$. On peut d'ores et d\'ej\`a remarquer que le m\^{e}me raisonnement appliqu\'e \`a $\frac{1}{N}\,\partial_{t_j} \Psi = \mathbf{M}_j\Psi$ conduit \`a :
\beq
\Psi \sim \widetilde{\Psi}^{(j)}\, \exp\Big\{N\sum_{a = 1}^{d} \Big(\int^x\!\! {}^{a}\underline{y}_j(x')\dd x'\Big)\mathbf{E}_{a,a}\Big\} \nn
\eeq
avec $\underline{y}_j$ d\'etermin\'e par :
\beq
\underline{\mathcal{E}}_j(\mathbf{t})\: : \: \mathrm{det}\big(\underline{y}_j - \mathbf{M}_j^{(0)}(x,\mathbf{t})\big) = 0 \nn
\eeq
L'existence de syst\`emes diff\'erentiels compatibles, donc la possibilit\'e de faire des calculs avec chacun des syst\`emes, pose des contraintes fortes sur les singularit\'es en $x$ des coefficients du d\'eveloppement asymptotique de $\Psi$.

\subsection{Corr\'elateurs et \'equations de boucles}
\label{sec:Shy}
Motiv\'es par les propri\'et\'es observ\'ees dans la chaine \`a deux matrices (un cas particulier de syst\`emes int\'egrables), Berg\`ere et Eynard \cite{BEdet} ont propos\'e de d\'efinir des \textbf{corr\'elateurs} \`a $n$ points $\mathcal{W}_n(x_1,\ldots,x_n)$ pour tout syst\`eme de Lax, par des formules d\'eterminantales\label{dett2} :
\bea
\mathcal{W}_1({}^{a}x) & = & \lim_{x' \rightarrow x} \Big(\mathbf{K}_{aa}(x,x') - \frac{1}{x - x'}\Big) \nn \\
\mathcal{W}_2({}^{a_1}x_1,{}^{a_2}x_2) & = & - \mathbf{K}_{a_1a_2}(x_1,x_2)\mathbf{K}_{a_2a_1}(x_2,x_1) - \frac{\delta_{a_1,a_2}}{(x_1 - x_2)^2} \nn \\
\mathcal{W}_n({}^{a_1}x_1,\ldots,{}^{a_n}x_n) & = & (-1)^{n + 1} \sum_{\sigma\,\,\mathrm{cycles}\,\mathrm{de}\,\mathfrak{S}_n} \prod_{i = 1}^{n} \mathbf{K}_{a_ia_{\sigma(i)}}(x_i,x_{\sigma(i)}) \nn
\eea
Les \textbf{corr\'elateurs non connexes} sont simplement les d\'eterminants :
\beq
\overline{\mathcal{W}}_n({}^{a_1}x_1,\ldots,{}^{a_n}x_n) = "\mathrm{det}"\big(\mathbf{K}_{a_ia_j}(x_i,x_j)\big)_{1 \leq i,j \leq n} \nn
\eeq
$"\mathrm{det}"$ signifie que l'on remplace chaque occurence de $\mathbf{K}_{aa}(x,x)$ (qui n'a pas de sens \`a cause de la singularit\'e de $\mathbf{K}$ \`a points co\"{i}ncidants) par $\mathcal{W}_1({}^{a}x)$, et chaque occurence de $\mathbf{K}_{a_1a_2}(x_1,x_2)\mathbf{K}_{a_2a_1}(x_2,x_1)$ (qui elle aurait un sens) par $-\mathcal{W}_2({}^{a_1}x_1,{}^{a_2}x_2)$.

Apr\`es un petit calcul :
\beq
\mathcal{W}_1({}^{a}x) =  -N\big(\Psi^{-1}(x)\mathbf{L}(x)\Psi(x)\big)_{aa} \nn
\eeq
et une simple comparaison avec la formule de Jimbo, Miwa, Ueno pour la fonction $\tau$ (\'{E}qn.~\ref{eq:Jibu}) montre que :
\beq
\mathcal{W}_1({}^{a}x) = \frac{1}{N}\,D_{(p;a)}^{(x)}\ln \tau - N(\partial_x \mathbf{T})_{aa} \nn
\eeq
Autrement dit,  les variations de $\tau$ le long des flots $\partial_{t_{p;a;j}}$ sont encod\'ees dans le d\'eveloppement de Taylor\footnote{C'est en fait un d\'eveloppement en s\'erie de Laurent, dont la partie n\'egative est trivialement connue. Ce ph\'enom\`ene de d\'ecalage pour $\mathcal{W}_1$ et $\mathcal{W}_2$ se manifestait d\'ej\`a dans les mod\`eles de matrices.} de $\mathcal{W}_1({}^{a}x)$ lorsque $x$ approche le p\^{o}le $p$. La propri\'et\'e d'autor\'eplication du noyau $\mathbf{K}$ permet ensuite d'\'etendre ce r\'esultat aux autres corr\'elateurs :
\bea
\mathcal{W}_2({}^{a_1}x_1,{}^{a_2}x_2) & = & N^{-2}\,D_{(p;a_1)}^{(x_1)}\,D_{(p;a_2)}^{(x_2)} \ln \tau - \frac{\delta_{a_1,a_2}}{(x_1 - x_2)^2} \nn \\
\label{eq:buds}\mathcal{W}_n({}^{a_1}x_1,\ldots,{}^{a_n}x_n) & = & N^{-n}\,D_{(p_1;a_1)}^{(x_1)}\cdots D_{(p_n;a_n)}^{(x_n)} \ln \tau
\eea
au sens d'une \'egalit\'e des d\'eveloppement de Taylor\footnote{Pour $\mathcal{W}_2$, voir la note pr\'ec\'edente.} lorsque $(x_1,\ldots,x_n) \rightarrow (p_1,\ldots,p_n)$. Ces formules peuvent aussi servir de d\'efinition des corr\'elateurs.

Ces objets sont particuli\`erement int\'eressants car ils v\'erifient des \'equations de boucles (comparer avec le \S~\ref{sec:heqb}).\label{heqb2}

\vspace{0.2cm}

\noindent $\diamond\,$ \emph{\'{E}quation de boucle lin\'eaire}
\beq
\label{eq:B1} \frac{1}{N}\sum_{a = 1}^{d} \mathcal{W}_1({}^{a}x) = - \Tr \mathbf{L}(x)
\eeq

\noindent $\diamond\,$ \emph{\'{E}quation de boucle quadratique}
\beq
\label{eq:B2}\frac{1}{N^2}\sum_{1 \leq a < b \leq d} \mathcal{W}_2({}^{a}x,{}^{b}x) + \mathcal{W}_1({}^{a}x)\mathcal{W}_1({}^{b}x) = \frac{1}{2}\big[\big(\Tr \mathbf{L}(x))^2 - \Tr \mathbf{L}^2(x)\big]
\eeq

Le point remarquable dans ces \'equations est que le membre de droite est une fonction rationnelle de $x$ dont la position des p\^{o}les est connue, alors que le membre de gauche vient d'une fonction multivalu\'ee en $x$. Puisque l'on a l'\'{E}qn.~\ref{eq:buds}, les applications successives de l'op\'erateur d'insertion g\'en\`erent l'ensemble des \'equations de boucles pour les $\mathcal{W}_n$.

\subsection{Lien avec la r\'ecurrence topologique}
\label{sec:contr}

Supposons que les $\mathcal{W}_n$ ait un d\'eveloppement asymptotique lorsque $N \rightarrow +\infty$ du type "d\'eveloppement topologique" :
\beq
\label{eq:juk}\mathcal{W}_n = \sum_{g \geq 0} N^{2 - 2g}\mathcal{W}_n^{(g)}
\eeq
Si l'on ins\`ere ce d\'eveloppement dans les relations que l'on vient d'obtenir, on trouve que $\mathcal{W}_n^{(g)}$ satisfait la hi\'erarchie d'\'equations de boucles du \S~\ref{sec:heqb}.
La r\'ecurrence topologique en donne l'unique solution qui n'a de singularit\'es qu'aux points de branchement de $\mathcal{E}$. D'o\`u le th\'eor\`eme suivant :
\begin{theo}
\label{thsu}Supposons :

\vspace{0.1cm}

\noindent $(i)\,$ $\mathbf{L}$ et $\mathbf{M}$ ont un d\'eveloppement asymptotique en puissances de $1/N$ :
 \beq
 \mathbf{L}(x,\mathbf{t}) = \sum_{k \geq 0} N^{-k}\,\mathbf{L}^{[[k]]}(x,\mathbf{t}),\quad \mathbf{M}(x,\mathbf{t}) = \sum_{k \geq 0} N^{-k}\,\mathbf{M}^{[[k]]}(x,\mathbf{t}) \nn
 \eeq

\vspace{0.1cm}

\noindent $(ii)\,$ $\widetilde{\Psi}$ d\'efini par l'\'{E}qn.~\ref{eq:nh} admet une limite quand $N \rightarrow +\infty$.

Cela assure que $\widetilde{\Psi}$ admet un d\'eveloppement asymptotique :
\beq
\widetilde{\Psi}(x,\mathbf{t}) = \sum_{k \geq 0} N^{-k}\,\widetilde{\Psi}^{[[k]]}(x,\mathbf{t}) \nn
\eeq

Ajoutons d'autres hypoth\`eses :

\vspace{0.1cm}

\noindent $(iii)\,$ Pour tout $b \in \{1,\ldots,d\}$, il existe une fonction $\psi_b^{[[k]]}(\cdot,\mathbf{t})$ m\'eromorphe sur $\mathcal{E}$, d\'efinie par $\widetilde{\Psi}_{a,b}^{[[k]]}(x,\mathbf{t})$ dans le feuillet $a \in \{1,\ldots,d\}$, dont les singularit\'es \'eventuelles se situent aux points de branchement de la courbe spectrale semiclassique ou aux p\^{o}les de $\mathbf{L}$ ou de $\mathbf{M}$.

\vspace{0.1cm}

\noindent $(iv)\,$ $\mathcal{W}_1$ admet un d\'eveloppement asymptotique de la forme $\sum_{g \geq 0} N^{1 - 2g}\mathcal{W}_1^{(g)}$.

\vspace{0.1cm}

Alors,  $\mathcal{W}_n$ et $\ln \tau$ admettent un d\'eveloppement topologique :
\beq
\ln \tau = \sum_{g \geq 0} N^{2 - 2g}\,F^{(g)}[\mathcal{S}],\qquad \mathcal{W}_n = \sum_{g \geq 0} N^{2 - 2g - n}\,\omega_n^{(g)}[\mathcal{S}] \nn
\eeq
Les coefficients sont donn\'es par la r\'ecurrence topologique appliqu\'ee \`a la courbe spectrale \label{bjs2}semiclassique $\mathcal{S} = [\mathcal{E},x,y,B]$ avec noyau de Bergman $B(z_1,z_2)$ d\'efini lorsque $z_1$ est dans le feuillet $a_1$ et $z_2$ dans le feuillet $a_2$ par :
\beq
B(z_1,z_2) = \dd x(z_1)\dd x(z_2)\,\mathcal{W}_2^{(0)}\big({}^{a_1}x(z_1),{}^{a_2}x(z_2)\big) + \delta_{a_1,a_2}\,\frac{\dd x_1\dd x_2}{(x_1 - x_2)^2} \nn
\eeq
\end{theo}

\subsubsection{Remarques}

\label{rq}Les travaux de \cite{BEdet} n'utilisaient qu'un op\'erateur d'insertion formel. Pour ces auteurs, il \'etait alors n\'ecessaire d'\'etablir les \'equations de boucles pour chaque $n$, conduisant \`a une preuve assez technique et valable uniquement pour les syst\`emes $2 \times 2$. De plus, ils devaient supposer l'existence d'un d\'eveloppement asymptotique pour tous les $\mathcal{W}_n$ afin d'avoir le r\'esultat du Th\'eor\`eme~\ref{thsu}. Je me suis rendu compte ensuite, gr\^{a}ce \`a \cite{BBT}, qu'il existe toujours un op\'erateur d'insertion r\'ealis\'e par des vrais flots (\'{E}qn.~\ref{eq:inser}). Cela permet d'\'etendre les r\'esultats de \cite{BEdet} aux syst\`emes $d \times d$ en toute g\'en\'eralit\'e, comme nous venons de le pr\'esenter.

Dans un exemple pratique, il faut surtout v\'erifier que $\psi_{b}^{[[k]]}$ n'a des singularit\'es qu'aux points de branchement de $\mathcal{E}$ (i.e. les z\'eros de $\dd x$ sur $\mathcal{E}$), et que $\mathcal{W}_1$ a un d\'eveloppement en puissances impaires de $1/N$. Il suffit pour cela d'\'etudier le d\'eveloppement asymptotique de $\big(\widetilde{\Psi}_{a,b}\big)_{1 \leq b \leq d}$, qui est une famille de solutions ind\'ependantes d'une \'equation diff\'erentielle d'ordre $d$ (autant dans la variable $x$ que dans les variables $t_j$), dont les coefficients sont des fonctions rationnelles en $x$.
Cette d\'emarche a \'et\'e d\'etaill\'ee pour les syst\`emes $2 \times 2$ dans mon article \cite{BETW} avec application au syst\`eme de Lax associ\'e \`a l'\'equation de Painlev\'e II, et sera d\'etaill\'ee dans un prochain article pour les syst\`emes $d \times d$ avec application aux mod\`eles minimaux $(p,q)$.

Il peut arriver que $\widetilde{\Psi}$ n'ait pas de limite lorsque $N \rightarrow +\infty$. En effet, de mani\`ere g\'en\'erique lorsque $\mathcal{E}$ est une courbe de genre $\mathfrak{g} > 0$, on s'attend \`a la pr\'esence de fonctions th\^{e}ta variant rapidement avec $N$, \`a l'ordre dominant et \`a tous les ordres du d\'eveloppement asymptotique (cf.~\S~\ref{sec:mono}). En principe, le Th\'eor\`eme~\ref{thsu} devrait se g\'en\'eraliser : il suffit d'identifier le comportement dominant de $\widetilde{\Psi}$ lorsque $N \rightarrow +\infty$, et de sommer comme au \S~\ref{sec:fraco} sur les fractions de remplissage de $\mathcal{E}$ de sorte \`a reproduire ce comportement dominant. Il y aura toutefois un travail d'analyse non n\'egligeable pour justifier un tel r\'esultat. Ce probl\`eme rejoint celui de l'\'etude de l'asymptotique dans les chaines de matrices.

\subsection{Applications}

\subsubsection{Hi\'erarchie de Painlev\'e I}
\label{hiePIA}
Berg\`ere et Eynard ont appliqu\'e cette m\'ethode \`a la hi\'erarchie de Painlev\'e I dans leur article \cite{BEdet}. La courbe spectrale semiclassique est\label{PIs} :
\beq
\mathcal{S}_{\mathrm{PI}}\::\: \left\{\begin{array}{l} y = -\sqrt{x + 2u^{[[0]]}}\Big(\sum_{k = 0}^{m} \widetilde{t}_k \sum_{j = 0}^k x^{k - j}\,\mathrm{C}_{2j}^{j}\,\big(-u^{[[0]]}/2\big)^{j}\Big) \\
B(z_1,z_2) = \frac{\dd z_1\dd z_2}{(z_1 - z_2)^2}\qquad z = \sqrt{x + 2u^{[[0]]}} \end{array}\right.\nn
\eeq
\`{A} la famille de solutions\footnote{Nous laissons de c\^{o}t\'e la question de l'existence et unicit\'e d'une telle solution.} du syst\`eme d\'efini par les \'{E}qns.~\ref{eq:sysy} et \ref{eq:sysy2} qui se comportent comme :
{\small \bea
&& \lim_{N \rightarrow \infty} u(s,\widetilde{\mathbf{t}}) = u^{[[0]]}(s,\widetilde{\mathbf{t}}),\qquad \sum_{k = 0}^{m} \mathrm{C}_{2k + 1}^{k}\,\widetilde{t}_k\,\big(-u^{[[0]]}/2\big)^{2k} = \frac{s}{4} \nn \\
&& \Psi(x,s,\widetilde{\mathbf{t}}) = \widetilde{\Psi}(x,s,\widetilde{\mathbf{t}})\,\exp\Big\{N\int_{-2u^{[[0]]}}^{x} y(x')\dd x'(\mathbf{E}_{1,1} - \mathbf{E}_{2,2})\Big\} \nn
\eea}
$\!\!\!$est associ\'ee une fonction $\tau(s,\mathbf{t})$. Ils v\'erifient les conditions $(i)$, $(ii)$, $(iii)$, et montrent que $\mathcal{W}_n$ a un d\'eveloppement a priori en $\sum_{k \geq 0} N^{n - 2k} \mathcal{W}_n^{[[k]]}$. En supposant que ce d\'eveloppement commence \`a l'ordre $N^{2 - n}$, ils concluent que $\ln \tau$ a un d\'eveloppement asymptotique :
\beq
\ln \tau = \mathrm{cte} + \sum_{g \geq 0} N^{2 - 2g}\,\mathcal{F}^{(g)}[\mathcal{S}_{\mathrm{PI}}] \nn
\eeq
\label{courbeK3}Remarquons que $\mathcal{S}_{\mathrm{PI}}$ est un cas particulier de la courbe spectrale de Kontsevich, donc les $\mathcal{F}^{(g)}[\mathcal{S}_{\mathrm{PI}}]$ pour $g \geq 2$ ont une expression en fonction des nombres d'intersections dans $\overline{\mathcal{M}}_g$ (cf.~\ref{sec:interFg}).

\subsubsection{Hi\'erarchie de Painlev\'e II}
\label{hiePIIA}
R\'ecemment, Cafasso et Marchal \cite{CM10} ont fait de m\^{e}me pour la hi\'erarchie de Painlev\'e II. Sa courbe spectrale semiclassique est\label{PIIs} :
\beq
\mathcal{S}_{\mathrm{PII}}\::\: \left\{\begin{array}{l} y = \sqrt{(u^{[[0]]})^2 - x^2}\Big(\sum_{k = 0}^{m} \widetilde{t}_k\,\sum_{j = 0}^{k - 1} \mathrm{C}_{2j}^{j} \,x^{2(k - j) - 1}\, \big(u^{[[0]]}/2\big)^{2j}\Big) \\ B(z_1,z_2) = \frac{\dd z_1\dd z_2}{(z_1 - z_2)^2}\qquad z = \frac{1 + \sqrt{x^2 - (u^{[[0]]})^2}}{u^{[[0]]}} \end{array}\right. \nn
\eeq
\`{A} la famille de solutions du syst\`eme d\'efini par les \'{E}qns.~\ref{eq:h1} et \ref{eq:h2} qui se comporte comme :
{\small \bea
&& \lim_{N \rightarrow \infty} u(s,\widetilde{\mathbf{t}}) = u^{[[0]]}(s,\widetilde{\mathbf{t}}),\qquad \sum_{k = 1}^{m} \mathrm{C}_{2k}^k\,\widetilde{t}_k\,\big(u^{[[0]]}/2\big)^{2k} = - s \nn \\
&& \Psi(x,s,\widetilde{\mathbf{t}}) = \widetilde{\Psi}(x,s,\widetilde{\mathbf{t}})\,\exp\Big\{N\int_{u^{[[0]]}}^{x} y(x')\dd x' (\mathbf{E}_{1,1} - \mathbf{E}_{2,2})\Big\} \nn
\eea}
$\!\!\!$est associ\'ee une fonction $\tau(s,\widetilde{\mathbf{t}})$. Toujours en supposant que $\mathcal{W}_n$ commence en $N^{2 - n}$, ils concluent :
\beq
\ln \tau = \mathrm{cte} + \sum_{g \geq 0} N^{2 - 2g}\,\mathcal{F}^{(g)}[\mathcal{S}_{\mathrm{PII}}] \nn
\eeq

L'argument que j'ai propos\'e conduit \`a l'\'{E}qn.~\ref{eq:buds} et compl\`ete ces travaux.

\subsubsection{Application \`a la loi de Tracy-Widom}
\label{sec:TWTW}
Dans l'article \cite{BETWc}, j'ai montr\'e que toutes les conditions \'etaient r\'eunies pour appliquer le Th\'eor\`eme~\ref{thsu} au syst\`eme de Lax dont l'\'equation de Painlev\'e II est l'\'equation de compatibilit\'e (\'{E}qn.~\ref{eq:poq}), et dont on recherche la solution $\Psi$ telle que :
\beq
\det\Psi \equiv 1,\qquad \Psi(x,s) = \widetilde{\Psi}_{\infty}(x,s)\,e^{i\big(\frac{4}{3}x^3 + sx\big)(\mathbf{E}_{1,1} - \mathbf{E}_{2,2})} \nn
\eeq
lorsque $|X| \rightarrow \infty$ dans le secteur angulaire $[-\pi/3,0]$.

La premi\`ere \'etape consiste \`a introduire \`a la main un param\`etre $N$. Si l'on souhaite que chaque d\'eriv\'ee $\partial_s$ soit d'ordre $1/N$ tandis que tous les autres termes restent d'ordre $1$, on est conduit \`a poser :
\beq
s = N^{2/3}\,S\,,\qquad x = N^{1/3}\,X\,,\qquad q(s) = N^{1/3}\,Q(S) \nn
\eeq
Ce qui nous am\`ene \`a un syst\`eme de Lax mis en forme comme en \'{E}qn.~\ref{eq:probN}, avec :
{\small \beq
\mathbf{L} = \left(\begin{array}{cc} -4iX^2 -i(S + 2Q^2(S)) & 4XQ(S) + \frac{2iQ'(S)}{N} \\ 4XQ(S) - \frac{2iQ'(S)}{N} & 4iX^2 + i(S + 2Q^2(S))\end{array}\right),\quad \mathbf{M} = \left(\begin{array}{cc} -iX & Q(S) \\ Q(S) & iX \end{array}\right) \nn
\eeq}
$\!\!\!$et l'\'equation de compatibilit\'e :
\beq
\label{eq:juml}\frac{1}{N^2}\,\partial_S^2 Q(S) = 2Q^3(S) + SQ(S)
\eeq
Cette mise en forme est pertinente si le membre de gauche tend vraiment vers $0$ lorsque $N \rightarrow +\infty$. Quand $A = 0$, c'est le cas lorsque $q$ est la solution d'Hastings-McLeod (\'{E}qn.~\ref{eq:jud}), pourvu que $S < 0$ : $\lim_{N \rightarrow \infty} Q(S) = \sqrt{-S/2}$. De l'\'{E}qn.~\ref{eq:juml}, on d\'eduit que $Q$ admet un d\'eveloppement asymptotique :
\beq
Q(S) = \sqrt{\frac{-S}{2}} + \sum_{g \geq 1} N^{-2g}\,Q^{(g)}(S) \nn
\eeq
avec des coefficients que l'on peut calculer r\'ecursivement. Donc $\mathbf{L}$ et $\mathbf{M}$ admettent un d\'eveloppement en $1/N$, et $N \rightarrow +\infty$ correspond au r\'egime $s \rightarrow -\infty$.

La courbe spectrale semiclassique de ce syst\`eme de Lax est \label{PIIs2}:
\beq
\label{eq:cSS}\mathcal{E}\::\: Y^2 = -16X^2\Big(X^2 + \frac{S}{2}\Big)
\eeq
C'est une courbe hyperelliptique : il y a deux feuillets, et l'on passe de l'un \`a l'autre en choisissant l'autre d\'etermination de la racine. Elle est de genre $0$, puisqu'elle admet un param\'etrage rationnel :
\beq
X(\zeta) = \sqrt{\frac{-S}{8}}\Big(\zeta + \frac{1}{\zeta}\Big),\quad Y(\zeta) = \frac{S}{2}\Big(\zeta^2 - \frac{1}{\zeta^2}\Big) \nn
\eeq
Les deux points de branchement $X = \pm \sqrt{-S/2}$ correspondent \`a $\zeta = \pm 1$. Il y a un unique noyau de Bergman qui est une forme m\'eromorphe sur $\mathcal{E}$ :
\beq
B(\zeta_1,\zeta_2) = \frac{\dd\zeta_1\dd\zeta_2}{(\zeta_1 - \zeta_2)^2} \nn
\eeq
On note $\mathcal{S}_S$ la courbe spectrale $[\mathcal{E},x,y,B]$.

Les points $(ii)$, $(iii)$ et $(iv)$ sont v\'erifi\'es en \'etudiant l'\'equation de type Schr\"{o}dinger satisfaite par $\Psi_{ab}$. Il apparait que $\Psi_{ab}(X,S)$, ordre par ordre en $1/N$, ne peut d\'evelopper de singularit\'es en $X$ qu'aux z\'eros communs de $Y(X)$ et de $\underline{Y}(X)$ d\'efini par :
\beq
\mathrm{det}\big(\underline{Y} - \mathbf{M}^{[[0]]}(X)\big) = 0 \quad \Leftrightarrow \quad \underline{Y}^2 = -\Big(X^2 + \frac{S}{2}\Big) \nn
\eeq
Par cons\'equent, il n'y a pas de singularit\'e en $X = 0$, seules les singularit\'es aux points de branchement $X = \pm \sqrt{-S/2}$ se manifestent, i.e. $(iii)$ est v\'erifi\'e. Enfin, il existe une sym\'etrie :
\beq
\big(\Psi_{-N}^{t}\big)^{-1}(X,S) = \Psi_N(X,S) \nn
\eeq
valable pour tout $N$ fini qui assure que $\mathcal{W}_1$ est bien impair en $N$, d'o\`u $(iv)$.

\begin{figure}
\begin{center}
\includegraphics[width=0.9\textwidth]{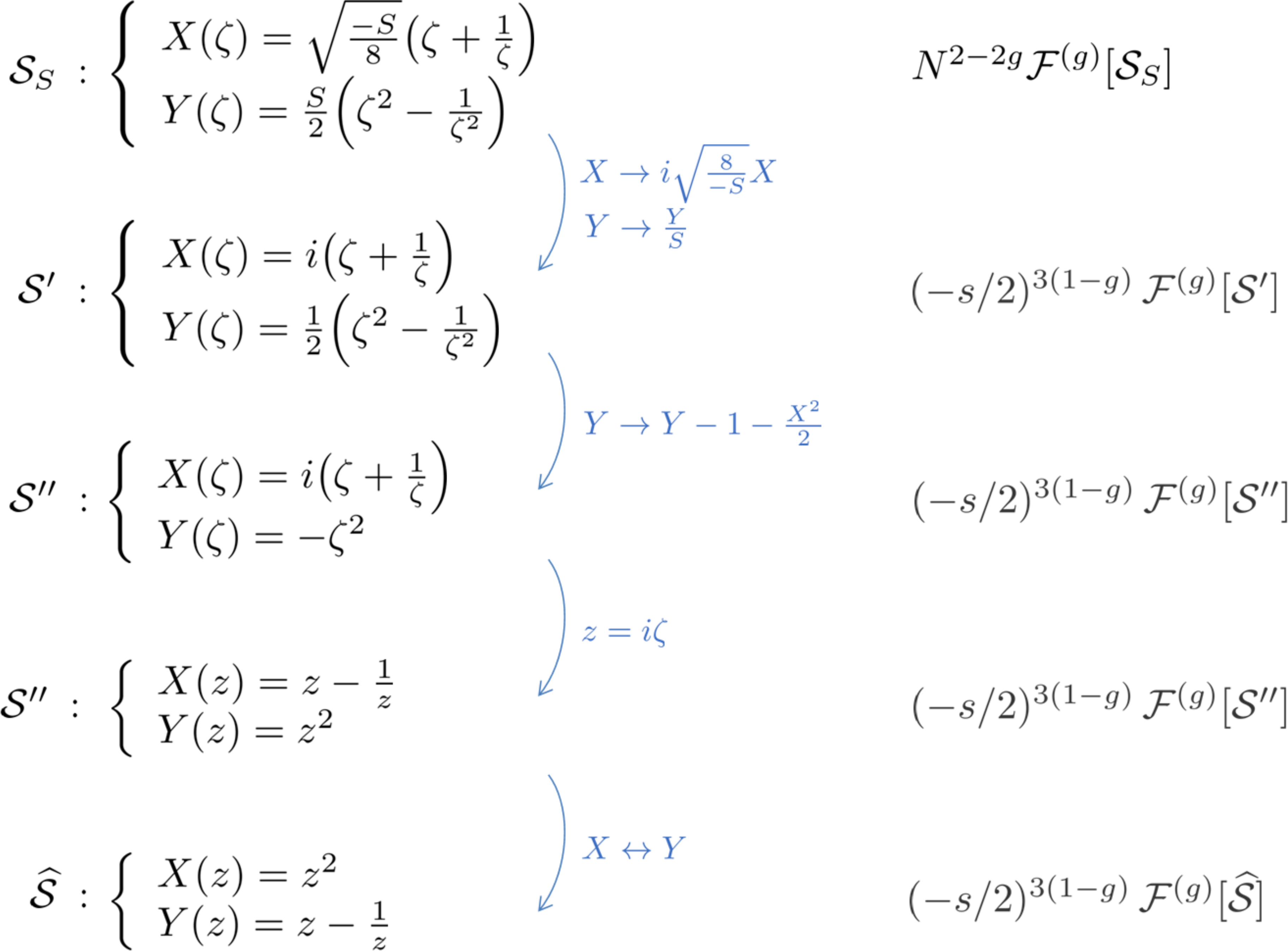}
\caption{\label{fig:sympchain} Relation entre la courbe spectrale semiclassique $\mathcal{S}_{S}$ du syst\`eme de Lax associ\'e \`a Painlev\'e II, et la courbe spectrale universelle $\widehat{\mathcal{S}}$ d\'ecrivant la transition point de branchement dur $\rightarrow$ point de branchement mou dans le cas g\'en\'erique.}
\end{center}
\end{figure}

D'apr\`es le Th\'eor\`eme~\ref{thsu}, $\ln \tau(S)$ admet le d\'eveloppement asymptotique lorsque $s \rightarrow -\infty$
\beq
\ln \tau(S) = \mathrm{cte} + \sum_{g \geq 0} N^{2 - 2g}\,\mathcal{F}^{(g)}[\mathcal{S}_S] \nn
\eeq
Par dilatation et transformations symplectiques, $\mathcal{S}_{S}$ se ram\`ene \`a la courbe \label{TWaa}spectrale $\widehat{\mathcal{S}} = [\mathbb{C},\widehat{x},\widehat{y},\widehat{B}]$ trouv\'ee au \S~\ref{sec:conn} (Fig.~\ref{fig:sympchain}) :
\beq
\widehat{x} = z^2,\quad \widehat{y} = z - \frac{1}{z},\quad \widehat{B}(z_1,z_2) = \frac{\dd z_1\dd z_2}{(z_1 - z_2)^2} \nn
\eeq
En utilisant les propri\'et\'es des invariants symplectiques, on obtient
\beq
\ln \tau(S) = \mathrm{cte} + \sum_{g \geq 0} (-s/2)^{3(1 - g)}\,\mathcal{F}^{(g)}[\widehat{\mathcal{S}}] \nn
\eeq
o\`u $(-s/2)^{0}$ doit \^{e}tre remplac\'e par $\ln |-s/2|$. D'apr\`es le Thm.~\ref{theu}, cela signifie que les invariants symplectiques de $\widehat{\mathcal{S}}$ sont les coefficients du d\'eveloppement asymptotique de la queue gauche de la loi de Tracy-Widom des matrices hermitiennes. Nous avons donc prouv\'e rigoureusement l'\'{E}qn.~\ref{eq:TWde} dans le cas $\beta = 1$.

Cette m\'ethode ne peut \^{e}tre utilis\'ee pour \'etablir \ref{eq:TWde} pour d'autres valeurs de $\beta$, m\^{e}me $\beta = 1/2$ ou $2$, car il n'existe pas de structure "int\'egrable" connue conduisant \`a la d\'eformation $\beta$ des \'equations de boucles. Autrement dit, partant d'un syst\`eme de Lax ou d'une d\'eformation de ce syst\`eme, on ne sait pas d\'efinir un objet int\'eressant $\widetilde{W}_1$ qui d\'epende de $\beta$ et satisfasse la d\'eformation $\beta$ (\'{E}qn.~\ref{eq:eqdg}) de l'\'equation de boucle quadratique (\'{E}qn.~\ref{eq:B2}).

\section{Construction d'un syst\`eme int\'egrable \`a partir d'une courbe spectrale}
\label{sec:consoK}

Dans un article c\'el\`ebre \cite{Kri77}, Krichever a montr\'e comment la g\'eom\'etrie d'une surface de Riemann compacte $\Sigma$ donne lieu \`a des solutions de la hi\'erarchie KP.
Cette classe de solutions, dites \textbf{alg\'ebro-g\'eom\'etriques}, est tr\`es particuli\`ere : $\Psi(x,\mathbf{t})$ peut \^{e}tre d\'efini comme fonction analytique univalu\'ee d'une variable $x$ vivant sur $\Sigma$. Outre les p\^{o}les, les seules singularit\'es qui apparaissent sont des points de branchement, d'ordre born\'e. Cela signifie que le groupe engendr\'e par les matrices de monodromie de $\Psi(x,\mathbf{t})$ est fini. Nous allons rappeler la construction alg\'ebro-g\'eom\'etrique au \S~\ref{sec:ALG}, en la reformulant d'une mani\`ere un peu diff\'erente de la litt\'erature sur les syst\`emes int\'egrables.

En g\'en\'eral, le groupe de monodromie d'une solution $\Psi(x,\mathbf{t})$ du probl\`eme lin\'eaire d'un syst\`eme int\'egrable est infini. Pour construire des solutions plus g\'en\'erales, on peut chercher \`a perturber une solution alg\'ebro-g\'eom\'etrique avec un petit param\`etre $1/N$, en autorisant l'ordre des singularit\'es de type "point de branchement" \`a cro\^{i}tre avec l'ordre en $1/N$. C'est typiquement ce qui arriverait si l'on avait une solution $\Psi(x,\mathbf{t})$ qui \'etait d\'efinie sur une surface de Riemann $\Sigma(\mathbf{t})$ qui \'evolue lentement (\`a une \'echelle de temps $\sim N$). Lorsque c'est le cas, on dira que $1/N$ le \textbf{param\`etre de dispersion}, et que $\Psi(x,\mathbf{t})$ est la solution\label{dispo} du probl\`eme lin\'eaire d'un \textbf{syst\`eme int\'egrable dispersif autour de $\Sigma$}. Tous les syst\`emes int\'egrables rencontr\'es jusqu'\`a pr\'esent \'etaient \label{semi2}des exemples de syst\`emes int\'egrables dispersifs autour de leur courbe spectrale semiclassique qui \'etait de genre $0$. L'enjeu du \S~\ref{sec:mono} est la construction d'un syst\`eme int\'egrable dispersif autour de n'importe quelle courbe $\Sigma$ de genre $\mathfrak{g}$. Plusieurs tentatives existent dans cette direction, notamment par Krichever \cite{Kri92}, et plus r\'ecemment Takasaki et Takebe \cite{TT}. L'int\'er\^{e}t de notre approche est de proposer des formules explicites pour organiser le d\'eveloppement asymptotique lorsque $N \rightarrow \infty$.

Cette partie esquisse le contenu d'un article en pr\'eparation avec Bertrand Eynard. J'indiquerai seulement l'id\'ee des d\'emonstrations, et soulignerai comme conjectures les points que nous n'avons pas encore d\'emontr\'es. Dans les premi\`eres parties de ce chapitre, nous partions d'un syst\`eme int\'egrable, et lui associions des notions comme celle de fonction tau, noyau int\'egrable, \ldots{} Ici, nous allons adopter une d\'emarche inverse : nous allons d'abord d\'efinir certains objets g\'eom\'etriques, puis nous allons rechercher le syst\`eme int\'egrable (s'il y en a un) dont ils apportent la solution. Pour cela, nous ferons souvent appel aux outils de g\'eom\'etrie complexe pr\'esent\'es dans l'Annexe~\ref{app:geomcx}. Pour \'eviter les confusions, nous allons maintenant d\'esigner la fonction tau par $\mathcal{T}$, et r\'eserver la lettre $\tau$ pour la matrice des p\'eriodes d'une surface de Riemann.

\subsection{Construction alg\'ebro-g\'eom\'etrique}
\label{sec:ALG}
Soit $\Sigma$ une surface de Riemann de genre $\mathfrak{g} > 0$, $X$ une fonction m\'eromorphe sur $\Sigma$, $(\mathcal{A}_{\mathfrak{h}},\mathcal{B}_{\mathfrak{h}})$ une base symplectique de cycles sur $\Sigma$, et $o \in \Sigma$ un point base. Ces donn\'ees additionnelles d\'efinissent une base de l'espace des formes m\'eromorphes sur $\Sigma$. Les \textbf{temps} de la hi\'erarchie seront les cooordonn\'ees $t_{p;j}$ :
\beq
\sigma(z) = \sum_{p} \Big(t_{p;0}\,\mathop{\dd S}_{o,p}(z) + \sum_{j \geq 1} t_{p;j}\,\Omega_{p;j}(z)\Big) + \sum_{\mathfrak{h} = 1}^{\mathfrak{g}} 2i\pi\,\epsilon_{\mathfrak{h}}\,\dd u_{\mathfrak{h}}(z) \nn
\eeq
Les fractions de remplissage $\epsilon_{\mathfrak{h}}$ \label{ffra3}ne joueront qu'un r\^{o}le spectateur, elles sont fix\'ees une fois pour toutes. \`{A} toute forme m\'eromorphe $\sigma$, on associe une forme $\chi$ d'int\'egrale nulle sur les $\mathcal{A}$-cycles, et le vecteur $\zeta[\sigma] = \big(\zeta_{\mathfrak{h}}[\sigma]\big)_{1 \leq \mathfrak{h} \leq \mathfrak{g}}$ de ses int\'egrales sur les $\mathcal{B}$-cycles :
\bea
\chi(z) & = & \sigma(z) - \sum_{\mathfrak{h} = 1}^{\mathfrak{g}} 2i\pi\epsilon_{\mathfrak{h}}\,\dd u_{\mathfrak{h}}(z) \nn \\
\zeta_{\mathfrak{h}}[\sigma] & = & \frac{1}{2i\pi}\oint_{\mathcal{B}_{\mathfrak{h}}} \chi = \frac{1}{2i\pi} \oint_{(\mathcal{B} - \tau\mathcal{A})_{\mathfrak{h}}} \!\!\!\!\!\!\!\!\!\sigma \nn
\eea
o\`u $\tau$ est la matrice des p\'eriodes de $\Sigma$. On d\'efinit une courbe spectrale en posant $\mathcal{S}_{\sigma} = [\Sigma,X,-\frac{\sigma}{\dd X},B]$, o\`u $B$ est le noyau de Bergman sur $\Sigma$ d'int\'egrale nulle sur les $\mathcal{A}$-cycles.

\subsubsection{Fonction tau}

Commen\c{c}ons par d\'efinir une fonction bilin\'eaire des temps $\widetilde{\mathcal{F}}^{(0)}[\sigma]$ :
\bea
\widetilde{\mathcal{F}}^{(0)}[\sigma] & = & \sum_{I,J} \frac{1}{2}\,t_It_J\,\iint_{\Omega^*_I \times \Omega^*_J} \!\!\!\!\!\!\!\! B(z_1,z_2) + \frac{1}{2}\sum_{\mathfrak{h},\mathfrak{h}' = 1}^{\mathfrak{g}} 2i\pi\,\epsilon_{\mathfrak{h}}\epsilon_{\mathfrak{h}'}\,\tau_{\mathfrak{h}\mathfrak{h}'} \nn
\eea
Nous annon\c{c}ons alors que :
\beq
\label{eq:taualg} \mathcal{T}\big[\sigma\big] = e^{\widetilde{\mathcal{F}}^{(0)}[\sigma]}\,\theta\big(\zeta[\sigma] + \mathbf{c}\big)
\eeq
est une fonction tau, ce qui est justifi\'e par les propositions suivantes.

\begin{prop}
\label{PP1} $\mathcal{T}$ v\'erifie une version globale de l'\'equation de Hirota :
\beq
\mathcal{T}\big[\sigma\big]\,\mathcal{T}\big[\sigma + \mathop{\dd S}_{z_1,z_2} + \mathop{\dd S}_{z_3,z_4}\big] = \mathcal{T}\big[\sigma + \mathop{\dd S}_{z_1,z_2}\big]\,\mathcal{T}\big[\sigma + \mathop{\dd S}_{z_3,z_4}\big] - \mathcal{T}\big[\sigma + \mathop{\dd S}_{z_1,z_4}\big]\,\mathcal{T}\big[\sigma + \mathop{\dd S}_{z_3,z_2}\big] \nn
\eeq
\end{prop}
Cette \'equation est exactement la r\'e\'ecriture pour $\mathcal{T}$ de l'identit\'e de Fay pour la fonction th\^{e}ta (\'{E}qn.~\ref{eq:fays}), avec $\mathbf{w} = \zeta[\sigma]$. Il suffit pour cela de remarquer que la transformation $\sigma \rightarrow \sigma + \mathop{\dd S}_{z_i,z_j}$ agit comme :
\bea
\zeta[\sigma] & \rightarrow & \zeta[\sigma] + \mathbf{u}(z_j) - \mathbf{u}(z_i) \nn \\
\widetilde{\mathcal{F}}^{(0)}[\sigma] & \rightarrow & \widetilde{\mathcal{F}}^{(0)}[\sigma] + \Big(\int_{z_i}^{z_j} \sigma\Big)  - \ln\Big(E(z_j,z_i)\sqrt{\dd x(z_i)\dd x(z_j)}\Big) \nn
\eea
$E$ est la forme primaire :
\beq
E(z_j,z_i) = \frac{\theta(\mathbf{u}(z_j) - \mathbf{u}(z_i) + \mathbf{c})}{\sqrt{\dd h_{\mathbf{c}}(z_j)\,\dd h_{\mathbf{c}}(z_i)}},\quad \dd h_{\mathbf{c}}(z) = \sum_{\mathfrak{h} = 1}^{\mathfrak{g}} (\partial_{u_{\mathfrak{h}}}\theta)(\mathbf{c})\,\dd u_{\mathfrak{h}}(z) \nn
\eeq

\label{opin2}On peut d\'efinir un op\'erateur d'insertion $\delta_{z_0}$, qui est la d\'eriv\'ee $\dd x(z_0)\frac{\partial}{\partial\varepsilon}\big|_{\varepsilon = 0}$ le long de la d\'eformation $\sigma_{\varepsilon}(z) = \sigma(z) + \varepsilon\,\frac{B(z_0,z)}{\dd x(z_0)}$. Cette d\'eformation peut aussi \^{e}tre obtenue en ajoutant deux p\^{o}les simples, en $z_0$ et $z_0'$, et en laissant $z_0'$ tendre vers $z_0$ :
\beq
\mathop{\dd S}_{z_0,z_0'}(z) = \int_{z_0}^{z_0'} \!\!B(z,\cdot) \sim \big(\xi(z_0') - \xi(z_0)\big) \frac{B(z_0,z)}{\dd \xi(z_0)} \nn
\eeq
dans toute coordonn\'ee locale $\xi$ au voisinage de $z_0$. Lorsque $z_0$ est dans une r\'egion o\`u $\xi_p$ (la coordonn\'ee locale autour d'un p\^{o}le $p$ de $\sigma$) est encore une coordonn\'ee locale, on a \'egalement la repr\'esentation :
\beq
\delta_z = \sum_{j \geq 1} \xi_p^{j - 1}(z_0)\,\partial_{t_{p;j}} \nn
\eeq

Par cons\'equent, si on laisse $z_3$ et $z_4$ coalescer vers un point $z$ :
\begin{prop}
\label{PP2} $\mathcal{T}$ v\'erifie une version infinit\'esimale de l'\'equation de Hirota :
\beq
\mathcal{T}\big[\sigma\big]\,(\delta_{z} \mathcal{T})\big[\sigma + \mathop{\dd S}_{z_1,z_2}\big]  - (\delta_z\mathcal{T})\big[\sigma\big]\,\mathcal{T}\big[\sigma + \mathop{\dd S}_{z_1,z_2}\big] =  - \mathcal{T}\big[\sigma + \mathop{\dd S}_{z_1,z}\big]\,\mathcal{T}\big[\sigma + \mathop{\dd S}_{z,z_2}\big] \nn
\eeq
\end{prop}

Signalons sans donner de d\'etails que les Propositions~\ref{PP1} et \ref{PP2} sont \'equivalentes.

\subsubsection{Noyau int\'egrable spinoriel}
\label{spino}
Nous allons introduire un \textbf{noyau int\'egrable spinoriel} $\psi(z_1,z_2)$, qui est du point de vue g\'eom\'etrique un objet plus fondamental\footnote{C'est sur ce point que la pr\'esentation diff\`ere, \`a notre connaissance, de la litt\'erature.} que le noyau int\'egrable. Il est d\'efini \`a partir d'une \label{Sato22}formule de Sato :
\bea
\psi(z_1,z_2) & = & \sqrt{\dd x(z_1)\dd x(z_2)}\,\frac{\mathcal{T}\big[\sigma + \mathop{\dd S}_{z_2,z_1}\big]}{\mathcal{T}\big[\sigma\big]} \nn \\
 & = & \frac{\exp\Big(\int_{z_2}^{z_1} \sigma\Big)}{E(z_1,z_2)}\,\frac{\theta\big(\zeta[\sigma] + \mathbf{u}(z_1) - \mathbf{u}(z_2) + \mathbf{c}\big)}{\theta\big(\zeta[\sigma] + \mathbf{c}\big)} \nn
\eea
La d\'ependance de $\psi(z_1,z_2)$ en $\sigma$ est sous-entendue pour simplifier les notations. C'est bien la formule de Sato, puisque si l'on choisit un p\^{o}le $p_0$ de $\sigma$ et $z_1,z_2$ dans une r\'egion o\`u $\xi_{p_0}$ est une coordonn\'ee locale, la transformation $\sigma \rightarrow \sigma + \mathop{\dd S}_{z_2,z_1}$ se repr\'esente par :
\beq
t_{p_0;j} \rightarrow t_{p_0;j} + \frac{\xi_{p_0}^j(z_1)}{j} - \frac{\xi_{p_0}^j(z_2)}{j}\,\,\,\mathrm{si}\,\,j \geq 1\qquad t_{p;j} \rightarrow t_{p;j}\,\,\mathrm{sinon} \nn
\eeq

\noindent $\psi(z_1,z_2)$ est l'unique objet avec les propri\'et\'es analytiques suivantes :

\vspace{0.2cm}

\noindent $\diamond\,$ $\psi(z_1,z_2)$ est un spineur en $z_1$ et en $z_2$. Autrement dit, lors d'un changement de coordonn\'ee locale $\xi$, il se transforme comme $\sqrt{\dd \xi(z_1)\,\dd\xi(z_2)}$.

\vspace{0.2cm}

\noindent $\diamond\,$ $\psi(z_1,z_2)$ est d\'efini globalement sur $\Sigma \times \Sigma$, i.e. n'a pas de monodromie lorsque $z_1$ (ou $z_2$) parcourt un $\mathcal{A}$-cycle ou un $\mathcal{B}$-cycle.

\vspace{0.2cm}

\noindent $\diamond\,$ $\psi$ admet un p\^{o}le simple avec coefficient $1$ \`a points co\"{i}ncidants, i.e. dans toute coordonn\'ee locale $\xi$ :
\beq
\psi(z_1,z_2) \sim \frac{\sqrt{\dd\xi(z_1)\,\dd\xi(z_2)}}{\xi(z_1) - \xi(z_2)}\,,\qquad z_1 \rightarrow z_2 \nn
\eeq

\vspace{0.2cm}

\noindent $\diamond\,$ $\psi$ a une singularit\'e essentielle aux p\^{o}les $p$ de $\sigma$, de sorte que $\psi(z_1,z_2)\exp\big(-\int^{z_2}_{z_1} \sigma\big) \in O(1)$ lorsque $z_1$ ou $z_2$ approchent $p$.

\vspace{0.2cm}

$\psi(z_1,z_2)\psi(z_2,z_3)$ est une forme diff\'erentielle m\'eromorphe en la variable $z_2$, qui a un p\^{o}le simple avec r\'esidu $-1$ lorsque $z_2 \rightarrow z_1$, et avec r\'esidu $1$ lorsque $z_2 \rightarrow z_3$. En utilisant le fait que c'est aussi un spineur en $z_1$ et $z_3$, inchang\'e lorsque $z_1$ (ou $z_3$) parcourt un $\mathcal{A}$-cycle ou un $\mathcal{B}$-cycle, on peut identifier :
\begin{prop} \label{poposu} Relation de passage dans $\Sigma$ :
\bea
\psi(z_1,z_2)\psi(z_2,z_3) & = & \psi(z_1,z_3)\Big(\mathop{\dd S}_{z_1,z_3}(z_2) + \sum_{\mathfrak{h} = 1}^{\mathfrak{g}} 2i\pi\,\alpha_{\mathfrak{h}}(z_1,z_3)\,\dd u_{\mathfrak{h}}(z_2)\Big) \nn \\
\alpha_{\mathfrak{h}}(z_1,z_3) & = & (\partial_{u_{\mathfrak{h}}}\ln \theta)\big(\mathbf{u}(z_1) - \mathbf{u}(z_3) + \zeta[\sigma] + \mathbf{c}\big) - (\partial_{u_{\mathfrak{h}}}\ln\theta)\big(\zeta[\sigma] + \mathbf{c}\big) \nn
\eea
\end{prop}
\noindent Par ailleurs, un calcul direct de $-\delta_{z_2}\psi(z_1,z_3)$ aboutit exactement au membre de droite dans la Prop.~\ref{poposu}. D'o\`u :
\begin{prop}
\label{PPA} $\psi$ a une propri\'et\'e d'autor\'eplication :
\beq
\label{eq:autore} (\delta_{z_2}\psi)(z_1,z_3)= -\psi(z_1,z_2)\psi(z_2,z_3)
\eeq
\end{prop}
En fait, ces calculs reviennent \`a d\'emontrer la version infinit\'esimale de l'identit\'e de Fay : la Proposition~\ref{PP2} est simplement la r\'e\'ecriture de la Proposition~\ref{PPA} en termes de $\mathcal{T}$.
Cette Proposition~\ref{PPA} est le r\'esultat cl\'e pour construire un syst\`eme int\'egrable. Pour calculer l'action de $\partial_{t_I}$ sur $\psi(z_1,z_3)$, il suffit d'int\'egrer l'\'{E}qn.~\ref{eq:autore} contre le cycle g\'en\'eralis\'e $z_2 \in \Omega_I^*$.

\subsubsection{Noyau int\'egrable}

Notons $d$, le degr\'e de la fonction $X$ sur $\Sigma$. Le noyau int\'egrable $\mathbf{K}(x_1,x_2)$ est la matrice $d \times d$ obtenue en \'evaluant $\psi(z_1,z_2)$ aux diff\'erents points ${}^{a_i}x_i \in \Sigma$ tels que $X({}^{a_i}x_i) = x_i \in \widehat{\mathbb{C}}$ :
\beq
\label{eq:defKK} \mathbf{K}(x_1,x_2) = \big(\psi({}^{a_1}x_1,{}^{a_2}x_2)\big)_{1 \leq a_1,a_2 \leq d}
\eeq
Tous les \'enonc\'es du paragraphe pr\'ec\'edent ont une contrepartie pour le noyau int\'egrable. Ainsi, le produit matriciel $\mathbf{K}(x_1,x_2)\mathbf{K}(x_2,x_3)$ est une fonction rationnelle de $x_2$ (car on a somm\'e sur toutes les pr\'eimages de $x_2$), que l'on peut identifier en \'etudiant son comportement aux p\^{o}les $x_2 = x_1$ et $x_2 = x_3$. Ou, de mani\`ere \'equivalente, on peut sommer dans la Proposition~\ref{poposu} sur toutes les pr\'eimages $z_2 = {}^{a_2}x_2$ de $x_2$. Quelque soit la mani\`ere, le r\'esultat est :
\begin{prop}
\label{poposu2} Relation de passage dans $\widehat{\mathbb{C}}$ :
\beq
\mathbf{K}(x_1,x_2)\mathbf{K}(x_2,x_3) = \frac{x_3 - x_1}{(x_2 - x_1)(x_2 - x_3)}\,\mathbf{K}(x_1,x_3) \nn
\eeq
\end{prop}
\begin{cor}
\label{poposu3} $\mathbf{K}(x_1,x_2)$ est une matrice inversible, et :
\beq
\mathbf{K}(x_1,x_2)^{-1} = -(x_1 - x_2)^2\,\mathbf{K}(x_2,x_1) \nn
\eeq
\end{cor}
De m\^{e}me, $\mathbf{K}$ a une propri\'et\'e d'autor\'eplication qui se d\'eduit de la Proposition~\ref{PPA}.

\subsubsection{Lien avec les fonctions de Baker-Akhiezer}
\label{bsbs}
Dans la litt\'erature sur les syst\`emes int\'egrables, on rencontre plus fr\'equemment les \textbf{fonctions de Baker-Akhiezer}, qui sont les uniques fonctions analytiques d'une seule variable $z \in \Sigma$, avec les m\^{e}mes singularit\'es essentielles, et $\mathfrak{g}$ p\^{o}les $p_1,\ldots,p_{\mathfrak{g}}$ de position fix\'ee. Ce type de fonction a \'et\'e introduit pour la premi\`ere fois par Akhiezer \cite{Akh} lorsque $\Sigma$ est de genre $1$. En genre quelconque, elles s'\'ecrivent :
\beq
\psi_{\mathrm{BA}}(z) = \exp\Big(\int_o^{z} \sigma\Big)\,\frac{\theta\big(\mathbf{u}(z) - \sum_{\mathfrak{h} = 1}^{\mathfrak{g}}\mathbf{u}(p_{\mathfrak{h}}) - \mathbf{k} + \zeta[\sigma]\big)}{\theta\big(\mathbf{u}(z) - \sum_{\mathfrak{h} = 1}^{\mathfrak{g}} \mathbf{u}(p_{\mathfrak{h}}) - \mathbf{k}\big)}  \nn
\eeq
\label{vcc}o\`u $\mathbf{k}$ est un vecteur des constantes de Riemann. On peut les obtenir en sp\'ecialisant le noyau spinoriel int\'egrable :
\beq
\label{eq:psisi}\psi_{\mathrm{BA}}(z) = \mathrm{cte}\,\frac{\psi(z,p_1)}{\sqrt{\dd h_{\mathbf{c}}(z_1)\dd h_{\mathbf{c}}(p_1)}}
\eeq
En effet, si $p_1$ est donn\'e et $\mathbf{c}$ est une caract\'eristique impaire non singuli\`ere, $F(z) = \theta(\mathbf{u}(z) - \mathbf{u}(p_1) + \mathbf{c})$ a exactement $\mathfrak{g}$ z\'eros \label{invJJJ}lorsque $z$ parcourt $\Sigma$, dont le point $p_1$ lui-m\^{e}me. D'apr\`es le th\'eor\`eme d'inversion de Jacobi (Thm.~\ref{thJaco}), il existe toujours $q_1,\ldots,q_{\mathfrak{g}} \in \Sigma$ tels que :
\beq
\label{eq:relsiu} -\mathbf{u}(p_1) + \mathbf{c} = -\mathcal{K} - \sum_{\mathfrak{h} = 1}^{\mathfrak{g}} \mathbf{u}(q_{\mathfrak{h}})
\eeq
Et d'apr\`es la description des z\'eros de la fonction th\^{e}ta, les $q_{\mathfrak{h}}$ doivent \^{e}tre pr\'ecis\'ement les z\'eros de $F(z)$ : quitte \`a permuter, on peut supposer $p_1 = q_{1}$. La substitution de la relation~\ref{eq:relsiu} dans la d\'efinition de $\psi$ conduit \`a l'\'{E}qn.~\ref{eq:psisi}.

\subsubsection{Reconstruction du probl\`eme lin\'eaire}
\label{sec:su2}
D\'efinissons des matrices $d \times d$, qui d\'ependent des temps via la forme $\sigma$ :
\bea
\mathbf{M}_{x_0|I}(x) & = & \frac{\partial}{\partial t_I} \mathbf{K}(x_0,x)\,\cdot\mathbf{K}^{-1}(x_0,x) \nn \\
\mathbf{M}_{x_0|x}(x) & = & \Big(\frac{\partial}{\partial x} + \frac{\mathbf{1}}{x - x_1}\Big)\mathbf{K}(x_0,x)\,\cdot\mathbf{K}^{-1}(x_0,x) \nn
\eea
Leurs entr\'ees sont des fonctions rationnelles de $x$, car le produit matriciel r\'ealise une somme sur les diff\'erentes pr\'eimages de $x$. $\Psi$ est donc solution de syst\`emes diff\'erentiels compatibles \`a coefficients rationnels :
\bea
\label{eq:ujp1}\partial_{t_I} \mathbf{K}(x_0,x)  & = & \mathbf{M}_{x_0|I}(x) \mathbf{K}(x_0,x) \\
\partial_x \mathbf{K}(x_0,x) & = & \Big(\frac{\mathbf{1}}{x - x_0} + \mathbf{M}_{x_0|x}(x)\Big)\mathbf{K}(x_0,x) \nn
\eea
Un point important est que la position et le degr\'e maximal des p\^{o}les ne d\'ependent pas du temps. Plus pr\'ecis\'ement :

 \vspace{0.2cm}

 \noindent $\diamond\,$ Si $X(p) \neq \infty$, $\mathbf{M}_{x_0|p;j}(x)$ a un p\^{o}le en $x = X(p)$ de degr\'e $j$, et $\mathbf{M}_{x_0|x}(x)$ a un p\^{o}le de degr\'e $j_{\mathrm{max}}$ (le degr\'e du p\^{o}le $p$ dans $\sigma$).

 \vspace{0.2cm}

\noindent $\diamond\,$ Si $X(p) = \infty$, en notant $d_{\infty}$ le maximum des degr\'es des p\^{o}les de $X$. $\mathbf{M}_{x_0|p;j}$ a un p\^{o}le en $X = \infty$ de degr\'e $1 + \lfloor\frac{j - 1}{d_\infty}\rfloor$, et $\mathbf{M}_{x_0|x}(x)$ a un p\^{o}le de degr\'e $1 + \lfloor \frac{j_{\mathrm{max}} - 1}{d_\infty} \rfloor$.

\noindent Pour toute fonction m\'eromorphe $Y$ sur $\Sigma$, ind\'ependante des temps, on d\'efinit :
\beq
\label{eq:Ldef} \mathbf{L}_{x_0}(x) = \mathbf{K}(x_0,x)\cdot\mathrm{diag}\big(Y({}^{1}x),\ldots,Y({}^{d}x)\big)\cdot\mathbf{K}^{-1}(x_0,x)
\eeq
C'est encore une matrice $d \times d$ \`a coefficients rationnels, qui v\'erifie :
\beq
\partial_{t_I} \mathbf{L}_{x_0}(x) = [\mathbf{M}_{x_0|I}(x),\mathbf{L}_{x_0}(x)] \nn
\eeq
Il est manifeste, d'apr\`es l'\'{E}qn.~\ref{eq:Ldef}, que les valeurs propres de $\mathbf{L}_{x_0}(x)$ sont ind\'ependantes du temps : on parle d'\textbf{\'evolution isospectrale}. Il y a exactement $\mathfrak{g}$ quantit\'es conserv\'ees ind\'ependantes, qui correspondent aux fractions de remplissages $\epsilon_{\mathfrak{h}}$.

Il y a une fa\c{c}on de sp\'ecialiser $x_1 = \infty$, ou bien $x_2 = \infty$, dans $\Psi(x_1,x_2)$, afin de d\'efinir une solution $\Psi(x)$ (ou bien $\Phi(x)$) du probl\`eme lin\'eaire, \`a partir de laquelle on peut reconstruire le noyau int\'egrable $\mathbf{K}(x_1,x_2)$. Nous ne l'\'evoquerons pas ici. Le \label{sjeu}syst\`eme int\'egrable que l'on vient de d\'ecrire est bien un cas particulier de \textbf{syst\`eme isomonodromique}, puisque les monodromies de $\mathbf{K}$ sont fix\'ees par les donn\'ees de $\Sigma$ et $X\,:\,\Sigma \rightarrow \widehat{\mathbb{C}}$, qui sont ind\'ependantes du temps. Comme nous l'avons dit, dans ce cas, le groupe engendr\'e par les monodromies est fini.

\subsubsection{Corr\'elateurs}
\label{coco0}
On d\'efinit les corr\'elateurs $\mathcal{W}_n$ et les corr\'elateurs non connexes $\overline{\mathcal{W}}_n$ comme les formes diff\'erentielles\footnote{Pour $\mathcal{W}_1$ et $\mathcal{W}_2$, cette d\'efinition diff\`ere de celle du \S~\ref{sec:Shy} par un d\'ecalage.} :
\bea
\mathcal{W}_n(z_1,\ldots,z_n) & = & \delta_{z_1}\cdots\delta_{z_n} \ln \mathcal{T}\big[\sigma\big] \nn \\
\overline{\mathcal{W}}_n(z_1,\ldots,z_n) & = & \frac{1}{\mathcal{T}}\,\delta_{z_1}\cdots\delta_{z_n} \mathcal{T}\big[\sigma\big] \nn
\eea
Par exemple, le corr\'elateur \`a un point vaut :
\beq
\mathcal{W}_1(z) = \sigma(z) + (\partial_{u_{\mathfrak{h}}} \ln \theta)\big(\zeta[\sigma] + \mathbf{c}\big)\,\dd u_{\mathfrak{h}}(z) \nn
\eeq

\begin{prop}
\label{dede} Les corr\'elateurs v\'erifient des formules d\'eterminantales. Pour les corr\'elateurs connexes :
\bea
\mathcal{W}_1(z) & = & \sigma(z) + \lim_{z' \rightarrow z} \Big(\psi(z,z')e^{-\int_{z'}^{z} \sigma} - \frac{\sqrt{\dd x(z)\,\dd x(z')}}{x(z) - x(z')}\Big)  \nn \\
\mathcal{W}_2(z_1,z_2) & = & -\psi(z_1,z_2)\psi(z_2,z_1) \nn \\
\mathcal{W}_n(z_1,\ldots,z_n) & = & (-1)^{n + 1} \sum_{\sigma\,\,\textrm{cycles}\,\,\textrm{de}\,\,\mathfrak{S}_n} \prod_{i = 1}^{n} \psi(z_i,z_{\sigma(i)}) \nn
\eea
et pour les corr\'elateurs non connexes :
\beq
\overline{\mathcal{W}}_n(z_1,\ldots,z_n) = "\mathop{\mathrm{det}}_{1 \leq i,j \leq n}" \big(\psi(z_i,z_j)\big) \nn
\eeq
o\`u $"\mathrm{det}"$ d\'esigne le d\'eterminant o\`u l'on remplace chaque occurence de $\psi(z_i,z_i)$ (qui n'a pas de sens) par $\mathcal{W}_1(z_i)$.
\end{prop}
La formule pour $\mathcal{W}_1$ est le r\'esultat d'un petit calcul. Puis, on obtient les formules d\'eterminantales pour les autres corr\'elateurs gr\^{a}ce \`a l'autor\'eplication de $\psi$ (\'{E}qn.~\ref{eq:autore}).

Dans cette construction alg\'ebro-g\'eom\'etrique, les corr\'elateurs ont des propri\'et\'es qui refl\`etent celles des fonctions th\^{e}ta, en particulier l'identit\'e de Fay (Eqns.~\ref{eq:fays} et \ref{eq:fays2}). Elle se traduit exactement par :
\begin{footnotesize}
\beq
\mathop{\mathrm{det}}_{1 \leq i,j \leq n}\big(\psi(z_i,z_j')\big) = (-1)^{n + 1}\,\Big[\frac{\prod_{1 \leq i<j \leq n} E(z_i,z_j)\,E(z_i',z_j')}{\prod_{1 \leq i,j \leq n} E(z_i,z_j')}\Big]\,\frac{\theta\Big(\sum_{i = 1}^n \big[\mathbf{u}(z_i) - \mathbf{u}(z_i')\big] + \zeta[\sigma] + \mathbf{c}\Big)}{\theta\big(\zeta[\sigma] + \mathbf{c}\big)} \nn
\eeq
\end{footnotesize}
$\!\!\!$Ainsi, $\overline{\mathcal{W}}_n$ encode les d\'eg\'en\'erescences $z_i' \rightarrow z_i$ de cette formule.

\subsection{D\'eformation dispersive}
\label{sec:mono}

Soit $\mathcal{S} = [\Sigma,X,Y,B]$ une courbe spectrale de genre $\mathfrak{g}$, $(\mathcal{A}_{\mathfrak{h}},\mathcal{B}_{\mathfrak{h}})_{\mathfrak{h}}$ une base symplectique de cycles, et $\mu,\nu \in \mathbb{C}^{\mathfrak{g}}$. Pour cette construction, nous allons utiliser le formalisme de la r\'ecurrence topologique (Chapitre~\ref{chap:toporec}), et en particulier les propri\'et\'es de g\'eom\'etrie speciale (\S~\ref{sec:geomspec}).
Les fractions de remplissage $\epsilon_{\mathfrak{h}} = -\frac{1}{2i\pi}\int_{\mathcal{A}_{\mathfrak{h}}} Y\dd X$ auront un r\^{o}le sp\'ecial. On note $\mathcal{S}_{\epsilon}$, la courbe spectrale $\mathcal{S}$ \`a fractions de remplissage fix\'ees. \label{ffra4}En ajoutant une forme holomorphe \`a $Y\dd X$, on peut atteindre n'importe quel vecteur de fractions de remplissage, sans toucher \`a la structure conforme de $\Sigma$. Les temps seront les coefficients des d\'eformations m\'eromorphes de $Y\dd X$ \`a une vitesse $1/N$ :
\beq
\delta(Y\dd X) = \frac{1}{N} \sum_{p} \sum_{j \geq 1} (\delta t_{p;j})\Omega_{p;j} \nn
\eeq
Contrairement \`a ce qui se passe dans la construction alg\'ebro-g\'eom\'etrique, la structure conforme \label{tauto2}de $\Sigma$ change avec les flots. Cela se traduit par la variation de la matrice des p\'eriodes :
\bea
\tau_{\mathfrak{h}\mathfrak{h}'} & = & \frac{1}{2i\pi}\,\frac{\partial^2 \mathcal{F}^{(0)}[\mathcal{S}]}{\partial \epsilon_{\mathfrak{h}}\partial \epsilon_{\mathfrak{h}'}} = \frac{1}{2i\pi}\oint_{\mathcal{B}_{\mathfrak{h}}}\oint_{\mathcal{B}_{\mathfrak{h}'}} B \nn \\
\partial_{t_I} \tau_{\mathfrak{h}\mathfrak{h}'} & = & \frac{1}{2i\pi}\oint_{\mathcal{B}_{\mathfrak{h}}}\oint_{\mathcal{B}_{\mathfrak{h}'}}\int_{\Omega^*_I} \omega_3^{(0)}[\mathcal{S}] \nn
\eea
et la variation de l'application d'Abel :
\bea
\mathbf{u}(z_0) & = & \frac{1}{2i\pi}\oint_{\mathcal{B}} B(z_0,\cdot) \nn \\
(\partial_{t_I} \mathbf{u})(z_0) & = & \frac{1}{2i\pi}\oint_{\mathcal{B}}\oint_{\Omega^*_I} \omega_3^{(0)}[\mathcal{S}](z_0,\cdot,\cdot) \nn
\eea
Toutes ces d\'eformations peuvent s'obtenir \`a partir de celle consistant \`a ajouter d'un p\^{o}le double $B(z,\cdot)$. Au niveau infinit\'esimal, le flot correspondant est l'op\'erateur d'insertion $\delta_z$. Par exemple :
\bea
\delta_z \tau_{\mathfrak{h}\mathfrak{h}'} & = & \frac{1}{2i\pi}\oint_{\mathcal{B}_{\mathfrak{h}}}\oint_{\mathcal{B}_{\mathfrak{h}'}} \omega_3^{(0)}[\mathcal{S}](z,\cdot,\cdot)  \nn \\
\delta_z \mathbf{u}(z_0) & = & \frac{1}{2i\pi}\oint_{\mathcal{B}} \omega_3^{(0)}[\mathcal{S}](z,z_0,\cdot) \nn
\eea
et l'expression pour $\omega_3^{(0)}$ figure \`a l'\'{E}qn.~\ref{eq:W30v}.

\subsubsection{Fonction tau}

Pour tout $\epsilon \in \mathbb{C}^{\mathfrak{g}}$, la r\'ecurrence topologique associe \`a $\mathcal{S}_{\epsilon}$ une fonction de partition\label{fonp3} :
\beq
Z[\mathcal{S}_{\epsilon}] = \exp\Big\{\sum_{g \geq 0} N^{2 - 2g}\,\mathcal{F}^{(g)}[\mathcal{S}_{\epsilon}]\Big\} \nn
\eeq
qui est solution des \'equations de boucles (\S~\ref{sec:heqb}) :
\bea
\sum_{a = 1}^{d} \delta_{{}^{a}x} Z & = & Z\,\sum_{a = 1}^{d} N\,Y\dd X({}^{a}x) \nn \\
\label{eq:eq000} \sum_{1 \leq a < b \leq d} \delta_{{}^{a}x}\delta_{{}^{b}x} Z & = &  Z\,Q(x)(\dd x)^2
\eea
o\`u $Q$ d\'esigne une fonction r\'eguli\`ere aux points de branchement. Ce sont des \'equations lin\'eaires sur $Z$. Toute combinaison lin\'eaire de solutions est encore une solution. En particulier :
\beq
\label{eq:hyh} \mathcal{T}_{(\mu,\nu)}[\mathcal{S}] \equiv \sum_{\mathbf{p} \in \mathbb{Z}^{\mathfrak{g}}} e^{2i\pi \nu \cdot \mathbf{p}}\,Z[\mathcal{S}_{\mu + \mathbf{p}/N}]
\eeq
est aussi solution des \'equations de boucles, et c'est notre candidat pour une fonction tau\footnote{Elle ne d\'epend pas des fractions de remplissage initiales de $\mathcal{S}$, mais seulement de $\mu$ et $\nu$ modulo $\mathbb{Z}^{\mathfrak{g}}$.}.

On peut se demander si la somme converge. La formule~\ref{eq:resul} donne un sens pr\'ecis \`a cette d\'efinition, en tant que s\'erie asymptotique lorsque $N \rightarrow \infty$ :
{\small \bea
&& \label{eq:blunta} \mathcal{T}_{(\mu,\nu)}[\mathcal{S}] = \exp\Big(N^2\,\mathcal{F}^{(0)}[\mathcal{S}_{\mu}] + \mathcal{F}^{(1)}[\mathcal{S}_{\mu}]\Big) \\
&& \times\,\Big\{\sum_{r = 0}^{\infty} \frac{1}{r!} \sum_{\substack{g_j \geq 0,\,\,\ell_j \geq 0 \\ 2 - 2g_j - \ell_j < 0}}\,N^{\sum_{j} (2 - 2g_j - \ell_j)} \Big[\prod_{j = 1}^{r} \frac{\mathcal{F}^{(g);(\ell_j)}[\mathcal{S}_{\mu}]}{\ell_j !}\cdot\frac{\nabla_{\mathbf{w}}^{\otimes \ell_j}}{(2i\pi)^{\ell_j}}\Big]\Big\}\Theta_{(\mu,\nu)}(\mathbf{w} = \mathbf{w}^*|\tau) \nn
\eea}
$\!\!\!$o\`u l'on a introduit :
{\small \bea
\Theta_{(\mu,\nu)}(\mathbf{w}|\tau) & = & \sum_{\mathbf{p} \in \mathbb{Z}^{\mathfrak{g}}} e^{2i\pi\nu\cdot N \mu + 2i\pi \mathbf{w}\cdot(\mathbf{p} - \mu) + i\pi(\mathbf{p} - \mu - N\epsilon)\cdot\tau(\mathbf{p} - \mu - N\epsilon)} \nn \\
& = & e^{i\pi N^2 \mu\cdot \tau \mu + 2i\pi (\nu - \mathbf{w})\cdot N\mu }\,\theta\big(\mathbf{w} - N\tau(\mu) + \nu|\tau\big) \nn \\
\mathbf{w}^* & = & \frac{N\,\big(\mathcal{F}^{(0)}\big)'[\mathcal{S}_{\mu}]}{2i\pi} + \nu \nn
\eea}
$\!\!\!$et l'on rappelle que $\mathcal{F}^{(g);(\ell)}[\mathcal{S}_{\mu}]$ est une notation pour le tenseur des d\'eriv\'ees $\ell^{\textrm{\,\`{e}me}}$ de $\mathcal{F}^{(g)}[\mathcal{S}_{\epsilon}]$ par rapport
aux fractions de remplissage, \'evalu\'e en $\epsilon = \mu$. Aux premiers ordres, avec la notation $\Theta^{(\ell)} = (2i\pi)^{-\ell}\big(\nabla_{\mathbf{w}}^{\otimes \ell} \Theta_{(\mu,\nu)}\big)(\mathbf{w} = \mathbf{w}^*|\tau)$ que l'on a d\'ej\`a rencontr\'ee :
{\small \bea
\label{eq:taupm} \mathcal{T} & = & \textcolor{misval}{\exp\Big(N^2\mathcal{F}^{(0)}}  + \mathcal{F}^{(1)}\Big) \\
& & \times \left\{\textcolor{misval}{\Theta} + \frac{1}{N}\Big[\big(\mathcal{F}^{(1)}\big)'\,\Theta' + \frac{\big(\mathcal{F}^{(0)}\big)'''}{6}\,\Theta''' \Big] \right. \nn \\
 && + \frac{1}{N^2}\Big[\mathcal{F}^{(2)}\Theta + \frac{\big[\big(\mathcal{F}^{(1)}\big)'\big]^2}{2}\,\Theta'' + \frac{\big(\mathcal{F}^{(1)}\big)''}{2}\,\Theta'' + \frac{\big(\mathcal{F}^{(0)}\big)''''}{24}\,\Theta'''' \nn \\
 && \left. + \frac{\big(\mathcal{F}^{(0)}\big)'''\,\big(\mathcal{F}^{(1)}\big)'}{6}\,\Theta'''' + \frac{\big[\big(\mathcal{F}^{(0)}\big)'''\big]^2}{72}\,\Theta^{(6)}\Big] + o(1/N^2)\right\} \nn
\eea}
$\!\!\!$Pour abr\'eger les formules, les r\'ef\'erences \`a $(\mu,\nu)$, $\mathcal{S}_{\mu}$, $\mathbf{w}^*$ seront dor\'enavant sous-entendues. \`{A} l'ordre dominant (en surbrillance), on retrouve\footnote{\`{A} un facteur constant pr\`es, et quitte \`a sp\'ecialiser $\mu + \tau(\nu)$ \`a une caract\'eristique impaire $\mathbf{c}$.} la valeur de la fonction tau alg\'ebro-g\'eom\'etrique (\'{E}qn.~\ref{eq:taualg}), avec $\sigma = -N\,Y\dd X$.

Cette d\'efinition est inspir\'ee des syst\`emes int\'egrables venant des mod\`eles de matrices hermitiennes : \label{fonp4}leur fonction de partition co\"{i}ncide avec la fonction tau. L'\'{E}qn~\ref{eq:blunta} est l'asymptotique \`a $N$ grand de la fonction de partition d\'eriv\'ee heuristiquement dans \cite{Ecv} et r\'eexpliqu\'ee \`a la Section~\ref{sec:pluscoup}. Par ailleurs, l'\'{E}qn.~\ref{eq:hyh} r\'ealise l'analogue \label{Whi}d'une \textbf{moyenne de Whitham} \cite{BBT}. Je pense que le syst\`eme int\'egrable que nous cherchons \`a construire co\"{i}ncide avec la hi\'erarchie de Whitham d\'ecrite par Krichever \cite{Kri92}. Si cela se v\'erifiait, nos formules apporteraient des solutions explicites, \`a tous les ordres, pour la hi\'erarchie de Whitham.

\subsubsection{Noyau int\'egrable spinoriel}

Comme pr\'ec\'edemment, la formule de Sato tient lieu de d\'efinition pour le noyau spinoriel int\'egrable. On note $\mathcal{S} + [z_1] - [z_2]$, la courbe spectrale $\mathcal{S}$ o\`u l'on a remplac\'e $Y\dd X$ par $Y\dd X + \frac{1}{N}\,\mathop{\dd S}_{z_1,z_2}$, et :
\beq
\label{eq:sSS}\psi_{\mathcal{S}}(z_1,z_2) = \sqrt{\dd x(z_1)\dd x(z_2)}\,\frac{\mathcal{T}_{(\mu,\nu)}\big[\mathcal{S} + [z_1] - [z_2]\big]}{\mathcal{T}_{(\mu,\nu)}[\mathcal{S}]}
\eeq
L'indice se r\'ef\'erant \`a la courbe spectrale $\mathcal{S}$ sera sous-entendu lorsqu'il n'y a pas d'ambig\"{u}it\'e. Le membre de droite se calcule \`a partir de l'\'{E}qn.~\ref{eq:blunta} en utilisant les propri\'et\'es de g\'eom\'etrie sp\'eciale (\S~\ref{sec:geomspec}). Si l'on introduit $\Theta^{(\ell)}_{12} = (2i\pi)^{-\ell}\big(\nabla_{\mathbf{w}}^{\otimes \ell}\,\Theta_{(\mu,\nu)}\big)(\mathbf{u}(z_1) - \mathbf{u}(z_2) + \mathbf{w}^*|\tau)$, les deux premiers ordres s'\'ecrivent :
\begin{footnotesize}\bea
\psi(z_1,z_2) & = & \frac{\exp\Big(-N\int_{z_2}^{z_1} Y\dd X\Big)}{E(z_1,z_2)}\left\{\frac{\Theta_{12}}{\Theta} + \frac{1}{N}\left[\frac{\Theta_{12}}{\Theta}\int_{z_2}^{z_1} \!\!\omega_1^{(1)} + \frac{1}{6}\,\frac{\Theta_{12}}{\Theta}\int_{z_2}^{z_1}\!\!\int_{z_2}^{z_1}\!\!\int_{z_2}^{z_1}\!\! \omega_3^{(0)} \right.\right. \nn \\
&& \left. + \frac{1}{2}\,\frac{\Theta_{12}'}{\Theta}\oint_{\mathcal{B}}\int_{z_2}^{z_1}\int_{z_2}^{z_1} \omega_3^{(0)} + \frac{1}{2}\oint_{\mathcal{B}}\!\oint_{\mathcal{B}}\!\int_{z_2}^{z_1}\!\!\omega_3^{(0)}\,\frac{\Theta''_{12}}{\Theta} \right. \nn \\
 && \left.+ \big(\mathcal{F}^{(1)}\big)'\,\Big(\frac{\Theta'_{12}}{\Theta} - \frac{\Theta_{12}\,\Theta'}{\Theta^2}\Big) + \frac{\big(\mathcal{F}^{(0)}\big)'''}{6}\,\Big(\frac{\Theta'''_{12}}{\Theta} -  \frac{\Theta_{12}\,\Theta'''}{\Theta^2}\Big) +\,o(1/N) \right\} \nn
\eea
\end{footnotesize}

Cet objet a les propri\'et\'es suivantes :

\vspace{0.2cm}

\noindent $\diamond\,$ $\psi$ est un spineur d\'efini globalement sur $\Sigma \times \Sigma$.

\vspace{0.2cm}

\noindent $\diamond\,$ $\psi$ a un p\^{o}le simple avec coefficient $1$ \`a points co\"{i}ncidants, i.e. dans toute coordonn\'ee locale $\xi$ :
\beq
\psi(z_1,z_2) \sim \frac{\sqrt{\dd\xi(z_1)\,\dd\xi(z_2)}}{\xi_p(z_1) - \xi_p(z_2)}\,,\qquad z_1 \rightarrow z_2 \nn
\eeq

\noindent $\diamond\,$ $\psi$ a une singularit\'e essentielle aux p\^{o}les de $y\dd x$, de sorte que $\psi(z_1,z_2)\exp\Big(N\int_{z_2}^{z_1} Y\dd X\Big) \in O(1)$ lorsque $z_1$ ou $z_2$ approchent un p\^{o}le de $Y\dd X$.

\vspace{0.2cm}

\noindent $\diamond\,$ Les autres singularit\'es de $\psi$ sont a priori des p\^{o}les aux points de branchement, ordre par ordre en $1/N$.

\vspace{0.2cm}
\label{conjd}
\begin{conj}
\label{eqly}$\psi$ a une propri\'et\'e d'autor\'eplication :
\beq
\frac{1}{N}\,(\delta_z \psi)(z_1,z_2) = -\psi(z_1,z)\psi(z,z_2) \nn
\eeq
\end{conj}
Les deux membres sont clairement des formes diff\'erentielles de $z$, qui ont les m\^{e}mes p\^{o}les en $z = z_1$ et $z = z_2$. Ils ne peuvent diff\'erer que par une forme m\'eromorphe ayant des p\^{o}les aux points de branchement, et notre conjecture dit que celle-ci est nulle. Cela peut \^{e}tre reformul\'e de fa\c{c}on plus technique. \`{A} partir de la repr\'esentation :
\beq
\frac{1}{N}\,(\delta_z \psi)_{\mathcal{S}}(z_1,z_2) =  \Res_{z' \rightarrow z} \psi_{\mathcal{S}}(z,z')\psi_{\mathcal{S} + [z_1] - [z_2]}(z',z) \nn
\eeq
on peut d\'eplacer les contours et \'ecrire :
\bea
\frac{1}{N}\,(\delta_z \psi)_{\mathcal{S}}(z_1,z_2) & = & -\Big(\Res_{z' \rightarrow z_2} + \sum_{i} \Res_{z' \rightarrow a_i}\Big) \psi_{\mathcal{S}}(z,z')\psi_{\mathcal{S} + [z_1] - [z_2]}(z',z) \nn \\
& = & -\psi_{\mathcal{S}}(z_1,z)\psi_{\mathcal{S}}(z,z_2) - \sum_{i} \Res_{z' \rightarrow a_i} \psi_{\mathcal{S}}(z,z')\psi_{\mathcal{S} + [z_1] - [z_2]}(z',z) \nn
\eea
Il faudrait donc d\'emontrer que cette somme de r\'esidus aux points de branchement est nulle, ordre par ordre dans le d\'ev\'eloppement asymptotique lorsque $N \rightarrow \infty$.

Cette conjecture a \'et\'e v\'erifi\'ee jusqu'\`a l'ordre $O(1/N)$ par un calcul direct. Malgr\'e plusieurs tentatives, \`a l'heure de l'\'ecriture, nous n'avons pas de m\'ethode g\'en\'erale pour la prouver. On sait au moins que toutes ses cons\'equences sont vraies pour toutes les courbes spectrales qui venant des mod\`eles de matrices int\'egrables, notamment du mod\`ele \`a une matrice hermitienne (qui \label{ropb2}atteignent toutes les courbes hyperelliptiques) et de la chaine \`a deux matrices hermitiennes. Cependant, cela ne permet pas d'atteindre toutes les courbes spectrales alg\'ebriques.

\subsubsection{\'{E}quations de Hirota et identit\'e de Fay}
\label{fsyq}
En termes de la fonction $\mathcal{T}$, la Conjecture~\ref{eqly} s'\'ecrit :
\begin{conj}
\label{conjs}$\mathcal{T}$ v\'erifie une version infinit\'esimale de l'\'equation de Hirota :
{\small\beq
 \mathcal{T}\big[\mathcal{S}\big]\,(\delta_{z} \mathcal{T})\big[\mathcal{S} + [z_1] - [z_2]\big]  - (\delta_z\mathcal{T})[\mathcal{S}]\,\mathcal{T}\big[\mathcal{S} + [z_1] - [z_2]\big] =  - N\, \mathcal{T}\big[\mathcal{S} + [z_1] - [z]\big]\,\mathcal{T}\big[\mathcal{S} + [z] - [z_2]\big] \nn
\eeq}
\end{conj}
Enfin, nous proposons aussi une conjecture globale qui implique la pr\'ec\'edente (mais que l'on pense en fait \'equivalente) :
\begin{conj}
\label{conjt}$\mathcal{T}$ v\'erifie une version globale de l'\'equation de Hirota globale :
\bea
&& \mathcal{T}\big[\mathcal{S}\big]\,\mathcal{T}\big[\mathcal{S} + [z_1] - [z_2] + [z_3] - [z_4]\big] \nn \\
& = & \mathcal{T}\big[\mathcal{S} + [z_1] - [z_2]\big]\,\mathcal{T}\big[\mathcal{S} + [z_3] - [z_4]\big] - \mathcal{T}\big[\mathcal{S} + [z_1] - [z_4]\big]\,\mathcal{T}\big[\mathcal{S} + [z_3] - [z_2]\big] \nn
\eea
\end{conj}

La conjecture~\ref{conjs} est une identit\'e de Fay diff\'erentielle, et \ref{conjt} est une identit\'e de Fay.

\subsubsection{Corr\'elateurs}
\label{coco02}
On d\'efinit \`a nouveau des corr\'elateurs $\mathcal{W}_n$ et des corr\'elateurs non connexes $\overline{\mathcal{W}}_n$ par :
\bea
\mathcal{W}_n(z_1,\ldots,z_n) & = & N^{-n}\delta_{z_1}\cdots\delta_{z_n} \ln \mathcal{T}\big[\sigma\big] \nn \\
\mathcal{W}_n(z_1,\ldots,z_n) & = & \frac{1}{\mathcal{T}}\,N^{-n}\,\delta_{z_1}\cdots\delta_{z_n} \mathcal{T}\big[\sigma\big] \nn
\eea
On peut calculer les premiers ordres en partant de l'\'{E}qn.~\ref{eq:taupm}.
\begin{footnotesize}
\bea
\mathcal{W}_1(z) & = & -N\,Y\dd X(z) + \frac{\Theta'}{\Theta}\,2i\pi\,\dd \mathbf{u}(z) \nn \\
& & \frac{1}{N}\left\{\omega_1^{(1)}(z) + i\pi\delta_z\tau\,\frac{\Theta''}{\Theta} + \Big[\big(\mathcal{F}^{(1)}\big)'\Big(\frac{\Theta''}{\Theta} - \frac{\Theta'^2}{\Theta^2}\Big) + \frac{\big(\mathcal{F}^{(0)}\big)'''}{6}\Big(\frac{\Theta''''}{\Theta} - \frac{\Theta'''\Theta'}{\Theta^2}\Big)\Big]2i\pi\dd \mathbf{u}(z) \right\} \nn \\
&& + o(1/N) \nn
\eea
\end{footnotesize}
$\!\!\!$Puis, en appliquant une nouvelle fois l'op\'erateur d'insertion :
\begin{footnotesize}
\bea
\mathcal{W}_2(z_1,z_2) & = & B(z_1,z_2) + \Big(\frac{\Theta''}{\Theta} - \frac{\Theta'^2}{\Theta^2}\Big) 2i\pi\dd\mathbf{u}(z_1)\otimes 2i\pi\dd\mathbf{u}(z_2) \nn \\
& & + \frac{1}{N}\left\{\int_{\mathcal{B}} \omega_3^{(0)}(\cdot,z_1,z_2)\,\frac{\Theta'}{\Theta} + \Big(i\pi\delta_{z_1}\tau\,2i\pi\dd\mathbf{u}(z_2) + 2i\pi\dd\mathbf{u}(z_1)\,i\pi\delta_{z_2}\tau\Big)\,\Big(\frac{\Theta''}{\Theta}\Big)' \right.\nn \\
& & \left. \Big[\big(\mathcal{F}^{(1)}\big)'\Big(\frac{\Theta''}{\Theta} - \frac{\Theta'^2}{\Theta^2}\Big)' + \frac{\big(\mathcal{F}^{(0)}\big)'''}{6}\Big(\frac{\Theta''''}{\Theta} - \frac{\Theta'''\Theta'}{\Theta^2}\Big)'\Big]2i\pi\dd\mathbf{u}(z_1)\otimes 2i\pi\dd\mathbf{u}(z_2) \right\} + o(1/N) \nn
\eea
\end{footnotesize}
$\!\!\!$Les contractions des tenseurs sont sous-entendues dans ces expressions, ainsi qu'un facteur $\frac{1}{2i\pi}$ par d\'erivation de fonction th\^{e}ta.

Par construction, les corr\'elateurs ont deux propri\'et\'es importantes. D'une part, ils v\'erifient des \'equations de boucles (cf. \'{E}qn.~\ref{eq:eq000}).
D'autre part, en utilisant un d\'eveloppement de Taylor pour r\'e\'ecrire la formule de Sato (\'{E}qn.~\ref{eq:sSS}), il \label{esoo}existe une formule exponentielle qui reconstruit $\psi$ \`a partir des $\mathcal{W}_n$ :
\beq
\psi(z_1,z_2) = \frac{\exp\Big(-N\int_{z_2}^{z_1} Y \dd X\Big)}{E(z_1,z_2)}\,\exp\Big(\sum_{n \geq 1} \int_{z_2}^{z_1}\cdots\int_{z_2}^{z_1} \widetilde{\mathcal{W}}_n\Big) \nn
\eeq
o\`u $\widetilde{\mathcal{W}}_1 = \mathcal{W}_1 - N\,\omega_1^{(0)}$, $\widetilde{\mathcal{W}}_2 = \mathcal{W}_2 - \omega_2^{(0)}$, et $\widetilde{\mathcal{W}}_n = \mathcal{W}_n$ pour $n \geq 3$. Enfin, un petit calcul montre :
\beq
\mathcal{W}_1(z) = -N\,Y\dd X(z) + \lim_{z' \rightarrow z} \Big(\psi(z,z')e^{N\int_{z'}^{z} Y\dd X} - \frac{\sqrt{\dd x(z)\,\dd x(z')}}{x(z) - x(z')}\Big)  \nn
\eeq

Si la conjecture~\ref{PPA} est vraie, on d\'eduit automatiquement que :
\vspace{0.2cm}

\noindent $\diamond\,$ $\mathcal{W}_n$ et $\overline{\mathcal{W}}_n$ satisfont des \label{dede2}formules d\'eterminantales (cf. Proposition~\ref{dede}).

\vspace{0.2cm}

\noindent $\diamond\,$ Le noyau\label{eq:noyu3} int\'egrable $\mathbf{K}(x_1,x_2)$ d\'efini comme \`a l'\'{E}qn.~\ref{eq:defKK} satisfait une relation de passage dans $\widehat{\mathbb{C}}$ (cf. Proposition~\ref{poposu2}).

\vspace{0.2cm}

\noindent $\diamond\,$ $\mathbf{K}(x_1,x_2)$ \label{sec:su3}est solution d'un probl\`eme lin\'eaire \`a coefficients rationnels (cf. \'{E}qn.~\ref{eq:ujp1}), o\`u les matrices $\mathbf{M}$ n'ont de p\^{o}les qu'\`a ceux de $Y\dd X$ (en particulier, n'ont pas de p\^{o}les aux points de branchement).

Cela justifierait que notre construction d\'ecrit bien un syst\`eme int\'egrable dispersif, dont $[\Sigma,X,Y]$ est la courbe spectrale semiclassique.

\section{Conclusion}
\label{sec:cnuc}
Une r\'eponse \`a la question suivante pourrait constituer une approche indirecte de la Conjecture~\ref{PPA} :
\begin{prob}
Peut-on r\'ealiser toute courbe plane alg\'ebrique $[\Sigma,X,Y]$ comme la courbe spectrale semiclassique d'un syst\`eme diff\'erentiel :
\beq
\frac{1}{N}\,\partial_x \Psi(x,\mathbf{t}) = \mathbf{L}(x,\mathbf{t})\,\Psi(x,\mathbf{t}),\qquad \frac{1}{N}\,\partial_{t_j} \Psi(x,\mathbf{t}) = \mathbf{M}(x,\mathbf{t})\Psi(x,\mathbf{t})  \nn
\eeq
o\`u $\mathbf{L}$ et $\mathbf{M}$ sont des matrices $d \times d$ :
 \begin{itemize}
 \item[$\diamond$] qui admettent une limite $\mathbf{L}^{[[0]]}$ et $\mathbf{M}^{[[0]]}$ lorsque $N \rightarrow \infty$ ;
 \item[$\diamond$] qui admettent un d\'eveloppement asymptotique lorsque $N \rightarrow \infty$ ;
 \item[$\diamond$] qui sont ordre par ordre des fonctions rationnelles en $x$, dont les p\^{o}les sont ceux de $\mathbf{L}^{[[0]]}$ (i.e. les projections en $X$ des p\^{o}les de $Y\dd X$) ?
\end{itemize}
\end{prob}

Dans la partie~\ref{sec:enjo}, nous avons expliqu\'e comment associer des \'equations de boucles \`a un syst\`eme int\'egrable. Dans la partie~\ref{sec:consoK}, nous avons cherch\'e r\'eciproquement \`a associer un syst\`eme int\'egrable aux \'equations de boucles pour une courbe spectrale alg\'ebrique. Si nos conjectures se v\'erifiaient, cela signifierait sch\'ematiquement :
\vspace{0.2cm}
\begin{center}
Int\'egrabilit\'e classique $\quad \Leftrightarrow \quad$ \'{E}quations de boucles
\end{center}

Il serait tr\`es int\'eressant d'\'etudier les cons\'equences d'une telle \'equivalence pour la th\'eorie des vari\'et\'es de \label{Frolon}Frobenius \cite{Dubrovin}, qui sous-tend certaines hi\'erarchies int\'egrables.
Dans cette th\'eorie, la construction des objets associ\'es au syst\`eme int\'egrable, \`a tout ordre dans le param\`etre dispersif, est un probl\`eme r\'eput\'e difficile. Une premi\`ere \'etape serait de comparer les premiers ordres de la fonction tau pr\'edit par l'\'{E}qn.~\ref{eq:hyh}, \`a des r\'esultats connus du c\^{o}t\'e vari\'et\'e de Frobenius. Il y a en tout cas un dictionnaire \`a \'etablir avec les m\'ethodes de r\'ecurrence topologique.

\newpage
\thispagestyle{empty}
\phantom{bbk}

\newpage

\chapter{Int\'egrales formelles de matrices et combinatoire}
\label{chap:formel}
\thispagestyle{plain}
\vspace{-1.5cm}

\rule{\textwidth}{1.5mm}
\addtolength{\baselineskip}{0.20\baselineskip}

\vspace{1.4cm}
{\textsf{Les probl\`emes de physique statistique sur r\'eseau al\'eatoire, i.e. de combinatoire de cartes (d\'ecor\'ees ou non), peuvent \^{e}tre repr\'esent\'es \`a l'aide d'int\'egrales formelles de matrices de taille $N \times N$. Chaque carte est compt\'ee modulo ses automorphismes, avec un poids $\propto N^{\chi}$, o\`u $\chi = 2 - 2g - n$ est la caract\'eristique d'Euler pour une carte de genre $g$ \`a $n$ bords. J'illustre cette m\'ethode pour un mod\`ele de boucles sur r\'eseau al\'eatoire, dont le mod\`ele $\On$ trivalent est un cas particulier. Le calcul des observables avec un nombre fini de changements de conditions de bord, se ram\`ene au calcul des observables avec conditions de bord uniformes. Ces derni\`eres sont solution d'\'equations de Schwinger-Dyson dans le mod\`ele de matrices, qui sont \'equivalentes \`a des relations combinatoires (i.e. d\'emontrables par une approche bijective). J'explique ensuite bri\`evement, dans le cas du mod\`ele $\On$ trivalent, la r\'esolution de ces \'equations par une r\'ecurrence topologique, et quelques cons\'equences pour l'asymptotique des grandes cartes. Ces r\'esultats sont d\'etaill\'es dans l'article \cite{BEOn}.}}

\addtolength{\baselineskip}{-0.20\baselineskip}

\section{Int\'egrales formelles de matrices}
\label{sec:foerm}
Dans ce chapitre, je vais parler de l'\'enum\'eration de cartes, sans d\'efinir math\'ematiquement ce qu'est une "carte". Je renvoie par exemple \`a \cite{ChapuyThese,Ebook}, ou \`a des ouvrages plus anciens et sp\'ecialis\'es \cite{Berge}. Je choisis plut\^{o}t de partir d'un mod\`ele formel de matrices assez g\'en\'eral, et de montrer selon la m\'ethode introduite par \cite{BIPZ}, quels objets combinatoires il compte. Ceux-ci correspondent bien \`a la notion de \textbf{carte topologique} qui a \'et\'e retenue en math\'ematiques. Cette partie est un peu technique mais ne contient pas de r\'esultat profond : j'essaie de justifier pas \`a pas les hypoth\`eses simplificatrices qui permettront d'\'etudier efficacement des probl\`emes de combinatoire gr\^{a}ce aux \'equations de Schwinger-Dyson. J'explique notamment les pr\'ecautions \`a prendre pour qu'un mod\`ele formel de matrices ait un d\'eveloppement topologique.

\subsection{D\'efinitions}

Nous\label{musi} avons d\'ej\`a rencontr\'e la notion d'int\'egrales formelles de matrices en introduction, ce n'est qu'un exemple de th\'eorie des champs perturbative (partie~\ref{sec:BIPZ}). Soit $\mathbf{X} = (X_\alpha)_{\alpha \in \mathcal{I}}$ une famille de matrices hermitiennes de taille $N \times N$. On se donne une mesure d'int\'egration :
\bea
\dd\nu(\mathbf{X}) & = & \dd\nu_0(\mathbf{X})\,\exp\Big\{\frac{N}{t}\,\mathcal{U}(\mathbf{X})\Big\} \nn \\
\label{eq:muodef}\dd\nu_0(\mathbf{X}) & = & \Big[\prod_{\alpha \in \mathcal{I}}\dd X_\alpha\Big]\,\exp\Big\{-\frac{N}{2t}\sum_{\alpha,a,b}\sum_{\beta,c,d} K_{\alpha|ab;\beta|cd} X_{\alpha|a,b}X_{\beta|cd}\Big\}
\eea
o\`u $\mathcal{U}(\mathbf{X})$ est une somme finie de mon\^{o}mes de degr\'e $\geq 3$ dans les entr\'ees $X_{\alpha|ab}$. Apr\`es plusieurs \'etapes qui seront explicit\'ees, il est possible de d\'efinir comme s\'eries formelles en $t$,

\vspace{0.2cm}

\noindent  $\diamond\,$ la\label{fonp5} \textbf{fonction de partition} :
\beq
Z = \int_{(\mathcal{H}_N)^{|\mathcal{I}|}}\!\!\dd\nu(\mathbf{X}) \nn
\eeq

\vspace{0.2cm}

\noindent $\diamond\,$ l'\textbf{\'energie libre} $F = \ln Z$\label{eneli2}.

\vspace{0.2cm}

\noindent $\diamond\,$ pour toute observable $\mathcal{P}(\mathbf{X})$ polyn\^{o}miale dans les entr\'ees $X_{\alpha|ab}$, la valeur moyenne :
\beq
\big\langle \mathcal{P}(\mathbf{X})\big\rangle = \frac{1}{Z}\int_{(\mathcal{H}_N)^{|\mathcal{I}|}}\!\!\dd\nu(\mathbf{X})\,\mathcal{P}(\mathbf{X}) \nn
\eeq

\vspace{0.2cm}

\noindent $\diamond\,$ pour toute famille d'observables $\mathcal{P}_j(\mathbf{X})$ \label{cum2}polyn\^{o}miales, les cumulants :
\beq
\Big\langle \prod_{j} \mathcal{P}_j(\mathbf{X})\Big\rangle_c = \Big[\prod_{j}\partial_{\epsilon_j = 0}\Big]\ln \big\langle e^{\sum_{j} \epsilon_j\,\mathcal{P}_j(\mathbf{X})}\big\rangle \nn
\eeq

\vspace{0.2cm}

Comme la mesure $\nu_0$ est gaussienne, on dispose du th\'eor\`eme de Wick pour calculer ses \label{momen2}moments. Si l'on note $\langle\cdots\rangle_0$ les valeurs moyennes pour $\nu_0$ :
\beq
\Big\langle \prod_{j = 1}^n X_{\alpha_j|a_j,b_j} \Big\rangle_0 = \sum_{\sigma \in \mathfrak{S}_n\,/\,\sigma^2 = \mathrm{id}} \prod_{j} \big\langle X_{\alpha_j|a_j,b_j}X_{\alpha_{\sigma(j)}|a_{\sigma(j)},b_{\sigma(j)}}\big\rangle_0 \nn
\eeq
et la valeur \label{appa3}des appariements est donn\'ee par l'inverse de la matrice de covariance :
\beq
\big\langle X_{\alpha|ab}X_{\beta|cd} \big\rangle_0 = \frac{t}{N}\,(K^{-1})_{\alpha|ab;\beta|cd} \nn
\eeq
Les moments d'ordre $2n$ sont donc proportionnels \`a $t^n$, tandis que les moments d'ordre impair sont nuls. D\'efinissons :
\beq
Z \equiv 1 + \sum_{m = 1}^{\infty} \frac{1}{m!}\,\frac{N^m}{t^m}\Big\langle\mathcal{U}^m(\mathbf{X})\Big\rangle_0 \nn
\eeq
Comme $\mathcal{U}(\mathbf{X})$ est une somme finie de mon\^{o}mes de degr\'e $\geq 3$, $\Big\langle\mathcal{U}^m(\mathbf{X})\Big\rangle_0$ est une somme finie de termes de degr\'e minimal $t^{\lceil 3m/2 \rceil}$. Cela assure que le coefficient de $t^v$ dans $Z$ est nul si $v < 0$, vaut $1$ si $v = 0$, et est une somme finie lorsque $v \geq 1$. Donc, $Z$ et $1/Z$ sont bien des s\'eries formelles\footnote{sous-entendu en puissances positives de $t$} de $t$. On en d\'eduit que pour toute observable $\mathcal{P}(\mathbf{X})$ polyn\^{o}miale :
\beq
\Big\langle \mathcal{P}(\mathbf{X})\Big\rangle \equiv \frac{1}{Z}\sum_{m = 0}^{\infty} \frac{1}{m!}\,\frac{N^m}{t^m}\Big\langle\mathcal{P}(\mathbf{X})\,\mathcal{U}^m(\mathbf{X})\Big\rangle_0 \nn
\eeq
est aussi une s\'erie formelle en $t$. Enfin, les cumulants sont des fonctions polyn\^{o}miales des moments, donc sont aussi des s\'eries formelles en $t$.

\subsection{Le d\'eveloppement topologique}
\label{sec:ezgt}
Les d\'efinitions pr\'ec\'edentes sont valables d\`es que $\dd\nu_0$ fait de la famille des $X_{\alpha|ab}$ un vecteur gaussien centr\'e. Nous allons \^{e}tre plus sp\'ecifiques en ne consid\'erant que les mesures $\dd\nu(\mathbf{X})$ et les observables $\mathcal{P}(\mathbf{X})$ qui sont invariantes sous l'action globale du groupe $\mathrm{U}(N)$ sur $(X_\alpha)_{\alpha \in \mathcal{I}}$, car ce sont ces mod\`eles qui ont une bonne interpr\'etation combinatoire en termes de cartes. La d\'erivation de cette combinatoire se d\'ecompose en plusieurs \'etapes.

\subsubsection{Hypoth\`eses techniques}

Nous allons supposer que $S_0$ dans $\dd\nu_0 = \dd \mathbf{X}\,e^{-S_0[\mathbf{X}]}$ ne contient pas de termes $\Tr X_{\alpha}\cdot\Tr X_{\beta}$. Alors, il est toujours possible de se ramener par des combinaisons lin\'eaires de $\mathbf{X}_\alpha$ \`a :
\beq
\dd\nu_0 = \prod_{\alpha \in \mathcal{I}} \dd X_\alpha\exp\Big\{-\frac{N}{t}\Tr\frac{X_\alpha^2}{2\z_\alpha}\Big\} \nn
\eeq
Les appariements valent :
\beq
\label{eq:regl}\big\langle X_{\alpha|ab}X_{\beta|cd}\big\rangle_0 = \frac{t}{N}\,\z_\alpha\,\delta_{\alpha\beta}\delta_{ad}\delta_{bc}
\eeq
Ceci sert uniquement \`a simplifier notre discussion.

Nous allons aussi faire une hypoth\`ese sur $\mathcal{U}$ importante pour la suite. D\'efinissons un \textbf{mot cyclique} dans l'alphabet $\mathcal{I}$ comme une suite finie $\mathfrak{m} \in \mathcal{I}^{\mathbb{Z}_r}$, modulo une succession de permutations cycliques $(\mathfrak{m}(i))_i  \sim (\mathfrak{m}(i + 1))_i$ et de commutations $(\mathfrak{m}(i))_i \sim (\mathfrak{m}(-i))_i$. On note $|\mathrm{Aut}\,\mathfrak{m}|$, le nombre d'\'el\'ements de la classe d'\'equivalence de $\mathfrak{m}$, et :
\beq
 T_{\mathfrak{m}}(\mathbf{X}) = \Tr X_{\mathfrak{m}(1)}X_{\mathfrak{m}(2)}\cdots X_{\mathfrak{m}(r)} \nn
\eeq
Les termes que nous autorisons dans $\mathcal{U}$ sont index\'es par un ensemble avec r\'ep\'etitions $L = \{\mathfrak{m}_1^{p_1},\ldots,\mathfrak{m}_s^{p_s}\}$ de mots cycliques \emph{\`a plus de trois lettres} sur l'alphabet $\mathcal{I}$. Nous allons \'ecrire :
\beq
\label{eq:ugfd}\frac{N}{t}\,\mathcal{U}(\mathbf{X}) = \sum_{L} g_{L}\,\prod_{\mathfrak{m} \in L} \frac{\big(\frac{N}{t}\,T_{\mathfrak{m}}(\mathbf{X})\big)^{p_{\mathfrak{m} \in L}}}{(p_{\mathfrak{m} \in L})!\,|\mathrm{Aut}\,\mathfrak{m}|^{p_{\mathfrak{m} \in L}}}
\eeq
En r\'ep\'etant un argument pr\'ec\'edent, on remarque que l'introduction de cette d\'ependance en $t$ dans $\mathcal{U}$ pr\'eserve le caract\`ere de s\'erie formelle en $t$ de la fonction de partition et des valeurs moyennes d'observables.

Les observables invariantes $\mathrm{U}(N)$ dans ce mod\`ele sont les fonctions des traces de produits \emph{arbitraires} de $X_{\alpha}$. Par la suite, nous utiliserons le mot \textbf{observable} pour d\'esigner aussi bien $\mathcal{P}(\mathbf{X})$ que sa valeur moyenne.

\subsubsection{Diagrammatique}

Il est commode de repr\'esenter un \'el\'ement de matrice hermitienne $X_{\alpha|ab}$ par un couple segment entrant (indice $a$)/segment sortant (indice $b$), avec une "couleur" $\alpha$. On peut alors lire un appariement (\'{E}qn.~\ref{eq:regl}) par une mise bout-\`a-bout de segments de m\^{e}me orientation et de m\^{e}me couleur, les fl\`eches entrantes pointant vers la zone d'interaction entre les deux matrices. Avec cette r\`egle, chaque ligne garde son indice matriciel $a \in \{1,\ldots,N\}$.

\begin{figure}[h!]
\begin{center}
\includegraphics[width = \textwidth]{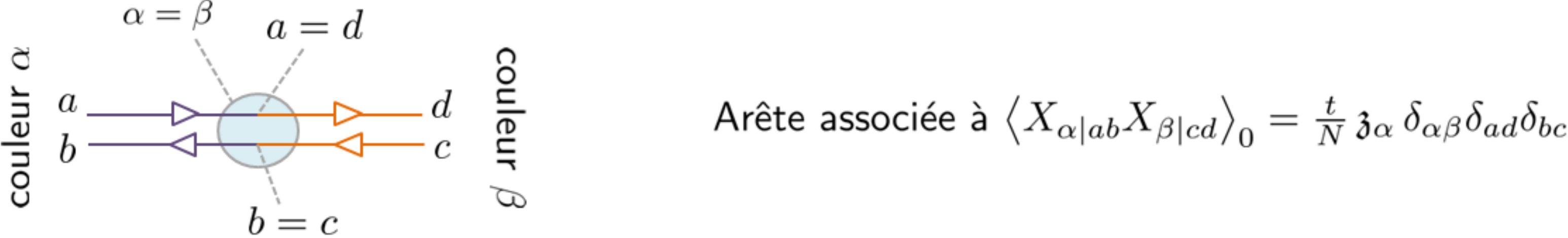}
\end{center}
\end{figure}

Dans un premier temps, on souhaite calculer :
\beq
\label{eq:aoerez}\big\langle \textcolor{misval}{T_{\mathfrak{m}_1}(\mathbf{X}) T_{\mathfrak{m}_2}(\mathbf{X}) \cdots}\, T_{\mathfrak{m'}_1}(\mathbf{X})T_{\mathfrak{m'}_2}(\mathbf{X})\cdots\big\rangle_0
\eeq
Naturellement, $T_{\mathfrak{m}}(\mathbf{X})$ se repr\'esente en disposant cycliquement autour d'un sommet les couples de segments entrants/sortants (les fl\'eches entrantes pointent vers le sommet) associ\'es \`a $X_{\mathfrak{m}(1)|a_1 b_1}$, $X_{\mathfrak{m}(2)|a_2 b_2}$, \ldots{} La trace impose $b_i = a_{i + 1}$, donc on va relier un segment entrant (indice $a_{i + 1}$) au segment sortant imm\'ediatement pr\'ec\'edent (indice $b_i$). Les indices vivant sur chaque ligne sont libres. On remarque que la ligne associ\'ee \`a $a_i$ entre avec une couleur $\mathfrak{m}(i)$ et ressort avec une couleur $\mathfrak{m}(i - 1)$ qui peut \^{e}tre diff\'erente.

\begin{figure}[h!]
\begin{center}
\includegraphics[width = \textwidth]{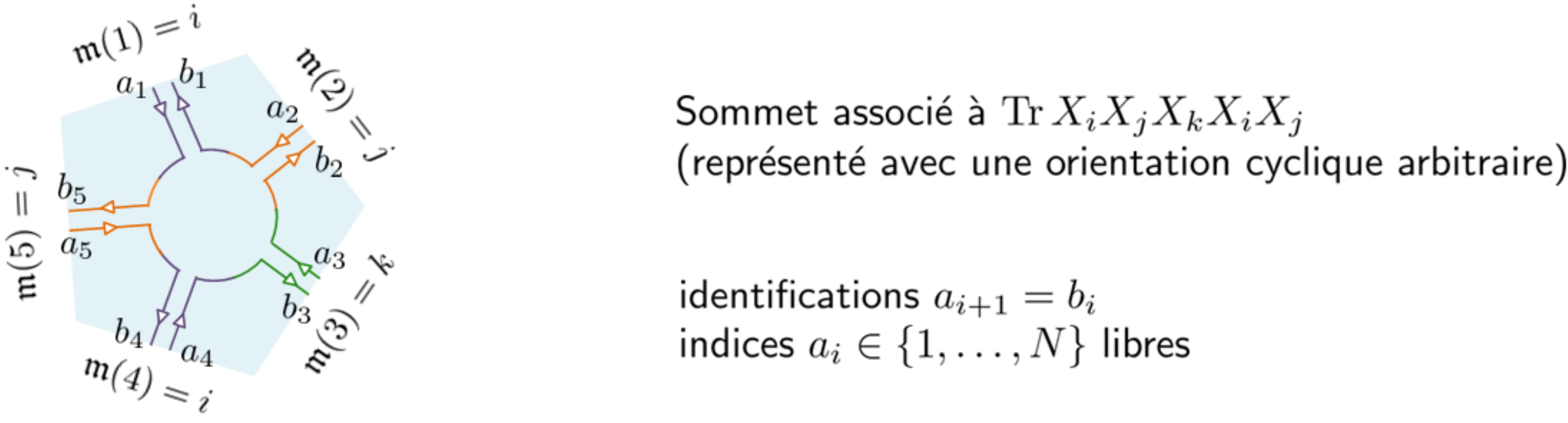}
\end{center}
\end{figure}

Le th\'eor\`eme de Wick dit que \'{E}qn.~\ref{eq:aoerez} est la somme sur toutes les "configurations" o\`u les lignes sont deux \`a deux appari\'ees. Les "configurations" sont donc d\'ecrites par des \textbf{graphes \'epais} $\mathcal{G}$, \label{epsi}o\`u les sommets sont les $T_{\mathfrak{m}}$, les ar\^{e}tes sont les appariements et portent des couleurs $\alpha \in \mathcal{I}$. Les lignes (bords des ar\^{e}tes) doivent former des cycles ferm\'es $\gamma$ sur lesquels circule un indice $a_{\gamma} \in \{1,\ldots,N\}$. Ils sont naturellement orient\'es par le sens des fl\`eches. Le dual $\mathcal{G}^*$ d'un graphe \'epais $\mathcal{G}$ d\'efinit ce que l'on appellera une \textbf{carte}. Intuitivement, c'est un recollement de faces polygonales le long de leur c\^{o}t\'es, ou encore une surface discr\`ete orientable (l'orientation vient du sens des fl\`eches le long des cycles de $\mathcal{G}$). Nous allons adopter le vocabulaire des cartes par la suite : en l'absence de pr\'ecision, "sommets", "ar\^{e}tes" et "faces" feront r\'ef\'erence \`a $\mathcal{G}^*$.

\begin{figure}[h!]
\begin{center}
\includegraphics[width=0.9\textwidth]{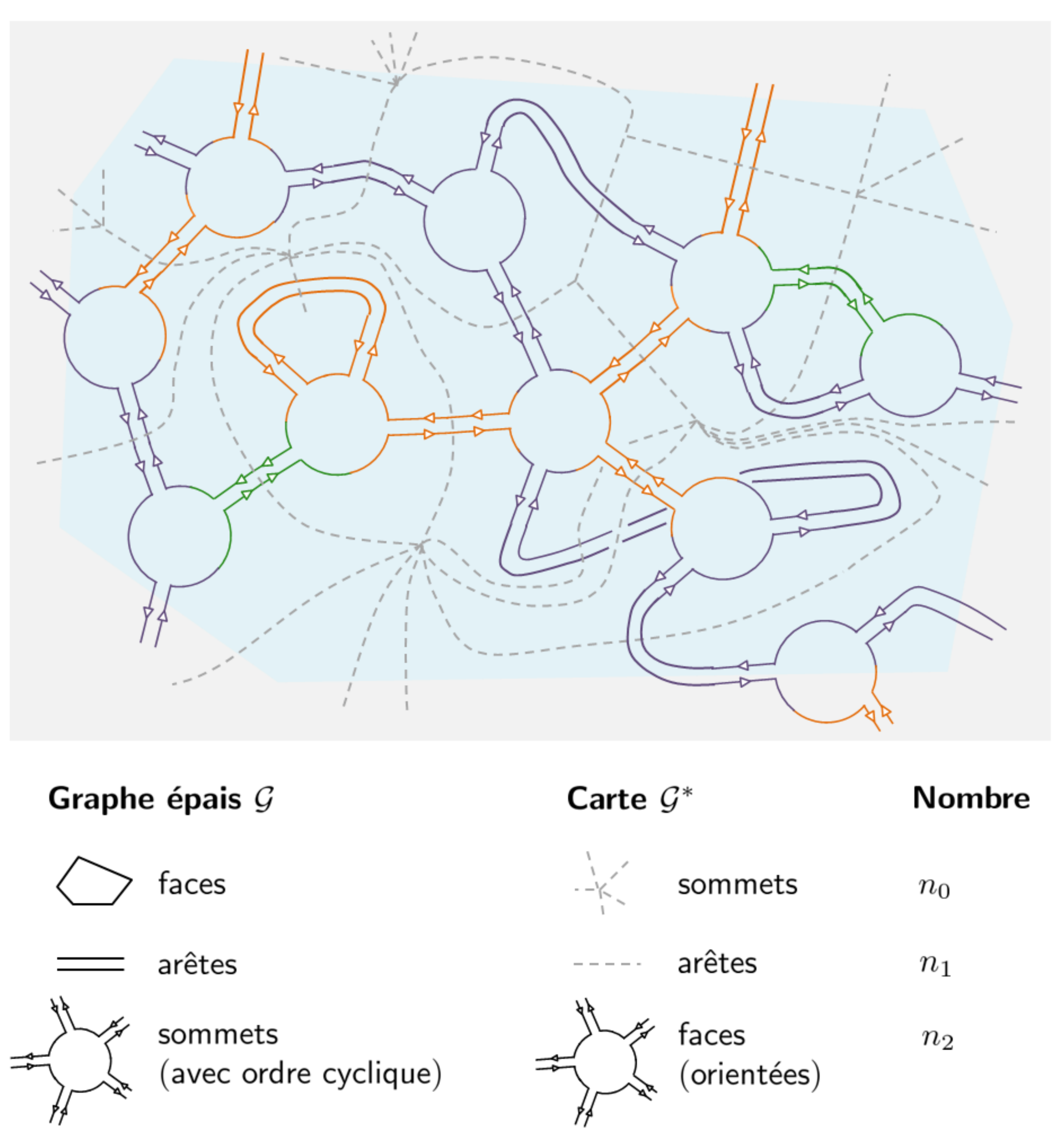}
\caption{\label{fig:Cartegenerale} Portion d'une carte construite par appariements. Attention : le th\'eor\`eme de Wick autorise les recollements de deux c\^{o}t\'es appartenant \`a la m\^{e}me face de $\mathcal{G}^*$, donc ce que l'on appelle "face polygonale" peut \^{e}tre un polygone un peu singulier. On remarquera une \textbf{trisection}, i.e. un croisement, qui implique que la carte sera de genre $g > 0$.}
\end{center}
\end{figure}

Dans un second temps, on souhaite calculer :
\bea
&& Z\cdot\Big\langle \textcolor{misval}{\prod_{j = 1}^n \Tr \frac{1}{x_j - X_{\mathfrak{m}_j(1)}\cdots X_{\mathfrak{m}_j(r_j)}}} \Big\rangle \nn \\
\label{eq:carj} & = & \sum_{(p_j)_j \geq 0}  \textcolor{misval}{\Big[\prod_{j = 1}^n \frac{1}{x_j^{p_j + 1}}\Big]}\,\Big\langle \prod_{j = 1}^n \textcolor{misval}{\Tr \big(X_{\mathfrak{m}_j(1)}\cdots X_{\mathfrak{m}_j(r_j)}\big)^{p_j}}\,e^{(N/t)\mathcal{U}(\mathbf{X})}\Big\rangle_0
\eea
o\`u les $x_j$ sont des param\`etres formels proches de $\infty$. C'est encore une somme sur des cartes $\mathcal{G}^*$. Pour les d\'ecrire, il faut maintenant distinguer les \textbf{faces (internes)} issues du d\'eveloppement de $e^{(N/t)\mathcal{U}}$, des \textbf{faces marqu\'ees} apparaissant en surbrillance. Dans l'expression pr\'ec\'edente, les faces marqu\'ees sont associ\'ees aux mots cycliques $\mathfrak{m}_j\cdots\mathfrak{m}_j$ concat\'en\'es $p_j$ fois, elles ont un p\'erim\`etre $r_jp_j$. On les repr\'esente avec une ar\^{e}te marqu\'ee et une orientation, i.e. on choisit (arbitrairement) un repr\'esentant du mot $\mathfrak{m}_j\cdots\mathfrak{m}_j$. Les faces marqu\'ees sont naturellement num\'erot\'ees (par un indice $j \in \{1,\ldots,n\}$), ce qui n'est pas le cas des faces internes. Pour d\'ecrire une "configuration" $\pi$ intervenant dans le calcul de \'{E}qn.~\ref{eq:carj}, il est n\'ecessaire de num\'eroter temporairement les polygones internes $\mathfrak{m}'_1,\mathfrak{m}'_2,\ldots$, et de num\'eroter leurs c\^{o}t\'es, disons dans un ordre cyclique (i.e. de fixer un repr\'esentant de chaque $\mathfrak{m}'_i$). $\pi$ est alors d\'etermin\'e par la liste des paires de c\^{o}t\'es que l'on veut identifier. Ceci fait, on obtient, apr\`es l'oubli de tous les num\'eros temporaires, un graphe \'epais $\mathcal{G}(\pi)$ ou une carte $\mathcal{G}^*(\pi)$.

Notons $\mathrm{Aut}\,\pi$ le sous-groupe du groupe $G$ des renum\'erotations des sommets et de leurs demi-ar\^{e}tes, laissant $\mathcal{G}^*(\pi)$ (avec tous ses marquages permanents) invariant. $|\mathrm{Aut}\,(\mathcal{G}(\pi))| = |G|/|\mathrm{Aut}\,\pi|$ est le nombre d'automorphismes du graphe $\mathcal{G}$, aussi appel\'e \textbf{facteur de sym\'etrie}.

\begin{figure}[h!]
\begin{center}
\hspace{-0.8cm}\begin{minipage}[c]{0.45\linewidth}
\raisebox{-5cm}{\includegraphics[width = \textwidth]{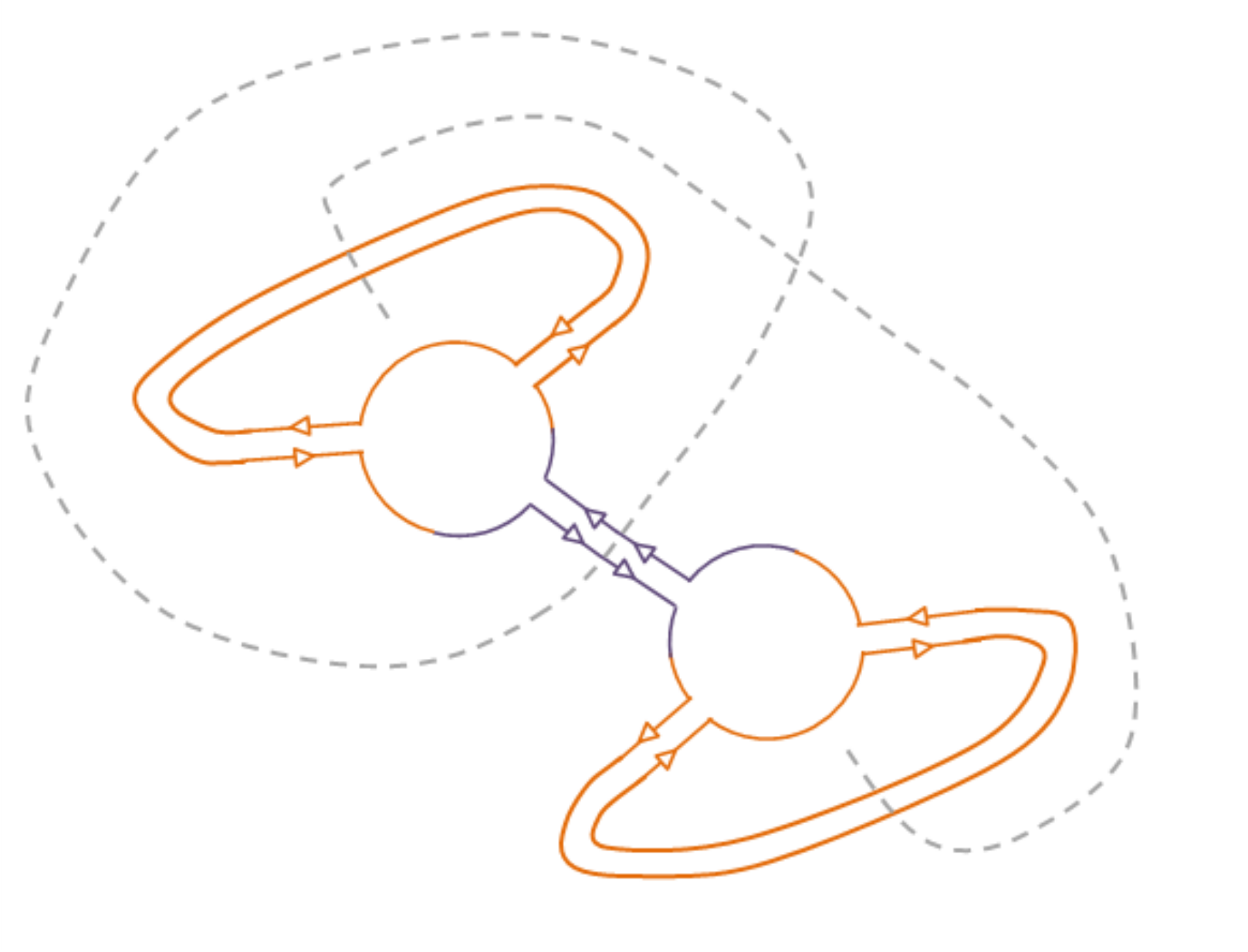}}
\hfill \end{minipage}\begin{minipage}[c]{0.47\linewidth}
 \caption{\label{fig:cartesym} Exemple de carte avec un facteur de sym\'etrie non trivial : on peut \'echanger les deux anses de m\^{e}me couleur, d'o\`u $|\mathrm{Aut}\,\mathcal{G}| = 2$.}
\end{minipage}
\end{center}
\end{figure}

Maintenant, $Z\cdot\big\langle\textcolor{misval}{\prod_{j = 1}^n\Tr \frac{1}{x_j - X_{\mathfrak{m}_j(1)}\cdots X_{\mathfrak{m}_j(r_j)}}}\big\rangle$ calcul\'e comme une somme sur les configurations $\pi$ fait apparaitre $|\mathrm{Aut}\,\pi|$ fois le m\^{e}me graphe $\mathcal{G}$. On peut donc la r\'e\'ecrire comme une s\'erie g\'en\'eratrice de toutes les cartes $\mathcal{G}^*$ qui contiennent au moins les faces marqu\'ees :
\beq
Z\cdot\Big\langle \textcolor{misval}{\prod_{j = 1}^n\Tr \frac{1}{x_j - X_{\mathfrak{m}_j(1)}\cdots X_{\mathfrak{m}_j(r_j)}}} \Big\rangle = \sum_{\mathcal{G}_0^* = \mathcal{G}^*(\pi)} |\mathrm{Aut}\,\pi|\,w_0(\mathcal{G}_0^*) \nn
\eeq
La normalisation au d\'enominateur de chaque terme de $\mathcal{U}$ (\'{E}qn.~\ref{eq:ugfd}) a \'et\'e choisie pr\'ecis\'ement pour reconstituer un facteur $\frac{1}{|G|}$ dans $w_0(\mathcal{G}_0^*)$. D'o\`u :
\beq
\label{eq:erogui}Z\cdot\Big\langle\textcolor{misval}{\prod_{j = 1}^n\Tr \frac{1}{x_j - X_{\mathfrak{m}_j(1)}\cdots X_{\mathfrak{m}_j(r_j)}}} \Big\rangle = \sum_{\substack{\mathcal{G}^*\,\textrm{\`a}\,k\,\textrm{faces}\,\textrm{marqu\'ees} \\ \textrm{de}\,\textrm{type}\,\mathfrak{m}_1,\ldots,\mathfrak{m}_k}} \frac{w(\mathcal{G}^*)}{|\mathrm{Aut}\,\mathcal{G}^*|}
\eeq
En particulier, lorsque $n = 0$ :
\beq
\label{eq:reogu}Z = \sum_{\mathcal{G}^*\,\textrm{sans}\,\textrm{face}\,\textrm{marqu\'ee}} \frac{w(\mathcal{G}^*)}{|\mathrm{Aut}\,\mathcal{G}^*|}
\eeq
Nous allons calculer ce poids, qui se d\'ecompose naturellement en $w(\mathcal{G}^*) = w_{\mathrm{topo}}(\mathcal{G}^*)\,\widehat{w}(\mathcal{G}^*)$, o\`u $w_{\mathrm{topo}}$ contiendra toute l'information sur la topologie de $\mathcal{G}^*$, et $\widehat{w}(\mathcal{G}^*)$ sera ind\'ependant de $N$.

\subsubsection{$w_{\mathrm{topo}}$ et regroupements par topologie}
\label{regr}
Soit $\mathcal{G}$ un graphe \'epais et $\mathcal{G}^*$ la carte associ\'ee. On note $n_2$ le nombre de faces polygonales de $\mathcal{G}^*$ (sommets internes de $\mathcal{G}$), $n_1$ le nombre d'ar\^{e}tes de $\mathcal{G}^*$ (le nombre d'appariements dans $\mathcal{G}$) et $n_0$ le nombre de sommets de $\mathcal{G}^*$ (le nombre de lignes formant des cycles dans $\mathcal{G}$). La d\'ependance en $N$ et $t$ du poids est tr\`es simple. Chaque ligne ferm\'ee dans $\mathcal{G}$ porte un indice matriciel $a \in \{1,\ldots,N\}$ sur lequel on somme, donc contribue \`a un facteur $N$. Et, par construction, chaque appariement vient avec un facteur $t/N$ (\'{E}qn.~\ref{eq:regl}), et chaque sommet de $\mathcal{G}$ avec un facteur $N/t$ (\'{E}qn.~\ref{eq:ugfd}). D'o\`u :
\beq
w(\mathcal{G}^*) \propto \Big(\frac{N}{t}\Big)^{\chi}\,t^{n_0},\qquad \chi = n_2 - n_1 + n_0 \nn
\eeq
$\chi$ est la caract\'eristique d'Euler de $\mathcal{G}^*$ ou de $\mathcal{G}$. Si l'on note, pour chaque composante connexe $\mathcal{G}_i^*$ de $\mathcal{G}^*$, $g_i$ son genre et $k_i$ son nombre de faces marqu\'ees,  elle vaut $\chi = \sum_{i} (2 - 2g_i - k_i)$. $\{(g_i,k_i)\quad i\}$ caract\'erise enti\`erement la topologie de la carte. Pour la terminologie, une carte connexe de genre $0$ est dite \textbf{planaire}, et une carte connexe avec une face marqu\'ee est dite \textbf{enracin\'ee}\footnote{Noter qu'une carte enracin\'ee n'a pas d'automorphismes : $|\mathrm{Aut}\,\mathcal{G}^*| = 1$.}. Par d\'efinition :
\beq
w_{\mathrm{topo}}(\mathcal{G}^*) = \Big(\frac{N}{t}\Big)^{\chi} \nn
\eeq

A priori dans un mod\`ele donn\'e, les sommets de $\mathcal{G}^*$ ont une valence arbitraire (la longueur des cycles dans $\mathcal{G}$ est arbitraire). Il n'est donc pas \'evident que l'ensemble $\mathbb{M}(\chi,n_0)$ des cartes \`a caract\'eristique d'Euler $\chi$ et nombre de sommets $n_0$ fix\'es, soit un ensemble fini. Notons $v_{\mathfrak{m}}$ la valence d'une face polygonale associ\'ee \`a $T_{\mathfrak{m}}$, et $n_2(\mathfrak{m})$ le nombre de telles faces dans $\mathcal{G}^*$. Le nombre d'ar\^{e}tes s'\'ecrit :
\beq
n_1 = \frac{1}{2}\sum_{\mathfrak{m}} v_{\mathfrak{m}}\,n_2(\mathfrak{m}) \nn
\eeq
D'o\`u :
\beq
\label{eq:eio}-\chi + n_0 = \sum_{\mathfrak{m}} \Big(\frac{v_{\mathfrak{m}}}{2} - 1\Big)\,n_2(\mathfrak{m}) \geq \frac{n_2}{2}
\eeq
avec l'in\'egalit\'e venant de l'hypoth\`ese technique faite sur les termes de $\mathcal{U}(\mathbf{X})$. On trouve que les cartes de $\mathbb{M}(\chi,n_0)$ sont obtenues par le recollement d'au plus $(-2\chi + 2n_0)$ faces polygonales, dont le nombre de c\^{o}t\'es est born\'e puisque $\mathcal{U}$ contient un nombre fini de termes. Donc, $\mathbb{M}(\chi,n_0)$ est fini.

Pour toute observable $\mathcal{O}$ qui s'\'ecrit :
\beq
\mathcal{O} = \sum_{\mathcal{G}^*}  \frac{w(\mathcal{G}^*)}{|\mathrm{Aut}\,\mathcal{G}^*|} \nn
\eeq
cela permet de d\'efinir $\mathcal{O}^{[[\chi]]}$ en regroupant les topologies :
\beq
\mathcal{O}^{[[\chi]]} = \sum_{\substack{\mathcal{G}^*\,\,\textrm{de}\,\,\textrm{caract\'eristique} \\ \textrm{d'Euler}\,\,\chi}} \frac{\widehat{w}(\mathcal{G}^*)}{|\mathrm{Aut}\,\mathcal{G}^*|} \nn
\eeq
en tant que s\'erie formelle en $t$. De plus, on a une \'egalit\'e entre s\'eries formelles :
\beq
\mathcal{O} = \sum_{\chi} \Big(\frac{N}{t}\Big)^{\chi}\,\mathcal{O}^{[[\chi]]} \nn
\eeq
i.e. \`a un ordre donn\'e $t^{k}$, il n'y a qu'un nombre fini de $\chi$ qui contribuent dans le membre de droite, et l'\'egalit\'e est valable.

\subsubsection{Poids $\widehat{w}(\mathcal{G})$ et mod\`eles \`a poids multiplicatifs}
\label{mukt}
Int\'eressons-nous maintenant \`a la d\'ependance du poids dans les param\`etres du mod\`ele. $\widehat{w}(\mathcal{G}^*)$ d\'epend du nombre $n_0(\mathfrak{m})$ de sommets internes de type $T_{\mathfrak{m}}$, et du nombre $n_1(\alpha)$ d'ar\^{e}tes de couleur $\alpha$ qui sont pr\'esents dans $\mathcal{G}^*$. Il y a aussi par construction un pr\'efacteur coupl\'e au p\'erim\`etre des faces marqu\'ees.
\beq
 \widehat{w}(\mathcal{G}^*) = \prod_{j} \frac{1}{x_j^{p_j + 1}}\,\prod_{\alpha \in \mathcal{I}} \z_\alpha^{n_1(\alpha)}\,\Big(\sum_{(L_k,r_k)_k}^* \prod_{k} g_{L_k}^{r_k}\Big) \nn
\eeq
$\sum^{*}$ porte sur toutes les familles finies constitu\'ee d'ensembles avec r\'ep\'etitions $L_k$ de mots cycliques de longueur $\geq 3$, et de multiplicit\'es $r_k \in \mathbb{N}$, telles que $n_0(\mathfrak{m}) = \sum_k r_k\cdot p_{\mathfrak{m} \in L_k}$ pour tout mot $\mathfrak{m}$. Cette somme est simplement le pr\'efacteur de $\prod_{\mathfrak{m}} T_{\mathfrak{m}}^{n_0(\mathfrak{m})}$ dans le d\'eveloppement de $e^{(N/t)\mathcal{U}}$. C'est en g\'en\'eral une fonction compliqu\'ee des $n_0(\mathfrak{m})$, qui n'est pas multiplicative. Par cons\'equent, si $\mathcal{G}^*$ a pour composantes connexes $\mathcal{G}^*_1,\ldots,\mathcal{G}^*_r$, en g\'en\'eral :
\beq
\label{eq:cos}\widehat{w}(\mathcal{G}^*) \neq \widehat{w}(\mathcal{G}_1^*)\cdots\widehat{w}(\mathcal{G}_r^*)
\eeq

On parle d'un \textbf{mod\`ele \`a poids multiplicatifs} lorsqu'il y a \'egalit\'e dans l'\'{E}qn.~\ref{eq:cos} pour $\mathcal{G}^*$ quelconque. Cela arrive seulement lorsque $\mathcal{U}(\mathbf{X})$ ne contient pas de produits de traces :
\beq
\label{eq:gfjk}\frac{N}{t}\,\mathcal{U}(\mathbf{X}) = \frac{N}{t}\sum_{\mathfrak{m}} \frac{t_{\mathfrak{m}}}{|\mathrm{Aut}\,\mathfrak{m}|}\,T_{\mathfrak{m}}(\mathbf{X})
\eeq
Dans ce cas :
\beq
\widehat{w}(\mathcal{G}^*) = \prod_{j} \frac{1}{x_j^{p_j + 1}} \prod_{\alpha \in \mathcal{I}} \z_\alpha^{n_1(\alpha)} \cdot \prod_{\mathfrak{m}} t_{\mathfrak{m}}^{n_0(\mathfrak{m})} \nn
\eeq
Par la suite, je vais limiter la discussion aux mod\`eles \`a poids multiplicatifs. Des mod\`eles \`a poids non multiplicatifs, comme le mod\`ele formel \`a une matrice :
\beq
\frac{N}{t}\,\mathcal{U}(X) = \frac{N}{t}\,\Tr V_1(X) + \Big(\frac{N}{t}\,V_2(X)\Big)^2 \nn
\eeq
avec $V_1$ polyn\^{o}me et $V_2$ polyn\^{o}me de bas degr\'e, ont \'et\'e \'etudi\'es dans la litt\'erature. Ils d\'ecrivent des cartes branch\'ees en arbre. La limite des grandes cartes planaires est alors d\'ecrite par la classe d'universalit\'e des polym\`eres branch\'es, ou des classes de polym\`eres branch\'es en comp\'etition avec les mod\`eles minimaux $(p,q)$ de la gravit\'e quantique. \`{A} ma connaissance, les topologies $g \geq 1$ ont \'et\'e peu \'etudi\'ees pour ces mod\`eles.

\subsubsection{Fonctions connexes}
\label{devo4}
La fonction de partition $Z$ est une somme sur les cartes sans face marqu\'ee. Dans un mod\`ele \`a poids multiplicatifs, un argument \'el\'ementaire montre que l'\'energie libre $F = \ln Z$ est la somme sur toutes les cartes connexes sans faces marqu\'ees. En regroupant les topologies, on peut d\'efinir $F^{(g)}$ comme la somme sur toutes les cartes connexes de genre $g$ et sans face marqu\'ee, et l'on a :
\beq
\ln Z = \sum_{g \geq 0} \Big(\frac{N}{t}\Big)^{2 - 2g}\,F^{(g)} \nn
\eeq
Toujours dans un mod\`ele \`a poids multiplicatifs, $\big\langle \prod_{j = 1}^n \Tr \frac{1}{x_j - X_{\mathfrak{m}_j(1)}\cdots X_{\mathfrak{m}_j(r_j)}} \big\rangle$ est la somme sur les cartes \`a $n$ faces marqu\'ees, de type $\mathfrak{m}_1,\ldots,\mathfrak{m}_n$, de sorte que toute composante connexe contienne au moins une face marqu\'ee. En regroupant les topologies :
\beq
\Big\langle \prod_{j = 1}^{n} \Tr \frac{1}{x_j - X_{\mathfrak{m}_j(1)\cdots\mathfrak{m}_j(r_j)}}\Big\rangle = \sum_{\chi \leq n} \Big(\frac{N}{t}\Big)^{\chi}\,\Big\langle \prod_{j = 1}^{n} \Tr \frac{1}{x_j - X_{\mathfrak{m}_j(1)\cdots\mathfrak{m}_j(r_j)}}\Big\rangle^{[[\chi]]} \nn
\eeq
On peut aussi d\'efinir le cumulant $\big\langle \cdots \big\rangle_c$, qui calcule la m\^{e}me somme restreinte aux cartes connexes. L\`a encore, on peut regrouper les topologies :
{\small \beq
\Big\langle \prod_{j = 1}^{n} \Tr \frac{1}{x_j - X_{\mathfrak{m}_j(1)}\cdots X_{\mathfrak{m}_j(r_j)}}  \Big\rangle_c = \sum_{g \geq 0} \Big(\frac{N}{t}\Big)^{2 - 2g - n}\,\Big\langle \prod_{j = 1}^n \Tr \frac{1}{x_j - X_{\mathfrak{m}_j(1)}\cdots X_{\mathfrak{m}_j(r_j)}} \Big\rangle_c^{(g)} \nn
\eeq}
$\!\!\!$Pour les cartes connexes, nous utilisons une notation avec le genre $g \in \mathbb{N}$ en exposant ${}^{(g)}$, plut\^{o}t que ${}^{[[\chi = 2 - 2g - n]]}$.

\subsubsection{Conclusion}

Pour r\'esumer, nous avons justifi\'e, pour les mod\`eles formels \`a poids multiplicatifs, que l'\'energie libre et les cumulants (et les moments) ont un d\'eveloppement topologique, dont les termes s'interpr\`etent comme des s\'eries g\'en\'eratrices de cartes de genre fix\'e. On a aussi introduit des param\`etres formels $x_j$ coupl\'es multiplicativement aux p\'erim\`etres des faces marqu\'ees.

Remarquons que les \'equations de Schwinger-Dyson (partie~\ref{sec:eqboucl1}) seront valables aussi bien pour les mod\`eles formels que convergents. Et, dans les mod\`eles formels, on a automatiquement l'existence d'un d\'eveloppement topologique de type \'{E}qn.~\ref{eq:dev}, avec $g_s = t/N$. La nature de ce d\'eveloppement est un peu particuli\`ere : chaque s\'erie g\'en\'eratrice est une s\'erie formelle en $t$, et \`a un ordre donn\'e $t^{k}$, il y a un nombre fini de termes qui ont la m\^{e}me puissance de $g_s$, et que l'on peut regrouper. Finalement, on pourra ins\'erer ce d\'eveloppement dans les \'equations de Schwinger-Dyson, et chercher \`a r\'esoudre la hi\'erarchie ainsi obtenue afin de calculer les s\'eries g\'en\'eratrices de cartes.

\subsection{Propri\'et\'es analytiques des s\'eries g\'en\'eratrices}
\label{sec:anaan}
Les s\'eries g\'en\'eratrices de cartes sont d\'efinies initialement comme des s\'eries formelles (en $t$, en $x_j$), mais elles ont des rayons de convergence non nuls car le nombre de cartes cro\^{i}t au plus g\'eom\'etriquement avec le nombre de faces. Ceci peut \^{e}tre justifi\'e en deux temps.

\begin{figure}[h!]
\begin{center}
\includegraphics[width=0.9\textwidth]{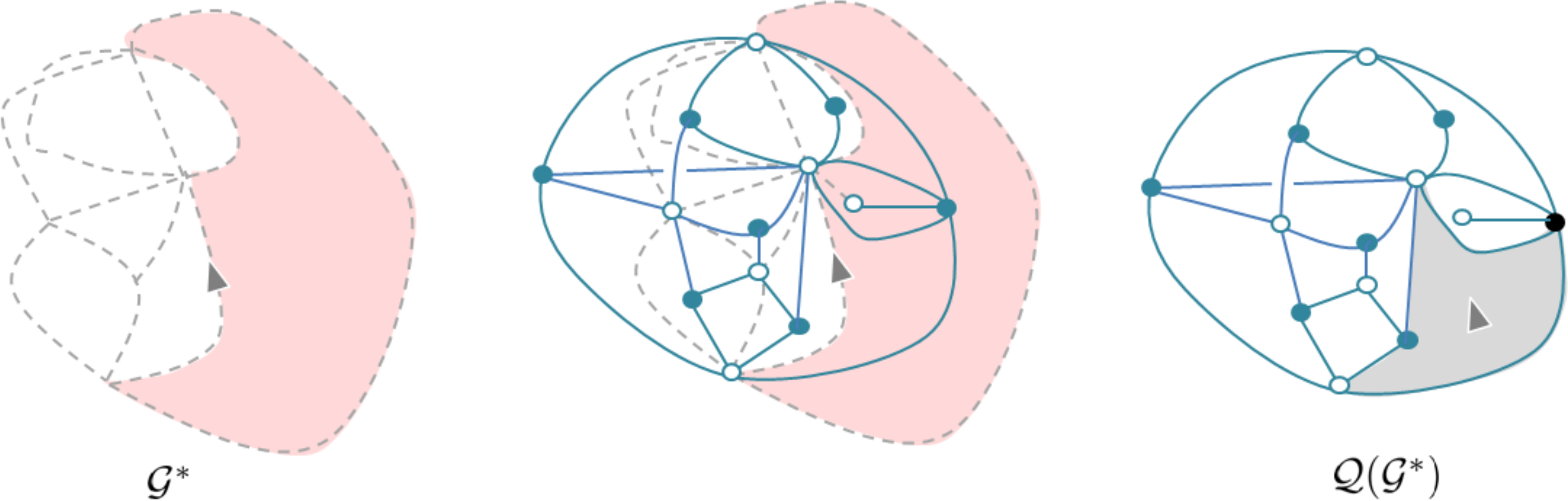}
\caption{\label{fig:Qcarte} Soit une carte $\mathcal{G}^*$ quelconque, $n_2,n_1,n_0$ ses nombres de faces, d'ar\^{e}tes et de sommets. On lui associe une quadrangulation bipartite $\mathcal{Q}_{\mathcal{G}^*}$ dont : les sommets $\circ$ sont aux sommets de $\mathcal{G}^*$ ; les sommets $\bullet$ aux centres des faces de $\mathcal{G}^*$ ; deux sommets $\circ$ appartiennent \`a la m\^{e}me face ssi ils sont reli\'es par une ar\^{e}te dans $\mathcal{G}^*$. $\mathcal{Q}_{\mathcal{G}^*}$ a $n_2' = n_1$ faces, $n_1' = 2n_1$ ar\^{e}tes, et $n_0' = n_0 + n_2$ sommets, donc $\chi(\mathcal{G}^*) = \chi(\mathcal{Q}_{\mathcal{G}^*})$. On transf\`ere aussi sur les faces de $\mathcal{Q}_{\mathcal{G}^*}$ les couleurs des ar\^{e}tes de $\mathcal{G}^*$, et l'on marque le sommet $\bullet$ de $\mathcal{Q}_{\mathcal{G}^*}$ correspondant \`a la face marqu\'ee. R\'eciproquement, il est possible de reconstruire $\mathcal{G}^*$ en partant de $\mathcal{Q}_{\mathcal{G}^*}$.}
\end{center}
\end{figure}

Les cartes consid\'er\'ees par Tutte et ses successeurs sont engendr\'ees par un mod\`ele formel \`a une matrice :
\beq
\label{eq:gfdq} Z = \int \mathrm{d}M\,e^{-(N/t)\Tr V(M)},\qquad V(M) = \frac{M^2}{2} - \sum_{j} t_j\,\frac{M^j}{j}
\eeq
Tutte a \'etabli \cite{Tutte2} que le nombre de quadrangulations planaires, enracin\'ees, \`a $n_2$ faces est :
\beq
\mathcal{Q}_{n_2} = \frac{2\cdot 3^{n_2}}{n_2 + 2}\,\textrm{\textsf{Cat}}(n_2) \mathop{\sim}_{n_2 \rightarrow \infty} \frac{2}{\sqrt{\pi}}\,\frac{12^{n_2}}{{n_2}^{5/2}} \nn
\eeq
o\`u \textsf{Cat}$(n) = \frac{(2n)!}{n!(n + 1)!}$ est le $n^{\textrm{\`eme}}$ nombre de Catalan. Comme il existe une bijection entre cartes et quadrangulations bipartites (Fig.~\ref{fig:Qcarte}) et que toute quadrangulation planaire est bipartite, c'est aussi le nombre de cartes planaires enracin\'ees \`a $n_2$ ar\^{e}tes. Il faut ajouter que nos hypoth\`eses techniques (toutes les faces internes ont un p\'erim\`etre $\geq 3$, tandis que les faces externes ont un p\'erim\`etre $\geq 1$ quelconque) impliquent que les cartes planaires enracin\'ees g\'en\'er\'ees dans l'\'{E}qn.~\ref{eq:gfdq} ne forment qu'un sous-ensemble des cartes planaires enracin\'ees de \cite{Tutte2}.
De ce r\'esultat, on peut \'egalement d\'eduire une croissance g\'eom\'etrique du nombre de cartes enracin\'ees de genre $g$ \`a $n_1$ ar\^{e}tes, soit en se r\'ef\'erant aux travaux ult\'erieurs de Bender et Canfield \cite{BenCan}, soit en \'etudiant directement le d\'eveloppement topologique des \'equations de Schwinger-Dyson du mod\`ele formel \`a une matrice. Enfin, ce r\'esultat est pr\'eserv\'e si l'on ajoute des faces marqu\'ees.

Les mod\`eles multiplicatifs plus g\'en\'eraux engendrent des cartes o\`u les ar\^{e}tes portent une couleur $\alpha \in \mathcal{I}$. On peut encore associer \`a toute carte $\mathcal{G}^*$ \`a $n_1$ ar\^{e}tes, une quadrangulation bipartite $\mathcal{Q}_{\mathcal{G}^*}$ \`a $n_1$ faces, et en d\'ecorant $\mathcal{Q}_{\mathcal{G}^*}$ convenablement, il est possible de remonter de fa\c{c}on unique \`a $\mathcal{G}^*$ (Fig.~\ref{fig:Qcarte}). De plus, cette association pr\'eserve la caract\'eristique d'Euler. Donc, le nombre de telles cartes \`a genre fix\'e sera un $O\big((12\cdot |\mathcal{I}|)^{n_1}\big)$.

Enfin, si le nombre de sommets $n_0$ et la caract\'eristique d'Euler de $\mathcal{G}^*$ sont fix\'es, le nombre d'ar\^{e}tes $n_1$ est born\'e par $(-2\chi + 2n_0)$ (\'{E}qn.~\ref{eq:eio}). Donc $|\mathbb{M}(\chi,n_0)|$ cro\^{i}t au plus g\'eom\'etriquement avec $n_0$. Par cons\'equent,
\beq
\Big\langle \prod_{j = 1}^{n} \Tr \frac{1}{x_j - X_{\mathfrak{m}_j(1)}\cdots X_{\mathfrak{m}_j(r_j)}}  \Big\rangle_c^{(g)} \nn
\eeq
a\label{rayy} un rayon de convergence non nul, c'est une fonction analytique pour $|t| < t^*$, et pour $|x_j| > R(t)$. Cela permet par la suite d'utiliser les outils puissants de l'analyse complexe. Toute relation entre s\'eries g\'en\'eratrices sera valable dans tout le domaine d'analyticit\'e pour, et donnera m\^{e}me des informations sur le domaine maximal d'analyticit\'e. L'\'etude des singularit\'es en $t$ donnera naturellement acc\`es \`a l'\'enum\'eration des grandes cartes.

\subsection{Limitations inh\'erentes aux mod\`eles de matrices}

Comme l'on ne peut associer de param\`etres libres qu'aux faces marqu\'ees, les observables des mod\`eles formels de matrices ne peuvent pas donner directement acc\`es \`a des informations locales sur la carte al\'eatoire, ni \`a des informations m\'etriques. Cependant, il existe des relations combinatoires (dont le vaste champ n'est qu'en partie explor\'e) entre des quantit\'es locales int\'eressantes, et les observables coupl\'ees au p\'erim\`etre des faces marqu\'ees. R\'ecemment, J\'er\'emie Bouttier et Emmanuel Guitter \cite{BouGui} ont d\'ecouvert que la statistique de la distance g\'eod\'esique dans les cartes planaires (sans d\'ecoration, i.e. engendr\'ees par \'{E}qn.~\ref{eq:gfdq}) \'etait encod\'ee dans le d\'eveloppement en fraction continue de $\big\langle\Tr \frac{1}{x - M}\big\rangle^{(0)}$. \`{A} l'avenir, il serait int\'eressant d'exploiter de telles m\'ethodes pour les cartes d\'ecor\'ees, par exemple pour pr\'edire l'exposant \'echelle de la distance g\'eod\'esique entre deux points d'une grande carte. Celui-ci est inconnu d\'ej\`a pour le mod\`ele de boucles le plus simple (le mod\`ele $\On$ trivalent, partie~\ref{sec:OnON}).

\section{Les mod\`eles de boucles}
\label{sec:bouton}
\subsection{Pourquoi les mod\`eles de boucles ?}
\label{sec:modbu}
Les mod\`eles de boucles consistent \`a compter des configurations de chemins auto\'evitants vivant sur les faces d'un r\'eseau \`a deux dimensions $\mathcal{G}^*$ (fixe ou al\'eatoire). Ces chemins peuvent relier deux extr\'emit\'es (que l'on appellera \textbf{m\`eches}) appartenant aux bords, ou bien former des boucles. Dans les mod\`eles $\mathcal{O}(\n_1)\times\cdots\times\mathcal{O}(\n_D)$, il y a $D$ types de boucles, et chaque boucle de type $d$ est compt\'ee avec un poids $\n_d$. On peut donner une orientation aux boucles si on le souhaite : en \'ecrivant $\n_d = e^{i\pi\mathfrak{b}_d} + e^{-i\pi\mathfrak{b}_d}$ dans le mod\`ele pr\'ec\'edent, la contribution d'une boucle qui a tourn\'e $\ell_+$ fois \`a droite et $\ell_-$ fois \`a gauche sera le pr\'efacteur de $e^{i\pi\mathfrak{b}_d(\ell_+ - \ell_-)}$.

Pour des valeurs g\'en\'eriques des $\n_d$, le poids des configurations est \textbf{non local}. Cela signifie que le poids $\n_d$ donn\'e globalement \`a une boucle de type $d$, a priori \'etendue sur $\mathcal{G}^*$, ne peut \^{e}tre redistribu\'e uniquement sur les sommets, ar\^{e}tes et faces de $\mathcal{G}^*$. La physique statistique dans les mod\`eles non locaux contient certainement de nouveaux effets par rapport \`a celle des mod\`eles locaux.

\subsubsection{Sur un r\'eseau fixe}

\begin{figure}
\begin{center}
\includegraphics[width = 0.8\textwidth]{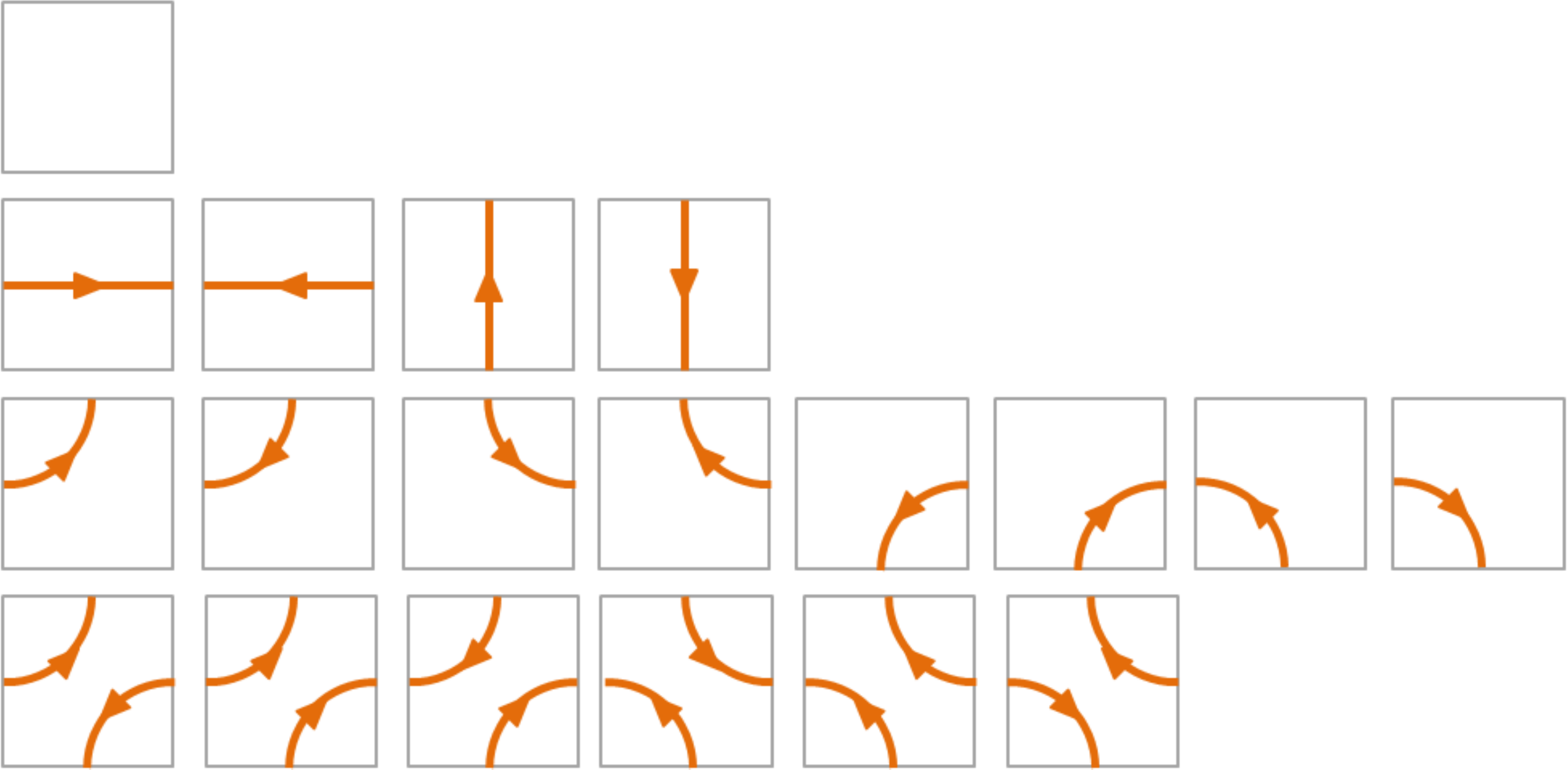}
\caption{\label{fig:vertex} Les briques du mod\`ele \`a dix-neuf vertex d'Izergin et Korepin.}
\end{center}
\end{figure}

\begin{figure}[h!]
\begin{center}
\includegraphics[width=\textwidth]{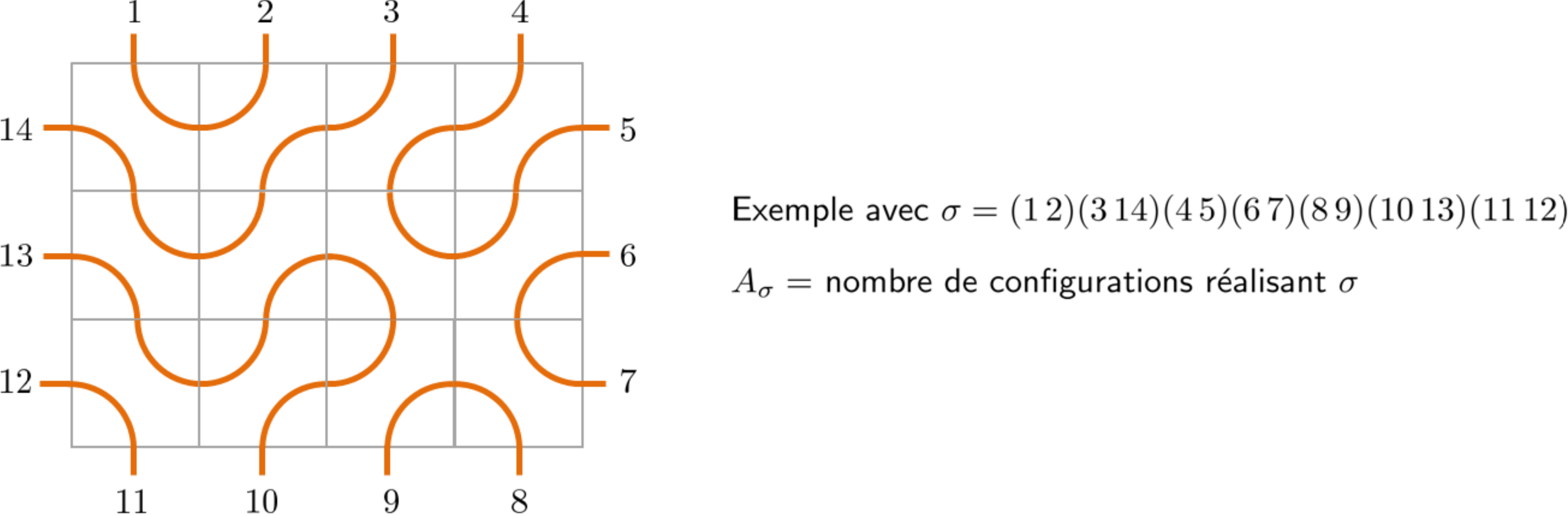}
\caption{\label{fig:Razumov} Si on note $2L$ le nombre de m\`eches, $\sum_{\sigma} A_{\sigma}$ est aussi le nombre de matrices \`a signes altern\'es de taille $L \times L$. La conjecture (maintenant, le th\'eor\`eme) de Razumov-Stroganov indique que $\sum A_{\sigma}|\sigma\rangle$ est le vecteur propre du fondamental de $\mathbf{H} = -\sum_{i = 1}^{2L} \mathbf{e}_i$, normalis\'e \`a $A_{(1\,2)(3\,4)\cdots(2L - 1\,2L)} = 1$. $\mathbf{e}_i$ sont les g\'en\'erateurs de l'alg\`ebre de Temperley-Lieb (cf. Fig~\ref{fig:TL}).}
\end{center}
\end{figure}

\begin{figure}[h]
\begin{center}
\includegraphics[width=\textwidth]{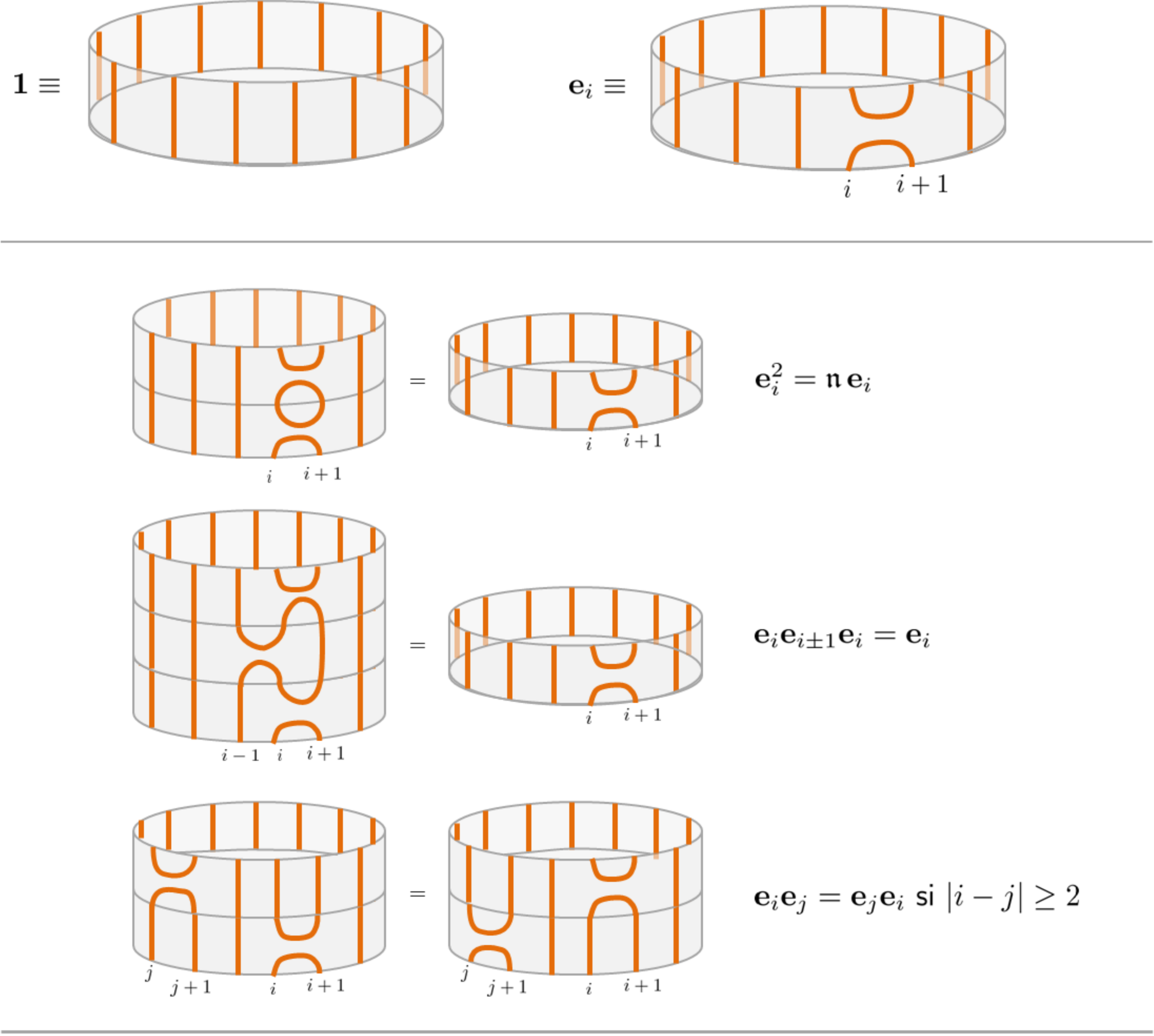}
\caption{\label{fig:TL} G\'en\'erateurs et relations de l'\textbf{alg\`ebre de Temperley-Lieb} avec $n$ m\`eches, \`a conditions de bord p\'eriodiques. Avec les produits des $\mathbf{e}_i$. TL forme un espace vectoriel engendr\'e par les syst\`emes de liens planaires sur le cylindre, donc est de dimension $n\cdot\textsf{Cat}(n - 1) = \frac{(2n - 2)!}{(n - 1)!^2}$. Pour les mod\`eles de boucles en genre sup\'erieur, c'est plut\^{o}t l'\textbf{alg\`ebre de Brauer} \label{Brauer} qui est pertinente : on ajoute un g\'en\'erateur qui \'echange les lignes $i$ et $i + 1$. Les appariements quelconques de $2n$ sites (\'eventuellement avec croisements sur le cylindre) en forment une base vectorielle, de dimension $\frac{(2n)!}{n!^2}$.}
\end{center}
\end{figure}

Les mod\`eles de boucles ont d'abord \'et\'e \'etudi\'es sur un r\'eseau fixe et r\'egulier, essentiellement sur le r\'eseau carr\'e ou le nid d'abeille. Le mod\`ele $\On$ a une longue histoire en physique statistique et en combinatoire. Lorsque toutes les faces du r\'eseau sont occup\'ees par des boucles, il est dual au mod\`ele de Potts \`a $\mathfrak{q} = \mathfrak{n^2}$ couleurs, et les boucles sont les fronti\`eres des amas de spins de m\^{e}me couleur. Lorsque les boucles sont interpr\'et\'ees comme lignes de niveaux et $\mathfrak{n} = -2\cos[2\pi/(\mathfrak{m} + 1)]$, c'est un mod\`ele de hauteurs al\'eatoires appel\'e RSOS (\textit{Restricted Solid on Solid}). En prenant des d\'eriv\'ees par rapport \`a $\n$, \'evalu\'ees \`a $\n =0$, le mod\`ele $\On$ capture la physique des polym\`eres, i.e. des marches al\'eatoires auto\'evitantes. Le mod\`ele $\On$ est aussi est reli\'e \`a des mod\`eles de vertex int\'egrables quantiques, i.e. solubles par ansatz de Bethe, comme le mod\`ele \`a six vertex (si les boucles occupent toutes les faces d'un r\'eseau carr\'e) ou le mod\`ele \`a dix-neuf vertex (sans cette contrainte, sur le r\'eseau carr\'e, Fig.~\ref{fig:vertex}). Le mod\`ele $\mathcal{O}(\mathfrak{1})$ est reli\'e \`a l'\'enum\'eration des matrices \`a signes altern\'es introduites par Mills et Robbins, \ldots{} La conjecture de Razumov-Stroganov\label{RazumovS}, d\'emontr\'ee en 2010 par Luigi Cantini et Andrea Sportiello \cite{Cantini}, est une belle illustration de cette ubiquit\'e (Fig.~\ref{fig:Razumov}) : dans un mod\`ele $\mathcal{O}(\mathfrak{1})$ sur r\'eseau carr\'e, o\`u tout carr\'e est occup\'e par un chemin, la probabilit\'e d'observer un appariement \label{msu}donn\'e de m\`eches $\sigma$ est proportionnelle aux composantes du vecteur propre de l'\'etat fondamental d'une chaine de spins $1/2$ de type XXZ. En g\'en\'eral, les observables avec changements de conditions de bord sont organis\'ees en une repr\'esentation de l'alg\`ebre de \label{TL} Temperley-Lieb $TL(q)$ (Fig.~\ref{fig:TL}) ou de ses variantes, avec $\n = - \mathfrak{q} - \mathfrak{q}^{-1}$.

\`{A} la limite continue, le mod\`ele $\On$ sur r\'eseau r\'egulier carr\'e, devrait \^{e}tre d\'ecrit par une \label{cc} th\'eorie conforme \cite{PagesJaunesOn} de charge centrale $\mathfrak{c}$ :
 \beq
 \mathfrak{c} = 1 - 6\big(\widetilde{\mathfrak{b}}^{1/2} - \widetilde{\mathfrak{b}}^{-1/2}\big)^2,\qquad \n = -2\cos(\pi\widetilde{\mathfrak{b}}) \nn
 \eeq
C'est une conjecture pour les math\'ematiciens\footnote{C'est un sujet de recherche tr\`es actif, avec des r\'esultats r\'ecents notamment pour $\n = 1$ (le mod\`ele d'Ising\label{Isi}). On pourra consulter la revue de S.~Smirnov \cite{Smir} pour un \'etat des progr\`es datant de 2006 ainsi que des r\'ef\'erences, ainsi que la description plus technique \cite{Smir2}.}, et un fait plausible largement utilis\'e par les physiciens pour faire des pr\'edictions, qui n'a jamais \'et\'e pris en d\'efaut par les r\'esultats exacts ou les simulations num\'eriques (pour les exposants, les r\`egles de fusion d'op\'erateurs, \ldots). Inversement, la limite continue des mod\`eles de boucles est particuli\`erement int\'eressante pour \'etudier, au moins heuristiquement, une vaste classe de th\'eories conformes et leurs conditions de bord \cite{JacobSen,Jerothese}. Cela inclut la s\'erie des \label{minimo2}mod\`eles minimaux unitaires $(m + 1,m)$ (lorsque $\widetilde{\mathfrak{b}} = (m + 1)/m$) et la s\'erie des mod\`eles minimaux $(p,q)$ (lorsque $\widetilde{\mathfrak{b}} = p/q$), mais ces th\'eories conformes sont par ailleurs tr\`es bien connues. En revanche, le cas $\widetilde{\mathfrak{b}} \notin \mathbb{Q}$ permet en principe d'atteindre les th\'eories conformes logarithmiques, dont la compr\'ehension est un enjeu actuel. On observe par exemple de fortes similarit\'es entre, la th\'eorie des repr\'esentations de  $TL(\mathfrak{q})$ et ses variantes qui conduit \`a la d\'efinition de mod\`eles  $\On$ ou $\mathcal{O}(\n_1)\times\mathcal{O}(\n_2)$, et celle de l'alg\`ebre de Virasoro (\`a l'{\oe}uvre dans la th\'eorie conforme).

\subsubsection{Sur une carte al\'eatoire}

Pour les physiciens, les cartes al\'eatoires (sans boucles) sont bien d\'ecrites \`a la limite continue par \label{Liouli}la \textbf{th\'eorie de Liouville}, qui est la seule th\'eorie des champs connue qui capture la notion de "gravit\'e quantique \`a deux dimensions". Naturellement, les cartes al\'eatoires dot\'ees de configurations de boucles devraient \^{e}tre d\'ecrites \`a la limite continue par la th\'eorie de Liouville coupl\'ee \`a une th\'eorie conforme. Dans ce cadre, il existe une correspondance appel\'ee \textbf{habillage gravitationnel} entre les observables d'une th\'eorie conforme de charge centrale $\mathfrak{c}$, et les observables de la m\^{e}me th\'eorie conforme coupl\'ee \`a la th\'eorie de Liouville. Knizhnik, Polyakov et A.B.~Zamolodchikov \cite{KPZ} pr\'edisent une relation entre dimension conforme $\Delta$ d'un op\'erateur $\Phi_{\Delta}$ (donnant un exposant critique sur r\'eseau fixe) et dimension conforme $\widetilde{\Delta}$ de l'op\'erateur habill\'e $\widetilde{\Phi_{\Delta}}$ (donnant un exposant\label{disss} critique pour les cartes al\'eatoires) :
\beq
\widetilde{\Delta} = \frac{\Delta(\Delta - \gamma_{\mathrm{str}})}{1 - \gamma_{\mathrm{str}}},\qquad \mathfrak{c} = 1 - 6\,\frac{\gamma_{\mathrm{str}}^2}{1 - \gamma_{\mathrm{str}}} \nn
\eeq

Les int\'egrales formelles de matrices donnent une technique d'\'etude syst\'ematique des mod\`eles de boucles sur cartes al\'eatoires. Le plus simple techniquement est le mod\`ele $\On$ trivalent, introduit en 1989 par Ivan Kostov \cite{KosOn}. C'est principalement sa limite continue qui a \'et\'e \'etudi\'ee par Kostov et bien d'autres : les points critiques\label{crti2} ont \'et\'e \'etudi\'es dans \cite{KosStau}, l'\'etude des conditions de bords parall\`element aux travaux de Jacobsen et Saleur sur le r\'eseau fixe \cite{JacobSen} a \'et\'e faite dans \cite{Kosbord}, le diagramme de phase et les flots de renormalisation pour les conditions de bords ont \'et\'e \'etudi\'es dans la th\`ese de Jean-\'{E}mile Bourgine \cite{TheseJE}, \ldots{} Le va-et-vient entre limite continue du mod\`ele de matrice $\On$, et th\'eorie de Liouville coupl\'ee \`a une th\'eorie conforme, ont contribu\'e \`a une meilleure compr\'ehension de chaque c\^{o}t\'e.

Dans l'optique de cette th\`ese, les mod\`eles de boucles sur cartes al\'eatoires sont \'egalement int\'eressants avant la limite continue. Pour la combinatoire, mais surtout pour leurs relations conjectur\'ees avec des alg\`ebres de type Temperley-Lieb et des questions d'int\'egrabilit\'e quantique. \`{A} terme, on aimerait transposer aux mod\`eles sur carte al\'eatoire, l'arsenal de structures math\'ematiques existant sur le r\'eseau fixe de taille finie. Plus modestement ici, l'objectif est de calculer le maximum d'observables dans les mod\`eles de boucles sur carte al\'eatoire sans prendre de limite continue. Dans les paragraphes suivants, je vais montrer pour le mod\`ele de boucles g\'en\'eral, quelles sont les \'equations \`a r\'esoudre et o\`u se trouve actuellement la fronti\`ere des connaissances. Cette pr\'esentation est inspir\'ee de travaux en cours, qui ne sont pas encore publi\'es. Le mod\`ele $\On$ trivalent fera l'objet de la partie~\ref{sec:OnON}.

\subsection{D\'efinitions}

Le mod\`ele formel de matrices qui engendre des cartes al\'eatoires portant des boucles auto\'evitantes est :
\bea
\label{eq:ms} \dd\nu_0 & = & \dd M \prod_{\alpha = 1}^{\n} \dd A_\alpha\,\exp\Big\{-\frac{N}{t}\Tr\Big(\frac{M^2}{2} + \sum_{\alpha = 1}^{\n} \frac{A_\alpha^2}{2\z_\alpha}\Big)\Big\}\\
\frac{N}{t}\,\mathcal{U} & = & \frac{N}{t}\Tr\Big(\sum_{j = 3}^{j_{\mathrm{max}}} \frac{t_j}{j}\,M^j + \sum_{\alpha = 1}^{\n}\sum_{k = 3}^{k_{\mathrm{max}}}\sum_{r = 1}^{\lfloor k/2 \rfloor - 1} \frac{g_{\alpha;k,r}}{1 + \delta_{r,k - 2 - r}}\,A_{\alpha}M^{r}A_{\alpha}M^{k - 2 - r} \Big) \nn
\eea
$\mathbf{X} = (M,A_1,\ldots,A_{\mathfrak{n}})$ sont des matrices hermitiennes. La matrice $M$ a un r\^{o}le particulier. Nous encodons son potentiel et le terme de couplage en posant :
 \bea
\label{eq:defV} V(\lambda) & =&  \frac{\lambda^2}{2} - \sum_{j = 3}^{j_{\mathrm{max}}} \frac{t_j}{j}\,\lambda^j  \\
 \label{eq:aoao} K_\alpha(\lambda,\mu) & = & \frac{1}{\z_\alpha} - \sum_{r_1,r_2} g_{\alpha;r_1 + r_2 + 2,\mathrm{min}(r_1,r_2)}\,\lambda^{r_1}\cdot\mu^{r_2}
 \eea

\begin{figure}[h]
\begin{center}
\hspace{-0.8cm} \includegraphics[width = 1.15\textwidth]{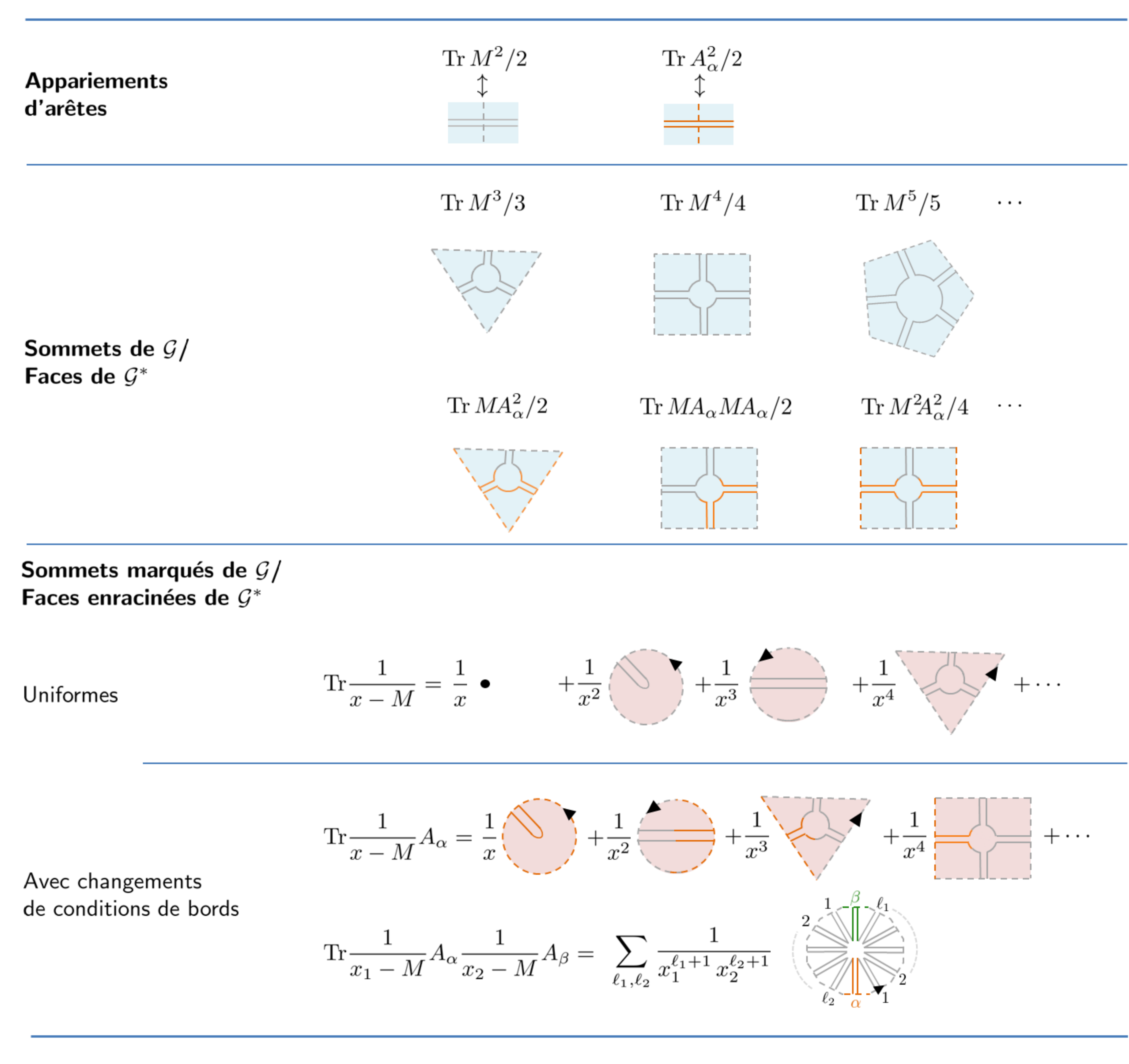}
\caption{\label{fig:briqbou} Briques \'el\'ementaires des cartes g\'en\'er\'ees par les mod\`eles de boucles}
\end{center}
\end{figure}

Attardons-nous sur les briques \'el\'ementaires des cartes dans ce mod\`ele (Fig.~\ref{fig:briqbou}). Les $M^j$ engendrent des $j$gones "standard" avec poids $t_j$. Les $\Tr(M^{r_1}A_{\alpha}M^{r_2}A_{\alpha})$ engendrent des $(r_1 + r_2 + 2)$gones avec deux c\^{o}t\'es sp\'eciaux : on peut les repr\'esenter comme des $k$gones ($k = r_1 + r_2 + 2$) travers\'es par un chemin de couleur $\alpha$ entre deux c\^{o}t\'es s\'epar\'es par exactement $r_1$ c\^{o}t\'es standard. Lors du recollement des polygones, ces chemins sont mis bout-\`a-bout, et doivent se fermer ou avoir leur extr\'emit\'es sur des c\^{o}t\'es de couleur $\alpha$ appartenant \`a une face marqu\'ee. Par construction, un chemin garde toujours sa couleur (pas de terme crois\'e $\Tr A_\alpha\cdots A_\beta\cdots$), et ne peut ni s'intersecter ni intersecter un autre chemin (pas de termes de degr\'e $\geq 3$ en $A_\alpha$ dans $\mathcal{U}$). Si $\ell_\alpha$ est la longueur totale des chemins de couleur $\alpha$, $n_2(\alpha;k,r)$ le nombre de polyg\^{o}nes de type $M^{r}A_{\alpha}M^{k - 2 - r}A_{\alpha}$, $n_2(j)$ le nombre de $j$gones standard, le poids est :
\beq
\prod_{j = 3}^{j_{\mathrm{max}}} t_j^{n_2(j)}\prod_{\alpha = 1}^{\n} \z_\alpha^{\ell_\alpha}\,\prod_{k = 3}^{k_{\mathrm{max}}}\prod_{r = 1}^{\lfloor k/2 \rfloor - 1} g_{\alpha;k,r}^{n_2(\alpha;k,r)} \nn
\eeq

On appellera \textbf{bords}, les faces marqu\'ees d'une carte. On appelle \textbf{observables (\`a conditions de bord) uniformes} les s\'eries g\'en\'eratrices $\big\langle \mathcal{P}(M)\big\rangle$. Elles g\'en\`erent des cartes o\`u sont dessin\'es des chemins qui ne peuvent que former des \textbf{boucles}. Les \label{coco3}corr\'elateurs de la matrice $M$ fournissent une base des observables uniformes :
\beq
W_n(x_1,\ldots,x_n) = \Big\langle \prod_{j = 1}^n \Tr \frac{1}{x_j - M}\Big\rangle_c  = \sum_{g \geq 0} \Big(\frac{N}{t}\Big)^{2 - 2g - n}\,W_n^{(g)}(x_1,\ldots,x_n) \nn
\eeq
$W_n^{(g)}(x_1,\ldots,x_n)$ \'enum\`ere les cartes $\mathcal{G}^*$ de genre $g$, avec $n$ faces marqu\'ees, et si l'on note $p_j$ est le p\'erim\`etre de la face $j$, elles sont compt\'ees avec le poids :
\beq
\label{eq:pods}\widehat{w}(\mathcal{G}^*) = \prod_{j = 1}^n \frac{1}{x_j^{p_j + 1}}\,\prod_{j = 3}^{j_{\mathrm{max}}} t_j^{n_2(j)}\prod_{\alpha = 1}^{\n} \z_\alpha^{\ell_\alpha}\,\prod_{k = 1}^{k_{\mathrm{max}}}\prod_{r = 1}^{\lfloor k/2 \rfloor - 1} g_{\alpha;k,r}^{n_2(\alpha;k,r)}
\eeq
En revanche, sur les cartes g\'en\'er\'ees par \label{nonu}des \textbf{observables \`a bords non uniformes}, les chemins de couleur $\alpha$ ont la possibilit\'e de former des boucles, ou de relier deux ar\^{e}tes de couleur $\alpha$ appartenant \`a un bord. On dira que ces ar\^{e}tes de couleur $\alpha$ sur le bord portent une \textbf{m\`eche} de couleur $\alpha$. Les m\`eches sont reli\'ees par un chemin trac\'e sur la carte, qui d\'efinit donc un appariement. On cherchera en Annexe~\ref{app:bornu} \`a calculer les observables o\`u les m\`eches sont en nombre fini, sont appari\'ees suivant un motif fix\'e, et o\`u les cartes sont pond\'er\'ees selon l'\'{E}qn.~\ref{eq:pods}, o\`u $x_j$ est coupl\'e \`a la distance $p_j$ entre deux m\`eches le long d'un m\^{e}me bord.

\begin{figure}[h]
\begin{center}
\includegraphics[width = \textwidth]{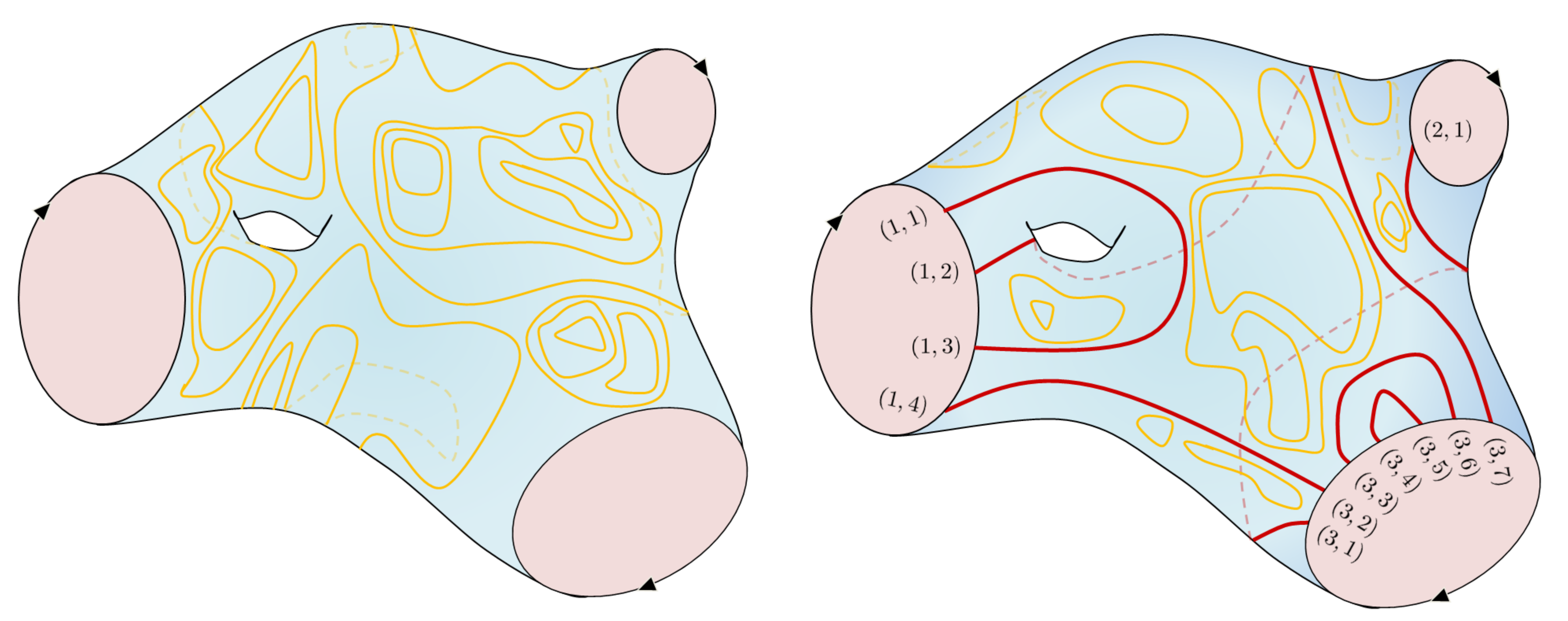}
\caption{\label{fig:Obson} \`{A} gauche : observable uniforme. \`{A} droite : observable \`a bords non uniformes, o\`u les m\`eches sont num\'erot\'ees cycliquement.}
\end{center}
\end{figure}

Le mod\`ele \'{E}qn.~\ref{eq:ms} est \'egalement int\'eressant lorsque certains $A_\alpha$ jouent le m\^{e}me r\^{o}le. Imaginons que $\mathcal{I}$ soit partitionn\'e en ensembles $\mathcal{I}_1,\ldots,\mathcal{I}_D$, tels que $\dd\nu$ soit sym\'etrique par \'echanges des $A_\alpha$ pour $\alpha$ \`a l'int\'erieur de $\mathcal{I}_d$. On peut donc d\'efinir des cartes $\mathcal{G}$ qui ne diff\'erencient plus les boucles de couleur $\alpha \in \mathcal{I}_d$. Les observables sont des s\'eries g\'en\'eratrices de ces cartes, compt\'ees avec un poids \'{E}qn.~\ref{eq:pods} multipli\'e par $\prod_{d} \n_d^{b(d)}$, o\`u $b(d)$ est le nombre de boucles de type $\mathcal{I}_d$, et $\n_d = |\mathcal{I}_d|$. Le probl\`eme d'\'enum\'eration de cartes avec des boucles de type $d = 1,\ldots,D$ et un poids $\n_d$ par boucle de type $d$, a un sens toute valeur complexe finie du param\`etre $\n_d$. Il est commun\'ement appel\'e \textbf{mod\`ele} $\mathcal{O}(\mathfrak{n}_1)\times\cdots\mathcal{O}(\mathfrak{n}_D)$. Lorsque les $\n_d$ sont des entiers positifs, il se repr\'esente par une int\'egrale formelle sur $1 + \sum_{d} \n_d$ matrices. Sinon, les observables de ce mod\`ele co\"{i}ncident avec le prolongement analytique des observables du mod\`ele de matrices. En particulier, toute \'equation valable dans le mod\`ele de matrices et analytique en $\n_d$ sera automatiquement valable pour le probl\`eme combinatoire o\`u $\n_d$ a une valeur complexe quelconque.

D'apr\`es \S~\ref{sec:anaan}, pour $t$ fix\'e tel que $|t| < t^*$, toutes ces observables sont des fonctions holomorphes de chaque variable $x_j$ au moins dans $\mathbb{C}\setminus\mathcal{D}(0;R(t))$, o\`u $R(t)$ est assez grand, et nous pouvons prendre $R(t) \rightarrow 0$ lorsque $t \rightarrow 0$.

\subsection{Int\'egration sur les matrices $A_\alpha$}
\label{sec:AAAS}
Il est facile d'int\'egrer sur les matrices $A_{\alpha}$ car leur poids est gaussien. C'est tr\`es simple pour les observables uniformes $\big\langle \mathcal{P}(M) \big\rangle$, le r\'esultat s'\'ecrit $\big\langle\!\!\big\langle \mathcal{Q}(M)\big\rangle\!\!\big\rangle$, o\`u $\big\langle\!\!\big\langle\cdots\big\rangle\!\!\big\rangle$ est la valeur moyenne par rapport \`a la mesure induite :
\beq
\dd\nu(M) = \Big[\prod_{\alpha = 1}^{\n}\frac{2\pi\,t\z_\alpha}{N}\Big]^{N^2/2}\,\frac{\dd M\,e^{-\frac{N}{t}\Tr V(M)}}{\prod_{\alpha = 1}^{\n}\sqrt{\mathrm{Det}\,\widehat{K}_\alpha(\mathbf{G}_M,\mathbf{D}_M)}} \nn
\eeq
$\mathbf{G}_M$ ou $\mathbf{D}_M$ sont les op\'erateurs de multiplication \`a gauche ou \`a droite par $M$, qui agissent sur l'espace des matrices hermitiennes. $\widehat{K}_\alpha(\mathbf{G}_M,\mathbf{D}_M)$ est donn\'e par l'\'{E}qn.~\ref{eq:aoao} o\`u le produit usuel $\cdot$ est remplac\'e par le produit tensoriel $\otimes$, et le terme constant par $\mathrm{Id}/\z_\alpha$.

Cette expression est plus lisible si l'on diagonalise $M = U\mathrm{diag}(\lambda_1,\ldots,\lambda_N)U^{\dagger}$ d\`es le d\'epart, et si l'on int\`egre sur $U \in \mathrm{U}(N)$ :
\bea
\label{eq:gaaa}\dd\nu(\lambda) & = & C_N\,\Big[\prod_{i = 1}^N \dd \lambda_i\,e^{-\frac{N}{t}V(\lambda_i)}\Big] \,\frac{|\Delta(\lambda)|^2}{\prod_{\alpha = 1}^{\mathfrak{n}}\prod_{i,j = 1}^N \sqrt{K_\alpha(\lambda_i,\lambda_j)}} \\
C_N & = & \Big[\prod_{\alpha = 1}^{\n} \frac{2\pi\,t\z_\alpha}{N}\Big]^{N^2/2}\,\frac{\mathrm{vol}(\mathrm{U}(N))}{N!\,(2\pi)^N} \nn
\eea
Il est l\'egitime de se demander si ces manipulations sont autoris\'ees \'etant donn\'e le caract\`ere formel du mod\`ele. Il me semble que oui. J'ai choisi de d\'efinir le mod\`ele avec une variable formelle $t$ vivant pr\`es de $0$ et des vraies variables $\z_{\alpha} > 0$ et $t_j,g_{\alpha;k,r} \in \mathbb{C}$ pour des raisons p\'edagogiques. Mais par homog\'en\'eit\'e, on peut reproduire les \'etapes du \S~\ref{sec:ezgt} si l'on d\'eclare $\z_{\alpha}, t_j,g_{\alpha;k,r}$ comme des variables formelles vivant pr\`es de $0$, et $t > 0$ comme une vraie variable. Dans ce cadre, l'\'{E}qn.~\ref{eq:gaaa} est valide, et les\label{rayy2} observables ont en fait un rayon de convergence non nul par rapport \`a chacun des param\`etres formels. Il n'y a pas d'ambig\"{u}it\'e venant d'\'eventuels z\'eros de $K_{\alpha}(\lambda_i,\lambda_j)$, car les puissances n\'egatives de $K_{\alpha} = 1/\z_{\alpha} + \cdots$ sont bien d\'efinies en tant que s\'eries formelles de $\z_{\alpha} \rightarrow 0$. De plus, dans les mod\`eles ou il y a $\n_{d}$ $A_{\alpha}$ qui jouent le m\^{e}me r\^{o}le, le prolongement analytique \`a $\n_d$ complexe est atteint en \'etudiant la mesure :
\bea
\label{eq:dss} \dd\nu & = & C_N\,\Big[\prod_{i = 1}^N \dd \lambda_i\,e^{-\frac{N}{t}V(\lambda_i)}\Big]\,|\Delta(\lambda)|^2\,\prod_{i,j = 1}^N K(\lambda_i,\lambda_j)  \\
K(\lambda,\mu) & = & \prod_{d} K_d(\lambda,\mu)^{-\n_d/2} \nn
\eea

Rappelons que\label{crti3} $W_n^{(g)}(x_1,\ldots,x_n)$ est analytique dans $\mathbb{C}\setminus\mathcal{D}(0;R(t))$. Cela implique, pour tous les calculs \`a partir $\dd\nu$, que tout se passe comme si les valeurs propres $\lambda_i$ vivaient seulement sur le segment $[-R(t),R(t)] \subseteq \mathcal{D}(0;R(t))$. Nous verrons avec le Thm.~\ref{theojj} que les \'equations de Schwinger-Dyson impliquent que $W_n^{(g)}(x_1,\ldots,x_n)$ est en fait holomorphe lorsque $x_j \in \mathbb{C}\setminus[a(t),b(t)]$, avec $[a(t),b(t)]$ un segment ind\'ependant de $g$ et $n$. Autrement dit, tout se passe comme si les valeurs propres vivaient uniquement sur $[a(t),b(t)]$. Les extr\'emit\'es $a(t)$ et $b(t)$ d\'ependent de $t$ et de tous les param\`etres du mod\`ele\footnote{Pour les param\`etres qui \'etaient formels, on demande que leur valeur reste dans le domaine d'analyticit\'e.}. Lorsque $t$ est proche de $0$,  $[a,b]$ est un petit segment autour de $\{0\}$. En augmentant $t$ ou en faisant varier les param\`etres du mod\`ele, on peut agrandir progressivement $[a(t),b(t)]$. Manifestement, il se passe quelque chose de sp\'ecial lorsque $\{|K(\lambda,\mu)|\quad (\lambda,\mu)\in [a(t),b(t)]\}$ atteint $0$ ou $\infty$ pour la premi\`ere fois. Heuristiquement, on s'attend \`a ce qu'un tel point donne lieu \`a une singularit\'e caract\'eristique des mod\`eles de boucles, i.e. qui n'est pas d\'ecrite par la gravit\'e pure. Ce point critique devrait \^{e}tre associ\'e \`a une limite de grandes cartes avec des boucles macroscopiques. Bien s\^{u}r, on peut \'egalement ajuster les coefficients de $V$ pour d\'ecrire un point multicritique qui est une comp\'etition entre le pr\'ec\'edent et la gravit\'e pure.

Les observables \`a bords non uniformes contiennent des $A_\alpha$, donc le th\'eor\`eme\label{Wick3} de Wick entre en jeu lorsque l'on veut int\'egrer sur les matrices $A_{\alpha}$. Finalement, ces observables s'\'ecrivent toujours $\big\langle\!\!\big\langle \mathcal{Q}(M) \big\rangle\!\!\big\rangle$, i.e. peuvent se d\'eduire de la seule connaissance des observables uniformes. Ce calcul demande l'introduction de nouvelles notations, il sera pr\'esent\'e dans le cas d'un nombre fini de m\`eches en Annexe~\ref{app:bornu}. \`{A} toute famille de $2L$ m\`eches dispos\'ees le long de $k$ bords, et \`a tout appariement abstrait $\pi$ entre ces m\`eches, on peut ainsi associer la s\'erie g\'en\'eratrice $H_{\pi}^{(g)}$ des cartes\label{mp} de genre $g$ o\`u est trac\'e un syst\`eme de liens auto\'evitants r\'ealisant $\sigma$.

\subsection{\'{E}quations de Schwinger-Dyson}
\label{sec:eqju}
\subsubsection{D\'erivation des \'equations}

\begin{figure}[h]
\begin{center}
\includegraphics[width = 0.95\textwidth]{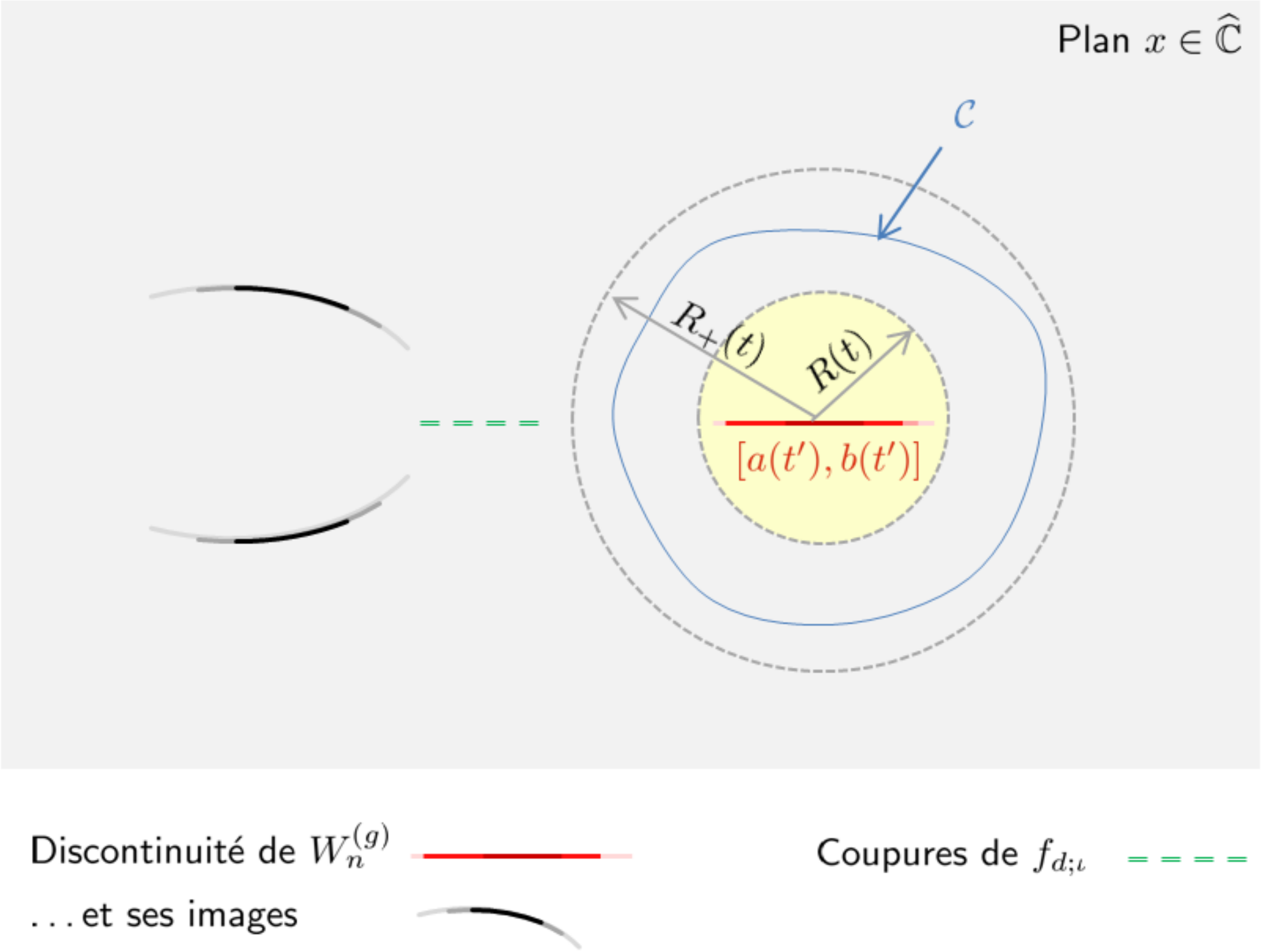}
\caption{\label{fig:lemma1cut} Pour $t$ assez petit, si $t' \in [0,t]$, $W_n^{(g)}(x,x_I)$ est une s\'erie convergente au moins pour $|x| > R(t)$. En fait, la seule singularit\'e pour $x \in \mathbb{C}$ de $W_n^{(g)}(x,x_I)$ est une discontinuit\'e sur un segment $[a(t),b(t)]$, qui est ind\'ependant de $n$ et $g$. Le lemme "une coupure" est valide jusqu'\`a un certain $t = t^*$ qui correspond : soit \`a la prise en d\'efaut de $W_1^{(0)}(x + i0) - W_1^{(0)}(x - i0) \propto \sqrt{x - a_i(t)}$ (points critiques de type "gravit\'e"), soit \`a la collision de plusieurs coupures au sens large (points critiques typiques d'un mod\`ele de boucles), soit \`a une combinaison des deux ph\'enom\`enes.}
\end{center}
\end{figure}

Les mod\`eles formels de matrices peuvent \^{e}tre \'etudi\'es \`a nouveau gr\^{a}ce aux \'equations de Schwinger-Dyson. Pour les mod\`eles de boucles, on peut se concentrer sur les observables uniformes, et \'etudier directement la mesure \'{E}qn.~\ref{eq:dss} obtenue apr\`es int\'egration des $A_{\alpha}$. L'invariance de $Z = \int \dd\nu $ sous le changement de variable infinit\'esimal $\lambda_i \rightarrow \lambda_i + \varepsilon\,\frac{1}{x - \lambda_i}$ se traduit par :
\bea
\Big\langle\!\!\Big\langle \Big(\sum_{i = 1}^N \frac{1}{x - \lambda_i}\Big)^2 - \frac{N}{t}\sum_{i = 1}^N \frac{V'(\lambda_i)}{x - \lambda_i} && \nn \\
+ \sum_{i,j = 1}^N \Big(\frac{(\partial_1 \ln K)(\lambda_i,\lambda_j)}{x - \lambda_i} + \frac{(\partial_2 \ln K)(\lambda_i,\lambda_j)}{x - \lambda_j}\Big)\Big\rangle\!\!\Big\rangle & = & 0 \nn
\eea
La deuxi\`eme ligne est propre aux mod\`eles de boucles, on pourra comparer \`a l'\'equation de Schwinger-Dyson (\'{E}qn.~\ref{eq:masterloop}) du mod\`ele \`a une matrice d\'efini en \'{E}qn.~\ref{eq:1mm}.
En \'echangeant les indices muets $i$ et $j$, les deux termes sont \'egaux. On peut \'ecrire cette \'equation en termes des corr\'elateurs :
\bea
W_2(x,x) + \big(W_1(x)\big)^2 + \frac{N}{t}\big(- V'(x)W_1(x) + P_1(x)\big) && \nn \\
 + 2\oint_{\mathcal{C}}\frac{\dd\lambda}{2i\pi}\oint_{\mathcal{C}}\frac{\dd\mu}{2i\pi}\,\frac{(\partial_1\ln K)(\lambda,\mu)}{x - \lambda}\big[W_2(\lambda,\mu) + W_1(\lambda)W_1(\mu)\big] & = & 0 \nn
\eea
$\mathcal{C}$ est un contour entourant les singularit\'es des $W_1^{(g)}$ et $W_2^{(g)}$ mais pas le point $x$. Pour le moment, nous ignorons la position de ces singularit\'es, nous savons seulement qu'elles se r\'eduisent \`a $\{0\}$ lorsque $t \rightarrow 0$. Nous supposerons dans un premier temps $t$ assez petit, afin que le rayon de convergence grossier $R(t)$ que l'on a \'etabli pour tous les $W_n^{(g)}(x_1,\ldots,x_n)$, soit tel que
 \beq
 \inf_{(\lambda,\mu) \in \mathcal{D}(0;R_+(t))} |K(\lambda,\mu)| > 0,\quad \inf_{(\lambda,\mu) \in \mathcal{D}(0;R_+(t))} |K^{-1}(\lambda,\mu)| > 0 \nn
 \eeq
 pour un $R_+(t) > R(t)$. Le choix de contour $\mathcal{C}(0;[R(t) + R_+(t)]/2)$ convient en pla\c{c}ant $x \in \mathbb{C}\setminus\mathcal{D}(0;[R(t) + R_+(t)]/2)$ (Fig.~\ref{fig:lemma1cut}).

$K(\lambda,\mu)$ est sym\'etrique\footnote{Le calcul des observables invariantes $\mathrm{U}(N)$, avec la mesure $\dd\nu$ donn\'ee par l'\'{E}qn.~\ref{eq:dss}, ne d\'epend que de la fonction $\widetilde{K}(\lambda,\nu) = \sqrt{K(\lambda,\mu)K(\mu,\lambda)}$. Si la fonction $K$ n'est pas sym\'etrique, on peut toujours la remplacer par $\widetilde{K}$.}, et dans les mod\`eles de boucles, c'est un produit de polyn\^{o}mes sym\'etriques \`a deux variables $K_d(\lambda,\mu)$ (cf. \'{E}qn.~\ref{eq:aoao}), \'eventuellement \'elev\'es \`a des puissances\footnote{$K_d(0,0) = 1/\z_d$ est r\'eel strictement positif, donc pour $t$ assez petit, $K_d(\lambda,\mu)$ reste dans $\mathbb{C}\setminus\mathbb{R}_-$ lorsque $(\lambda,\mu)$ parcourt $\mathcal{C}\times\mathcal{C}$ : il n'y a pas d'ambig\"{u}it\'e \`a \'elever $K_d$ a une puissance non enti\`ere.}  $-\n_d/2\in \mathbb{C}$. Il est commode de les factoriser :
\beq
K_d(\lambda,\mu) = e^{E(\mu)}\,c_d(\lambda)\,\prod_{\iota} (\mu - f_{d;\iota}(\lambda))^{m_{d;\iota}} \nn
\eeq
Si $K_d$ n'est pas un polyn\^{o}me, il peut admettre une infinit\'e de z\'eros, et il faut autoriser la pr\'esence d'une fonction sans z\'eros en pr\'efacteur. Lorsque $K_d$ est un polyn\^{o}me, $E$ est une constante. Par cons\'equent :
\beq
(\partial_1 \ln K)(\lambda,\mu) = \sum_{d} \frac{\n_{d}}{2}\Big(\frac{c'_d(\lambda)}{c_d(\lambda)} + \sum_{\iota} \frac{m_{d;\iota}\,f_{d;\iota}'(\lambda)}{\mu - f_{d;\iota}(\lambda)}\Big)  \nn
\eeq
En d\'epla\c{c}ant le contour $\mathcal{C}$ vers $\infty$, on collecte les r\'esidus aux p\^{o}les simples \`a $\mu = f_{d;\iota}(\lambda)$ :
\bea
\label{eq:gdr} W_2(x,x) + \big(W_1(x)\big)^2 + \frac{N}{t}\big(-\widetilde{V}'(x)W_1(x) + P_1(x)\big)\phantom{p\^{o}les} && \\
 - \sum_{d,\iota} \n_d\,m_{d;\iota} \oint_{\mathcal{C}} \frac{\dd\lambda}{2i\pi}\,\frac{f'_{d;\iota}(\lambda)}{x - \lambda}\big[W_2(\lambda,f_{d;\iota}(\lambda)) + W_1(\lambda)W_1(f_{d;\iota}(\lambda))\big] & = & 0 \nn
\eea
avec :
\bea
\widetilde{V}(x) & = & V(x) + t \sum_{d} \n_d\,\ln c_d(x) \\
P_1(x) & = & \oint_{\mathcal{C}} \frac{\dd\lambda}{2i\pi}\,\frac{\widetilde{V}'(\lambda) - \widetilde{V}'(x)}{\lambda - x}\,W_1(\lambda) \nn
\eea
Ce que l'on r\'e\'ecrit
\bea
\label{eq:hu} W_2(x,x) - \sum_{d,\iota} \n_d\,m_{d;\iota}\,f'_{d;\iota}(x)\,W_2(x,f_{d;\iota}(x)) \phantom{-\frac{N}{t}\,\widetilde{V}'(x)W_1(x)\,} && \\
+ \big(W_1(x)\big)^2 - \sum_{d,\iota} \n_d\,m_{d;\iota}\,f'_{d;\iota}(x)\,W_1(x)\,W_1(f_{d;\iota}(x)) - \frac{N}{t}\,\widetilde{V}'(x)\,W_1(x) & = & Q_1(x) \nn
\eea
$Q_1(x)$ est une fonction de $x$ qui est r\'eguli\`ere aux singularit\'es des $W_n$ :
{\small \beq
 Q_1(x) =  \frac{N}{t}\,P_1(x) + \sum_{d,\iota} \n_d\,m_{d,\iota}\oint_{\mathcal{C}}\frac{\dd\lambda}{2i\pi}\,\frac{\overline{W}_2(\lambda,f_{d;\iota}(\lambda))f'_{d;\iota}(\lambda) - \overline{W}_2(x,f_{d;\iota}(x))f'_{d;\iota}(x)}{x - \lambda} \nn
\eeq}
$\!\!\!$o\`u $\overline{W}_2(\lambda,\mu) = W_2(\lambda,\mu) + W_1(\lambda)W_1(\mu)$ est un corr\'elateur non connexe (\'{E}qn.~\ref{eq:nonon1}).

Les \'equations de Schwinger-Dyson pour $W_n$ s'en d\'eduisent, en laissant agir l'op\'erateur d'insertion (partie~\ref{sec:correlateurs}). On donne seulement le r\'esultat :
{\footnotesize \bea
&& W_{n + 1}(x,x,x_I) - \sum_{d,\iota} \n_d\,m_{d;\iota}\,f'_{d;\iota}(x)\,W_{n + 1}(x,f_{d;\iota}(x),x_I)  \nn \\
&& \sum_{J \subseteq I}\Big(W_{|J| + 1}(x,x_J)\,W_{n - |J|}(x,x_{I\setminus J}) -  \sum_{d;\iota} \n_d\,m_{d;\iota}\,f'_{d;\iota}(x)\,W_{|J| + 1}(x,x_{J})\,W_{n - |J|}(f_{d;\iota}(x),x_{I\setminus J})\Big)\nn \\
\label{eq:mastj}  && + \sum_{i \in I} \frac{W_{n - 1}(x,x_{I\setminus\{i\}})}{(x - x_i)^2} - \frac{N}{t}\,\widetilde{V}'(x)\,W_n(x) = Q_n(x,x_I)
\eea}
$\!\!\!$o\`u $Q_n(x,x_I)$ est une fonction r\'eguli\`ere lorsque $x$ s'approche des singularit\'es de $W_n(x,x_I)$, qui s'\'ecrit explicitement :
{\footnotesize \bea
Q_n(x,x_I) & = & \sum_{i \in I} \frac{\dd}{\dd x_i}\Big(\frac{W_{n - 1}(x_I)}{x - x_i}\Big) + \sum_{d,\iota} \n_{d}\,m_{d;\iota}\,\oint_{\mathcal{C}}\frac{\dd\lambda}{2i\pi}\frac{1}{x - \lambda} \nn \\
&& \left\{f'_{d;\iota}(\lambda)\Big(W_{n + 1}(\lambda,f_{d;\iota}(\lambda),x_I)  + \sum_{J \subseteq I} W_{|J| + 1}(\lambda,x_{J})\,W_{n - |J|}(f_{d;\iota}(\lambda),x_{I\setminus J})\Big)\right. \nn \\
&& \left. - f'_{d;\iota}(x)\Big(W_{n + 1}(x,f_{d;\iota}(x),x_I)  + \sum_{J \subseteq I} W_{|J| + 1}(x,x_{J})\,W_{n - |J|}(f_{d;\iota}(x),x_{I\setminus J})\Big)\right\} \nn
 \eea}

\subsubsection{Le lemme "une coupure"}
\label{leqk}
L'\'{E}qn.~\ref{eq:gdr} est une \'equation quadratique pour la s\'erie g\'en\'eratrice des cartes planaires enracin\'ees $W_1^{(0)}$ :
\beq
\big(W_1^{(0)}(x)\big)^2 - W_1^{(0)}(x)\Big[\widetilde{V}'(x) + \sum_{d,\iota} \n_d\,m_{d;\iota}\,f'_{d;\iota}(x)\,W_1^{(0)}(f_{d;\iota}(x))\Big] = Q_1^{(0)}(x) \nn
\eeq
On reconnait lorsque $\n_d \equiv 0$ l'\'equation quadratique de Tutte pour les cartes sans boucles \cite{TutteQ}.  Nous recherchons la solution qui se comporte en $1/x$ quand $x \rightarrow \infty$ :
\bea
W_1^{(0)}(x) & = & \frac{1}{2}\Big(B(x) - \sqrt{\Delta(x)}\Big) \nn \\
B(x) & = & \widetilde{V}'(x) + \sum_{d,\iota} \n_d\,m_{d;\iota}\,f'_{d;\iota}(x)W_{1}(f_{d;\iota}(x)),\qquad \Delta(x) = B^2(x) + Q_1^{(0)}(x) \nn
\eea
Comme $B(x)$ et $\Delta(x)$ sont r\'eguliers dans $\mathcal{D}(0;R(t))$, la seule singularit\'e de $W_1^{(0)}(x)$ dans ce disque est une discontinuit\'e sur le lieu de $\Delta(x) \in \mathbb{R}_-$.
\`{A} $t= 0$, $W_1^{(0)} \equiv 0$ et $\Delta(x) = V'(x)^2$ poss\`ede un z\'ero double en $x = 0$. Par analyticit\'e, pour $t > 0$ assez petit, $\Delta(x) = V'(x)^2 + O(t)$, donc ce z\'ero double se scinde en deux z\'eros simples $a(t)$ et $b(t)$, qui ont un d\'eveloppement en s\'erie formelle de $\sqrt{t}$, et tels que $a(t) + b(t) = O(t)$. Ils sont n\'ecessairement r\'eels, puisque les corr\'elateurs sont holomorphes sur $\mathbb{C}$, priv\'e de l'axe r\'eel $\mathbb{R}$ o\`u vivent a priori les valeurs propres $\lambda_i$.

Pour tous les autres $W_n^{(g)}$, les \'equations de Schwinger-Dyson (\'{E}qn.~\ref{eq:mastj}) sont lin\'eaires en $W_n^{(g)}(x)$. $W_n^{(g)}(f_{d;\iota}(x))$ intervient aussi, mais on sait d\'ej\`a que cette quantit\'e est holomorphe pour $x \in \mathcal{D}(0;R(t))$, et on peut la traiter comme une inconnue qui ne contribuera pas aux singularit\'es de $W_n^{(g)}$. Finalement, $W_n^{(g)}$ ne d\'eveloppe pas d'autres singularit\'es que celles initialement pr\'esentes dans $W_1^{(0)}$ :

\begin{theo} \label{theojj} (Lemme "une coupure"). Pour $t > 0$ assez petit, il existe $a(t)$ et $b(t)$, qui sont des s\'eries formelles en $\sqrt{t}$, telles que $a(t) + b(t) \in O(t)$, de sorte que $W_n^{(g)}(x_1,\ldots,x_n)$ soit holomorphe pour $x_i \in\mathbb{C}\setminus[a(t),b(t)]$, et ait une discontinuit\'e pour $x_i$ de part et d'autre de $[a(t),b(t)]$. De plus :
\bea
W_1^{(0)}(x + i0) - W_1^{(0)}(x - i0) & \mathop{\propto}_{x \rightarrow a(t)^+} & \sqrt{x - a(t)} \nn \\
 & \mathop{\propto}_{x \rightarrow b(t)^-} & \sqrt{x - b(t)} \nn
\eea
\end{theo}

C'est la propri\'et\'e analytique cl\'e, qui distingue la solution combinatoire qui nous int\'eresse parmi l'ensemble des solutions des \'equations de Schwinger-Dyson.

\subsubsection{R\'eapparition des \'{e}quations de boucles}
\label{heqb3}
En calculant la discontinuit\'e des \'equations de Schwinger-Dyson sur $[a(t),b(t)]$, on arrive par r\'ecurrence \`a montrer, pour tout $n,g$, $x_1 \in [a(t),b(t)]$, et $x_I \in \big(\mathbb{C}\setminus[a(t),b(t)]\big)^{n - 1}$ :
\bea
&& W_n^{(g)}(x_1 + i0,x_I) + W_n^{(g)}(x_1 - i0,x_I) - \sum_{d,\iota} \n_d\,m_{d;\iota}\,f'_{d;\iota}(x_1)\,W_n^{(g)}(f_{d;\iota}(x_1),x_I) \nn \\
\label{eq:pmpm} & = & \delta_{g,0}\Big(\delta_{n,1}\widetilde{V}'(x_1) - \frac{\delta_{n,2}}{(x_1 - x_2)^2}\Big)
\eea
Les objets naturels sont les formes diff\'erentielles :
\beq
\mathcal{W}_n^{(g)}(x_1,\ldots,x_n) = W_n^{(g)}(x_1,\ldots,x_n)\dd x_1\cdots\dd x_n \nn
\eeq
qui absorbent les facteurs $f_{d;\iota}'(x)$ dans l'\'{E}qn.~\ref{eq:pmpm} :
\bea
&& \mathcal{W}_n^{(g)}(x_1 + i0,x_I) + \mathcal{W}_n^{(g)}(x_1 - i0,x_I) - \sum_{d,\iota} \n_d\,m_{d;\iota}\,f'_{d;\iota}(x_1)\,W_n^{(g)}(f_{d;\iota}(x_1),x_I) \nn \\
& = & \delta_{g,0}\Big(\delta_{n,1}\dd \widetilde{V}(x_1) - \delta_{n,2}\frac{\dd x_1\dd x_2}{(x_1 - x_2)^2}\Big) \nn
\eea
On obtient une \'equation lin\'eaire \`a coefficients constants, qui ressemble beaucoup \`a la hi\'erarchie d'\'equations de boucles lin\'eaires (\'{E}qn.~\ref{eq:VVVV1}). De m\^{e}me, les \'equations de Schwinger-Dyson sous la forme \'{E}qn.~\ref{eq:hu} et \ref{eq:mastj}, \'ecrites pour $\mathcal{W}_n^{(g)}$, ressemblent fortement \`a la hi\'erarchie d'\'equations de boucles quadratiques (\'{E}qn.~\ref{eq:VVVV2}). Ceci fait \'echo \`a la g\'en\'eralisation de la notion d'\'equation de boucles\footnote{Historiquement, les \'equations de \underline{boucles} ont re\c{c}u ce nom par r\'ef\'erence au bord des surfaces discr\`etes que g\'en\`ere $\big\langle\Tr \underbrace{M\cdots M}_{p\,\mathrm{fois}}\big\rangle$, qui est une \underline{boucle} de longueur $p$. Cela est \'etranger aux \textit{boucles} al\'eatoires auto\'evitantes qui sont dessin\'ees sur les surfaces discr\`etes g\'en\'er\'ees par les mod\`eles de \textit{boucles}.} \'evoqu\'ee au \S~\ref{sec:defsq}.

Ces progr\`es ont \'et\'e faits peu avant l'\'ecriture, et le projet n'est pas encore achev\'e. La difficult\'e est de d\'eterminer $W_1^{(0)}$ et $W_2^{(0)}$ satisfaisant l'\'equation lin\'eaire~\ref{eq:pmpm} pour des $K(\lambda,\mu)$ les plus g\'en\'eraux, i.e. de d\'eterminer la courbe spectrale. Cela passe par la r\'esolution de l'\'equation lin\'eaire homog\`ene :
\beq
\label{eq:equalin}\forall x \in \gamma^{(0)},\qquad \phi(x + i0) + \phi(x - i0) - \sum_{d,\iota} \n_d\,f'_{d;\iota}(x)\,\phi(x) = 0
\eeq
pour une fonction $\phi$ dont on connait les propri\'et\'es analytiques sur $\mathbb{C}\setminus\gamma^{(0)}$. Pour la combinatoire, $\gamma^{(0)}$ est un segment $[a,b]$ de l'axe r\'eel, mais on peut chercher d'autres types de solutions de l'\'{E}qn.~\ref{eq:equalin}, o\`u $\gamma$ est une collection d'arcs du plan complexe. Dans un premier temps, on supposera que la \textbf{vraie coupure} $\gamma$ ne \label{ohu} rencontre ni ses \textbf{coupures images} $f_{d;\iota}(\gamma)$, ni les discontinuit\'es de $f_{d;\iota}$ que l'on peut appeler \textbf{coupures fant\^{o}mes}. Un point critique sera probablement atteint lorsque les coupures (au sens large) commencent \`a se rencontrer, autrement dit lorsque la courbe spectrale devient singuli\`ere.

Si cette \'etape est franchie, avec une notion sous-jacente de g\'eom\'etrie complexe et de feuillets, je \label{resss}pense qu'il sera ais\'e de montrer que tous les autres $\mathcal{W}_n^{(g)}$ sont donn\'es par une formule des r\'esidus, avec un formalisme de r\'ecurrence topologique (Chapitre~\ref{chap:toporec}), et toutes les propri\'et\'es qui y sont associ\'ees.

\section{Le mod\`ele $\On$ trivalent}
\label{sec:OnON}
\subsection{Pr\'esentation}

Ce programme a \'et\'e compl\'et\'e dans mon article \cite{BEOn} pour le mod\`ele $\On$ trivalent introduit par Kostov \cite{KosOn} :
\beq
\label{eq:Onmat}\dd\nu = \dd M \dd A_1 \cdots \dd A_{\mathfrak{n}}\,\exp\Big\{-\frac{N}{t}\Tr\Big(V(M) + \sum_{\alpha = 1}^{\n} \frac{1}{\z}A_{\alpha}^2  - M A_\alpha^2\Big)\Big\}
\eeq
ou bien, apr\`es int\'egration des $A_{\alpha}$ :
\beq
\dd\nu = C_N\,\Big[\prod_{i = 1}^N \dd \lambda_i\,e^{-\frac{N}{t}V(\lambda_i)}\Big]\,\frac{|\Delta(\lambda)|^2}{\prod_{i,j = 1}^N \big(1/\z - (\lambda_i + \lambda_j)\big)^{\n/2}} \nn
\eeq
Il engendre des cartes constitu\'ees de $k$gones ($k \geq 3$), o\`u les chemins vivent seulement sur des triangles, et chaque boucle est compt\'ee avec un poids $\n$ (Fig.~\ref{fig:carteon}). Le mod\`ele combinatoire est parfaitement d\'efini pour toute valeur de $\n \in \mathbb{C}$. Il est bon garder \`a l'esprit les mod\`eles populaires en physique statistique qui lui sont reli\'es :

\begin{figure}[h]
\begin{center}
\includegraphics[width = \textwidth]{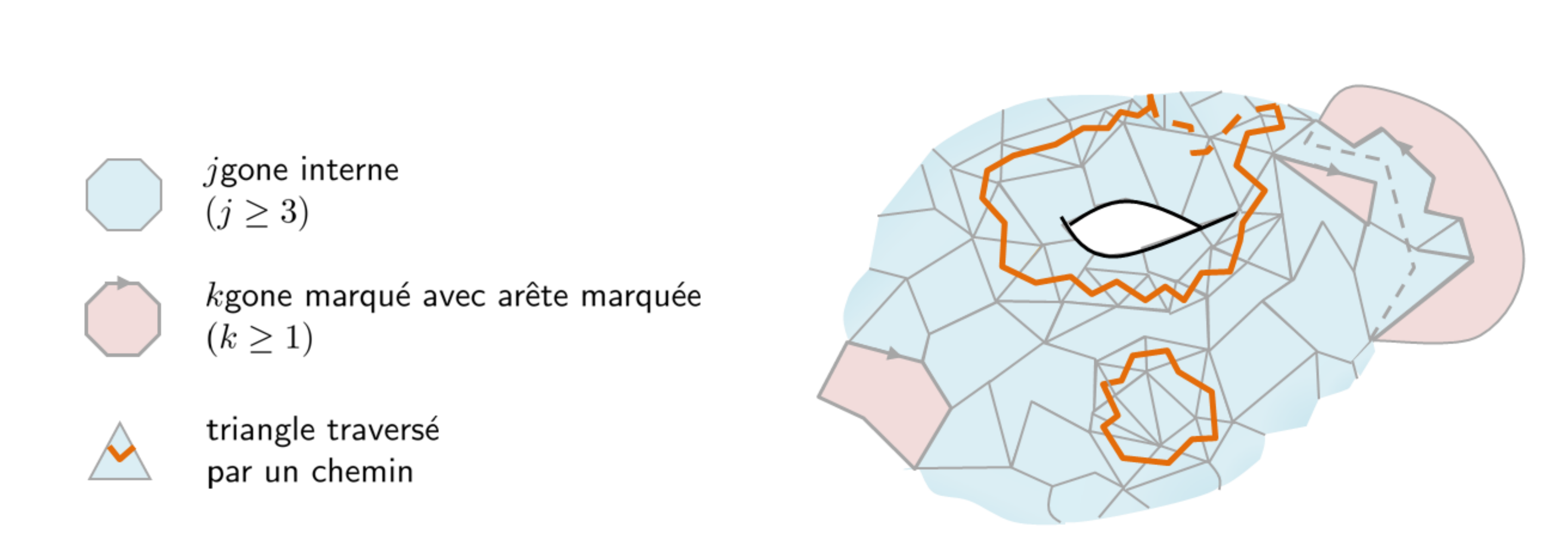}
\caption{\label{fig:carteon} Les briques pour les cartes du mod\`ele $\On$, et un exemple de carte de genre $g = 1$, avec $n = 3$ faces marqu\'ees.}
\end{center}
\end{figure}

\vspace{0.2cm}

\begin{itemize}
\item[$\diamond$] Le mod\`ele $\mathcal{O}(\mathfrak{1})$ trivalent avec $V(M) = \frac{M^2}{2} - \frac{t_3\,M^3}{3}$ est exactement dual \`a un mod\`ele \`a deux matrices, qui est un mod\`ele d'Ising \label{Isi2}sur triangulations al\'eatoires \cite{KazakovO1}.
\item[$\diamond$] Le mod\`ele $\On$ pour $\n \rightarrow 0$ d\'ecrit en principe la physique des polym\`eres dans un milieu al\'eatoire \cite{DuK}.
\item[$\diamond$] Le mod\`ele $\mathcal{O}(-\mathfrak{2})$ trivalent est exactement dual \`a un mod\`ele de for\^{e}ts couvrantes sur triangulations al\'eatoires. Cette relation s'\'etend \`a des cartes al\'eatoires quelconques, modulo le passage \`a un mod\`ele $\On$ avec un noyau $K(\lambda,\mu)$ quelconque (\'{E}qn.~\ref{eq:dss}) \cite{Sporti}.
\end{itemize}

\vspace{0.2cm}

Du point de vue des \'equations de boucles, le mod\`ele $\On$ trivalent est techniquement le plus simple : $K(\lambda,\mu)$ est une fonction affine, donc n'a qu'un seul z\'ero $\mu = f(\lambda)$, qui est fonction affine de $\lambda$. L'\'equation lin\'eaire :
\beq
\label{eq:gfdq0}\forall x \in [a,b],\qquad \phi(x + i0) + \phi(x - i0) + \n\,\phi(1/\z - x) = 0
\eeq
a \'et\'e r\'esolue pour $\n$ quelconque\footnote{La solution est connue depuis longtemps dans le cas $\n = -2\cos(\pi{}p/q)$, o\`u $p/q$ est une fraction irr\'eductible. Elle s'exprime uniquement en termes de racines carr\'ees et de racines $q^{\textrm{\`eme}}$. On pourra par exemple consulter \cite{EZJ1}.} dans la th\`ese de Bertrand Eynard \cite{TheseBE}, et lors de ses travaux avec Charlotte Kristjansen \cite{EKOn,EKOn2} Cela leur a permis de trouver $W_1^{(0)}(x)$, $W_2^{(0)}(x)$, et en principe les s\'eries g\'en\'eratrices pour les autres topologies gr\^{a}ce \`a la m\'ethode des moments \label{kujj}initialement d\'evelopp\'ee par Ambj{\o}rn et al. \cite{ACM92}.

Mes travaux ont consist\'e \`a reformuler cette solution dans le cadre de la g\'eom\'etrie complexe, puis \`a d\'emontrer que tous les $W_n^{(g)}$ sont bien donn\'es par une r\'ecurrence topologique. Par rapport \`a la m\'ethode des moments, l'avantage de la r\'ecurrence topologique est d'\'elucider compl\`etement le caract\`ere g\'eom\'etrique du probl\`eme. Elle donne aussi un algorithme de calcul plus facile \`a mettre en {\oe}uvre.

Pour simplifier, je vais supposer $\n \in ]-2,2[$. On peut naturellement \'etudier les cas $\n = \pm 2$ en prenant des limites $\n \rightarrow \pm 2$. Nous n'aborderons pas le cas $\n \notin [-2,2]$, o\`u la nature des solutions de l'\'{E}qn.~\ref{eq:gfdq0} est un peu modifi\'ee \cite{EKOn2}.

\subsubsection{Interpr\'etation combinatoire des \'equations de Schwinger-Dyson}
\label{SDD2}
Les \'equations de Schwinger-Dyson traduisent des relations entre s\'eries g\'en\'eratrices de cartes. En g\'en\'eral, elles sont \'equivalentes \`a des bijections que l'on peut construire en \^{o}tant une ar\^{e}te marqu\'ee dans une carte. Cependant, en manipulant les \'equations de Schwinger-Dyson, on peut arriver \`a des \'equations dont la signification bijective n'est pas tr\`es claire. C'est le cas de l'\'{E}qn.~\ref{eq:gdr}, obtenue apr\`es avoir d\'eplac\'e des contours dans le plan complexe. Le mod\`ele $\On$ est \`a la fois assez simple et \'elabor\'e pour illustrer comment on peut parvenir, par une approche bijective, \`a :
\begin{footnotesize}
\bea
\label{eq:EQN3} W_2^{(g - 1)}(x_1,x_1) + \n\,W_2^{(g - 1)}(x_1,1/\z - x_1) + W_2^{(g - 1)}(1/\z - x_1,1/\z - x_1) && \\
+ \Big[\sum_{h = 0}^{g} W_1^{(h)}(x_1)\,W_1^{(g - h)}(x_1) + \mathfrak{n}\,W_1^{(h)}(x_1)\,W_1^{(g - h)}(1/\z - x_1) && \nn \\
+ W_1^{(h)}(1/\z - x_1)\,W_1^{(g - h)}(1/\z - x_1)\Big] -\big[V'\,W_1^{(g)}\big]_{-}(x_1) - \big[V'\,W_1^{(g)}\big]_{-}(1/\z - x_1) & = & 0 \nn
\eea
\end{footnotesize}
$\!\!\!$qui est le d\'eveloppement topologique de l'\'{E}qn.~\ref{eq:hu}, o\`u l'on identifie :
\begin{footnotesize}
\bea
Q_1^{(g)}(x_1) & = & -W_2^{(g - 1)}(1/\z - x_1,1/\z - x_1) - \sum_{h = 0}^{(g)} W_1^{(h)}(1/\z - x_1)\,W_1^{(g - h)}(1/\z - x_1) \nn \\
&& - \big[V'\,W_1^{(g)}\big]_{+}(x_1) + \big[V'\,W_1^{(g)}\big]_{-}(1/\z - x_1) \nn
\eea
\end{footnotesize}

La m\'ethode consiste toujours \`a \'etablir une relation de r\'ecurrence entre s\'eries g\'en\'eratrices de cartes, en d\'ecrivant les cartes obtenues en \^{o}tant l'ar\^{e}te marqu\'ee d'une face marqu\'ee.

\begin{figure}[h!]
\begin{center}
\includegraphics[width=1.15\textwidth]{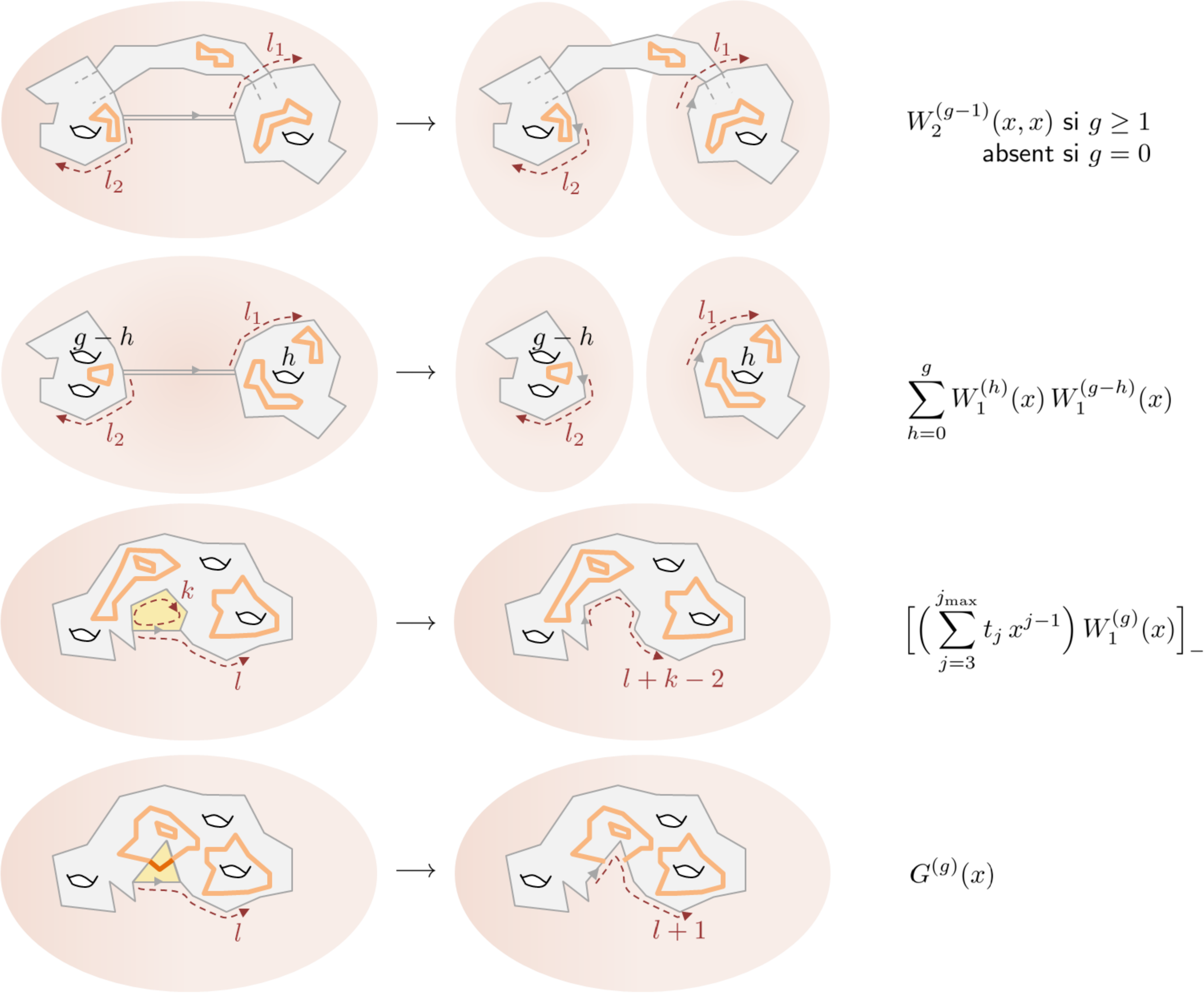}
\caption{\label{fig:recOn1} La s\'erie g\'en\'eratrice des cartes enracin\'ees, non r\'eduites \`a un point, de genre $g$, auxquelles on a \^{o}t\'e l'ar\^{e}te marqu\'ee de la face marqu\'ee, est $x\,W_1^{(g)}(x) - t$. Mais c'est aussi $W_2^{(g - 1)}(x,x)  + \Big(\sum_{h = 0}^{g} W_1^{(h)}(x)\,W_1^{(g - h)}(x)\Big) + \Big[\big(x - V'(x)\big)\,W_1^{(g)}(x)\Big]_{-} + \mathfrak{n}\,G^{(g)}(x)$. Le facteur $\mathfrak{n}$ indique le nombre de couleurs possibles pour la boucle rencontr\'ee dans le quatri\`eme cas. $\big[\cdots\big]_{-}$ d\'esigne la partie du d\'eveloppement en s\'erie de Laurent ne contenant que des puissances strictement n\'egatives de $x$. On rappelle que $V'(x) = x - \sum_{j \geq 3} t_j x^{j - 1}$ (cf. \'{E}qn.~\ref{eq:defV}). Alors, le d\'ecalage $\big[x\,W_1^{(g)}(x)\big]_{-} = x\,W_1^{(g)}(x) - \delta_{g,0}\,t$ compense exactement le membre de gauche, d'o\`u l'\'{E}qn.~\ref{eq:EQN1}.}
\end{center}
\end{figure}

Pour le mod\`ele $\On$, ce proc\'ed\'e est appliqu\'e dans un premier temps aux cartes enracin\'ees de genre $g$ \`a la Fig.~\ref{fig:recOn1}. Il fait apparaitre des cartes avec deux m\`eches\label{msu2} adjacentes, dont la s\'erie g\'en\'eratrice est :
\beq
G^{(g)}(x) = \Big\langle \Tr\Big(\frac{1}{x - M}\,A_{1}^2\Big)\Big\rangle^{(g)}_c \nn
\eeq
Le r\'esultat est :
\bea
\label{eq:EQN1} W_2^{(g - 1)}(x,x) + \sum_{h = 0}^{g} W_1^{(h)}(x)\,W_1^{(g - h)}(x) &&  \\
- \big[V'(x)\,W_1^{(g)}(x)\big]_{-} + \n\,G^{(g)}(x) & = & 0 \nn
\eea

\begin{figure}[h!]
\begin{center}
\includegraphics[width=1.15\textwidth]{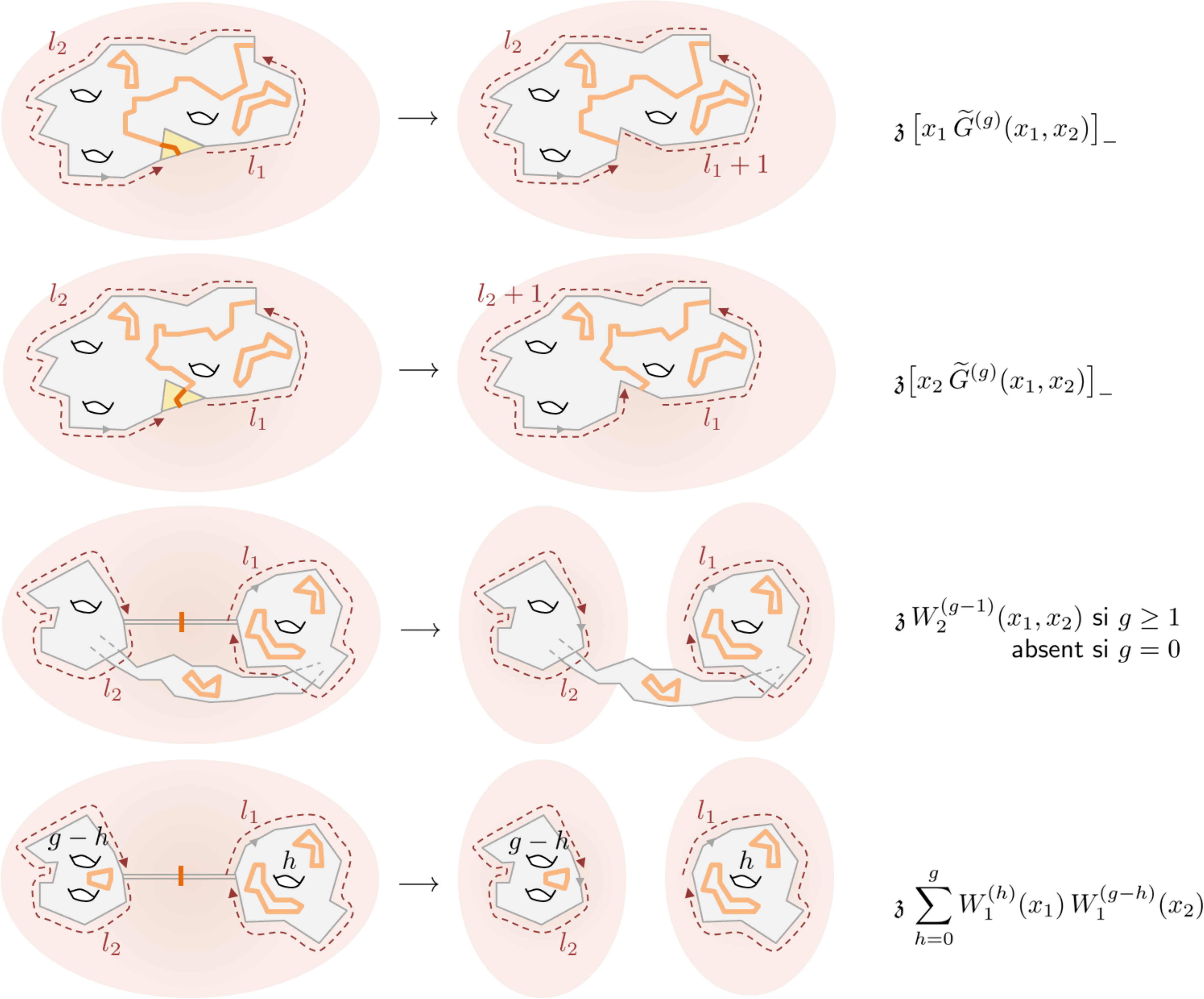}
\caption{\label{fig:recOn2} La s\'erie g\'en\'eratrice des cartes de genre $g$, avec une face marqu\'ee et deux m\`eches de couleur fix\'ee, est $\widetilde{G}^{(g)}(x_1,x_2)$. En \^{o}tant l'ar\^{e}te o\`u se trouve la m\`eche de d\'epart, on obtient une bijection entre cartes qui implique une \'egalit\'e de s\'eries g\'en\'eratrices $\widetilde{G}^{(g)}(x_1,x_2) = \z \big[(x_1 + x_2)\widetilde{G}(x_1,x_2)\big]_{-} + \z\,W_2^{(g - 1)}(x_1,x_2) + \z\,\sum_{h = 0}^{g} W_1^{(h)}(x_1)\,W_1^{(g - h)}(x_2)$. Il faut remarquer $\big[x\widetilde{G}^{(g)}(x,x_i)\big]_{-} = x\widetilde{G}^{(g)}(x,x_i) - G(x_i)$ pour arriver \`a l'\'{E}qn.~\ref{eq:EQN2}.}
\end{center}
\end{figure}

Il faut trouver une autre relation pour \'eliminer $G(x)$. Le m\^{e}me proc\'ed\'e est appliqu\'e \`a cette fin aux cartes enracin\'ees de genre $g$, dont la face marqu\'ee poss\`ede deux m\`eches, \`a la Fig.~\ref{fig:recOn2}. Leur s\'erie g\'en\'eratrice d\'ependant de deux variables, $x_1$ et $x_2$, coupl\'ees aux distances s\'eparant les deux m\`eches le long de la face marqu\'ee :
\beq
\widetilde{G}^{(g)}(x_1,x_2) = \Big\langle \Tr\Big(\frac{1}{x_1 - M}\,A_1\,\frac{1}{x_2 - M}\,A_1\Big)\Big\rangle^{(g)}_c \nn
\eeq
Le r\'esultat est :
\begin{footnotesize}
\beq
\label{eq:EQN2} W_2^{(g - 1)}(x_1,x_2) + \sum_{h = 0}^{g} W_1^{(h)}(x_1)\,W_1^{(g - h)}(x_2) + (x_1 + x_2 - 1/\z)\widetilde{G}(x_1,x_2) = G(x_1) + G(x_2)
\eeq
\end{footnotesize}
$\!\!\!$En sp\'ecialisant $x_2 = 1/\z - x_1$ dans l'\'{E}qn.~\ref{eq:EQN2}, on obtient $G(x_1) + G(1/\z - x_1)$ en termes d'observables uniformes uniquement. On en d\'eduit l'\'{E}qn.~\ref{eq:EQN3} en additionnant l'\'{E}qn.~\ref{eq:EQN1} pour les valeurs $x = x_1$ et $x = 1/\z - x_1$. Ces changements de variables sont autoris\'es car ils pr\'eservent le caract\`ere de s\'erie formelle en $1/x$ et en $\z$ des observables $G^{(g)}$ et $W_n^{(h)}$.

\hspace{-0.5cm}

\hspace{-0.5cm}

Comme l'interpr\'etation combinatoire de l'op\'erateur d'insertion $\partial/\partial V(x_i)$ est limpide (on consid\`ere des cartes ayant une face marqu\'ee suppl\'ementaire, de p\'erim\`etre coupl\'e \`a la variable $x_i$), cela donne une preuve combinatoire compl\`ete des \'equations de Schwinger-Dyson du mod\`ele $\On$.

\subsection{S\'eries g\'en\'eratrices des topologies instables}
\label{nonu2}
\subsubsection{R\'esolution de l'\'equation lin\'eaire}

\begin{figure}[h!]
\begin{center}
\includegraphics[width=0.9\textwidth]{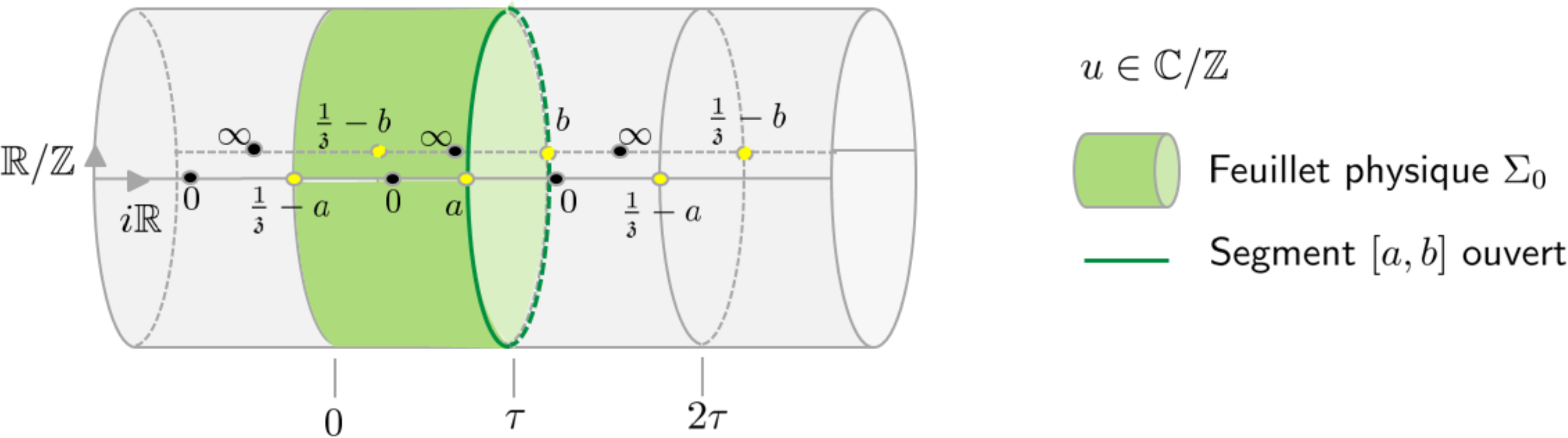}
\caption{\label{fig:CourbeS} Le plan complexe $u \in \mathbb{C}$, avec identifications $u \sim u + m$ ($m \in \mathbb{Z}$). On a indiqu\'e les valeurs de $x$ pour les points remarquables. La sym\'etrie $x \mapsto 1/\z - x$ du param\'etrage \'{E}qn.~\ref{eq:parmu} assure que $\tau$ est imaginaire pur. De plus, $\tau$ satisfait la contrainte de positivit\'e $\mathrm{Im}\,\tau > 0$.}
\end{center}
\end{figure}

$W_1^{(0)}$ et $W_2^{(0)}$ satisfont, pour $x \in [a,b]$ :
\bea
\label{eq:apq}W_1^{(0)}(x + i0) + W_1^{(0)}(x - i0) + \n W_1^{(0)}(1/\z - x) & = & V'(x)  \\
 W_2^{(0)}(x_1 + i0,x_2) + W_2^{(0)}(x_1 - i0,x_2) + \n\,W_2^{(0)}(1/\z - x_1,x_2) & = & -\frac{1}{(x_1 - x_2)^2} \nn \\
\label{eq:apq2} & &
\eea
Ces \'equations sont lin\'eaires. Comme le second membre est une fonction r\'eguli\`ere sur $[a,b]$, elles ont une solution particuli\`ere facile :
\bea
\label{eq:apoix}(W_1^{(0)})_{\mathrm{P}}(x) & = & \frac{2 V'(x) - \n V'(1/\z -x)}{4 - \n^2} \\
\label{eq:apoix2}(W_2^{(0)})_{\mathrm{P}}(x_1,x_2) & = & \frac{1}{4 - \n^2}\Big(\frac{-2}{(x_1 - x_2)^2} + \frac{\n}{(x_1  + x_2 - 1/\z)^2}\Big)
\eea
Les quantit\'es $(W_n^{(0)})_{\mathrm{H}} = W_n^{(0)} - (W_n^{(0)})_{\mathrm{P}}$ sont solutions de l'\'equation homog\`ene~\ref{eq:gfdq0}. Rappelons l'id\'ee de la r\'esolution de l'\'{E}qn.~\ref{eq:gfdq0}. Si l'on cherche \`a exprimer la solution g\'en\'erale de l'\'{E}qn.~\ref{eq:gfdq0} sans autre information, on sera naturellement conduit \`a consid\'erer des fonctions qui ont une discontinuit\'e sur $\gamma = [a,b]$ et sur son image $\gamma' = [1/\z - b,1/\z - a]$. Il est judicieux d'introduire un param\'etrage qui "ouvre" ces deux coupures \label{coucou3} :
\beq
\label{eq:parmu} u(x) = \frac{ib}{2K(k')}\int_{1/\z - a}^{x} \frac{\mathrm{d}\xi}{\sqrt{(\xi - a)(\xi + a - 1/\z)(\xi - b)(\xi + b - 1/\z)}}
\eeq
o\`u $K(k')$ est l'int\'egrale elliptique compl\`ete \'evalu\'ee en $k' = \sqrt{1 - a^2/b^2}$. $u$ n'est pas une fonction univalu\'ee. La d\'etermination principale envoie $x \in \mathbb{C}\setminus (\gamma\cup\gamma')$ sur le domaine $u \in \Sigma_0 = [-1/2,1/2[ + ]0,\tau[$, et la fonction r\'eciproque $x(u)$ peut \^{e}tre \'etendue analytiquement pour $u \in \mathbb{C}$ (Fig.~\ref{fig:CourbeS}) en une fonction qui satisfait :
\beq
x(u) = x(u + 1),\qquad x(u) = x(u + 2\tau),\qquad x(u) = -x(u), \qquad x(u) = x(u + 1) \nn
\eeq
$\tau$ est un nombre complexe de partie imaginaire strictement positive, c'est ici le \textbf{module} de la surface de Riemann :
\beq
\mathcal{E} =\big\{(s,x)\in \widehat{\mathbb{C}}^2\quad s^2 = (x - a)(x + a - 1/\z)(x - b)(x + b - 1/\z)\big\} \simeq \mathbb{C}/(\mathbb{Z}\oplus\tau\mathbb{Z}) \nn
\eeq
Le domaine\label{fph2} $\Sigma_0$ est appel\'e \textbf{feuillet physique}. L'op\'eration miroir $x \mapsto 1/\z - x$ devient $u \mapsto \tau - u$ dans le feuillet physique avec le nouveau param\'etrage.

Les solutions de l'\'{E}qn.~\ref{eq:gfdq0} qui sont des fonctions analytiques sur $\mathbb{C}\setminus [a,b]$ s'\'ecrivent alors $\phi(x) = \Phi(u(x))\,u'(x)$, avec $\Phi$ une fonction qui se prolonge analytiquement sur $\mathbb{C}$ et v\'erifie :
\bea
\label{eq:phu}&& \left\{\begin{array}{l} \Phi(u + 1) = \Phi(u) \\ \Phi(u - 2\tau) - \n\,\Phi(u - \tau) + \Phi(u) = 0 \end{array}\right.  \\
\label{eq:phyp}&& \,\,\,\Phi(u) = -\Phi(-u)
\eea
Ces \'equations nous disent que $\Phi$ n'est pas univalu\'ee sur la surface de Riemann $\mathcal{E}$, car elle admet une monodromie non triviale autour de $\gamma = [a,b] \subseteq \mathcal{E}$.
La contrainte d'imparit\'e peut \^{e}tre impos\'ee a posteriori. Si l'on se concentre sur l'\'{E}qn.~\ref{eq:phu}, l'ensemble des solutions $\mathrm{E}$ est un espace vectoriel de dimension $2$ sur le corps des \textbf{fonctions elliptiques}, i.e. p\'eriodiques de p\'eriode $1$ et $\tau$ et m\'eromorphes sur $\mathbb{C}$. Toute solution s'\'ecrit comme combinaison lin\'eaire (\`a coefficients "fonctions elliptiques") de fonctions prenant des phases $e^{\pm i\pi\mathfrak{b}}$ \`a chaque translation par $\tau$
\beq
\Phi \in \mathrm{E}_{\pm} \quad \Leftrightarrow \quad \Phi(u + 1) = \Phi(u),\qquad \Phi(u + \tau) = e^{\pm i \pi \mathfrak{b}} \Phi(u) \nn
\eeq
avec $\n = 2\cos(\pi\mathfrak{b})$. Tout d'abord, il existe de telles fonctions. Il suffit d'en \label{the}construire une, dans $\mathrm{E}_+$, et cela peut \^{e}tre fait avec un ratio de fonctions th\^{e}ta ou une s\'erie infinie\footnote{Comme la s\'erie ne converge pas absolument, il faut pr\'eciser que $\sum_{m \in \mathbb{Z}}$ est une notation pour $\Big(\sum_{m \geq 0} + \sum_{m \leq -1}\Big)$.} :
\bea
\zeta_{\mathfrak{b}}(u) & = &  \sum_{m \in \mathbb{Z}} e^{-i\pi m\mathfrak{b}}\,\pi\,\mathrm{cotan}\,\pi(u + m\tau) \nn \\
& = & -\frac{\vartheta_1'\big(0|\tau\big)}{\vartheta_1\big(\mathfrak{b}/2|\tau\big)}\,\frac{\vartheta_1\big(u - \mathfrak{b}/2|\tau\big)}{\vartheta_1\big(u|\tau\big)} \nn
\eea
Celle-ci est dans $\mathrm{E}_+$ et a des p\^{o}les simples lorsque $u = 0\,\,\mathrm{mod}\,\mathbb{Z}\oplus\tau\mathbb{Z}$, tels que :
\beq
\zeta_{\mathfrak{b}}(u) \mathop{\sim}_{u \rightarrow 0} \frac{1}{u} \nn
\eeq

La th\'eorie de ces fonctions\label{psu} \textbf{pseudoelliptiques} (les fonctions de $E_{\pm}$) est analogue \`a celle des fonctions elliptiques esquiss\'ee en Annexe~\ref{app:geomcx}. Il n'y a pas de fonction pseudoelliptique holomorphe, autre que la fonction constante\footnote{Cela n'est vrai que pour $\mathfrak{b} \in \mathbb{R}$, i.e. pour $\n \in [-2,2]$. Autrement dit, quand $e^{\pm i\pi\mathfrak{b}}$ est une phase. $z\mapsto e^{\pm 2i\pi z}$ est un exemple de fonction enti\`ere, $1$ p\'eriodique et prenant un facteur constant de module $\neq 1$ apr\`es translation $z \mapsto z + \tau$, puisque $\mathrm{Im}\,\tau > 0$.}. Une fonction pseudoelliptique est caract\'eris\'ee par son comportement aux p\^{o}les, ou par la position de ses p\^{o}les et de ses z\'eros. On peut identifier l'analogue des fonctions de Weierstra{\ss} $\wp$ (p\^{o}le double sans r\'esidu), $\zeta$ (p\^{o}le simple), $\sigma$ (singularit\'e logarithmique), \label{Wei2}et des fonctions $\theta$ de Jacobi. Ce sont les briques qui permettent de repr\'esenter de diverses mani\`eres toute fonction pseudoelliptique, par exemple avec des formules de Cauchy.

Pour les fonctions solutions de l'\'{E}qn.~\ref{eq:phu}-\ref{eq:phyp}, le noyau de Cauchy appropri\'e est :
\bea
\mathcal{G}(u_0,u) & = & \frac{1}{2}\,\frac{1}{e^{i\pi\mathfrak{b}} - e^{-i\pi\mathfrak{b}}}\Big[-e^{-i\pi\mathfrak{b}}\zeta_{\mathfrak{b}}(u_0 + u) + e^{i\pi\mathfrak{b}}\zeta_{\mathfrak{b}}(u_0 - u)  \nn \\
&& \phantom{\frac{1}{2}\,\frac{1}{e^{i\pi\mathfrak{b}} - e^{-i\pi\mathfrak{b}}}} + e^{-i\pi\mathfrak{b}}\zeta_{\mathfrak{b}}(-u_0 + u) -  e^{i\pi\mathfrak{b}}\zeta_{\mathfrak{b}}(-u_0 - u)\Big] \nn \\
\label{eq:Chachy} &&
\eea
et la repr\'esentation d'une fonction $\Phi$ m\'eromorphe s'\'ecrit :
\beq
\label{eq:AQ}\Phi(u_0) = \sum_{\alpha} \epsilon_{\alpha}\,\Res_{u \rightarrow u(\alpha)} \dd u\,\mathcal{G}(u_0,u)\,\Phi(u)
\eeq
C'est une somme de r\'esidus aux p\^{o}les de $\Phi$ dans le feuillet physique. Lorsque $\alpha \in \{a,b,1/\z - a,1/\z - b\}$, $\epsilon_{\alpha} = 1$, et pour tous les autres p\^{o}les, $\epsilon_{\alpha} = 2$. Cette formule est correcte stricto sensu lorsque $\Phi(u)$ n'a pas de r\'esidus \`a $u = u(a),u(b),u(1/\z - a),u(1/\z - b)$. Pour d\'eterminer $\Phi$, il suffit donc de d\'eterminer son comportement aux p\^{o}les dans le feuillet physique.

La m\'ethode d'Eynard et Kristjansen donne des r\'esultats "plus explicites" dans la variable $x$. Deux remarques permettent de faire le lien avec leurs expressions. Tout d'abord, si $f$ est une fonction pseudoelliptique, alors $f'(u)/f(u)$ est une fonction elliptique. Ensuite, il est possible de montrer que $h$ est une fonction elliptique de p\'eriode $1$ et $\tau$ ssi $H(x(u)) = h(u)$ s'\'ecrit :
\beq
\label{eq:elli}H(x) = A(x^2) + x\,s(x)\,B(x^2)
\eeq
o\`u $A$ et $B$ des fractions rationnelles paires, et $s(x) = \sqrt{(x - a)(x - b)(x + a - 1/\z)(x + b - 1/\z)}$. Cela permet toujours d'exprimer $\ln f(u(x))$ comme une int\'egrale elliptique. Dans \cite{EKOn,EKOn2}, Eynard et Kristjansen exhibent dans un premier temps une base de $\mathrm{E}_{\pm}$. Puis, ils d\'ecomposent $\Phi(u(x))$, la solution recherch\'ee de l'\'equation lin\'eaire, sur cette base, avec des coefficients qui sont des fonctions elliptiques \'ecrites selon l'\'{E}qn.~\ref{eq:elli}. Enfin, ils d\'eterminent les fractions rationnelles paires inconnues \`a partir des propri\'et\'es de $\phi(x)$ au voisinage de ses singularit\'es.

\subsubsection{Les disques : $W_1^{(0)}$}

$(W_1^{(0)})_{\mathrm{H}}$ est la solution de l'\'{E}qn.~\ref{eq:gfdq0} caract\'eris\'ee par les propri\'et\'es analytiques :
\bea
(W_1^{(0)})_{\mathrm{H}}(x) = -\Big(\frac{2V'(x) - \n V'(1/\z -x)}{4 - \n^2}\Big) + t/x + o(t/x)\,\, && x \rightarrow \infty \nn \\
(W_1^{(0)})_{\mathrm{H}}(x + i0) - (W_1^{(0)})_{\mathrm{H}}(x - i0) \propto \sqrt{x - a_i}\phantom{t/x + o(t/x)} \,\, && x \rightarrow a_i \in \{a,b\} \nn
\eea
Autrement dit, si l'on note $\mathcal{W}_1^{(0)}(u)$ le prolongement analytique de $(\overline{W}_1^{(0)})_{\mathrm{H}}(x(u))\,x'(u)$ \`a $u \in \mathbb{C}$, il n'a de singularit\'e dans le feuillet physique qu'en $u = u(\infty)$.

On peut appliquer directement la formule \'{E}qn.~\ref{eq:AQ}. Il est commode de param\'etrer le potentiel, en fonction de son comportement au voisinage de $\infty$ :
\bea
V'(x(u))x'(u) & = & \frac{2K(k')}{ib}\,s(x)\,\Big(x - \sum_{j = 3}^{j_{\mathrm{max}}} t_j\,x^{j - 1}\Big) \nn \\
& = & -\frac{1}{2}\sum_{j = 1}^{j_{\mathrm{max}} + 2} \frac{(2 + (-1)^j\n)\,v_j}{(u - u(\infty))^j} + O(1) \nn
\eea
En utilisant les propri\'et\'es de $\mathcal{G}(u_0,u)$, le r\'esultat est :
{\small \beq
\label{eq:Wms} \mathcal{W}_1^{(0)}(u) =  - 2t\mathcal{G}(u_0,u(\infty)) + \sum_{j = 1}^{j_{\mathrm{max}} + 1} v_{j + 1}\,\frac{1}{j!}\,(\partial_2^{j}\mathcal{G})(u_0,u(\infty))
\eeq}
$\!\!\!$On remonte facilement \`a $W_1^{(0)}$. La position des bords $a$ et $b$ est d\'etermin\'ee par la consistance de l'\'{E}qn.~\ref{eq:Wms} avec :
\beq
W_1^{(0)}(x + i0) - W_1^{(0)}(x - i0) \mathop{\propto}_{x \rightarrow a} \sqrt{x - a},\quad W_1^{(0)}(x + i0) - W_1^{(0)}(x - i0) \mathop{\propto}_{x \rightarrow b} \sqrt{x - b} \nn
\eeq
i.e. par deux conditions d'annulation au bord de la discontinuit\'e. $(a,b)$ doit \^{e}tre l'unique solution de ce syst\`eme telle que $a < b$ et  $a(t) + b(t) \propto \sqrt{t}$ lorsque $t \rightarrow 0$.

La m\'ethode suivie par Eynard et Kristjansen donne une expression diff\'erente de l'\'{E}qn.~\ref{eq:Wms}, mais \'equivalente et que nous \'ecrivons :
\beq
\label{eq:Tritri} W_1^{(0)}(x) = \frac{1}{2i\pi}\oint_{\mathcal{C}([a,b])} \frac{\dd \xi\,V'(\xi)}{x - \xi}\,T(\xi,x)
\eeq
$T(\xi,x)$ est un noyau construit \`a partir d'un choix de fonctions de base dans $E_{\pm}$, dont l'expression est donn\'ee dans \cite{BEOn}. Cette formule int\'egrale g\'en\'eralise celle de Tricomi \cite{Tricomi} :
\beq
W_1^{(0)}(x)\big|_{\n = 0} = \frac{1}{2i\pi}\oint_{\mathcal{C}([a,b])} \frac{\dd \xi\,V'(\xi)}{x - \xi}\,\sqrt{\frac{(x - a)(x - b)}{(\xi - a)(\xi - b)}} \nn
\eeq

\subsubsection{Les cylindres : $W_2^{(0)}$}

On note $\mathcal{W}_2^{(0)}(u_1,u_2)$, le prolongement analytique de $(W_2^{(0)})_{\mathrm{H}}(x(u_1),x(u_2))x'(u_1)x'(u_2)$ \`a $(u_1,u_2) \in \mathbb{C}^2$. C'est une fonction m\'eromorphe, qui a des p\^{o}les doubles sans r\'esidus lorsque $x(u_1) = x(u_2)$ et $x(u_1) = 1/\z - x(u_2)$ :
\bea
(W_2^{(0)})_{\mathrm{H}}(x_1,x_2) = \frac{2}{4 - \n^2}\,\frac{1}{(x_1 - x_2)^2}  + O(1) && x_1 \rightarrow x_2 \nn \\
\phantom{(W_2^{(0)})_{\mathrm{H}}(x_1,x_2)} = \frac{-\n}{4 - \n^2}\,\frac{1}{(x_1 + x_2 - 1/\z)^2} + O(1) && x_1 \rightarrow 1/\z - x_2 \nn
\eea
 $\mathcal{W}_2^{(0)}(u_1,u_2)$ n'admet aucune autre singularit\'e. On peut donc la construire par combinaison lin\'eaire (\`a coefficients constants) de fonctions de Weierstra{\ss} pseudoelliptiques $\wp_{\mathfrak{b}}(u) = -\zeta_{\mathfrak{b}}'(u)$ :
\bea
&& \mathcal{W}_2^{(0)}(u_1,u_2) \nn \\
& = & \frac{\dd u_1\dd u_2}{4 - \n^2}\Big\{-\wp_{\mathfrak{b}}(u_1 + u_2) + \wp_{\mathfrak{b}}(u_1 - u_2) + \wp_{\mathfrak{b}}(-u_1 + u_2) - \wp_{\mathfrak{b}}(-u_1 - u_2)\Big\}\nn \\
& = & \dd u_1\dd u_2 \sum_{m \in \mathbb{Z}} 2\pi^2\,\frac{\cos(\pi\mathfrak{b}m)}{4 - \n^2}\Big\{\frac{1}{\sin^2 (u_1 - u_2 + m\tau)} - \frac{1}{\sin^2(u_1 + u_2  + m\tau)}\Big\} \nn \\
\label{eq:wsp} &&
\eea
On remonte facilement \`a $W_2^{(0)}$. L\`a encore, Eynard et Kristjansen trouvent une expression diff\'erente, avec des fonctions qui d\'ependent s\'epar\'ement de $x_1$ et $x_2$. Elle est bien s\^{u}r \'equivalente : $\wp_{\mathfrak{b}}$ satisfait, comme la fonction de Weierstra{\ss} usuelle, des formules d'addition des arguments, reliant $\wp_{\mathfrak{b}}(u + u_0)$, $\wp_{\mathfrak{b}}(u)$ et $\wp_{\mathfrak{b}}(u_0)$. L'avantage de l'\'{E}qn.~\ref{eq:wsp} est de bien montrer que $\mathcal{W}_2^{(0)}$ ne d\'epend que des distances  $d(x_1;x_2) = u(x_1) - u(x_2)$ et $d(x_1;1/\z - x_2) = u(x_1) + u(x_2) - \tau$, une fois mesur\'ees dans la bonne coordonn\'ee $u$. Cela g\'en\'eralise un ph\'enom\`ene bien connu dans le mod\`ele \`a une matrice.

Dans les deux cas, on observe que $W_2^{(0)}$ est une fonction universelle : toute la d\'ependance dans les param\`etres du mod\`ele se r\'esume \`a la position des extr\'emit\'es des coupures (i.e de $a$, $b$ et du param\`etre $1/\z$). Cela \'etait aussi attendu.

\subsubsection{Les sph\`eres : $F^{(0)}$}

Les s\'eries g\'en\'eratrices de cartes sans bord se d\'eduisent par int\'egration des s\'eries g\'en\'eratrices des cartes \`a bords. Dans le cas de $F^{(0)}$, cela se manifeste par la relation d'homog\'en\'eit\'e :
\beq
F^{(0)} = \frac{t}{2}\Res_{x \rightarrow \infty} \dd x\,V'(x)W_1^{(0)}(x) + \frac{t}{2}\frac{\partial F^{(0)}}{\partial t} \nn
\eeq
et la relation de g\'eom\'etrie sp\'eciale :
\beq
\frac{\partial^2 F^{(0)}}{\partial t_{j_1}\partial t_{j_2}} = \Res_{x_1 \rightarrow \infty} \dd x_1\,x_1^{j_1}\,\Res_{x_2 \rightarrow \infty} \dd x_2\,x_2^{j_2}\,W_2^{(0)}(x_1,x_2) \nn
\eeq
On donne seulement le r\'esultat de l'int\'egration :
\bea
\label{eq:Fko} F^{(0)} & = & \frac{t}{2}\Res_{x \rightarrow \infty} \dd x\,V'(x)\,W_1^{(0)}(x) \\
& & + \frac{t}{2}\,\lim_{\epsilon \rightarrow 0}\Big((2 - \n)t\ln x(u(\infty) + \epsilon) - V\big(x(u(\infty) + \epsilon\big) - 2\int^{u(b)}_{u(\infty) + \epsilon} \dd u\,Y(u)\Big) \nn
\eea
o\`u $Y(u) = -\frac{1}{2}\big[\mathcal{W}_1^{(0)}(u) + \mathcal{W}_1^{(0)}(\overline{u})\big]$. L'\'{E}qn.~\ref{eq:Fko} est la g\'en\'eralisation pour le mod\`ele $\On$ du pr\'epotentiel d\'efini \`a l'\'{E}qn~\ref{eq:F0def}. On peut pousser plus loin ces formules pour calculer $F^{(0)}$ (en utilisant l'\'{E}qn.~\ref{eq:Wms}) ou  $\partial_t^2 F^{(0)}$ (en utilisant l'\'{E}qn.~\ref{eq:wsp}). Quelques unes sont donn\'ees dans \cite{BEOn}. Ces r\'esultats sont nouveaux par rapport \`a \cite{EKOn,EKOn2}.

\subsubsection{Les tores : $F^{(1)}$}

Du point de vue des mod\`eles de matrices, $F^{(1)}$ est la fonction la plus difficile \`a calculer. Nous avons montr\'e qu'une formule similaire \`a l'\'{E}qn.~\ref{eq:F1def} existe, avec une fonction $\mathcal{T}_{B,x}$ d\'efinie \`a partir de $W_2^{(0)}$ par une d\'eformation de l'\'{E}qn.~\ref{eq:TauBerg}. Cela n'est pas tr\`es explicite.

Dans les mod\`eles de boucles sur r\'eseau fixe, la fonction de partition sur le tore\footnote{Il va de soi que le tore $\mathbb{C}/(\mathbb{Z}\oplus\tau\mathbb{Z})$ intervenant dans le param\'etrage $u$ n'a rien \`a avoir avec la topologie des cartes de ce paragraphe.} est une quantit\'e importante, qui se calcule exactement en termes de caract\`eres de Temperley-Lieb (avant la limite continue) ou de Virasoro (\`a la limite continue). Tout progr\`es pour le calcul du $F^{(1)}$ serait int\'eressant en vue d'une comparaison avec les r\'esultats sur r\'eseau fixe. On pourrait \`a l'avenir exploiter un argument heuristique bas\'e sur une int\'egrale de chemins, qui relie notre $F^{(1)}$ \`a un d\'eterminant de Fredholm\label{Fredo3}.

\subsection{S\'eries g\'en\'eratrices des topologies stables}

\subsubsection{R\'esultat principal}

Il est naturel, dans le mod\`ele formel de matrices, de consid\'erer les fonctions $\mathcal{W}_n^{(g)}(u_1,\ldots,u_n)$, qui sont les formes diff\'erentielles $W_n^{(g)}(x_1,\ldots,x_n)\dd x_1\cdots\dd x_n$ lues avec la coordonn\'ee $u$, avec un d\'ecalage pour $(n,g) = (1,0)$ et $(2,0)$ (\'{E}qn.~\ref{eq:apoix} et \ref{eq:apoix2}). Elles s'\'etendent analytiquement en fonctions m\'eromorphes pour $(u_i)_i \in \mathbb{C}^n$. Une cons\'equence imm\'ediate du lemme "une coupure" et des \'equations de Schwinger-Dyson, est que les $\mathcal{W}_n^{(g)}(u_1,\ldots,u_n)$ pour $2 - 2g - n < 0$ n'ont de p\^{o}les dans le feuillet physique que lorsque $u_i = u(a)$ ou $u(b)$.

D\'efinissons une courbe spectrale $\mathcal{S} = [\Sigma,x,y,B]$ pour le mod\`ele $\On$.

\vspace{0.2cm}

\noindent $\diamond\,$ $\Sigma = \mathbb{C}/\mathbb{Z}$ est un cylindre. $u$ est une coordonn\'ee globale sur $\Sigma$.

\vspace{0.2cm}

\noindent $\diamond\,$ $u \mapsto x(u)$ est une fonction m\'eromorphe sur $\Sigma$. Parmi les points de branchement de $x$, on s\'electionne seulement ceux qui se trouvent dans le feuillet \label{invu}physique : $a_i \in \{u(a),u(b)\}$. On d\'efinit une involution (qui est ici globale) par $\overline{u} = 2\tau - u - 1$.

\vspace{0.2cm}

\noindent $\diamond\,$ $u \mapsto y(u) = \mathcal{W}_1^{(0)}(u)/x'(u)$ d\'efinit une fonction m\'eromorphe sur $\Sigma$. Lorsque $\mathfrak{b} \notin \mathbb{Q}$, la relation entre $x$ et $y$ est non alg\'ebrique.

\vspace{0.2cm}

\begin{figure}[h!]
\begin{center}
\begin{minipage}[c]{0.54\linewidth}
\raisebox{-2.7cm}{\includegraphics[width=\textwidth]{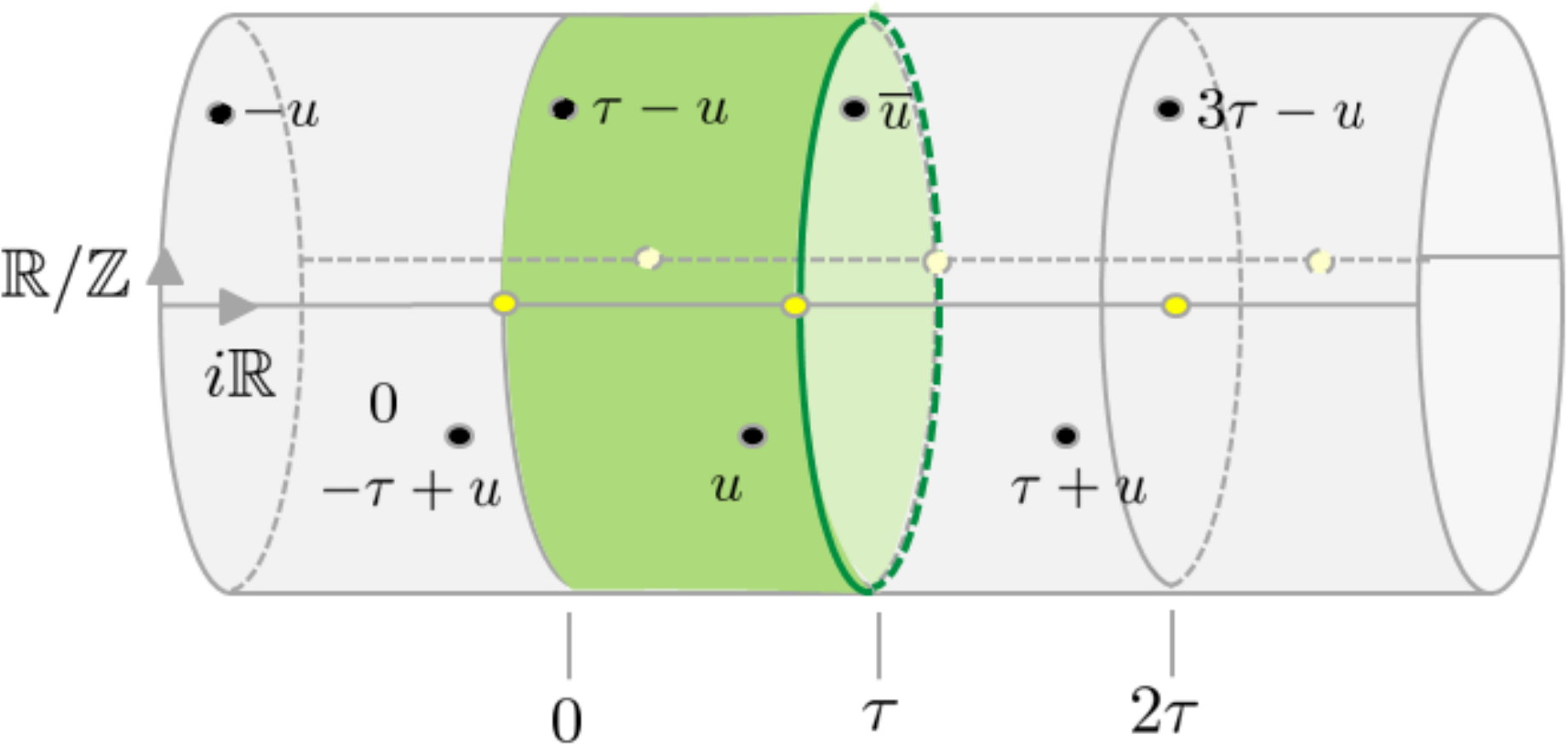}}
\end{minipage}
\hfill \begin{minipage}{0.355\linewidth}
\caption{\label{fig:CourbeS2} $u$ et leurs images $\pm u + m\tau$ ($m \in \mathbb{Z}$). Les points jaunes (qui se projettent sur $a,b,1/\z - a,1/\z - b$) co\"{i}ncident avec l'une de leurs images, et cela est \`a l'origine de l'absence de facteur $2$ dans la formule de Cauchy (\'{E}qn.~\ref{eq:AQ}). $B(u_1,u_2)$ n'a pas de p\^{o}les autres que $u_1 = u_2$ lorsque $u_1,u_2$ sont dans le feuillet physique.}
\end{minipage}
\end{center}
\end{figure}

\noindent $\diamond\,$ On d\'efinit un pseudo noyau de Bergman :
\bea
\label{eq:BBBQ} B(u_1,u_2) & = & \dd u_1\dd u_2\Big\{\mathcal{W}_2^{(0)}(u_1,u_2) \\
&&  + \frac{2 - \n^2}{4 - \n^2}\,\frac{x'(u_1)x'(u_2)}{(x(u_1) - x(u_2))^2} + \frac{\n}{4 - \n^2}\,\frac{x'(u_1)x'(u_2)}{(x(u_1) + x(u_2) - 1/\z)^2}\Big\} \nn
\eea
$B$ a bien un p\^{o}le double avec coefficient $1$ \`a points co\"{i}ncidants et pas de p\^{o}le $u_1 = \tau - u_2$  et $u_1 = \overline{u_2}$ (c'est la raison de ce d\'ecalage). Mais ce n'est pas stricto sensu un noyau de Bergman, car il admet des p\^{o}les suppl\'ementaires, lorsque $u_1$ atteint les autres points $\pm u_2 + m\tau$, $m \in \mathbb{Z}$ (cf. Fig.~\ref{fig:CourbeS2}).

\begin{theo}
Pour $\chi = 2 - 2g - n < 0$, les s\'eries g\'en\'eratrices des cartes de genre $g$ \`a $n$ bords sont donn\'ees par la r\'ecurrence topologique appliqu\'ee \`a la courbe spectrale $\mathcal{S}$.
\beq
\mathcal{W}_n^{(g)}(u_1,\ldots,u_n)\dd u_1\cdots\dd u_n = \mathcal{\omega}_n^{(g)}[\mathcal{S}](u_1,\ldots,u_n) \nn
\eeq
\noindent Autrement dit, on a la formule\label{ressss} des r\'esidus (\S~\ref{sec:axiom}) :
{\small \beq
\omega_n^{(g)}(u,u_I) = \Res_{u \rightarrow u(a),u(b)}\dd u\,\mathcal{K}(u_0,u)\Big[\mathcal{W}_{n + 1}^{(g - 1)}(u,\overline{u},u_I) + \sum_{\substack{J \subseteq I \\ \,0 \leq h \leq g}}^{'} \mathcal{W}_{|J| + 1}^{(h)}(u,u_J)\,\mathcal{W}_{n - |J|}^{(g - h)}(\overline{u},u_{I\setminus J}) \Big] \nn
\eeq \small}
$\omega_2^{(0)} = B$ est reli\'e \`a $\mathcal{W}_2^{(0)}$ par la relation~\ref{eq:BBBQ}. Le noyau de r\'ecurrence est :
\beq
\mathcal{K}(u_0,u) = -\frac{1}{2}\frac{\int_{\overline{u}}^{u} \omega_2^{(0)}(u,v)}{\big(y(u) - y(\overline{u})\big)\dd x(u)} \nn
\eeq
$\sum^{'}$ signifie que l'on exclut de la somme tous les termes ou $\mathcal{W}_1^{(0)}$ apparait.
\end{theo}

Les autres propri\'et\'es de la r\'ecurrence topologique sont aussi valides, ce sont essentiellement des cons\'equences de cette formule, et je ne les red\'erive pas ici. Je mentionne seulement l'\'equation de dilatation qui permet de calculer les s\'eries g\'en\'eratrices des cartes sans bord :

\begin{theo}
Pour $g \geq 2$ :
\beq
F^{(g)} = \frac{1}{2 - 2g}\,\Res_{u \rightarrow u(a),u(b)}\dd u\Big(\int^{u} \!\! y(v)\dd v\Big)\,\mathcal{W}_1^{(g)}(u) \nn
\eeq
\end{theo}

Ainsi, la solution du mod\`ele de matrice $\On$ ne se distingue pas de celle du mod\`ele \`a une matrice, une fois la courbe spectrale identifi\'ee. M\^{e}me la relation~\ref{eq:W2H} entre le corr\'elateur $W_2^{(0)}(x_1,x_2)$ et le noyau de Bergman est encore valide :
\beq
W_2^{(0)}(x_1,x_2)\dd x_1\dd x_2 = B(x_1,x_2) - \frac{\dd x_1\dd x_2}{(x_1 - x_2)^2} \nn
\eeq

\subsubsection{Preuve}

Nous allons utiliser les \'equations de Schwinger-Dyson et la formule de Cauchy (\'{E}qn.~\ref{eq:AQ}), pour arriver exactement \`a la formule des r\'esidus. Le r\^{o}le de l'involution $u \mapsto \overline{u} = 2\tau - u - 1$ est essentiel\footnote{Toutes les fonctions que l'on consid\`ere sont $1$ p\'eriodiques en $u$, donc le $-1$ dans $\overline{u}$ n'est pas important.}. Il est facile d'\^{e}tre \'egar\'e lorsque l'on fait ce calcul la premi\`ere fois, \`a cause des d\'ecalages de $W_1^{(0)}$ et $W_2^{(0)}$. C'est pourquoi je choisis de d\'etailler cette preuve.

\vspace{0.2cm}

\noindent $\diamond\,$ On pose :
 \bea
 -2Y(u) & = & \big[2\,W_1^{(0)}(x(u)) + \n\,W_1^{(0)}(1/\z - x(u)) - V'(x(u))\big]x'(u) \nn \\
 & = & \mathcal{W}_1^{(0)}(u) + \mathcal{W}_1^{(0)}(\overline{u}) \nn
 \eea
et l'on va d\'efinir $\widetilde{\mathcal{W}}_2^{(0)}(u_1,u_2) = \frac{B(u_1,u_2)}{\dd u_1\dd u_2}$, et $\widetilde{\mathcal{W}}_{n}^{(g)} = \mathcal{W}_n^{(g)}$ dans tous les autres cas. Les \'equations de Schwinger-Dyson~\ref{eq:mastj}, pour $2 - 2g - n < 0$, s'\'ecrivent toujours :
\beq
\label{eq:abeau}\mathcal{E}_n^{(g)}(u,u_I) - 2Y(u)\widetilde{\mathcal{W}}_n^{(g)}(u,u_I) = \mathcal{R}_n^{(g)}(u,u_I)
\eeq
avec $\mathcal{R}_n^{(g)}(u,u_I)$ une fonction r\'eguli\`ere lorsque $u \rightarrow u(a)$ ou $u(b)$. Les cas $(n,g) = (1,1)$ et $(3,0)$ doivent \^{e}tre trait\'es s\'epar\'ement, \`a cause du d\'ecalage :
\beq
W_2^{(0)}(x(u_1),x(u_2))x'(u_1)x'(u_2) = \widetilde{\mathcal{W}}_2^{(0)}(u_1,u_2) - \frac{x'(u_1)x'(u_2)}{(x(u_1) - x(u_2))^2} \nn
\eeq

\vspace{0.2cm}

\noindent $\diamond\,$ Lorsque $2 - 2g - n \leq -2$ :
{\small
\bea
&& \mathcal{E}_n^{(g)}(u,u_I) \nn \\
& = & \widetilde{\mathcal{W}}_{n + 1}^{(g - 1)}(u,u,u_I) + \n\,\widetilde{\mathcal{W}}_{n + 1}^{(g - 1)}(u,\tau - u,u_I) \nn \\
&& + \sum_{\substack{J \subseteq I \\ 0 \leq h \leq g}}^{'} \widetilde{\mathcal{W}}_{|J| + 1}^{(h)}(u,u_J)\big(\widetilde{\mathcal{W}}_{n - |J|}^{(g - h)}(u,u_{I\setminus J}) + \n\widetilde{\mathcal{W}}_{n - |J|}^{(g - h)}(\tau - u,u_{I\setminus J})\big) \nn \\
&& + \sum_{i \in I} \Big\{-\frac{x'(u)x'(u_i)}{(x(u) - x(u_i))^2}\,\big[2\widetilde{\mathcal{W}}_{n - 1}^{(g)}(u,u_{I\setminus\{i\}}) + \n\widetilde{\mathcal{W}}_{n - 1}^{(g)}(\tau - u,u_{I\setminus\{i\}})\big]  \nn \\
 & &  + \sum_{i \in I} - \frac{\n\,x'(u)x'(u_i)}{(x(u) + x(u_i) - 1/\z)^2}\,\widetilde{\mathcal{W}}_{n - 1}^{(g)}(u,u_{I\setminus\{i\}}) + \sum_{i \in I} \frac{x'(u)x'(u_i)}{(x(u) - x(u_i))^2}\,\mathcal{W}_{n - 1}^{(g)}(u,u_{I\setminus\{i\}}) \Big\}\nn \\
& = & \mathcal{W}_{n + 1}^{(g - 1)}(u,\overline{u},u_{I}) + \sum^{'} \mathcal{W}_{|J| + 1}^{(h)}(u,u_J)\mathcal{W}_{n - |J|}^{(g - h)}(\overline{u},u_{I\setminus J}) \nn \\
\label{eq:eql} & & - \sum_{i \in I}\Big(\frac{x'(u)x'(u_i)}{(x(u) - x(u_i))^2} + \frac{\n\,x'(u)x'(u_i)}{(x(u) + x(u_i) - 1/\z)^2}\Big)\widetilde{\mathcal{W}}_{n - 1}^{(g)}(u,u_I)
\eea}

\vspace{0.2cm}

\noindent $\diamond\,$ Lorsque $(n,g) = (1,1)$, l'\'{E}qn~\ref{eq:eql} ne peut pas \^{e}tre correcte car $\mathcal{W}_2^{(0)}(u,v)$ a un p\^{o}le lorsque $u \rightarrow \overline{v}$. En fait :
{\small \bea
\mathcal{E}_1^{(1)}(u) & = & \lim_{v \rightarrow u} \Big\{\mathcal{W}_2^{(0)}(u,v) + (\mathcal{W}_2^{(0)})_{\mathrm{P}}(u,v) + \n\Big[\mathcal{W}_2^{(0)}(u,\tau - v) + (\mathcal{W}_2^{(0)})_{\mathrm{P}}(u,\tau - v)\big]\Big\} \nn \\
& = & -\frac{\n}{4 -\n^2}\frac{(x'(u))^2}{4\,(x(u))^2} + \lim_{v \rightarrow u} \Big\{\mathcal{W}_2^{(0)}(u,\overline{v}) + \frac{\n^2 - 2}{4 - \n^2}\,\frac{x'(u)x'(v)}{(x(u) - x(v))^2}\Big\} \nn \\
& = & \widetilde{\mathcal{W}}_2^{(0)}(u,\overline{u}) \nn
\eea}
$\!\!\!$avec la d\'efinition de $B(u,v)$ adopt\'ee \`a l'\'{E}qn.~\ref{eq:BBBQ}.

\vspace{0.2cm}

\noindent $\diamond\,$ Lorsque $(n,g) = (3,0)$, il y a des termes suppl\'ementaires dans l'\'{E}qn.~\ref{eq:eql}, qui sont des produits de deux $\frac{x'(u)x'(u_i)}{(x(u) - x(u_i))^2}$ ou $\frac{x'(u)x'(u_i)}{(x(u) + x(u_i) - 1/\z)^2}$.  Comme il sont r\'eguliers lorsque $u \rightarrow u(a),u(b)$, on peut les inclure dans $\mathcal{R}_n^{(g)}(u,u_1,u_2)$, et l'\'{E}qn.~\ref{eq:eql} est encore valide.

\vspace{0.2cm}

\noindent $\diamond\,$ Nous pouvons maintenant utiliser la formule de Cauchy (\'{E}qn.~\ref{eq:Chachy}) :
\beq
\mathcal{W}_n^{(g)}(u_0,u_I) = \Res_{u \rightarrow u(a),u(b)} \dd u\,\mathcal{G}(u_0,u)\,\mathcal{W}_n^{(g)}(u,u_I)  \nn
\eeq
et y ins\'erer l'\'equation de Schwinger-Dyson sous la forme \'{E}qn.~\ref{eq:abeau}.
\beq
\mathcal{W}_n^{(g)}(u_0,u_I) = \Res_{u \rightarrow u(a),u(b)} \dd u\,\frac{\mathcal{G}(u_0,u)}{2Y(u)}\big[\mathcal{E}_n^{(g)}(u,u_I) - \mathcal{R}_n^{(g)}(u,u_I) \big] \nn
\eeq
Puisque $\mathcal{G}(u_0,u) = -\mathcal{G}(u_0,\overline{u})$, $\mathcal{G}$ s'annule lorsque $u \rightarrow u(a),u(b)$. Par cons\'equent :
\beq
\mathcal{K}(u_0,u) \equiv \frac{\mathcal{G}(u_0,u)}{2Y(u)} \nn
\eeq
est r\'egulier lorsque $u \rightarrow u(a),u(b)$, et le terme impliquant $\mathcal{R}_n^{(g)}$ ne contribue pas au r\'esidu :
\beq
\mathcal{W}_n^{(g)}(u_0,u) = \Res_{u \rightarrow u(a),u(b)} \dd u\,\mathcal{K}(u_0,u)\,\mathcal{E}_n^{(g)}(u,u_I) \nn
\eeq

\vspace{0.2cm}

\noindent $\diamond\,$ Puisque $\dd u\,\mathcal{K}(u_0,u) = \dd \overline{u}\,\mathcal{K}(\overline{u},u_0)$, on peut sym\'etriser :
\beq
\mathcal{W}_n^{(g)}(u_0,u) = \Res_{u \rightarrow u(a),u(b)} \dd u\,\mathcal{K}(u_0,u)\,\frac{1}{2}\big[\mathcal{E}_n^{(g)}(u,u_I) + \mathcal{E}_{n}^{(g)}(\overline{u},u_I)\big] \nn
\eeq
On peut d\'ecomposer :
\beq
\frac{1}{2}\big[\mathcal{E}_n^{(g)}(u,u_I) + \mathcal{E}_n^{(g)}(\overline{u},u_I)\big] = \widetilde{\mathcal{E}}_n^{(g)}(u,u_I) + \Delta_{n}^{(g)}(u,u_I) \nn
\eeq
avec :
{\small\bea
\widetilde{\mathcal{E}}_n^{(g)}(u,u_I) & = & \widetilde{\mathcal{W}}_{n + 1}^{(g - 1)}(u,\overline{u},u_{I}) + \sum^{'} \widetilde{\mathcal{W}}_{|J| + 1}^{(h)}(u,u_J)\widetilde{\mathcal{W}}_{n - |J|}^{(g - h)}(\overline{u},u_{I\setminus J}) \nn \\
\Delta_{n}^{(g)}(u,u_I) & = & \n \sum_{i \in I} \Big(\frac{x'(u)x'(u_i)}{(x(u) - x(u_i))^2} + \frac{\n\,x'(u)x'(u_i)}{(x(u) + x(u_i) - 1/\z)^2}\Big)\widetilde{\mathcal{W}}_{n - 1}^{(g)}(\tau - u,u_{I\setminus\{i\}}) \nn \\
&& + \delta_{g,0}\delta_{n,3}\,\big[\textrm{r\'egulier}\big] \nn
\eea}
$\!\!\!$M\^{e}me si l'on n'a pas explicit\'e le terme produit provenant du d\'ecalage lorsque $(g,n) = (3,0)$, $\Delta_{n}^{(g)}$ est r\'egulier lorsque $u \rightarrow u(a),u(b)$. Par cons\'equent, on trouve la formule des r\'esidus, pour $2 - 2g - n < 0$ :
\beq
\widetilde{\mathcal{W}}_n^{(g)}(u_0,u_I) = \Res_{u \rightarrow u(a),u(b)} \mathcal{K}(u_0,u)\,\widetilde{\mathcal{E}}_{n}^{(g)}(u,u_I) \nn
\eeq

\vspace{0.2cm}

\noindent $\diamond\,$ En comparant l'expression de $\mathcal{W}_2^{(0)}$ (\'{E}qn.~\ref{eq:wsp}) et celle du noyau de Cauchy (\'{E}qn.~\ref{eq:Chachy}), et en rappelant que $\wp_{\mathfrak{b}} = -\zeta_{\mathfrak{b}}'$, on peut \'etablir :
\beq
\mathcal{G}(u_0,u) = \frac{1}{2}\int_{\overline{u}}^{u} \mathcal{W}_2^{(0)}(u_0,v)\dd v\nn
\eeq
On peut ajouter \`a $\mathcal{W}_2^{(0)}(u_0,v)$ toute fonction impaire de $v$ sous l'involution sans changer l'int\'egrale, en particulier on remplacer $\mathcal{W}_2^{(0)}$ par $\widetilde{\mathcal{W}}_2^{(0)} = B$ (\'{E}qn.~\ref{eq:BBBQ}). D'o\`u la formule :
\beq
\mathcal{K}(u_0,u) = -\frac{1}{2}\frac{\int_{\overline{u}}^{u} B(u_0,v)\dd v}{\mathcal{W}_1^{(0)}(u) + \mathcal{W}_1^{(0)}(\overline{u})} \nn
\eeq

\subsection{Limite des grandes cartes}
\label{gcart}
Les points critiques propres au mod\`ele $\On$ sont ceux o\`u $[a,b]$ rencontre la \label{ohu2} coupure image $[1/\z - b,1/\z - a]$, i.e. lorsque $a \rightarrow 1/2\z$ ou $b = +\infty$. La limite de $\mathcal{W}_1^{(0)}$ a \'et\'e extensivement \'etudi\'ee par Kostov et Staudacher \cite{KosStau}, en \'etudiant l'\'equation lin\'eaire directement \`a $a = 1/2\z$ ou $b = +\infty$. On peut retrouver leurs r\'esultats en prenant la limite dans divers r\'egimes de l'\'{E}qn.~\ref{eq:Tritri} ou \ref{eq:Wms}, sachant que $a \rightarrow 0$ correspond \`a $\tau \rightarrow 0$.

Si l'on fixe $\mathfrak{b} \in ]0,1[$ tel que $\n = 2\cos\pi\mathfrak{b}$, il y a autant de phases possibles que de d\'eterminations de $\widetilde{\mathfrak{b}} \in \mathbf{R}_+^*$ tel que $\n = - 2\cos(\pi\widetilde{\mathfrak{b}})$. La r\'esolution de la courbe spectrale est :
\beq
\left\{\begin{array}{c} x(u) \sim  (a - 1/2\z)\,x^*(z) \\ y(u) \sim (a - 1/2\z)^{\widetilde{\mathfrak{b}}}\,y^*(z) \end{array} \right.,\qquad \mathcal{W}_2^{(0)}(u_1,u_2)\dd u_1\dd u_2 \sim (\mathcal{W}_2^{(0)})^*(z_1,z_2)\dd z_1\dd z_2  \nn
\eeq
avec une courbe plane limite non alg\'ebrique :
\beq
x^*(z) = \ch z,\quad y^*(z) = \sum_{j \in \mathbb{Z}} c_j\,\sh\big[(\mathfrak{b} + 2j)z + 2i\pi\mathfrak{b}\big] \nn
\eeq
telle que $y^*(x^*) \propto (x^*)^{\widetilde{\mathfrak{g}}}$ lorsque $x^* \rightarrow \infty$. Par ailleurs :
\bea
(\mathcal{W}_2^{(0)})^*(z_1,z_2) & = & \frac{2}{4 - \n^2}\Bigg\{\frac{\ch\big[(1 - \mathfrak{b})(z_1 - z_2)\big]}{(z_1 - z_2)^2} - \frac{(1 - \mathfrak{b})\sh\big[(1 - \mathfrak{b})(z_1 - z_2)\big]}{z_1 - z_2} \nn \\
&& - \frac{\ch\big[(1 - \mathfrak{b})(z_1 + z_2) - 2i\pi\mathfrak{b}\big]}{(z_1 + z_2)^2} + \frac{\sh\big[(1 - \mathfrak{b})(z_1 + z_2) - 2i\pi\mathfrak{b}\big]}{(z_1 + z_2)}\Bigg\} \nn
\eea
et pour la limite du pseudo noyau de Bergman :
\bea
\frac{B^*(z_1,z_2)}{\dd z_1\dd z_2} & = & \mathcal{W}_2^{(0)}(z_1,z_2) + \frac{1}{4}\Bigg\{\frac{2 - \n^2}{4 - \n^2}\Bigg[\frac{1}{\sh{}^2[(z_1 - z_2)/2]} - \frac{1}{\sh{}^2[(z_1 + z_2)/2]}\Bigg] \nn \\
& & + \frac{\n}{4 - \n^2}\Bigg[\frac{1}{\ch{}^2[(z_1 - z_2)/2]} - \frac{1}{\ch{}^2[(z_1 + z_2)/2]}\Bigg]\Bigg\} \nn
\eea
Cela d\'efinit une courbe spectrale limite $\mathcal{S}^* = [\mathbb{C},x^*,y^*,B^*]$ avec un point de branchement \`a $z = 0$.

Ces points critiques sont atteints \`a certaines valeurs $t \rightarrow (t^*)^{-}$, et l'approche se fait \`a une vitesse :
\beq
(a - 1/2\z) \propto C\,(1 - t/t^*)^{\frac{1}{\widetilde{\mathfrak{b}} + 1 - \mathfrak{b}}} \nn
\eeq
L'exposant de \label{susc}susceptibilit\'e $\gamma_{\mathrm{str}}$ s'en d\'eduit :
\beq
\frac{\partial^2 F^{(0)}}{\partial t^2} \propto (1 - t/t^*)^{-\gamma_{\mathrm{str}}},\qquad \gamma_{\mathrm{str}} = -\frac{2\,\mathfrak{b}}{\widetilde{\mathfrak{b}} + 1 - \mathfrak{b}} \nn
\eeq
Comme ces exposants\label{exposs} sont connus, nous n'allons pas nous attarder sur ces r\'esultats.

Comme la r\'ecurrence topologique passe bien \`a la limite singuli\`ere de courbes spectrales, nous avons aussi d\'emontr\'e, pour les s\'eries g\'en\'eratrices de cartes :
\bea
W_n^{(g)}(x(z_1),\ldots,x(z_n)) & \sim & C^{(\widetilde{\mathfrak{b}} + 1)(2 - 2g - n) - n}\,(1 - t/t^*)^{\frac{(\widetilde{\mathfrak{b}} + 1)(2 - 2g - n) - n}{\widetilde{\mathfrak{b}} + 1 - \mathfrak{b}}}\,\frac{\omega_n^{(g)}[\mathcal{S}^*](z_1,\ldots,z_n)}{\dd x^*(z_1)\cdots\dd x^*(z_n)} \nn \\
F^{(g)} & \sim & C^{(\widetilde{\mathfrak{b}} + 1)(2 - 2g)}\,(1 - t/t^*)^{\frac{(\widetilde{\mathfrak{b}} + 1)}{\widetilde{\mathfrak{b}} + 1 - \mathfrak{b}}\,(2 - 2g)}\,\mathcal{F}^{(g)}[\mathcal{S}^*] \nn
\eea
avec les subtilit\'es habituelles (d\'ecalages, parties non singuli\`eres) pour les topologies instables ($2 - 2g - n > 0$). Ces exposants et leur d\'ependance dans la topologie est attendue du c\^{o}t\'e de la th\'eorie \label{cc2} conforme de charge centrale :
\beq
\mathfrak{c} = 1 - 6\big(\widetilde{\mathfrak{b}}^{1/2} - \widetilde{\mathfrak{b}}^{-1/2}\big) \nn
\eeq
coupl\'ee \`a la th\'eorie de Liouville : c'est la relation d'\'echelle KPZ\label{KPZZ} (pour le param\`etre d'\'echelle $a \rightarrow 0$) \cite{KPZ}. Elle apparait comme une cons\'equence de la relation de Gauss-Bonnet pour une feuille d'univers de caract\'eristique d'Euler $\chi$ : $\int \mathrm{courbure} = 4\pi\chi$. R\'ecemment, pour les cartes sans boucles, Guillaume Chapuy \cite[Chapitre 2]{ChapuyThese} a trouv\'e une interpr\'etation combinatoire de cette d\'ependance en $g$ : la relation d'\'echelle garde trace de la probabilit\'e de trouver $2g$ trisections dans une carte, qui la rendent de genre $g$. Son argument se g\'en\'eralise probablement aux cartes du mod\`ele $\On$, et aux autres mod\`eles de boucles.

\newpage
\thispagestyle{empty}
\phantom{bbk}

\newpage
\thispagestyle{empty}
\phantom{bbk}

\newpage

\chapter{Cordes topologiques et conjecture BKMP}
\label{chap:cordes}
\thispagestyle{plain}
\vspace{-1.5cm}

\rule{\textwidth}{1.5mm}
\addtolength{\baselineskip}{0.20\baselineskip}

\vspace{1.4cm}
{\textsf{Les invariants de Gromov-Witten d'une vari\'et\'e de Calabi-Yau torique $\mathfrak{X}$ peuvent \^{e}tre calcul\'es efficacement de plusieurs mani\`eres : par des techniques de localisation (relations avec la th\'eorie de l'intersection de $\overline{\mathcal{M}}_{g,n}$), par des sommes sur des partitions (relations avec des probl\`emes combinatoires et de th\'eorie des repr\'esentations du groupe sym\'etrique), par des hi\'erarchies d'\'equations diff\'erentielles int\'egrables. En 2007, Bouchard, Klemm, Mari\~{n}o et Pasquetti ont conjectur\'e une nouvelle m\'ethode, bas\'ee sur la r\'ecurrence topologique appliqu\'ee \`a la courbe miroir de $\mathfrak{X}$. L'application la plus simple de cette conjecture donne une fa\c{c}on de calculer les nombres de Hurwitz simples gr\^{a}ce \`a la r\'ecurrence topologique appliqu\'ee \`a la courbe de Lambert $e^{x} = ye^{-y}$, et cela a \'et\'e \'etabli dans \cite{BEMS}. Des progr\`es, auxquels je n'ai pas contribu\'e, ont \'et\'e faits depuis en direction de la conjecture BKMP g\'en\'erale, mais il y a encore beaucoup \`a comprendre. L'objectif de cette partie est d'expliquer, \`a un niveau \'el\'ementaire, l'int\'er\^{e}t des m\'ethodes du Chapitre~\ref{chap:toporec} pour la th\'eorie des cordes topologiques.}}

\addtolength{\baselineskip}{-0.20\baselineskip}

\section{Notions de th\'eorie des cordes}
\label{sec:nono}
Commen\c{c}ons par une introduction simplifi\'ee \`a la th\'eorie des cordes dont nous aurons besoin.

\subsection{Pourquoi les espaces de Calabi-Yau ?}
\label{CYs}
Une \textbf{th\'eorie des cordes} est une th\'eorie des champs o\`u les configurations sont des plongements de surfaces de Riemann $\Sigma_{g,n}$ (les \textbf{feuilles d'univers}) dans un espace cible $\mathfrak{Y}$. Nous n'allons pas aborder de telles d\'efinitions par la th\'eorie des champs. Disons seulement que, en g\'en\'eral, l'on s'int\'eresse aux th\'eories dites \textbf{sans anomalie}, o\`u la valeur des observables ne d\'epend pas du choix de param\'etrage de $\Sigma_{g,n}$ et $\mathfrak{Y}$ utilis\'e pour les calculer. Cela impose des conditions purement g\'eom\'etriques sur $\mathfrak{Y}$. La th\'eorie des supercordes, avec $\mathfrak{Y}$ un espace riemannien et de dimension$_{\mathbb{R}}$ $10$ et qui admet un spineur constant, fournit une classe de th\'eories sans anomalie.
Les physiciens se sont tout de suite demand\'es : si l'on \'ecrit $\mathfrak{Y} = \mathfrak{X} \times M_4$, o\`u $M_4$ est une vari\'et\'e de dimension $3 + 1$ qui joue le r\^{o}le d'espace-temps (par exemple $M_4 \simeq \mathbb{R}^{4}$), quelle est la th\'eorie effective sur $M_4$ ? Pour cela, il faut que $\mathfrak{X}$ soit une vari\'et\'e de dimension$_{\mathbb{R}}$ $6$, sur laquelle il existe un spineur constant.
Les configurations qui extr\'emisent l'action des supercordes sur $\mathfrak{X}$ sont celles o\`u la m\'etrique $\mathsf{g}$ de $\mathfrak{Y}$ satisfait les \'equations d'Einstein :
\beq
\label{eq:ric} \mathsf{Ric}_{\mu\nu} - \frac{1}{2}\,\mathsf{R}\,\mathsf{g}_{\mu\nu} =  \frac{8\pi G_N}{c^4}\,\mathsf{T}_{\mu\nu}
\eeq
En l'absence de mati\`ere, le tenseur \'energie-impulsion $\mathsf{T}_{\mu\nu}$ est nul. Si l'on recherche les solutions o\`u $(\mathfrak{Y},\mathsf{g})$ est un produit direct $(\mathfrak{X} \times M_4,\mathsf{g}_{\mathfrak{X}} + \mathsf{g}_4)$, $(M_4,\mathsf{g}_4)$ et $(\mathfrak{X},\mathsf{g}_{\mathfrak{X}})$ sont n\'ecessairement solution de l'\'{E}qn.~\ref{eq:ric} en dimension$_{\mathbb{R}}$ $4$ et $6$ respectivement.

Une restriction suppl\'ementaire consiste \`a ne s'int\'eresser qu'\`a des\label{Kahl} \textbf{vari\'et\'es de K\"{a}hler} $(\mathfrak{X},\omega)$. Ce sont les vari\'et\'es complexes $\mathfrak{X}$ dont la m\'etrique $\mathsf{g}_{\mathfrak{X}}$ est induite par un \'el\'ement $\omega \in H^{1,1}(\mathfrak{X})$, autrement dit
\beq
\omega = \sum_{i,j} \mathsf{g}_{i,j}\dd z_i \otimes \dd\overline{z_{j}} \nn
\eeq
lu dans un syst\`eme de coordonn\'ees locales holomorphes $(z_i)_i$. Les \textbf{vari\'et\'es de Calabi-Yau} (CY en abr\'eg\'e) sont pr\'ecis\'ement d\'efinies comme les vari\'et\'es de K\"{a}hler qui admettent un spineur constant. Elles ont plusieurs autres d\'efinitions \'equivalentes :

\vspace{0.2cm}

\noindent \emph{D\'efinition originale} $\diamond\,$  Il existe une forme volume sur $\mathfrak{X}$ qui ne s'annule jamais.

\vspace{0.2cm}

\noindent \emph{Conjecture de Calabi, d\'emontr\'ee par Yau} $\diamond\,$ Il existe une \label{Calabi} m\'etrique sur $\mathfrak{X}$ de courbure de Ricci nulle : $\textsf{Ric} \equiv 0$. En particulier, la courbure scalaire $R = \Tr \mathsf{Ric}$ est nulle.

\vspace{0.2cm}

Les CY de dimension$_{\mathbb{C}}$ $3$ fournissent donc un cadre naturel pour \'etudier la th\'eorie des supercordes sans anomalie dans un espace du type $\mathfrak{X} \times M_4$.

\subsection{Th\'eories des cordes de type A et B}
\label{sec:thAB}
La th\'eorie des supercordes sans anomalie associ\'ee \`a $\mathfrak{X}$ existe en deux versions, type IIA et IIB, qui diff\`erent par la r\'ealisation de la supersym\'etrie. L'intensit\'e globale des interactions entre cordes est fix\'ee par un param\`etre $g_s$, coupl\'e \`a la topologie des feuilles d'univers. En th\'eorie topologique des cordes, seules les observables qui ne d\'ependent que de la topologie vont nous int\'eresser. On dit que ces observables appartiennent au \textbf{secteur topologique}, \textbf{ouvert} ou \textbf{ferm\'e} selon la pr\'esence ou non de bords dans les feuilles d'univers. Dans le secteur ferm\'e, les quantit\'es \`a calculer sont :
\begin{itemize}
\item[$\diamond$] \label{eneli3}l'\textbf{\'energie libre de genre $g$}, $F_g(\mathfrak{X})$.
\item[$\diamond$] l'\'energie libre $F(g_s,\mathfrak{X}) = \sum_{g \geq 0} g_s^{2g - 2} F_g(\mathfrak{X})$.
\item[$\diamond$] la\label{fonp6} \textbf{fonction de partition} $Z(g_s,\mathfrak{X}) = e^{F(g_s,\mathfrak{X})}$.
\end{itemize}
Comme en thermodynamique, les observables du secteur ouvert sont reli\'ees aux variations de ces quantit\'es par rapport \`a certains param\`etres dont d\'epend la g\'eom\'etrie de $\mathfrak{X}$.
Pour les th\'eories des cordes, l'\'evaluation de ces observables est une premi\`ere \'etape et dicte les couplages entre les autres secteurs. Pour les math\'ematiques, ces observables sont int\'eressantes si on les voit comme des s\'eries g\'en\'eratrices en $g_s$, car elles contiennent des informations sur la g\'eom\'etrie de $\mathfrak{X}$.

\subsubsection{Dans le mod\`ele B}

\label{Kahl2} C'est la structure de vari\'et\'e complexe de $\mathfrak{X}$ qui compte. Il est muni d'une forme de K\"{a}hler $\omega \in H^{1,1}(\mathfrak{X})$, qui d\'efinit une forme volume holomorphe $\Omega \in H^{3,0}(\mathfrak{X})$. Les observables du secteur ferm\'e encodent les int\'egrales de $\Omega$ sur des cycles de $\mathfrak{X}$. L'\'energie libre de genre $0$ est bien d\'efinie par :
\beq
\frac{\partial F_0}{\partial T_I} = \oint_{\mathcal{A}_I} \Omega,\qquad T_I = \oint_{\mathcal{B}_I} \Omega \nn
\eeq
o\`u $(\mathcal{A}_I,\mathcal{B}_I)_I$ est une base de $H_3(\mathfrak{X},\mathbb{Z})$ telle que :
\beq
\mathcal{A}_I\cap\mathcal{B}_J = \delta_{IJ},\quad \mathcal{A}_I \cap \mathcal{A}_J = 0,\quad \mathcal{B}_I \cap \mathcal{B}_J = 0 \nn
\eeq
L'\'energie libre de genre $1$ est d\'efinie \`a partir de la \textbf{torsion de Ray-Singer} $\mathcal{T}_{\mathrm{RS}}(\mathfrak{X})$, qui est un invariant de vari\'et\'e symplectique :
\beq
F_1 = \frac{1}{2}\ln\mathcal{T}_{\textrm{RS}} = \frac{1}{2}\ln\Big[\prod_{p = 0}^{\mathrm{dim}_{\mathbb{C}}\,\mathfrak{X}} \big(\mathrm{det}\,\Delta_{(p,0)}\big)^{p(-1)^p}\Big] \nn
\eeq
$\Delta_{(p,0)}$ est le laplacien agissant sur les $p$ formes holomorphes, et le d\'eterminant est r\'egularis\'e selon une proc\'edure classique en th\'eorie spectrale \cite{Voros}. Les $F_g$ pour $g \geq 2$, ou les observables du secteur ouvert, n'avaient pas de d\'efinition rigoureuse et assez \label{BKMPZ} g\'en\'erale jusqu'aux propositions de Bouchard, Klemm, Mari\~{n}o et Pasquetti \cite{BKMP} (cf. \S~\ref{sec:BKMP}). Toutefois, si elles existent, on souhaite qu'elles v\'erifient des \'equations \textbf{d'anomalie holomorphe}, \label{anoholo} d\'eriv\'ees \cite{BCOV2} via la th\'eorie des champs qui repr\'esente le mod\`ele B. Ces \'equations dictent la d\'ependance en $\overline{t_I}$ des $F_g$, mais en g\'en\'eral ne les d\'eterminent pas compl\`etement.

\subsubsection{Dans le mod\`ele A}
\label{modA}
\begin{figure}
\begin{center}
\includegraphics[width=0.9\textwidth]{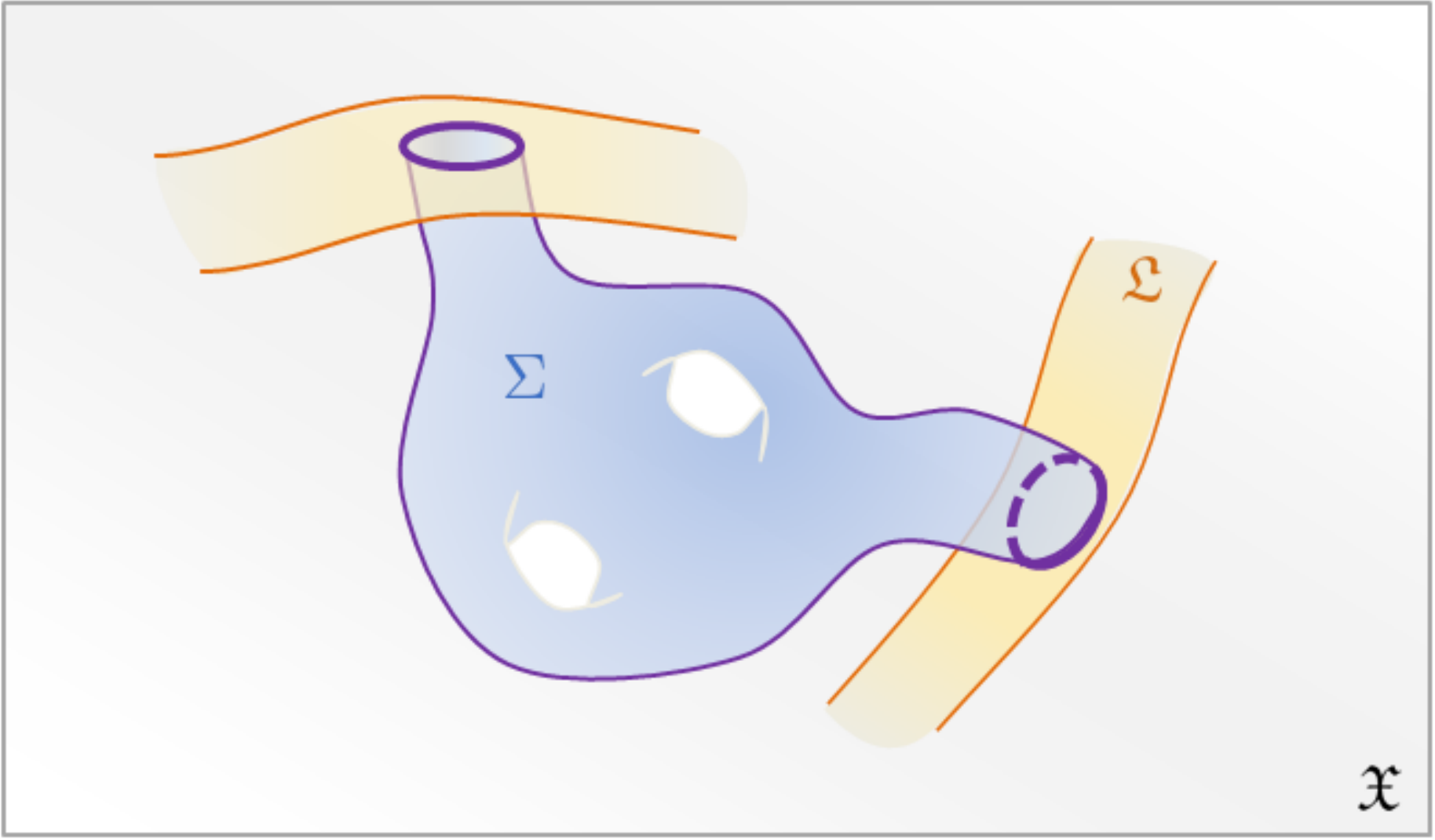}
\caption{\label{fig:GWdessin} Les invariants de Gromov-Witten comptent le nombre de fa\c{c}ons de plonger une feuille d'univers $\Sigma$ dans $\mathfrak{X}$ en respectant certaines conditions aux bords.}
\end{center}
\end{figure}

C'est la structure de vari\'et\'e symplectique de $\mathfrak{X}$ qui compte. Les observables \'enum\`erent les surfaces de Riemann $\Sigma_{g,n}$ de genre $g$ \`a $n$ bords, plong\'ees dans $\mathfrak{X}$ avec un degr\'e donn\'e $\vec{\beta} \in H_2(\mathfrak{X},\mathbb{Z})$ et s'enroulant sur certaines surfaces $\mathfrak{L} \subseteq \mathfrak{X}$ (Fig.~\ref{fig:GWdessin}). Les $\mathfrak{L}$ autoris\'es sont prescrits par la structure symplectique sur $\mathfrak{X}$. Ces nombres sont des invariants g\'eom\'etriques de $\mathfrak{X}$. Il existe deux approches pour les d\'efinir, suivant le point de vue adopt\'e sur $\phi = \Sigma_{g,n} \hookrightarrow \mathfrak{X}$.

\vspace{0.2cm}

\noindent \emph{DT} $\diamond\,$ $\phi$ est d\'efini implicitement par une \'equation dans $\mathfrak{X}$. Il existe une mesure de comptage naturelle sur l'ensemble $\mathcal{I}_n(\mathfrak{X},\vec{\beta})$ de telles \'equations. Les \textbf{invariants de Donaldson-Thomas} sont d\'efinis par son int\'egrale sur\label{Thomas} $\mathcal{I}_n(\mathfrak{X},\vec{\beta})$. On les rassemble dans une fonction de partition $Z_{\mathrm{DT}}$, qui est par construction une s\'erie en puissance de $e^{-g_s}$, o\`u $g_s$ vit naturellement pr\`es de $\infty$.

 \vspace{0.2cm}

 \noindent \emph{GW} $\diamond\,$ $\phi$ est d\'efini par un param\'etrage dans $\mathfrak{X}$. L'ensemble de ces param\'etrages co\"{i}ncide \label{cini}avec l'espace $\mathcal{M}_{g,n}(\mathfrak{X},\vec{\beta})$ introduit dans les parties~\ref{sec:introgeom} et \ref{sec:cordes1}. Il existe une bonne mesure de comptage, not\'ee $[\mathbf{1}]$, sur cet espace, et les nombres
\beq
  N_{g,n}(\mathfrak{X},\vec{\beta}) = \int_{\overline{\mathcal{M}}_{g,n}(\mathfrak{X},\vec{\beta})}[\mathbf{1}] \nn
\eeq
sont appel\'es \textbf{invariants de Gromov-Witten}. \label{TGW}On les rassemble dans une s\'erie g\'en\'eratrice en puissances de $g_s$, qui vit naturellement pr\`es de $0$. En particulier dans le secteur ferm\'e :
\beq
F_g(\mathfrak{X}) = \sum_{\vec{\beta} \in H_2(\mathfrak{X},\mathbb{Z})} N_{g,0}(\mathfrak{X},\beta)\,e^{-\vec{t}\cdot\vec{\beta}}  \nn
\eeq
et l'on note $Z_{\mathrm{GW}}$ la fonction de partition. Un comptage de dimension montre que beaucoup de ces invariants sont nuls lorsque $\mathfrak{X}$ est de dimension$_{\mathbb{C}} > 3$. Cela concentre l'int\'er\^{e}t, dans un premier temps, sur les espaces $\mathfrak{X}$ de basse dimension$_{\mathbb{C}}$ $0,1,2$ et $3$.

Il a \'et\'e d\'emontr\'e \cite{GWDT} que $Z_{\mathrm{GW}}$ et $Z_{\mathrm{DT}}$ sont \'egaux modulo une identification ad\'equate des param\`etres des s\'eries g\'en\'eratrices. Cela signifie qu'il existe une fonction de partition $Z$ pour le mod\`ele $A$, de sorte que le d\'eveloppement asymptotique \`a $g_s \rightarrow \infty$ de $Z$ engendre les invariants DT, et le d\'eveloppement \`a $g_s \rightarrow 0$ de $\ln Z$ les invariants GW. Nous allons surtout parler de l'aspect Gromov-Witten.

\vspace{0.3cm}

Nous supposerons dor\'enavant que $\mathfrak{X}$ est un CY de dimension$_{\mathbb{C}}$ $3$ (CY$_3$ en abr\'eg\'e). Pour\label{toto} la famille des CY$_3$ qui admettent une sym\'etrie torique $\mathbb{T}_3$, des techniques de calcul assez g\'en\'erales ont \'et\'e d\'evelopp\'ees \`a la fois pour les mod\`eles A et B. Ce n'est pas \'etranger au fait que la classification des CY$_3$ toriques est \label{orbi}tr\`es simple, et s'\'etend aussi au cas des orbivari\'et\'es. Le paragraphe \S~\ref{sec:classCY} r\'esume cette classification et introduit plusieurs notions utiles \`a la partie~\ref{sec:calcGW}. En revanche, le programme de classification des CY$_3$ n'est pas encore achev\'e. Il existe des travaux sur la th\'eorie topologique des cordes dans d'autres familles connues de $\mathfrak{X}$ (qui ont ou pas la propri\'et\'e de Calabi-Yau), mais cela sort du cadre de cette th\`ese.

\subsection{Sym\'etrie miroir}
\label{miros}
Il existe une \textbf{sym\'etrie miroir} disant que le mod\`ele de type $A$ sur un espace $\mathfrak{X}$ est dual au mod\`ele de type $B$ sur un \textbf{espace miroir} $\widetilde{\mathfrak{X}}$. C'est une conjecture profonde en math\'ematiques, et dont la formulation pr\'ecise demande un investissement en g\'eom\'etrie alg\'ebrique bien au-del\`a des objectifs de cette th\`ese. On pourra consulter le programme original de Kontsevich \cite{Kont94}, et nous n'allons pas faire \'etat des progr\`es r\'ecents pour y r\'epondre. Le livre \cite{CoxKatz} d\'ecrit le ph\'enom\`ene avec de nombreux exemples.

La sym\'etrie miroir a \'et\'e justifi\'ee par les physiciens Hori et Vafa \cite{Mirror}, disons pour le moins, lorsque $\mathfrak{X}$ est un CY$_3$. Son miroir $\widetilde{\mathfrak{X}}$ est \'egalement un CY$_3$ que l'on sait construire explicitement. Ici, nous adopterons un point de vue purement calculatoire : sous hypoth\`eses, nous disposerons de d\'efinitions algorithmiques pour certaines observables des mod\`eles A et B, et nous verrons quelles propri\'et\'es elles impliquent. La sym\'etrie miroir pr\'edit une correspondance pr\'ecise entre observables, souvent v\'erifiable num\'eriquement, et permet de transposer (en conjecture) des id\'ees d\'evelopp\'ees pour le mod\`ele A vers le mod\`ele B et inversement. La conjecture BKMP en est une manifestation (\S~\ref{sec:BKMP}).

\subsection{Description des CY toriques}
\label{sec:classCY}

Cette pr\'esentation sera minimale, et s'inspire de celles employ\'ees dans les articles d'Aganagic, Klemm, Mari\~{n}o, \ldots \cite{TopoV,MarinoV,BKMP}. Son but est surtout de permettre au lecteur d'identifier les objets que l'on manipule, en s'appuyant d\`es que possible sur des dessins et une intuition. La plupart des justifications seront laiss\'ees de c\^{o}t\'e. Un amateur qui aimerait avoir plus de prises sur les g\'eom\'etries toriques pourra consulter l'article p\'edagogique \cite{Closset}.

\subsubsection{Les CY$_3$ toriques $\mathfrak{X}$ \ldots}
\label{orbi3}
\`{A} l'exception de quelques cas particuliers, tous les CY$_3$ toriques peuvent \^{e}tre construits comme des quotients\footnote{ou des recollements de ces quotients. En fait, les espaces $\mathfrak{X}$ que l'on consid\`ere ici ne sont pas des vari\'et\'es lisses. On travaille plut\^{o}t avec des \textbf{sch\'emas}, i.e. des atlas d'anneaux constitu\'es des polyn\^{o}mes dans les coordonn\'ees locales, qui satisfont des conditions de recollement. Le point de vue des sch\'emas a l'avantage de d\'ecrire sans heurts les points o\`u $\mathfrak{X}$ n'est pas lisse, par exemple au voisinage desquels $\mathfrak{X}$ est une orbivari\'et\'e. Les "CY$_3$ toriques" que l'on consid\`ere en pratique sont parfois appel\'es \textbf{CY$_3$ toriques formels} dans la litt\'erature.} :
\beq
\label{eq:X} \mathfrak{X} = \Big\{\vec{\zeta} \in \mathbb{C}^{r + 3}\qquad \sum_{\alpha = 1}^{r + 3} Q_j^{(\alpha)}\,|\zeta_{\alpha}|^2 = t_j \Big\}/\sim
\eeq
avec les identifications $(\zeta_{\alpha})_{\alpha} \sim \big(\lambda^{Q_j^{(\alpha)}}\zeta_{\alpha}\big)_{\alpha}$ pour tout $\lambda \in \mathbb{C}^*$. Les $t_j$ sont appel\'es \textbf{param\`etres de K\"{a}hler}, \label{Kahh}ils forment un syst\`eme de coordonn\'ees r\'eelles sur $H^{1,1}(\mathfrak{X})$. Les vecteurs $\vec{Q}_j \in \mathbb{Z}^{r + 3}$ sont appel\'es \textbf{charges}. L'espace $\mathfrak{X}$ est \label{CYs2}bien de Calabi-Yau ssi :
\beq
\label{eq:condCY} \forall j \in \{1,\ldots,r\}\qquad \sum_{\alpha = 1}^{r + 3} Q_j^{(\alpha)} = 0
\eeq
$\mathfrak{X}$ n'est pas n\'ecessairement une vari\'et\'e lisse, mais la description \'{E}qn.~\ref{eq:X} est assez explicite pour que l'on comprenne ses singularit\'es.

\begin{figure}[h!]
\begin{center}
\hspace{-0.8cm}\includegraphics[width=1.1\textwidth]{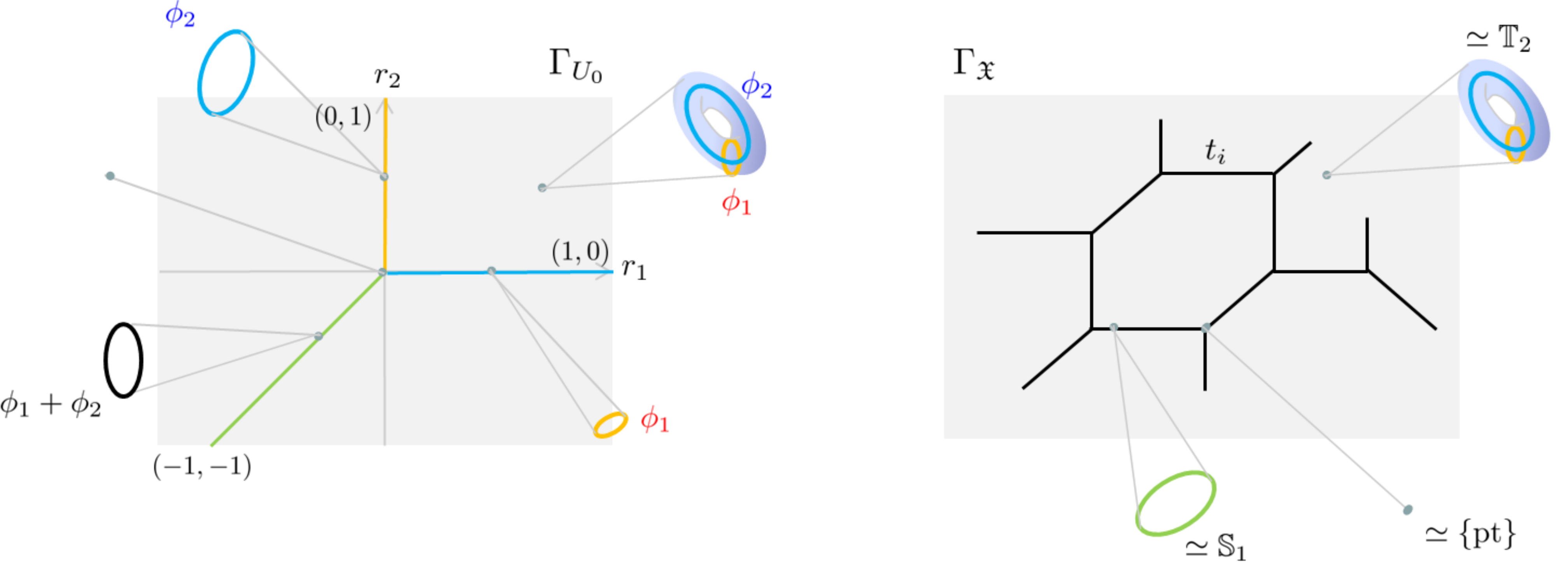}
\caption{\label{fig:patchCYU} \`{A} gauche, lieu $\Gamma_{U_0}$ des singularit\'es de la fibre $\mathbb{T}_2$ dans le plan $s = 0$. \`{A} droite, le diagramme torique $\Gamma_{\mathfrak{X}}$ est le recollement des $\Gamma_{U_i}$, d\'efinis avec la fibration \'{E}qn.~\ref{eq:coordq}, pour un atlas de cartes $(U_i,\Phi_i)$ sur $\mathfrak{X}$. Les param\`etres de K\"{a}hler $t_i$ apparaissent comme les longueurs des ar\^{e}tes de $\Gamma_{\mathfrak{X}}$.}
\end{center}
\end{figure}

Si l'on choisit initialement un syst\`eme de coordonn\'ees locales $\Phi_0 = (z_{1},z_{2},z_{3})$ valable sur un ouvert $U_0 \subseteq \mathfrak{X}$, on peut r\'ealiser $U$ comme un fibr\'e en tores $\mathbb{T}_2$ sur $\mathbb{R}^2 \times \mathbb{C}$. (Fig.~\ref{fig:patchCYU}, \`a gauche). Par exemple, en introduisant $(r_1,r_2) \in \mathbb{R}^2$ et $s \in \mathbb{C}$ par :
\bea
\label{eq:coordq} && r_{1} = |z_1|^2 - |z_3|^2,\qquad r_{2} = |z_2|^2 - |z_3|^2 \\
&& s = \zeta_1\cdots\zeta_{r + 3} \nn
\eea
et en d\'efinissant l'action de $(e^{i\phi_1},e^{i\phi_2}) \in \mathbb{T}_2$ par :
\beq
(z_1,z_2,z_3) \rightarrow (e^{i\phi_1}\,z_{1},e^{i\phi_2}\,z_{2},e^{-i(\phi_1 + \phi_2)}\,z_{3}) \nn
\eeq
La condition~\ref{eq:condCY} implique que l'action $(\mathbb{C}^*)^r$ des charges $(\vec{Q}_j)_{1 \leq j \leq r}$ laissent la coordonn\'ee $s$ invariante, ce qui permet de projeter \`a $s = 0$ dans tout syst\`eme de coordonn\'ees sans perdre d'information.

Si l'on souhaite d\'ecrire les fibres \`a $r_1, r_2$ constants dans une autre r\'egion $U' \subseteq \mathfrak{X}$ avec un autre syst\`eme de coordonn\'ees $\Phi' = (z_1',z_2',z_3')$, il faut utiliser l'\'{E}qn.~\ref{eq:X} pour exprimer $r_1$ et $r_2$ en fonction des $z'_{j}$. On peut parcourir ainsi tout un atlas de $\mathfrak{X}$. M\^{e}me si les choix de $\Phi_0$ et de la fibration \'{E}qn.~\ref{eq:coordq} sont arbitraires, la g\'eom\'etrie de $\mathfrak{X}$ est enti\`erement caract\'eris\'ee par le lieu $\Gamma_{\mathfrak{X}} \subseteq \mathbb{R}^2$ o\`u la fibre $\mathbb{T}_2$ est singuli\`ere. C'est en g\'en\'eral un graphe trivalent dont les ar\^{e}tes ont une pente fractionnaire (Fig.~\ref{fig:patchCYU}, \`a droite). R\'eciproquement, la donn\'ee d'un tel graphe permet de reconstruire un CY$_3$ par le recollement appropri\'e de syst\`emes de coordonn\'ees $\Phi_i$ associ\'es \`a chaque vertex de $\Gamma_{\mathfrak{X}}$.

$\Gamma_{\mathfrak{X}}$ est appel\'e le \textbf{diagramme torique} de $\mathfrak{X}$. La g\'eom\'etrie de $\mathfrak{X}$ est encod\'ee de fa\c{c}on transparente dans le dual $(\Gamma_{\mathfrak{X}})^*$ : les sommets correspondent \`a des points de $\mathfrak{X}$, les ar\^{e}tes \`a des courbes $C$ invariantes sous l'action torique, et les faces donnent des relations entre les classes d'homologie $[C]$ de courbes invariantes toriques (elles sont prescrites par les charges $\vec{Q}_i$). Pour chacune de ces courbes, on peut d\'efinir un param\`etre de K\"{a}hler complexe $t_{[C]} = \int_{[C]} \omega$, et ces relations indiquent que tous les $t_{[C]}$ ne sont pas ind\'ependants.

\subsubsection{\ldots et leurs miroirs $\widetilde{\mathfrak{X}}$}

{A} tout CY$_3$ torique $\mathfrak{X}$ est associ\'e un espace \textbf{miroir} $\widetilde{\mathfrak{X}}$, qui est aussi un CY$_3$ torique :
\beq
\label{eq:relX}\widetilde{\mathfrak{X}} = \left\{(\vec{\zeta},u,v) \in \mathbb{C}^{r + 3}\times\mathbb{C}^2 \qquad \begin{array}{l} \sum_{\alpha = 1}^{r + 3} \zeta_{\alpha} = 0 \\ \sum_{\alpha = 1}^{r + 3} Q_i^{(\alpha)}\zeta_{\alpha} = \tau_i \\  \sum_{\alpha = 1}^{r + 3} e^{-\zeta_{\alpha}} = e^{u + v} \end{array}\right\}
\eeq
$\Re \tau_i = t_i$ co\"{i}ncident avec les param\`etres de K\"{a}hler de $\mathfrak{X}$. Modulo un choix de coordonn\'ees $(x,y)$ fonctions des $\zeta_{\alpha}$, on munit $\widetilde{\mathfrak{X}}$ de la forme volume holomorphe $\Omega = \dd u\wedge \dd x \wedge \dd y$.

\begin{figure}[h!]
\begin{center}
\includegraphics[width=0.8\textwidth]{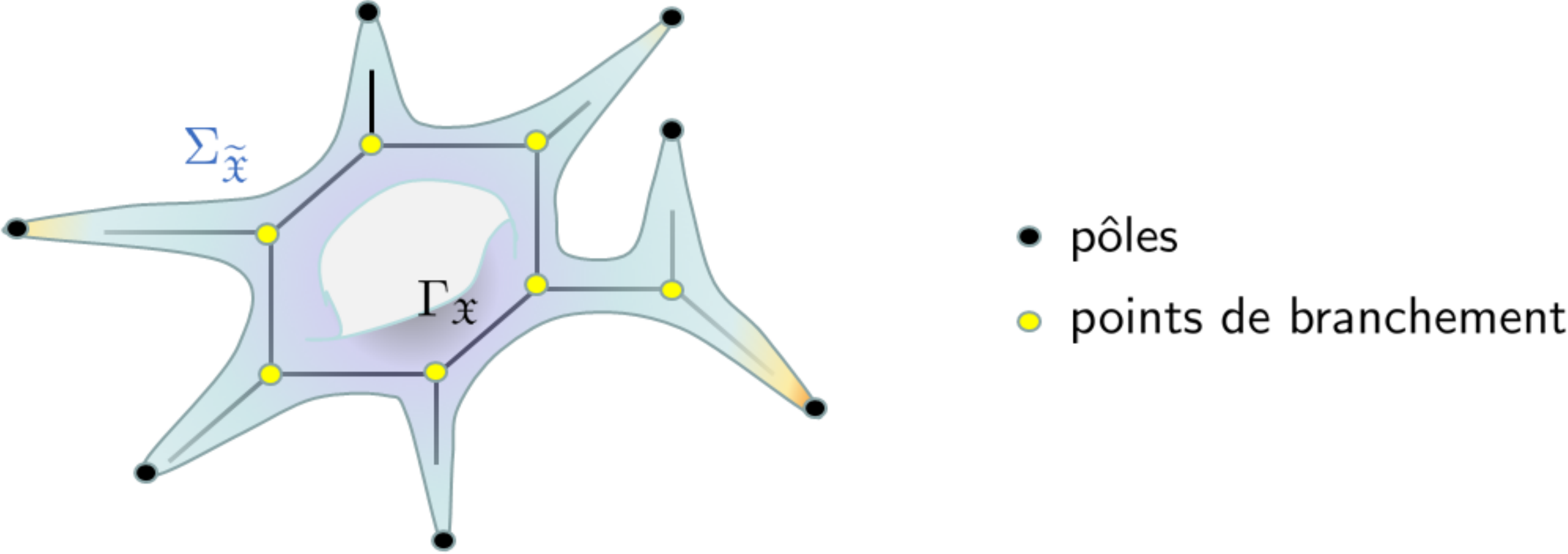}
\caption{\label{fig:Cmiroir} $[\Sigma_{\widetilde{\mathfrak{X}}},x,y]$ s'interpr\`ete comme le diagramme torique $\Gamma_{\mathfrak{X}}$ \'epaissi. On peut alors facilement d\'eterminer son genre $\mathfrak{g}$ (nombre de faces de $\Gamma_{\mathfrak{X}}$), ses points de branchement $a_i$ (qui sont en bijection avec les sommets de $\Gamma_{\mathfrak{X}}$), et les p\^{o}les $x = y = \infty$ (correspondant aux extr\'emit\'es de $\Gamma_{\mathfrak{X}}$).}
\end{center}
\end{figure}

Il est commode d'utiliser des coordonn\'ees vivant dans $\mathbb{C}^*$ : $U = e^{u}$, $V = e^{v}$, $Z_\alpha = e^{-\zeta_{\alpha}}$. On peut voir $\mathfrak{X}$ comme un fibr\'e en hyperboles, o\`u les fibres ont pour \'equation $UV = \mathrm{cte}$. Le lieu $U = V = 0$ o\`u ces fibres deviennent singuli\`eres est appel\'e\label{cm} \textbf{courbe miroir} et not\'e $\Sigma_{\widetilde{\mathfrak{X}}}$ (Fig.~\ref{fig:Cmiroir}). Son \'equation se d\'eduit de :
 \beq
 \sum_{\alpha = 1}^{r + 3} Z_{\alpha} = 0 \nn
 \eeq
 \`a partir des relations d\'efinissant $\widetilde{\mathfrak{X}}$ (\'{E}qn.~\ref{eq:relX}). Elle est toujours de la forme :
\beq
\Sigma_{\widetilde{\mathfrak{X}}} = \big\{(x,y) \in \mathbb{C}^2\qquad \mathrm{Pol}(e^{x},e^{y}) = 0\big\} \nn
\eeq

\subsubsection{L'application miroir}
\label{amiro}
Il existe une \textbf{application miroir}, qui donne la correspondance entre les param\`etres de K\"{a}hler $t_i$ de $\mathfrak{X}$ (mod\`ele A) et les p\'eriodes $T_I$ de $\widetilde{\mathfrak{X}}$, de sorte que la sym\'etrie miroir soit valide dans le secteur ferm\'e :
\beq
F^{g}(\mathfrak{X};t_i)\big|_{\textrm{mod\`{e}le\,\,A}} = F^{g}(\widetilde{\mathfrak{X}},T_I)\big|_{\textrm{mod\`{e}le\,\,B}} \nn
\eeq
En fait, gr\^{a}ce aux sym\'etries toriques, les p\'eriodes $T_I$ sur les cycles de dimension$_{\mathbb{R}}$ $3$ de $\widetilde{\mathfrak{X}}$ se r\'eduisent aux p\'eriodes de $y\dd x$ sur des cycles $(\mathcal{A}'_I,\mathcal{B}_I')$ de la courbe miroir :
\beq
\frac{\partial F^0}{\partial T_I} = \oint_{\mathcal{B}'_I} y\dd x,\qquad T_I = \frac{1}{2i\pi}\oint_{\mathcal{A}_I'} y\dd x \nn
\eeq

\begin{figure}[h!]
\begin{center}
\includegraphics[width=\textwidth]{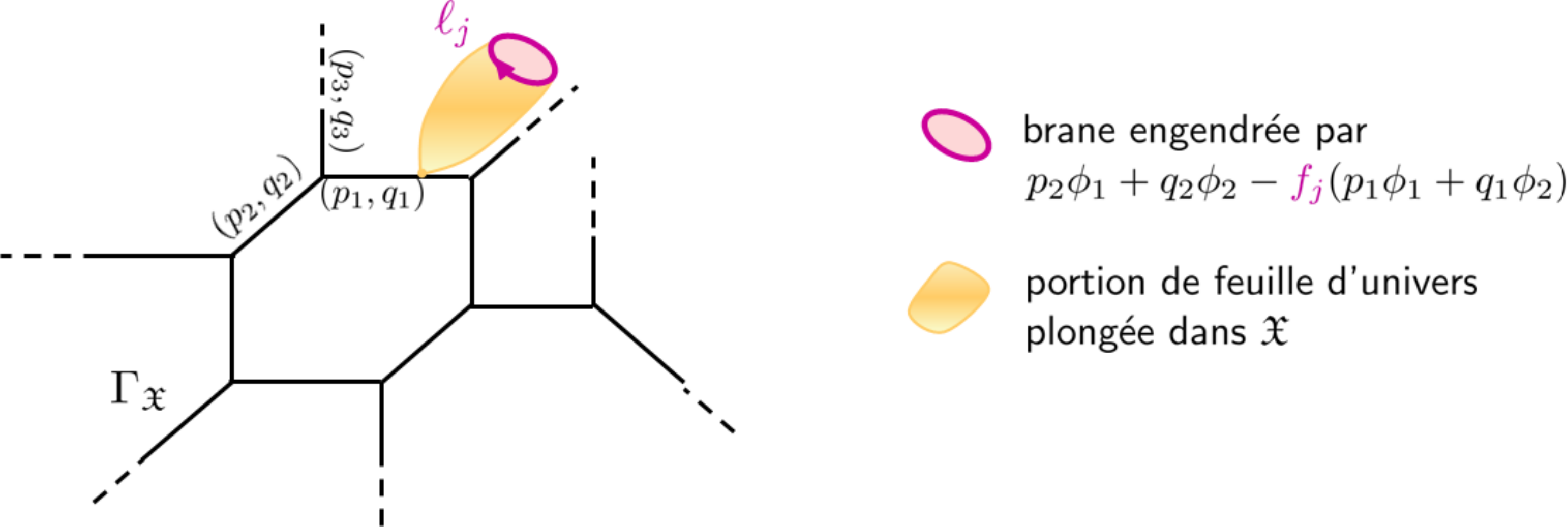}
\caption{\label{fig:framing} Branes dans le mod\`ele A. On a indiqu\'e les pentes $(p,q)$ des ar\^{e}tes, telles que la fibre sur ces ar\^{e}tes est engendr\'ee par l'action de $p\phi_1 + q\phi_2$.}
\end{center}
\end{figure}

Avant de parler de sym\'etrie miroir dans le secteur ouvert, il faut pr\'eciser dans le mod\`ele A sur quelles surfaces $\mathfrak{L}_j \subseteq \mathfrak{X}$ s'enroulent les bords de la feuille d'univers $\Sigma_{g,n}$. \label{baba} G\'en\'eriquement, ces bords sont appel\'es $\textbf{branes}$. Les $\mathfrak{L}_j$ qui se proposent naturellement comme support des branes sont des surfaces lagrangiennes\label{Kahl3}, i.e.\footnote{Contrairement au secteur ferm\'e qui apparemment n'utilise que la structure de vari\'et\'e complexe de $\mathfrak{X}$, le secteur ouvert du mod\`ele A fait intervenir la structure de K\"{a}hler sur $\mathfrak{X}$.} $\omega|_{\mathfrak{L}_j} \equiv 0$, qui vivent sur les ar\^{e}tes de $\Gamma_{\mathfrak{X}}$. Leur position est fix\'ee par un entier $f_j$, dont le nom anglais \textit{framing} \label{frami}n'a pas de traduction en fran\c{c}ais. Intuitivement, il sp\'ecifie le plongement de $\mathfrak{L}_j$ dans la fibre $\mathbb{T}_2$ (Fig.~\ref{fig:framing}). Les observables ouvertes de genre $g$, \`a $n$ bords du mod\`ele A sont alors les s\'eries g\'en\'eratrices :
\beq
\Phi_{g,n}(\mathfrak{X};z_1,\ldots,z_n)|_{\textrm{mod\`{e}le}\,\mathrm{A}} = \sum_{\vec{\beta},\ell_1,\ldots,\ell_n}  \Big[\prod_{j = 1}^n z_j^{\ell_j}\Big]\,N_{g,n}(\mathfrak{X},\vec{\beta})\,e^{-\vec{t}\cdot\vec{\beta}} \nn
\eeq
o\`u $\ell_j \in \mathbb{Z}$ est le nombre d'enroulements de la $j^{\textrm{\`{e}me}}$ brane autour de $\mathfrak{L}_j$.

Il y a aussi un secteur ouvert dans le mod\`ele B : intuitivement, ajouter des branes revient \`a ajouter des singularit\'es dans la structure complexe de $\widetilde{\mathfrak{X}}$. Cela se traduit naturellement par l'ajout de p\^{o}les dans $y\dd x$. On comprend alors que la position de la $j^{\textrm{\`{e}me}}$ brane est d\'etermin\'ee par un point $\xi \in \Sigma_{\widetilde{\mathfrak{X}}}$ associ\'e au choix d'une coordonn\'ee locale $x$. A priori, les observables ouvertes $\Phi_{g,n}(z_1,\ldots,z_n)|_{\textrm{mod\`{e}le}\,\,\mathrm{B}}$ d\'ependent du choix du param\'etrage $(x,y)$ de $\Sigma_{\widetilde{\mathfrak{X}}}$. Il existe une \textbf{application miroir "ouverte"} qui, \`a partir d'une configuration de brane donn\'ee dans le mod\`ele A, prescrit le bon choix $(x,y)$ qui est fonction des $t_i$, de sorte que la sym\'etrie miroir soit valide. On peut changer la configuration de brane \'etudi\'ee gr\^{a}ce \`a des transformations $(x,y) \rightarrow (ax + by,cx + dy)$ appropri\'ees. En particulier, les changements de \textit{framing} sont impl\'ement\'es par $(x,y) \rightarrow (x - fy,y)$ pour $f \in \mathbb{Z}$, tandis que les autres \label{ininini}transformations de $\mathrm{SL}_2(\mathbb{Z})$ permettent de d\'eplacer une brane d'une ar\^{e}te de $\Gamma_{\mathfrak{X}}$ vers une autre. On peut reconnaitre dans ce $\mathrm{SL}_2(\mathbb{Z})$ un sous-groupe des transformations symplectiques de  $[\Sigma_{\widetilde{\mathfrak{X}}},x,y]$ introduites au \S~\ref{sec:defc}.

S'il faut \^{e}tre conscient de l'existence de ces applications miroirs, leur construction ne sera pas discut\'ee ici. Il suffit de se r\'ef\'erer \`a la litt\'erature lorsque l'on souhaite travailler sur des exemples pr\'ecis.

\subsection{Exemples}
\label{sec:exCY}
\noindent $\diamond\,$ $\mathfrak{X} = \mathbb{C}^3$. Le diagramme torique est constitu\'e par un seul vertex, et la courbe miroir a pour \'equation :
\beq
\Sigma_{\widetilde{\mathbb{C}^3}} \: : \: 1 + e^{x} + e^{y} = 0 \nn
\eeq
Apr\`es changement de r\'ef\'erentiel $(x',y') = (x +fy,y)$, l'\'equation de la courbe miroir devient :
\beq
(\Sigma_{\widetilde{\mathbb{C}^3}})_{f}\: : \: e^{fy'} + e^{x'} + e^{y'(f + 1)} = 0 \nn
\eeq
C'est une courbe de genre $0$, sur laquelle $y$ d\'efinit une coordonn\'ee globale, et qui a un unique point de branchement $a$, \`a la position $Y(a) = e^{y(a)} = -\frac{f}{f + 1}$. Lorsque l'on grossit le voisinage de $a$ dans la limite $f \rightarrow \infty$ selon :
\beq
e^{x} = e^{x(a)}e^{\widehat{x}},\qquad e^{y} = e^{y(a)}e^{(1 - \widehat{y})/f} \nn
\eeq
\label{Lambert2}on obtient la courbe de Lambert :
\beq
(\Sigma_{\widetilde{\mathbb{C}^3}})_{\infty}\: : \: \widehat{y}e^{-\widehat{y}} = e^{\widehat{x}} \nn
\eeq

\vspace{0.2cm}

\begin{figure}[h!]
\begin{center}
\includegraphics[width=\textwidth]{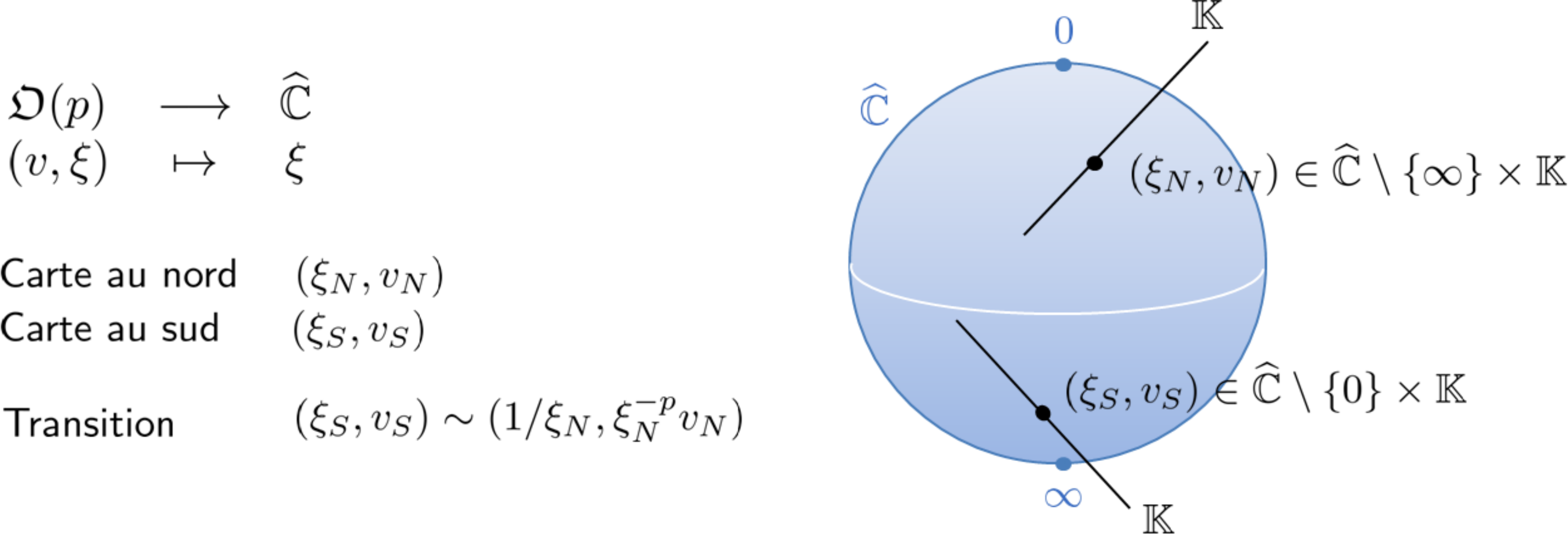}
\caption{\label{fig:ofibre} Ici $\mathbb{K} = \mathbb{C}$. $\mathfrak{O}(p) \rightarrow \widehat{\mathbb{C}}$ est un fibr\'e vectoriel de rang$_{\mathbb{K}}$ $1$ construit \`a partir de deux cartes $(\xi_N,v_N)$ et $(\xi_S,v_S)$. L'application de transition entre ces deux cartes d\'epend de $p$, et intrique le comportement de l'espace vectoriel fibre (ici $\mathbb{K}$) et la position sur $\widehat{\mathbb{C}}$.}
\end{center}
\end{figure}

\vspace{0.2cm}

\noindent $\diamond\,$ Si $\Sigma$ est une surface de Riemann, on trouve des CY$_3$ parmi les fibr\'es vectoriels de rang$_{\mathbb{C}}$ $2$ au-dessus de $\Sigma$. De m\^{e}me, si $S$ est une vari\'et\'e complexe de dimension$_{\mathbb{C}}$ $2$, on trouve des CY$_3$ parmi les fibr\'es vectoriels de rang$_{\mathbb{C}}$ $1$ au-dessus de $S$. De tels espaces sont appel\'es \textbf{g\'eom\'etries locales}.

\vspace{0.2cm}

\noindent $\diamond\,$ Les CY$_3$ bas\'es sur $\Sigma = \widehat{\mathbb{C}}$ sont tous du type $\mathfrak{X}_p = \mathfrak{O}(p) \oplus \mathfrak{O}(-p - 2)$. La figure~\ref{fig:ofibre} explique cette notation. Lorsque $p \neq 1$, ces espaces ne peuvent \^{e}tre r\'ealis\'es\footnote{Je remercie Marcos Mari{\~{n}}o pour cette pr\'ecision.}. par un quotient du type Eqn.~\ref{eq:X}. Il existe tout de m\^{e}me une notion de courbe miroir, que l'on peut d\'eterminer \`a partir du vertex topologique \cite{MariXp}, mais que nous n'allons pas expliciter. Son expression est par exemple donn\'ee dans \cite{Ep}.

\begin{figure}[h!]
\begin{center}
\includegraphics[width = 0.9\textwidth]{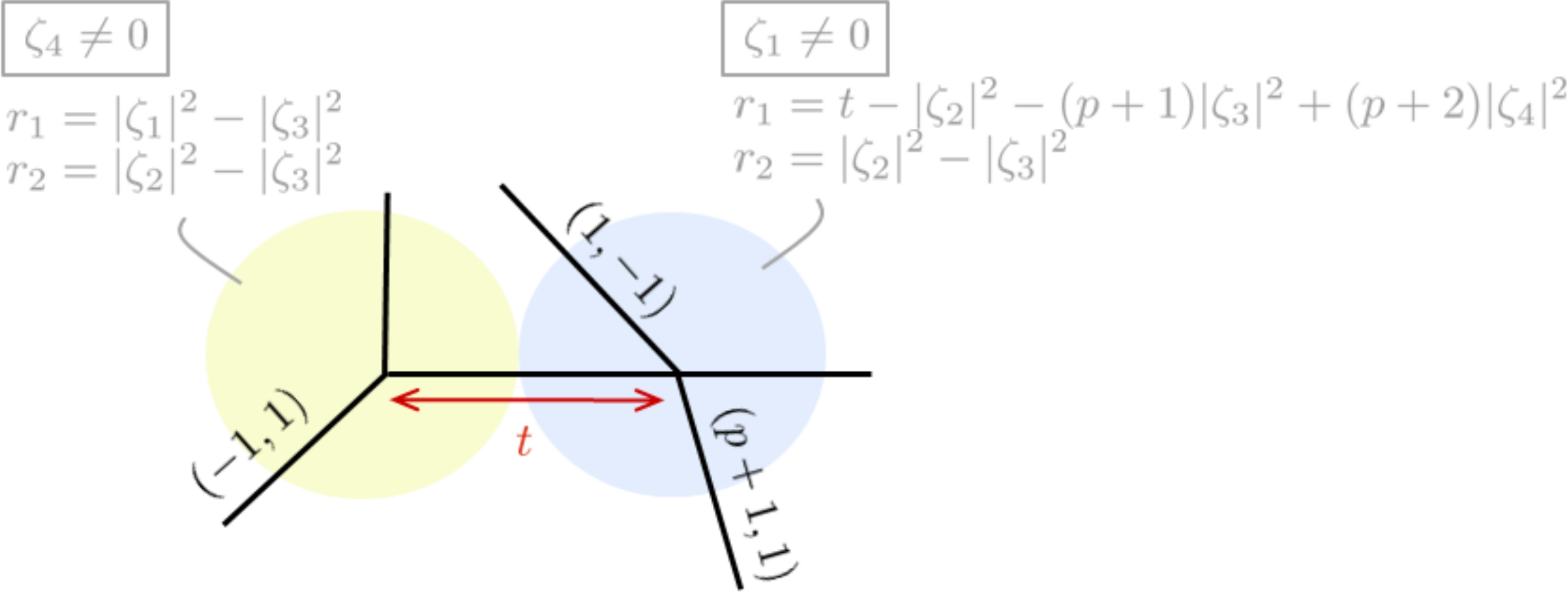}
\caption{\label{fig:torXp} Le diagramme torique de $\mathfrak{O}(p) \oplus \mathfrak{O}(- p - 2) \rightarrow \widehat{\mathbb{C}}$. On a indiqu\'e sur les ar\^{e}tes les vecteurs directeurs qui ne sont pas \'evidents. L'ar\^{e}te centrale est un fibr\'e en $\mathbb{S}_1$ au-dessus d'un segment, dont la fibre se r\'eduit \`a un point aux deux extr\'emit\'es : c'est le $\widehat{\mathbb{C}}$ duquel on est parti, plong\'e dans $\mathfrak{X}_p$ sous la forme $\big\{(\zeta_1,\zeta_2) \in \mathbb{C}^2 \quad |\zeta_1|^2  + |\zeta_2|^2 = t\big\}$. Le param\`etre de K\"{a}hler $t$ est directement li\'e au volume de cette sph\`ere.}
\end{center}
\end{figure}

\vspace{0.2cm}

\noindent $\diamond\,$ On appelle \textbf{r\'esolution du c\^{o}ne} la famille d'espaces $\mathfrak{X}_{p = -1} = \mathfrak{O}(-1) \oplus \mathfrak{O}(-1) \rightarrow \widehat{\mathbb{C}}$ de l'exemple pr\'ec\'edent.
\beq
\mathfrak{X}_{-1} = \big\{\vec{\zeta} \in \mathbb{C}^{4}\quad |\zeta_1|^2 + |\zeta_{2}|^2 - |\zeta_3|^2 - |\zeta^4|^2 = t\big\}/\mathbb{C}^* \nn
\eeq
$\mathfrak{X}_{-1}$ "extrapole de mani\`ere lisse" la singularit\'e conique $Z^2 - Z'^2 = t \rightarrow 0$, d'o\`u son nom. La courbe miroir, dans un param\'etrage adapt\'e \`a une brane sur une ar\^{e}te externe de $\Gamma_{\mathfrak{X}_{-1}}$, a pour \'equation :
\beq
\Sigma_{\widetilde{\mathfrak{X}_{-1}}}\: : \: 1 + e^{-x} + e^{-y} + e^{-\tau}e^{-x + y} = 0 \nn
\eeq

\vspace{0.2cm}

\begin{figure}[h!]
\begin{center}
\includegraphics[width=\textwidth]{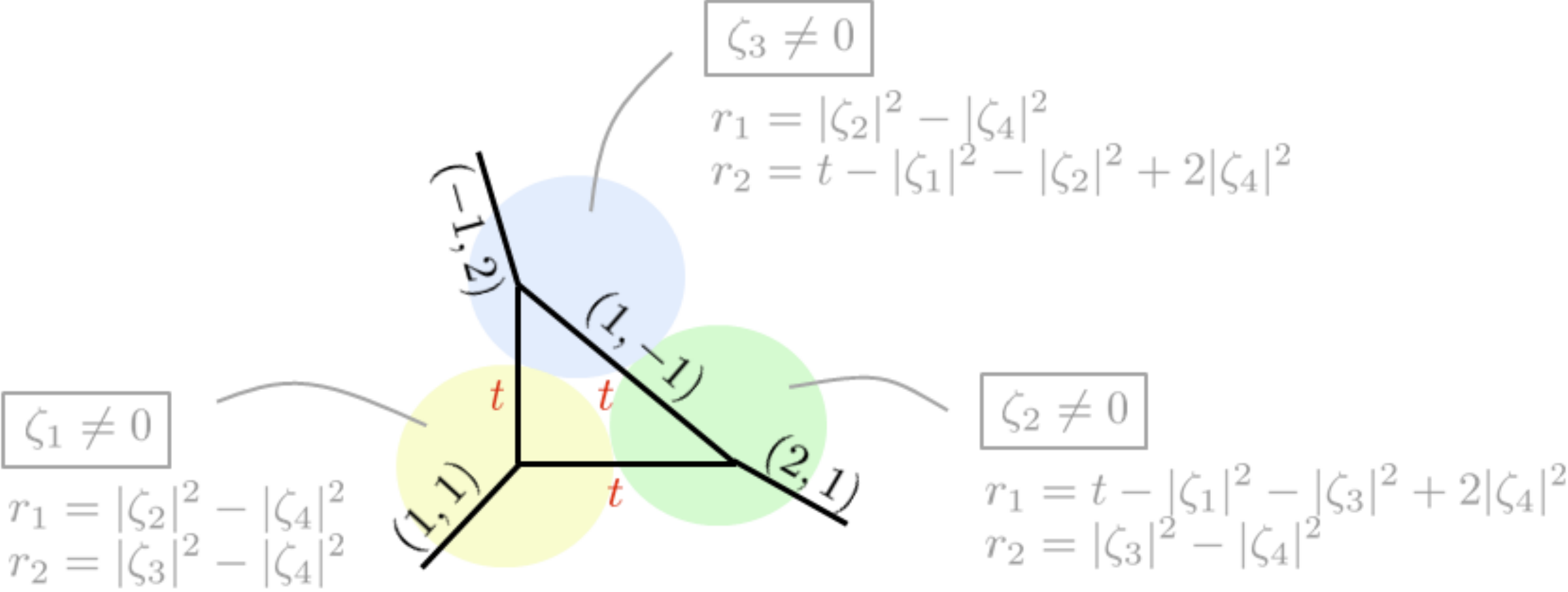}
\caption{\label{fig:torP2} Le diagramme torique de l'espace $\mathbb{P}_2$ local. Il diff\`ere de $\mathfrak{X}_{-1}$ par la pr\'esence d'une troisi\`eme carte, i.e. d'une identification suppl\'ementaire dans $\mathfrak{X}_{-1}$.}
\end{center}
\end{figure}

\noindent $\diamond\,$ Le seul fibr\'e de rang$_{\mathbb{C}}$ $1$ qui permet de d\'efinir un espace de Calabi-Yau bas\'e sur l'espace projectif $S = \mathbb{P}_2(\mathbb{C})$, est du type $\mathfrak{O}(-3) \rightarrow \mathbb{P}_2(\mathbb{C})$. Il est r\'ealis\'e par :
\beq
\mathfrak{X} = \big\{\vec{\zeta} \in \mathbb{C}^4 \quad |\zeta_1|^2 + |\zeta_2|^2 + |\zeta_3|^2 - 3|\zeta_4|^2 = t\big\}/\mathbb{C}^* \nn
\eeq
recouvert par les trois cartes $\zeta_i \neq 0$ ($i \in \{1,2,3\}$). L'action de $\mathbb{C}^*$ est engendr\'ee par la charge $\vec{Q} = (1,1,1,-3)$. $\mathfrak{X}$ est appel\'e \textbf{espace $\mathbb{P}_2$ local}. Son diagramme torique est repr\'esent\'e \`a la figure~\ref{fig:torP2}. Une \'equation pour la courbe miroir est :
\beq
(\Sigma_{\widetilde{\mathfrak{X}}})\: : \: e^{-2y} + e^{-y} + e^{-(x + y)} + e^{-\tau}e^{-3x} = 0 \nn
\eeq
avec $e^{-x} = Z_4/Z_2$ et $e^{-y} = Z_3/Z_2$. On admet que ce param\'etrage $(x,y)$ est adapt\'e pour d\'ecrire une brane positionn\'ee sur une ar\^{e}te externe de $\Gamma_{\mathfrak{X}}$.

\section{R\'ecurrence topologique et th\'eorie topologique des cordes}
\label{sec:calcGW}

Nous allons pr\'esenter quelques aspects du calcul des invariants de Gromov-Witten dans les CY$_3$ toriques, o\`u l'introduction r\'ecente de techniques de r\'ecurrence topologique a permis quelques progr\`es.

\subsection{Localisation et vertex topologique}
\label{sec:topover}
Dans un CY$_3$ torique, la sym\'etrie $\mathbb{T}_3$ permet de \textbf{localiser} les int\'egrales sur $\overline{\mathcal{M}}_{g,n}(\mathfrak{X},\vec{\beta})$, i.e. de les exprimer comme une somme discr\`ete, avec certains poids, d'int\'egrales sur $\overline{\mathcal{M}}_{g,n}$. Les termes de cette somme sont appel\'es \textbf{instantons} : chaque configuration $\Sigma^* \hookrightarrow \mathfrak{X}$ invariante sous l'action de $\mathbb{T}_3$ donne lieu \`a un instanton. D'o\`u l'importance du diagramme torique : de telles surfaces de Riemann $\Sigma^*$ doivent s'appuyer sur les ar\^{e}tes de $\Gamma_{\mathfrak{X}}$, et sont engendr\'ees par la fibre $\mathbb{S}_1$. Si ce n'\'etait pas le cas, on disposerait au moins d'une action non triviale de $\mathbb{T}_2$, dont au moins un sous-groupe $\mathbb{S}_1$ ferait tourner $\Sigma^*$ dans $\mathfrak{X}$. En g\'en\'eral, les instantons sont index\'es par une suite de partitions $(\alpha,\beta,\ldots)$ vivant sur les ar\^{e}tes de $\Gamma_{\mathfrak{X}}$. Ainsi, la d\'ependance dans la g\'eom\'etrie de $\mathfrak{X}$ ne se trouve plus dans l'espace d'int\'egration, mais dans les poids statistiques d'un mod\`ele de partitions al\'eatoires.

\vspace{0.2cm}

Le r\'esultat est la formule de Mari\~{n}o-Vafa exprimant\label{MVV} la fonction de partition du mod\`ele A comme une somme sur des instantons. Leur poids font intervenir :

\vspace{0.2cm}

\noindent $\diamond\,$ une amplitude d\'ependant de deux partitions $(\mu,\nu)$, donn\'ee par les \textbf{nombres de Hurwitz doubles} dont nous reparlerons au \S~\ref{sec:Hurwitz} :
\beq
\label{eq:Hdoubleg} \mathcal{H}_{\mu,\nu}(g_s) = \frac{1}{|\mathrm{Aut}\,\mu|\,|\mathrm{Aut}\,\nu|} \sum_{\alpha\,\vdash\,N} e^{g_s f_{\alpha}(C_2)}\,\chi_{\alpha}(\mu)\,\chi_{\alpha}(\nu)
\eeq
Ici, $|\mu| = |\nu| = N$ est le nombre de boites des partitions, $\alpha \vdash N$ d\'esigne une partition de $N$ boites, qui d\'efinit \'egalement une repr\'esentation unitaire irr\'eductible du groupe sym\'etrique $\mathfrak{S}_N$. $f_{\alpha}(C_2)$ est le Casimir \label{Casi}quadratique associ\'e \`a $\alpha$ (\'{E}qn.~\ref{eq:erratum}), et $\chi_{\alpha}(\mu)$ est le caract\`ere de cette repr\'esentation \'evalu\'e sur la classe de conjugaison de $\mathfrak{S}_N$ associ\'ee \`a $\mu$.

\vspace{0.2cm}

\noindent $\diamond\,$ une amplitude d\'ependant de trois partitions $(\mu,\nu,\rho)$ \label{Hodge2} donn\'ee par les \textbf{int\'egrales de Hodge triples} :
{\small \beq
I_{\mu,\nu,\rho}(f) = \int_{\overline{\mathcal{M}}_{g,\ell_{\mathrm{tot}}}} \frac{\Lambda_g^{\vee}(1)\,\Lambda_g^{\vee}(f)\,\Lambda_g^{\vee}(- f - 1)}{\prod_i(1 - \mu_i \psi_i)(1 - \nu_i \psi_{i + \ell(\mu)}/f)(1 + \rho_i \psi_{i + \ell(\mu) + \ell(\nu)}/(f + 1))}\nn
\eeq}
$\!\!\!\ell_{\mathrm{tot}} = \ell(\mu) + \ell(\nu) + \ell(\rho)$ est la longueur totale des partitions. $\Lambda_g^{\vee}(f) = \sum_{h = 0}^{g} (-f)^{g - h} \,\lambda_h$ \label{Hodge} est appel\'ee \textbf{classe de Hodge}. Les classes $\lambda_h$ sont d\'efinies \cite{Mum} pour $h \in \{0,\ldots,g\}$, et se comportent comme des \'el\'ements de volume $2h$-dimensionnels dans $\overline{\mathcal{M}}_{g,n}$ (et $\lambda_0 = 1$). Elles peuvent s'exprimer en fonction des \label{frami2}classes $\psi$ et $\kappa_r$. Dans la formule ci-dessus, $f$ est un entier de \textit{framing}.

\vspace{0.2cm}

\begin{figure}[h!]
\begin{center}
\includegraphics[width=0.71\textwidth]{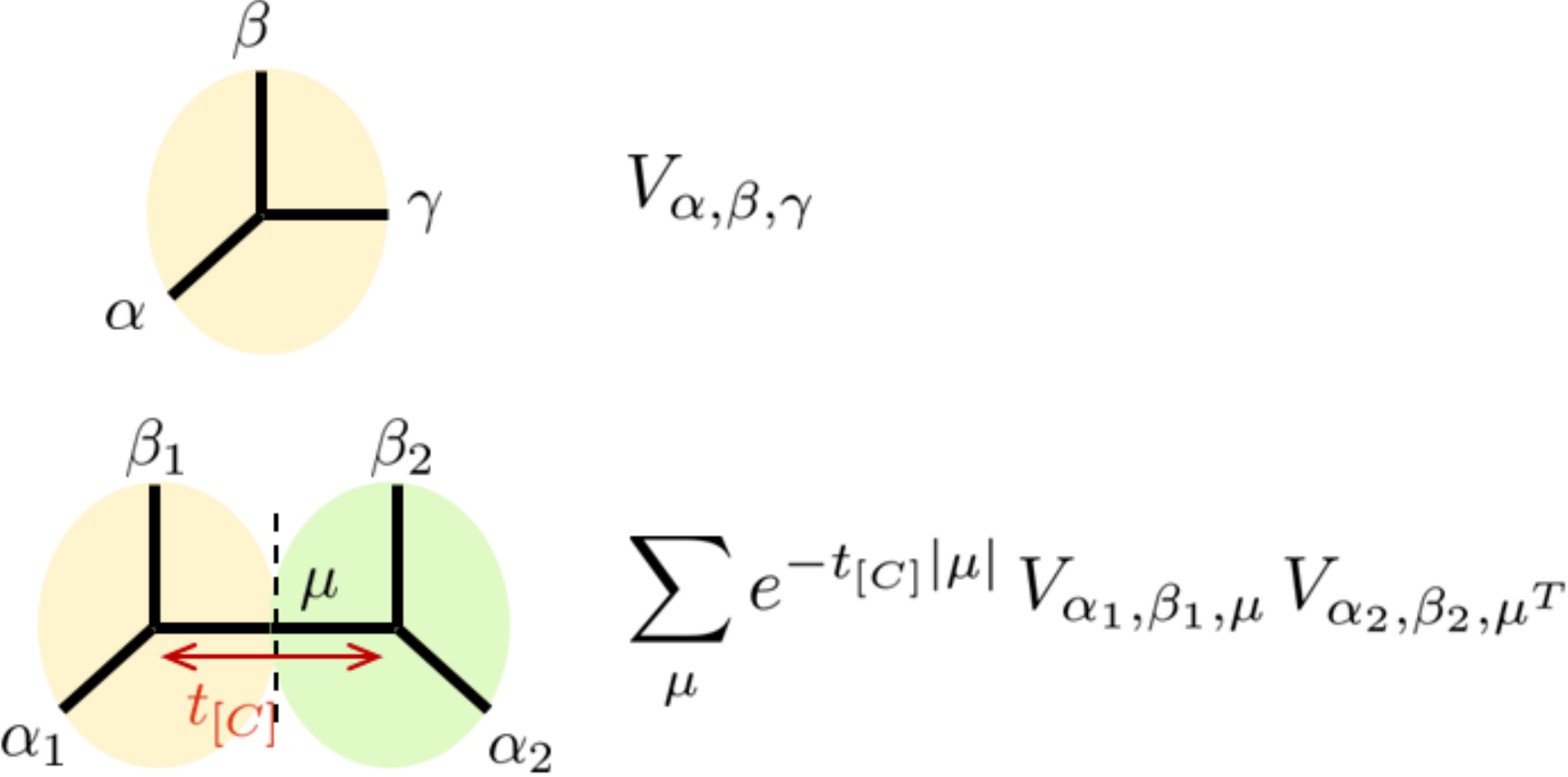}
\caption{\label{fig:vertexa} Le vertex topologique et l'algorithme de recollement.}
\end{center}
\end{figure}

Il est important de souligner que les poids des instantons ne d\'ependent que la g\'eom\'etrie locale de $\mathfrak{X}$ : on peut d\'ecouper $\Sigma^*$ suivant chaque ar\^{e}te de $\Gamma_{\mathfrak{X}}$, \'etudier s\'epar\'ement la contribution de $\Sigma^*$ dans chaque carte $\mathbb{C}^3$ associ\'ee \`a un sommet de $\Gamma_{\mathfrak{X}}$, et les recoller en sommant sur tous les d\'ecoupages possibles (Fig.~\ref{fig:vertexa}). La contribution d'un sommet \label{baba2} avec la configuration de brane la plus g\'en\'erale est appel\'ee \textbf{vertex topologique}, et not\'e $V_{\alpha,\beta,\gamma}$. Plusieurs formules \'equivalentes permettent de calculer le vertex topologique en tant que s\'erie en puissances de $e^{-g_s}$. Ce sont toujours des sommes sur des partitions, que \cite{TopoV} expriment \`a partir de $I_{\alpha,\beta,\gamma}(f)$. \cite{Cristal} donne une jolie interpr\'etation de $V_{\alpha,\beta,\gamma}$, par un mod\`ele statistique de grandes partitions\label{uuu} planes s'appuyant sur $(\alpha,\beta,\gamma)$.

Historiquement, le vertex topologique et l'algorithme de recollement ont \'et\'e d\'ecouverts en physique \cite{TopoVO}. En fait, il existe une dualit\'e bien document\'ee \cite{Witten} entre  $(a)$ certains invariants de n{\oe}uds de Chern-Simons \label{CS10} pour le groupe $\mathrm{U}(N)$ sur un espace $X$ de dimension$_{\mathbb{R}}$ $3$, et $(b)$ le secteur ouvert du mod\`ele A sur $\mathfrak{X} = T^*X$ o\`u les branes s'enroulent sur une vari\'et\'e \label{Frlo}lagrangienne. Cette dualit\'e vaut \`a tout ordre dans l'asymptotique $N = \frac{t}{ig_s} \rightarrow \infty$, et a men\'e \`a la conjecture de la formule de Mari\~{n}o-Vafa \cite{MarinoVk}, d\'emontr\'ee dans \cite{ZhouMV}.  Les auteurs de \cite{TopoVO} ont d\'ecrit pr\'ecis\'ement comment transposer les id\'ees correctes dans la th\'eorie de Chern-Simons, pour reproduire toutes les caract\'eristiques attendues du secteur ouvert du mod\`ele A dans un CY$_3$ torique quelconque. Gr\^{a}ce \`a la localisation, ces pr\'edictions ont \'et\'e prouv\'ees pour l'essentiel dans \cite{TopoV}. Le lecteur pourra consulter \cite{MLiuHodge} pour plus d'informations sur les int\'egrales de Hodge et leur lien avec les invariants de Gromov-Witten.

\subsection{La conjecture BKMP}

\label{sec:BKMP}

La th\'eorie du vertex topologique calcule $Z$ en puissances de $e^{-g_s}$, i.e. du c\^{o}t\'e Donaldson-Thomas. \label{Tomas}On aimerait \'egalement avoir une m\'ethode directe pour calculer les invariants de Gromov-Witten, i.e. $F_g$, le coefficient de $g_s^{2g - 2}$ dans $\ln Z$. C'est l'objet de la conjecture de Bouchard, Klemm, Mari\~{n}o et Pasquetti \cite{BKMP}.
\begin{conj}\label{cong:K}
Soit $\mathfrak{X}$ est un CY$_3$ torique. On d\'efinit une courbe spectrale $\mathcal{S}_{\mathfrak{X}} = [\Sigma_{\widetilde{\mathfrak{X}}},x,y,B]$ \`a partir de la courbe miroir et de l'application miroir ouverte. Alors, les s\'eries g\'en\'eratrices d'invariants de Gromov-Witten de $\mathfrak{X}$ sont donn\'ees par la r\'ecurrence topologique du Chapitre~\ref{chap:toporec} appliqu\'ee \`a $\mathcal{S}_{\mathfrak{X}}$. Les invariants symplectiques encodent le secteur ferm\'e :
\beq
F_g(\mathfrak{X}) = \mathcal{F}^{(g)}[\mathcal{S}_{\mathfrak{X}}] \nn
\eeq
Les primitives :
\beq
\Phi_n^{(g)} = \int^{\xi_1}\!\!\cdots\int^{\xi_n} \omega_n^{(g)}[\mathcal{S}_{\mathfrak{X}}](\xi_1',\ldots,\xi_n') \nn
\eeq
encodent le secteur ouvert lorsqu'elles sont d\'evelopp\'ees en s\'erie de Taylor en $z(\xi_i)$, o\`u $z$ est une coordonn\'ee locale sur $\Sigma_{\widetilde{\mathfrak{X}}}$ prescrite par l'application miroir.
\end{conj}

Les formules de r\'esidus (\'{E}qn.~\ref{eq:res}) et de dilatation (\'{E}qn.~\ref{eq:dilat}) donnent un algorithme efficace pour calculer ces quantit\'es, qui h\'eritent automatiquement de toutes les propri\'et\'es de la r\'ecurrence topologique. Celles-ci ressemblent fortement aux caract\'eristiques que l'on attend des observables des mod\`eles A et B.

\vspace{0.2cm}

\noindent $\diamond\,$ Dans le secteur ouvert, les $\omega_n^{(g)}$ d\'ependent du choix du param\'etrage $(x,y)$, \`a des transformations d'\'equivalence faible pr\`es. Dans le secteur ferm\'e, $\mathcal{F}^{(g)}$ est inchang\'e sous des transformations symplectiques de $(x,y)$. Cela est \label{baba3} consistant avec l'id\'ee que l'on peut changer les configurations de branes par des transformations symplectiques de $(x,y)$. En particulier, les $\omega_n^{(g)}$ pour $n \geq 1$ d\'ependent du choix d'un \textit{framing}, \label{frami3}mais pas les $\mathcal{F}^{(g)}$.

\vspace{0.2cm}

\noindent $\diamond\,$ Les $\mathcal{F}^{(g)}$ et $\omega_n^{(g)}$ ont les propri\'et\'es de \label{sgu}g\'eom\'etrie sp\'eciale. C'est ce que l'on attend des observables du mod\`ele B.

\vspace{0.2cm}

\noindent $\diamond\,$ Le choix d'un noyau de Bergman $B$ invariant modulaire rend les $\mathcal{F}^{(g)}$ et $\omega_n^{(g)}$ invariants modulaires, mais leur donne une d\'ependance non holomorphe en les p\'eriodes sur $\Sigma_{\widetilde{\mathfrak{X}}}$. Cette d\'ependance est toutefois r\'egie par des \'equations d'anomalie holomorphe \cite{EMO}. Pour un autre choix de noyau de Bergman, d\'ependant de mani\`ere holomorphe des p\'eriodes, $\mathcal{F}^{(g)}$ et $\omega_n^{(g)}$ d\'epend de mani\`ere holomorphe des p\'eriodes, et ont \label{modula}des transformations modulaires que l'on sait expliciter. C'est aussi le ph\'enom\`ene \`a l'\oe{}uvre dans le mod\`ele B.

\vspace{0.2cm}

La conjecture BKMP a \'et\'e motiv\'ee par ces analogies. Elle est bien test\'ee num\'eriquement par comparaison avec la m\'ethode du vertex topologique\label{sech}. En fait, \cite{BKMP} proposent $\mathcal{F}^{(g)}[\mathcal{S}_{\mathfrak{X}}]$ et $\omega_n^{(g)}[\mathcal{S}_{\mathfrak{X}}]$ comme d\'efinition des secteurs ouvert et ferm\'e du mod\`ele B dans $\widetilde{\mathfrak{X}}$.

\subsection{La conjecture de Bouchard et Mari\~{n}o}
\label{BMM}
Dans le cas le plus simple $\mathfrak{X} = \mathbb{C}^3$ avec \textit{framing} $f \rightarrow \infty$, les observables du mod\`ele A encodent les int\'egrales de Hodge simple. \label{Hodge3}
\beq
\label{eq:Hosimple} H_{g;\mu} = \prod_{i = 1}^{n} \frac{\mu_i^{\mu_i}}{\mu_i!} \int_{\overline{\mathcal{M}}_{g,n}} \frac{\Lambda^{\vee}_g(1)}{\prod_{i = 1}^n (1 - \mu_i \psi_i)}
\eeq
D\'efinissons la s\'erie g\'en\'eratrice
\beq
\label{eq:Hgnci} H_{g,n}(t;\mathbf{v}) = \sum_{\mu\:/\:\ell(\mu) = n} t^{|\mu|}\,\mu_1\cdots\mu_n\,M_{\mu}(\mathbf{v})\,H_{g;\mu}
\eeq
o\`u les $M_{\mu}$ sont les mon\^{o}mes sym\'etriques  :
\beq
M_{\mu}(v_1,\ldots,v_n) = \frac{1}{|\mathrm{Aut}\,\mu|}\sum_{\sigma \in \mathfrak{S}_n} \prod_{i = 1}^n v_{\sigma(i)}^{\mu_i} \nn
\eeq
et $\mathbf{v} = (v_1,\ldots,v_n)$ est une suite de variables proches de $0$. Bouchard et Mari\~{n}o \cite{BMconj} ont d\'eriv\'e une application de la conjecture~\ref{cong:K} qui permet de les calculer. La courbe miroir de $\mathbb{C}^3$ dans la limite $f \rightarrow \infty$ a pour \'equation $x = -y + \ln(y/t)$. $y$ est une coordonn\'ee globale sur cette courbe, et on choisit le noyau de Bergman :
\beq
B(y_1,y_2) = \frac{\dd y_1\dd y_2}{(y_1 - y_2)^2} \nn
\eeq
pour d\'efinir la courbe spectrale $\mathcal{S}_{\mathrm{Lambert}}$.

\begin{conj}
\label{conjBM} (Bouchard et Mari{\~{n}}o)
\beq
H_{g,n}(t;\mathbf{v}) = \frac{\omega_n^{(g)}[\mathcal{S}_{\mathrm{Lambert}}](z_1,\ldots,z_n)}{\dd x(z_1)\cdots\dd x(z_n)} \nn
\eeq
o\`u $z_i$ est l'unique point de $\Sigma_{\widetilde{\mathbb{C}^3}}$ tel que $x(z_i) = \ln v_i$, et $y(z_i) \rightarrow 0$ lorsque $v_i \rightarrow 0$.
\end{conj}

Les int\'egrales de Hodge simple ont aussi une interpr\'etation combinatoire : ce sont des nombres de Hurwitz simples, que nous allons d\'efinir maintenant. Cette formule donne un algorithme efficace pour les calculer, genre par genre. Elle pr\'esente un int\'er\^{e}t ind\'ependamment du contexte de la th\'eorie des cordes o\`u elle est apparue.

\subsection{Les nombres de Hurwitz}
\label{sec:Hurwitz}

\begin{figure}[h!]
\begin{center}
\includegraphics[width=0.74\textwidth]{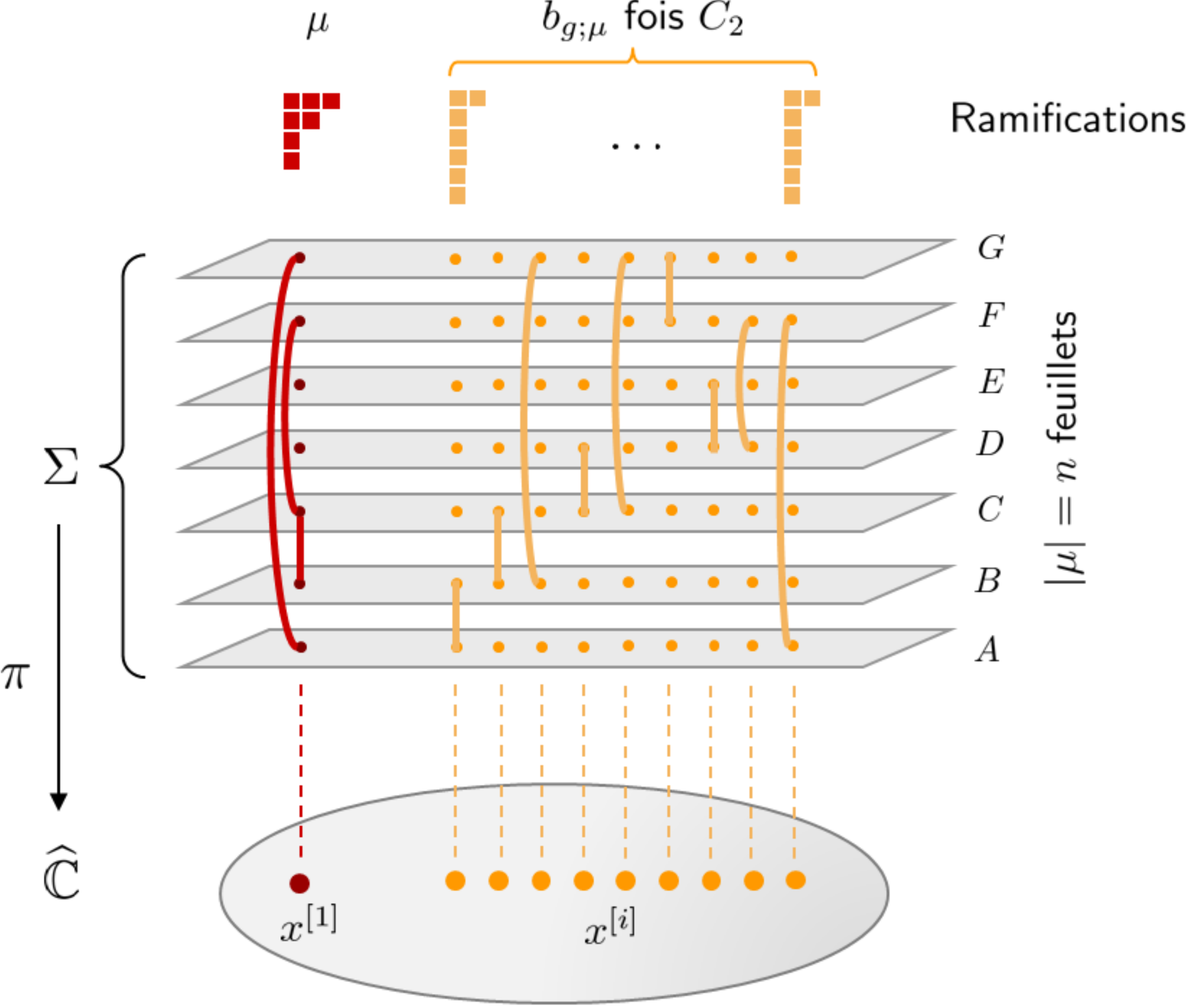}
\caption{\label{fig:Hur} Un exemple de recouvrement de $\widehat{\mathbb{C}}$, compt\'e dans les nombres de Hurwitz simples. $\Sigma$ est l'espace topologique constitu\'e de $n$ copies de $\widehat{\mathbb{C}}$ (les feuillets), apr\`es les identifications de points indiqu\'ees sur le dessin. Ici, $n = 7$, $b_{g;\mu} = 9$, $\ell(\mu) = 4$ donc $\Sigma$ est une surface de genre $0$.}
\end{center}
\end{figure}

Par d\'efinition, les \textbf{nombres de Hurwitz} $h(\mu^{[1]},\ldots,\mu^{[r]})$ comptent le nombre de recouvrements de la sph\`ere de Riemann par une surface de Riemann $\Sigma$ (\'eventuellement non connexe), \`a \'equivalence topologique pr\`es (Fig.~\ref{fig:Hur}).
 \beq
\label{eq:formq} h(\mu^{[1]},\ldots,\mu^{[r]}) = \sum_{\substack{[\pi\::\:\Sigma \rightarrow \widehat{\mathbb{C}}] \\ \mathrm{ramifications}\,\,\mu^{[1]},\ldots,\mu^{[r]}}} \frac{1}{|\mathrm{Aut}\,\pi|}
 \eeq
Le degr\'e de $\pi$ est le nombre de pr\'eimages (compt\'ees avec multiplicit\'es) d'un point $x \in \widehat{\mathbb{C}}$, et il ne d\'epend pas de $x$. Tous les $x$ ont exactement $n = \mathrm{deg}\,\pi$ pr\'eimages distinctes, sauf un nombre fini d'entre eux $x^{[1]},\ldots,x^{[r]} \in \widehat{\mathbb{C}}$. $\mu^{[i]}$ est une partition de $n$ boites, qui prescrit les degr\'es de ramification des pr\'eimages de $x_i$. Autrement dit, au voisinage d'une pr\'eimage $z^{[i]}_j$ de $x^{[i]}$ :
\beq
\pi(z) - \pi(z^{[i]}_j) \propto (z - z^{[i]}_j)^{\mu^{[i]}_j} \nn
\eeq
o\`u l'on a implicitement utilis\'e des coordonn\'ees locales dans $\widehat{\mathbb{C}}$ et $\Sigma$. On note $h_g(\mu^{[1]},\ldots,\mu^{[r]})$ le nombre de recouvrements par une surface connexe de genre $g$. $g$ est en fait contraint par le profil des ramifications, d'apr\`es la formule de Riemann-Hurwitz\label{RHU} :
\beq
2 - 2g = (2 - r)n + \sum_{i = 1}^r \ell(\mu^{[i]}) \nn
\eeq

Enum\'erer ces recouvrements revient \`a compter dans $\mathfrak{S}_n$ les d\'ecompositions $\sigma^{[1]}\circ \cdots \circ\sigma^{[r]} = \mathrm{id}$ de sorte que la classe de \label{sysys}conjugaison de $\sigma^{[i]}$ soit $\mu^{[i]}$, pond\'er\'ees par leur facteur de sym\'etrie. C'est un probl\`eme r\'esolu il y a longtemps\footnote{Lorsqu'il n'y a que des points de branchement simples, Hurwitz a reli\'e l'\'enum\'eration des recouvrements aux d\'ecompositions dans le groupe sym\'etrique \cite{Hurwitz0}, et a ensuite \'etabli la formule~\ref{eq:formq} \cite{Hurwitz1}.} en termes des caract\`eres $\chi_{\alpha}$ du groupe sym\'etrique :
\beq
\label{eq:formU} h(\mu^{[1]},\ldots,\mu^{[r]}) =  \sum_{\alpha\,\vdash\,n} \Big(\frac{\mathrm{dim}\,\alpha}{n!}\Big)^2 \prod_{i = 1}^{r} f_{\alpha}(\mu^{[i]}),\qquad f_{\alpha}(\mu) = \frac{n!}{\mathrm{dim\,\alpha}}\,\frac{\chi_{\alpha}(\mu)}{|\mathrm{Aut}\,\mu|}
\eeq
qui est une somme sur les partitions \`a $n$ boites.

On parle de \textbf{nombres de Hurwitz simples} lorsque tous les points de branchement\label{branch3} sont simples\footnote{Rappelons que $x^{[i]} \in \widehat{\mathbb{C}}$ est \textbf{point de branchement simple} si $\dd\pi(z)$ a exactement un z\'ero $z_0 \in \Sigma_g$, qui est simple.}, sauf un point de branchement $\mu$ arbitraire. \`{A} genre fix\'e, cela impose exactement :
\beq
b_{g;\mu} = 2g - 2 + n + \ell(\mu) \nn
\eeq
 points de branchement simple. La classe de conjugaison qui les d\'ecrit est celle d'une transposition, on la note $C_{2}$. $f_{\alpha}(C_2)$ co\"{i}ncide avec le \textbf{Casimir quadratique} de la repr\'esentation index\'ee par $\alpha$ :
\beq
\label{eq:erratum} f_{\alpha}(C_{2}) = \sum_{i = 1}^{\ell(\alpha)} \alpha_i\Big(\frac{\alpha_i}{2} - i + \frac{1}{2}\Big)
\eeq
La formule \label{ELSV}ELSV \cite{ELSV} relie les nombres de Hurwitz simples d\'efinis par la formule~\ref{eq:formU} aux int\'egrales de Hodge simple\footnote{La formule d\'ecouverte par ELSV est un cas particulier de la formule de Mari{\~{n}}o-Vafa.} de l'\'{E}qn.~\ref{eq:Hosimple} :
\beq
\label{eq:ELSV} H_{g,\mu} = \frac{|\mathrm{Aut}\,\mu|}{b_{g;\mu}!}\,h_{g}(\mu,\underbrace{C_2,\ldots,C_2}_{b_{g;\mu}\,\,\textrm{fois}})
\eeq

On parle de \textbf{nombres de Hurwitz doubles} lorsque tous les points de branchement sont simples, sauf deux points de branchement $\mu$ et $\nu$ arbitraires. Cette fois-ci, il y a exactement :
\beq
b_{g;\mu,\nu} = 2g - 2  + \ell(\mu) + \ell(\nu) \nn
\eeq
points de branchement simple. Cela donne une intepr\'etation combinatoire \`a la quantit\'e d\'efinie en \'{E}qn.~\ref{eq:Hdoubleg} :
\beq
\mathcal{H}_{\mu,\nu}(g_s) = \sum_{b \geq 0} \frac{g_s^b}{b!}\,|\mathrm{Aut}\,\mu|\,|\mathrm{Aut}\,\nu|\,h(\mu,\nu,\underbrace{C_2,\ldots,C_2}_{b\,\,\textrm{fois}}) \nn
\eeq

\section{R\'esultats r\'ecents}
\label{sec:zeroa}
Venons-en concr\`etement aux strat\'egies qui permettent d'aller dans le sens de la conjecture~\ref{cong:K}.

\subsection{Utilisation des mod\`eles de matrices}

Une premi\`ere approche est d'essayer de repr\'esenter la fonction de partition $Z$, qui s'\'ecrit comme sommes sur des partitions\footnote{Collision entre le nom consacr\'e "fonction de partition" en physique statistique, et le nom "partition" pour les suites $\lambda = (\lambda_1 \geq \lambda_2 \geq \ldots \geq \lambda_{\ell} \geq 0)$.} ou des partitions planes, par une chaine de matrices hermitiennes avec champ ext\'erieur (ou tout mod\`ele plus simple). Nous \label{devo5}savons alors que la r\'ecurrence topologique s'applique, notamment $\ln Z = \sum_{g \geq 0} g_s^{2g - 2}\,\mathcal{F}^{(g)}[\mathcal{S}]$, pourvu qu'un tel d\'eveloppement existe et que l'on ait identifi\'e la courbe spectrale $\mathcal{S}$.

La repr\'esentation par un mod\`ele de matrices est la partie astucieuse, mais facile car purement alg\'ebrique.

\vspace{0.2cm}

\noindent $\diamond\,$ On se restreint \`a des partitions de longueur $N$ fix\'ee, en excluant les partitions de longueur $> N$, et en\label{eq:yyy} compl\'etant celles de longueur $\ell < N$ par $N - \ell$ z\'eros.

\begin{figure}[h!]
\begin{minipage}[l]{0.59\linewidth}
\vspace{0.5cm}
\noindent $\diamond\,$ Les partitions $\mu$ sont en bijection avec des suites strictement d\'ecroissantes d'entiers $h_i(\mu) = \mu_i - i + \mathrm{cte}$, et l'on peut sym\'etriser :
{\small \beq
\sum_{\mu} w(\mathbf{h}(\mu),\bullet) = \frac{1}{N!}\sum_{h_1}\cdots\sum_{h_N} w(\mathbf{h},\bullet) \nn
\eeq}
\end{minipage}
\hfill \begin{minipage}[c]{0.4\linewidth}
\raisebox{-2cm}{\includegraphics[width=\textwidth]{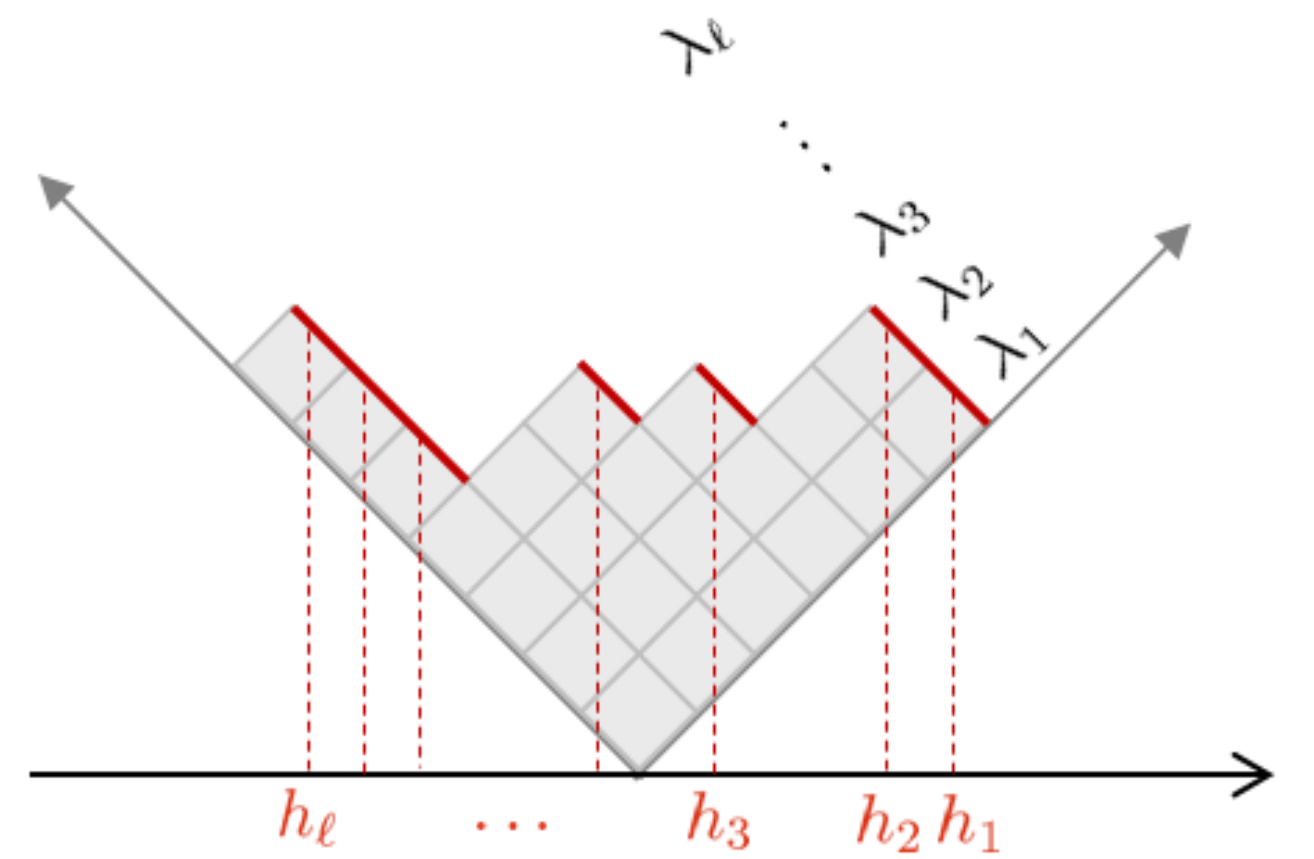}}
\end{minipage}
\end{figure}

\noindent $\diamond\,$ Les poids sur les partitions qui interviennent en th\'eorie topologique des cordes d\'erivent de notions de la th\'eorie des repr\'esentations de $\mathfrak{S}_n$. Ils d\'ependent analytiquement des $h_i$, et s'annulent lorsque deux $h_i$ sont \'egaux. La somme sur des entiers peut donc se repr\'esenter comme une int\'egrale de contour contre une fonction $f$ ayant des p\^{o}les simples avec r\'esidu $1$ en ces points :
    \beq
    \frac{1}{N!}\sum_{h_1}\cdots\sum_{h_N} w(\mathbf{h},\bullet) = \frac{1}{N!}\oint_{\mathcal{C}_1}\frac{\dd h_1\,f(h_1)}{2i\pi}\cdots\oint_{\mathcal{C}_N}\frac{\dd h_N\,f(h_N)}{2i\pi}\, w(\mathbf{h},\bullet) \nn
    \eeq

\vspace{0.2cm}

\noindent $\diamond\,$ Souvent, $w(\mathbf{h},\bullet)$ contient des d\'eterminants de \label{Vande3}Vandermonde $\Delta(u(\mathbf{h})) = \big[\prod_{1 \leq i < j \leq N} u(h_i) - u(h_j)\big]$. $u(h_i) = h_i$ et $u(h_i) = e^{-g_sh_i}$ sont des exemples rencontr\'es fr\'equemment. Cela sugg\`ere d'associer, \`a chaque partition, une matrice hermitienne $M = U\mathrm{diag}(u(h_1),\ldots,u(h_N))U^{\dagger}$.

\vspace{0.2cm}

\noindent $\diamond\,$ Enfin, quitte \`a introduire des matrices al\'eatoires auxiliaires, ou des matrices fix\'ees qui jouent le r\^{o}le de champ externe, on cherche \`a reconnaitre $Z$ comme l'int\'egrale d'une mesure d\'erivant d'une chaine de matrices. La formule d'Harish-Chandra (\'{E}qn.~\ref{eq:HarishChandra}) et la repr\'esentation du d\'eterminant de Cauchy par une int\'egrale de matrice (\'{E}qn.~\ref{eq:Cauint}) sont tr\`es utiles pour cette \'etape.\label{detCb}

\vspace{0.2cm}

D\`es lors, la proc\'edure du \S~\ref{sec:bout} s'applique :  il reste \`a prouver que $\ln Z$ et les corr\'elateurs ont un d\'eveloppement du type \ref{eq:dev} dans le petit param\`etre $g_s$, et \`a d\'eterminer la courbe spectrale $\mathcal{S}$ qui est solution de l'\'equation de boucle maitresse (\'{E}qn.~\ref{eq:probRHW10}). C'est la partie technique, dont la difficult\'e varie suivant les probl\`emes. Souvent, en pratique, \`a des transformations d'\'equivalence faible pr\`es, cette courbe spectrale ne d\'epend plus de $N$ au-del\`a d'un $N_0$ assez grand. C'est une manifestation du ph\'enom\`ene du cercle\label{arc} arctique \'evoqu\'e \`a la partie~\ref{sec:introuniv}. Empiriquement, s'il est assez facile de deviner une courbe spectrale solution, il est parfois difficile de prouver que c'est l'unique solution.

\subsection{Preuve de la conjecture de Bouchard et Mari\~{n}o}
\label{BMM2}
Ces id\'ees ont \'et\'e appliqu\'ees dans mon article \cite{BEMS} pour d\'emontrer la conjecture~\ref{conjBM}. Pour $\mathfrak{X} = \mathbb{C}^3$ \`a la limite de \textit{framing} infini, la fonction de partition s'\'ecrit :
\beq
\label{eq:jk} Z \propto \exp\Big\{\sum_{g \geq 0} g_s^{2g - 2} \sum_{\mu} t^{|\mu|}\Big[\prod_{i = 1}^{\ell(\mu)} p_{\mu_i}\Big]\,H_{g;\mu}\Big\}
\eeq
C'est une s\'erie g\'en\'eratrice d\'ependant d'une suite infinie de variables $p_1,\ldots,p_N,\ldots$ vivant pr\`es de $0$, qui permettent d'encoder les degr\'es des branchements. On peut toujours r\'ealiser $p_1,\ldots,p_N$ comme sommes de puissances de $N$ variables $v_1,\ldots,v_N$, i.e. $p_k = g_s \sum_{i = 1}^N v_i^k$. En suivant la proc\'edure pr\'ec\'edente, la s\'erie $Z_N$ d\'efinie par l'\'{E}qn.~\ref{eq:jk} restreinte \`a $\ell(\mu) \leq N$, se repr\'esente comme une int\'egrale formelle \`a une matrice hermitienne, de taille $N \times N$, dans un champ externe $\mathbf{R} = \mathrm{diag}(\ln v_1,\ldots,\ln v_N)$ :
\beq
\label{eq:momat} Z \propto \int_{\mathcal{H}(\Gamma^N)} \!\!\!\!\!\!\dd M e^{-\frac{1}{g_s}\Tr(V(M) - M\mathbf{R})}
\eeq
Le potentiel vaut :
\beq
V = -\frac{x^2}{2} + g_s(N - 1/2)x + x\ln(g_s/t) + i\pi x - g_s\ln\Gamma(-x/g_s) \nn
\eeq
Par construction (cf. \'{E}qn.~\ref{eq:jk}), $\ln Z$ a un d\'eveloppement topologique en $g_s$, et on peut aussi justifier que cela est aussi vrai des corr\'elateurs. Puisque $V$ d\'epend de $g_s$, la courbe spectrale d\'epend ici de $g_s$ (on la note $\mathcal{S}(g_s)$) et doit satisfaire l'\'equation de boucle maitresse (\'{E}qn.~\ref{eq:fdogih}). Comme nous avons affaire \`a une int\'egrale formelle, $\mathcal{S}(g_s)$ lorsque $g_s \rightarrow 0$ doit \^{e}tre une perturbation de la courbe spectrale d'un mod\`ele gaussien, qui est de genre $0$. Cet argument intuitif est justifi\'e dans \cite{BEMS}, d'o\`u l'on d\'eduit que $\mathcal{S}(g_s)$ admet un param\'etrage ($x(z),y(z)$) par une variable $z \in \mathbb{C}$. Cette information adjointe \`a l'\'equation de boucle maitresse d\'etermine une unique courbe spectrale $\mathcal{S}(g_s)$, que l'on sait exhiber. Par cons\'equent :
\beq
Z \propto \exp\Big\{\sum_{g \geq 0} g_s^{2g - 2}\,F_g\Big\},\qquad F_g = \mathcal{F}^{(g)}[\mathcal{S}(g_s)] \nn
\eeq
La derni\`ere \'etape de la preuve consiste \`a remarquer :
\beq
H_{g,n}(t;\mathbf{v}) = \lim_{g_s \rightarrow 0} \frac{1}{g_s^n}\,\frac{\partial^n F_g}{\partial v_1\cdots \partial v_n}\Big|_{v_{n + 1} = \cdots = v_N = 0} \nn
\eeq
Les propri\'et\'es de g\'eom\'etrie sp\'eciale (cf. \S~\ref{sec:geomspec}) de la r\'ecurrence topologique permettent de calculer cette d\'eriv\'ee $n^{\textrm{\`{e}me}}$ en fonction de $\omega_n^{(g)}[\mathcal{S}(g_s)]$. On trouve que le cycle dual de $\partial_{v_i}$ lit simplement une forme diff\'erentielle dans la coordonn\'ee $x$. Cette relation passe bien \`a la limite $g_s \rightarrow 0$ :
\beq
H_{g,n}(t;\mathbf{v}) = \frac{\omega_n^{(g)}[\mathcal{S}(g_s = 0)](z_1,\ldots,z_n)}{\dd x(z_1)\cdots \dd x(z_n)}\qquad z_i = \ln v_i \nn
\eeq
\label{Lambert3}Enfin, $\mathcal{S}(g_s = 0)$ est exactement la courbe spectrale de Lambert pr\'edite par Bouchard et Mari\~{n}o.

\subsection{Vers la conjecture BKMP}
\label{vBLM}
Pour un CY$_3$ torique quelconque $\mathfrak{X}$, le dual $\Gamma_{\mathfrak{X}}^*$ peut \^{e}tre construit \`a partir d'un r\'eseau $\mathfrak{R}_{P\times Q}$ rectangulaire de taille $P \times Q$ assez grande apr\`es une succession de \textbf{renversements} de faces, et l'envoi de certains param\`etres de K\"{a}hler vers l'infini (ce qui efface la partie du diagramme \`a "distance infinie"). Les invariants de Gromov-Witten ne changent pas lors des renversements, pourvu que l'on suive la trace des param\`etres de K\"{a}hler $t_{[C]}$ \`a chaque \'etape. Il suffit donc de les \'etudier pour les CY$_3$ toriques $\mathfrak{X}_{P\times Q}$ associ\'es \`a $\mathfrak{R}_{P\times Q}$. Eynard, Kashani-Poor et Marchal \cite{EKM1} ont trouv\'e que $Z(\mathfrak{X}_{P\times Q})$ donn\'e par le vertex topologique comme une somme sur une famille de partitions, se repr\'esente comme une chaine de matrices hermitiennes, avec un champ externe \`a chaque extr\'emit\'e (Fig.~\ref{eq:ZchainBKMP}). Ils ont ensuite montr\'e \cite{EKM2} que la courbe miroir $\Sigma_{\widetilde{\mathfrak{X}}_{P\times Q}}$ \'etait une solution de l'\'equation de boucle maitresse (\'{E}qn.~\ref{eq:probRHW10}).

\begin{landscape}
\addtolength{\footskip}{30pt}
\addtolength{\linewidth}{50pt}
\begin{figure}
\begin{center}
\raisebox{-3.5cm}{\includegraphics[width=1.43\textwidth]{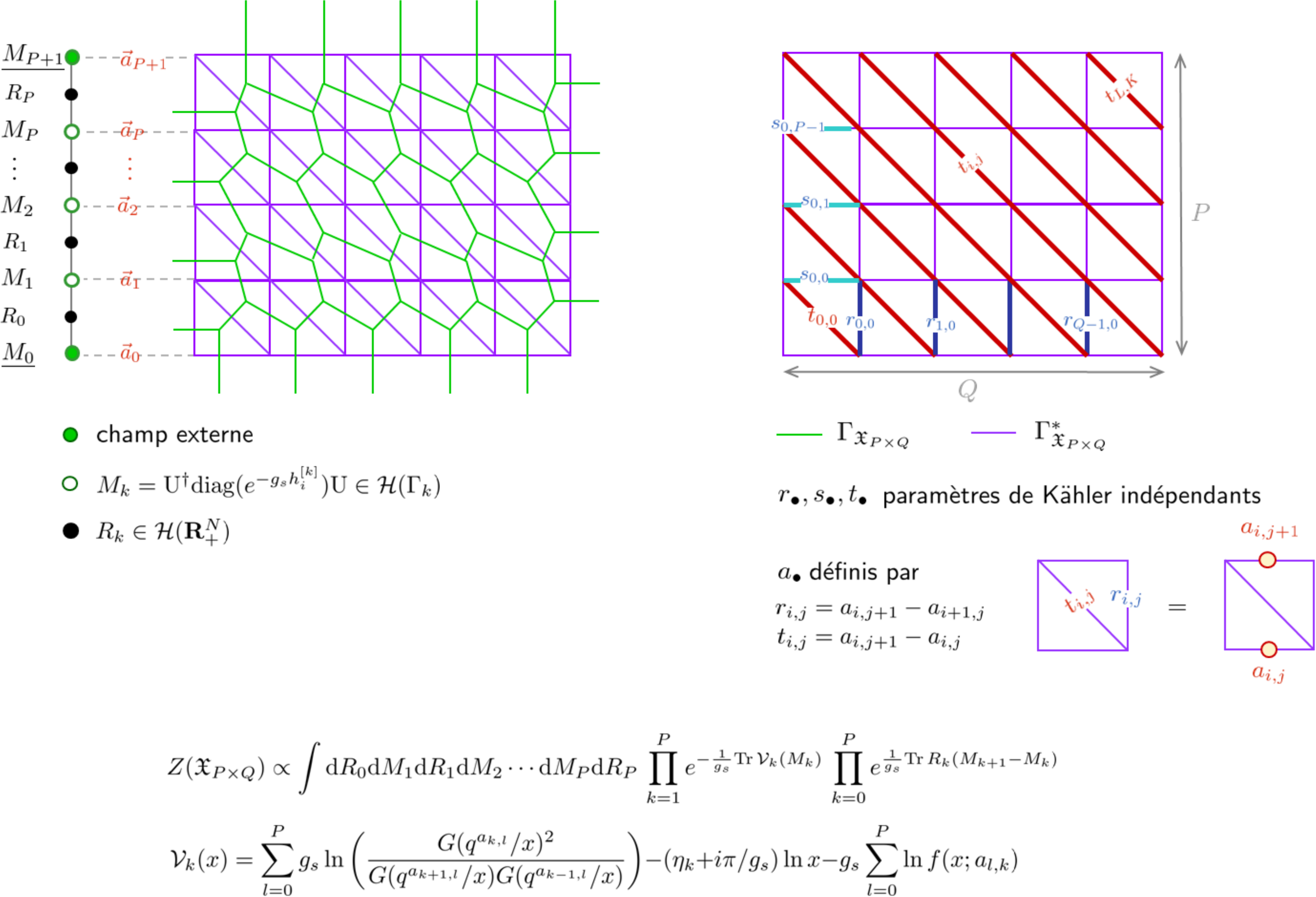}}
\caption{\label{eq:ZchainBKMP} Le mod\`ele de matrices de \cite{EKM1} pour $\mathfrak{X}_{P \times Q}$. $G_q(x) = \prod_{n = 1}^{\infty}(1 - q^n/x)$ est le dilogarithme quantique. Les $\eta_k$ d\'ependent des param\`etres de K\"{a}hler $s_k$.}
\end{center}
\end{figure}
\end{landscape}

Il faudrait compl\'eter deux points pour en d\'eduire une preuve de la conjecture~\ref{cong:K}.

\vspace{0.2cm}

\noindent $\diamond\,$ Le vertex topologique construit $Z$ comme une s\'erie en puissances de $e^{-g_s}$, lorsque $g_s \rightarrow \infty$ (cf. \S~\ref{sec:topover}). Gr\^{a}ce \`a la correspondance GW/DT \'etablie par \cite{GWDT} (cf. \S~\ref{sec:thAB}), on sait que $\ln Z$ a bien un d\'eveloppement en puissance de $g_s$ lorsque $g_s \rightarrow 0$. Il faudrait montrer \label{coco34}que c'est aussi le cas des corr\'elateurs $W_{n;p} = \big\langle \prod_{i = 1}^n \Tr \frac{1}{x_i - M_p}\big\rangle_c$ et des corr\'elateurs mixtes. Ce sont des travaux en cours, auxquels je participe. Nous favorisons pour l'instant des techniques de polyn\^{o}mes orthogonaux pour \'etudier le mod\`ele de matrices de la Fig.~\ref{eq:ZchainBKMP}. \`{A} l'heure de l'\'ecriture, la fa\c{c}on dont $W_{n;p}$ encode des observables ouvertes du mod\`ele A n'est pas encore tr\`es claire.

\vspace{0.2cm}

\noindent $\diamond\,$ Il faudrait \label{cm2} d\'emontrer que la courbe miroir est l'\textit{unique} courbe spectrale solution de l'\'equation de boucle maitresse qui \textit{minimise} le pr\'epotentiel (cf. \S~\ref{sec:bout}). Cette \'etape parait assez difficile.

\subsection{Autres m\'ethodes}

\subsubsection{Relations combinatoires et \'equations de boucles}
\label{heqb4}
Les nombres de Hurwitz simples (\S~\ref{sec:Hurwitz}) v\'erifient des relations combinatoires r\'ecursives, appel\'ees \'equations \textit{cut-and-join}, red\'ecouvertes et \'etudi\'ees par Goulden, Jackson \cite{GJ} et Vakil \cite{VakilThese}. Elles sont illustr\'ees \`a la Fig.~\ref{fig:cutjoin}. Leur structure est tr\`es similaire \`a celles des \'equations de boucles (Chapitre~\ref{chap:toporec}) ou des relations combinatoires r\'ecursives que v\'erifient les cartes (Chapitre~\ref{chap:formel}) : il y a deux types de termes, selon l'effet d'un coup de ciseau dans une surface $\Sigma_g$ qui recouvre $\widehat{\mathbb{C}}$. Eynard, Mulase et Safnuk ont montr\'e dans \cite{EMS} que les \'equations \textit{cut-and-join} sont en fait \'equivalentes aux \'equations de boucles pour la courbe spectrale de Lambert. Ceci donne une seconde preuve de la conjecture de Bouchard et Mari\~{n}o, sans recourir \`a la repr\'esentation par un mod\`ele de matrices. Il existe \'egalement des \'equations \textit{cut-and-join} pour les int\'egrales de Hodge \label{Hodge4} triple, d\'ependant d'un \textit{framing} $f$. G\'en\'eralisant \cite{EMS}, Zhou a montr\'e dans \cite{Zhou1,Zhou2} qu'elles sont \'equivalentes aux \'equations de boucles pour la courbe spectrale miroir de $\mathbb{C}^3$, dont l'\'equation est $e^{fy} + e^{x} + e^{y(f + 1)} = 0$ (cf. \S~\ref{sec:exCY}). Il a ainsi obtenu une preuve de la conjecture BKMP lorsque $\mathfrak{X} = \mathbb{C}^3$. Le recollement de plusieurs cartes $\mathbb{C}^3$ pour atteindre la conjecture BKMP g\'en\'erale devrait \^{e}tre possible en suivant ces techniques. C'est un travail entrepris notamment par Nicolas Orantin.

\begin{landscape}
\addtolength{\footskip}{30pt}
\addtolength{\linewidth}{50pt}
\begin{figure}
\begin{center}
\raisebox{-3cm}{\includegraphics[width = 1.5\textwidth]{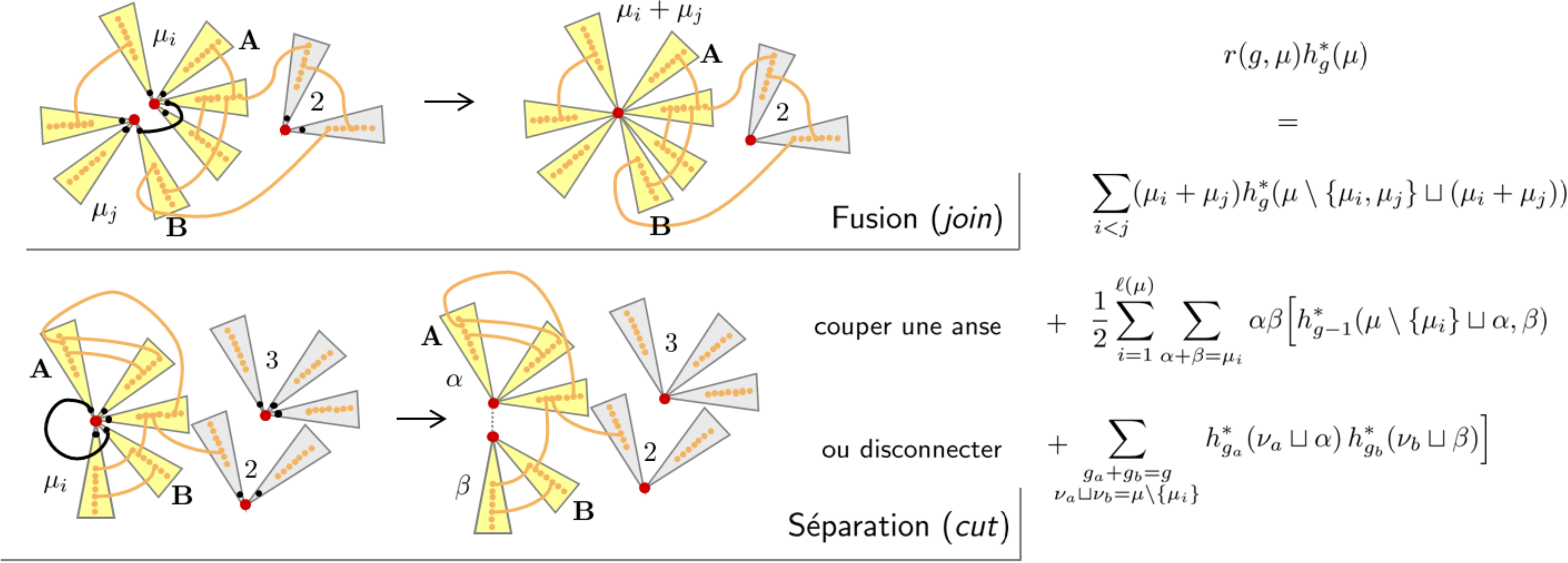}}
\caption[]{\label{fig:cutjoin} Partons d'un recouvrement de genre $g$ de $\widehat{\mathbb{C}}$ qui admet un point de branchement non trivial $x^{[1]}$, et $b_{g;\mu}$ points de branchement simple $x^{[2]},x^{[3]},\ldots$, et o\`u les feuillets sont num\'erot\'es. Le dessin r\'esume ce qu'il se passe lorsque l'on fait tendre $x^{[2]}$ vers $x^{[1]}$. $x^{[2]}$ est un point de branchement simple, qui provient de l'identification de deux points $x^{[2]}_{\mathbf{A}}$ et $x^{[2]}_{\mathbf{B}}$ appartenant \`a des feuillets \textbf{A} et \textbf{B} au-dessus de $x^{[2]}$. Notons \'egalement $x^{[1]}_{\mathbf{A}}$ et $x^{[1]}_{\mathbf{B}}$ les points de $\Sigma$ qui sont dans les feuillets \textbf{A} et \textbf{B} au-dessus de $x^{[1]}$. Si $x^{[1]}_{\mathbf{A}} \neq x^{[1]}_{\mathbf{B}}$, la limite $x^{[2]} \rightarrow x^{[1]}$ est une situation o\`u la ligne de $\mu$ o\`u figure \textbf{A} et celle o\`u figure \textbf{B} sont fusionn\'ees en une seule. L'effet sur $\Sigma$ est d'identifier les feuillets \textbf{A} et \textbf{B} au-dessus de $x^{[1]}$, i.e. o\`u les deux lignes. Si $x^{[1]}_{\mathbf{A}} = x^{[1]}_{\mathbf{B}}$, la limite $x^{[2]} \rightarrow x^{[1]}$ est une situation o\`u l'on a retir\'e la ligne de $\mu$ contenant \textbf{A} et \textbf{B}, pour la remplacer par une ligne o\`u figure \textbf{A} et une ligne o\`u figure \textbf{B}. L'effet sur $\Sigma$ est de s\'eparer les feuillets \textbf{A} et \textbf{B} au-dessus de $\Sigma$, ce qui lui coupe une anse ou bien la disconnecte. Cette proc\'edure construit un nouveau recouvrement de $\widehat{\mathbb{C}}$ qui poss\`ede $b - 1$ points de branchement simple. Pour faire le chemin inverse, il faut tenir compte des possibilit\'es de renum\'erotation des feuillets. Le r\'esultat est l'\'equation \textit{cut-and-join} pour les nombres de Hurwitz simples $h^*_{g}(\mu) \equiv |\mathrm{Aut}\,\mu|\,h_g(\mu,\underbrace{C_2,\ldots,C_2}_{b_{g;\mu}\,\,\mathrm{fois}})$.}
\end{center}
\end{figure}
\end{landscape}

\subsubsection{Nombres d'intersections dans $\overline{\mathcal{M}}_{g,n}$}
\label{cls2}
Nous avons vu au \S~\ref{sec:interFg} que les $\omega_n^{(g)}[\mathcal{S}_{\mathrm{K}}]$ pour les courbes spectrales $\mathcal{S}_{\mathrm{K}} = [\Sigma,x,y,B_{\mathrm{K}}]$ admettant un point de branchement mou $x = a$, et o\`u le noyau de Bergman est :
\beq
\label{eq:Bergm} B_{\mathrm{K}}(x_1,x_2) = \frac{\dd\sqrt{x_1 - a}\,\dd\sqrt{x_2 - a}}{\big(\sqrt{x_1 - a} - \sqrt{x_2 - a}\big)^2}
\eeq
sont des s\'eries g\'en\'eratrices de nombres d'intersections. D\`es que $2 - 2g - n < 0$ :
\bea
\omega_n^{(g)}[\mathcal{S}](z_1,\ldots,z_n) & = & \sum_{d_1 + \cdots + d_n = d_{g,n}} \Big[\prod_{i = 1}^n \dd \varsigma_{d_i}(z_i)\Big] \int_{\overline{\mathcal{M}}_{g,n}} \!\!\!\! K^{\vee}[\mathcal{S}] \prod_{i = 1}^n \psi_i^{d_i} \nn \\
\label{eq:ELSq} \dd \varsigma_{k} & = &  (2k + 1)!!\,\frac{\dd \sqrt{x - a}}{(x - a)^{k + 1}}
\eea
$K^{\vee}[\mathcal{S}]$ est une s\'erie g\'en\'eratrice de classes $\kappa_r$ d\'efinie par les \'{E}qns.~\ref{eq:hujs} et \ref{eq:Kvee}.

Il devrait exister une relation similaire lorsque l'on choisit un autre noyau de Bergman. De plus, il devrait \^{e}tre possible de faire des "recollements" pour arriver \`a des courbes spectrales \`a plusieurs points de branchement. Actuellement, ce recollement est bien connu au niveau de la fonction de partition $Z$ \cite{Orantin1}, et il faudrait comprendre ce qu'il advient au niveau des fonctions de corr\'elations et de l'\'{E}qn.~\ref{eq:ELSq}. Ce sont des travaux en cours.

Par exemple, la courbe spectrale calculant les nombres de Hurwitz est :
\beq
\mathcal{S}\: : \: x(y) = -y + \ln(y/t),\qquad B(y_1,y_2) = \frac{\dd y_1\,\dd y_2}{(y_1 - y_2)^2} \nn
\eeq
Le noyau de Bergman est diff\'erent de $B_{\mathrm{K}}$, donc l'\'{E}qn.~\ref{eq:ELSq} ne s'applique pas. Cependant, on sait de fa\c{c}on indirecte, par la formule ELSV (\'{E}qn.~\ref{eq:ELSV}) :
\bea
\omega_n^{(g)}[\mathcal{S}](y_1,\ldots,y_n) & = & \sum_{d_1 + \cdots + d_n = d_{g,n}} \Big[\prod_{i = 1}^n \dd\varsigma_{d_i}(y_i)\Big] \int_{\overline{\mathcal{M}}_{g,n}} \!\!\!\!\Lambda_g^{\vee}(1)
 \prod_{i = 1}^n \psi_i^{d_i} \nn \\
 \dd\varsigma_k(y) & = & \sum_{m = 1}^{\infty} \frac{m^{m + 1 + k}}{m!}\,x^{m - 1} \dd x  \nn
 \eea
\label{ELSVSS}et $\Lambda^{\vee}(1)$ selon Mumford est une s\'erie g\'en\'eratrice de classes $\psi_i$ et $\kappa_r$. Une formule g\'en\'eralisant l'\'{E}qn.~\ref{eq:ELSq} pour un noyau de Bergman quelconque permettrait de mieux comprendre l'apparition des int\'egrales de Hodge du point de vue de la r\'ecurrence topologique. Dans le cas de la courbe de Lambert, on en d\'eduirait sans doute une nouvelle preuve de la formule ELSV.

\subsubsection{Hi\'erarchies int\'egrables}

Les observables de th\'eorie topologique des cordes sont solutions de syst\`emes d'\'equations diff\'erentielles int\'egrables. Une autre voie \`a explorer pour d\'emontrer la conjecture BKMP serait de les relier aux syst\`emes int\'egrables venant des \'equations de boucles, dont on a \'evoqu\'e la construction au chapitre~\ref{chap:int}. Les travaux r\'ecents d'Andrea Brini \cite{BriniA}, o\`u sont d\'ecrits pr\'ecis\'ement ces hi\'erarchies int\'egrables et l'op\'erateur d'insertion dans le secteur ouvert, seraient un bon point de d\'epart.

\section{Perspectives}
\label{sec:drop}
La plupart des aspects des th\'eories topologiques des cordes peuvent \^{e}tre unifi\'es du point de vue de la r\'ecurrence topologique. Il n'est donc pas \'etonnant d'y rencontrer, outre les propri\'et\'es que l'on vient d'\'evoquer, des syst\`emes int\'egrables, des nombres d'intersections dans $\overline{\mathcal{M}}_{g,n}$, des mod\`eles de matrices, des sommes sur les partitions ou des partitions planes, \ldots{} Les \'equations de boucles/contraintes de Virasoro, dont la r\'ecurrence topologique fournit les solutions, sont un d\'enominateur commun \`a tous ces probl\`emes. Pour rendre ce point de vue complet, il faudrait mieux comprendre la relation g\'en\'erale entre r\'ecurrence topologique et l'intersection de classes tautologiques dans $\overline{\mathcal{M}}_{g,n}$ (i.e. les int\'egrales m\^{e}lant classes $\psi_i$, $\kappa_r$, et $\lambda_h$). La fin de ce chapitre \'evoque des questions r\'ecentes, auxquels ces outils seront sans doute utiles dans le futur, autant du point de vue technique que de la compr\'ehension des ph\'enom\`enes math\'ematiques en jeu.

\subsubsection{Mod\`eles de matrices g\'en\'eralis\'es}
\label{shys}
Actuellement, on ne sait appliquer la r\'ecurrence topologique en g\'en\'eral qu'aux int\'egrales de matrices du type :
\beq
Z_{\textrm{cha\^{i}ne}} \propto \int \Big[\prod_{i} \dd M_i\Big]\,\exp\Big\{-\frac{1}{g_s}\Tr\Big(\sum_{i} V_i(M_i) + c_{i,i+1}\,M_i M_{i + 1}\Big)\Big\} \nn
\eeq
Notamment, cela suppose que la quantit\'e en exposant ne contienne pas de produits de traces, et que les valeurs de propres de $M_i$ subissent toutes le m\^{e}me potentiel $V_i$. Autrement dit, $w(\mathbf{h},\bullet)$, outre les quantit\'es absorb\'ees dans les mesures $\dd M_i$, ne doit pas de facteurs $e^{\sum_{i,j} h_ih_j}$, $e^{\sum_{i,j,k} h_ih_jh_k}$, \ldots{} ni de $e^{\sum_i f(i)h_i}$. Dans le mod\`ele de \cite{EKM1}, ces deux types de facteurs sont pr\'esents au d\'ebut du calcul. Toutefois, ils disparaissent \`a la fin gr\^{a}ce aux relations entre param\`etres de K\"{a}hler impos\'ees par la condition de Calabi-Yau. Si l'on peut toujours\label{CYs3} imaginer d'autres g\'eom\'etries d\'efinies \`a partir de $\Gamma_{\mathfrak{X}}$ avec des param\`etres de K\"{a}hler $t_{[C]}$ tous ind\'ependants, les g\'eom\'etries de Calabi-Yau se distinguent par leur simplicit\'e du point de vue du mod\`ele de matrices.

\beq
\label{eq:gfdqa}\int_{\Gamma} \Big[\prod_{i = 1}^N \dd\lambda_i w(\lambda_i)\Big]\,\Big[\prod_{1 \leq i < j \leq N} K_{\mathrm{sing}}(\lambda_i,\lambda_j)\Big]\,\Big[\prod_{i,j = 1}^N K_{\mathrm{reg}}(\lambda_i,\lambda_j)\Big]
\eeq
peut \^{e}tre consid\'er\'e comme un  \textbf{mod\`ele de matrice g\'en\'eralis\'e}. Ce type d'int\'egrales et sa g\'en\'eralisation "\`a plusieurs matrices" se manifeste dans plusieurs probl\`emes int\'eressants. Nous les avons d\'ej\`a rencontr\'ees dans les mod\`eles de boucles sur des surfaces al\'eatoires au Chapitre~\ref{chap:formel} (\'{E}qn.~\ref{eq:gaaa}). On les retrouve souvent dans le contexte de la physique des hautes \'energies. Voici quelques exemples :

\vspace{0.2cm}

\noindent $\diamond\,$ La th\'eorie de Chern-Simons \label{CS11} sur les espaces de Seifert $X(p_1/q_1,\ldots,p_K/q_K)$, qui est duale au mod\`ele A sur $T^*X(p_1/q_1,\ldots,p_K/q_K)$. Ici, $p_1,\ldots,p_K$ sont des entiers deux \`a deux premiers, et $q_i$ est un entier premier avec $p_i$. La fonction de partition s'\'ecrit \cite{MarinoSei} :
{\small \beq
Z \propto \int_{\mathbb{R}^N} \Big[\prod_{i = 1}^N \dd \lambda_i e^{-\frac{\lambda_i^2}{2g_s} + \big(- 2(K - 2)(N - 1) + \sum_{k = 1}^K (N - 1)/p_k - t_k\big) \lambda_i}\Big]\,\big(\Delta(e^{-\lambda})\big)^2\,\prod_{k = 1}^{K} \frac{\Delta(e^{-\lambda/p_k})}{\Delta(e^{-\lambda})} \nn
\eeq}
$\!\!\!$Dans le cas particulier des espaces lenticulaires $L(p,q) = X(p/q)$, la courbe spectrale de ce mod\`ele est connue \cite{Halma}.

\vspace{0.2cm}

\noindent $\diamond\,$ La fonction de partition \label{fonp7}introduite par Nekrasov pour les th\'eories de jauge $\mathcal{N} = 2$ superconformes \cite{Nekra}. Dans le cadre des dualit\'es entre th\'eories de cordes et th\'eories de jauge \`a $4$, $5$ voire $6$ dimensions$_{\mathbb{R}}$, cette fonction de Nekrasov r\'ealise une d\'eformation \`a deux param\`etres $(\epsilon_1,\epsilon_2)$ de la somme sur les instantons du mod\`ele A. \label{sukj}En notant $\beta = -\epsilon_1/\epsilon_2$, Su{\l}kowski \cite{Sulko} la r\'e\'ecrit :
{\footnotesize \bea
Z & = & \sum_{\mu^{[1]},\ldots,\mu^{[r]}} e^{-g_s\sum_{j} |\mu^{[j]}|} w_{\mathrm{vect}}\,w_{\mathrm{(anti)fond}} \nn \\
w_{\mathrm{vect}}& = & \prod_{(j,i) \neq (j',i')} \frac{G(\mu^{[i]}_j - \mu^{[i']}_{j'} + \beta(i' - i) + b^{[j]} - b^{[j']} + \beta)}{G(\mu^{[i]}_j - \mu^{[i']}_{j'} + \beta(i' - i) + b^{[j]} - b^{[j']})}\,\frac{G(\beta(i' - i) + b^{[j]} - b^{[j']})}{G(\beta(i' - i) + b^{[j]} - b^{[j']} + \beta)} \nn \\
w_{\mathrm{fond}} & = & \prod_{j} \prod_{k = 1}^{N_f} \prod_{i = 1}^{\ell(\mu^{[j]})} \frac{G(\mu^{[j]}_i + b^{[j]} - M_k - i\beta + 1)}{G(b^{[j]} - M_k - i\beta + 1)} \nn
\eea}
$\!\!\!$Pour d\'ecrire rapidement cette formule\footnote{Nous avons \'evit\'e d'inclure des champs de Chern-Simons, qui conduisent \`a un poids $w_i^{[j]}(\mu^{[j]}_i)$ qui d\'ependent explicitement de $i$, ce qui complique encore l'analyse.} : $w_{\mathrm{vect}}$ est un poids associ\'e \`a des champs de mati\`ere dans la repr\'esentation fondamentale de $\mathrm{U}(N)$, $w_{\mathrm{(anti)fond}}$ est associ\'e \`a $N_f$ multiplets de masses $m_k = \epsilon_2 M_k$ vivant dans la repr\'esentation (anti)fondamentale de $\mathrm{U}(N)$. Selon la dimension$_{\mathbb{R}}$ de la th\'eorie de jauge :
\beq
G_{\mathrm{4d}}(x) = \Gamma(x),\qquad G_{\mathrm{5d}}(x) = \frac{1}{\prod_{n = 0}^{\infty} (1 - e^{\epsilon_2(n + x)})} \nn
\eeq
Cela conduit \`a un mod\`ele de matrice g\'en\'eralis\'e o\`u les facteurs $(\lambda - \mu)$ du d\'eterminant de Vandermonde sont remplac\'es par une fonction plus compliqu\'ee $K_{\mathrm{sing}}(\lambda,\mu) = G(\lambda - \mu)$, singuli\`ere lorsque $\lambda \rightarrow \mu$. \cite{Sulko} en a d\'etermin\'e la courbe spectrale seulement dans la limite $\epsilon_2 \rightarrow 0$, o\`u $K_{\mathrm{sing}}$ se r\'eduit \`a un d\'eterminant de Vandermonde.

\vspace{0.2cm}

\noindent $\diamond\,$ La conjecture \label{AGTss} d'Alday-Gaiotto-Tachikawa \cite{AGTc} \'enonce une relation entre ces th\'eories de jauge et la th\'eorie conforme de Liouville de charge centrale \label{cc3} :
 \beq
\mathfrak{c} = 1 + 6\Big(\mathfrak{b}_0 + \frac{1}{\mathfrak{b}_0}\Big)^2,\qquad \mathfrak{b}_0 = \sqrt{\epsilon_1/\epsilon_2} \nn
 \eeq
 Elle est \`a l'origine d'une activit\'e de recherche intense ces deux derni\`eres ann\'ees. Mentionnons seulement deux points. Du c\^{o}t\'e CFT, Bonelli et al. \cite{Bonelli} ont propos\'e un mod\`ele de matrice g\'en\'eralis\'e calculant les corr\'elateurs de la th\'eorie de Liouville sur une surface de Riemann $\Sigma_{g,n}$ de genre $g$ \`a $n$ points marqu\'es. Il fait intervenir, \`a la place du d\'eterminant de Vandermonde, la forme primaire $K_{\mathrm{sing}}(\lambda,\mu) = E(\lambda,\mu)$, i.e. une \label{the2}fonction th\^{e}ta de genre $g$ qui s'annule \`a points co\"{i}ncidants (\'{E}qn.~\ref{eq:prima}). Dans le cas $g = 1$ par exemple, $\Sigma_{1,n} \simeq \mathbb{C}/(\mathbb{Z} \oplus \tau\mathbb{Z})$, avec des points marqu\'es $z_1,\ldots,z_n \in \mathbb{C}$ :
 \beq
 Z_{g = 1,n} \propto \int_{\Gamma} \Big[\prod_{i = 1}^N \dd \lambda_i\,e^{-\frac{N\beta}{g_s}\big(8i\pi^2p \lambda_i + \sum_{k = 1}^n \mu_k \ln\vartheta_1(\lambda_i - z_k|\tau)\big)}\Big]\,\prod_{1 \leq i < j \leq N} \vartheta_1(\lambda_i - \lambda_j|\tau)^{2\beta} \nn
 \eeq
Le nombre $N$ de valeurs propres et l'exposant $\beta$ sont li\'es aux param\`etres de la th\'eorie de Liouville :
 \beq
 N = \frac{\sum_{k = 1}^n m_k}{\mathfrak{b}_0g_s},\qquad \beta = -\mathfrak{b}_0^2 \nn
 \eeq
et $\mu_i$ est la masse partielle plac\'ee au point $z_i$. Les auteurs de \cite{Bonelli} ont trouv\'e la courbe spectrale de ce mod\`ele. Du c\^{o}t\'e th\'eorie de jauge, ont \'et\'e identifi\'es depuis longtemps \cite{ShataNek} des syst\`emes int\'egrables o\`u la projection $x$ ne vit pas sur $\widehat{\mathbb{C}}$ mais sur une surface de Riemann $\Sigma_{0;g}$ de genre $g > 0$ : on troque une matrice de Lax $\mathbf{L}(x)$ rationnelle en $x$, pour une fonction m\'eromorphe sur $\Sigma_{0;g}$. Ce sont des exemples de\label{Hitchin} \textbf{syst\`emes de Hitchin}. Il serait int\'eressant de d\'evelopper l'analogue de la th\'eorie du chapitre~\ref{chap:int} pour ces syst\`emes int\'egrables : est-elle ou non fondamentalement diff\'erente ?

\vspace{0.2cm}

Lorsque les m\'ethodes \'ebauch\'ees au \S~\ref{sec:eqju} pour \'etudier les mod\`eles de matrices g\'en\'eralis\'es via leurs \'equations de boucles auront \'et\'e approfondies, j'esp\`ere qu'elles seront applicables \`a ces probl\`emes. L'enjeu est bien s\^{u}r de calculer toutes les corrections en $g_s$, \`a partir de la courbe spectrale $g_s \rightarrow 0$. On s'attend \`a ce que la r\'eponse soit donn\'ee par une r\'ecurrence topologique, et b\'en\'eficie de toutes ses propri\'et\'es.



\newpage
\thispagestyle{empty}
\phantom{bbk}

\newpage

\chapter{Conclusion}
\label{eq:conclu}
Toutes les cons\'equences des m\'ethodes bas\'ees sur les \'equations de boucles, et leur solution via la r\'ecurrence topologique, n'ont pas \'et\'e exploit\'ees. Quand elles s'appliquent, ces techniques offrent un algorithme de calcul efficace et permettent d'\'etablir des propri\'et\'es structurelles des quantit\'es que l'on calcule. Les cartes al\'eatoires (avec ou sans boucles), les nombres de Hurwitz simples, le d\'eveloppement asymptotique des int\'egrales de matrices, et plus g\'en\'eralement des fonctions tau des syst\`emes int\'egrables, \ldots{} rentrent dans ce cadre. Un grand chantier reste ouvert pour \'etablir les relations conjectur\'ees actuellement, avec les invariants de Gromov-Witten et la th\'eorie topologique des cordes, les nombres de Hurwitz doubles, la fonction de partition de Nekrasov, les syst\`emes de Hitchin, les th\'eories conformes coupl\'ees \`a la th\'eorie de Liouville, \ldots{} D'autres relations avec l'int\'egrabilit\'e quantique ou la th\'eorie des n{\oe}uds sont plus incertaines. La liste des applications potentielles n'est sans doute pas close.

J'ai l'intention de me concentrer sur trois probl\`emes principaux dans les mois qui viennent.

\vspace{0.2cm}

$\diamond\,$ La r\'esolution des \'equations de boucles des mod\`eles de matrices g\'en\'eralis\'es. J'ai quelques id\'ees pour une m\'ethode g\'en\'erale de r\'esolution, la partie difficile \'etant de r\'esoudre l'\'equation maitresse. Cela aurait de nombreuses applications en th\'eorie des n{\oe}uds (mod\`eles de matrices venant des th\'eories de Chern-Simons), en th\'eorie topologique des cordes, en th\'eorie de Liouville coupl\'ee \`a la CFT, \ldots{} En physique statistique, cela conduirait probablement \`a de nouveaux types de points critiques, lorsque les coupures images ou/et les coupures fant\^{o}mes se rejoignent de diff\'erentes mani\`eres.
Ces points critiques sont pertinents par exemple dans les mod\`eles de boucles sur r\'eseau al\'eatoire.

\vspace{0.2cm}

$\diamond\,$ Le d\'eveloppement asymptotique des mod\`eles de matrices (g\'en\'eralis\'es), en particulier une justification rigoureuse de la s\'erie th\^{e}ta (\'{E}qn.~\ref{eq:resul}) \`a partir des m\'ethodes d'\'equations de boucles. Une premi\`ere \'etape pourrait consister \`a \'etudier les mod\`eles int\'egrables, comme le mod\`ele \`a deux matrices, ou les syst\`emes de Lax de taille $d \times d$.

\vspace{0.2cm}

$\diamond\,$ La statistique des chemins dans le mod\`ele $\On$. J'ai commenc\'e \`a travailler avec Bertrand Eynard sur le processus al\'eatoire constitu\'e par un chemin de longueur $l$ dans une carte al\'eatoire du mod\`ele $\On$, par des m\'ethodes purement combinatoires. Le seul pr\'eliminaire est le calcul des observables uniformes qui est complet d'apr\`es mon article \cite{BEOn}. Nous avons des r\'esultats exacts sur le r\'eseau al\'eatoire de taille finie. Cependant, prendre la limite continue est techniquement difficile. Nous esp\'erons retrouver des caract\'eristiques d'un processus al\'eatoire invariant conforme (puisque l'on a moyenn\'e en quelque sorte sur tous les r\'eseaux possibles), i.e. d'un SLE. J'ai choisi de ne pas le pr\'esenter car ce travail n'est pas abouti. L'objectif est, \`a plus long terme, une compr\'ehension fine des cartes al\'eatoires (d\'ecor\'ees ou non) et des processus al\'eatoires qui vivent dessus, afin de justifier rigoureusement leur description en physique par des th\'eories conformes.

\newpage
\thispagestyle{empty}
\phantom{bbk}

\newpage

\appendix
\numberwithin{equation}{chapter}
\chapter{G\'eom\'etrie complexe sur les surfaces de Riemann compactes}
\label{app:geomcx}
\thispagestyle{plain}
\vspace{-1.5cm}
\rule{\textwidth}{1.5mm}
\vspace{2.5cm}

\addtolength{\baselineskip}{0.20\baselineskip}

Quelques faits \'el\'ementaires sur les surfaces de Riemann, utilis\'es dans le cours du texte, sont rappel\'es sans d\'emonstration dans cette annexe. Parmi les r\'ef\'erences remarquables, signalons : \cite{FarkasKra} pour l'aspect "g\'eom\'etrie diff\'erentielle" ; le livre de fonctions sp\'eciales \cite{GuoWang} pour l'aspect "fonctions elliptiques" ; le cours de Marco Bertola \cite{BertoCours}, qui est agr\'eable \`a lire et est un bon compromis entre d\'etails math\'ematiques et aspects pratiques utilis\'es dans cette th\`ese.

\section{G\'eom\'etrie}

\subsubsection{Topologie}
\label{topot}
\begin{figure}[h!]
\begin{center}
\includegraphics[width=\textwidth]{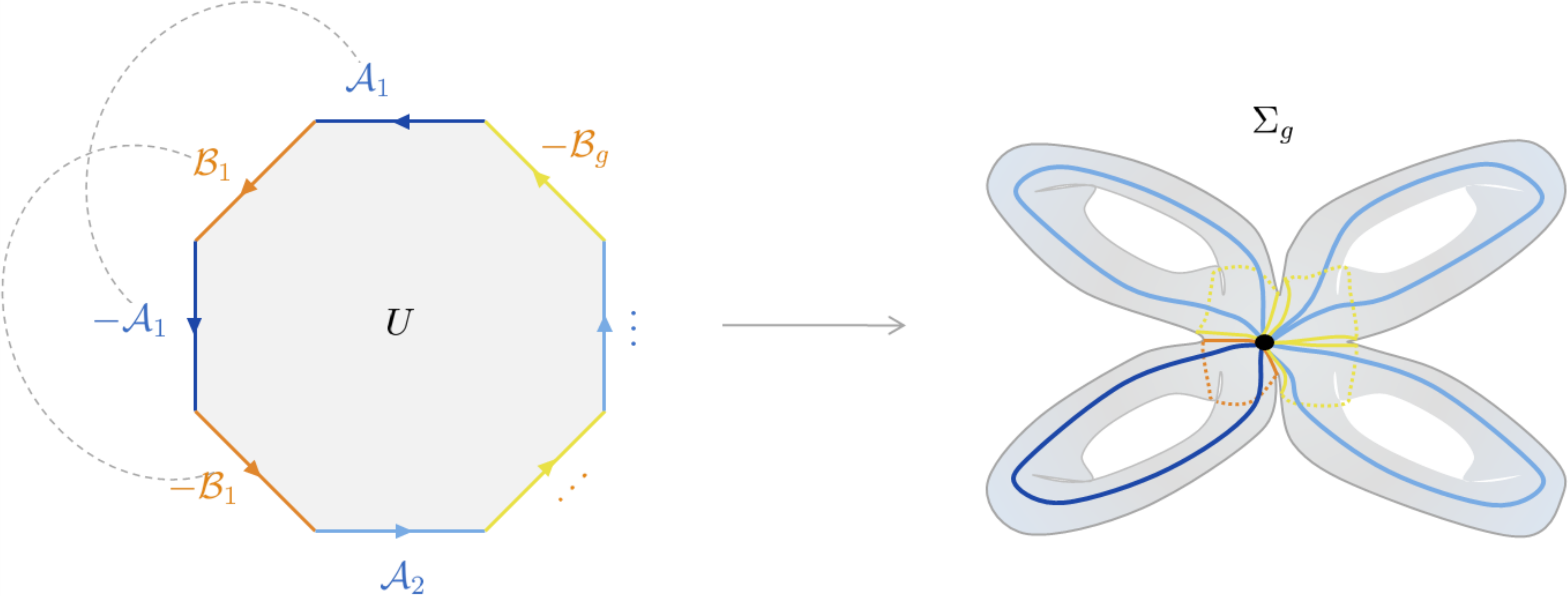}
\caption{\label{fig:4ggone} En identifiant les c\^{o}t\'es dans un $4\mathfrak{g}$-gone $\mathcal{A}_i\mathcal{B}_1(-\mathcal{A}_1)(-\mathcal{B}_1)\mathcal{A}_2\mathcal{B}_2(-\mathcal{A}_2)(-\mathcal{B}_2)\cdots(-\mathcal{B}_{\mathfrak{g}})$, on obtient une surface de genre $\mathfrak{g}$. La famille de cycles $(\mathcal{A}_{\mathfrak{h}},\mathcal{B}_{\mathfrak{h}})_{\mathfrak{h}}$ forme alors une base symplectique de $\Sigma_{\mathfrak{g}}$.}
\end{center}
\end{figure}

Les surfaces orientables \`a $\mathfrak{g}$ anses $\Sigma_{\mathfrak{g}}$ sont obtenues topologiquement par identification de c\^{o}t\'es dans un $4\mathfrak{g}$-gone (Fig.~\ref{fig:4ggone}).
\label{cycsym4} Son groupe d'homologie est $H_1(\Sigma_{\mathfrak{g}},\mathbb{Z})$ est isomorphe \`a $\mathbb{Z}^{2\mathfrak{g}}$ : elle poss\`ede une base de $2\mathfrak{g}$ cycles non contractibles. Ils peuvent \^{e}tre arrang\'es de sorte que :
\beq
\label{eq:orthosp} \mathcal{A}_{\mathfrak{h}} \cap \mathcal{A}_{\mathfrak{h}'} = 0,\quad \mathcal{A}_{\mathfrak{h}} \cap \mathcal{B}_{\mathfrak{h}'} = \delta_{\mathfrak{h},\mathfrak{h}'},\quad \mathcal{B}_{\mathfrak{h}} \cap \mathcal{B}_{\mathfrak{h}'} = 0
\eeq
Une base de cycles $(\mathcal{A}_{\mathfrak{h}},\mathcal{B}_{\mathfrak{h}})_{1 \leq \mathfrak{h} \leq \mathfrak{g}}$ qui v\'erifie ces relations est dite \textbf{symplectique}. Les changements de base symplectique sont param\'etr\'es par le groupe des matrices \`a coefficients entiers qui pr\'eservent l'\'{E}qn.~\ref{eq:orthosp}, i.e. $\mathrm{Sp}(2\mathfrak{g},\mathbb{Z})$.

$U = \Sigma_{\mathfrak{g}}\setminus\bigcup_{\mathfrak{h}}(\mathcal{A}_{\mathfrak{h}}\cup\mathcal{B}_{\mathfrak{h}})$ est un \textbf{domaine fondamental} de $\Sigma_{\mathfrak{g}}$, i.e. un ouvert simplement connexe maximal.

\subsubsection{Formes diff\'erentielles}

\label{hol}
Si $\omega$ est une forme diff\'erentielle m\'eromorphe non nulle sur $\Sigma_{\mathfrak{g}}$, notons $q_1,\ldots,q_r$ ses z\'eros et $p_1,\ldots,p_s$ ses p\^{o}les compt\'es avec multiplicit\'es. $\omega$ v\'erifie les deux relations fondamentales :
\beq
\label{eq:popol} \sum_{l = 1}^{s} \Res_{z \rightarrow p_l} \omega(z) = 0 \nn,\qquad \sum_{j = 1}^{r} \mathbf{u}(q_j) - \sum_{l = 1}^{s} \mathbf{u}(p_l) \in \mathbb{Z}^{\mathfrak{g}} \oplus \tau(\mathbb{Z}^{\mathfrak{g}})
\eeq
Si $\omega_1$ et $\omega_2$ sont deux formes diff\'erentielles, et $\omega_1$ n'a pas de r\'esidus, elles v\'erifient l'\textbf{identit\'e bilin\'eaire de Riemann} :\label{bii}
\beq
\sum_{p} \Res_{z \rightarrow p} \Big(\int^{z}_o \!\!\omega_1\Big)\omega_2(z) = \frac{1}{2i\pi}\sum_{\mathfrak{h} = 1}^{\mathfrak{g}}\Big(\oint_{\mathcal{A}_{\mathfrak{h}}} \!\!\omega_1\Big)\Big(\oint_{\mathcal{B}_{\mathfrak{h}}}\!\! \omega_2\Big) - \Big(\oint_{\mathcal{B}_{\mathfrak{h}}} \!\!\omega_1\Big)\Big(\oint_{\mathcal{A}_{\mathfrak{h}}} \!\!\omega_2\Big) \nn
\eeq
Il est possible d'en d\'eduire que l'espace des formes holomorphes sur $\Sigma_{\mathfrak{g}}$ est engendr\'e par les $\dd u_{\mathfrak{h}}$, donc est de dimension$_{\mathbb{C}}$ $\mathfrak{g}$. Celles-ci forment une base duale \`a $(\mathcal{A}_{\mathfrak{h}})_{1 \leq \mathfrak{h} \leq \mathfrak{g}}$ :
\beq
\label{eq:qj1} \oint_{\mathcal{A}_{\mathfrak{h}}} \dd u_{\mathfrak{h}'} = \delta_{\mathfrak{h},\mathfrak{h}'}
\eeq
Les int\'egrales sur les $\mathcal{B}$-cycles d\'efinissent la \textbf{matrice des p\'eriodes} :
\beq
\label{eq:qj2} \tau_{\mathfrak{h}\mathfrak{h}'} = \frac{1}{2i\pi}\oint_{\mathcal{B}_{\mathfrak{h}}} \dd u_{\mathfrak{h}'}
\eeq
Une propri\'et\'e fondamentale est que $\tau$ est une matrice \textbf{sym\'etrique}, et $\mathrm{Im}\,\tau$ est \textbf{d\'efinie positive}. Ceci est une cons\'equence de la compacit\'e et l'orientabilit\'e de $\Sigma_{\mathfrak{g}}$.

Soit $o \in \Sigma_{\mathfrak{g}}$ un point arbitraire. Les primitives des $\dd u_{\mathfrak{h}}$ sont des fonctions holomorphes sur le domaine fondamental, mais ne sont pas univalu\'ees sur $\Sigma_{\mathfrak{g}}$ au vu de l'\'{E}qn.~\ref{eq:qj1} et \ref{eq:qj2}. On peut cependant d\'efinir l'\textbf{application d'Abel} :
\beq
\begin{array}{rcl} u\::\:\Sigma_{\mathfrak{g}} & \rightarrow & \mathbb{C}^{\mathfrak{g}}/\big(\mathbb{Z}^{\mathfrak{g}} \oplus \tau(\mathbb{Z}^{\mathfrak{g}})\big) \\ \phantom{u\::\:} z & \mapsto & \big[\int_o^z \dd u_1,\ldots,\int_o^z \dd u_{\mathfrak{g}}\big] \end{array} \nn
\eeq
$J_{\Sigma} = \mathbb{C}^{\mathfrak{g}}/\big(\mathbb{Z}^{\mathfrak{g}} \oplus \tau(\mathbb{Z}^{\mathfrak{g}})\big)$ est la \textbf{jacobienne} de $\Sigma_{\mathfrak{g}}$. Lorsque $\mathfrak{g} = 1$, $\mathbf{u}$ est un isomorphisme de surfaces de Riemann. Si $\mathfrak{g} \geq 2$, $\mathbf{u}$ est un plongement mais ne peut plus \^{e}tre surjective pour des raisons de dimension. Toutefois, on a la propri\'et\'e importante :
\begin{theo}
\label{thJaco}Th\'eor\`eme d'inversion de Jacobi.
\beq
\forall \mathbf{w} \in J_{\Sigma},\,\,\exists p_1,\ldots,p_{\mathfrak{g}} \in \Sigma_{\mathfrak{g}} \quad \mathbf{w} = \sum_{\mathfrak{h} = 1}^{\mathfrak{g}} \mathbf{u}(p_{\mathfrak{h}}) \nn
\eeq
\end{theo}

\section{Fonction th\^{e}ta}
\label{the3}
\subsubsection{D\'efinition g\'en\'erale}

Dans cette annexe, nous choisissons d'appeler \textbf{module}, toute matrice \`a entr\'ees complexes, sym\'etrique et de partie imaginaire d\'efinie positive. Pour tout module $\tau$ de taille $g \times g$, la \textbf{fonction th\^{e}ta de Riemann} est d\'efinie par :
\beq
\theta(\mathbf{w}|\tau) = \sum_{\mathbf{p} \in \mathbb{Z}^{\mathfrak{g}}} e^{i\pi\mathbf{p}\cdot \tau\mathbf{p} + 2i\pi \mathbf{w}\cdot\mathbf{p}} \nn
\eeq
C'est une fonction enti\`ere sur $\mathbb{C}^{\mathfrak{g}}$, p\'eriodique dans la direction $\mathbb{Z}^{\mathfrak{g}}$, et qui prend une phase lin\'eaire dans la direction $\tau(\mathbb{Z}^{\mathfrak{g}})$ :
\beq
\label{eq:deacl} \forall \mathbf{m},\mathbf{n} \in \mathbb{Z}^{\mathfrak{g}},\quad \theta(\mathbf{w} + \mathbf{m} + \tau\mathbf{n}) = e^{-i\pi(2\mathbf{w} + \tau\mathbf{n})\cdot \mathbf{n}}\,\theta(\mathbf{w}|\tau)
\eeq
Elle v\'erifie une \'equation de diffusion :
\beq
\frac{\partial \theta}{\partial \tau_{\mathfrak{h}\mathfrak{h}'}} = \frac{1}{4i\pi}\frac{\partial^2\theta}{\partial w_{\mathfrak{h}} \partial w_{\mathfrak{h'}}},\qquad \theta(\mathbf{w}|\tau = 0) = \delta(\mathbf{w}) \nn
\eeq
$\theta$ peut donc \^{e}tre vue comme un noyau de la chaleur sur l'espace $\mathbb{C}^{\mathfrak{g}}$ avec certaines conditions de p\'eriodicit\'e. En genre $\mathfrak{g} = 1$, $\theta$ s'annule uniquement \`a $w = \frac{1 + \tau}{2}$ modulo $\mathbb{Z}\oplus\tau\mathbb{Z}$. En genre $\mathfrak{g} \geq 2$, la situation est plus compliqu\'ee et sera d\'ecrite dans le paragraphe suivant. Il y a toujours des z\'eros \'evidents aux vecteurs demi-entiers $\mathbf{c} = \frac{1}{2}(\mathbf{m} + \tau\mathbf{n})$ tel que $\mathbf{m}\cdot\mathbf{n} \in 2\mathbb{Z} + 1$, que l'on appelle \textbf{caract\'eristiques impaires}.

Il est justifi\'e \`a la partie~\ref{sec:introgeom} que la dimension$_{\mathbb{C}}$ maximale d'une famille de surfaces de Riemann de genre $\mathfrak{g} \geq 2$, est $d_{\mathfrak{g},0} = 3\mathfrak{g} - 3$. Or, un module de taille $\mathfrak{g} \times \mathfrak{g}$ contient $d'_{\mathfrak{g}} = \frac{\mathfrak{g}(\mathfrak{g} + 1)}{2}$ param\`etres libres. Lorsque $\mathfrak{g} \geq 4$, $d_{\mathfrak{g},0} < d_{\mathfrak{g}}'$, donc il existe toujours des \label{tauto3}modules $\tau$ qui ne sont pas des matrices de p\'eriodes d'une surface de Riemann.

\subsubsection{Fonctions th\^{e}ta issues d'une surface de Riemann}

Lorsque $\tau$ est bien une matrice de p\'eriodes pour une surface de Riemann $\Sigma_{\mathfrak{g}}$, la fonction $\theta(\cdot|\tau)$ poss\`ede d'autres propri\'et\'es. On omet d'\'ecrire la d\'ependance en $\tau$ dans ce qui suit.

\vspace{0.2cm}

\noindent \emph{Z\'eros de $\theta$} $\diamond\,$ Il existe \label{vcc2}un \textbf{vecteur des constantes de Riemann} $\mathbf{k}$, tel que, $\theta(\mathbf{w}|\tau) = 0$ ssi il existe $\mathfrak{g} - 1$ points $z_j \in \Sigma_{\mathfrak{g}}$ tels que $\mathbf{w} = \sum_{j = 1}^{\mathfrak{g} - 1} \mathbf{u}(z_j) + \mathbf{k}$. Ce vecteur $\mathbf{k}$ d\'epend du choix du point $o$.

\vspace{0.2cm}

\noindent \emph{Identit\'e de Fay} $\diamond\,$ La fonction th\^{e}ta v\'erifie une propri\'et\'e d'addition des arguments, appel\'ee \textbf{identit\'e de Fay} \cite{Fay}. Cette addition se fait naturellement dans la jacobienne : si $z_j,z_l \in \Sigma_\mathfrak{g}$, notons $\mathbf{u}_{jl} = \mathbf{u}(z_j) - \mathbf{u}(z_l)$. Soit $\mathbf{w} \in \mathbb{C}^{\mathfrak{g}}$ et $\mathbf{c}$ une caract\'eristique impaire. Il y a plusieurs fa\c{c}ons d'\'ecrire l'identit\'e de Fay, voici la forme qui nous sera utile :
\begin{footnotesize}
\bea
\label{eq:fays} && \theta(\mathbf{w} + \mathbf{c})\,\theta(\mathbf{u}_{21} + \mathbf{u}_{43} + \mathbf{w} + \mathbf{c})\,\frac{E(z_2,z_4)\,E(z_1,z_3)}{E(z_2,z_3)\,E(z_1,z_4)}\,\frac{1}{E(z_2,z_1)\,E(z_4,z_3)} \\
& = & \frac{\theta(\mathbf{w} + \mathbf{u}_{21} + \mathbf{c})}{E(z_2,z_1)}\,\frac{\theta(\mathbf{w} + \mathbf{u}_{43} + \mathbf{c})}{E(z_4,z_3)} - \frac{\theta(\mathbf{w} + \mathbf{u}_{23} + \mathbf{c})}{E(z_2,z_3)}\,\frac{\theta(\mathbf{w} + \mathbf{u}_{41} + \mathbf{c})}{E(z_4,z_1)} \nn
\eea
\end{footnotesize}
$\!\!\!$o\`u $E$ est la forme primaire d\'efinie un peu plus loin, \`a l'\'{E}qn.~\ref{eq:prima}. Cette identit\'e a de nombreuses cons\'equences, et est intimement li\'ee \`a l'existence de syst\`emes int\'egrables bas\'es sur $\Sigma_{\mathfrak{g}}$ (Chapitre~\ref{chap:int}). Elle est \'equivalente \`a l'\'equation de\label{Hiro} Hirota que v\'erifie la fonction tau de la famille des solutions alg\'ebro-g\'eom\'etriques (\S~\ref{sec:ALG}). Elle se g\'en\'eralise facilement :
\begin{footnotesize} \bea
\label{eq:fays2} && \mathop{\mathrm{det}}_{1 \leq i,j \leq n} \Big(\frac{\theta(\mathbf{u}(z_i) - \mathbf{u}(z'_j) + \mathbf{w} + \mathbf{c})}{\theta(\mathbf{w} + \mathbf{c})\,E(z_i,z_j')}\Big) \\
 & = & (-1)^{n + 1}\,\Big[\frac{\prod_{1 \leq i < j \leq n} E(z_i,z_j)\,E(z_i',z_j')}{\prod_{1 \leq i,j \leq n} E(z_i,z_j')}\Big]\,\frac{\theta\Big(\sum_{i = 1}^n \big[\mathbf{u}(z_i) - \mathbf{u}(z_i')\big] + \mathbf{w} + \mathbf{c}\Big)}{\theta(\mathbf{w} + \mathbf{c})} \nn
\eea
\end{footnotesize}
$\!\!\!$Des limites \`a points co\"{i}ncidants de l'\'{E}qn.~\ref{eq:fays} conduisent \`a des \'equations diff\'erentielles non lin\'eaires venant d'une hi\'erarchie int\'egrable.

\subsubsection{Fonctions analytiques sur $\Sigma_g$}

Une caract\'eristique impaire est dite \textbf{non singuli\`ere} si la fonction $\theta(\mathbf{u}(z_1) - \mathbf{u}(z_2) + \mathbf{c})$ n'est pas identiquement nulle pour $z_1,z_2 \in \Sigma_{\mathfrak{g}}$. Il existe des caract\'eristiques impaires non singuli\`eres, et l'on en choisit une. Soit $f$ une fonction m\'eromorphe sur $\Sigma_{\mathfrak{g}}$, qui admet des z\'eros en $q_1,\ldots,q_r$ et des p\^{o}les en $p_1,\ldots,p_s$ (compt\'es avec multiplicit\'es). Il existe une constante $C > 0$ telle que :
\beq
\label{eq:thte} f(z) = C\,\frac{\prod_{j = 1}^{r} \theta(\mathbf{u}(z) - \mathbf{u}(q_j) + \mathbf{c})}{\prod_{l = 1}^{s} \theta(\mathbf{u}(z) - \mathbf{u}(p_l) + \mathbf{c})}
\eeq
Les relations \ref{eq:popol} assurent que le membre de droite restent inchang\'e lorsque $z \rightarrow z + \mathcal{B}_{\mathfrak{h}}$. On dit que les ratios de l'\'{E}qn.~\ref{eq:thte} sont \textbf{bien \'equilibr\'es}.

On appelle \textbf{fonctions pseudom\'eromorphes} des fonctions qui ne sont pas univalu\'ees sur $\Sigma_{\mathfrak{g}}$, mais prennent des phases lorsque l'on translate $z \rightarrow z + \mathcal{A}_{\mathfrak{h}}$ ou $\mathcal{B}_{\mathfrak{h}}$. Les ratios mal \'equilibr\'es de fonctions $\theta$ permettent de construire ce type de fonctions.

\section{Formes m\'eromorphes}
\label{sec:meroe}
On se fixe une caract\'eristique impaire non singuli\`ere $\mathbf{c}$.

\subsubsection{Noyau de Bergman}

Il existe un unique objet $B(z_1,z_2)$, qui est une forme diff\'erentielle en $z_1$ et $z_2 \in \Sigma_{\mathfrak{g}}$, tel que :

\vspace{0.2cm}

\noindent $\diamond\,$ La seule singularit\'e de $B$ est un p\^{o}le double sans r\'esidu \`a points co\"{i}ncidants. I.e. dans toute coordonn\'ee locale $\xi$ :
\beq
B(z_1,z_2) = \frac{\dd \xi(z_1)\dd\xi(z_2)}{\big(\xi(z_1) - \xi(z_2)\big)^2} + O(1),\quad z_1 \rightarrow z_2 \nn
\eeq

\noindent $\diamond\,$ $\oint_{z_1 \in \mathcal{A}_{\mathfrak{h}}} B(z_1,\cdot) = 0$.

\vspace{0.2cm}

De plus, cet objet est sym\'etrique en $z_1$ et $z_2$ : c'est un noyau de Bergman pour $\Sigma_{\mathfrak{g}}$. Son int\'egrale sur les $\mathcal{B}$-cycles redonne les formes holomorphes :
\beq
\oint_{z_1 \in \mathcal{B}_{\mathfrak{h}}} B(z_1,z_2) = 2i\pi\,\dd u_{\mathfrak{h}}(z_2),\qquad \oint_{z_1\in\mathcal{B}_{\mathfrak{h}}}\oint_{z_2\in\mathcal{B}_{\mathfrak{h}'}} B(z_1,z_2) = 2i\pi\,\tau_{\mathfrak{h}\mathfrak{h}'} \nn
\eeq
Il se construit explicitement :
\beq
\label{eq:Berg}B(z_1,z_2) = \dd_{z_1}\dd_{z_2}\ln \theta(\mathbf{u}(z_1) - \mathbf{u}(z_2) + \mathbf{c}|\tau)
\eeq
qui est ind\'ependant du choix de $\mathbf{c}$.

\subsubsection{Forme primaire}

Lorsque $\mathbf{c}$ est une caract\'eristique impaire non singuli\`ere, $\dd h_{\mathbf{c}}(z) = \sum_{\mathfrak{h} = 1}^{\mathfrak{g}} \partial_{u_{\mathfrak{h}}}\theta(\mathbf{c})\dd u_h(z)$ est une forme holomorphe qui a exactement $\mathfrak{g}$ z\'eros, qui sont doubles.
\beq
\label{eq:prima} E(z_1,z_2) = \frac{\theta(\mathbf{u}(z_1) - \mathbf{u}(z_2) + \mathbf{c})}{\sqrt{\dd h_{\mathbf{c}}(z_1)\dd h_{\mathbf{c}}(z_2)}}
\eeq
est un objet m\'eromorphe sur le domaine fondamental de $\Sigma_{\mathfrak{g}}$, qui se comporte comme l'inverse d'un spineur en $z_1$ et $z_2$, et admet un z\'ero simple \`a points co\"{i}ncidants. $E$ n'est pas d\'efini de fa\c{c}on univalente sur $\Sigma_{\mathfrak{g}}$, il prend une phase lorsque $z \rightarrow z + \mathcal{B}_{\mathfrak{h}}$ comme la fonction th\^{e}ta. En fait, $E$ ne d\'epend pas de $\mathbf{c}$, et on l'appelle \textbf{forme primaire}. Elle v\'erifie $E(z_2,z_1) = -E(z_1,z_2)$, et :
\bea
\label{eq:formpi} && \exp\Big\{\int_{z_1}^{z_2}\!\!\int_{z_3}^{z_4} B\Big\} = \frac{E(z_2,z_4)E(z_1,z_3)}{E(z_2,z_3)E(z_1,z_4)} \\
&& \exp\Big\{\frac{1}{2}\int_{z_1}^{z_2} \!\!\int_{z_1}^{z_2}\Big(B(z,z')- \frac{\dd x(z)\dd x(z')}{\big(x(z) - x(z')\big)^2}\Big)\Big\} =  \frac{1}{\sqrt{\dd x(z_1)\dd x(z_2)}}\,\frac{x_2 - x_1}{E(z_2,z_1)} \nn
\eea

\subsubsection{Noyau de Cauchy}

\`{A} cause de l'\'{E}qn.~\ref{eq:popol}, une forme diff\'erentielle m\'eromorphe ne peut admettre un unique p\^{o}le simple. On peut n\'eanmoins construire une forme diff\'erentielle $\dd S_{z_1,z_2}(z)$ m\'eromorphe sur $\Sigma_{\mathfrak{g}}$, qui admet un p\^{o}le simple avec r\'esidu $-1$ en $z = z_1$, et un p\^{o}le simple avec r\'esidu $1$ en $z = z_2$ :
\beq
\mathop{\dd S}_{z_1,z_2}(z) = \int_{z_1}^{z_2} B(z,\cdot) \nn
\eeq
Comme corolaire de l'identit\'e bilin\'eaire de Riemann, on trouve une formule de Cauchy pour repr\'esenter par des int\'egrales de contours n'importe quelle forme diff\'erentielle m\'eromophe $\Omega$ sur $\Sigma_{\mathfrak{g}}$, dont on note $p_i$ les p\^{o}les :
\beq
\Omega(z_0) = \sum_{i} \Res_{z \rightarrow p_i} \mathop{\dd S}_{o,z}(z_0)\,\Omega(z) + \sum_{\mathfrak{h} = 1}^{\mathfrak{g}}\Big(\oint_{\mathcal{A}_{\mathfrak{h}}} \!\!\Omega\Big)\dd u_{\mathfrak{h}}(z_0) \nn
\eeq

\subsubsection{P\^{o}les multiples}

Soit $p \in \Sigma_{\mathfrak{g}}$, et $\xi$ une coordonn\'ee locale telle que $\xi(p) = 0$. Si $\Omega$ est une forme diff\'erentielle m\'eromorphe avec un p\^{o}le de degr\'e $d_p + 1$ en $p$, les coefficients dans le d\'eveloppement de Laurent :
\beq
\Omega(z) \mathop{=}_{z \rightarrow p} \sum_{j = 0}^{d_p} t_{p;j}\,\frac{\dd\xi(z)}{\xi(z)^{j + 1}} + O(1),\qquad t_{p;j} = \Res_{z \rightarrow p} \xi(z)^{j}\,\Omega(z) \nn
\eeq
d\'ependent du choix de $\xi$.

Inversement, on peut chercher \`a construire des formes diff\'erentielles m\'eromorphes $\Omega_{p;j}$ telles que $t_{q;l} = \delta_{j,l}\delta_{p,q}$. Elles sont d\'efinies \`a l'ajout d'une forme holomorphe pr\`es. Si l'on impose \`a $\Omega_{p;j}$ d'avoir des int\'egrales nulles sur les $\mathcal{A}$-cycles, il y a une unique solution, qui s'\'ecrit :
\beq
\Omega_{p;j}(z_0) = \Res_{z \rightarrow p} \xi^{-j}(z)\,B(z_0,z) \nn
\eeq

\subsubsection{Base de formes m\'eromorphes}

\begin{figure}[h!]
\begin{center}
\includegraphics[width=1.05\textwidth]{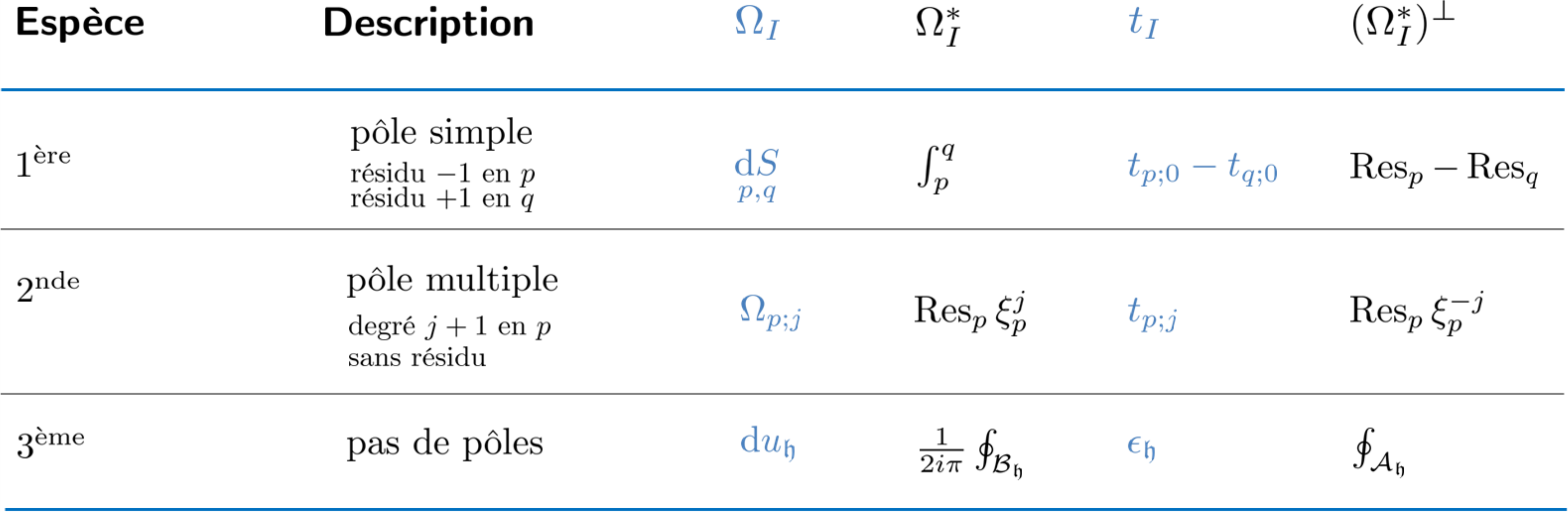}
\caption{\label{fig:cl} R\'ecapitulatif des formes m\'eromorphes sur les surfaces de Riemann. On a indiqu\'e la classification standard en "esp\`eces".}
\end{center}
\end{figure}

Soit $x$ une fonction m\'eromorphe sur $\Sigma_{\mathfrak{g}}$. Elle d\'efinit de fa\c{c}on canonique des coordonn\'ees locales $\xi_p$ au voisinage de tout point $p$, telles que $\xi_p(p) = 0$ :
\begin{itemize}
\item[$\diamond$] Si $x$ a un p\^{o}le de degr\'e $d_p$ en $p$, on pose $\xi_p(z) = \big(x(z)\big)^{-1/d_p}$.
\item[$\diamond$] Si $x$ est r\'eguli\`ere en $p$, et d'ordre $m_p$ (i.e. $\dd x$ a un z\'ero d'ordre $m_p - 1$ en $p$), on pose $\xi_p(z) = \big(x(z) - x(p)\big)^{1/m_p}$.
\end{itemize}
Par cons\'equent, elle d\'efinit une base de formes m\'eromorphes sur $\Sigma_{\mathfrak{g}}$, et toute forme m\'eromorphe $\Omega$ sur $\Sigma_{\mathfrak{g}}$ se d\'ecompose de mani\`ere unique :
\beq
\label{eq:altern}\Omega(z_0) = \sum_{p} \Big(t_{p;0}\,\mathop{\dd S}_{o,p}(z_0) + \sum_{j \geq 1} t_{p;j}\,\Omega_{p;j}(z_0)\Big) + \sum_{\mathfrak{h} = 1}^{\mathfrak{g}} 2i\pi\,\epsilon_{\mathfrak{h}}\,\dd u_{\mathfrak{h}}(z_0)
\eeq
Les \textbf{temps} $t_{I}$ et les formes diff\'erentielles de la base $\Omega_{I}$ sont toujours de la forme
\beq
t_I = \int_{(\Omega_I^*)^{\bot}}\!\!\!\! \omega,\qquad \Omega_I(z_0) = \int_{z \in \Omega^*_I} \!\!\!\! B(z_0,z) \nn
\eeq
o\`u $\Omega_I^*$ et $(\Omega_I^*)^{\bot}$ sont des cycles g\'en\'eralis\'es, i.e. la donn\'ee d'un contour dans $\Sigma_{\mathfrak{g}}$ et d'une fonction d\'efinie sur ce contour (Fig.~\ref{fig:cl}).
Cette d\'ecomposition ne d\'epend pas du point $o$ car la somme des r\'esidus de $\Omega$ est nulle : $\sum_{p} t_{p;0} = 0$.

L'ensemble des fonctions m\'eromorphes non nulles est un corps. La repr\'esentation d'une fonction m\'eromorphe $f$ comme ratio de fonctions th\^{e}ta (\'{E}qn.~\ref{eq:thte}) souligne la structure de groupe, tandis que la repr\'esentation utilisant les primitives des $\Omega_I$ (primitive de \'{E}qn.~\ref{eq:altern} appliqu\'ee \`a $\dd f$) met en \'evidence la structure d'anneau. Cette derni\`ere est pratique pour l'int\'egration ou la d\'erivation. En r\'esum\'e, pour identifier une fonction ou une forme m\'eromorphe, il suffit d'identifier ses p\^{o}les et z\'eros, ou son comportement aux p\^{o}les, et de reproduire cette structure gr\^{a}ce \`a des fonctions bien connues.

\section{Transformations modulaires}
\label{modula2}
Rassemblons les cycles dans des vecteurs colonnes, et les formes holomorphes dans des vecteurs lignes. On note $\mathbf{X}^{t}$, la transpos\'ee de $\mathbf{X}$. Les changements de base symplectique de cycles, aussi appel\'es \textbf{transformations modulaires}, s'\'ecrivent :
\beq
\left(\begin{array}{c} \mathcal{A}' \\ \mathcal{B}' \end{array}\right) = \left(\begin{array}{cc} \alpha & \beta \\ \gamma & \delta \end{array}\right)\left(\begin{array}{c} \mathcal{A} \\ \mathcal{B}\end{array}\right),\qquad \left\{\begin{array}{l} \alpha^{t}\,\beta - \beta^{t}\,\alpha = 0 \\ \alpha^{t}\,\delta - \beta^{t}\,\gamma = \mathbf{1}_{\mathfrak{g} \times \mathfrak{g}} \\ \gamma^{t}\,\delta - \delta^{t}\,\gamma = 0 \end{array}\right. \nn
\eeq
Notons que l'inverse de $\mathbf{M} = \left(\begin{array}{cc} \alpha & \beta \\ \gamma & \delta \end{array}\right)$ s'\'ecrit $\mathbf{M}^{-1} = \left(\begin{array}{cc} \delta^{t} & -\gamma^{t} \\ -\beta^{t} & \alpha^{t}\end{array}\right)$.

\noindent Cette transformation se r\'epercute sur tous les objets introduits jusqu'\`a pr\'esent.

\vspace{0.2cm}

\noindent $\diamond\,$ Les formes holomorphes $\dd \mathbf{u}' = (\alpha + \beta\tau)^{-1} \dd \mathbf{u}$.

\vspace{0.2cm}

\noindent $\diamond\,$ La matrice des p\'eriodes $\tau' = (\alpha + \beta\tau)^{-1}(\gamma + \delta\tau)$.

\vspace{0.2cm}

\noindent $\diamond\,$ Le noyau de Bergman $B'(z_1,z_2) = B(z_1,z_2) + 2i\pi\dd \mathbf{u}(z_1)\,\kappa\,\dd \mathbf{u}(z_2)^{t}$, avec la matrice sym\'etrique :
\beq
\kappa = -(\alpha + \beta\tau)^{-1}\beta\tau \nn
\eeq

\vspace{0.2cm}

\noindent $\diamond\,$ La fonction th\^{e}ta devient \cite{MumTata} :
\beq
\theta(\mathbf{w}'|\tau') = \eta_{\mathbf{M}}\,\big(\mathrm{det}(\alpha + \beta\tau)\big)^{1/2}\,e^{-i\pi\mathbf{w}\cdot\tau\mathbf{w}}\theta(\mathbf{w}|\tau) \nn
\eeq
$\eta_{\mathbf{M}}$ est une phase d\'ependant seulement de $\mathbf{M}$, et les arguments $\mathbf{w}$ et $\mathbf{w}'$ sont reli\'es par :
\beq
\mathbf{w} = (\alpha + \beta\tau)\big[\mathbf{w}' - \mathrm{diag}(\gamma^{t}\,\delta)/2 - \tau\,\mathrm{diag}(\alpha^{t}\,\beta)/2\big] \nn
\eeq

\section{Surfaces de Riemann elliptiques}

Les courbes spectrales de genre $1$ se rencontrent dans le mod\`ele \`a une matrice dans un r\'egime \`a deux coupures (partie~\ref{sec:pluscoup}), ou comme outil interm\'ediaire pour le mod\`ele $\On$ trivalent (partie~\ref{sec:OnON}). Lorsque $\mathfrak{g} = 1$, $\Sigma_{\mathfrak{1}}$ est en bijection conforme avec $\mathbb{C}/\mathbb{Z} \oplus \tau\mathbb{Z}$ via l'application d'Abel. $\mathcal{A}_1 = [0,1]$ et $\mathcal{B}_{1} = [0,\tau]$ est la base symplectique de cycles correspondante. Il n'y a qu'une seule caract\'eristique impaire, $c = \frac{1 + \tau}{2}$, et elle est non singuli\`ere.

Cette bijection rend la g\'eom\'etrie sur $\Sigma_{\mathfrak{1}}$ plus maniable. Les fonctions m\'eromorphes sur $\Sigma_{\mathfrak{g}}$ sont les fonctions m\'eromorphes sur $\mathbb{C}$ qui sont $1$ et $\tau$ p\'eriodiques. Ces fonctions bip\'eriodiques sont aussi appel\'ees \textbf{fonctions elliptiques}, car la fonction $\mathrm{sn}$ qui intervient pour calculer \label{elli}l'abscisse curviligne sur une ellipse poss\`ede ce type de propri\'et\'e.
Une fonction elliptique sans p\^{o}les est born\'ee sur $\mathbb{C}$, donc constante : $\dd u$ est la seule forme holomorphe \`a une constante multiplicative pr\`es, comme l'on s'y attend. Il y a plusieurs jeux de fonctions elliptiques sp\'eciales, \`a partir desquelles toute fonction elliptique peut s'exprimer. Celui impliquant les fonctions de Jacobi $\mathrm{sn}$, $\mathrm{cn}$ et $\mathrm{dn}$ ne sera pas pr\'esent\'e ici.

Les fonctions\label{elli2} \textbf{pseudoelliptiques} du chapitre~\ref{chap:formel} sont des fonctions $f$ p\'eriodiques dans la direction $\mathbb{Z}$, et qui prennent une phase dans la direction $\tau\mathbb{Z}$ :
\beq
f(w + 1) = f(w),\qquad f(w + \tau) = e^{i\pi\mathfrak{b}} f(w) \nn
\eeq

\subsubsection{Fonction th\^{e}ta}

La fonction $\vartheta_1$ de Jacobi est une variante de la fonction th\^{e}ta de Riemann. Pour $\mathrm{Im}\,\tau > 0$ :
\bea
\label{eq:serith}\vartheta_1(w|\tau) & \equiv & i \sum_{p \in \mathbb{Z}} (-1)^p\,e^{i\pi\tau(p - 1/2)^2 + i\pi(2p - 1)w} \\
& = &  -ie^{i\pi\tau/4}\,e^{-i\pi w}\,\theta(w + (1 + \tau)/2|\tau) \nn
\eea
$\vartheta_1(w|\tau)$ est une fonction impaire de $w$, en particulier elle s'annule en $w = 0$. Ses propri\'et\'es par translation sont :
\beq
\vartheta_1(w + 1|\tau) = -\vartheta_1(w|\tau),\qquad \vartheta_1(w + \tau) = -e^{i\pi(2w - \tau)}\vartheta_1(w|\tau) \nn
\eeq
On dispose \'egalement de la repr\'esentation en \textbf{produit triple de Jacobi} :
\beq
\label{eq:jcbo} \vartheta_1(w|\tau) = 2e^{i\pi\tau/4}\,\sin(\pi w)\,\prod_{p = 1}^{\infty} (1 - e^{2i\pi\tau p}) (1 - e^{2i\pi(w + \tau p)})(1 - e^{2i\pi(-w + \tau p)})
 \eeq

Les transformations modulaires agissent sur $\tau$ par homographies \`a coefficients entiers :
\beq
\tau' = \frac{a\tau + b}{c\tau + d}\,,\qquad \left(\begin{array}{cc} c & d \\ a & b \end{array}\right) \in \mathrm{PSL}_2(\mathbb{Z})\nn
\eeq
La r\`egle de transformation de $\vartheta_1$ est assez simple, notamment pour $\tau' = -1/\tau$ :
\beq
\label{eq:modo} \vartheta_1(w|\tau) = i\,(-i\tau)^{-1/2}\,e^{-i\pi w^2/\tau}\vartheta_1(\tau'w|\tau')
\eeq
Dans la limite $|\tau| \rightarrow \infty$, $\vartheta_1$ se r\'eduit \`a une ou deux exponentielles qui dominent la s\'erie~\ref{eq:serith}. La limite $|\tau| \rightarrow 0$ s'en d\'eduit gr\^{a}ce \`a la transformation modulaire~\ref{eq:modo}.

Si $f$ est une fonction m\'eromorphe dont on connait les p\^{o}les et les z\'eros, on peut l'exprimer comme un ratio bien \'equilibr\'e de fonctions $\vartheta_1$. Les ratios mal \'equilibr\'es permettent de repr\'esenter les fonctions pseudoelliptiques du Chapitre~\ref{chap:formel}. Par exemple,
\beq
f(u) = C\,\frac{\prod_{j = 1}^{r} \vartheta_1(u - v_j)}{\prod_{l = 1}^{s} \vartheta_1(u - w_l)} \nn
\eeq
admet des z\'eros aux $v_j$, des p\^{o}les aux $w_l$, et :
\beq
f(u + 1) = f(u),\quad f(u + \tau) = (-1)^{s - r}\,e^{i\pi(s - r)u}\,e^{i\pi\big((s - r)\tau + \sum_{k} w_k - \sum_{j} v_j\big)}\,f(u) \nn
\eeq

\subsubsection{Fonctions de Weierstra{\ss}}
\label{Wei3}
La fonction $\wp$ de Weierstra{\ss} est d\'efinie par :
\bea
\wp(w|\tau) & \equiv & \frac{1}{w^2} + \sum_{p,q \in \mathbb{Z}^2\setminus\{(0,0)\}} \frac{1}{(w + p + \tau q)^2} - \frac{1}{(p + \tau q)^2}  \nn \\
& = & -C_0(\tau) + \sum_{p \in \mathbb{Z}} \frac{\pi^2}{\sin^2\pi(w + p\tau)} \nn \\
C_0(\tau) & = & \frac{\pi^2}{3} + \sum_{p \in \mathbb{Z}^*} \frac{\pi^2}{\sin^2\pi p \tau} \nn
\eea
C'est une fonction elliptique, paire, et qui poss\`ede un p\^{o}le double $\wp(w) \sim w^{-2}$ lorsque $w \rightarrow 0$. Elle co\"{i}ncide \`a une constante pr\`es avec le noyau de Bergman $B$, d'int\'egrale nulle sur le cycle $\mathcal{A} = [0,1]$ :
\beq
B(u_1,u_2) = \dd u_1\dd u_2\big[\wp(u_1 - u_2|\tau) + C_0(\tau)\big] \nn
\eeq
En comparant avec l'\'{E}qn.~\ref{eq:Berg}, on obtient une autre repr\'esentation de la fonction de Weierstra{\ss} :
\beq
\label{eq:djua}\wp(w|\tau) = -\partial_{w}^2\ln\vartheta_1\big(w + (1 + \tau)/2|\tau\big) + \mathrm{cte}
\eeq
On peut \'etablir l'\'equation diff\'erentielle :
\beq
\label{eq:diffs}\big(\wp'(w)\big)^2 = 4\wp(w)^3 - g_2\,\wp(w) - g_3
\eeq
Les deux membres sont des fonctions elliptiques, il suffit de comparer leur comportement au voisinage de l'unique p\^{o}le $w = 0$ pour d\'emontrer l'\'{E}qn.~\ref{eq:diffs}. On identifie au passage les constantes :
\beq
g_2 = \sum_{(p,q) \in \mathbb{Z}^{2}\setminus\{(0,0)\}} \frac{60}{(p + q\tau)^4},\qquad g_3 = \sum_{(p,q) \in \mathbb{Z}^{2}\setminus\{(0,0)\}} \frac{120}{(p + q\tau)^6} \nn
\eeq
$\wp'(w)$ admet exactement trois z\'eros modulo $\mathbb{Z}\oplus\tau\mathbb{Z}$, qui sont situ\'es aux points $w\in \{\frac{1}{2},\frac{\tau}{2},\frac{1 + \tau}{2}\}$.
Pour illustration, on peut ais\'ement exprimer $\wp$ et $\wp'$ comme ratio \'equilibr\'e de fonctions th\^{e}ta :
{\small \bea
\wp(w|\tau) - \wp(1/2|\tau) & = & \Big[\frac{\vartheta_1'(0|\tau)}{\vartheta_1(1/2|\tau)}\,\frac{\vartheta_1(w + 1/2|\tau)}{\vartheta_1(w|\tau)}\Big]^2 \nn \\
\wp'(w|\tau) & = & C\,e^{2i\pi w}\,\frac{\vartheta_1(w + 1/2|\tau)\vartheta_1(w + \tau/2|\tau)\vartheta_1(w + (1 + \tau)/2|\tau)}{\vartheta_1^3(w|\tau)} \nn
\eea}
$\!\!\!$La base de formes m\'eromorphes $\Omega_{p;j}$ s'exprime en fonction de la base alternative :
\beq
\widetilde{\Omega}_{v_j,k}=\frac{(-1)^k}{k!}\,(\partial_u^{k - 1}\wp)(u - v_j|\tau)\,\dd u\qquad (k \geq 1) \nn
\eeq
constitu\'ee de $\wp$ et de ses d\'eriv\'ees.

La fonction $\zeta$ de Weierstra{\ss} est d\'efinie par :
\bea
\zeta(w|\tau) & \equiv & \frac{1}{w} + \int^{w}_0 \!\!\dd v\Big(\frac{1}{v^2} - \wp(v|\tau)\Big) \nn \\
& = & C_0(\tau)w + \pi\mathrm{cotan}(\pi w) +  \sum_{p \in \mathbb{Z}^*} \pi\,\big[\mathrm{cotan}\,\pi(w + p\tau) - \mathrm{cotan}\,\pi p \tau\big]  \nn
\eea
C'est une fonction telle que $\zeta'(w) = -\wp(w)$, elle poss\`ede un p\^{o}le simple avec r\'esidu $1$ en $w = 0$. Elle est reli\'ee naturellement au noyau de Cauchy :
\beq
\mathop{\dd S}_{u_2,u_1}(u) = C_0(\tau)(u_1 - u_2) + \zeta(u - u_2|\tau) - \zeta(u - u_1|\tau) \nn
\eeq
Les int\'egrales de cycles de $B$ impliquent que $\zeta$ n'est pas une fonction univalu\'ee sur $\Sigma_1$ :
\beq
\zeta(u + 1|\tau) - \zeta(u|\tau) = C_0(\tau),\qquad \zeta(u + \tau|\tau) - \zeta(u|\tau) = \tau C_0(\tau) - 2i\pi \nn
\eeq

La fonction $\sigma$ de Weierstra{\ss} est d\'efinie par :
\bea
\sigma(w|\tau) & = & w\,\exp\Big\{\int^{w}_0 \!\!\dd v\Big(\zeta(v|\tau) - \frac{1}{v}\Big)\Big\} \nn \\
& = &  \pi\,\sin(\pi w)\,\prod_{p = 1}^{\infty} \frac{\sin\pi(w + p\tau)\,\sin\pi(-w + p\tau)}{\sin^2\pi p\tau} \nn
\eea
Elle v\'erifie $(\ln\sigma)'(w) = \zeta(w)$. L'\'{E}qn.~\ref{eq:djua} permet de la relier \`a la fonction th\^{e}ta :
\beq
\frac{\vartheta_1(w|\tau)}{\vartheta_1'(0|\tau)} = e^{C_0(\tau) w^2/2}\,\sigma(w|\tau) \nn
\eeq
Ceci conduit \`a l'expression de $\vartheta_1$ comme produit triple de Jacobi (\'{E}qn.~\ref{eq:jcbo}).

\newpage
\thispagestyle{empty}
\phantom{bbk}

\newpage

\chapter{Observables \`a bords non uniformes dans les mod\`eles de boucles}
\label{app:bornu}
\thispagestyle{plain}
\vspace{-1.5cm}
\rule{\textwidth}{1.5mm}
\vspace{1.25cm}

Dans le cadre du mod\`ele de boucles d\'efini par la mesure :
\bea
\dd\nu & = & \frac{1}{Z}\,\dd M \prod_{\alpha = 1}^{\n} \dd A_\alpha\,\exp\Big\{-\frac{N}{t}\Big(\Tr\frac{M^2}{2} + \sum_{\alpha = 1}^{\mathfrak{n}} \frac{1}{2}\,\Tr K_{\alpha}(A_{\alpha},A_{\alpha})\Big)\Big\} \nn
\eea
nous allons calculer les observables admettant $k$ bords avec un nombre fini de m\`eches dans une configuration arbitraire. Lorsque le $i^{\textrm{\`{e}me}}$ bord n'a pas de m\`eche, son p\'erim\`etre est coupl\'e \`a une seule variable $x^{i}_1$. Lorsqu'il poss\`ede $r^i$ m\`eches, on les indexe en suivant l'ordre cyclique par $({}^{i}_1),\ldots,({}^{i}_{r^i})$, et l'on note $\alpha^{i}_j$ leur couleur. La longueur de l'intervalle entre $({}^{i}_{j})$ et $({}^{i}_{j + 1})$ est coupl\'ee \`a une variable $x^i_j$. On note $\mathsf{B}$ l'ensemble des bords, et $\mathsf{B}_{0}$ l'ensemble des bords sans m\`eches.

L'observable en question est :
\bea
\overline{H}  & = & \Big\langle \prod_{i \in \mathsf{B}_0} \Tr\,\frac{1}{x^i_1 - M}\,\prod_{i \in \mathsf{B}\setminus\mathsf{B}_0} \Tr\,\Big(A_{\alpha^{i}_1}\,\frac{1}{x^{i}_1 - M}\cdots A_{\alpha^{i}_{r^i}}\,\frac{1}{x^{i}_{r^i} - M}\Big) \Big\rangle \nn \\
\label{eq:defss} &&
\eea
On peut aussi d\'efinir une observable connexe, et regrouper les topologies :
\bea
\overline{H} & = & \Big\langle \prod_{i \in \mathsf{B}_0} \Tr\,\frac{1}{x^i_1 - M}\,\prod_{i \in \mathsf{B}\setminus\mathsf{B}_0} \Tr\,\Big(A_{\alpha^{i}_1}\,\frac{1}{x^{i}_1 - M}\cdots A_{\alpha^{i}_{r^i}}\,\frac{1}{x^{i}_{r^i} - M}\Big) \Big\rangle_c \nn \\
& = & \sum_{g \geq 0} \Big(\frac{N}{t}\Big)^{2 - 2g -k}\,H^{(g)} \nn
\eea

\section{Observables non connexes}

L'int\'egration sur les matrices $A_{\alpha}$ est possible car, \`a matrice $M = U\cdot\mathrm{diag}(\lambda_1,\ldots,\lambda_N)\cdot U^{\dagger}$ fix\'ee, elles sont gaussiennes, avec pour matrice de covariance :
\beq
\label{eq:concona}\contraction[1ex]{}{A_{\alpha|ab}}{\,}{A_{\beta|cd}}
A_{\alpha|ab}\,A_{\beta|cd} = \frac{t}{N}\,\frac{\delta_{\alpha,\beta}\,\delta_{a,d}\,\delta_{b,c}}{K_{\alpha}(\lambda_a,\lambda_b)} \nn
\eeq
Par application du th\'eor\`eme de Wick :
\beq
\overline{H} = \sum_{\pi} \overline{H}_{\pi} \nn
\eeq
La\label{appa} somme porte sur les appariements de m\`eches $\pi$ pr\'eservant la couleur $\alpha^{i}_{j} = \alpha_{\pi({}^{i}_{j})}$. Pour fixer les id\'ees, on dira que $L = \{({}^{i}_{j}),({}^{i'}_{j'})\}$ avec $\pi({}^{i}_j) = ({}^{i'}_{j'})$ est un \textbf{lien}, et que $\pi$ est un \textbf{syst\`eme de liens}.  $\overline{H}_{\pi}$ compte les cartes sur lesquelles sont trac\'es des chemins auto\'evitants qui r\'ealisent le \textbf{syst\`eme de liens} $\pi$.

La m\`eche $({}^{i}_{j})$ a des indices matriciels que l'on indique en \'ecrivant $A_{\alpha^{i}_j|a^{i}_{j - 1}a^{i}_{j}}$. Lors de l'appariement entre $({}^{i}_j)$ et $\pi({}^{i}_j)$ :
{\small \bea
&& \contraction[2ex]{\!\!\!\!\!\!\cdots\,\,\frac{1}{x^{i}_{j - 1} - \lambda_{a^{i}_{j - 1}}}\,}{(A_{\alpha^{i}_{j}}|a^{i}_{j - 1}a^{i}_{j})}{\frac{1}{x^{i}_{j} - \lambda_{a^{i}_j}}\,\,\cdots\,\,\frac{1}{x^{i'}_{j' - 1} - \lambda_{a^{i'}_{j' - 1}}}\,}{(A_{\alpha^{i'}_{j'}|a^{i'}_{j' - 1}a^{i'}_{j'}})}
\!\!\!\!\!\!\cdots\,\,\frac{1}{x^{i}_{j - 1} - \lambda_{a^{i}_{j - 1}}}\,(A_{\alpha^{i}_j|a^{i}_{j - 1}a^{i}_j})\,\frac{1}{x^{i}_{j} - \lambda_{a^{i}_j}}\,\,\cdots\,\,\frac{1}{x^{i'}_{j' - 1} - \lambda_{a^{i'}_{j' - 1}}}\,(A_{\alpha^{i'}_{j'}|a^{i'}_{j' - 1}a^{i'}_{j'}})\,\frac{1}{x^{i'}_{j'} - \lambda_{a^{i'}_{j'}}}\,\,\cdots \nn \\
&& \!\!\!\!\!\!\phantom{\cdots\,\,\frac{1}{x^{i}_{j - 1} - \lambda_{a^{i}_{j - 1}}}\,(A_{\alpha^{i}_j|a^{i}_{j - 1}a^{i}_{j}})\,\frac{1}{x^{i}_{j} - \lambda_{a^{i}_j}}\,\,\:} \downarrow \nn \\
&& \!\!\!\!\!\!\cdots\,\,\frac{1}{x^{i}_{j - 1} - \lambda_{b}}\,\frac{1}{x^{i'}_{j'} - \lambda_{b}}\,\,\Big[\frac{t}{N}\,\frac{\delta_{b,a^{i}_{j - 1}}\delta_{b,a^{i'}_{j'}}\,\delta_{b',a^{i}_j}\delta_{b' ,a^{i'}_{j' - 1}}}{K(\lambda_{a},\lambda_{b})}\Big]\,\,\frac{1}{x^{i'}_{j' - 1} - \lambda_{b'}}\,\frac{1}{x^{i}_{j} - \lambda_{b'}}\,\,\cdots \nn
\eea}
$\!\!\!$Les\label{appa2} appariements successifs des matrices $A$ font apparaitre des cycles $\gamma$ qui sont parcourus par un m\^{e}me indice matriciel $b_{\gamma}$ :
\bea
\cdots \rightarrow (i,j - 1)  \rightarrow (i',j') \rightarrow \cdots \nn\\
\cdots \rightarrow (i',j' - 1) \rightarrow (i,j) \rightarrow \cdots \nn
\eea

Si l'on introduit $S$, l'op\'eration \textbf{successeur} $S({}^i_j) = ({}^{\,\,i}_{j + 1})$, ces cycles sont les cycles disjoints de la permutation\footnote{$S \circ \pi$ fait aussi l'affaire.} $\pi\circ S$. Il est commode d'ajouter de nouveaux cycles $\gamma_I$ pour chaque variable $x_I$ vivant sur un bord sans m\`eche. \`{A} chaque variable de bord $x_I$ (entre deux m\`eches $I$ et $S(I)$, ou sur un bord sans m\`eche), on peut donc associer le cycle $\gamma_I$ auquel $I$ appartient. Il faut ensuite sommer sur les indices $b_{\gamma} \in \{1,\ldots,N\}$ ind\'ependants, et int\'egrer sur les valeurs propres $\lambda_l$, avec la mesure induite (\'{E}qn.~\ref{eq:gaaa}) :
\beq
\label{eq:mesi}\dd\nu(\lambda) \propto \Big[\prod_{l = 1}^N \dd\lambda_l e^{-\frac{N}{t}V(\lambda_l)}\Big]\,\frac{|\Delta(\lambda)|^2}{\prod_{\alpha = 1}^{\mathfrak{n}}\prod_{1 \leq l, l' \leq N} \sqrt{K_{\alpha}(\lambda_l,\lambda_{l'})}}
\eeq
Pour chaque lien $L = \{m^{-},m^{+}\}$ de couleur $\alpha_L$, les m\`eches $m^{-}$ et $m^{+}$ appartiennent \`a des cycles not\'es $\gamma^{+}_{L}$ et $\gamma^{-}_L$ (qui sont peut-\^{e}tre identiques), et il existe dans $\overline{H}_{\pi}$ un facteur de couplage entre $\lambda_{b_{\gamma^{+}_L}}$ et $\lambda_{b_{\gamma^{-}_L}}$ via le noyau $K_{\alpha_L}^{-1}$ (cf. \'{E}qn.~\ref{eq:concona}). Ainsi :
\beq
\overline{H}_{\pi} = \Big\langle\!\!\Big\langle \sum_{(b_{\gamma})_{\gamma}} \Big[\prod_{I \in \mathsf{P}} \frac{1}{x_I - \lambda_{b_{\gamma_I}}}\Big]\Big[\prod_{L\,\,\mathrm{lien}\,\mathrm{de}\,\pi} \frac{1}{K_{\alpha_L}(\lambda_{b_{\gamma^{+}_L}},\lambda_{b_{\gamma^{-}_L}})}\Big] \Big\rangle\!\!\Big\rangle \nn
\eeq
avec $\langle\!\langle\cdots\rangle\!\rangle$ la valeur moyenne pour la mesure de l'\'{E}qn.~\ref{eq:mesi}, et $\mathsf{P}$ l'ensemble des variables de bord.

Ce r\'esultat peut \^{e}tre r\'e\'ecrit en fonction du corr\'elateur non connexe $\overline{W}_{n}(\xi_1,\ldots,\xi_n)$, o\`u $n$ est le nombre total de cycles. On rappelle que cette observable est holomorphe pour $\xi_i \in \mathbb{C}\setminus[a,b]$ et admet une discontinuit\'e sur le segment $[a,b]$ (Thm.~\ref{theojj}).
\bea
\label{eq:fdhyu}  \overline{H}_{\pi} & = & \Big(\frac{t}{N}\Big)^{|\mathsf{L}(\pi)|}\Big[\prod_{\gamma \in \mathsf{F}(\pi)}\oint_{\mathcal{C}([a,b])} \frac{\dd\xi_{\gamma}}{2i\pi}\Big]\,\overline{W}_{|\mathsf{F}(\pi)|}\big((\xi_\gamma)_{\gamma}\big)  \\
& & \phantom{\Big(\frac{t}{N}\Big)^{|\mathsf{L}(\pi)|}}\times \Big[\prod_{I \in \mathsf{P}} \frac{1}{x_I - \xi_{\gamma_{I}}}\Big]\Big[\prod_{L \in \mathsf{L}(\pi)} \frac{1}{K_{\alpha_L}(\xi_{\gamma_{L}^+},\xi_{\gamma_{L}^{-}})}\Big] \nn
\eea
$\mathsf{L}(\pi)$ d\'esigne l'ensemble des liens de $\pi$, $\mathsf{F}(\pi)$ est l'ensemble des cycles de $\mathrm{S}\circ \pi$ augment\'e de l'ensemble $\mathsf{B}_0$ des bords sans m\`eches. Cette formule a une interpr\'etation graphique (Fig.~\ref{fig:MixteOn}) : si l'on d\'ecoupe le long des liens une carte o\`u est r\'ealis\'e le syst\`eme de liens $\pi$, on obtient une surface discr\`ete avec $|\mathsf{F}(\pi)|$ \textbf{bords'}. Les variables $\xi_{\gamma}$ vivent sur ces bords'. Ils sont constitu\'es dans un ordre cyclique de l'ar\^{e}te portant $x_I$, du lien $I \rightsquigarrow (\pi\circ S)(I)$, de l'ar\^{e}te portant $x_{(S\circ \pi \circ S)(I)}$, du lien $(S\circ \pi \circ S) \rightsquigarrow (\pi\circ S\circ \pi \circ S)(I)$, \ldots{} Chaque lien $L$ est commun \`a deux bords (\'eventuellement identiques) index\'es par $\gamma^{\pm}_{L}$, et contribue par un facteur $K^{-1}_{\alpha_L}(\xi_{\gamma^{+}_L},\xi_{\gamma^{-}_L})$ \`a l'int\'egrand.

Nous avons donc r\'eduit le calcul de $\overline{H}_{\pi}$ \`a celui des observables uniformes.

\section{Observables connexes \ldots}

De la m\^{e}me fa\c{c}on :
\beq
H = \sum_{\pi} H_{\pi},\qquad H_{\pi} = \sum_{g \geq 0} \Big(\frac{N}{t}\Big)^{2 - 2g - k}\,H^{(g)}_{\pi} \nn
\eeq
et $H_{\pi}^{(g)}$ \'enum\`ere les cartes $\mathcal{G}^*$ connexes de genre $g$ sur lesquelles le syst\`eme de liens $\pi$ est r\'ealis\'e. Si $\pi$ est fix\'e, les $k$ bords initiaux se regroupent en \label{amas} \textbf{amas} $\mathsf{B} = \bigcup_{p = 1}^{s} \mathsf{B}_p$ : on peut passer d'un bord $i \in \mathsf{B}_p$ \`a un bord $i' \in \mathsf{B}_p$ en suivant uniquement les liens de $\pi$, mais pas \`a un bord $i'' \in \mathsf{B}_q$ si $q \neq p$.

\subsubsection{\ldots \`a un amas}

Parfois, $\pi$ assure la connexion entre tous les bords, i.e. $s = 1$ (cela suppose que tous les bords portent des m\`eches). Dans ce cas, la contrainte de connectivit\'e n'en est pas une : $H_{\pi} = \overline{H}_{\pi}$. En identifiant la plus grande puissance de $N$ dans la formule~\ref{eq:fdhyu}, on en d\'eduit le genre minimal d'une carte $\mathcal{G}^*$ sur laquelle $\pi$ peut \^{e}tre trac\'e :
\beq
\label{eq:genuo} g_{\mathrm{min}}(\pi) = 1 + \frac{|\mathsf{L}(\pi)| - |\mathsf{F}(\pi)| - |\mathsf{B}|}{2}
\eeq
C'est aussi le genre minimal d'une vari\'et\'e $\Sigma_g$ de dimension $2$ sur laquelle $\pi$ peut \^{e}tre trac\'e, puisque l'on peut toujours se ramener pour ce probl\`eme \`a des discr\'etisations (par exemple, des triangulations) de $\Sigma_g$. Les syst\`emes de liens pour lesquels $g_{\mathrm{min}}(\pi) = 0$ sont dits \textbf{planaires}, ce sont \label{TL2}eux qui forment une base des alg\`ebres de type Temperley-Lieb (Fig.~\ref{fig:TL}). Ils sont caract\'eris\'es par :
\beq
|\mathsf{F}(\pi)| = |\mathsf{L}(\pi)| - |\mathsf{B}| + 2 \nn
\eeq

En genre minimal, $W_n(\xi_1,\ldots,\xi_n)$ se factorise en $W_1^{(0)}(\xi_l)$, ce qui simplfie un peu l'expression :
\beq
H_{\pi}^{(g_{\mathrm{min}}(\pi))} = \Big[\prod_{\gamma \in \mathsf{F}(\pi)} \oint_{\mathcal{C}([a,b])} \frac{\dd\xi_{\gamma}}{2i\pi}\,W_1^{(0)}(\xi_{\gamma})\Big]\,\Big[\prod_{I \in \mathsf{P}} \frac{1}{x_I - \xi_{\gamma_{I}}}\Big]\Big[\prod_{L \in \mathsf{L}(\pi)} \frac{1}{K_{\alpha_L}(\xi_{\gamma_{L}^+},\xi_{\gamma_{L}^{-}})}\Big] \nn
\eeq

\subsubsection{\ldots \`a plusieurs amas}

Lorsque $s > 1$, le calcul de $H_{\pi}$ est tr\`es similaire \`a celui de $\overline{H}_{\pi}$. La seule diff\'erence est qu'il faut maintenant imposer la connectivit\'e entre les amas. Les cycles se regroupent aussi naturellement en amas, not\'es $\mathsf{F}_1(\pi),\ldots,\mathsf{F}_s(\pi)$, avec la r\`egle $\gamma \in \mathsf{F}_p(\pi)$ ssi un bord $i \in \mathsf{B}_p$ est repr\'esent\'e dans le cycle $\gamma$. Par cons\'equent :
\bea
\label{eq:expiu} H_{\pi} & = & \Big(\frac{t}{N}\Big)^{|\mathsf{L}(\pi)|}\Big[\prod_{\gamma \in \mathsf{F}(\pi)}\oint_{\mathcal{C}([a,b])} \frac{\dd\xi_{\gamma}}{2i\pi}\Big]\,W_{|\mathsf{F}(\pi)|;\pi\mathrm{-}c}\big((\xi_\gamma)_{\gamma}\big) \\
&& \phantom{\Big(\frac{t}{N}\Big)^{|\mathsf{L}(\pi)|}}\times\Big[\prod_{I \in \mathsf{P}} \frac{1}{x_I - \xi_{\gamma_{I}}}\Big]\Big[\prod_{L \in \mathsf{L}(\pi)} \frac{1}{K_{\alpha_L}(\xi_{\gamma_{L}^+},\xi_{\gamma_{L}^{-}})}\Big] \nn
\eea
avec le \textbf{corr\'elateur $\pi$-connexe}, qui est par d\'efinition le cumulant des observables d'amas $\mathcal{O}_p$ :
\beq
W_{n|\pi\mathrm{-}c} = \Big\langle\!\!\Big\langle \prod_{p = 1}^{s} \mathcal{O}_p \Big\rangle\!\!\Big\rangle_c \qquad \mathcal{O}_p = \prod_{\gamma \in \mathsf{F}_p(\pi)} \Big(\Tr\,\frac{1}{\xi_{\gamma} - M}\Big) \nn
\eeq
Il est remarquable que $H$ soit une fonction sym\'etrique des variables $x_I$ qui vivent sur un m\^{e}me cycle $\gamma$.

Notons $\pi_p$ le syst\`eme de liens $\pi$ restreint aux m\`eches de l'amas $\mathsf{B}_p$. Il peut \^{e}tre trac\'e sur une surface de genre minimal $g_{\mathrm{min}}(\pi_p)$. Par convention, $g_{\mathrm{min}}(\pi_p) \equiv 0$ si $\mathsf{B}_p$ est constitu\'e d'un seul bord qui n'a pas de m\`eche. L'inspection des puissances de $N$ maximales apparaissant dans l'Eqn~\ref{eq:expiu} montre alors que :
\beq
g_{\mathrm{min}}(\pi) = \sum_{p = 1}^s g_{\mathrm{min}}(\pi_p)  \nn
\eeq

\begin{landscape}
\addtolength{\footskip}{30pt}
\addtolength{\linewidth}{50pt}
\begin{figure}[h!]
\begin{center}
\includegraphics[width=1.57\textwidth]{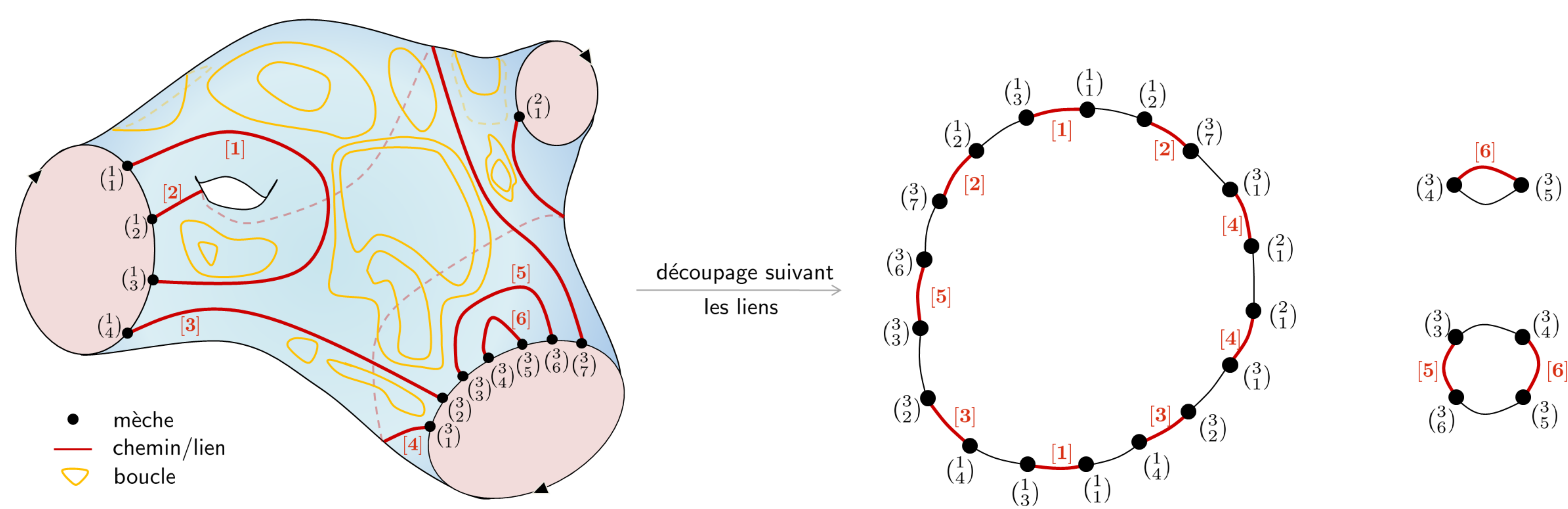}
\vspace{1cm}
\caption{\label{fig:MixteOn} Exemple d'observable \`a un amas (donc connexe), de genre $1$, avec $12 = 2|\mathsf{L}(\pi)|$ m\`eches dispos\'ees sur $|\mathsf{B}| = 3$ bords. Apr\`es d\'ecoupage, on obtient $|\mathsf{F}(\pi)| = 3$ faces, que l'on appelle bords' pour \'eviter la confusion avec les faces de $\mathcal{G}^*$ et les bords initiaux. Par cons\'equent, $g_{\mathrm{min}}(\pi) = 1$. $\pi$ ne peut \^{e}tre trac\'e sur une surface de genre $0$ : sur une surface de genre $0$, le lien $[\mathbf{2}]$ devrait croiser le lien $[\mathbf{1}]$ pour rejoindre le $2^{\textrm{\`{e}me}}$ bord.}
\end{center}
\end{figure}
\end{landscape}

\section{Exemples}

Les expressions~\ref{eq:fdhyu} et \ref{eq:expiu} peuvent \^{e}tre transform\'ees en d\'epla\c{c}ant les contours vers les p\^{o}les de l'int\'egrand, et \'eventuellement en utilisant les \'equations de Schwinger-Dyson (\S~\ref{sec:eqju}). Illustrons-le sur quelques exemples simples : dans le mod\`ele $\On$ trivalent, i.e. $K_{\alpha}(\lambda,\mu) = 1/\z - (\lambda + \mu)$ ind\'ependamment de $\alpha$, pour des observables \`a un seul bord, i.e. $|\mathsf{B}| = 1$, et en genre minimal.

\vspace{0.1cm}

\noindent \begin{minipage}[l]{0.28\linewidth}
\emph{Un seul lien} $\diamond\,$
\end{minipage}
\begin{minipage}[c]{0.4\linewidth}
\includegraphics[width=\textwidth]{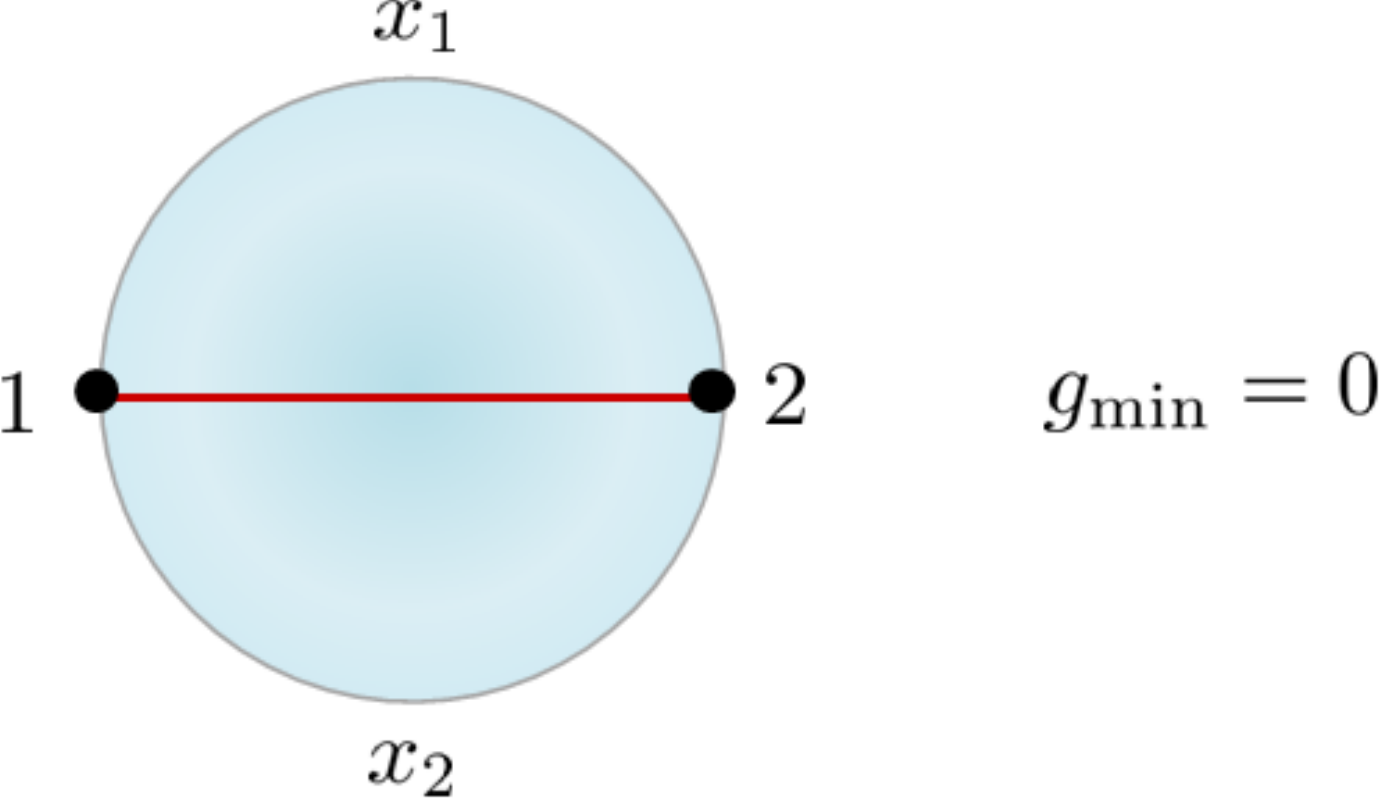}
\end{minipage} \hfill

\vspace{-0.25cm}

\begin{footnotesize} \bea
H_{(12)}^{(0)} & = & \oint_{\mathcal{C}([a,b])^2} \frac{\dd\xi}{2i\pi}\,\frac{\dd\eta}{2i\pi}\,\frac{1}{x_1 - \xi}\,\frac{1}{x_2 - \eta}\,\frac{1}{1/\z - (\xi + \eta)}\,\overline{W}_2(\xi,\eta) \nn \\
& = & \oint_{\mathcal{C}([a,b])} \frac{\dd \xi}{2i\pi}\,\frac{1}{x_1 - \xi}\,\frac{1}{1/\z - (x_2 + \xi)}\Big(\overline{W}_2(\xi,x_2) - \overline{W}_2(\xi,1/\z - \xi)\Big) \nn \\
& \mathop{=}^{\!\!\!\!\!\!{}^{\mathrm{SD}}} & \oint_{\mathcal{C}([a,b])} \frac{\dd \xi}{2i\pi}\,\frac{1}{x_1 - \xi}\,\frac{1}{1/\z - (x_2 + \xi)}\Big\{\overline{W}_2(\xi,x_2)  + \nn \\
&& \mathfrak{n}^{-1}\Big(\overline{W}_2(\xi,\xi) - [V'W_1]_{-}(\xi) + \overline{W}_2(1/\z - \xi,1/\z - \xi) - [V'W_1]_{-}(1/\z - \xi)\Big)\Big\} \nn \\
& \,\,\,\mathop{=}^{\!\!\!\!\!\!\!\!\!\!{}^{1\,\textrm{coup.}}} & \oint_{\mathcal{C}([a,b])} \frac{\dd \xi}{2i\pi}\,\frac{1}{x_1 - \xi}\,\frac{1}{1/\z - (x_2 + \xi)}\Big\{\overline{W}_2(\xi,x_2) + \mathfrak{n}^{-1}\Big(\overline{W}_2(\xi,\xi) - [V'W_1]_{-}(\xi)\Big)\Big\} \nn \\
& = & \frac{1}{1/\z - (x_1 + x_2)}\Big\{\overline{W}_2(x_1,x_2) - \overline{W}_2(1/\z - x_2,x_2) + \nn \\
&& \mathfrak{n}^{-1}\Big(\overline{W}_2(x_1,x_1) - [V'W_1]_{-}(x_1) - \overline{W}_2(1/\z - x_2,1/\z - x_2) + [V'W_1]_{-}(1/\z - x_2)\Big)\Big\} \nn \\
& \mathop{=}^{\!\!\!\!\!\!{}^{\mathrm{SD}}} & \frac{\overline{W}_2(x_1,x_2) + \mathfrak{n}^{-1}\Big(\overline{W}_2(x_1,x_1) + \overline{W}_2(x_2,x_2) - [V'W_1]_{-}(x_1) - [V'W_1]_{-}(x_2)\Big)}{1/\z - (x_1 + x_2)} \nn
\eea
\end{footnotesize}

\noindent \emph{Un seul bord'} $\diamond\,$ Lorsque $|\mathsf{F}(\pi)| = 1$, $\pi$ apparie $2l$ m\`eches de sorte que $g_{\mathrm{min}}(\pi) = l/2$ (cela impose \`a $l$ d'\^{e}tre pair). $H_{\pi}$ ne d\'epend pas d'un tel $\pi$, car les variables $x_1,\ldots,x_{2l}$ jouent des r\^{o}les compl\`etement sym\'etriques :
\bea
H_l^{(l/2)} & = & \oint_{\mathcal{C}([a,b])} \frac{\dd \xi}{2i\pi}\,\frac{W_1(\xi)}{(1/\z - 2\xi)^{l}}\,\Big[\prod_{j = 1}^{2l} \frac{1}{x_j - \xi}\Big] \nn \\
& = & \frac{(-1)^{l - 1}}{2^{l}\,(l - 1)!}\Big[\frac{\dd^{l - 1} W_1}{\dd x^{l - 1}}\Big](1/2\z) + \sum_{j = 1}^{2l} \frac{W_1(x_j)}{(1/\z - 2x_j)^l \prod_{j' \neq j} (x_{j'} - x_j)} \nn
\eea

\begin{figure}[h!]
\begin{center}
\includegraphics[width=\textwidth]{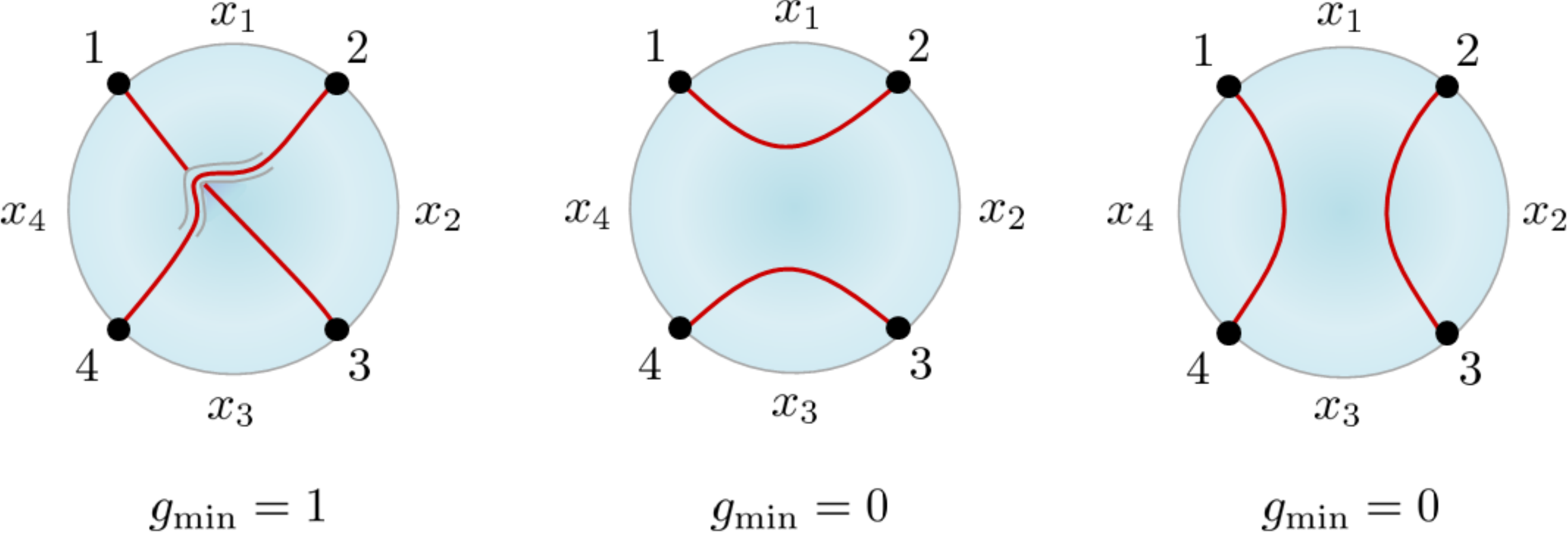}
\caption{\label{fig:2lin} Syst\`emes de liens entre quatre m\`eches dispos\'ees sur un bord.}
\end{center}
\end{figure}

\noindent \emph{Deux liens} $\diamond\,$ Pour $\pi = {(13)(24)}$, $|\mathsf{F}(\pi)| = 1$ et l'on se trouve dans le cas pr\'ec\'edent. Les deux autres possibilit\'es $\pi = (12)(34)$ et $\pi' = (14)(23)$ sont des syst\`emes de liens planaires, et admettent $|\mathsf{F}(\pi)| = 3$ bords'. $\pi$ et $\pi'$ se d\'eduisent l'un de l'autre par permutation circulaire des variables $(x_1,x_2,x_3,x_4)$ (Fig.~\ref{fig:2lin}).

\vspace{-0.25cm}

\begin{footnotesize}
\bea
H_{(12)(34)}^{(0)} & = & \oint_{\mathcal{C}([a,b])^3} \frac{\dd \xi}{2i\pi}\,\frac{\dd \eta}{2i\pi}\,\frac{\dd \sigma}{2i\pi}\,\frac{\overline{W}_3(\xi,\eta,\sigma)}{(x_1 - \xi)(x_2 - \sigma)(x_3 - \eta)(x_4 - \sigma)}\,\frac{1}{1/\z - (\xi + \sigma)}\,\frac{1}{1/\z - (\eta + \sigma)} \nn \\
& = & \oint_{\mathcal{C}([a,b])} \frac{\dd \sigma}{2i\pi}\,\frac{1}{x_2 - \sigma}\,\frac{1}{x_4  - \sigma}\,\frac{1}{1/\z - (x_1 + \sigma)}\,\frac{1}{1/\z - (x_3 + \sigma)} \nn \\
&& \times\Big\{\overline{W}_3(x_1,\sigma,x_3) - \overline{W}_3(x_1,\sigma,1/\z - \sigma) - \overline{W}_3(1/\z - \sigma,\sigma,x_3) + \overline{W}_3(1/\z - \sigma,\sigma,1/\z - \sigma)\Big\} \nn
\eea
\end{footnotesize}
$\!\!\!$Les \'equations de Schwinger-Dyson permettent d'exprimer $\overline{W}_3(\sigma,1/\z - \sigma,x_j)$ et $\overline{W}_3(\sigma,1/\z - \sigma,1/\z - \sigma)$ en termes de quantit\'es qui sont (\'eventuellement) discontinues sur $[a,b]$ seulement. Le r\'esultat final s'\'ecrit :
\begin{scriptsize}
\bea
H_{(12)(34)}^{(0)} & = & \frac{1}{x_4 - x_2}\Bigg\{\frac{\overline{W}_3(x_1,x_2,x_3) + \mathfrak{n}^{-1}\Big(\overline{W}_3(x_2,x_2,x_1) - [V'\overline{W}_2(\cdot,x_1)]_{-}(x_2)\Big) - \mathfrak{n}^{-1}\big(x_1 \leftrightarrow x_3\big)}{(1/\z - (x_1 + x_2)\big)\big(1/\z - (x_3 + x_2)\big)} \nn \\
&& - \frac{\overline{W}_3(x_1,x_4,x_3) + \mathfrak{n}^{-1}\Big(\overline{W}_3(x_4,x_4,x_1) - [V'\overline{W}_2(\cdot,x_1)]_{-}(x_4)\Big) - \mathfrak{n}^{-1}\big(x_1 \leftrightarrow x_3\big)}{\big(1/\z - (x_1 + x_4)\big)\big(1/\z - (x_3 + x_4)\big)}\Bigg\} \nn \\
& & + \frac{1}{x_1 - x_3}\left\{\begin{array}{c} \overline{W}_3(x_1,1/\z - x_1,x_3) - \mathfrak{n}^{-1}\Big(\overline{W}_3(1/\z - x_1,1/\z - x_1,x_3) - [V'\overline{W}_2(\cdot,x_3)]_{-}(x_1)\Big)  \\ + \mathfrak{n}^{-1}\Big(\overline{W}_3(1/\z - x_1,1/\z - x_1,x_1) - [V'\overline{W}_2(\cdot,x_1)]_{-}(1/\z - x_1)\Big) \\ \hline \big(1/\z - (x_1 + x_2)\big)\big(1/\z - (x_1 + x_4)\big) \end{array}\right. \nn\\
&& \left. \raisebox{-0.3cm}{ - }\: \begin{array}{c} \overline{W}_3(x_3,1/\z - x_3,x_1) - \mathfrak{n}^{-1}\Big(\overline{W}_3(1/\z - x_3,1/\z - x_3,x_1) - [V'\overline{W}_2(\cdot,x_1)]_{-}(x_3)\Big) \\  + \mathfrak{n}^{-1}\Big(\overline{W}_3(1/\z - x_3,1/\z - x_3,x_3) - [V'\overline{W}_2(\cdot,x_3)]_{-}(1/\z - x_3)\Big) \\ \hline \big(1/\z - (x_3 + x_2)\big)\big(1/\z - (x_3 + x_4)\big) \end{array}\right\} \nn
\eea
\end{scriptsize}

\section{Questions}

Si l'on constate que l'on peut calculer explicitement les observables $H_{\pi}$ au cas par cas, la structure g\'en\'erale reste \`a \'etablir. Par exemple, il existe une action naturelle des g\'en\'erateurs $\mathbf{e}_m$ de l'alg\`ebre de Temperley-Lieb \`a $r^i$ m\`eches sur le cylindre (Fig.~\ref{fig:TL}), sur le $i^{\textrm{\`{e}me}}$ bord : elle transforme $\pi$, par composition, en un syst\`eme de liens $\pi' = \mathbf{e}_m\cdot \pi$. On aimerait d\'ecrire cette transformation au niveau de la repr\'esentation int\'egrale~\ref{eq:expiu}. On remarque par ailleurs, en calculant les int\'egrales de contours dans le mod\`ele $\On$ trivalent, que $H_{\pi}$ est toujours la somme d'une observable "irr\'eductible" qui d\'epend pleinement de tous les $x_I$, et de termes de complexit\'e inf\'erieure. Sans donner une d\'efinition g\'en\'erale, on aimerait identifier comme parties irr\'eductibles dans les exemples pr\'ec\'edents :
{\small
\bea
(H_{(12)}^{(0)})_{\mathrm{irr}} & = & \frac{\overline{W}_2(x_1,x_2)}{1/\z - (x_1 + x_2)} \nn \\
(H_{(12)(34)}^{(0)})_{\mathrm{irr}} & = & \frac{1}{x_4 - x_2}\left\{\frac{\overline{W}_3(x_1,x_2,x_3)}{\big(1/\z - (x_1 + x_2)\big)\big(1/\z - (x_3 + x_2)\big)} \right. \nn \\
&& \phantom{\frac{1}{x_4 - x_2}\,} \left. + \frac{\overline{W}_3(x_1,x_3,x_4)}{\big(1/\z - (x_1 + x_4)\big)\big(1/\z - (x_3 + x_4)\big)}\right\} \nn
\eea}
$\!\!\!$ $H_{\pi}$ tel que nous l'avons d\'efini semble ne pas \^{e}tre la bonne base d'observable. Il faudrait donc trouver la bonne d\'efinition qui remplacerait l'\'{E}qn.~\ref{eq:defss} et s\'electionnerait la partie irr\'eductible de $H$. Cela faciliterait l'identification de la structure alg\'ebrique des $H_{\pi}$.

Il serait instructif de comparer avec les r\'esultats connus pour la chaine \`a deux matrices hermitiennes. En effet, le mod\`ele :
\beq
 \dd\nu(M_1,M_2) = \dd M_1\,\dd M_2\,e^{-\frac{N}{t}\Tr\big(V(M_1) + V(M_2) - c_{12}\,M_1M_2\big)},\qquad \mathrm{deg}\,V = 3 \nn
\eeq
est dual au mod\`ele $\mathcal{O}(\mathfrak{1})$ trivalent sur triangulations al\'eatoires, en posant $M = M_1 + M_2$ et $A = M_1 - M_2 - \mathrm{cte}$. Dans les chaines \`a deux matrices hermitiennes, toutes les observables invariantes $\mathrm{U}(N)$ impliquant $M_1$ et $M_2$ ont \'et\'e calcul\'ees par Eynard et Orantin \cite{EOmixed}. Ils ont trouv\'e notamment \cite{EOfirst} que les observables planaires avec un bord non uniforme :
 \beq
H_{n}^{(0)}(x_1,y_1;\ldots;x_n,y_n) = \Big\langle \Tr\Big(\frac{1}{x_1 - M_1}\frac{1}{y_1 - M_2}\cdots \frac{1}{x_n - M_1}\frac{1}{y_n - M_2}\Big)\Big\rangle^{(0)} \nn
\eeq
se d\'ecomposent sur les syst\`emes de liens planaires $\pi$ :
\beq
H_{n}^{(0)}(x_1,y_1;\ldots;x_n,y_n) = \sum_{\pi} C_{n;\pi}^{(0)}(x_1,y_1;\ldots;x_n,y_n)\,\prod_{j= 1}^n H_{1}^{(0)}(x_j,y_{\pi(j)}) \nn
\eeq
$C_{n;\pi}^{(0)}$ sont des fractions rationnelles des $x_j$ et $y_j$, dont la construction ainsi que la structure alg\'ebrique est connue. Elles appartiennent \label{tru}au commutant d'une certaine famille de matrices de transfert de taille $n!\times n!$, d\'ependant de deux param\`etres spectraux $\xi$ et $\eta$ :
\bea
&& \mathcal{M}_{\pi,\pi'}(\mathbf{x},\mathbf{y};\xi,\eta) \equiv \prod_{j = 1}^{n}\Big(\delta_{\pi(j),\pi'(j)} + \frac{1}{(\xi - x_j)(\eta - x_{\pi(j)})}\Big) \nn \\
&& \forall \xi,\xi',\eta,\eta' \in \mathbb{C},\qquad [\mathcal{M}(\mathbf{x},\mathbf{y};\xi,\eta),\mathcal{M}(\mathbf{x},\mathbf{y};\xi',\eta')] = 0 \nn
\eea
$H_{n}^{(0)}$ est une s\'erie g\'en\'eratrice de cartes ayant la topologie d'un disque et dont les faces portent un mod\`ele\label{Isi3} d'Ising\footnote{Autrement dit, une face g\'en\'er\'ee par des matrices $M_1$ (resp. $M_2$) porte un $\ominus$ (resp. $\oplus$).}. En revanche, la d\'ecomposition sur les syst\`emes de liens planaires est purement alg\'ebrique, l'interpr\'etation combinatoire des termes $H_{n;\pi}^{(0)}$ pris isol\'ement n'est pas connue. Dans le mod\`ele $\On$, la situation est invers\'ee : on connait l'interpr\'etation combinatoire de $H_{\pi}$, mais pas la structure alg\'ebrique du r\'esultat. Cela appelle de nouveaux travaux pour \'eclairer les relations entre observables non uniformes dans le mod\`ele $\On$, celles de la chaine \`a deux matrices, leur interpr\'etation combinatoire, les propri\'et\'es d'int\'egrabilit\'e quantique en relation avec les matrices de transfert $\mathcal{M}$, et l'alg\`ebre de Temperley-Lieb. Le va-et-vient entre les deux mod\`eles et l'impl\'ementation de leur dualit\'e au niveau des observables sera probablement source d'inspirations.

Par ailleurs, l'universalit\'e de la formule~\ref{eq:expiu} est remarquable : elle est valable dans tous les mod\`eles de boucles, seul le noyau $K^{-1}$ d\'epend des d\'etails du mod\`ele.
On peut se demander \`a quel point la structure alg\'ebrique que l'on aimerait trouver d\'epend du noyau $K$, et si il n'y a pas d\'ej\`a de nombreuses propri\'et\'es li\'ees \`a Temperley-Lieb cach\'ees dans les fractions rationnelles de l'int\'egrand dans l'\'{E}qn.~\ref{eq:expiu}.





\newpage
\thispagestyle{empty}
\phantom{bbk}

\newpage

\chapter{Index}
\thispagestyle{plain}
\vspace{-1.5cm}
\rule{\textwidth}{1.5mm}

\vspace{2cm}

\label{inde}

{\small \textsf{
\begin{minipage}[l]{0.45\linewidth}
Alg\`ebre \\
\phantom{algi} de Brauer \phantom{a} \pageref{Brauer} \\
\phantom{algi} de Temperley-Lieb \phantom{a} \pageref{TL}, \pageref{TL2} \\
\textbf{Alg\'ebro-g\'eom\'etrique} \phantom{a} \pageref{sec:ALG} \phantom{a} \\
Amas \phantom{a} \pageref{amas}\\
Anomalie holomorphe \phantom{a} \pageref{anoholo} \\
Appariements \phantom{a}  \pageref{appa3}, \pageref{epsi}, \pageref{appa2}  \\
Application miroir \phantom{a} \pageref{amiro}, \pageref{sec:BKMP} \\
Autor\'eplication \phantom{a} \pageref{autorep}, \pageref{PPA}, \pageref{eqly} \\
Base \\
\phantom{algi} d'homologie de contours \phantom{a} \pageref{hypo3} \\
\phantom{algi} symplectique de cycle \phantom{a} \pageref{cycsym}, \pageref{cycsym2} \\
\phantom{algi} \pageref{cycsym3}, \pageref{sec:ALG}, \textbf{\pageref{cycsym4}} \\
Bord dur, bord mou \phantom{a} \pageref{bobo3}, \pageref{bobo2}  \\
Branes \phantom{a} \pageref{baba}, \pageref{baba2}, \pageref{baba3} \\
Cartes \phantom{a} \pageref{caca}, \pageref{sec:foerm} \\
Casimir quadratique \phantom{a} \pageref{Casi}, \textbf{\pageref{eq:erratum}} \\
Cercle arctique \phantom{a} \pageref{fig:okoun}, \pageref{arc} \\
Charge centrale \phantom{a} \pageref{cc}, \pageref{cc2}, \pageref{cc3} \\
Chern-Simons \phantom{a} \pageref{CS10}, \pageref{CS11}
\end{minipage}
\hfill \begin{minipage}[l]{0.45\linewidth}
Classes \\
\phantom{algi} $\kappa_r$ \phantom{a} \pageref{cls}, \pageref{kaka}, \pageref{cls2} \\
\phantom{algi} $\psi_i$ \phantom{a} \pageref{cls}, \pageref{psipsi}, \pageref{cls2} \\
\phantom{algi} $\lambda_g$ \phantom{a} \pageref{Hodge} \\
\phantom{algi} d'universalit\'e \phantom{a} \pageref{noyu}, \pageref{sec:introuniv}, \pageref{sec:conn} \\
Conjecture \\
\phantom{algi} AGT \phantom{a} \pageref{AGTss} \\
\phantom{algi} BGS \phantom{a} \pageref{boh} \\
\phantom{algi} BKMP \phantom{a} \pageref{BKMPZ}, \textbf{\pageref{sec:BKMP}} \\
\phantom{algi} de Bouchard et Mari\~{n}o \phantom{a} \textbf{\pageref{BMM}}, \pageref{BMM2} \\
\phantom{algi} de Calabi \phantom{a} \pageref{Calabi} \\
\phantom{algi} de Montgomery \phantom{a} \pageref{mon} \\
\phantom{algi} sur les syst\`emes dispersifs \phantom{a} \pageref{conjd} \\
\phantom{algi} de Razumov-Stroganov \phantom{a} \pageref{RazumovS}\\
\phantom{algi} de Witten \phantom{a} \pageref{WW} \\
Constantes \\
\phantom{algi} vecteur des constantes de Riemann \\
\phantom{algi} \pageref{vcc}, \textbf{\pageref{vcc2}} \\
\phantom{algi} de Tracy-Widom \phantom{a} \pageref{cteTW}, \pageref{eq:taube} \\
Convexit\'e \phantom{a} \pageref{convex}
\end{minipage}
\newpage
\begin{minipage}[l]{0.45\linewidth}
Corr\'elateurs \phantom{a} \textbf{\pageref{sec:correlateurs}}, \pageref{sec:Shy}, \pageref{coco0}, \pageref{coco02}, \pageref{coco3}, \pageref{coco34} \\
Coupures \phantom{a} \pageref{coucou}, \pageref{sec:pluscoup}, \pageref{Joukov}, \pageref{theojj}, \pageref{coucou3} \\
Coupures images, coupures fant\^{o}mes \phantom{a} \textbf{\pageref{ohu}}, \pageref{ohu2}\\
Courbe \\
\phantom{algi} de Boutroux \phantom{a} \pageref{boubou}, \pageref{boubou2} \\
\phantom{algi} hyperelliptique \phantom{a} \pageref{hyper}, \pageref{hyper2}, \pageref{hyper3}, \pageref{eq:cSS} \\
\phantom{algi} miroir \phantom{a} \pageref{cm}, \pageref{sec:exCY}, \pageref{sec:BKMP}, \pageref{cm2} \\
Courbe spectrale \\
\phantom{algi} d\'efinition \phantom{a} \pageref{sec:defc}, \pageref{sec:defsq} \\
\phantom{algi} d'Airy \phantom{a} \pageref{eq:Airy}, \pageref{Airy} \\
\phantom{algi} de Kontsevich \phantom{a} \pageref{eq:courbeK}, \pageref{courbeK2}, \pageref{courbeK3} \\
\phantom{algi} de Lambert \phantom{a} \pageref{Lambert}, \pageref{Lambert2}, \pageref{Lambert3} \\
\phantom{algi} de Painlev\'e I \phantom{a} \pageref{PIs} \\
\phantom{algi} de Painlev\'e II \phantom{a} \pageref{PIIs}, \pageref{PIIs2} \\
\phantom{algi} de Tracy-Widom \phantom{a} \pageref{TWaa2}, \pageref{TWaa}\\
\phantom{algi} semiclassique \phantom{a} \pageref{semi}, \pageref{semi2} \\
Cumulants \phantom{a} \pageref{cum}, \pageref{cum2}\\
D\'ecomposition \\
\phantom{algi} polaire \phantom{a} \pageref{decop}\\
\phantom{algi} de Strebel \phantom{a} \pageref{fig:Streb22} \\
Densit\'e de valeurs propres \phantom{a} \pageref{dens}, \pageref{dens2}, \pageref{dens3}, \pageref{dens4}, \pageref{dens5}, \pageref{dens6} \\
D\'eterminant \\
\phantom{algi} de Cauchy \phantom{a} \pageref{detCa}, \pageref{detCb}\\
\phantom{algi} de Fredholm \phantom{a} \pageref{Fredo}, \pageref{Fredo2}, \pageref{Fredo3}\\
\phantom{algi} de Vandermonde \phantom{a} \pageref{Vande}, \textbf{\pageref{Vande2}}, \pageref{Vande3} \\
D\'eveloppement \\
\phantom{algi} asymptotique complet \phantom{a} \pageref{sec:Amj}, \pageref{sec:pluscoup}, \\
\phantom{algi} \pageref{sec:contr}, \pageref{sec:mono} \\
\phantom{algi} topologique \phantom{a} \pageref{devo}, \pageref{devo6}, \pageref{devo3}, \pageref{eq:juk} \\
\phantom{algi} \pageref{thsu}, \pageref{devo4}, \pageref{devo5} \\
Diagramme torique \phantom{a} \textbf{\pageref{fig:patchCYU}}, \pageref{sec:topover}, \pageref{eq:ZchainBKMP} \\
Dispersif (syst\`eme int\'egrable) \phantom{a} \pageref{dispo}
\end{minipage}
\hfill \begin{minipage}[l]{0.45\linewidth}
Double limite d'\'echelle \phantom{a} \pageref{heurintro1}, \pageref{dls} \\
\'{E}nergie libre \phantom{a} \pageref{eneli}, \pageref{sec:1maj}, \pageref{orps}, \pageref{eneli2}, \pageref{eneli3}\\
Ensemble \\
\phantom{algi} gaussien \phantom{a} \pageref{2point}, \pageref{eq:SelG}, \pageref{dls} \\
\phantom{algi} de Laguerre \phantom{a} \pageref{eq:SelL}, \pageref{dls} \\
\'{E}quation \\
\phantom{algi} de boucle maitresse \phantom{a} \textbf{\pageref{eq:VVVV1}}, \pageref{eq:probRHW10} \\
\phantom{algi} de Hirota \phantom{a} \pageref{eq:Hirhi}, \pageref{eq:amto}, \pageref{PP1}, \pageref{conjt}, \\
\phantom{algi} \pageref{Hiro} \\
\phantom{algi} de Korteweg-de Vries \phantom{a} \pageref{KDV}, \pageref{KDV2}\\
\phantom{algi} de Kadomtsev-Petviashvili \phantom{a} \pageref{KPeq} \\
\phantom{algi} de Painlev\'e I \phantom{a} \pageref{API}, \pageref{eq:PIA} \\
\phantom{algi} de Painlev\'e II \phantom{a} \pageref{APII}, \pageref{eq:PII}\\
\'{E}quations \\
\phantom{algi} de boucles \phantom{a} \pageref{sec:heqb}, \pageref{heqb2}, \pageref{heqb3}, \pageref{heqb4} \\
\phantom{algi} \textit{cut-and-join} \phantom{a}  \pageref{heqb4} \\
\phantom{algi} de Schwinger-Dyson \phantom{a} \pageref{sec:eqboucl1}, \pageref{SDD}, \pageref{sec:eqju}, \\
\phantom{algi} \pageref{SDD2} \\
Espace des modules des surfaces \\
 de Riemann \phantom{a} \pageref{em}, \pageref{sec:interFg}, \pageref{cini} \\
Espaces \\
\phantom{algi} de Calabi-Yau \phantom{a} \pageref{CYs}, \pageref{CYs2}, \pageref{CYs3} \\
\phantom{algi} de K\"{a}hler \phantom{a} \pageref{Kahl}, \pageref{Kahl3} \\
Exposants critiques \phantom{a} \pageref{sec:diss}, \pageref{disss}, \pageref{exposs} \\
Feuillet physique \phantom{a} \pageref{fph}, \pageref{fig:Joukov}, \pageref{fph2} \\
Flots \phantom{a} \pageref{flot}, \pageref{RHH}, \pageref{sec:ALG}, \pageref{tauto2} \\
Fonction \\
\phantom{algi} de Barnes \phantom{a} \pageref{Seljk}, \textbf{\pageref{Seljk2}}, \pageref{eq:taube} \\
\phantom{algi} de Hastings-McLeod \phantom{a} \textbf{\pageref{eq:jud}}, \pageref{fhua}, \\
\phantom{algi} \pageref{theu}, \pageref{eq:juml} \\
\phantom{algi} $\wp$ de Weierstra{\ss} \phantom{a} \pageref{Wei}, \pageref{Wei2}, \textbf{\pageref{Wei3}} \\
\phantom{algi} de partition \phantom{a} \pageref{fonp}, \pageref{fonp2}, \pageref{fonp3}, \pageref{fonp4}, \\
\phantom{algi} \pageref{fonp5}, \pageref{fonp6} \\
\phantom{algi} de partition de Nekrasov \phantom{a} \pageref{fonp7} \\
\phantom{algi} tau \phantom{a} \pageref{thKont}, \pageref{eq:Jibu}, \pageref{eq:taualg}, \pageref{eq:hyh}
\end{minipage}
\newpage
\begin{minipage}[l]{0.45\linewidth}
Fonction \\
\phantom{algi} tau de Bergman \phantom{a} \pageref{eq:hy3} \\
\phantom{algi} th\^{e}ta \phantom{a} \pageref{theq}, \pageref{eq:taualg}, \pageref{the}, \pageref{the2}, \pageref{the3} \\
\phantom{algi} z\^{e}ta de Riemann \phantom{a} \pageref{Zetu2}, \pageref{zetu}  \\
Fonctions \\
\phantom{algi} de Baker-Akhiezer \phantom{a} \pageref{bsbs}\\
\phantom{algi} elliptiques \phantom{a} \pageref{eq:parmu}, \textbf{\pageref{elli}} \\
\phantom{algi} pseudoelliptiques \phantom{a} \pageref{psu}, \pageref{elli2}\\
\phantom{algi} quasi p\'eriodiques \phantom{a} \pageref{qausi}\\
Formes diff\'erentielles holomorphes \phantom{a} \pageref{sec:geomspec}, \textbf{\pageref{hol}} \\
Formes diff\'erentielles m\'eromorphes \phantom{a} \pageref{sec:geomspec}, \pageref{sec:ALG}, \pageref{hol}, \textbf{\pageref{sec:meroe}} \\
Formule \\
\phantom{algi} ELSV \phantom{a} \pageref{ELSV}, \pageref{ELSVSS}\\
\phantom{algi} exponentielle \phantom{a} \pageref{esoo}\\
\phantom{algi} de dilatation \phantom{a} \pageref{eq:dilat} \\
\phantom{algi} de Karlin-McGregor \phantom{a} \pageref{Karlin} \\
\phantom{algi} KPZ \phantom{a} \pageref{disss}\\
\phantom{algi} de Mari{\~{n}}o-Vafa \phantom{a} \pageref{MVV} \\
\phantom{algi} des r\'esidus \phantom{a} \pageref{eq:res}, \pageref{resss}, \pageref{ressss}\\
\phantom{algi} de Riemann-Hurwitz \phantom{a} \pageref{RHU} \\
\phantom{algi} de Sato \phantom{a} \pageref{eq:Sato}, \pageref{Sato22}, \pageref{eq:sSS} \\
\phantom{algi} variationnelle de Rauch \phantom{a} \pageref{eq:Rauch}, \pageref{Rauchh} \\
Formules d\'eterminantales \phantom{a} \pageref{dett}, \pageref{dett2}, \pageref{dede}, \pageref{dede2} \\
Fractions de remplissage \phantom{a} \pageref{ffra}, \pageref{ffra2}, \pageref{ffra3}, \pageref{ffra4}\\
\textit{Framing} \phantom{a} \pageref{frami}, \pageref{frami2}, \pageref{frami3}\\
\textbf{G\'eom\'etrie sp\'eciale} \phantom{a} \textbf{\pageref{sec:geomspec}}, \pageref{eq:sSS}, \pageref{sgu}  \\
Grandes d\'eviations \phantom{a} \pageref{fig:tra}, \textbf{\pageref{sec:ben}}\\
Graphes \'epais \phantom{a} \pageref{fig:Streb}, \pageref{epsi} \\
Gravit\'e quantique \phantom{a} \pageref{gravq}, \pageref{gravq2}, \pageref{Liouli} \\
Groupe sym\'etrique \phantom{a} \pageref{eq:Hdoubleg}, \pageref{eq:formU}\\
Hi\'erarchie KP \phantom{a} \pageref{KPKP}, \pageref{sec:consoK}
\end{minipage}
\hfill \begin{minipage}[l]{0.45\linewidth}
Hi\'erarchie \\
\phantom{algi} de Painlev\'e I \phantom{a} \pageref{hiePI}, \pageref{hiePIA}\\
\phantom{algi} de Painlev\'e II \phantom{a} \pageref{sec:PII}, \pageref{hiePIIA} \\
\phantom{algi} de Whitham \phantom{a} \pageref{Whi} \\
Identit\'e \\
\phantom{algi} bilin\'eaire de Riemann \phantom{a} \pageref{bii}\\
\phantom{algi} de Fay \phantom{a} \pageref{PP1}, \pageref{fsyq}, \pageref{eq:fays} \\
Instantons \phantom{a} \pageref{instan}, \pageref{sec:topover}, \pageref{fonp7} \\
Int\'egrales \phantom{a} \\
\phantom{algi} convergentes \phantom{a} \pageref{concons} \\
\phantom{algi} d'Harish-Chandra \phantom{a} \pageref{eq:HarishChandra}, \pageref{detCb} \\
\phantom{algi} de Hodge \phantom{a} \pageref{Hodge2}, \pageref{Hodge3}, \pageref{Hodge4} \\
\phantom{algi} de Selberg \phantom{a} \pageref{Seljk}, \pageref{Seljk2}, \pageref{Seljk3} \\
Interpr\'etation combinatoire \phantom{a} \pageref{fig:toporec}, \pageref{SDD2}, \pageref{heqb4} \\
Invariants \\
\phantom{algi} de Donaldson-Thomas \phantom{a} \pageref{Thomas}, \pageref{Tomas} \\
\phantom{algi} de Gromov-Witten \phantom{a} \pageref{TGWH}, \textbf{\pageref{TGW}}, \pageref{sec:BKMP} \\
Invariants symplectiques \phantom{a} \pageref{inini}, \pageref{Tomas} \\
Involution locale \phantom{a} \pageref{brancha}, \pageref{invu} \\
Isomonodromique \phantom{a} \pageref{sec:isomono}, \pageref{sejio}, \pageref{sjeu}\\
Isospectral \phantom{a} \pageref{sjeu} \\
Lemme "une coupure" \phantom{a} \pageref{leqk}\\
Limite \\
\phantom{algi} des grandes cartes \phantom{a} \pageref{gcart} \\
\phantom{algi} singuli\`ere de courbes spectrales \\
\phantom{algi} \textbf{\pageref{sec:limitos}}, \pageref{sec:conn}, \pageref{exposs} \\
Lois de Tracy-Widom \phantom{a} \textbf{\pageref{sec:largenty}}, \pageref{sec:conn}, \pageref{theu}, \pageref{sec:TWTW} \\
Marches al\'eatoires \phantom{a}  \pageref{sec:introuniv}\\
Matrice des p\'eriodes \phantom{a} \pageref{hyper2}, \pageref{tauto}, \pageref{tauto2}, \textbf{\pageref{eq:qj2}}, \pageref{tauto3} \\
Matrices \\
\phantom{algi} de transfert \phantom{a} \pageref{tru} \\
\phantom{algi} de Wigner \phantom{a} \pageref{Wignsu}
\end{minipage}
\newpage
\begin{minipage}[l]{0.45\linewidth}
M\`eche \phantom{a} \pageref{sec:modbu}, \pageref{msu}, \pageref{msu2}, \pageref{app:bornu} \\
Mesure empirique \phantom{a} \pageref{dens3} \\
M\'ethode des moments \phantom{a} \pageref{sec:boubou}, \pageref{kujj} \\
Mod\`ele \\
\phantom{algi} d'Ising \phantom{a} \pageref{Isi}, \pageref{Isi2}, \pageref{Isi3}\\
\phantom{algi} $\On$ \phantom{a} \pageref{Ontri}, \pageref{Ontri2}, \pageref{eq:Krk1}, \pageref{sec:modbu}, \textbf{\pageref{sec:OnON}} \\
\phantom{algi} \`a poids multiplicatifs \phantom{a} \pageref{mukt} \\
Mod\`eles minimaux $(p,q)$ \phantom{a} \pageref{sec:introuniv}, \pageref{minimo}, \pageref{minimo2}\\
Mod\`eles de matrice \\
\phantom{algi} \`a une matrice hermitienne \phantom{a} \pageref{sec:Mehtapoly}, \\
\phantom{algi} \textbf{\pageref{eq:1mm}}, \pageref{eq:dsgts}, \pageref{sec:1maj} \\
\phantom{algi} \`a une matrice version $\beta$ \phantom{a} \textbf{\pageref{eq:1mmbeta}}, \pageref{eq:Zbeta2}, \\
\phantom{algi} \pageref{eq:ansat}, \pageref{bebeb}, \pageref{bebeb2}, \pageref{dls}\\
\phantom{algi} \`a une matrice avec champ \\
 \phantom{algi} ext\'erieur \phantom{a} \textbf{\pageref{eq:1mmchpext}}, \pageref{chuk}, \pageref{eqoug}, \pageref{devo5} \\
\phantom{algi} g\'en\'eralis\'e \phantom{a} \textbf{\pageref{eq:intqh}}, \pageref{sec:defsq}, \pageref{sec:gene}, \pageref{eq:ms}, \pageref{shys}\\
\phantom{algi} chaine de matrices \phantom{a} \textbf{\pageref{sec:chain}}, \pageref{sec:devM}, \pageref{rpob}, \\
\phantom{algi} \pageref{ropb2}, \pageref{devo5}, \pageref{eq:ZchainBKMP} \\
Moments \phantom{a} \textbf{\pageref{sec:correlateurs}}, \pageref{momen}, \pageref{momen2} \\
Nombres \\
\phantom{algi} de Hurwitz simples \phantom{a} \pageref{sec:Hurwitz} \\
\phantom{algi} de Hurwitz doubles \phantom{a} \pageref{eq:Hdoubleg}\\
\phantom{algi} d'intersections \phantom{a} \textbf{\pageref{eq:psiclass}}, \pageref{sec:interFg}, \pageref{courbeK3}, \pageref{cls2} \\
Noyau \\
\phantom{algi} d'Airy \phantom{a} \pageref{aaa}, \pageref{dens5}, \pageref{theu} \\
\phantom{algi} de Bergman \phantom{a} \textbf{\pageref{Berg}}, \pageref{bjs}, \pageref{bjs2} \\
 \phantom{algi} pseudo noyau de Bergman \phantom{a} \pageref{eq:BBBQ} \\
\phantom{algi} sinus \phantom{a} \pageref{dett}, \pageref{aaa} \\
\phantom{algi} spinoriel int\'egrable \phantom{a}  \pageref{spino}, \pageref{eq:sSS}\\
\phantom{algi} int\'egrable \phantom{a} \pageref{noyu}, \pageref{noyu2}, \pageref{eq:defKK}, \pageref{eq:noyu3} \\
Observables \\
\phantom{algi} uniformes \phantom{a}  \pageref{coco3}, \pageref{nonu2} \\
\phantom{algi} \`a bords non uniformes \phantom{a} \pageref{nonu}, \pageref{app:bornu}\\
Op\'erateur d'insertion \phantom{a} \pageref{eq:Wnconnexe}, \pageref{eq:opin}, \pageref{eq:inser}, \pageref{rq}, \pageref{opin2}
\end{minipage}
\hfill \begin{minipage}[l]{0.45\linewidth}
Osculation \phantom{a} \pageref{fig:okoun} \\
Param\'etrage de Joukovski \phantom{a} \pageref{Joukov} \\
Param\`etres de K\"{a}hler \phantom{a} \pageref{Kahh}, \pageref{amiro}, \pageref{CYs3}\\
Partitions \phantom{a} \pageref{bbb}, \pageref{sec:topover}, \pageref{sec:Hurwitz}, \pageref{eq:yyy}, \pageref{sukj} \\
Partitions planes \phantom{a} \pageref{bbb}, \pageref{uuu} \\
Point \\
\phantom{algi} de branchement \phantom{a} \pageref{brancha}, \pageref{branch2}, \pageref{branch3} \\
\phantom{algi} de ramification \phantom{a} \pageref{brancha}, \pageref{sec:Hurwitz} \\
Points critiques \phantom{a} \pageref{crti}, \pageref{orhu2}, \pageref{crti2}, \pageref{crti3} \\
Polyn\^{o}mes \\
\phantom{algi} de Gelfand-Dikii \phantom{a} \pageref{GD}, \pageref{GD2} \\
\phantom{algi} (bi)orthogonaux \phantom{a} \pageref{sec:Mehtapoly}, \pageref{sec:intnin}, \pageref{rpob}, \\
\phantom{algi} \pageref{orhu}, \pageref{orhu2}, \pageref{coco34}\\
Potentiel chimique \phantom{a} \pageref{cycsym} \\
Pr\'epotentiel \phantom{a} \pageref{pre}, \pageref{dens3}, \pageref{eq:hy1}, \pageref{eq:Fko}\\
Probl\`eme \\
\phantom{algi} lin\'eaire \pageref{sec:su}, \pageref{eq:problin}, \pageref{sec:su2}, \pageref{sec:su3} \phantom{a} \\
\phantom{algi} de Riemann-Hilbert scalaire \phantom{a} \pageref{sec:heqb}, \\
 \phantom{algi} \pageref{jjjj}, \pageref{eq:equalin}, \pageref{eq:gfdq0} \\
\phantom{algi} de Riemann-Hilbert matriciel \phantom{a} \pageref{sec:1maj},  \\
\phantom{algi} \pageref{RHH}, \pageref{RHH2} \\
Quadrangulations \phantom{a} \pageref{eq:gfdq}\\
Rayon de convergence \phantom{a} \pageref{rayy}, \pageref{rayy2}, \pageref{leqk}\\
Relation \\
\phantom{algi} de dispersion \phantom{a} \pageref{eq:nh} \\
\phantom{algi} d'\'echelle KPZ \phantom{a} \pageref{KPZZ} \\
\phantom{algi} de passage \phantom{a} \pageref{poposu}, \pageref{eq:noyu3} \\
R\'esolution d'une courbe spectrale \phantom{a} \pageref{sec:limitos}, \pageref{eq:AVSS}, \pageref{gcart}, \pageref{Lambert2} \\
Secteur ferm\'e/ouvert \phantom{a} \pageref{sec:thAB} \\
Sym\'etrie \\
\phantom{algi} miroir \phantom{a} \pageref{miros}, \pageref{sec:BKMP} \\
\phantom{algi} torique \phantom{a} \pageref{toto}, \pageref{sec:classCY} \\
\phantom{algi} classes de sym\'etrie \phantom{a} \pageref{2point} \\
\phantom{algi} facteur de sym\'etrie \phantom{a} \pageref{fig:cartesym}, \pageref{sysys}
\end{minipage}
\newpage
\begin{minipage}[l]{0.45\linewidth}
Syst\`eme \\
\phantom{algi} de Hitchin \phantom{a} \pageref{Hitchin} \\
\phantom{algi} de Lax \phantom{a} \pageref{sec:su} \\
\phantom{algi} de liens \phantom{a} \pageref{fig:TL}, \pageref{mp}, \textbf{\pageref{appa}} \\
Transformations \\
\phantom{algi} d'\'equivalence faible \phantom{a} \pageref{faible} \\
\phantom{algi} modulaires \phantom{a} \pageref{modula}, \textbf{\pageref{modula2}}, \pageref{eq:modo} \\
\phantom{algi} de Schlesinger \phantom{a} \pageref{Schl} \\
\phantom{algi} symplectiques  \phantom{a} \textbf{\pageref{inini}}, \pageref{TWaa}, \pageref{ininini}, \pageref{baba3} \\
Topologie \\
\phantom{algi} des surfaces de Riemann \phantom{a} \pageref{topot}\\
\phantom{algi} stable/instable \phantom{a} \pageref{stbsu} \\
\phantom{algi} regroupement par topologie \phantom{a} \pageref{regr} \\
Th\'eor\`eme \\
\phantom{algi} central limite \phantom{a}  \pageref{thcl} \\
\phantom{algi} d'inversion de Jacobi \phantom{a} \pageref{invJJJ}, \textbf{\pageref{thJaco}} \\
\phantom{algi} de la limite forte de Szeg\"{o} \phantom{a} \pageref{eq:hy1} \\
\phantom{algi} de Wick \phantom{a} \pageref{Wick2}, \textbf{\pageref{momen2}}, \pageref{epsi}, \pageref{Wick3}, \pageref{eq:concona} \\
Th\'eorie \\
\phantom{algi} des champs perturbative \phantom{a} \pageref{sec:BIPZ}, \pageref{musi} \\
\phantom{algi} conforme \phantom{a} \pageref{confor}, \pageref{cc}, \pageref{KPZZ}, \pageref{AGTss}\\
\phantom{algi} des cordes de type A \phantom{a} \pageref{modA} \\
\phantom{algi} des cordes de type B \phantom{a} \pageref{Kahl2} \\
\phantom{algi} des cordes topologiques \phantom{a} \pageref{sec:thAB}\\
\phantom{algi} de Galois \phantom{a} \pageref{Galois} \\
\phantom{algi} de Liouville \phantom{a} \pageref{Liouli}, \pageref{KPZZ}, \pageref{cc3} \\
Universalit\'e locale \phantom{a} \pageref{sec:unn} \\
Vari\'et\'e \\
\phantom{algi} de Frobenius \phantom{a} \pageref{Frolon} \\
\phantom{algi} lagrangienne \phantom{a}  \pageref{Kahl3}, \pageref{Frlo} \\
Vertex topologique \phantom{a} \textbf{\pageref{sec:topover}}, \pageref{sech}, \pageref{vBLM}
\end{minipage}
\vfill
}}

\newpage

\chapter{Liste d'articles}
\thispagestyle{plain}
\vspace{-1.5cm}
\rule{\textwidth}{1.5mm}

\vspace{2cm}
\label{app:arti}

\subsubsection{Publi\'es dans une revue \`a comit\'e de lecture}

\noindent $\diamond\,$ \cite{BEMS} \emph{A matrix model for simple Hurwitz numbers, and topological recursion} \\
\noindent \phantom{$\diamond\,$ [BEMS10]} G.~Borot, B.~Eynard, M.~Mulase, B.~Safnuk \\
\noindent \phantom{$\diamond\,$ [BEMS10]} Journal of Geometry and Physics, Volume 61, Issue 2 \\
\noindent  \phantom{$\diamond\,$ [BEMS10]} F\'evrier 2011, pages 522-540, \href{http://arxiv.org/abs/0906.1206}{\textsf{math-ph/0906.1206}}

\vspace{0.2cm}

{\small \noindent Erratum p526.
\beq
s_{\lambda}(1,\ldots,1) \neq \mathrm{dim}\,\lambda,\quad \mathrm{dim}\,\lambda = |\lambda|!\,\frac{\Delta(\mathbf{h})}{\prod_{i = 1}^{N} h_i!} \nn
\eeq
De plus, la formule correcte pour le Casimir quadratique est
\beq
f_{\lambda}(C_2) = \sum_{i = 1}^{N} \lambda_i\Big(\frac{\lambda_i}{2} - i + \frac{1}{2}\Big) = \frac{1}{2}\sum_{i} h_i^2 - (N - 1/2)\sum_{i} h_i + \frac{N(N - 1)(2N - 1)}{6} \nn
\eeq
}

\vspace{0.2cm}

\noindent $\diamond\,$ \cite{BEOn} \emph{Enumeration of maps with self avoiding loops and the $\mathcal{O}(\mathfrak{n})$ model on random lattices of all topologies} \\
\noindent \phantom{$\diamond\,$ [BE11]} G.~Borot, B.~Eynard \\
\noindent \phantom{$\diamond\,$ [BE11]} Journal of Statistical Mechanics, Theory and Experiment \\
\noindent  \phantom{$\diamond\,$ [BE11]} \href{http://arxiv.org/abs/0910.5896}{\textsf{math-ph/0910.5896}}

\subsubsection{Soumis dans une revue \`a comit\'e de lecture}

\noindent $\diamond\,$ \cite{BEMN} \emph{Large deviations of the maximal eigenvalue of random matrices} \\
\noindent \phantom{$\diamond\,$ [BEMN10]} G.~Borot, B.~Eynard, C.~Nadal, S.~Majumdar \\
\noindent \phantom{$\diamond\,$ [BEMN10]} Journal of Statistical Physics \\
\noindent \phantom{$\diamond\,$ [BEMN10]} \href{http://arxiv.org/abs/1009.1945}{\textsf{math-ph/1009.1945}}

\vspace{0.2cm}

\noindent $\diamond\,$ \cite{BETW} \emph{Tracy-Widom GUE law and symplectic invariants} \\
\noindent \phantom{$\diamond\,$ [BG11]} G.~Borot, B.~Eynard \\
\noindent  \phantom{$\diamond\,$ [BG11]} Journal of Physics A: Mathematical and Theoretical, P01010 \\
\noindent \phantom{$\diamond\,$ [BG11]} \href{http://arxiv.org/abs/1011.1418}{\textsf{nlin.SI/1011.1418}}

\subsubsection{Pr\'epublication}

\noindent $\diamond\,$ \cite{BG11} \emph{Asymptotic expansion of $\beta$ matrix models in the one-cut regime} \\
\noindent \phantom{$\diamond\,$ [BEMS]} G.~Borot, A.~Guionnet \\
\noindent \phantom{$\diamond\,$ [BEMS]} \href{http://arxiv.org/abs/1107.1167}{\textsf{math.PR/1107.1167}}

\bibliographystyle{amsalpha}
\addtolength{\baselineskip}{-0.5\baselineskip}
\bibliography{Biblithese1}
\end{document}